\documentclass[epj,nopacs]{svjour}
\usepackage{atlasphysics} 
\usepackage{wrapfig}
\usepackage{graphicx}
\usepackage{mathptmx} 
\usepackage{amssymb}
\usepackage{helvet}
\usepackage{subfig}
\usepackage{epsfig}
\usepackage{multirow}
\usepackage{rotating}
\usepackage[switch]{lineno}
\usepackage{cite}
\usepackage{collref}

\usepackage{url}
\usepackage{hyperref}
\usepackage{stfloats} 
\usepackage{fixltx2e} 
\usepackage[mathcal]{eucal}

\newcommand{\antikt}{anti-$k_{t}$}
\newcommand{\Antikt}{Anti-$k_{t}$}
\newcommand{\asym}{\ensuremath{\mathcal{A}}}

\newcommand{\JES} {{\rm JES}}
\newcommand{\EMJES} {{\rm EM+JES}}
\newcommand{\GCWJES}{{\rm GCW+JES}}
\newcommand{\LCWJES}{{\rm LCW+JES}}
\newcommand{\EM}    {{\rm EM}}
\newcommand{\GCW}   {{\rm GCW}}
\newcommand{\LCW}   {{\rm LCW}}
\newcommand{\GSL}   {{\rm GSL}}
\newcommand{\GS}    {{\rm GS}}

\newcommand{\MTF}   {{\rm MTF}}
\newcommand{\MPF}   {{\rm MPF}}
\newcommand{\Etmiss}   {\ensuremath{{E}_{\mathrm{T}}^{\mathrm{miss}}}}
\newcommand{\vecEtmiss}{\ensuremath{\vec{E}_{\mathrm{T}}^{\mathrm{miss}}}}

\newcommand{\EoverP}    {\ensuremath{E/p}}
\newcommand{\Etrue}     {\ensuremath{E^{\rm truth}}}

\newcommand{\EcaloEM}   {\ensuremath{E^{\rm jet}_\EM}}
\newcommand{\Response}  {\ensuremath{\mathcal{R}}} 
\newcommand{\Rcalo}     {\ensuremath{\Response^{\rm jet}}} 
\newcommand{\RcaloEM}   {\ensuremath{\Rcalo_\EM}}
\newcommand{\RMPF}      {\ensuremath{\Response_\MPF}}
\newcommand{\EcaloCALIB}{\ensuremath{E^{\rm jet}}}
\newcommand{\RcaloCALIB}{\ensuremath{\mathcal{R}^{\rm jet}}}

\newcommand{\Rtrack}     {\ensuremath{\mathcal{R}^{\rm track  \; jet}}}

\newcommand{\rtrk}             {\ensuremath{r_{\rm trk}}}
\newcommand{\rtrackjet}        {\ensuremath{r^{\rm calo / track \; jet}}}
\newcommand{\rtrackjetiso}     {\ensuremath{\rtrackjet_{\rm iso}}}
\newcommand{\rtrackjetnoniso}  {\ensuremath{\rtrackjet_{\rm non-iso}}}
\newcommand{\rtrackjetisoratio}{\ensuremath{\rtrackjet_{\rm non-iso/iso}}}

\newcommand{\Rmin}     {\ensuremath{R_{\rm min}}}
\newcommand{\DeltaR}   {\ensuremath{\Delta R}}
\newcommand{\DeltaRdef}{\ensuremath{\DeltaR = \sqrt{(\Delta\eta)^{2} + (\Delta\phi)^{2}}}}

\newcommand{\etaDet}{\ensuremath{\eta_{\rm det}}}
\newcommand{\etatrk}{\ensuremath{\eta^{\rm track}}}

\newcommand{\etaRange}[2]{\ensuremath{{#1}\leq|\eta|<{#2}}}
\newcommand{\AetaRange}[1]{\ensuremath{|\eta|<{#1}}}

\newcommand{\gammajet}{\ensuremath{\gamma}\mbox{-}{\rm jet}}
\newcommand{\deltaphijetgamma}{\ensuremath{\Delta \phi_{{\rm jet}\mbox{-}\ensuremath{\gamma}}}}

\def\gtap{\raisebox{-.4ex}{\rlap{$\sim$}} \raisebox{.4ex}{$>$}}

\newcommand{\insitu}{{\textit{in situ}}}
\newcommand{\Insitu}{{\textit{In situ}}}

\newcommand{\etjet}  {\ensuremath{\et^{\mathrm{jet}}}}
\newcommand{\ptavg}  {\ensuremath{\pt^\mathrm{avg}}}
\newcommand{\ptjet}  {\ensuremath{\pt^\mathrm{jet}}}
\newcommand{\ptRecoil}{\ensuremath{\pt^\mathrm{Recoil}}}
\newcommand{\ptLeading}{\ensuremath{\pt^\mathrm{Leading}}}

\newcommand{\ptprobe}{\ensuremath{\pt^{\mathrm{probe}}}}
\newcommand{\ptref}  {\ensuremath{\pt^{\mathrm{ref}}}}
\newcommand{\ptRange}[2]{\ensuremath{{#1} \leq \ptjet<{#2} \GeV}}
\newcommand{\pttrue}  {\ensuremath{\pt^\mathrm{truth}}}
\newcommand{\pttruth} {\ensuremath{\pt^{\mathrm{truth}}}}
\newcommand{\pttrk}   {\ensuremath{\pt^{\mathrm{track}}}} 
\newcommand{\ptrk}    {\ensuremath{p^{\mathrm{track}}}} 
\newcommand{\pTtrkjet}{\ensuremath{p^{\mathrm{track \; jet}}_{\rm T}}}

\newcommand{\ptl}{\ensuremath{\pt^{\rm left}}}
\newcommand{\ptr}{\ensuremath{\pt^{\rm right}}}

\newcommand{\rapjet}{\ensuremath{y}}
\newcommand{\etajet}{\ensuremath{\eta}}
\newcommand{\phijet}{\ensuremath{\phi}}

\newcommand{\alpgen}  {{\sc Alpgen}}
\newcommand{\pythia}  {{\sc Pyth\-ia}}
\newcommand{\Perugia} {{\sc Perugia}}
\newcommand{\geant}   {G{\sc eant}4}
\newcommand{\herwig}  {{\sc Herwig}}
\newcommand{\herwigpp}{{\sc Herwig++}}
\newcommand{\jimmy}   {{\sc Jimmy}}

\newcommand{\radlength}{\ensuremath{X_0}}

\newcommand{\topo} {topo-clust\-er}
\newcommand{\Topo} {Topo-clust\-er}
\newcommand{\topos}{topo-clust\-ers}
\newcommand{\Topos}{Topo-clust\-ers}

\newcommand{\central}{$0.3$ \ensuremath{\leq |\eta| < } $0.8$}
\newcommand{\ecap}   {$2.1$ \ensuremath{\leq |\eta| < } $2.8$}
\newcommand{\forward}{$3.6$ \ensuremath{\leq |\eta| < } $4.5$}

\newcommand{\Npv}{\ensuremath{N_{{\rm PV}}}}
\newcommand{\Nref}{\ensuremath{\Npv^{\rm ref}}}

\newcommand{\bunchSp}{\ensuremath{\tau_{\rm bunch}}}

\newcommand{\ATLAS}   {ATLAS}
\newcommand{\Lone}    {\texttt{L1}}
\newcommand{\HLT}     {\texttt{HLT}}
\newcommand{\ID}      {\texttt{ID}}
\newcommand{\Pixel}   {\texttt{Pixel}}
\newcommand{\SCT}     {\texttt{SCT}}
\newcommand{\TRT}     {\texttt{TRT}}

\newcommand{\LHC}    {LHC}
\newcommand{\EMB}    {\texttt{EMB}}
\newcommand{\EME}    {\texttt{EME}}
\newcommand{\EMEC}   {\texttt{EMEC}}
\newcommand{\FCAL}   {\texttt{FCAL}}
\newcommand{\Cryo}   {\texttt{Cryo}}
\newcommand{\Gap}    {\texttt{Gap}}
\newcommand{\Scint}  {\texttt{Scint}}
\newcommand{\HEC}    {\texttt{HEC}}
\newcommand{\LAr}    {\texttt{LAr}}
\newcommand{\FCal}   {\texttt{FCal}}
\newcommand{\Tile}   {\texttt{Tile}}
\newcommand{\TileExt}{\texttt{TileExt}}
\newcommand{\TileBar}{\texttt{TileBar}}
\newcommand{\MBTS}   {\texttt{MBTS}}
\newcommand{\Presampler}   {\texttt{Presampler}}

\newcommand{\width}   {{\rm width}}

\newcommand{\ftile}{\ensuremath{f_{\Tile 0}}}
\newcommand{\fem}  {\ensuremath{f_{\LAr 3}}}
\newcommand{\fpres}{\ensuremath{f_{\rm PS}}}
\newcommand{\fhec} {\ensuremath{f_{\HEC 0}}}
\newcommand{\ffcal}{\ensuremath{f_{\FCal 1}}}

\newcommand{\btag   }{{\ensuremath{b}\mbox{\rm-tagging}}}
\newcommand{\btagged}{{\ensuremath{b}\mbox{\rm-tagged}}}
\newcommand{\bquark }{{\ensuremath{b}\mbox{\rm-quark}}}
\newcommand{\bquarks}{{\ensuremath{b}\mbox{\rm-quarks}}}
\newcommand{\bjet}{{\ensuremath{b}\mbox{-}{\rm jet}}}
\newcommand{\bjets}{{\ensuremath{b}\mbox{\rm-jets}}}
\newcommand{\mylumi}{ $38$~\ipb}

\makeatletter
\g@addto@macro\bfseries{\boldmath}
\makeatother
\makeatletter
\usepackage{makeidx}
\makeindex
\hypersetup{
  colorlinks=true,
  linkcolor=blue,
  citecolor=red,
  urlcolor=red,
  pdftitle={Jet energy scale and its systematic uncertainty 
    in proton-proton collisions at $\sqrt{s}=7$~TeV with ATLAS 2010 data}
  pdfauthor={ATLAS Collaboration},
  pdfpagemode={UseOutlines},
  bookmarksopen=true,
  bookmarksnumbered=true,
  pdfstartview={Fit}
}
\def\bfseries{\fontseries\bfdefault\selectfont\boldmath}
\def\itshape{\fontshape\itdefault\selectfont\let\mathrm=\mathit}
\begin{document}
%

\title{\vspace{-3.0cm}\flushleft{\normalsize\normalfont{CERN-PH-EP-2011-191}} \flushright{\vspace{-0.86cm}\normalsize\normalfont{Submitted to Eur. Phys. J. C}} \\[3.0cm] \flushleft{Jet energy measurement with the ATLAS detector in proton-proton collisions at $\mathrm{\sqrt{s}=7}$~TeV
}}
%
\author{The \ATLAS{} Collaboration}
\institute{}
%
%
%
%
%
\date{\today}
%
%
%
\newpage
%



%
\abstract{
The jet energy scale (\JES) and its systematic uncertainty are determined for jets measured  
with the \ATLAS{} detector at the \LHC{} in proton-proton collision data at a centre-of-mass energy of $\sqrt{s}=7$~\TeV{} 
corresponding to an integrated luminosity of \mylumi.
Jets are reconstructed with the \antikt{} algorithm with distance parameters $R=0.4$ or $R=0.6$.
Jet energy and angle corrections are determined from Monte Carlo simulations
to calibrate jets with transverse momenta $\pt \geq 20$~\GeV{} and pseudorapidities \AetaRange{4.5}.
The \JES{} systematic uncertainty is estimated 
using the single isolated hadron response measured \insitu{} and in test-beams, exploiting the
transverse momentum balance between central and forward jets in events with dijet
topologies and studying systematic variations in Monte Carlo simulations.
The \JES{} uncertainty is less  than $2.5 \%$ in the central calorimeter region
(\AetaRange{0.8}) for jets with $60 \le \pt <800$~\GeV, and is maximally $14\%$
for $\pt < 30$~\GeV{} in the most forward region \etaRange{3.2}{4.5}. 
The uncertainty for additional energy from multiple proton-proton collisions in the same bunch crossing
is less than $1.5\%$ per additional collision for jets with $\pt > 50$~\GeV{} after a dedicated
correction for this effect.
The \JES{} is validated for jet transverse momenta up to  $1$~\TeV{} to the level of a few percent 
using several \insitu{} techniques by comparing a well-known reference such as the recoiling
photon \pt, the sum of the transverse momenta of tracks associated to the jet, 
or a system of low-\pt{} jets recoiling against a high-\pt{} jet.
More sophisticated jet calibration schemes are presented 
based on calorimeter cell energy density weighting
or hadronic properties of jets, providing an improved jet energy resolution 
and a reduced flavour dependence of the jet response. 
The \JES{} systematic uncertainty determined from a combination of  \insitu{} techniques
are consistent with the one derived from single hadron response measurements over a wide kinematic range.
The nominal corrections and uncertainties are derived for isolated jets in an inclusive sample of high-\pt{} jets.
Special cases such as event topologies with close-by jets,
or selections of samples with an enhanced content of jets originating from light quarks, heavy quarks or gluons 
are also discussed and the corresponding uncertainties are determined.
\vspace{10.cm}
}

\authorrunning{\ATLAS{} collaboration} 
\titlerunning{Jet measurement with the \ATLAS{} detector}

\maketitle


%
%
\setcounter{tocdepth}{2}
\newpage
\tableofcontents
\newpage
\section{Introduction}
%
\begin{figure*}[htp!]
\begin{center}
\includegraphics[width=0.9\textwidth]{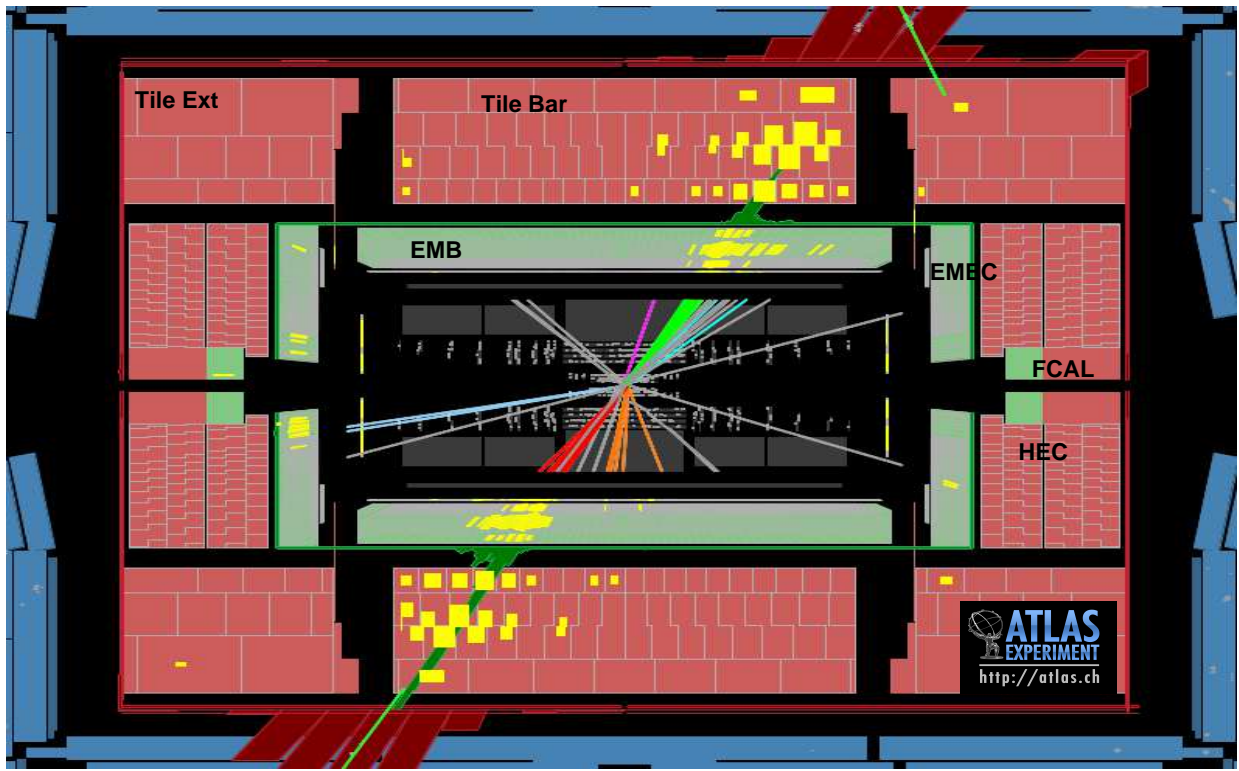} 
\caption{
Display of the central part of the \ATLAS{} detector in the $x$-$z$ view
showing the highest mass central dijet event collected during the $2010$ data taking period. 
The two leading jets have $\ptjet = 1.3$~\TeV{} with $\rapjet = -0.68$ and 
$\ptjet = 1.2$~\TeV{} with $\rapjet = 0.64$, respectively. 
The two leading jets have an invariant mass of approximately $3.1$~\TeV. 
The missing transverse energy in the event is $46$~\GeV.
The lines in the inner detector indicate the reconstructed particle trajectories.
The energy deposition in the calorimeter cells are displayed as light rectangles.
The size of the rectangles is proportional to the energy deposits.
The histograms attached to the \LAr{} and the \Tile{} calorimeter illustrate
the amount of deposited energy.
}
\label{fig:ATLASfigure30}
\end{center}
\end{figure*}

Collimated sprays of energetic hadrons, called jets, are
the dominant feature of high energy proton-proton interactions
at the Large Hadron Collider (\LHC) at CERN. In Quantum Chromodynamics (QCD)
jets are produced via the fragmentation of quarks and gluons.  
They are key ingredients for many physics measurements and for searches 
for new phenomena.

During the year $2010$ the \ATLAS{} detector collected  
proton-proton collision data at a centre-of-mass energy of $\sqrt{s}=7$~\TeV{}
corresponding to an integrated luminosity of \mylumi.
The uncertainty in the jet energy measurement is the dominant experimental 
uncertainty for numerous physics results, for example the cross-section measurement
of inclusive jets, dijets or multijets~\cite{atlasjet2010,atlasmultijet2011,Atlasjetshape,dijetdphi,Aad:2011jz}, 
as well as of vector bosons accompanied by jets~\cite{atlaswzjets2010,Aad:2011jn}, and new physics searches 
with jets in the final state~\cite{atlasexoticdijets2010,Atlasdimass,Aad:2011aj,Atlassusy1lepton,susyoleptons,Aad:2011ks}. 

Jets are observed as groups of topologically related energy deposits in the
\ATLAS{} calorimeters. 
They are reconstructed with the \antikt{} algorithm~\cite{Cacciari:2008gp}. 

Using a Monte Carlo (MC) simulation the observed jets are calibrated such that, on average,
the jet energy corresponds to that of the associated stable particles in the \ATLAS{} detector.
The calibration of the jet energy scale (\JES) 
should ensure the correct measurement of the average
energy across the whole detector and needs to be independent of additional events
produced in  proton-proton collisions at high luminosity compounding on the event
of interest. 

In this document, the jet calibration strategies adopted by the \ATLAS{} experiment
are outlined and studies to evaluate the uncertainties in the jet
energy measurement are presented. 
A first estimate of the \JES{} uncertainty, described in Ref.~\cite{atlasjet2010},
was based on information available before the first \LHC{} collisions. 
It also exploited transverse momentum balance in events
with only two jets at high transverse momenta (\pt).
A reduced uncertainty with respect to Ref.~\cite{atlasjet2010}  is presented that
is based on the increased knowledge of the detector performance obtained during the analysis 
of the first year of \ATLAS{} data taking.

\ATLAS{} has developed several jet calibration schemes~\cite{cscbook} with different levels of complexity
and different sensitivity to systematic effects, which 
are complementary in their contribution to the jet energy measurement.
Each calibration scheme starts from the measured calorimeter energy at the
electromagnetic (\EM) energy scale, which correctly measures the energy deposited by electromagnetic showers. 
In the simplest scheme (\EMJES)
the jet calibration is derived as a simple correction relating the calorimeter's response
to the true jet energy. 
More sophisticated schemes exploit the topology of the ca\-lo\-ri\-me\-ter 
energy depositions to correct for calorimeter non-compensation (nuclear energy losses, etc.) 
and other jet reconstruction effects. 

For the simple \EMJES{} calibration scheme based only on the \JES{} correction, the \JES{} uncertainty
can be determined from the single hadron response measurements in small data sets collected
\insitu{} or in test-beams.
With a large data set available the \JES{} uncertainty can also
be determined using the ratio of the jet transverse momentum to the momentum of a reference object
and by a comparison of the data to the Monte Carlo simulation.

Several techniques have been developed to directly determine the uncertainty on the jet energy
measurement \insitu.  The \JES{} uncertainty can be obtained by comparing the
jet energy to a well calibrated reference object.
A standard technique to probe the absolute jet energy scale, used also in earlier hadron collider experiments, 
is to measure the \pt{} balance between the jet and a well-measured object: 
a photon or a $Z$ boson. 
However, the currently limited data statistics imposes a limit on the \pt{} range that can be tested
with this technique. The \JES{} uncertainty on higher jet transverse momenta up to the \TeV-scale can be assessed
using the multijet balance technique where a recoil system of well-calibrated jets at lower \pt{} is balanced 
against a single jet at higher \pt. 
A complementary technique uses the total momentum
of the tracks associated to the jets as reference objects. While the resolution
of the jet energy measurement using tracks in jets is rather poor,
the mean jet energy can be determined to the precision of a few percent.

The standard jet calibration and the corresponding uncertainty on the energy measurement
are determined for isolated jets in an inclusive jet data sample. 
Additional uncertainties are evaluated for differences in the response
of jets induced by quarks or gluons and for special topologies with close-by jets. 

The outline of the paper is as follows.

First the \ATLAS{} detector (Section~\ref{sec:ATLAS}) is described.
An over\-view of the jet calibration procedures and the various calibration schemes
is given in Section~\ref{sec:overviewjetcalibration}.
The Monte Carlo simulation framework is introduced in Section~\ref{sec:MC}. 
The data samples, data quality assessment and event selection are described
in Section~\ref{sec:datasample}.
Then, the reconstruction (Section~\ref{sec:JetReco}), 
and the selection (Section~\ref{sec:JetSelection}) of jets are discussed.
The jet calibration method is outlined in Section~\ref{sec:JetCalib} %
which includes a prescription
to correct for the extra energy due to multiple proton-proton interactions (pile-up).

Section~\ref{sec:JESUncertainties} describes the sources of systematic 
uncertainties for the jet energy measurement and their estimation
using Monte Carlo simulations and collision data. 
Section~\ref{sec:insituvalidation} describes several \insitu{} techniques
used to validate these systematic uncertainties. 
Section~\ref{sec:calibTechnique} presents a technique to improve the resolution of the
energy measurements and to reduce the flavour response differences by exploiting the topology of the jets.
The systematic uncertainties %
associated with this technique are described in Section~\ref{sec:JESCalibGSC}.
The jet calibration schemes based on calorimeter cell energy weighting in jets
are introduced in Section~\ref{sec:JetCalibSchemes}, and the associated \JES{}
uncertainties are estimated from the \insitu{} techniques as described in Section~\ref{sec:JESUncertaintiescellweighting}.
Section \ref{sec:JesUncertaintycomparison} summarises the systematic uncertainties for all studied jet calibration schemes.

The jet reconstruction efficiency and its uncertainty is discussed in Section~\ref{sec:jetrecoeff}. 
The response uncertainty of non-isolated jets is investigated in Section~\ref{sec:closeby},
while Section~\ref{sec:quarkgluon} and Section~\ref{sec:gsc-quark-gluon} 
discuss response difference for jets originating from light quarks or gluons
and presents a method to determine, on average, the jet flavour content in a given data sample. 
In Section~\ref{sec:bjet} \JES{} uncertainties for jets where a heavy quark is identified
are investigated.
Finally, possible effects from lack of full calorimeter containment of jets with high transverse momentum are studied 
in Section~\ref{sec:punchthrough}. 
The overall conclusion is given in Section~\ref{sec:conclusions}.

%
\begin{figure}[ht!]
\begin{center}
\includegraphics[width=0.45\textwidth]{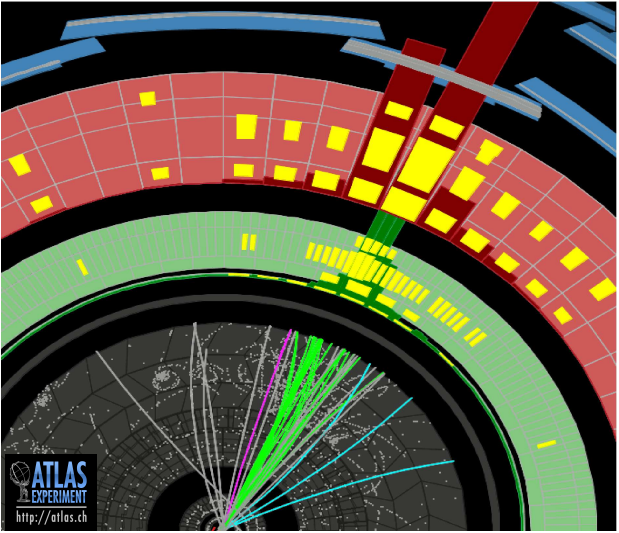} 
\caption{
Zoom of the $x$-$y$ view of the \ATLAS{} detector showing one of the high-\pt{}
jets of the event shown in Figure~\ref{fig:ATLASfigure30}. 
The energy depositions in the calorimeter cells are displayed as light rectangles.
The size of the rectangles is proportional to the energy deposits.
The dark histograms attached to the \LAr{} (\Tile{}) calorimeter illustrates the
amount of deposited energy. The lines in the \ID{} display the reconstructed tracks
originating from the interaction vertex.
}
\label{fig:ATLASfigure30zoom}
\end{center}
\end{figure}

\section{The ATLAS detector}
\label{sec:ATLAS}
\index{ATLAS Detector}   
The \ATLAS{} detector is a multi-purpose detector designed to observe 
particles produced in proton-proton and heavy ion collisions. 
A detailed description can be found in Ref.~\cite{DetectorPaper}.
The detector consists of an inner detector, sampling electromagnetic and hadronic calorimeters 
and muon chambers. 
Figure~\ref{fig:ATLASfigure30} shows a sketch of the detector outline together with
an event with two jets at high transverse momenta.

\index{ATLAS Inner Detector}   
The inner detector (\ID) is a tracking system immersed in a magnetic field of $2$~T provided by a solenoid
and covers a pseudorapidity\footnote{
The \ATLAS{} coordinate system is a right-handed system with the $x$-axis 
pointing to the centre of the \LHC{} ring and the $y$-axis pointing upwards. 
The polar angle $\theta$ is measured with respect to the \LHC{} beam-line.
The azimuthal angle $\phi$ is measured with respect to the $x$-axis.
The pseudorapidity $\eta$ is an approximation for rapidity $y$ in the high 
energy limit, and it is related to the polar angle $\theta$ 
as $\eta = -\ln \tan\frac{\theta}{2} $.
The rapidity is defined as $y = 0.5 \times {\rm ln} [ ( E + p_z )/( E - p_z ) ]$, 
where $E$ denotes the energy and $p_z$ is the component of the momentum along the beam direction.
Transverse momentum and energy are defined as 
$\pt = p \times \sin \theta$ and $ E_{\rm T} = E \times \sin \theta$, respectively.
} 
$|\eta| \lesssim 2.5$.
The \ID{} barrel region $|\eta| \lesssim 2$ consists of three layers of 
pixel detectors (\Pixel) close to the beam-pipe, 
four layers of double-sided silicon micro-strip detectors (\SCT) providing
eight hits per track at intermediate radii, and a transition radiation tracker (\TRT)
composed of straw tubes in the outer part providing $35$ hits per track.
At $|\eta| > 1$ the \ID{} endcap regions each provide three \Pixel{} discs and
nine \SCT{} discs perpendicular to the beam direction.
\index{pseudorapidity}
\index{pixel detector}    
\index{silicon micro strip detector (SCT)}     
\index{transition radiation tracker (TRT)}

\index{ LAr calorimeter}     
The liquid argon (\LAr) cal\-ori\-me\-ter 
is composed of sampling detectors with full azimuthal symmetry, housed
in one barrel and two endcap cryostats. A highly granular electromagnetic (\EM) 
cal\-ori\-me\-ter with
accordion-shaped electrodes and lead absorbers in liquid argon covers 
the pseudo\-rapidity range $|\eta| < 3.2$.
It contains a barrel part (\EMB{},  $|\eta| < 1.475$) and an endcap part 
(\EMEC{}, $1.375 \le |\eta| < 3.2$) 
each with three layers in depth
(from innermost to outermost \EMB 1, \EMB 2, \EMB 3 and \EMEC 1, \EMEC 2, \EMEC 3).
The middle layer has a $0.025 \times 0.025$ granularity in $\eta \times \phi$ space.
The innermost layer (strips) consists of cells with eight times finer
granularity in the $\eta$-direction and with $3$-times coarser granularity in the $\phi$ direction.
\index{ LAr calorimeter EMB} \index{ LAr calorimeter EMC}          

For $|\eta| < 1.8$, a presampler (\Presampler), consisting 
of an active \LAr{} layer is installed directly in front of the \EM{} calorimeters,
and provides a measurement of the energy lost before the calorimeter. 
\index{ LAr calorimeter presampler}

A copper-liquid argon had\-ronic endcap calorimeter 
(\HEC, $1.5 \le |\eta| < 3.2$) is located behind the \EMEC{}.
A copper/tungsten-liquid argon forward calorimeter (\FCal) covers the region closest to the beam 
at $3.1 \le |\eta| < 4.9$. 
The \HEC{} has four layers and the \FCAL{} has three layers. From innermost to outermost these are:
\HEC 0, \HEC 1, \HEC 2, \HEC 3 and \FCal 0, \FCal 1, \FCal 2.
Altogether, the \LAr{} calorimeters 
correspond to a total of $182,468$ readout cells, i.e. $97.2\%$ of the full \ATLAS{} calorimeter readout.
\index{ LAr calorimeter layers}

The hadronic \Tile{} calorimeter ($|\eta| < 1.7$) surrounding
the \LAr{} cryostats completes the \ATLAS{} calorimetry.
It consists of plastic scintillator tiles and steel absorbers 
covering $|\eta| < 0.8$ for the barrel and $0.8 \leq |\eta|<1.7$ for the extended barrel.
Radially, the hadronic \Tile{} calorimeter is segmented into three layers, 
approximately $1.4$, $3.9$ and $1.8$ interaction lengths thick at $\eta = 0$; 
the $\Delta\eta \times \Delta \phi$ segmentation is $0.1 \times 0.1$ ($0.2 \times 0.1$ in the last radial layer).
The last layer is used to catch the tails of the longitudinal shower development.
\index{Tile calorimeter}
\index{Tile calorimeter, barrel} \index{Tile calorimeter, extended barrel}
The three radial layers of the \Tile{} calorimeter will be referred to (from innermost to outermost) as
\Tile 0, \Tile 1, \Tile 2 \footnote{In the barrel, the \Tile{} layers will be called
\TileBar 0, \TileBar 1, \TileBar 2 and in 
the extended barrel \TileExt 0, \TileExt 1 and \TileExt 2.
}.

Between the barrel and the extended barrels there is a gap of about $60~\mathrm{cm}$, 
which is needed for the \ID{} and the \LAr{} services. 
Gap scintillators (\Gap)
covering the region $1.0 \le |\eta|  < 1.2$
are installed on the inner radial surface of the extended barrel modules in the 
region between the \Tile{} barrel and the extended barrel.
Crack scintillators (\Scint)
are located on the front of the \LAr{} endcap and cover the region
$1.2 \le |\eta| < 1.6$.
\index{Tile calorimeter, gap scintillators}

The muon spectrometer surrounds the \ATLAS{} calorimeter. A system of three large air-core
toroids, a barrel and two endcaps, generates a magnetic field in the
pseudorapidity range of $|\eta| < 2.7$. The muon spectrometer measures muon tracks 
with three layers of precision tracking chambers and is instrumented with separate trigger chambers.
\index{ muon spectrometer}

The trigger system for the \ATLAS{} detector consists of a
hardware-based Level~1~(\Lone) and a software-based higher level trigger~(\HLT) \cite{triggerperformance}.  
Jets are first identified at L1 using a sliding window algorithm from coarse granularity
calorimeter towers.  This is refined using jets reconstructed from
calorimeter cells in the \HLT.  
The lowest threshold inclusive jet trigger is fully efficient for jets with $\pt \gtrsim 60$~\GeV. 
Events with lower \pt{} jets are triggered by the
minimum bias trigger scintillators (\MBTS) mounted at each end of the
detector in front of the \LAr{} endcap calorimeter cryostats at
$|z| = \pm 3.56$~{\rm m}.
\index{trigger}

\section{Introduction to jet energy calibration methods}
\label{sec:overviewjetcalibration}
Hadronic jets used for \ATLAS{} physics analyses are reconstruct\-ed by
a jet algorithm starting from the energy depositions of electromagnetic
and hadronic showers in the calorimeters.
An example of a jet recorded by the \ATLAS{} detector and displayed
in the plane transverse to the beam line
is shown in Figure~\ref{fig:ATLASfigure30zoom}. 

The jet Lorentz four-momentum is reconstructed 
from the corrected energy and angles with respect to the primary event vertex.
For systematic studies and calibration purposes {\it track jets}
are built from charged particles using their momenta measured in the
inner detector.
Reference jets in Monte Carlo simulations ({\it truth jets}) are formed
from simulated stable particles using the same jet algorithm.

The jet energy calibration relates the jet energy measured with the 
\ATLAS{} calorimeter to the true energy of the corresponding jet
of stable particles entering the \ATLAS{} detector.

The jet calibration corrects for the following detector effects that affect the jet energy measurement:
\noindent
\begin{enumerate}
 \item {\bf Calorimeter non-compensation}: 
       partial measurement of the energy deposited by hadrons.
 \item {\bf Dead material}: 
       energy losses in inactive regions of the detector.
 \item {\bf Leakage}: 
        energy of particles reaching outside the calorimeters.
 \item {\bf Out of calorimeter jet cone}: 
       energy deposits of particles inside the truth jet entering
       the detector that are not included in the reconstructed jet.
 \item {\bf Noise thresholds and particle reconstruction efficiency}: 
       signal losses in the calorimeter clustering and jet reconstruction.
\end{enumerate}

Jets reconstructed in the calorimeter system are formed from calorimeter energy depositions
reconstructed at the {\it electromagnetic energy scale} (\EM) 
or from energy depositions that are corrected for the lower detector response to hadrons.
The \EM{} scale correctly reconstructs the energy deposited by particles
in an electromagnetic shower in the calorimeter. \index{electromagnetic energy scale}
This energy scale is established using test-beam measurements for
electrons in the barrel~\cite{ctb2004electronseoverp,ctb2004electrons, LArTB02uniformity, LArTB02linearity,Tile2002}
and the endcap calorimeters~\cite{Pinfold:2008zzb,EndcapTBelectronPion2002}.
The absolute calorimeter response to energy deposited via electromagnetic processes
was validated in the hadronic calori\-met\-ers using muons, both from 
test-beams \cite{Tile2002,LArTB02muons} and produced \insitu{} by cosmic rays \cite{TileReadiness}.
The energy scale of the electromagnetic calorimeters
is corrected using the invariant mass of $Z$ bosons produced 
in proton-proton collisions ($Z \rightarrow e^+ e^-$ events) \cite{Atlaselectronpaper}.
The correction for the lower response to hadrons is solely based on the topology of the energy 
depositions observed in the calorimeter.

In the simplest case the measured jet energy is corrected, on average, using Monte Carlo simulations, as follows:
\begin{equation}
\label{eq:CALIB}
E^{\rm jet}_{\rm calib} = E^{\rm jet}_{\rm meas} / \mathcal{F}_{{\rm calib}}( E^{\rm jet}_{\rm meas}),
\; {\rm with} \; E^{\rm jet}_{\rm meas} = \EcaloEM - {\cal O}(\Npv).
\end{equation}
The variable \EcaloEM{} is the calorimeter energy measured at the electromagnetic scale, 
$E^{\rm jet}_{\rm calib}$ is the calibrated jet energy
and $\mathcal{F}_{{\rm calib}}$ is the calibration function
that depends on the measured jet energy and is evaluated in small jet pseudo\-rapidity regions. 
The variable ${\cal O}(\Npv)$ denotes the correction for additional energy 
from multiple proton-proton interactions depending on the number of primary vertices (\Npv).

The simplest calibration scheme (called \EMJES) applies
the \JES{} corrections to jets reconstructed at the electromagnetic scale. 
This calibration scheme allows a simple evaluation of the systematic uncertainty 
from single hadron response measurements and systematic Monte Carlo variations.
This can be achieved with small data sets and is therefore suitable for early physics analyses. 

Other calibration schemes 
use additional cluster-by-cluster and/or jet-by-jet information to reduce some of the sources of 
fluctuations in the jet energy response, thereby improving
the jet energy resolution. For these calibration schemes the same jet calibration procedure
is applied as for the \EMJES{} calibration scheme, but the energy corrections are numerically smaller. 

The global calorimeter cell weighting (\GCW) calibration exploits the observation 
that electromagnetic showers in the calorimeter leave more compact energy depositions 
than hadr\-on\-ic showers with the same energy. 
Energy corrections are derived for each calorimeter cell within a jet, 
with the constraint that the jet energy resolution is minimised. 
The cell corrections account for all energy losses of a jet in the \ATLAS{} detector.
Since these corrections are only applicable to jets and not to energy depositions in general,
they are called ``global'' corrections.

The local cluster weighting (\LCW) calibration method first clusters together 
topologically connected calorimeter cells and classifies these clusters as either
electromagnetic or hadronic. Based on this classification energy corrections are derived from single pion 
Monte Carlo simulations. 
Dedicated corrections are derived for the effects of non-compensation, 
signal losses due to noise threshold effects, 
and energy lost in non-instrumented regions. 
They are applied to calorimeter clusters and are defined without reference to a jet definition.
They are therefore called ``local'' corrections.
Jets are then built from these calibrated clusters using a jet algorithm.

The final jet energy calibration (see Equation~\ref{eq:CALIB})
can be applied to \EM{} scale jets, with the resulting calibrated jets referred to as \EMJES,
or to \GCW{} and \LCW{} calibrated jets, with the resulting jets referred to as \GCWJES{} and \LCWJES{} jets.

A further jet calibration scheme, called global sequential (\GS) calibration, 
starts from jets calibrated with the \EMJES{} calibration and exploits the topology of the energy deposits
in the calorimeter to characterise fluctuations in the jet particle content
of the hadronic shower development. Correcting for such fluctuations can improve the
jet energy resolution.
The corrections are applied such that the mean jet energy is left unchanged.
The correction uses several jet properties and each correction is applied sequentially.
In particular, the longitudinal and transverse structure of the hadronic shower in the calorimeter
is exploited.

The simple \EMJES{} jet calibration scheme does not provide the best performance, but allows 
in the central detector region the most direct evaluation of the systematic uncertainties
from the calorimeter response to single isolated hadron measured \insitu{} and in test-beams
and from systematic variations of the Monte Carlo simulation. 
For the \GS{} the systematic
uncertainty is obtained by studying the response after applying the \GS{} 
calibration with respect to the  \EMJES{} calibration.
For the  \GCWJES{} and \LCWJES{} calibration schemes
the \JES{} uncertainty is determined from \insitu{} techniques.

For all calibration schemes the \JES{} uncertainty in the forward detector regions
is derived from the uncertainty in the central region using the transverse momentum balance
in events where only two jets are produced.


In the following, the calibrated calorimeter jet transverse momentum will be denoted as \ptjet{}, and
the jet pseudorapidity as $\etajet$.

\section{Monte Carlo simulation}
\label{sec:MC}
\subsection{Event generators}
\label{sec:EventGenerators}
The energy and direction of particles produced in proton-proton collisions
are simulated using various event generators. An over\-view of Monte Carlo
event generators for \LHC{} physics can be found in Ref.~\cite{mcforlhc}.
The samples using different event generators and theoretical models 
used are described below:
\begin{enumerate}
\item \pythia{} with the MC10 or AMBT1 tune:
\index{\pythia{}}  \index{ MC10 tune}   
The event generator \pythia{} \cite{pythia}
simulates non-diffractive proton-proton collisions using a $2 \to 2$ matrix element
at leading order in the strong coupling to model the hard subprocess,
and uses \pt-ordered parton showers to model additional radiation in the leading-logarithmic approximation \cite{pythiapartonshower}. 
Multiple parton interactions \cite{Sjostrand:2004ef}, as well as fragmentation and hadronisation 
based on the Lund string model \cite{lundstring} are also simulated. 
The proton parton distribution function (PDF) set used is 
the modified leading-order PDF set MRST LO* \cite{PDF-MRST}. 
The parameters used for tuning multiple parton interactions include  
charged particle spectra measured by \ATLAS{} in minimum bias collisions~\cite{mc10chargedparticles}, 
and are denoted as the \ATLAS{} MC10 tune\cite{MC10}. 
\index{ \pythia{} \Perugia2010 tune}   
\item The \Perugia2010 tune is an independent tune of \pythia{} with increased 
final state radiation to better reproduce 
the jet shapes and hadronic event shapes using {\sc LEP} and  {\sc Tevatron} data~\cite{Perugia2010}. 
In addition, parameters sensitive to the production of particles with strangeness and related to 
jet fragmentation have been adjusted.
\index{\herwig{} + \jimmy{}}   
\item \herwig+\jimmy{} 
uses a leading order  $2 \to 2$ matrix element supplemented 
with angular-ordered parton showers in the leading-logarithm approximation \cite{herwig3,herwig2,herwig}. 
The cluster model is used for the hadronisation  \cite{herwigclustermodel}.
Multiple parton interactions are modelled using \jimmy{}~\cite{jimmy}.
The model parameters of \herwig{}/\jimmy{} have been tuned to \ATLAS{} data (AUET1 tune) \cite{AUET1}.
The MRST LO* PDF set \cite{PDF-MRST} is used.
\index{\herwigpp{}}   
\item \herwigpp{} \cite{Herwigpp} is based on the event generator \herwig,       
but redesigned in the {\textit C++} programming language.
The generator contains a few modelling improvements.                    
It also uses angular-ordered parton showers, but with an updated 
evolution variable and a better phase space treatment.
Hadronisation is performed using the cluster model. 
The underlying event and soft inclusive interactions are described using a 
hard and soft multiple partonic interactions model \cite{HerwigppUI}.
The MRST LO* PDF set \cite{PDF-MRST} is used.
\index{\alpgen{}}   
\item \alpgen{}
is a tree level matrix-element generator for hard multi-parton processes ($2 \to n$) 
in hadronic collisions ~\cite{Alpgen}. 
It is interfaced to \herwig{} to produce parton showers in the leading-logarithmic 
approximation. Parton showers are matched to the matrix element with the MLM matching scheme~\cite{MLM}. 
For the hadronisation,  \herwig{} is used and soft multiple parton interactions are modelled 
using \jimmy \cite{jimmy} (with the \ATLAS{} MC09 tune \cite{MC09}).
The PDF set used is CTEQ6L1 \cite{PDF-CTEQ}.
\end{enumerate}
\subsection{Simulation of the ATLAS detector}
\label{sec:MCSIM}
\index{Geant}   
The \geant{} software toolkit~\cite{Geant4} within the \ATLAS{} simulation framework~\cite{simulation} 
propagates the generated particles through the \ATLAS{} detector and simulates their interactions with 
the detector material. 
The energy deposited by particles in the active detector material is 
converted into detector signals with the same format as the \ATLAS{} detector read-out.
The simulated detector signals are in turn reconstructed with the same reconstruction software
as used for the data. %

In \geant{} the model for the interaction of hadrons with the detector material
can be specified for various particle types and for various energy ranges.
For the simulation of hadronic interactions in the detector, the \geant{} set of 
processes called \texttt{QGSP\_BERT} is chosen~\cite{geanthadronic}. 
\index{\texttt{QGSP\_BERT}}   
In this set of processes, the 
Quark Gluon String model~\cite{QGS,QGSP2,QGSP3,QGSP4,QGSP5} is used for the 
fragmentation of the nucleus, 
and the Bertini cascade model~\cite{Bertini,Bertini1,Bertini2,Bertini3} for the description 
of the interactions of  hadrons in the nuclear medium. 

The \geant{} simulation and in particular the hadronic interaction model
for pions and protons, has been validated with test-beam measurements for the
barrel~\cite{Tile2002, Tile2002pionproton, CTB2004topology, CTB04pion, CTB2004vlepion} 
and endcap~\cite{EndcapTBelectronPion2002,Pinfold:2008zzb, Kiryunin:2006cm} calorimeters. 
Agreement within a few percent is found between simulation and data
for pion momenta between $2$~\GeV{} and $350$~\GeV.
\index{test-beam validation}   

Further tests have been carried out \insitu{} comparing the single hadron response, measured using isolated tracks
and identified single particles.
Agreement within a few percent is found for the inclusive measurement \cite{singleparticle900,atlassingleparticle2011}
and for identified pions and protons from the decay products of kaon and lambda particles
produced in proton-proton collisions at $7$~\TeV~\cite{ZachIdParticles}.
With this method particle momenta of pions and protons   
in the range from a few hundred \MeV{} to $6$~\GeV{} can be reached.
Good agreement between Monte Carlo simulation and data is found. 
\index{\insitu{} single hadron response}   
\subsection{Nominal Monte Carlo simulation samples}
\label{sec:NominalSample}
\index{ Monte Carlo simulation sample, nominal}   
The baseline (nominal) Monte Carlo sample used to derive the jet energy scale and to estimate the 
sources of its systematic uncertainty is a sample containing high-\pt{} jets produced via strong
interactions.
It is generated with the \pythia{} event generator with the MC10 tune (see Section~\ref{sec:EventGenerators}),
passed through the full \ATLAS{} detector simulation and is reconstructed as the data.

The \ATLAS{} detector geometry used in the simulation of the nominal sample reflects the
geometry of the detector as best known at the time of these studies. 
Studies of the material of the inner detector in front of the calorimeters have been performed using
secondary hadronic interactions~\cite{Aad:2011cxa}.
Additional information is obtained from studying photon conversions \cite{PhotonConversions900} and
the energy flow in minimum bias events \cite{MaterialBudgetMinBias7}.

\subsection{Simulated pile-up samples}
\label{sec:PileupMCSample}
For the study of multiple proton-proton interactions, two samples have been used, one for in-time and one for out-of-time pile-up. 
The first simulates additional proton-proton interactions per bunch crossing, 
while the second one also contains pile-up arising from bunches before or after
the bunch where the event of interest was triggered 
(for more details see Section~\ref{sec:datasample} and Section~\ref{sec:pileup}). 
The bunch configuration of LHC (organised in bunch trains) is also simulated.
The additional number of primary vertices in the in-time (bunch-train) pile-up sample is $1.7$ ($1.9$) on average.

\section{Data sample and event selection}
\label{sec:datasample}
\subsection{Data taking period and LHC conditions}
Proton-proton collisions at a centre-of-mass energy of
$\sqrt{s} = 7$~\TeV, recorded from March to October $2010$ are analysed. 
Only data with a fully functioning calorimeter and  inner detector are used. 
The data set corresponds to an integrated luminosity of \mylumi.
Due to different data quality requirements the integrated luminosity 
can differ for the various selections used in the \insitu{} technique analyses.

Several distinct periods of machine configuration and detector operation were present 
during the $2010$ data taking. 
As the \LHC{} commissioning progressed, changes in the beam optics and proton bunch parameters 
resulted in changes in the number of pile-up interactions per bunch crossing.
The spacing between the bunches was no less than $150$~{\rm ns}.

\begin{figure}
\begin{center}
  \includegraphics[angle=0,width=0.45\textwidth]{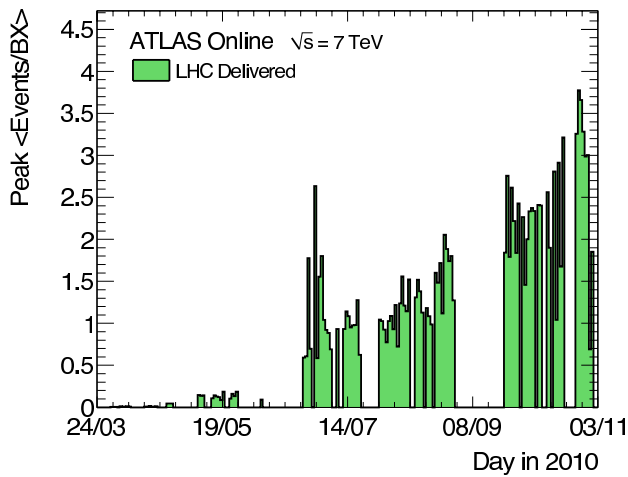}
        \caption{The peak number of interactions per bunch crossing (``BX'') as measured 
online by the \ATLAS{} luminosity detectors~\cite{Aad:2011dr}. 
}
        \label{fig:lumitimeline}
\end{center}
\end{figure}

Figure~\ref{fig:lumitimeline} shows the evolution of the 
maximum of the distribution of the number of interactions (peak) 
derived from the online luminosity 
measurement and assuming an inelastic proton-proton scattering cross section of $71.5$~{\rm mb}~\cite{Aad:2011dr}. 

The very first data were essentially devoid of multiple proton-proton interactions until the optics 
of the accelerator beam (specifically $\beta^{*}$) were changed in order to decrease the transverse size
of the beam and increase the luminosity\footnote{The parameter $\beta^{*}$ is the value of the 
$\beta$-function (the envelope of all trajectories of the beam particles)
at the collision point and smaller values of $\beta^{*}$ imply a smaller physical 
size of the beams and thus a higher instantaneous luminosity.}. 
This change alone raised the fraction of events with at least two observed interactions 
from less than $ 2\%$ to between $8 \%$ and $10 \%$ (May-June 2010). 

A further increase in the number of interactions
occurred when the number of protons per bunch ({\rm ppb}) 
was increased from approximately $5 - 9 \cdot 10^{10}$ to $1.15 \cdot 10^{11}$~{\rm ppb}. 
Since the number of proton-proton collisions per bunch crossing is proportional to the square of the bunch intensity, 
the fraction of events with pile-up increased to more than $50 \%$ for runs between June and September 2010. 

Finally, further increasing the beam intensity slowly raised the average number of interactions per bunch crossing to more 
than three by the end of the proton-proton run in November $2010$.

\subsection{Event selection }
Different triggers are used to select the data samples, in order to be maximally efficient over the 
entire jet \pt-range of interest. 
The dijet sample is selected using the hardware-based calorimeter jet triggers~\cite{jetTriggers,triggerperformance}, 
which are fully efficient for jets with $\ptjet > 60$~\GeV. 
For lower \ptjet{} a trigger based on the minimum bias trigger scintillators is used.
\index{dijet sample }

The multijet sample uses either the inclusive jet trigger or 
a trigger that requires at least two, three or more jets 
with  $\pt > 10$~\GeV\ at the \EM{} scale. 
These triggers are fully efficient for jets with $\ptjet > 80$~\GeV.
\index{multijet sample }

Each event is required to have a primary hard scattering vertex. 
A primary vertex is required to have at least five tracks $(N_{\rm pp}^{\rm tracks})$  
with a transverse momentum of $\pttrk > 150$~\MeV. 
The primary vertex associated to the event of interest (hard scattering vertex)
is the one with the highest associated transverse track momentum 
squared used in the vertex fit $\Sigma (\pt^{\rm track})^2$, 
where the sum runs over all tracks used in the vertex fit. 
This renders the contribution from fake vertices due to beam backgrounds to be negligible.
\index{primary vertex}

The \gammajet{} sample is selected using a photon trigger \cite{triggerperformance}
that is fully efficient for photons passing 
offline selections. The higher  threshold for the photon \pt{}
is $40$~\GeV{} and this trigger was not pre-scaled;
the lower threshold is $20$~\GeV{} and this trigger was pre-scaled at high luminosity. 
\index{$\gamma$-jet sample }

\subsection{Data quality assessment}
\index{data quality}
The \ATLAS{} data quality (DQ) selection is based upon inspection of a standard set of distributions 
that leads to a data quality assessment for each subdetector, usually segmented into barrel, forward and endcap regions, 
as well as for the trigger  and for each type of reconstructed physics object (jets, electrons, muons, etc.). 
Each subsystem sets its own DQ flags, which are recorded in a conditions database. 
Each analysis applies DQ selection criteria, and defines a set of luminosity blocks 
(each corresponds to approximately two minutes of data taking). 
The good luminosity blocks used are those not flagged for having issues affecting a relevant subdetector.

Events with minimum bias and calorimeter triggers were required to belong 
to specific runs and run periods in which the detector, 
trigger and reconstructed physics objects have passed a data quality assessment 
and are deemed suitable for physics analysis.

The primary systems of interest for this study are the electromagnetic and hadronic calorimeters, 
and the inner tracking detector for studies of the properties of tracks associated with jets.

\begin{table*}[ht!]
\begin{center}
\begin{tabular}{c|c|c}
\hline \hline
           & Loose & Medium \\
\hline
\HEC{} spikes & ($f_\HEC > 0.5$ and $\mid f_{\rm \HEC quality} \mid > 0.5 $) 
           &  Loose or \\
           & or $\mid E_{\rm neg} \mid > 60$~\GeV 
           & $f_{\HEC} > 1- \mid f_{\rm \HEC quality} \mid $  \\
\hline
Coherent  & $f_{\EM} > 0.95$ and $f_{{\rm quality}} > 0.8$  & Loose or \\
\EM{} noise  & and $\mid \etajet \mid < 2.8 $                      & $f_{\EM} > 0.9 $ and $f_{\rm quality}>0.8$ and $\mid \etajet \mid < 2.8$ \\
\hline
Non-collision & $\mid t_{\rm jet} \mid > 25$~{\rm ns} or  &  Loose or\\
background    & ($f_{\EM} < 0.05$ and $f_{\rm ch} < 0.05$ and $\mid \etajet \mid < 2 $)  & $\mid t_{\rm jet} \mid > 10$ {\rm ns} \\
              & or ($f_{\EM} < 0.05$ and $\mid \etajet \mid \geq 2 $) & or ($f_{\EM} < 0.05$ and $f_{\rm ch} < 0.1$ and $\mid \etajet \mid < 2 $)  \\
              & or ($f_{\rm max} > 0.99$ and $\mid \etajet \mid < 2 $)  & or ($f_{\EM} > 0.95$ and $f_{\rm ch} < 0.05$ and $\mid \etajet \mid < 2 $) \\
\hline \hline
\end{tabular}
\caption{Selection criteria used to reject fake jets and non-collision background.}
\label{cut_table}
\end{center}
\end{table*}

\section{Jet reconstruction}
\label{sec:JetReco}
In data and Monte Carlo simulation 
jets are reconstructed using the \antikt{} algorithm~\cite{Cacciari:2008gp} with distance 
parameters $R = 0.4$ or $R = 0.6$
using the {\sc FastJet} software \cite{Cacciari200657,Fastjet}.
\index{{\sc FastJet} software} \index{\antikt{} jet algorithm} 
The four-momentum recombination scheme is used.
Jet finding is done in $y$-$\phi$ coordinates, while
jet corrections and performance studies are often done in $\etajet$-$\phijet$ coordinates. 
The jet \pt{} reconstruction threshold is $\ptjet > 7$~\GeV.

In the following, only \antikt{} jets with distance parameter $R = 0.6$ are discussed in detail. 
The results for jets with $R = 0.4$ are similar, if not stated otherwise.

\subsection{Reconstructed calorimeter jets}
\label{sec:calorimeterjets}
The input to {\it calorimeter jets} can be topological calorimeter clusters
({\it \topos})~\cite{EndcapTBelectronPion2002,TopoClusters} or calorimeter
towers. Only \topos{} or towers with a positive energy are considered as
input to jet finding.  \index{calorimeter jets}
\subsubsection{Topological calorimeter clusters}
\label{sec:topocluster}
\index{Topological calorimeter clusters} Topological clusters are groups of
calorimeter cells that are designed to follow the shower
development taking advantage of the fine segmentation of the \ATLAS{}
calorimeters.  The \topo{} formation algorithm starts from a
\textit{seed} cell, whose signal-to-noise ($S/N$) ratio is above a threshold of $S/N = 4$.
The noise is estimated as the absolute value of the energy deposited in the calorimeter cell divided by
the RMS of the energy distribution measured in events triggered at random bunch crossings. 
Cells neighbouring the seed (or the cluster
being formed) that have a signal-to-noise ratio of at least $S/N = 2$
are included iteratively. Finally, all calorimeter cells
neighbouring the formed \topo{} are added.
The \topo{} algorithm efficiently suppresses the calorimeter noise.

The \topo{} algorithm also includes a splitting step in order to
optimise the separation of showers from different close-by particles:
All cells in a \topo{} are searched for local maxima in terms of energy content with
a threshold of $500$~\MeV. This means that the selected calorimeter cell has to be more energetic than any of its neighbours. 
The local maxima are then used as seeds for a new iteration of topological clustering, which splits
the original cluster into more \topos{}.

A \topo{} is defined to have an energy equal to the energy sum of all
the included calorimeter cells, zero mass and a reconstructed
direction calculated from the weighted averages of the
pseudorapidities and azimuthal angles of the constituent cells.
The weight used is the absolute cell energy and the positions of the cells
are relative to the nominal \ATLAS{} coordinate system.

\subsubsection{Calorimeter towers}
\label{sec:topotowers}
\index{Calorimeter towers} 
{\it Calorimeter towers} are static, $\Delta \eta \times \Delta \phi = 0.1 \times 0.1$,
grid elements built directly from calorimeter cells\footnote{
For the few calorimeter cells 
that are larger than the $\Delta \eta \times \Delta \phi = 0.1 \times 0.1$ 
(like in the last \Tile{} calorimeter layer and the \HEC{} inner wheel) 
or have a special geometry (like in the \FCAL), 
projective tower grid geometrical weights are defined that
specify the fraction of calorimeter cell energy to be attributed to a particular calorimeter tower.}. 

\ATLAS{} uses two types of calorimeter towers: with and without noise suppression.
Calorimeter towers based on all calorimeter cells are called
{\it non-noise-suppressed calorimeter towers} in the following.
\index{non-noise-suppressed calorimeter tower} 
Noise-suppressed towers make use of the  \topos{} algorithm, i.e.
only calorimeter cells that are included in \topos{} are used.
\index{Noise-suppressed calorimeter tower} 
Therefore, for a fixed geometrical area, noise-suppressed towers have the
same energy content as the \topos.

Both types of calorimeter towers have an energy equal to the energy sum of all 
included calorimeter cells. The formed Lorentz four-momentum has zero mass. 

\subsection{Reconstructed track jets}
\label{sec:trackjets}
\index{Track jets} 
Jets built from charged particle tracks originating from the primary hard scattering vertex
({\it track jets}) are used to define jets that are insensitive 
to the effects of pile-up and provide a stable reference to study close-by jet effects. 

Tracks with $\pt^{\rm track}>0.5$~\GeV{} and $|\eta|<2.5$ are selected. They are required to have 
at least one (six) hit(s) in the \Pixel{} (\SCT) detector. 
The transverse ($d_0$) and longitudinal ($z_0$) impact parameters of the tracks measured with
respect to the primary vertex are also required to be $|d_0|<1.5$~mm and $|z_0\sin\theta|<1.5$~mm, respectively.

The track jets must have at least two constituent tracks and a total 
transverse momentum of $\pTtrkjet>3$~\GeV. 
Since the tracking system has a coverage up to $|\eta| = 2.5$,
the performance studies of calorimeter jets is carried out in the range
\AetaRange{1.9} for $R = 0.6 $ and \AetaRange{2.1} for $R = 0.4$. 

\subsection{Monte Carlo truth jets and flavour association}
\label{sec:truthjets}
\index{Monte Carlo truth jets} 
Monte Carlo simulation {\it truth jets} are built from 
stable particles defined to have proper lifetimes longer than $10$~{\rm ps} 
excluding muons and neutrinos.

For certain studies, jets in the Monte Carlo simulation are additionally identified as jets initiated by light or heavy quarks 
or by gluons based on the generator event record.
The highest energy parton that points to the truth jet\footnote{With $\Delta R < 0.6$ for jets 
with $R = 0.6$ and $\Delta R < 0.4$ for jets with $R = 0.4$, 
where $\DeltaRdef$.} determines the flavour of the jet.
Using this method, only a small fraction of the jets ($ < 1 \%$ at low \pt{} and less at high \pt) 
could not be assigned a partonic flavour.
This definition is sufficient to study the flavour dependence of the jet response.
Any theoretical ambiguities of jet flavour assignment do not need to be addressed in the context
of a performance study.
\index{jet flavour}

\newpage
\begin{figure*}[htp!]
  \begin{center}
    \subfloat[         $|\etajet|<0.3$]{\includegraphics[width=70mm]{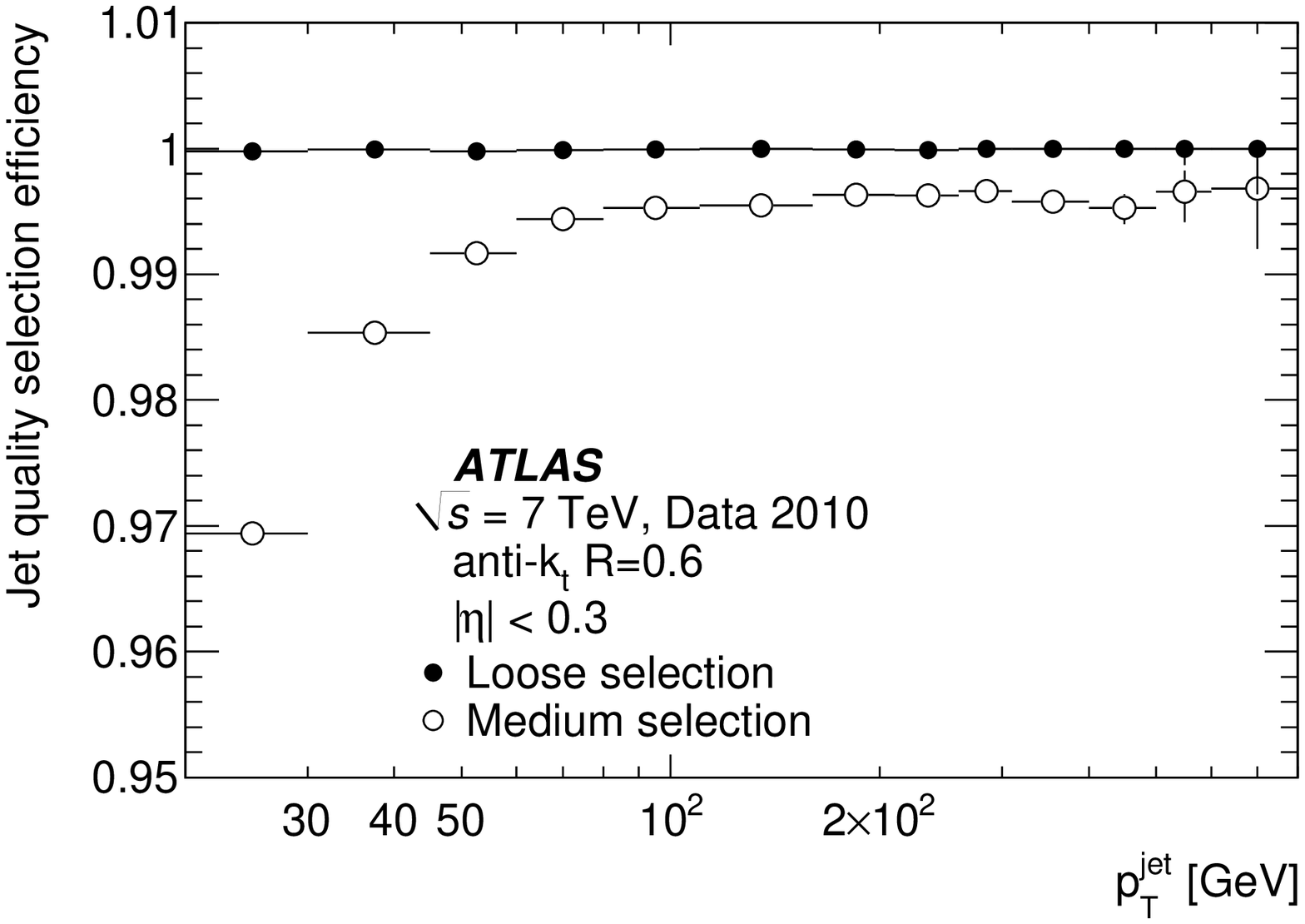}}
    \subfloat[$0.3 \leq |\etajet|<0.8$]{\includegraphics[width=70mm]{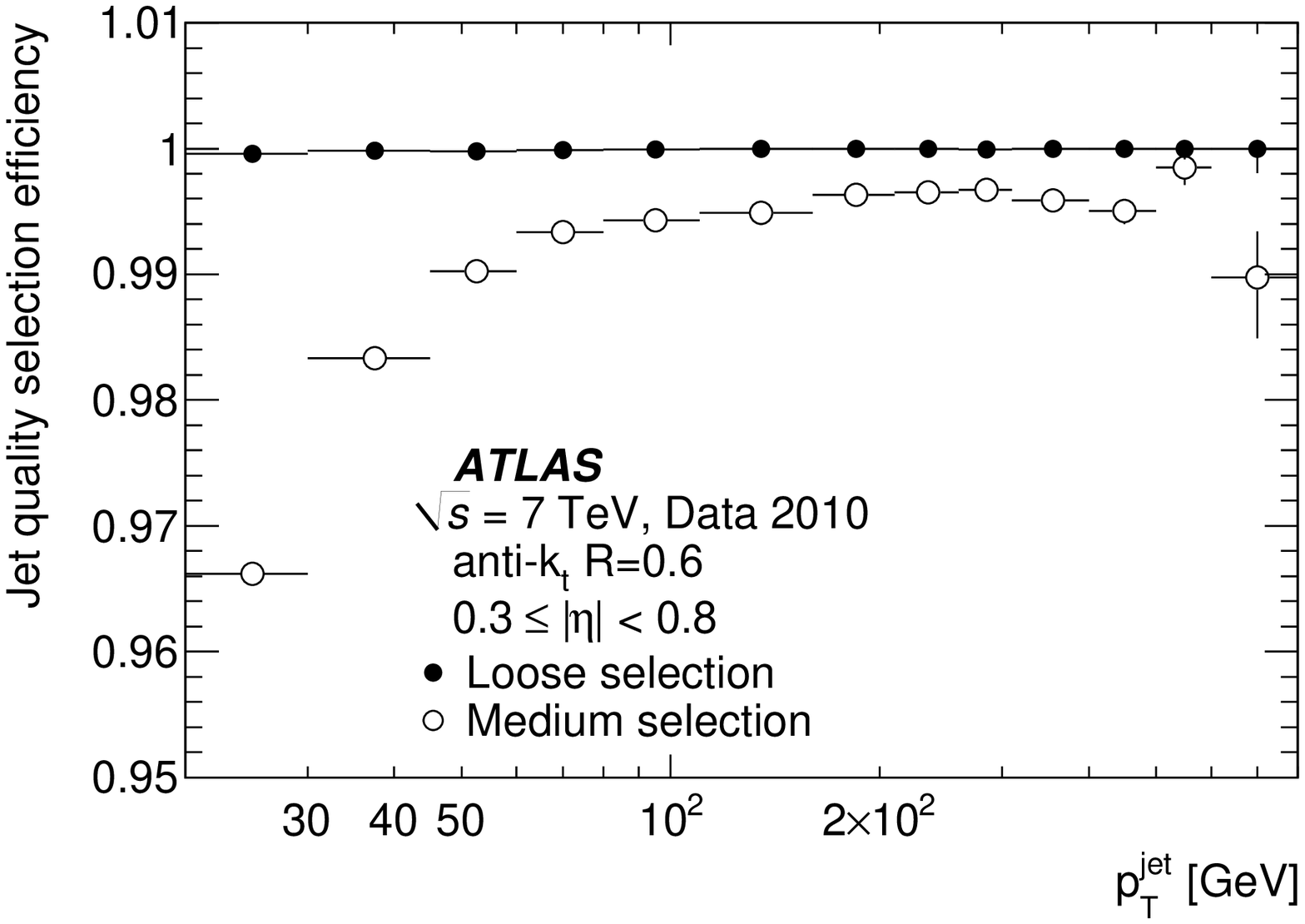}}\\
    \subfloat[$0.8 \leq |\etajet|<1.2$]{\includegraphics[width=70mm]{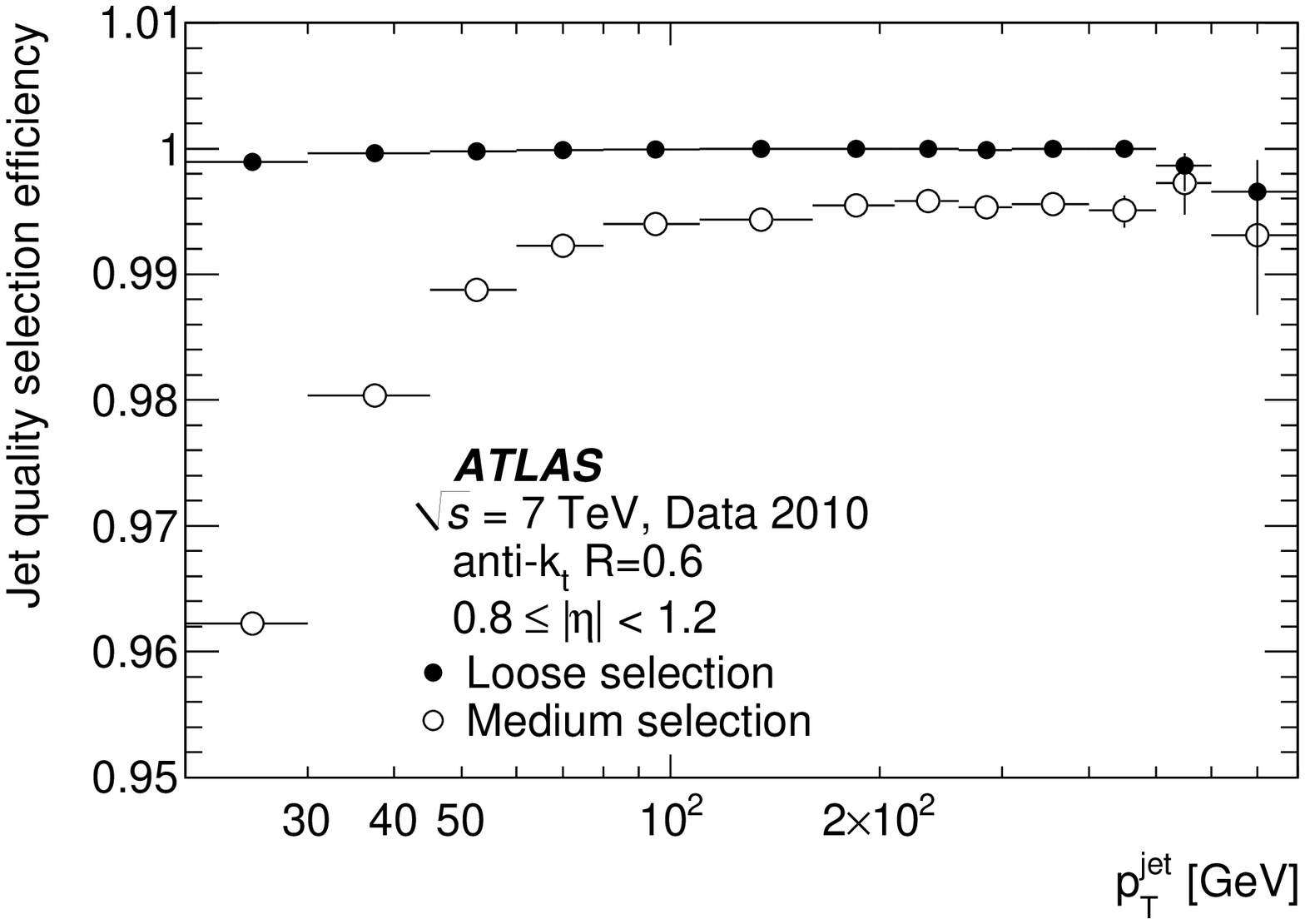}}
    \subfloat[$1.2 \leq |\etajet|<2.1$]{\includegraphics[width=70mm]{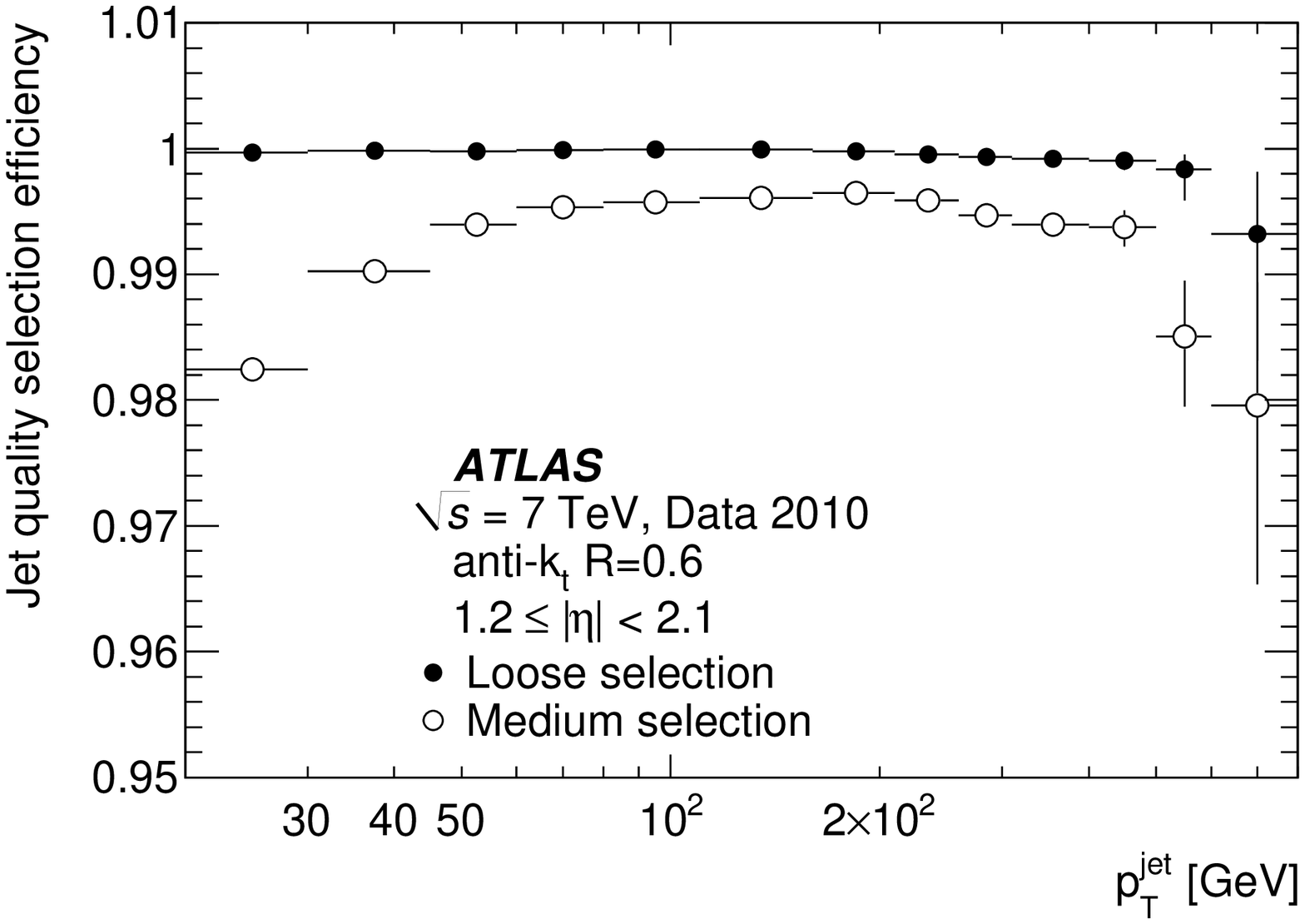}}\\
    \subfloat[$2.1 \leq |\etajet|<2.8$]{\includegraphics[width=70mm]{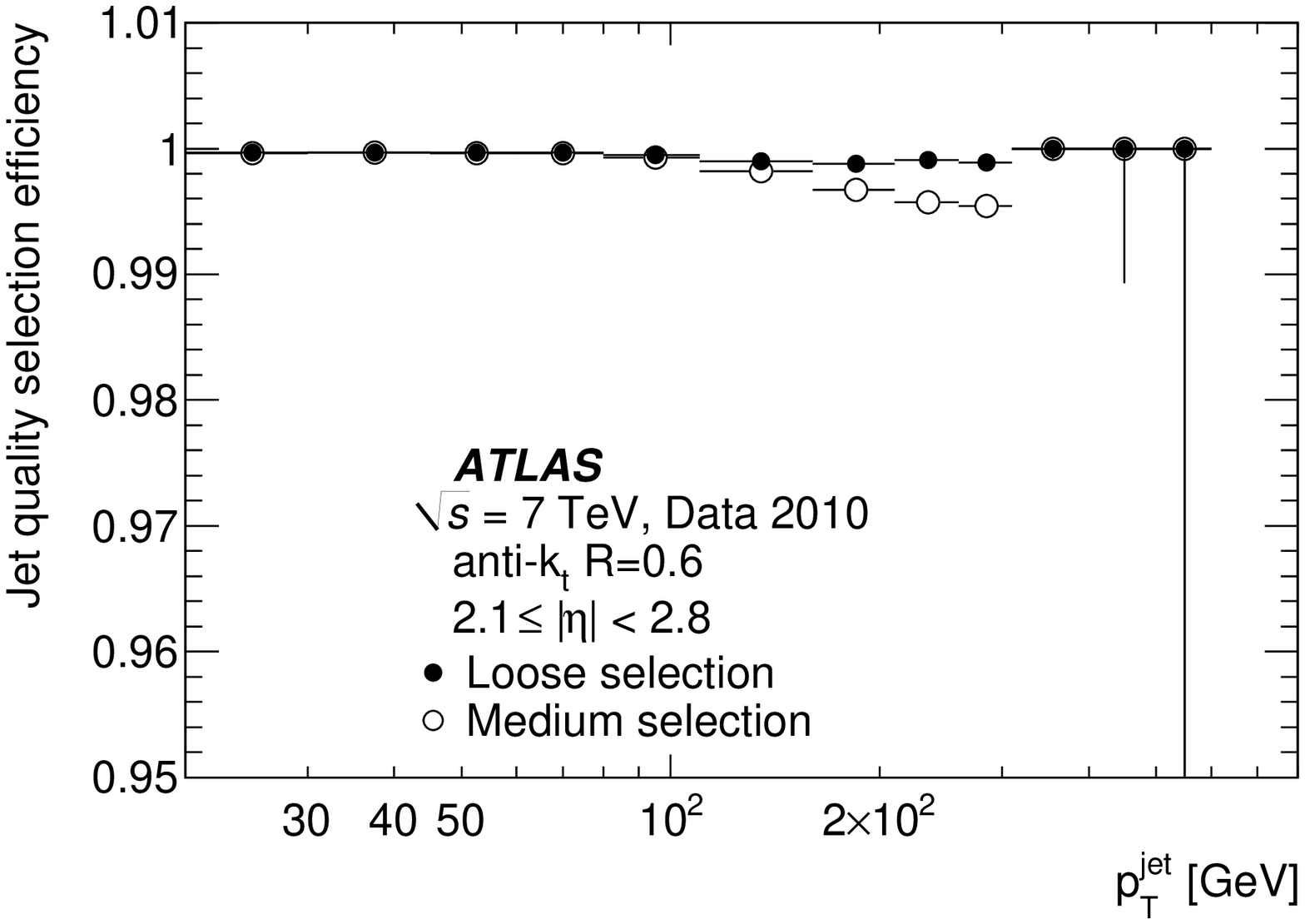}}
    \subfloat[$2.8 \leq |\etajet|<3.6$]{\includegraphics[width=70mm]{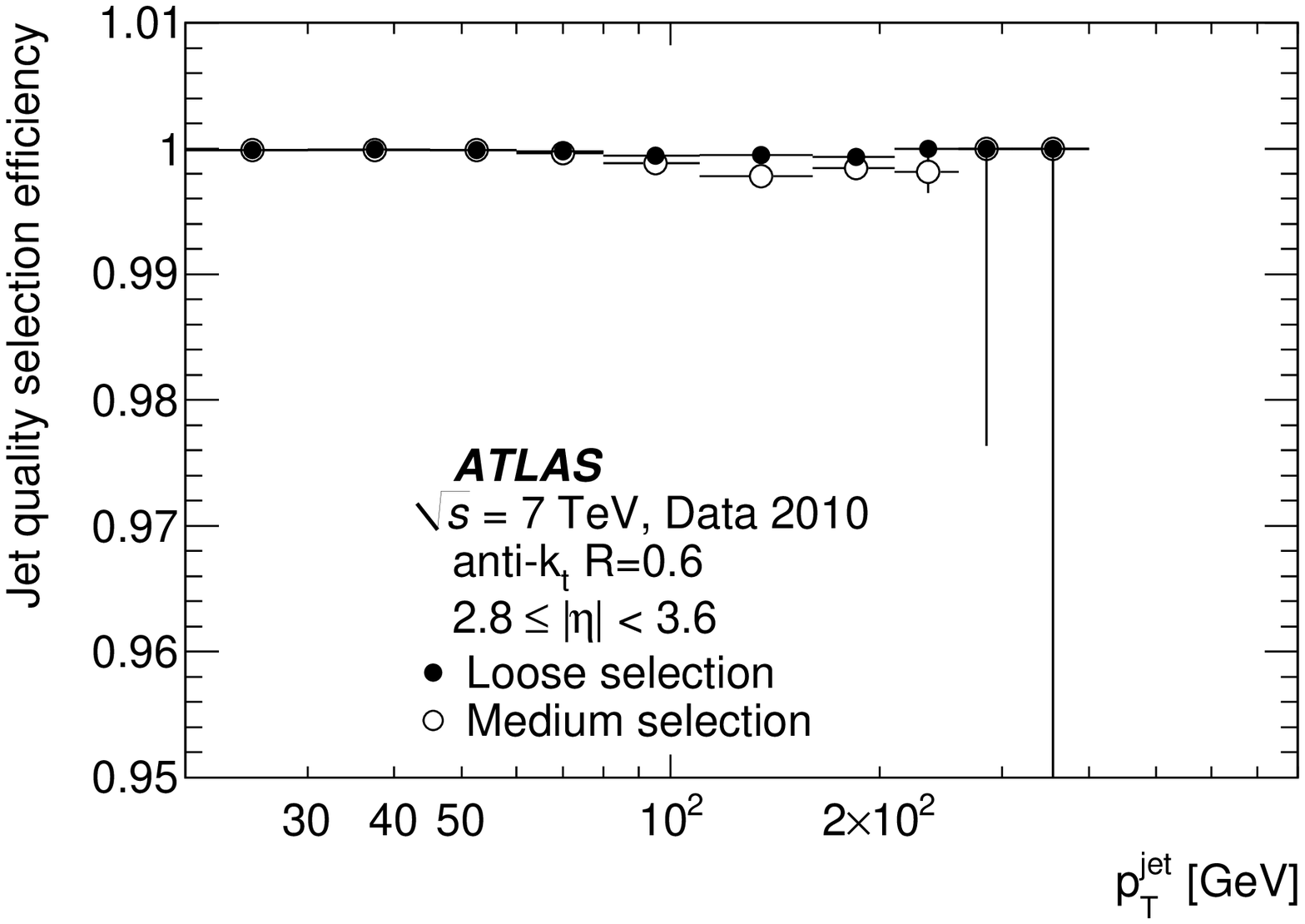}}\\
    \subfloat[$3.6 \leq |\etajet|<4.4$]{\includegraphics[width=70mm]{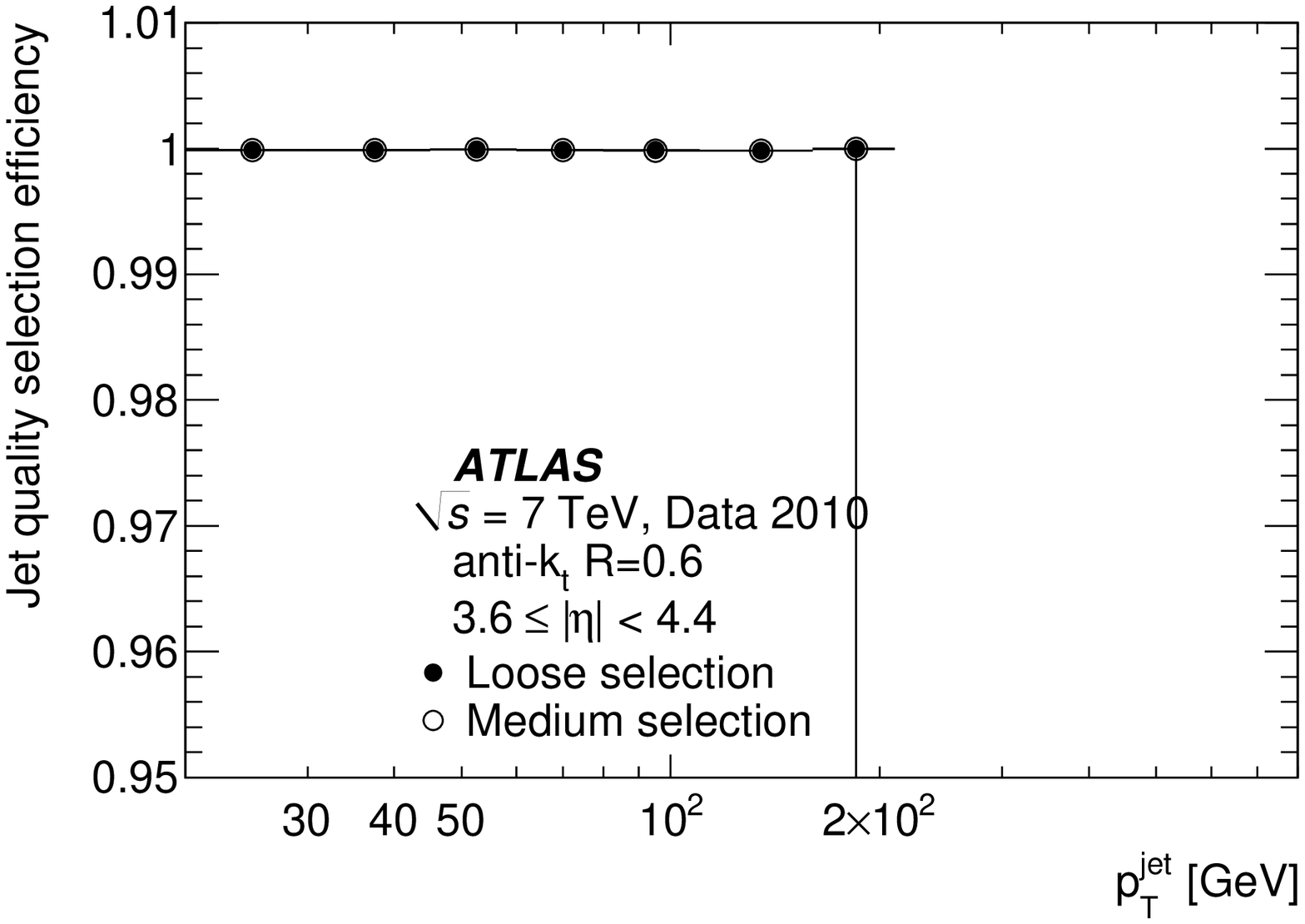}} \hfill\\
     \caption{Jet quality selection efficiency for \antikt{} jets with $R = 0.6$ measured with a tag-and-probe technique 
      as a function of \ptjet{} in bins of $\etajet$, for loose and medium selection criteria (see Table~\ref{cut_table}).
      Only statistical uncertainties are shown. In (e), (f), (g) the loose and medium results overlap.}
     \label{fig:jeteff}
  \end{center}
\end{figure*}

\section{Jet quality selection}
\label{sec:JetSelection}
Jets at high transverse momenta produced in proton-proton collisions must be distinguished 
from background jets not originating from hard scattering events.
The main backgrounds \index{jet backgrounds} 
are the following:
\noindent
\begin{enumerate}
\item Beam-gas events, where one proton of the beam collided with the residual gas within the beam pipe.
\item Beam-halo events, for example caused by interactions in the tertiary collimators
in the beam-line far away from the \ATLAS{} detector.
\item Cosmic ray muons overlapping in-time with collision events.
\item Large calorimeter noise.
\end{enumerate}

The criteria to efficiently reject jets arising from background are only applied to data.
They are discussed in the following sections. 

\subsection{Criteria to remove non-collision background}
\label{sec:JetSelectioncriteria}
\subsubsection{Noise in the calorimeters}
Two types of calorimeter noise are addressed:
\begin{enumerate}
\index{Sporadic noise bursts} 
\item {\bf Sporadic noise bursts} in the hadronic endcap calorimeter (\HEC), where a single noisy calorimeter 
cell contributes almost all of the jet energy. Jets reconstructed from these problematic cells 
are characterised by a large energy fraction in the \HEC{} calorimeter ($f_{\HEC}$) 
as well as a large fraction of the energy in calorimeter cells with poor signal shape quality\footnote{
The signal shape quality is obtained by comparing the measured pulse from the calorimeter
cell to the expected pulse shape.} ($f_{\rm \HEC quality}$). 
Due to the capacitive coupling between channels, the neighbouring calorimeter cells 
will have an apparent negative energy ($E_{\rm neg}$).
\index{Rare coherent noise} 
\item {\bf Rare coherent noise} in the electromagnetic calorimeter. 
Similarly, fake jets arising from this source are characterised 
by a large electromagnetic energy fraction ($f_{\rm EM}$)\footnote{The \EM{} fraction is defined
as the ratio of the energy deposited in the \EM{} calorimeter to the total energy.},
 and a large fraction of calorimeter cells with poor signal shape quality ($f_{\rm quality}$).
\index{Fake collision events:} 
\end{enumerate}

\subsubsection{Cosmic rays or non-collision background} 
Cosmic rays or non-collision backgrounds can induce events
where the jet candidates are not in-time with the beam collision. 
A cut on the jet time ($t_{\rm jet}$) is applied to reject these backgrounds.
The jet time is reconstructed from the energy deposition in the calorimeter 
by weighting the reconstructed time of calorimeter cells forming the jet
with the square of the cell energy. 
The calorimeter time is defined with respect to the event time recorded by the trigger. 

A cut on the $f_{\rm EM}$ is applied to make sure 
that the jet has some energy deposited in the calorimeter layer closest to the interaction region 
as expected for a jet originating from the nominal interaction point. 

Since a real jet is expected to have tracks, 
the $f_{\EM}$ cut is applied together with a cut on the minimal jet charged fraction ($f_{\rm ch}$), 
defined as the ratio of the scalar sum of the \pt{} of the tracks associated to the jet divided by the jet \pt{}, 
for jets within the tracking acceptance. 

A cut on the maximum energy fraction in any single calorimeter layer ($f_{\rm max}$) 
is applied to further reject non-collision background.

\subsubsection{Jet quality selections} 
Two quality selections are provided: 
\noindent
\begin{enumerate}
\index{loose jet cleaning} 
\item A {\bf loose selection} is designed with an efficiency above $99 \%$, 
that can be used in most of the \ATLAS{} physics analyses.
\index{medium jet cleaning} 
\item  A {\bf  medium selection} is designed for analyses that select jets at high transverse momentum, 
such as for jet cross-section measurements \cite{atlasjet2010}.
\end{enumerate}
A tight quality selection has been developed for the measurement of the jet quality selection efficiency 
described in Section~\ref{sec:effEval}, 
but is not used in physics analyses, since the medium jet quality selection is sufficient for removing fake jets.
The quality selection criteria used to identify and reject fake jets are listed in Table~\ref{cut_table}.

\subsection{Evaluation of the jet quality selection efficiency}
\label{sec:effEval}
\index{jet selection measurement} 
The criteria for the jet quality selection are optimised by studying samples with
good and fake jets classified by their amount
of missing transverse momentum significance\footnote{
The missing transverse momentum (\Etmiss) significance is defined as
$\Etmiss/\sqrt{\Sigma E_{\rm T}}$, where $\Sigma E_{\rm T}$ is the scalar sum of the transverse energies of
all energy deposits in the calorimeter.
}:
\begin{enumerate}
\item Good jets belong to events where the two leading jets have %
      $\ptjet > 20$~\GeV, and are back-to-back ($\Delta \phi_{\rm j-j} > 2.6$ radian) in the plane transverse to the beam, 
and with a small missing transverse momentum significance $\Etmiss/\sqrt{\Sigma E_{\rm T}} < 1$.
\item Fake jets belong to events with a high transverse momentum significance
$\Etmiss/\sqrt{\Sigma E_{\rm T}} > 3$ and with
a reconstructed jet back-to-back 
to the missing transverse momentum direction ($\Delta \phi_{\Etmiss -\rm j} > 2.6$ radian).
\end{enumerate}

As the jet quality selection criteria are only applied to data
an efficiency correction for data is determined.
This efficiency %
is measured using a tag-and-probe method in events 
with two jets at high transverse momentum. The reference jet ($\pt^{\rm ref}$)
is required to pass the tightened version of the jet quality selections, 
and to be back-to-back and well-balanced with the probe jet ($\pt^{\rm probe}$): 
\begin{equation}
(| \pt^{\rm probe} - \pt^{\rm ref}| / p^{\rm avg}_{\rm T} < 0.4), {\rm with} \; \; p^{\rm avg}_{\rm T} 
= ( p_{\rm T}^{\rm probe} + p_{\rm T}^{\rm ref} )/2.
\end{equation}
The jet quality selection criteria were then applied to the probe jets, 
measuring the fraction of jets passing as a function of $\etajet$ and $\ptjet$. 

The resulting efficiencies for jets with $R = 0.6$ for loose and medium selections 
applied to the probe jets are shown in Figure~\ref{fig:jeteff}. 
The tight selection of the reference jet was varied to study the systematic uncertainty.
The loose selection criteria are close to $100 \%$ efficient. 
In the forward region the medium selection criteria are also close to fully efficient.
In the central region they have an efficiency of $99 \%$ for $\ptjet > 50$~\GeV. 
For lower \pt{} jets of about $25$~\GeV{} an inefficiency of up to $3 - 4 \%$ is observed.

\subsection{Summary of the jet quality selection}
Quality selections used to reject fake jets %
with the \ATLAS{} detector have been developed.
Simple variables allow the removal of fake jets due to sporadic noise in the calorimeter 
or non-collision background at the analysis level, with an efficiency greater than $99 \%$
over a wide kinematic range.

\section{Jet energy calibration in the EM+JES scheme}
\label{sec:JetCalib}
\index{ jet energy calibration}
The simple \EMJES{} calibration scheme  applies
corrections as a function of the jet energy and pseudorapidity 
to jets reconstructed at the electromagnetic scale. 
\index{ jet energy calibration \EMJES}

The additional energy due to multiple proton-proton
collisions within the same bunch crossing (pile-up) is corrected before the hadronic
energy scale is restored, such that the derivation of the jet energy scale calibration 
is factorised and does not depend on the number of additional interactions measured. 

The \EMJES{} calibration scheme consists of three subsequent steps as
outlined below and detailed in the following subsections:
\begin{enumerate}
\item {\bf Pile-up correction}:
  The average additional energy due to additional proton-proton interactions 
  is subtracted from the energy measured in the calorimeters
  using correction constants obtained from \insitu{} measurements.
\item  {\bf Vertex correction}:
  The direction of the jet is corrected such that the jet originates from the primary vertex
  of the interaction instead of the geometrical centre of the detector.
\item {\bf Jet energy and direction correction}:
  The jet energy and direction as reconstructed in the calorimeters are corrected 
  using constants derived from the comparison of the kinematic observables of reconstructed 
jets and those from truth jets in Monte Carlo simulation.
\end{enumerate}

\begin{figure}[!ht]
\begin{center}
  \includegraphics[angle=0,width=0.45\textwidth]{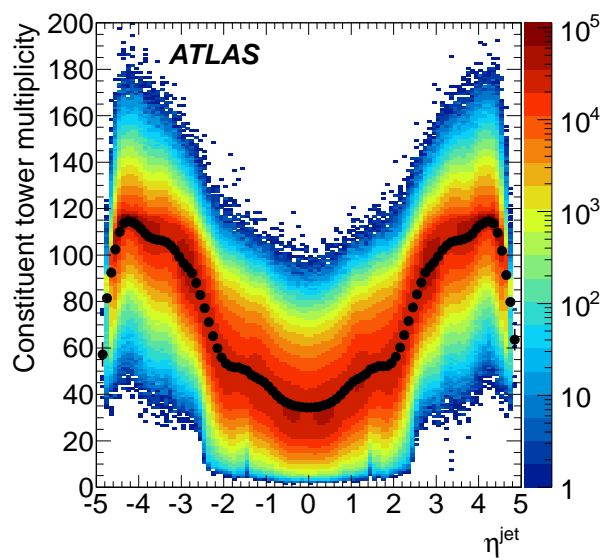}
        \caption{Distribution of the number constituent calorimeter towers as a function of the
 jet pseudorapidity for \antikt{} jets with $R=0.6$ and $\ptjet>7$~\GeV.
 The black dots indicate the average number of tower constituents.}
        \label{fig:towers}
\end{center}
\end{figure}
\begin{figure*}[!ht]
\begin{center}
\subfloat[Tower offset]{\label{fig:offset:tower}
\includegraphics[angle=0,width=0.49\textwidth]{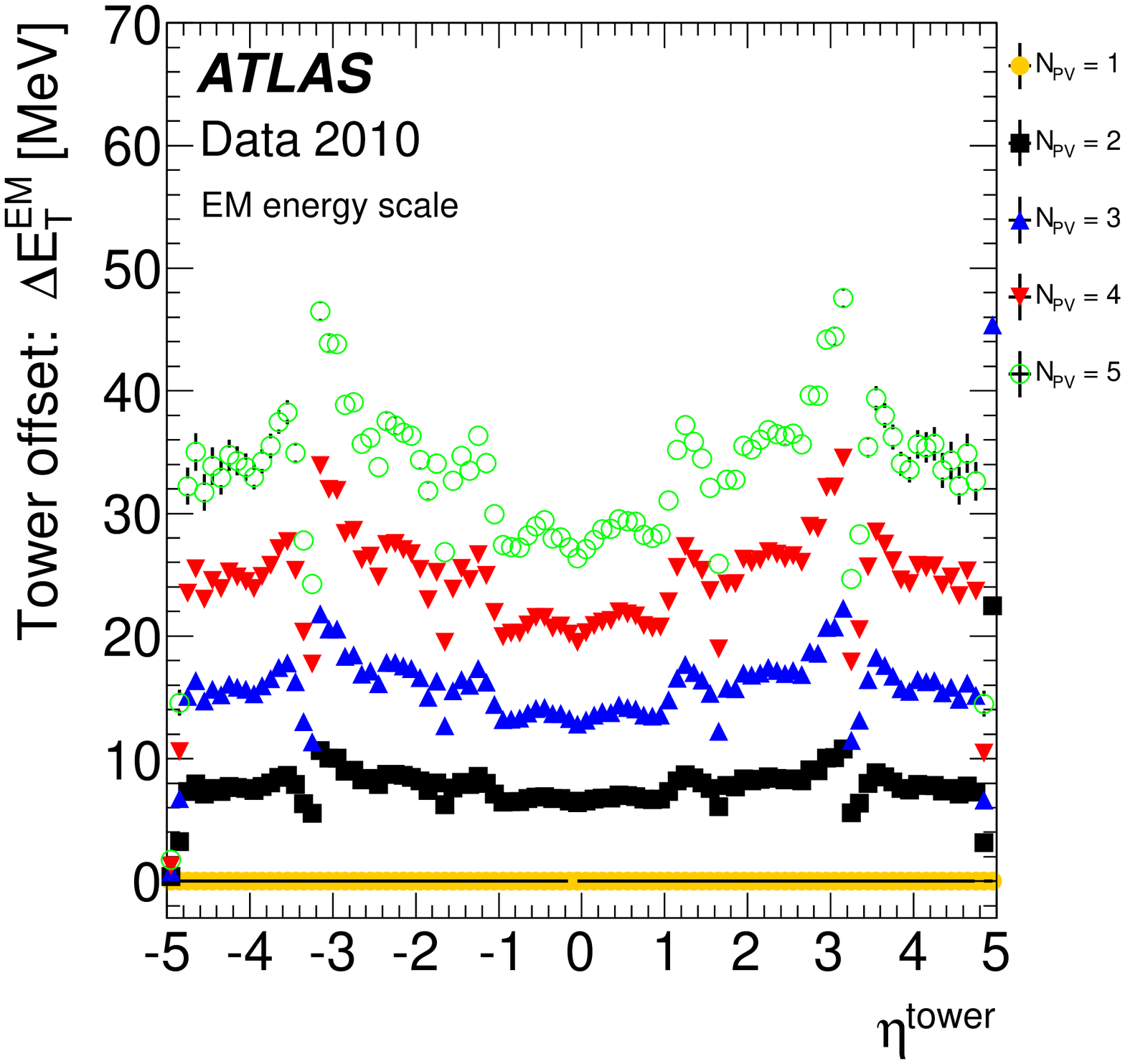}}
\hfill
\subfloat[Jet offset]{\label{fig:offset:jet}\includegraphics[angle=0,width=0.49\textwidth]{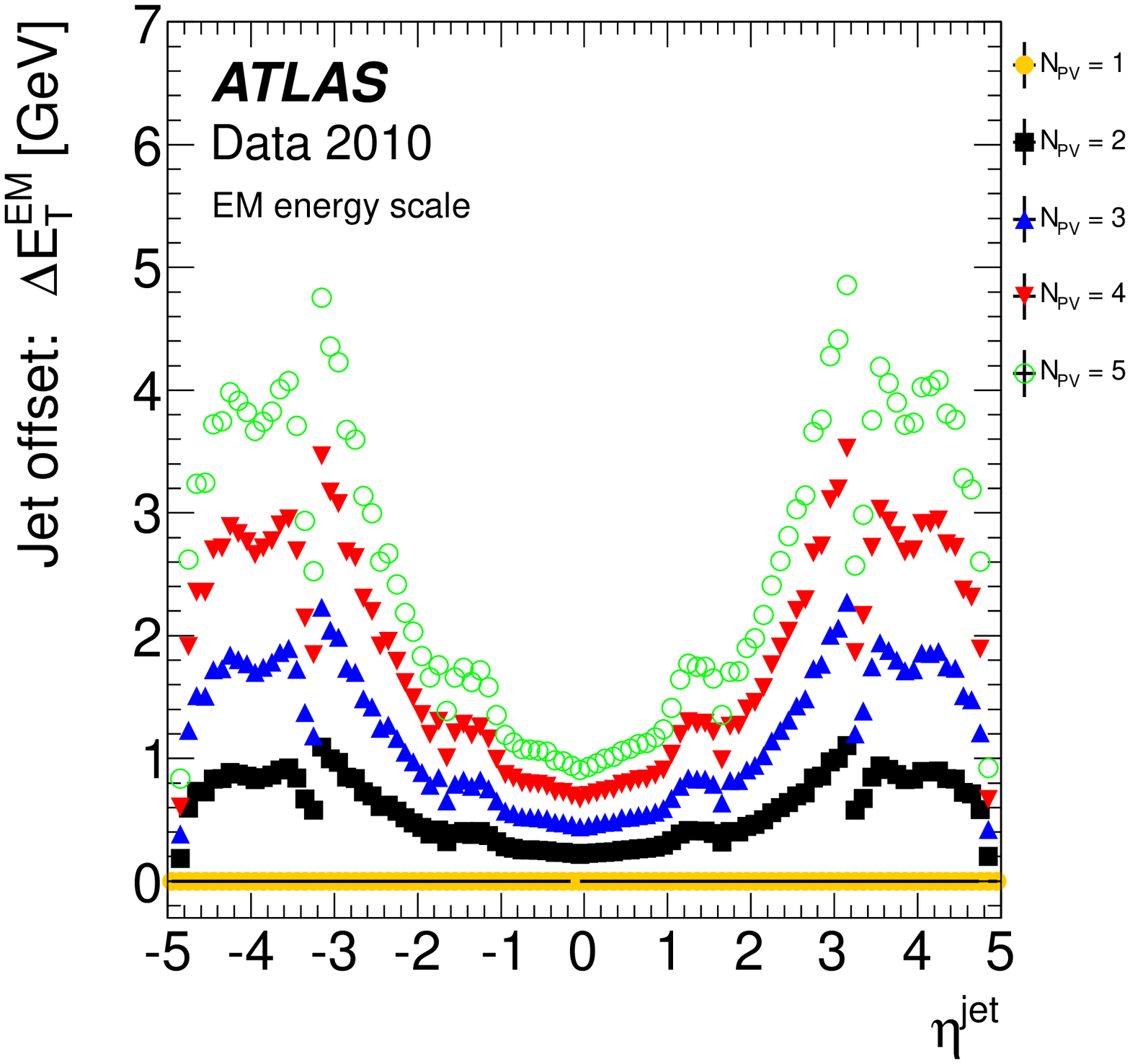}}
\caption{Tower offset (a) and jet offset (b)
at the \EM{} scale as a function of the tower or jet pseudorapidity in bins of
the number of reconstructed primary vertices. The jet offset is shown for \antikt{} jets with $R = 0.6$.
Only statistical uncertainties are shown. They are typically smaller than the marker size.
}
\label{fig:offset}
\end{center}
\end{figure*}
\begin{figure*}[!ht]
\begin{center}
\subfloat[Tower jets]{\label{fig:trkoffset:fittower}
\includegraphics[width=0.49\textwidth]{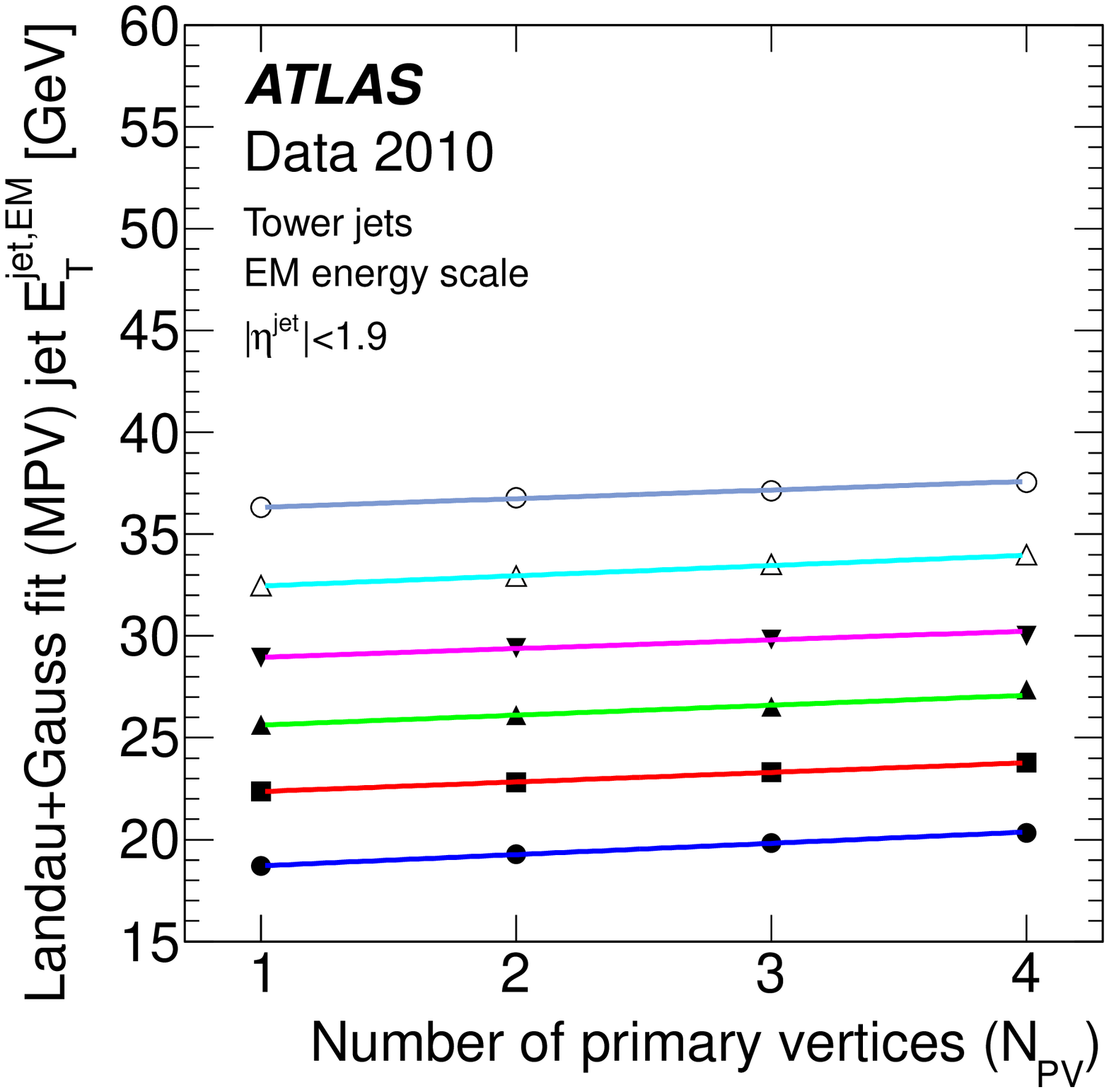}}
\subfloat[\Topo{} jets]{\label{fig:trkoffset:fittopo}
\includegraphics[width=0.49\textwidth]{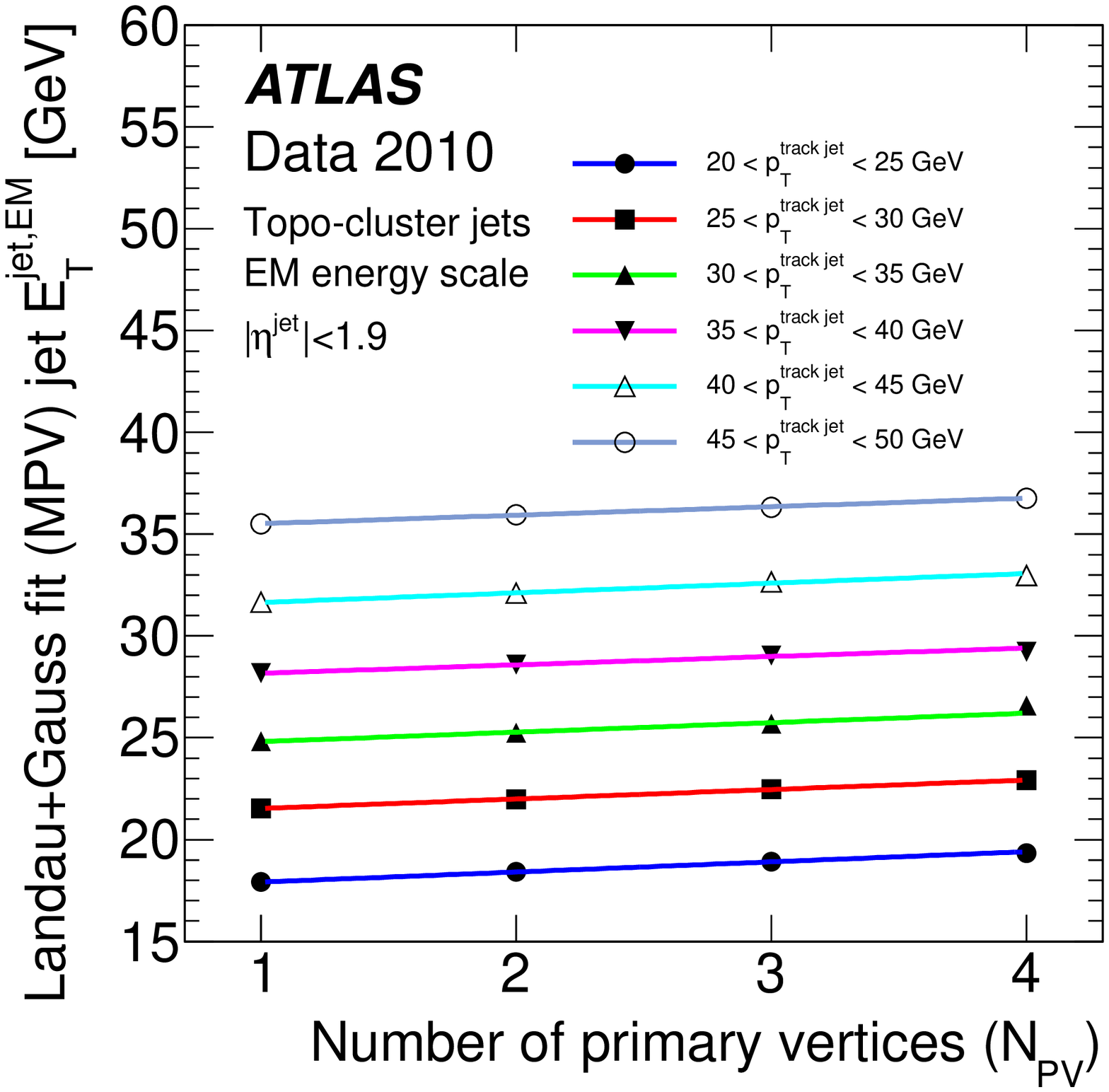}}
\caption{Transverse jet energy \etjet{} for calorimeter jets associated to track jets 
measured at the \EM{} scale using a Landau-Gauss fit
as a function of the reconstructed vertex multiplicity, \Npv,
in bins of \pTtrkjet{}.
Calorimeter jets are reconstructed at the \EM{} scale with calorimeter towers (a) and \topos{} (b) as inputs. 
Systematic uncertainties are not shown.
The statistical uncertainties from the fit results are smaller than the marker size.
}
\label{fig:calopt}
\end{center}
\end{figure*}
\begin{figure*}[htp]
\begin{center}
\subfloat[Tower jet offset (\EM{} scale)]{\label{fig:trkoffset:towerEM}       \includegraphics[width=0.44\textwidth]{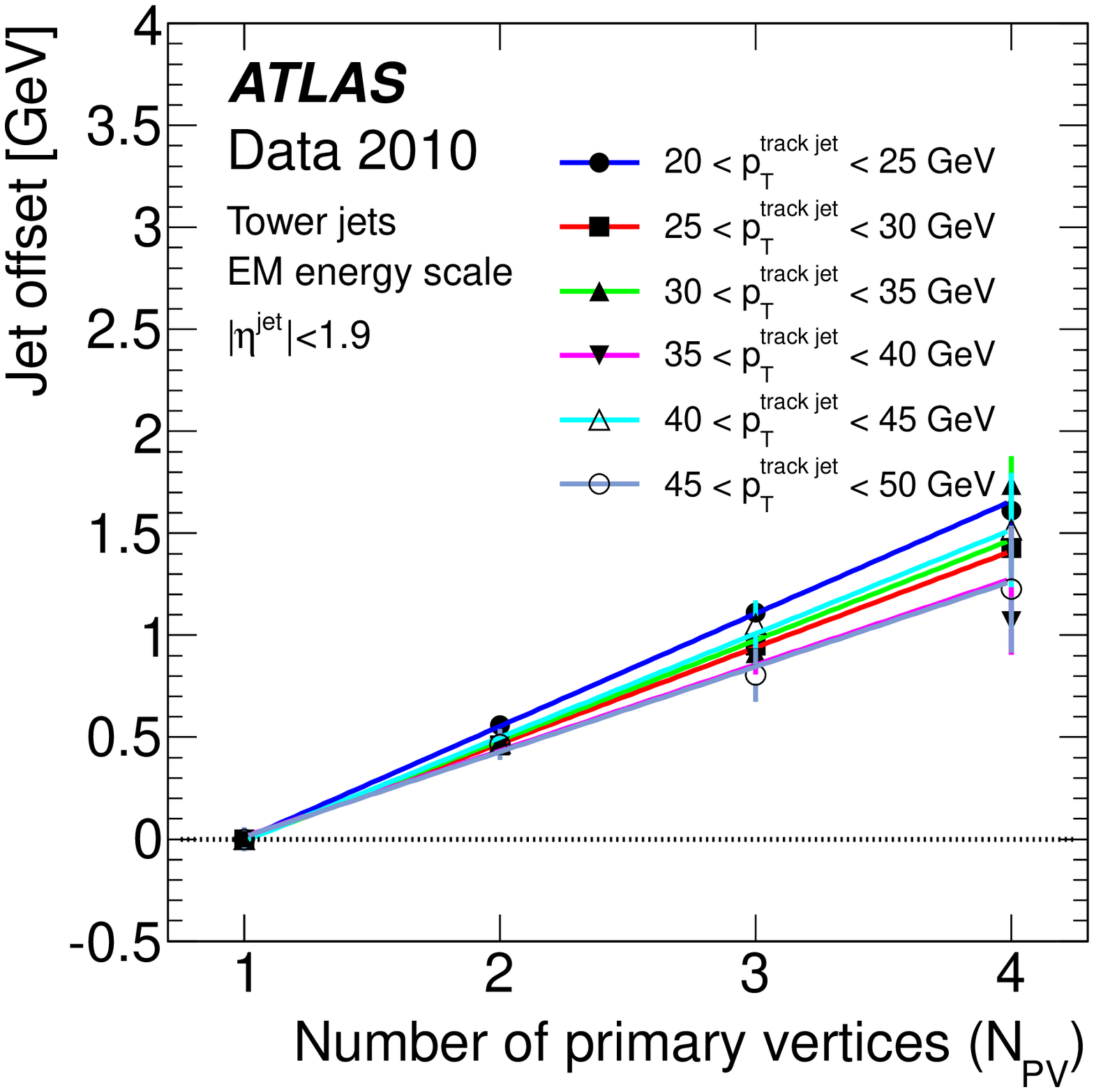}}
\hspace{1.cm}
\subfloat[\Topo{} jet offset (\EM{} scale)]{\label{fig:trkoffset:topoEM}      \includegraphics[width=0.44\textwidth]{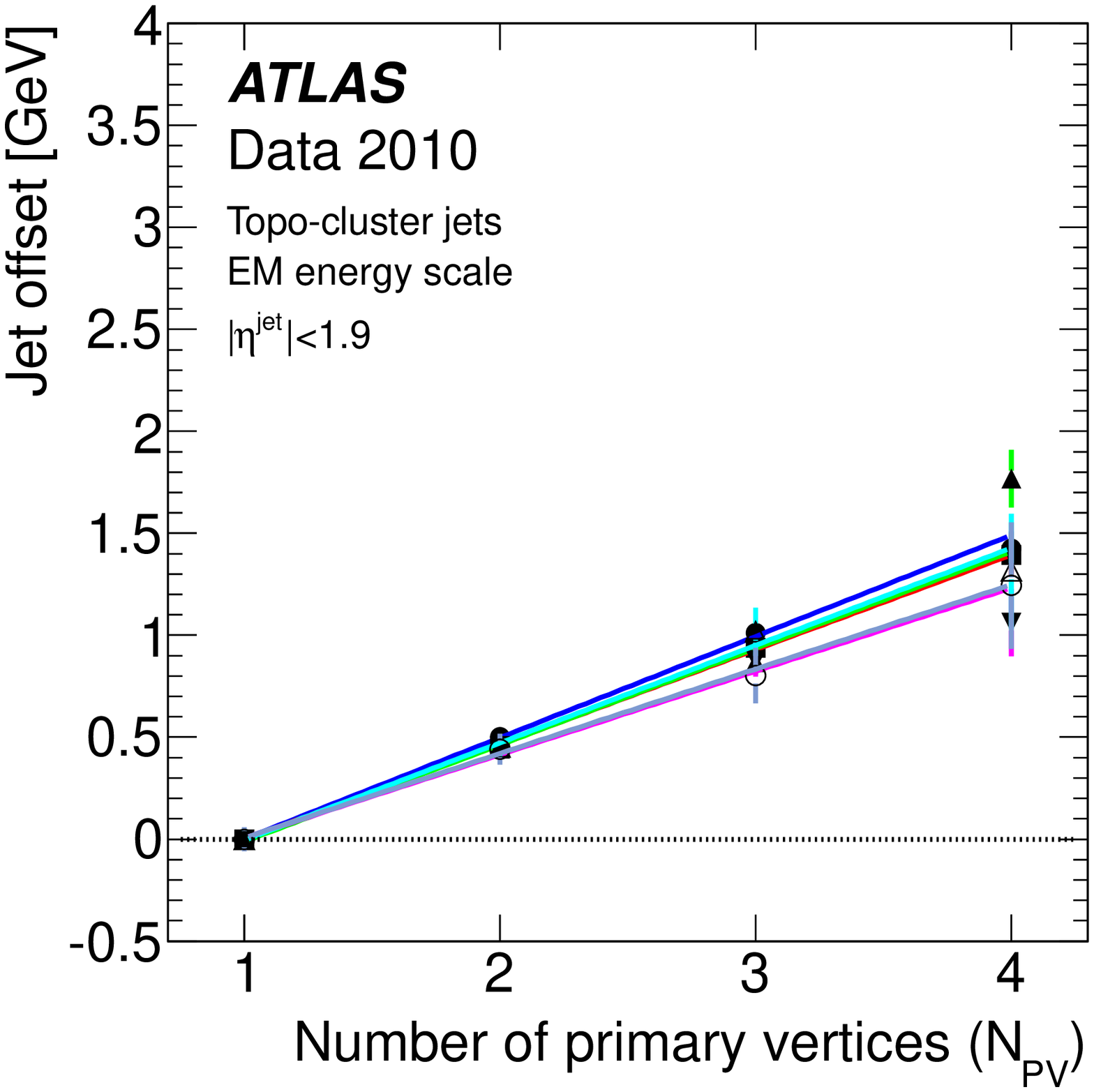}}\\
\subfloat[Tower jet offset (\EMJES{} scale)]{\label{fig:trkoffset:towerEMJES} \includegraphics[width=0.44\textwidth]{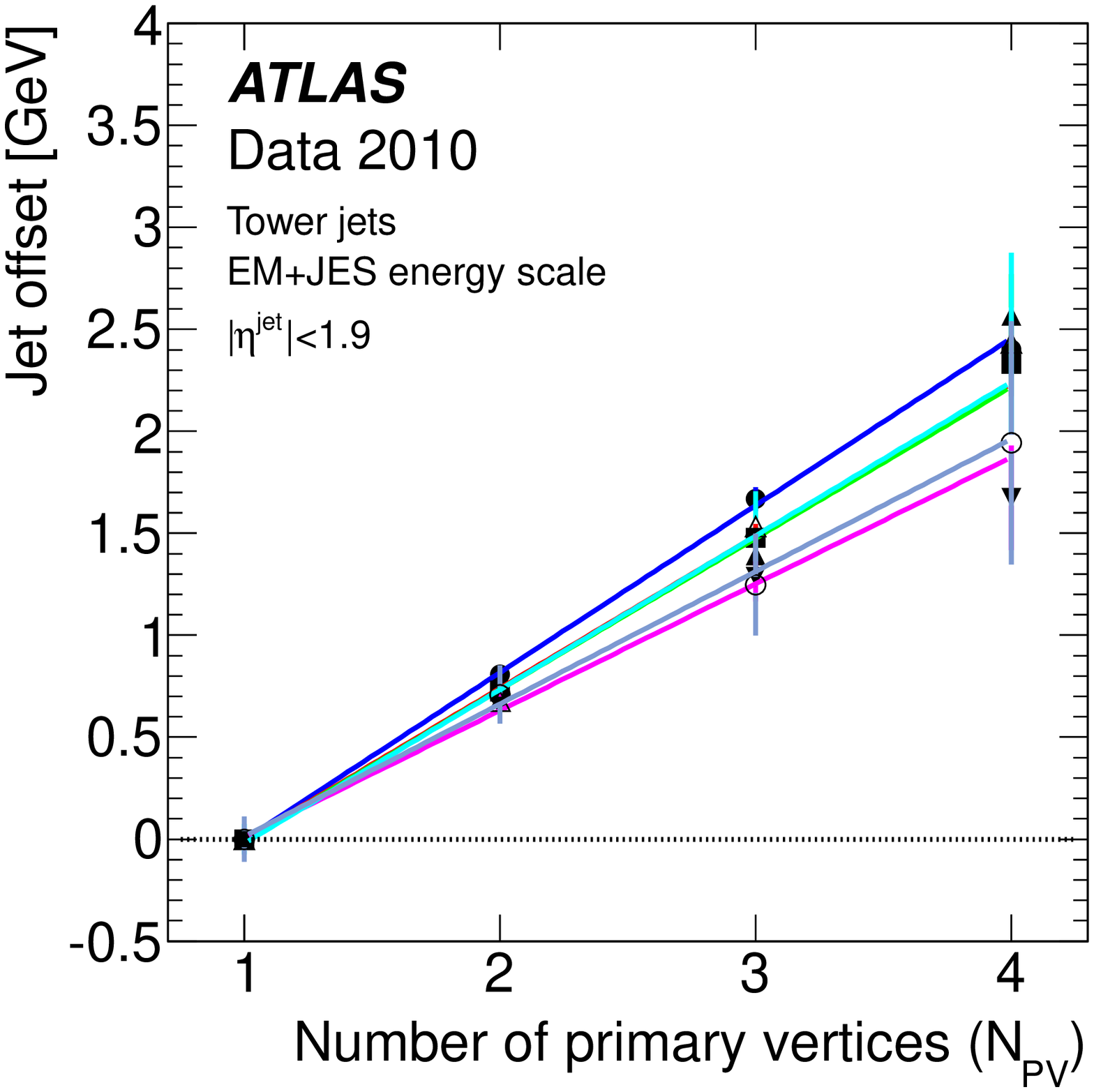}}
\hspace{1.cm}
\subfloat[\Topo{} jet offset (\EMJES{} scale)]{\label{fig:trkoffset:topoEMJES}\includegraphics[width=0.44\textwidth]{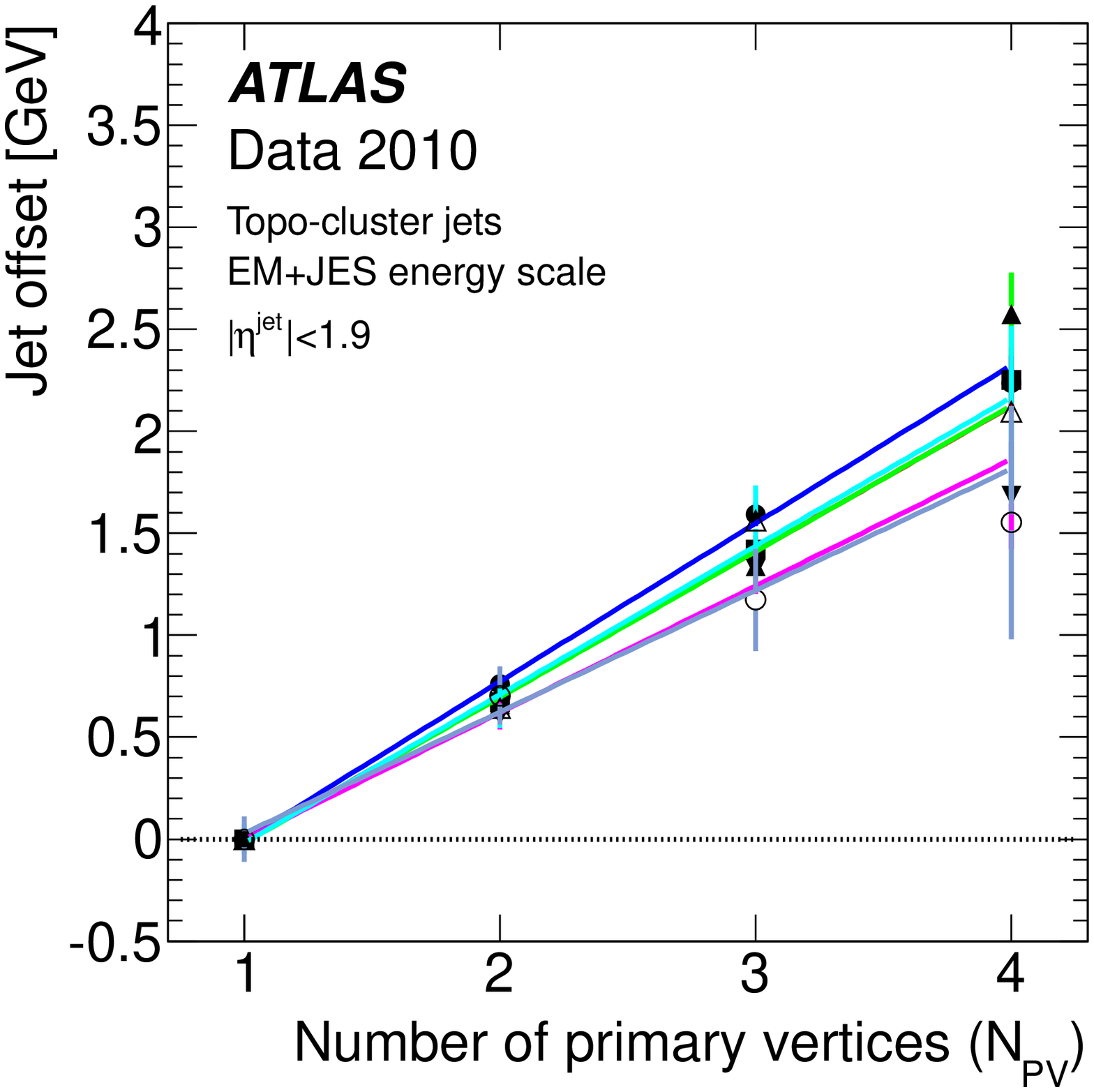}}
\caption{Jet offset as a function of the number of primary vertices 
for several ranges of \pTtrkjet{} values. The track jet offset is derived for  
calorimeter tower jets at the \EM{} scale (a), 
\topo{} jets at the \EM{} scale (b), 
calorimeter tower jets at the \EMJES{} scale (c), and 
\topo{} jets at the \EMJES{} scale (d). 
Only statistical uncertainties from the fit results are shown. }
       \label{fig:trkoffset}
\vfil
\end{center}
\end{figure*}

%
\subsection{Pile-up correction}
\label{sec:pileup} 
\index{Pile-up correction}
\subsubsection{Correction strategy}
The measured energy of reconstructed jets can be affected by contributions 
that do not originate from the hard scattering event of interest, but are instead 
produced by additional proton-proton collisions within 
the same bunch crossing. 
\index{multiple proton-proton collisions}
An offset correction for pile-up is derived from minimum bias data as a function of the number of
reconstructed primary vertices, \Npv, the jet pseudorapidity, $\etajet$,
and the bunch spacing.

This offset correction applied to the jet transverse energy (\et) at the \EM{} scale as the first step
of jet calibration can be written generically as:
\begin{equation}
  E^{\rm corrected}_{\rm T} = E^{\rm uncorrected}_{\rm T} - \mathcal{O}(\eta,\Npv,\bunchSp),
\end{equation}
where $\mathcal{O}(\eta,\Npv, \bunchSp)$  corrects for the jet offset due to pile-up.

Due to the varying underlying particle spectrum and the variation in the calorimeter geometry
the jet offset is derived as a function of the jet pseudo\-rapidity. 
The amount of in-time pile-up is parameterised by \Npv. 
The spacing between consecutive bunches, \bunchSp, is considered, 
because it can impact the amount by which collisions in previous bunch crossings affect
the jet energy measurement\footnote{The dependence on \bunchSp{} is 
explicitly allowed for due to the possibility of pile-up contributions 
from previous proton-proton bunch crossings for closely spaced bunches. 
This will be an important consideration for the 2011-2012 \LHC{} run as the number of bunches is increased 
and the spacing between consecutive bunches is reduced.}. 
\index{bunch spacing}

The jet offset correction is proportional to the number of constituent 
towers in a jet as a measure of the jet area.
For jets built directly from dynamically-sized topological clusters, for which no clear geometric definition 
is available, a model is used that describes the average area of a jet in terms of the 
equivalent number of constituent towers.

\subsubsection{Constituent tower multiplicity of jets}
\label{sec:tower-offset}
The multiplicity of calorimeter towers in jets depends on the internal jet composition and 
on  the presence of pile-up. %
The average tower multiplicity can be measured \insitu.
\index{tower multiplicity}

Figure~\ref{fig:towers} depicts the distribution of the 
constituent tower multiplicity for jets based on towers
with $\ptjet > 7$~\GeV{} as a function of the jet pseudo\-rapidity.
The average number of constituent towers is also indicated.
This distribution is governed by the change in physical size 
of calorimeter towers for a constant interval in pseudo\-rapidity, as well as by differences in the noise spectrum for the 
various calorimeters and sampling regions. 

\subsubsection{Pile-up offset for towers and jets}
\label{subsec:towerjetlevel}
\index{Tower-based offset correction for jets}
The calorimeter tower offset at the \EM{} scale is derived by measuring the average tower transverse energy 
for all towers in events with $\Npv=1,2,...N$ 
and comparing directly to events with $\Npv=\Nref=1$: 
\begin{equation}
  \mathcal{O}_{\rm tower}(\eta, \Npv)=\langle E^{\rm tower}_{\rm T}(\eta,\Npv)\rangle - \langle E^{\rm tower}_{\rm T}(\eta, \Nref)\rangle,
\end{equation}
where the angled brackets denote a statistical average over all events. 
The average is computed for events at each primary vertex multiplicity. 
For this measurement non-noise-suppressed calorimeter towers are used
(see Section \ref{sec:topotowers})
in order to remain sensitive to low energy depositions
that may not rise above noise threshold except inside of a jet.
The calorimeter tower offset is shown in Figure~\ref{fig:offset:tower} for $1\leq\Npv\leq 5$.

The tower offset can be extrapolated to an \EM{} scale jet offset using:
%
\begin{equation}
  \mathcal{O}_{\rm jet|tower}(\eta, \Npv)=\mathcal{O}_{\rm tower}(\eta, \Npv)
\; \cdot \;
A^{\rm jet},
\end{equation}
where $A^{\rm jet}$ is the jet area that, for jets built from calorimeter towers,
can be estimated from the constituent tower multiplicity,
$A^{\rm jet} = N^{\rm jet}_{\rm towers}$.
For jets built from \topos, the mean equivalent constituent tower multiplicity 
($A^{\rm jet} = \langle N^{\rm jet}_{\rm towers} \rangle$) 
is used\footnote{
The equivalent constituent tower multiplicity for jets based on \topos{}
is calculated from the location of the calorimeter cells of the constituent \topos{}
in the jet.}.
The small dependencies of the constituent multiplicity on \ptjet{} and \Npv{} are 
neglected in the correction, but incorporated as systematic uncertainties
(see Section \ref{sec:pileupSummary}). 

The jet offset for jets with $R=0.6$ is shown in Figure~\ref{fig:offset:jet}. 

\subsubsection{Track jet based validation and offset correction}
\label{sec:track-jet-offset}
\index{Tower-based offset at jet-based}
Track jets  
constructed from charged particles
originating from the primary hard-scattering vertex 
matched to the calorimeter jets provide a stable reference 
that can be used to measure the variation of the calorimeter \etjet{} as a function of \Npv. 
It is therefore possible to 
validate the tower-based offset correction  
and also to directly estimate the pile-up energy contribution to jets. 

As this method is only applicable to jets within the inner detector acceptance, it serves primarily 
as a cross-check for the tower-based method discussed above. 
It can also be used, however, to derive a dedicated offset correction 
that can be applied to jets at energy scales other than the electromagnetic energy scale.
Studying the variation of the offset correction as a function 
of \pTtrkjet{} can establish the systematic uncertainty of the pile-up correction. 

The criterion to match a track jet to a calorimeter jet with $R = 0.6$ is
\begin{equation}
  \Delta R({\rm jet, {\rm track jet}}) < 0.4,
\end{equation}
where \DeltaRdef. 
The offset is calculated by measuring the average calorimeter jet \etjet{}
as a function of \Npv{} and the transverse momentum of the matched track jet, \pTtrkjet:
%

\begin{equation}
  \mathcal{O}_{{\rm track \; jet}} = \langle\etjet(\Npv | \pTtrkjet )\rangle - \langle\etjet(\Nref | \pTtrkjet)\rangle.
\end{equation}
The reference $\Nref = 1 $ is used.

Figure~\ref{fig:calopt} shows the jet $E_{\rm T}$ as a function of \Npv{} for several bins in \pTtrkjet{}. 
Both tower and \topo{} jets at the electromagnetic scale are used.
The most probable value (MPV) of the calorimeter jet $E_{\rm T}$ is determined from a fit using a Landau distribution 
convolved with a Gaussian for each range of \pTtrkjet. 
A consistent offset of nearly $\mathcal{O}=0.5$~\GeV{} 
per vertex is found for $|\etajet|<1.9$. 
No systematic trend of the offset as a function of \pTtrkjet{} 
is observed.

Figure~\ref{fig:trkoffset} 
presents the jet-based offset correction as a function of \Npv{} derived with 
respect to $\Nref=1$ for tower and \topo{} based jet using the \EM{}
and the \EMJES{} scale.
As expected, the magnitude of the offset is higher after \EMJES{} calibration 
(see Figure~\ref{fig:trkoffset:towerEMJES} and Figure~\ref{fig:trkoffset:topoEMJES}), 
and the increase corresponds to the average jet energy correction
(see Section \ref{sec:EMJES}).

%
\begin{figure}[ht!]
  \centering
  \includegraphics[width=0.49\textwidth]{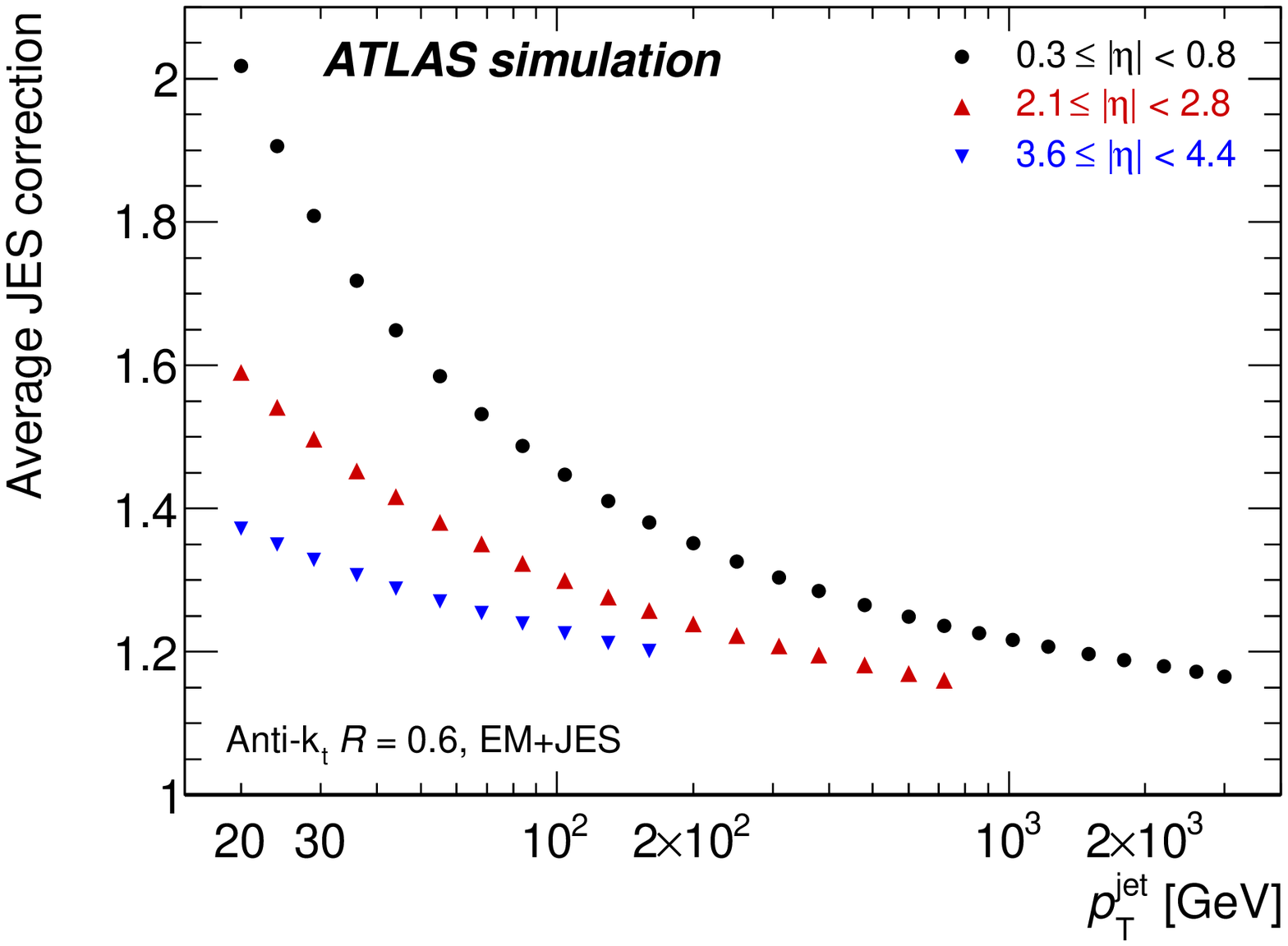}
  \caption{
    Average jet energy scale correction 
    as a function of the calibrated jet transverse momentum
    for three representative $\etajet$-intervals obtained from the nominal Monte Carlo simulation sample.
    The correction is only shown over the accessible kinematic range. 
  }
 \label{fig:JESvsPt}
\end{figure}

\begin{figure}
  \centering
  \includegraphics[width=0.49\textwidth]{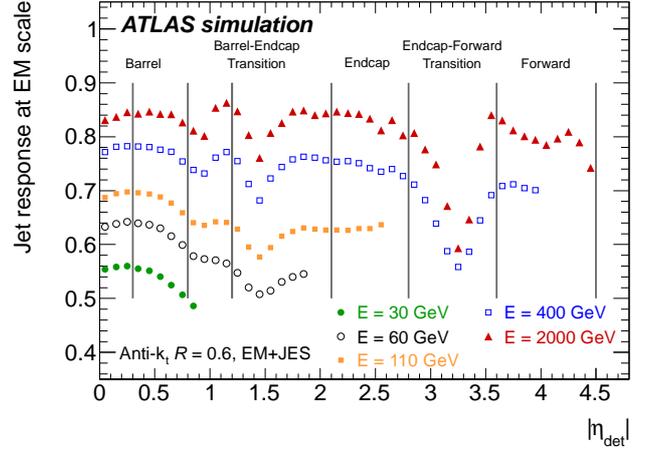}
  \caption{
    Average simulated jet response (\RcaloEM) at the electromagnetic scale in bins of
    \EMJES{} calibrated jet energy  and as a function of
    the detector pseudorapidity
    $\etaDet$. Also shown are the $\eta$-intervals used to evaluate 
    the \JES{} uncertainty (see Table~\ref{tab:etaregions}).
    The inverse of the response shown in each bin is equal to the
    average jet energy scale correction.} 
  \label{fig:EMJES_vs_eta}
\end{figure}
\begin{figure}[ht!]
  \centering
  \includegraphics[width=0.49\textwidth]{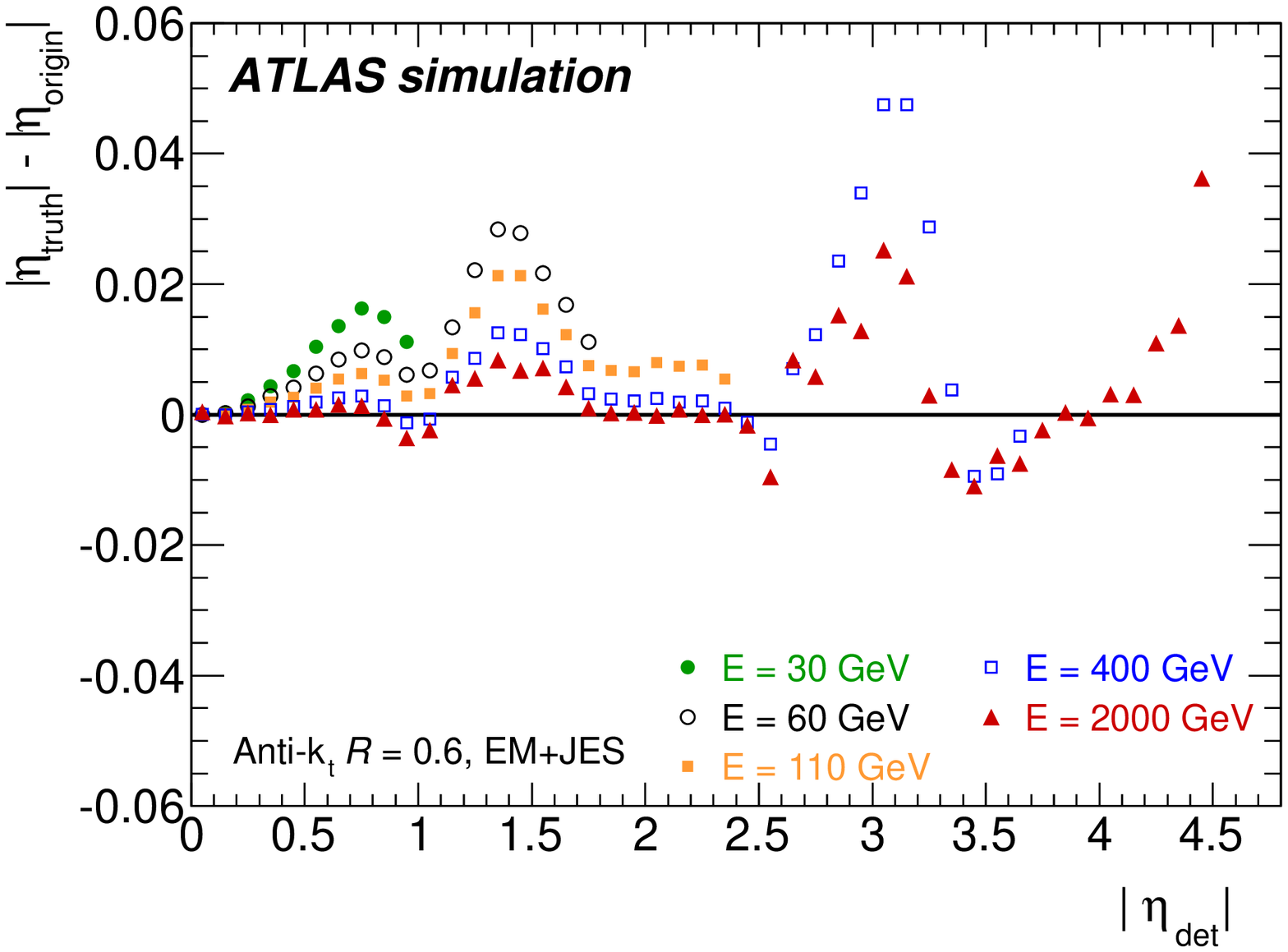}
  \caption{
    Difference between the jet
    pseudorapidity calculated using an origin correction and the true jet pseudorapidity 
    in bins of the calorimeter jet energy calibrated with the \EMJES{} scheme
    as a function of the detector pseudorapidity $|\etaDet|$.
  }
  \label{fig:EMJES_eta_corr}
\end{figure}

\subsection{Jet origin correction}
\label{sec:origin-corr}

Calorimeter jets are reconstructed using the geometrical centre of the \ATLAS{} detector as reference to calculate 
the direction of jets and their constituents (see Section~\ref{sec:JetReco}).
The jet four-momentum is corrected for each event such that
the direction of each \topo{} points back to the primary hard-scattering vertex.
The kinematic observables of each \topo{} are recalculated using 
the vector from the primary hard-scattering vertex to the \topo{} centroid
as its direction. The raw jet four-momentum is thereafter redefined as the
vector sum of the \topo{} four-momenta. The origin-corrected pseudorapidity is called $\eta_{\rm origin}$.
This correction improves the angular resolution and results in a small
improvement ($< 1\%$) in the jet \pt{} response. The jet energy is unaffected.

\subsection{Jet energy correction}
\label{sec:EMJES}
The final step of the \EMJES{} jet calibration restores the reconstructed jet energy
to the energy of the Monte Carlo truth jet. Since pile-up effects have already been
corrected for, the Monte Carlo samples used to derive the calibration
do not include multiple proton-proton interactions.

The calibration is derived using all
isolated calorimeter jets that have a matching isolated truth jet within $\Delta R = 0.3$.
Here, an isolated jet is defined as a jet having no other jet with
$\ptjet > 7$~\GeV{} within $\Delta R = 2.5 R$, 
where $R$ is the distance parameter of the jet algorithm.
A jet is defined to be isolated, if it is isolated with respect to the same jet type, i.e.
either a calorimeter or a truth jet.
\index{calibration of isolated jets}

The final jet energy scale calibration is first parametrised as a function of uncalibrated jet energy and $\etajet$.
Here the detector pseudorapidity is used rather than the origin-corrected $\eta$ (used by default in physics analyses),
since it more directly correspond to a region of the calorimeter. Energy is used rather than \pt{}, since
the calorimeter responds to energy, and the response curves can be directly compared to
expectation and between $\eta$ bins. The method to derive this calibration is detailed below.

The \EM-scale jet energy response 
\begin{equation}
\RcaloEM = \EcaloEM/E_{\rm  truth}^{\rm jet}
\end{equation}
for each pair of calorimeter and truth jets is measured in bins of the truth jet energy
$E_{\rm truth}^{\rm jet}$ and the calorimeter jet detector pseudorapidity $\etaDet$\footnote{Here, 
pseudorapidity refers to the original reconstructed jet before the origin correction.}.  
For each $(E_{\rm truth}^{\rm jet},\etaDet)$-bin, the averaged jet response 
$\left< \RcaloEM \right>$ is defined as the peak position of a Gaussian fit to the 
$\EcaloEM/E_{\rm  truth}^{\rm jet}$ distribution. 
In the same $(E_{\rm truth}^{\rm jet},\etaDet)$-bin, in addition, the average jet energy
response ($\left< \EcaloEM \right>$) is derived from the mean of the \EcaloEM{} distribution. 
%
For a given $\etaDet$-bin $k$, 
the jet response calibration function $\mathcal{F}_{{\rm calib},k}(\EcaloEM)$ 
is obtained using a fit of the
$(\left< \EcaloEM \right>_j,\left<\RcaloEM\right>_j)$ values for each $E_{\rm truth}^{\rm jet}$-bin $j$.

The fitting function is parameterised as:
\begin{equation}
  \mathcal{F}_{{\rm calib},k}( \EcaloEM) =
  \sum_{i=0}^{N_{\rm max}} a_i
  \left(\ln \EcaloEM\right)^i,
  \label{eq:Rjet}
\end{equation}
where $a_i$ are free parameters, and $N_{\rm max}$ is chosen
between $1$ and $6$ depending on the goodness of the fit.
\index{jet calibration, numerical inversion}

The final jet energy scale correction that relates the measured calorimeter jet energy 
to the true energy is then defined as $1/\mathcal{F}_{{\rm calib}}(E^{\rm calo}_{\EM})$ in the following:
\begin{equation}
  E^{\rm jet}_\EMJES = \frac{\EcaloEM}{\mathcal{F}_{\rm calib}( \EcaloEM ) |_{\etaDet}},
  \label{eq:JES}
\end{equation}
where $\mathcal{F}_{\rm calib}(\EcaloEM )|_{\etaDet}$ is
the jet response calibration function for the relevant $\etaDet$-bin~$k$.

The average jet energy scale correction $\left<1/\mathcal{F}_{{\rm calib},k}(E_{\rm calo}^{\rm EM})\right>$
is shown as a function of calibrated jet transverse momentum 
for three jet $\eta$-intervals  
in Figure~\ref{fig:JESvsPt}. 
In this and the following figures 
the correction is only shown over the accessible kinematic range,
i.e. values for jets above the kinematic limit are not shown. 

The calorimeter jet response \RcaloEM{} is shown for various
en\-er\-gy- and $\etaDet$-bins in Figure~\ref{fig:EMJES_vs_eta}. The values of the jet energy correction 
factors range from about $2.1$ at low jet energies in the central region to less 
than $1.2$ for high energy jets in the most forward region. 
\index{calorimeter jet response}
\index{jet origin and energy correction}

\subsection{Jet pseudorapdity correction}
After the jet origin and energy corrections
the origin-correc\-ted jet $\eta$ is further corrected for a bias due to
poorly instrumented regions of the calorimeter.
In these regions \topos{} are reconstructed with a lower energy 
with respect to better instrumented regions (see Figure~\ref{fig:EMJES_vs_eta}). 
This causes the jet direction to be biased towards the better instrumented calorimeter regions.
\index{jet direction correction}

The $\eta$-correction is derived as the average difference $\Delta\eta=\eta_{\rm truth}-\eta_{\rm origin}$ 
in $(E^{\rm truth},\etaDet)$-bins, and is parameterised as a function of the calibrated jet energy
$E^{\rm calo}_{\rm EM+JES}$ and the uncorrected $\etaDet$. 
The correction is very small ($\Delta\eta < 0.01$) for most regions of the calorimeter
but larger in the transition regions. 
The size of the bias is illustrated as a function of the detector pseudorapidity 
$|\etaDet|$ and \EMJES{} calibrated jet energy in
Figure~\ref{fig:EMJES_eta_corr}.

\section{Jet energy scale uncertainties for the EM+JES scheme}
\label{sec:JESUncertainties}
The \JES{} systematic uncertainty is derived combining information from the
single hadron response measured \insitu{} and 
single pion test-beam measurements, 
uncertainties on the amount of material of the \ATLAS{} detector, 
the description of the electronic noise, and 
the Monte Carlo modelling used in the event generation.  
Dedicated Monte Carlo simulation test samples are generated with different conditions with respect
to the nominal Monte Carlo sample described in Section~\ref{sec:NominalSample}. 
These variations are expected to provide an
estimate of the systematic effects contributing to the \JES{}{} uncertainty. 

The pseudorapidity bins used for the estimate of the \JES{} uncertainty divide
the \ATLAS{} detector in the eight $\eta$-regions specified in Table~\ref{tab:etaregions}
and Figure~\ref{fig:EMJES_vs_eta}.

\begin{table}[!ht]
\begin{center}
\begin{tabular}{c|c}
\hline
\hline
$\eta$ region & ATLAS detector regions \\
\hline
$         |\eta| < 0.3$ & Central Barrel\\ 
$0.3 \leq |\eta| < 0.8$ & \\ 
\hline
$0.8 \leq |\eta| < 1.2$ & Barrel-Endcap Transition\\
$1.2 \leq |\eta| < 2.1$ & \\
\hline
$2.1 \leq |\eta| < 2.8$ & Endcap \\
\hline
$2.8 \leq |\eta| < 3.2$ & Endcap-Forward Transition \\
$3.2 \leq |\eta| < 3.6$ & \\ 
\hline
$3.6 \leq |\eta| < 4.5$ & Forward \\ 
\hline
\hline
\end{tabular}
\end{center}
\caption{Detector regions used for the \JES{} uncertainty estimate.
\label{tab:etaregions}
}
\end{table}

The \JES{} systematic uncertainty for all jets with pseudorapidity $|\eta| > 0.8$ 
is determined using the \JES{} uncertainty for 
the central barrel region ($0.3 \leq |\eta| < 0.8$) as a baseline, with a contribution from the 
relative calibration of the jets with respect to the central barrel region. 
This choice is motivated by the good knowledge of the 
detector geometry in the central region, and by the use of pion response measurements
in the \ATLAS{} combined test-beam, which used a full slice of the \ATLAS{} barrel detector,
for the estimate of the calorimeter response uncertainties.
The region $0.3 \leq |\eta| < 0.8$ is the largest fully instrumented $|\eta|$ region considered
where combined test-beam results, used to estimate the calorimeter uncertainty, are available for the entire
pseudorapidity range.

\index{\JES{} systematic uncertainties}
This section describes the sources of systematic uncertainties and
their effect on the response of \EMJES{} calibrated jets. 
In Section~\ref{sec:Selection}, the selection of jets used to derive 
Monte Carlo based components of the \JES{} systematic uncertainty is discussed.
The contributions to the \JES{} systematics due to the following effects are then described:
\begin{enumerate}
\item \JES{} calibration method (Section~\ref{sec:nonclosure}).
\item Calorimeter response (Section~\ref{sec:SingleParticle}).
\item Detector simulation (Section~\ref{sec:DetectorUnc}).
\item Physics model and parameters employed in the Monte Carlo event generator (Section~\ref{sec:MCUnc}).
\item Relative calibration for jets with $|\eta|>0.8$ (Section~\ref{sec:etaintercalibration}).
\item Additional proton-proton collisions (pile-up) (Section~\ref{sec:pileupSummary}).
\end{enumerate}

Section~\ref{sec:JESSummary} discusses how the final uncertainties are calculated.
Additional uncertainties such as those for close-by jets are mentioned in Section~\ref{sec:JESUncertaintySpecialcase}
and discussed in more detail in Section~\ref{sec:closeby}.
\index{close-by jets}

\subsection{Jet response definition for the JES uncertainty evaluation}
\label{sec:Selection}
The components of the JES uncertainty derived from Monte Carlo samples are
obtained by studying the average calorimeter energy response of calibrated
jets. The average energy or \pt{} response, defined as 
\begin{equation}
\left< \Rcalo \right> = \left< \EcaloCALIB/\Etrue \right>
\; {\rm or} \; 
\left< \mathcal{R}(\ptjet) \right> = \left< \ptjet / \pttrue \right>, 
\end{equation}
is obtained as the peak position from a Gaussian fit to the distribution
of the ratio of the kinematic quantities for reconstructed and truth jets 
by matching isolated calorimeter jets to Monte Carlo truth
jets as described in Section~\ref{sec:EMJES}, but without the isolation cut for truth 
jets\footnote{The isolation cut for truth jets on the average jet response has a negligible 
impact on the average jet response given that truth jets are matched to isolated reconstructed jets.}.
This is done separately for the nominal and each of the alternative Monte
Carlo samples.
Only MC truth jets with $\pttrue{}>15$~\GeV, and calorimeter jets 
with $\ptjet{}>7$~\GeV{} after calibration, are considered. 
The calibrated response $\left< \Rcalo \right>$
is studied in bins of the truth jet transverse momentum \pttrue{}.
For each \pttrue-bin, an associated calibrated \pt{} value is calculated by multiplying the bin centre
with the average response.

The shifts between the Monte Carlo truth level \pttrue{} bin centres and the
reconstructed \ptjet{} bin centres are negligible with respect to the chosen \pt{} bin widths. 
Hence the average jet response can be obtained to a good approximation as a function of \ptjet.

\index{jet selection for uncertainty calculation}

\subsection{Uncertainty in the JES calibration}
\label{sec:nonclosure}
\begin{figure}[p]
 \centering
  \subfloat[\etaRange{0.3}{0.8}]{\includegraphics[width=0.45\textwidth]{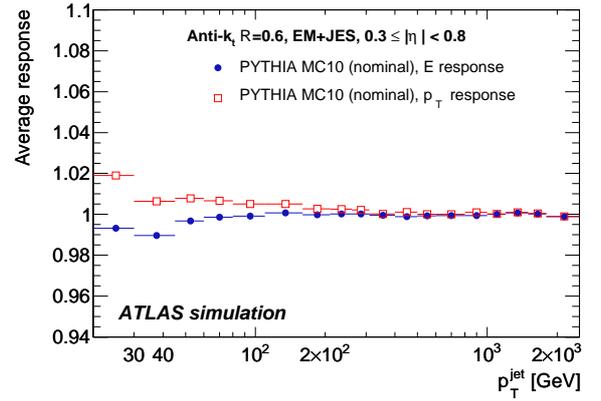}}\\
  \subfloat[\etaRange{2.1}{2.8}]{\includegraphics[width=0.45\textwidth]{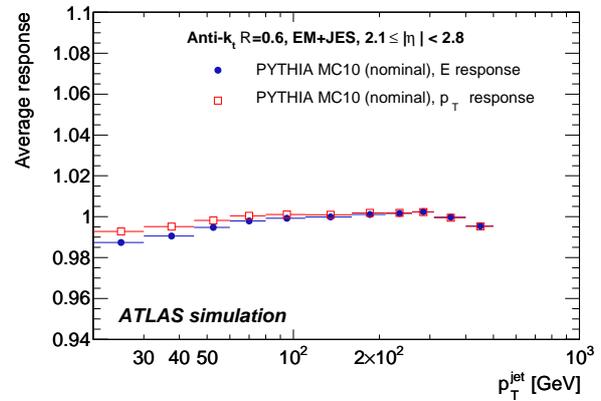}}\\
  \subfloat[\etaRange{3.6}{4.5}]{\includegraphics[width=0.45\textwidth]{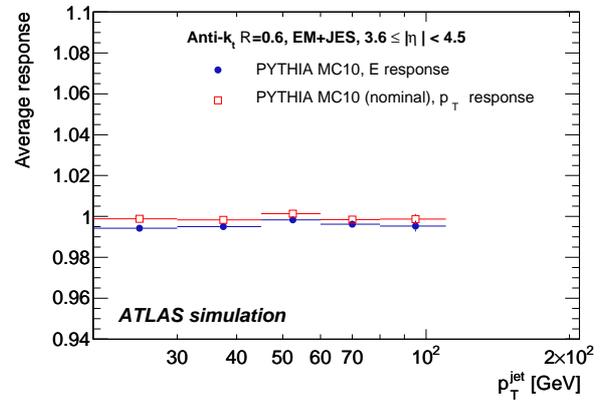}}
  \caption{Average simulated jet \pt{} response
(open squares) after the \EMJES{} calibration and jet energy response (full circles)
as a function of \ptjet{} for the nominal Monte Carlo sample for jets in the
central (a), endcap (b) and most forward (c) calorimeter regions.
Systematic uncertainties are not shown. Statistical uncertainties 
are smaller than the marker size.}
 \label{fig:NonClosure}
\end{figure}

After the jets in the nominal jet Monte Carlo simulation sample are calibrated
(see Section~\ref{sec:JetCalib}), 
the jet energy and \pt{} response still show slight deviations from unity at low \pt{} 
(non-closure). 
\index{jet calibration non-closure}
This can be seen in Figure~\ref{fig:NonClosure}, showing the jet response 
for \pt{} and energy as a function of \ptjet{} for the nominal Monte Carlo sample in the 
barrel (a) and endcap (b) and the most forward (c) regions for \antikt{} jets with $R = 0.6$.

Any deviation from unity  in the jet energy or \pt{} response after the application of the JES to the nominal Monte Carlo sample 
implies that the kinematic observables of the calibrated calorimeter jet are not restored to that of the corresponding truth jet
(non-closure). 
Besides approximations made when deriving the calibration (fit quality, parametrisation of calibration
curve), the non-closure is due to the application of the same correction factor for energy and transverse momentum.
Closure can therefore only be achieved if the reconstructed jet mass is close to the true jet mass. 
If this is not the case, such as for low \pt{} jets, restoring only the jet energy and pseudorapidity will
lead to a bias in the \pt{} calibration. The non-closure is also affected by jet resolution
and by details how the Monte Carlo samples are produced in order to cover the large kinematic
range in jet transverse momentum.

The systematic uncertainty due to the non-closure of the 
nominal JES calibration is taken as the larger deviation of the response 
in either energy or \pt{} from unity.
In the barrel region ($0.3 \leq |\eta| < 0.8$) this contribution amounts 
to about $2 \%$ at low \ptjet{} and 
less than $1 \%$ for $\ptjet{} > 30$~\GeV. In the endcap and forward regions, 
the closure is less than $1\%$ for $\ptjet{} > 20$~\GeV, 
and the energy response is within $1 \%$ for jets with transverse momentum above $30$~\GeV.
The deviation of the jet response from unity after calibration is taken as a
source of systematic uncertainty. 

For physics analysis the non-closure uncertainty only needs to be considered
when an absolute jet energy or transverse momentum is needed. 
For analyses where only the description of the data by the Monte Carlo
simulation is important, this uncertainty does not need to be considered.

\index{non-closure of jet calibration}

\begin{figure}[!htb]
  \centering
  \subfloat[Energy response]{\includegraphics[width=0.45\textwidth]{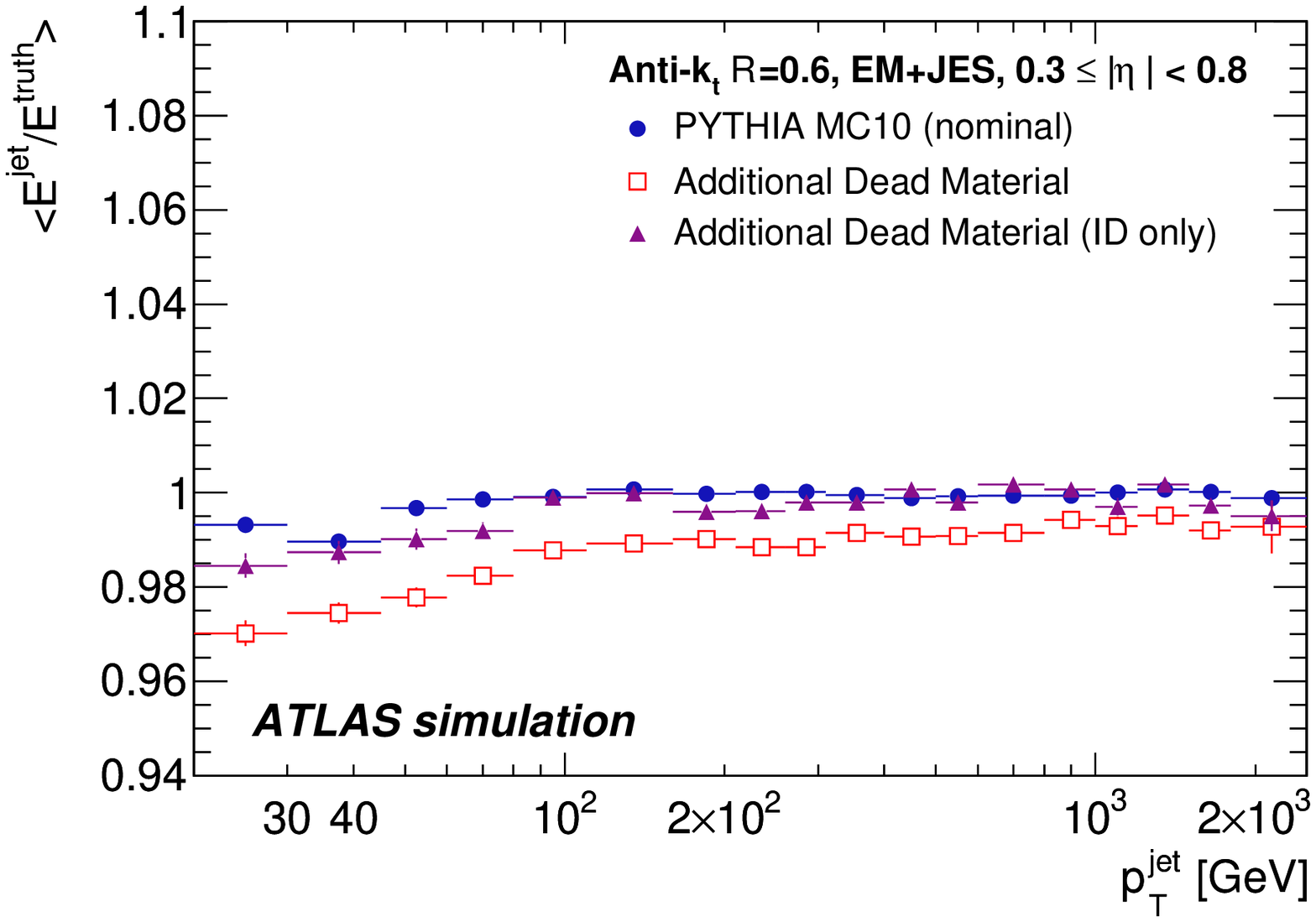}}\\
  \subfloat[Transverse momentum response]{\includegraphics[width=0.45\textwidth]{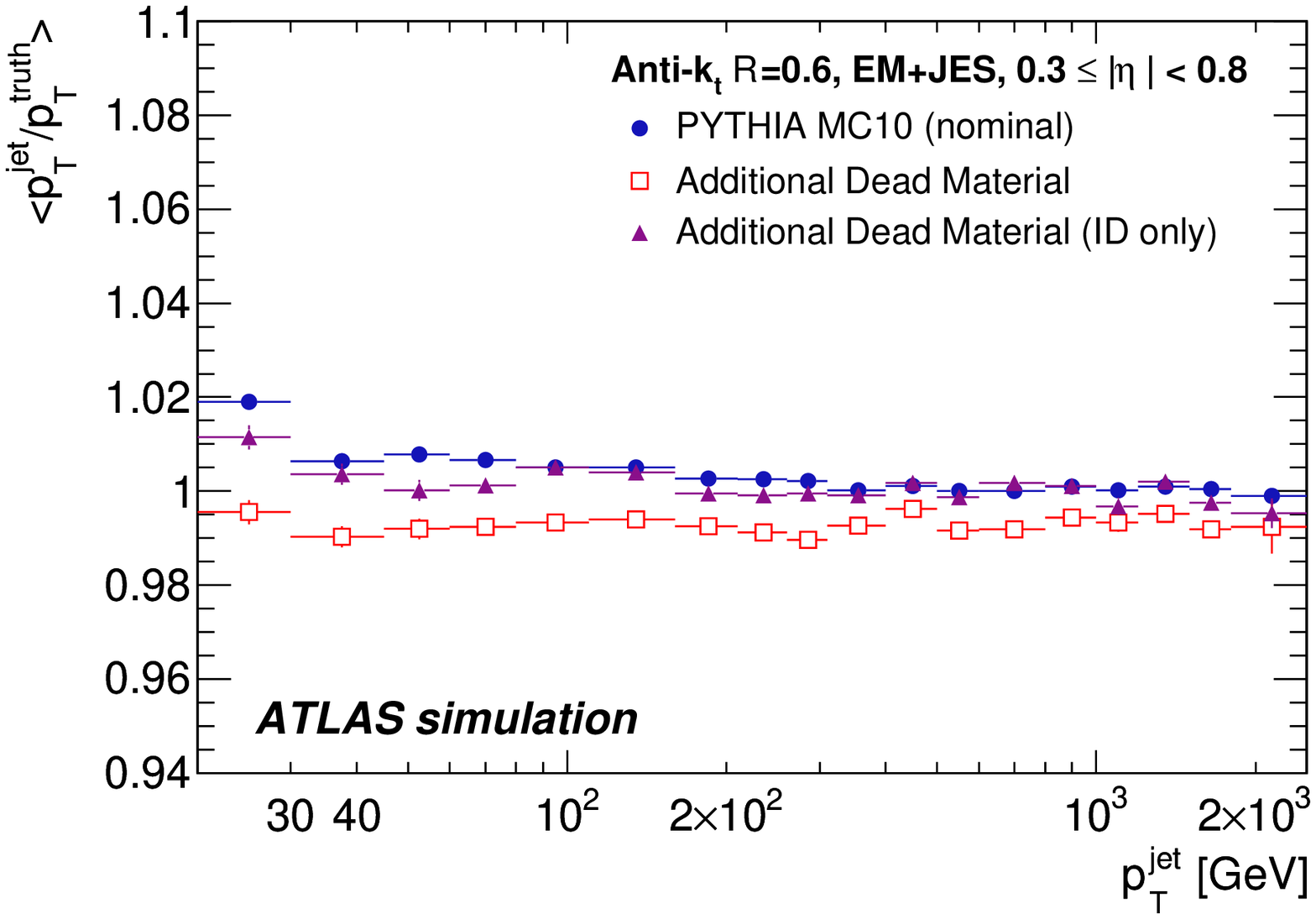}}
  \caption{Average simulated jet response in energy (a) and in \pt{} (b) 
    as a function of \ptjet{} in the central region (\central{}) in
    the case of additional dead material in the inner detector (full
    triangles) and in both the inner detector and the calorimeters
    (open squares). The response within the nominal Monte Carlo sample
    is shown for comparison (full circles).
    Only statistical uncertainties are shown.}
 \label{fig:DeadMaterial}
\end{figure}

\begin{figure}[!htb]
  \centering
  \subfloat[Energy response]{\includegraphics[width=0.45\textwidth]{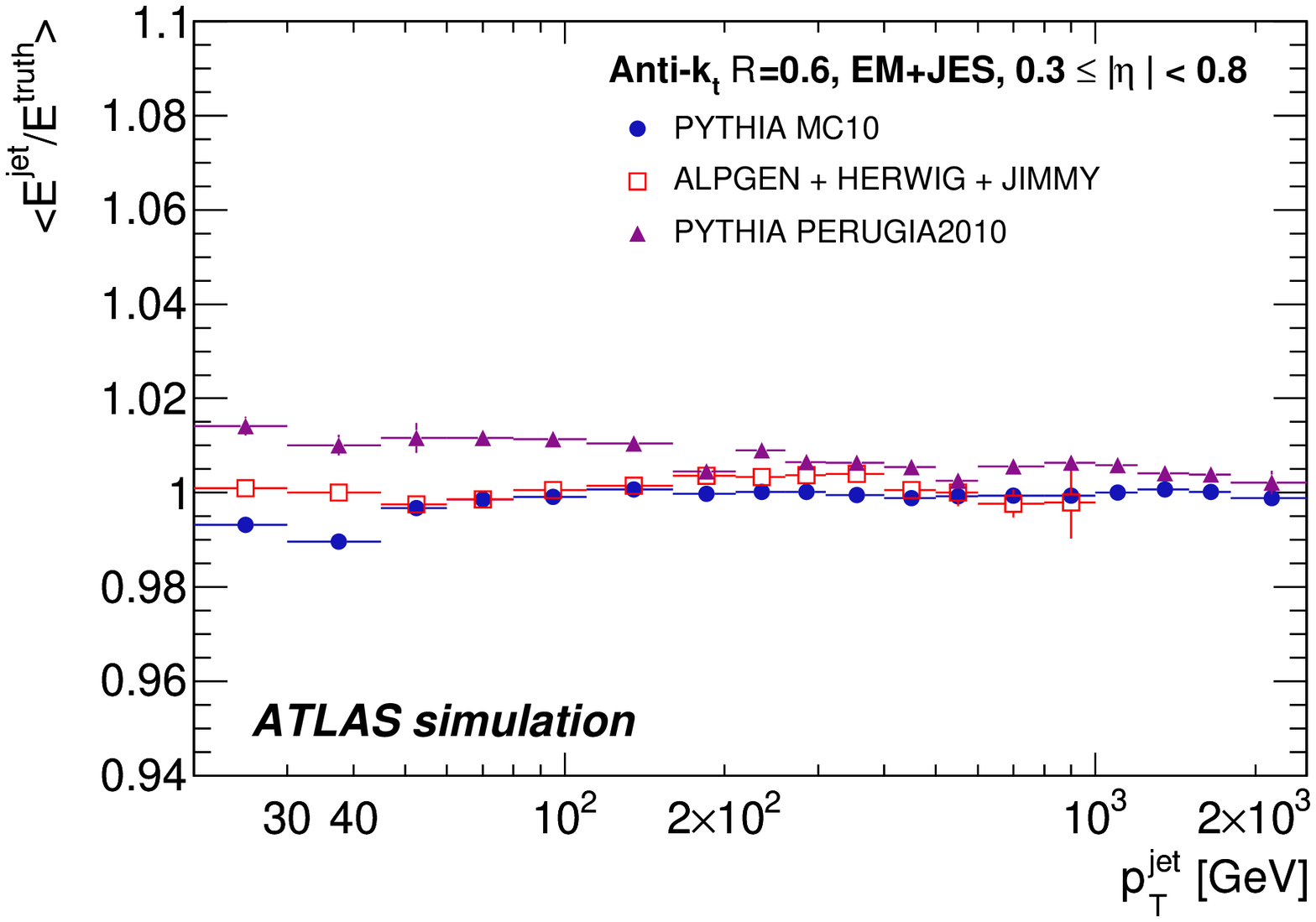}}\\
  \subfloat[Transverse  momentum response]{\includegraphics[width=0.45\textwidth]{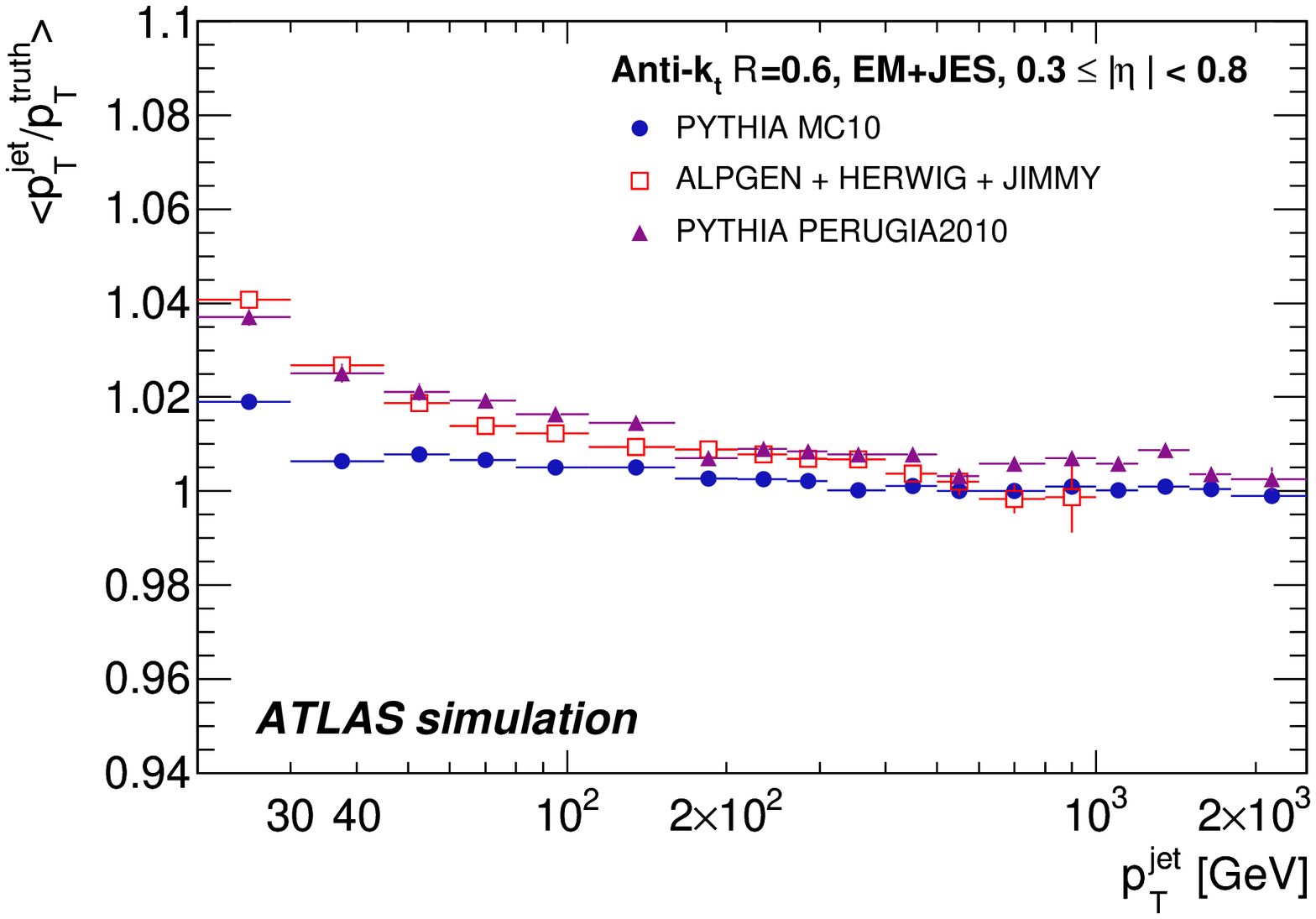}}
  \caption{
    Average simulated response in energy (a) and in \pt{} (b) as a function of \ptjet{} in the central region (\central{})
    for \alpgen+\herwig+\jimmy{} (open squares) and \pythia{} with the \Perugia2010 tune (full triangles). The
    response of the nominal Monte Carlo simulation sample is shown for comparison (full circles).
    Only statistical uncertainties are shown.}
  \label{fig:TheoResp}
\end{figure}

\subsection{Uncertainty on the calorimeter response}
\label{sec:SingleParticle}
The response and corresponding uncertainties for single particles interacting in the \ATLAS{} calorimeters 
can be used to derive the jet energy scale uncertainty in the central calorimeter region
as detailed in Ref.~\cite{singleparticle,atlassingleparticle2011}.

\index{calibration hits and particle identification}
In the \ATLAS{} simulation infrastructure the true calorimeter energy deposits in each calorimeter
cell can be traced to the particles generated in the collision. The uncertainty in the calorimeter 
response to jets can then be obtained from the response uncertainty in the individual particles constituting the jet.
The \insitu{} measurement of the single particle response detailed in Ref.~\cite{atlassingleparticle2011}
significantly reduces the uncertainty due to the limited knowledge of the exact detector geometry,
in particular that due to the presence of additional dead material, and
the modelling of the exact way particles interact in the detector.

\index{single hadron response measurement}
The following single particle response measurements are used:
\begin{enumerate}
\item The single hadron energy measured in a cone around an isolated track
  with respect to the track momentum (\EoverP) in the momentum range from $0.5 \le \ptrk < 20$~\GeV.
\item The pion response measurements performed in the $2004$ combined test-beam, 
  where a full slice of the \ATLAS{} detector was exposed to pion beams with momenta 
  between $20$~\GeV{} and $350$~\GeV~\cite{CTB04pion}. 
\end{enumerate}
Uncertainties for charged hadrons are estimated from these measurements
as detailed in Ref.~\cite{atlassingleparticle2011}.
Additional uncertainties are related to:
\begin{enumerate}
\item The calorimeter acceptance for low \pt{} particles that do not reach the calorimeter
      or are not reconstructed in a \topo{} due to the noise thresholds.
\item Calorimeter response to particles with $p > 400$~\GeV{} 
for which the uncertainty is conservatively estimated as $10 \%$, 
to account for possible calorimeter non-linearities and longitudinal leakage. 
\item The baseline absolute electromagnetic scale for the hadronic and electromagnetic calorimeters for particles 
      in the kinematic range not measured \insitu.
\item The calorimeter response to neutral hadrons is estimated by comparing various models in \geant .
      An uncertainty of $10 \%$ for particles with an energy $E< 3$~\GeV{} and $5 \%$ for higher energies is obtained.
\end{enumerate}

At high transverse momentum, the dominant contribution
to the calorimeter response uncertainties is due to particles 
with momenta covered by the test-beam.
In the pseudorapidity range $0 \leq |\eta| < 0.8$ the shift 
 of the relative jet energy scale expected from the single hadron response measurements in the test-beam
is up to $\approx 1 \%$, and the uncertainty on the shift is from $1 \%$ to $3 \%$. 
The total envelope (the shift added linearly to the uncertainty) 
of about $1.5 - 4\%$, depending on the jet transverse momentum,
is taken as the relative JES calorimeter uncertainty. 
The calorimeter uncertainty is shown in Figure~\ref{fig:FinalJES}.

\index{jet uncertainty from single particles}

\subsection{Uncertainties due to the detector simulation}
\label{sec:DetectorUnc}

\subsubsection{Calorimeter cell noise thresholds}
\label{sec:calonoise}
As described in Section~\ref{sec:topocluster}, \topos{} are constructed based
on the signal-to-noise ratio of calorimeter cells, where the {\it noise} is defined as
the RMS of the measured cell energy distribution
in events with no energy depositions from collision events.
Discrepancies between the simulated noise and the real
noise in data can lead to differences in the cluster shapes and to the
presence of fake \topos. %
For data, the noise can change over time\footnote{Time-dependent noise changes for single cells in data
are accounted for using regular measurements.},
while the noise RMS used in the simulation is fixed at the time of
the production of the simulated data sets. 
These effects  can lead to biases in the jet reconstruction and calibration, if  
the electronic noise injected in the Monte Carlo simulation does not reflect that data. 
\index{noise thresholds}

The effect of the calorimeter cell noise mis-modelling on the jet response
is estimated by reconstructing \topos{}, and thereafter jets, in Monte
Carlo using the noise RMS measured from data. The actual energy and
noise simulated in the Monte Carlo are left unchanged,
but the values of the thresholds used to include a given calorimeter
cell in a \topo{} are shifted according to the cell noise RMS measured in data.
The response for jets reconstructed with the modified noise thresholds
are compared with the response for jets reconstructed in exactly
the same sample using the default Monte Carlo noise thresholds. 

To further understand the effect of the noise thresholds on the jet response,
the noise thresholds were shifted. 
An increase of each calorimeter cell threshold by $7 \%$ in the Monte Carlo simulation
is found to give a similar shift in the jet response as using
the noise RMS from data. Raising and lowering the cell thresholds by $7 \%$
shows that the effect on the jet response
from varying the cell noise thresholds is symmetric. This allows
the use of the calorimeter cell noise thresholds derived from data as a 
representative sample to determine the jet energy scale uncertainty
and covers the cases when the data have either more or less noise than the simulation.

The maximal observed change in jet response is used to estimate the
uncertainty on the jet energy measurement due to the calorimeter cell noise modelling.
It is found to be below 3\% for the whole pseudorapidity range, and 
negligible for jets with transverse momenta above $45$~\GeV.
The uncertainties assigned to jets with transverse momenta below $45$~\GeV{} are:
\begin{itemize}
\item $1 \%$ and $2 \%$ for \ptRange{20}{30} for \antikt{} jets with $R = 0.4$ and $R = 0.6$ jets, respectively,
\item $1 \%$ for \ptRange{30}{45} for both $R$ values.
\end{itemize}

\subsubsection{Additional detector material}
\label{sec:dmmaterial}
\index{JES uncertainty due to dead material}
The jet energy scale is affected by possible deviations in the material description as
the jet energy scale calibration has been derived to restore
the energy lost assuming a geometry as simulated in the nominal Monte Carlo sample. 
Simulated detector geometries that include systematic variations
of the amount of material have been designed using test-beam measurements~\cite{ctb2004electrons},
in addition to $900$~\GeV{} and $7$~\TeV{} data~\cite{PhotonConversions900, MaterialBudgetMinBias7, MaterialBudgetKaons900, MaterialBudgetSecondary7}. The possible additional material amount is estimated from these \insitu{} measurements
and the {\it a priori} knowledge of the detector construction.
Specific Monte Carlo simulation samples have been produced using these distorted geometries.

In the case of uncertainties derived with \insitu{} techniques, 
such as those coming from the single hadron response measurements detailed in Section~\ref{sec:SingleParticle},
most of the effects on the jet response due to additional dead material 
do not apply, because \insitu{} measurements do not rely on any simulation where the material
could be misrepresented.
However, the quality criteria of the track selection for the single hadron response measurement
effectively only allow particles that have not interacted in the \Pixel{} and \SCT{} layers of the inner detector to be
included in the measurement.

Therefore the effect of dead material in these inner detector layers on the jet response
needs to be taken into account for particles in the momentum range of the single hadron response
measurement.
This is achieved using a specific Monte Carlo sample where the amount of material is systematically 
varied by adding $5 \%$ of material to the existing inner detector services \cite{Atlaselectronpaper}. 
The jet response in the two cases is shown in Figure~\ref{fig:DeadMaterial}.

Electrons, photons, and hadrons with momenta $p > 20$~\GeV{} are not included in the single hadron response 
measurements and therefore there is no estimate based on \insitu{} techniques
for the effect of any additional material in front of the calorimeters. 
This uncertainty is estimated using a dedicated Monte Carlo simulation sample 
where the overall detector material is systematically varied within the current 
uncertainties\cite{Atlaselectronpaper} on the detector geometry. 
The overall changes in the detector geometry include:
\begin{enumerate}
\item The increase in the inner detector material mentioned above.
\item An extra $0.1$ radiation length (\radlength{}) in the cryostat in front of the barrel of the electromagnetic calorimeter ($|\eta| < 1.5$).
\item An extra $0.05$ \radlength{} between the presampler and the first layer of the electromagnetic calorimeter.
\item An extra $0.1$ \radlength{} in the cryostat after the barrel of the electromagnetic calorimeter.
\item Extra material in the barrel-endcap transition region in the electromagnetic calorimeter ($1.37 < |\eta| < 1.52$).
\end{enumerate}

The uncertainty contribution due to the overall additional detector material is estimated 
by comparing the \EMJES{} jet response in the nominal Monte Carlo simulation
sample with the jet response in a Monte Carlo simulation sample with a distorted geometry 
(see Figure~\ref{fig:DeadMaterial}), and scaled by the average energy fraction of electrons, 
photons and high transverse momentum hadrons within a jet as a function of \pt.

\index{jet uncertainty detector simulation}

\subsection{Uncertainties due to the event modelling in Monte Carlo generators}
\label{sec:MCUnc}
The contributions to the JES uncertainty from the modelling of the fragmentation, the underlying event and 
other choices in the event modelling of the Monte Carlo event generator are obtained from samples based on
\alpgen+\herwig+\jimmy{} and the \pythia{} \Perugia2010 tune discussed in Section~\ref{sec:MC}.
\index{jet uncertainty Monte Carlo modelling}

By comparing the baseline \pythia{} Monte Carlo sample to the \pythia{} \Perugia2010 tune,
the effects of soft physics modelling  are tested. The \Perugia2010 tune
provides, in particular, a better description of the internal jet structure recently measured
with \ATLAS~\cite{Atlasjetshape}.
The \alpgen{}  Monte Carlo uses different theoretical models for all steps of the
event generation and therefore gives a reasonable estimate 
of the systematic variations. However, the possible
compensation of modelling effects that shift the jet response in opposite
directions cannot be excluded.

Figure~\ref{fig:TheoResp} shows the calibrated jet kinematic response for the two Monte Carlo generators and tunes used to
estimate the effect of the Monte Carlo theoretical model on the jet energy scale uncertainty.
The kinematic response for the nominal sample is shown for comparison. The ratio of the nominal response to 
that for each of the two samples is used to estimate 
the systematic uncertainty to the jet energy scale, and the procedure is further detailed in Section~\ref{sec:JESSummary}.

\index{jet uncertainty from Monte Carlo modelling}

\subsection{\Insitu{} intercalibration using events with dijet topologies}
\label{sec:etaintercalibration}
The response of the \ATLAS{} calorimeters to jets depends on the
jet direction, due to the different calorimeter technology and to the
varying amounts of dead material in front of the calorimeters.
A calibration is therefore needed to ensure a uniform 
calorimeter response to jets.  This can be achieved by applying
correction factors derived from Monte Carlo simulations. Such
corrections need to be validated \insitu{} 
given the non-compensating nature of the calorimeters in conjunction with the complex
 calorimeter geometry and material distribution.
  
The relative jet calorimeter response and its uncertainty
is studied by comparing the transverse momenta of a well-calib\-rated central jet and a jet
in the forward region in events with only two jets at 
high transverse momenta (dijets).
Such techniques have been applied in previous hadron collider
experiments\cite{ref:D0_MPF,cdf06}.

\subsubsection{Intercalibration method using a fixed central reference region}
\index{Central reference region method for intercalibration}
The traditional approach for $\eta$-intercalibration with dijet events is
to use a fixed central region of the calorimeters as the reference region. 
The relative calorimeter response to  jets in other calorimeter regions 
is then quantified by the \pt{} balance between the reference jet and the probe jet,
exploiting the fact that these jets are expected to have equal \pt{} 
due to transverse momentum conservation. 
The \pt{} balance can be characterised by the asymmetry \asym{},
defined as
\begin{equation}
  \label{eq:asym}
  \asym = \frac{\ptprobe - \ptref}{\ptavg},
\end{equation}
with $\ptavg = (\ptprobe + \ptref)/2$. The reference region is
chosen as the central region of the barrel: $|\eta|<0.8$. 
If both jets fall into the reference region, each jet is used, in
turn, as the reference jet.
As a consequence, the average asymmetry in the reference region will
be zero by construction.

The asymmetry is then used to measure an $\eta$-intercalibration factor $c$ for the probe jet, or its
response relative to the reference jet $1/c$, using
the relation
\begin{equation}
  \label{eq:rr}
  \frac{\ptprobe}{\ptref}
  =\frac{2+\asym}{2-\asym}
  =1/c.
\end{equation}

The asymmetry distribution is calculated in bins of jet $\etaDet$ and \ptavg:
The bins are labeled $i$ for each probe jet $\etaDet$ and $k$ for each \ptavg{}-bin. 
Intercalibration factors are calculated for each bin according to
Equation~(\ref{eq:rr}):
\begin{equation}
  \label{eq:c}
  c_{ik}=\frac{2-\left<\asym_{ik}\right>}{2+\left<\asym_{ik}\right>}, 
\end{equation}
where the $\left<\asym_{ik}\right>$ is the mean value of the asymmetry
distribution in each bin. The uncertainty on $\left<\asym_{ik}\right>$
is taken to be the RMS/$\sqrt{N}$ of each distribution,
where $N$ is the number of events per bin.

\subsubsection{Intercalibration using the matrix method}
\index{matrix method for eta intercalibration}
A disadvantage with the method outlined above is that all events are
required to have a jet in the central reference region. This results in a
significant loss of event statistics, especially in the forward region, where 
the dijet cross section drops steeply as the rapidity interval between the
jets increases. 
In order to use the full event statistics, the default method can be extended
by replacing the ``probe'' and ``reference'' jets by ``left'' and ``right'' jets
defined as $\eta^{\rm left}<\eta^{\rm right}$.
Equations~(\ref{eq:asym}) and~(\ref{eq:rr}) then become:
\begin{eqnarray}
  \label{eq:frederik}
  \asym = \frac{\ptl - \ptr}{\ptavg}   \; & \mathrm{and} & 
  \mathcal{R}_{l r}=
  \frac{\ptl}{\ptr}=
  \frac{c^{\rm right}}{c^{\rm left}}=
  \frac{2+\asym}{2-\asym},
\end{eqnarray}
where the term $\mathcal{R_{l r}}$ denotes the ratio of the responses,
and $c^{\rm left}$ and $c^{\rm right}$ are the $\eta$-intercalibration factors
for the left and right jets, respectively. 

In this approach there is a response ratio distribution,
$\mathcal{R}_{ijk}$, whose average value $\left < \mathcal{R}_{ijk} \right >$ is
evaluated for each $\eta^{\rm left}$-bin~$i$, $\eta^{\rm right}$-bin~$j$ and
\ptavg{}-bin~$k$. 
The relative correction factor $c_{ik}$ for a given jet $\eta$-bin $i$ and for a fixed
$\ptavg$-bin $k$, is obtained by minimising a matrix of linear equations:
\begin{eqnarray}
  S(c_{1k},...,c_{Nk}) = \hspace{4.5cm}  \nonumber \\
  \sum_{j=1}^{N}
  \sum_{i=1}^{j-1}
  \left(\frac{1}{\Delta \left< \mathcal{R}_{ijk} \right> }\left(c_{ik}
    \left< \mathcal{R}_{ijk} \right> -c_{jk}\right)\right)^2
  + X(c_{ik}),
  \label{eq:MM}
\end{eqnarray}
where $N$ denotes the number of $\eta$-bins, 
$\Delta \left< \mathcal{R}_{i j k} \right>$ is the statistical uncertainty of
$\left< \mathcal{R}_{i j k} \right>$  and the function $X(c_{ik})$ is used to quadratically
suppress deviations from unity of the average
corrections\footnote{$X(c_{ik})=K\left( N_{\rm bins}^{-1}
    \sum_{i=1}^{N_{\rm bins}} c_{ik}
    - 1\right)^{2}$ is defined with $K$ being a constant and $N_{\rm bins}$ being the number
of $\eta$-bins (number of indices~$i$). This term prevents the
minimisation from choosing the trivial solution: all $c_{ik}$ equal to zero. 
The value of the constant $K$ does not impact the solution as
long as it is sufficiently large ($K \approx 10^6$).}.
Note that if the jet response does not vary with $\etajet$, then the relative response will be unity 
for each $(\eta_{\rm left},\eta_{\rm right})$-bin combination (see Equation~\ref{eq:frederik}).
A perfect minimization $S=0$ is achieved when all correction factors equal unity.

The minimisation of Equation~\ref{eq:MM} is done separately for each
\ptavg-bin~$k$, and the resulting calibration factors $c_{ik}$ (for each
jet $\eta$-bin $i$) are scaled such that the average calibration
factor in the reference region $|\eta|<0.8$ equals unity.

\begin{figure}[htp!]
\centering
 \subfloat[$30 \leq \ptavg< 40$~\GeV{}]{\includegraphics[width=0.48\textwidth]{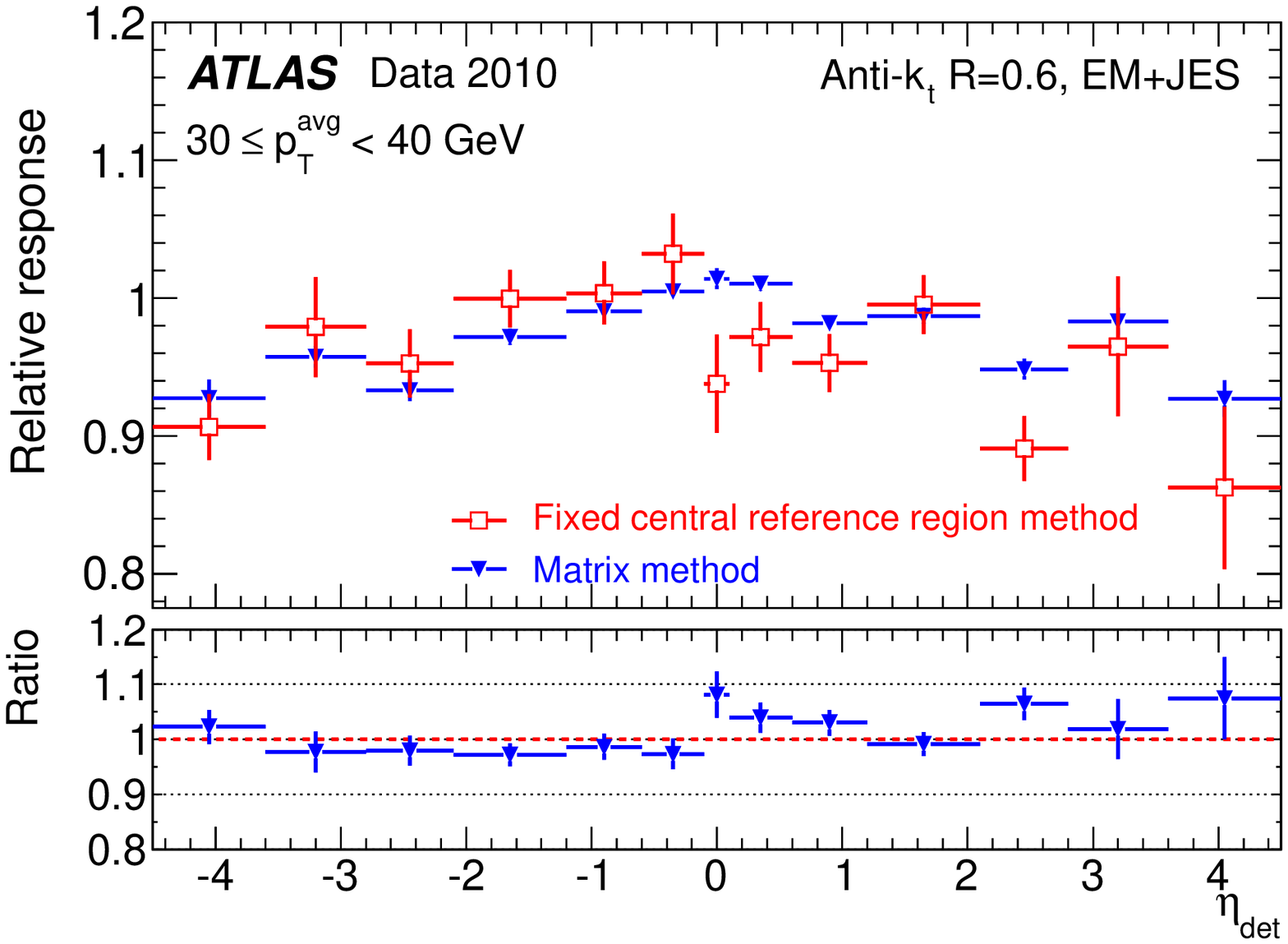}}\quad
 \subfloat[$60 \leq \ptavg< 80$~\GeV{}]{\includegraphics[width=0.48\textwidth]{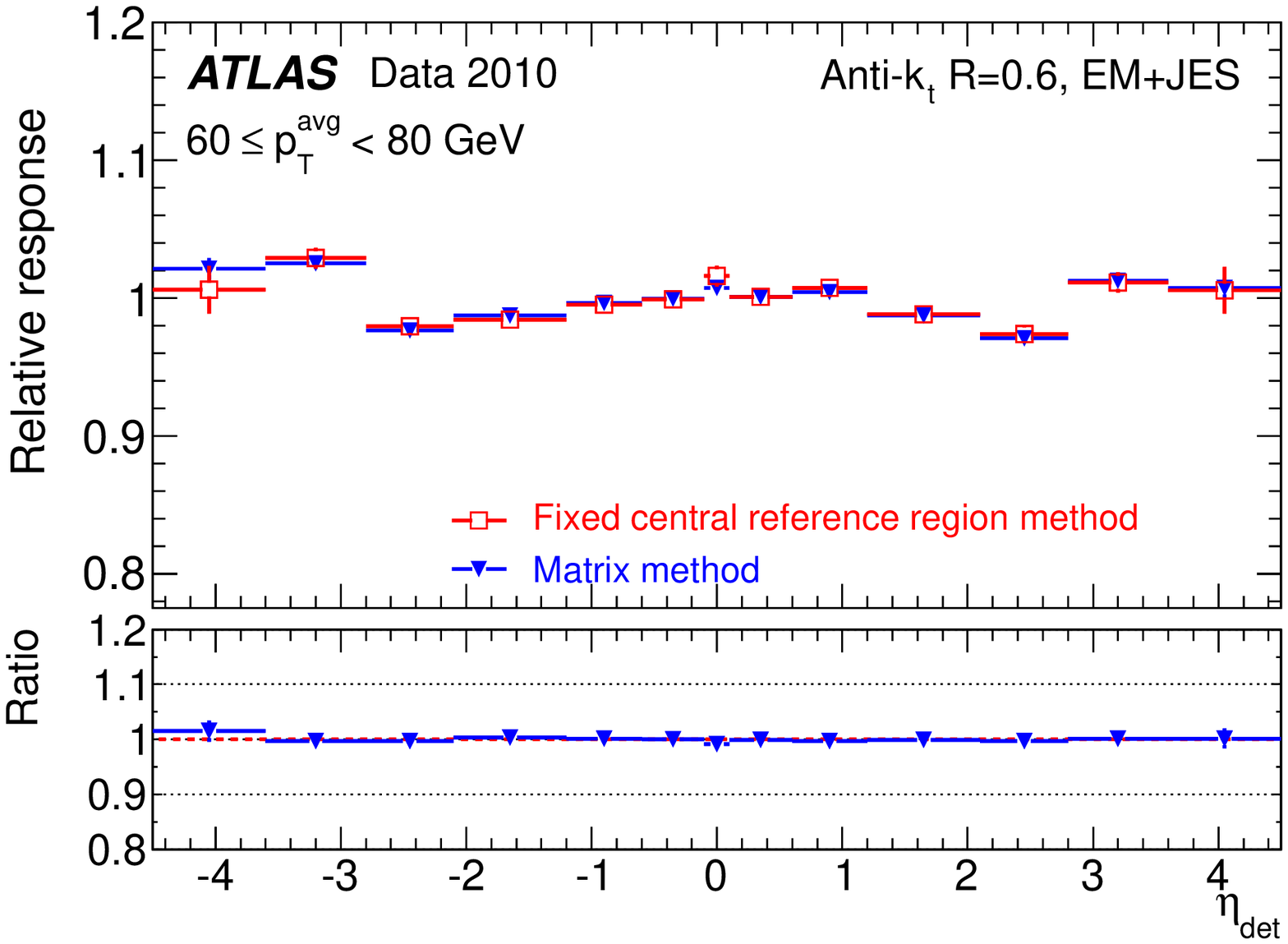}}
 \caption{Relative response of \antikt{} jets with $R = 0.6$ calibrated with the \EMJES{} scheme, $1/c$, 
  as a function of the pseudorapidity measured 
  using the matrix and fixed central reference region $\etajet$-intercalibration methods. 
  Results are presented for two bins of \ptavg{}: 
  $30 \leq \ptavg{} < 40$~\GeV{} measured in minimum bias data (a), and 
  $60 \leq \ptavg{} < 80$~\GeV{} measured in data collected using jet triggers (b).
  The lower part of the figures shows the ratio of the two methods.
  The central reference region is $0.1 \leq |\etajet|<0.6$. 
  Only statistical uncertainties are shown.
  \label{fig:comp} }
\end{figure}

\subsubsection{Selection of dijet events}
\label{sec:evsel}
\label{sec:etaintercalibrationselection}
\index{jet selection intercalibration}
Events are retained if there were at least two jets above the
jet reconstruction threshold of $\ptjet > 7$~\GeV. 
The event is rejected if either of the two leading jets did not satisfy 
the standard jet selection criteria (see Section~\ref{sec:JetSelection}). 

Events are required to satisfy a specific 
logic using one central and one forward jet trigger, 
which select events based on jet activity in the 
central ($|\eta|<3.2$) and forward ($|\eta|>3.2$) trigger regions, 
respectively \cite{triggerperformance}. 
The requirements are chosen such that the trigger efficiency, 
for a specific region of $\ptavg{}$, was greater than $99\%$ and approximately flat 
as a function of the pseudorapidity of the probe jet. 

To cover the region $\ptavg{}<45$~\GeV, events triggered by the minimum bias trigger scintillators
were used. 
To enhance events which have only two jets at high \pt, the following selection criteria 
are applied;
\begin{eqnarray}
  \ptavg>20~\GeV, \qquad 
  \Delta\phi({\rm j_{1}},{\rm j_{2}}) > 2.6~{\rm rad}, \qquad  \\ 
  p_{\rm T}^{}({\rm j_{3}}) < {\rm max}(0.15 \, \ptavg{}, 7~\GeV), \qquad \qquad
\end{eqnarray}
where $\mathrm{j}_{i}$ denotes the $i^{\rm th}$ highest \pt{} jet in the
event and $\Delta\phi({\rm j_{1}},{\rm j_{2}})$ is the azimuthal angle 
between the two leading jets. 

The lowest \ptavg{}-bins are likely to suffer from biases. 
At very low \ptavg, it is expected that this technique may not measure accurately the relative
response to jets, because the assumption of dijet balance at hadron level may start to fail.
First, there are residual low-$\pt$ jet effects since the selection criterion 
on the third jet, which is used to suppress the unbalancing effects of soft QCD radiation, 
is not as efficient due to the jet reconstruction threshold of $7$~\GeV.
Second, the jet reconstruction efficiency is worse for low-\pt{} jets. 
%

%
\begin{figure*}[htp!]
  \centering
  \subfloat[$20 \leq \ptavg < 30$~\GeV] {\includegraphics[width=0.48\textwidth]{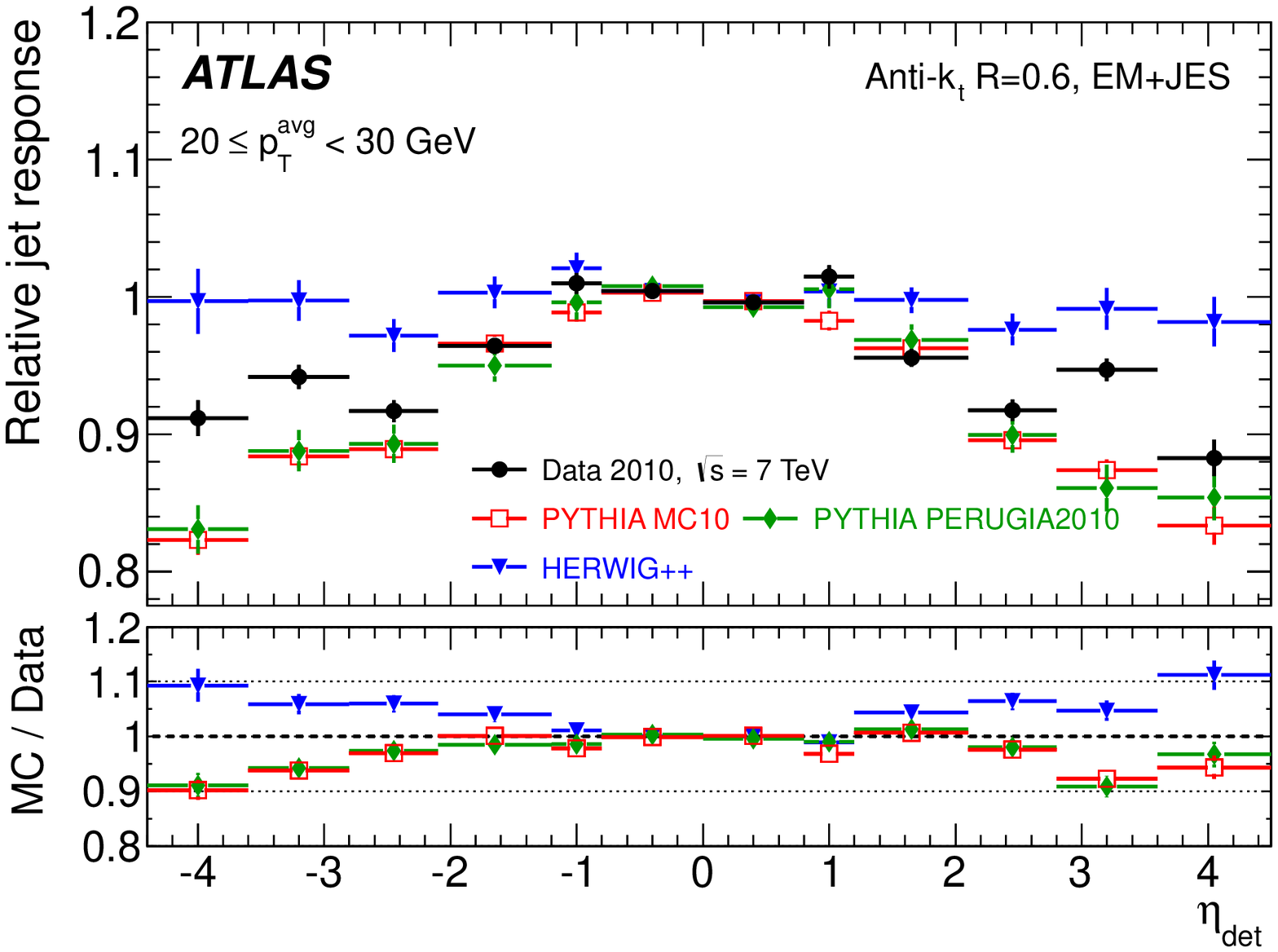}}\quad
  \subfloat[$30 \leq \ptavg < 45$~\GeV] {\includegraphics[width=0.48\textwidth]{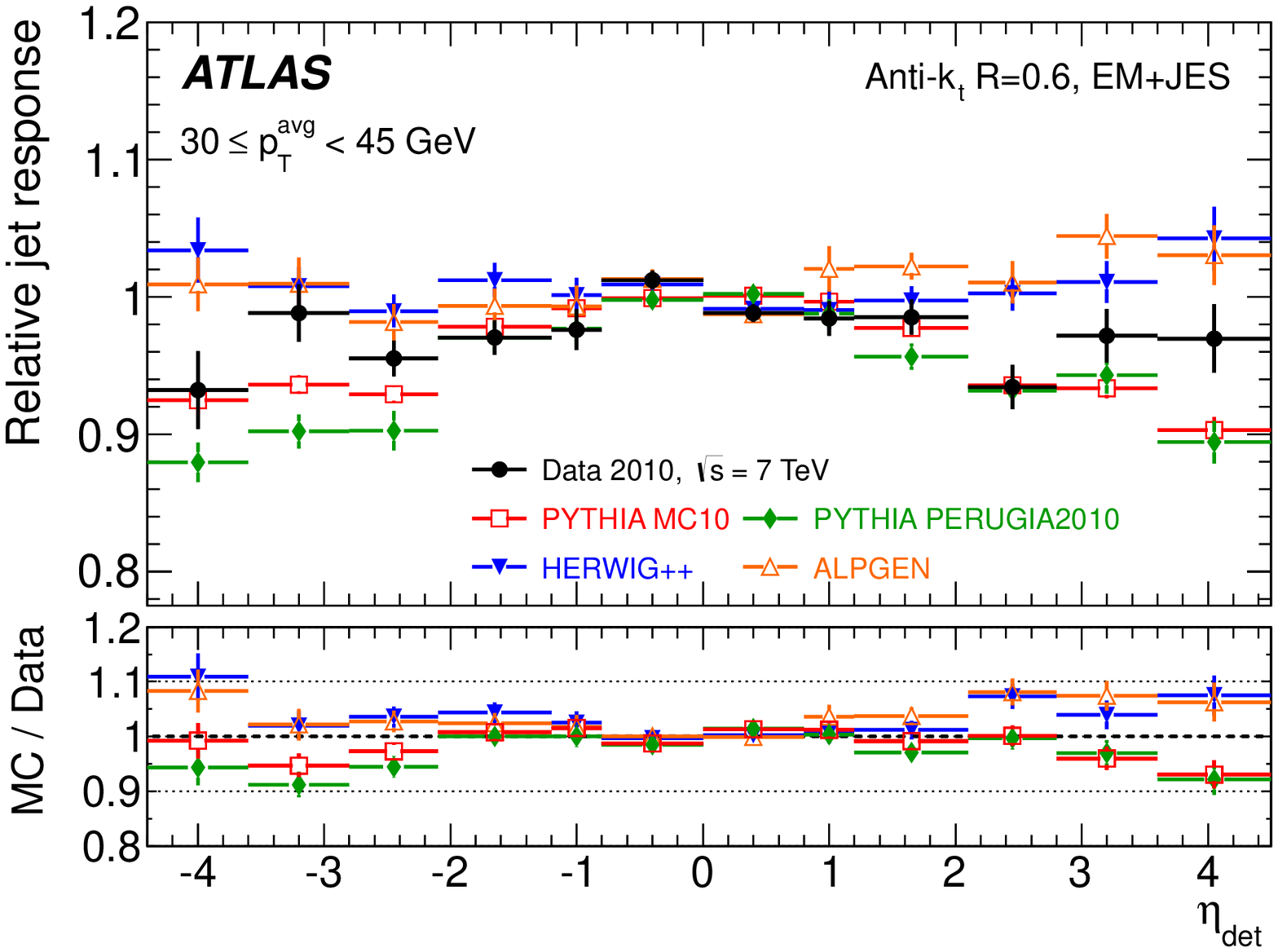}} \\
  \subfloat[$60 \leq \ptavg < 80$~\GeV] {\includegraphics[width=0.48\textwidth]{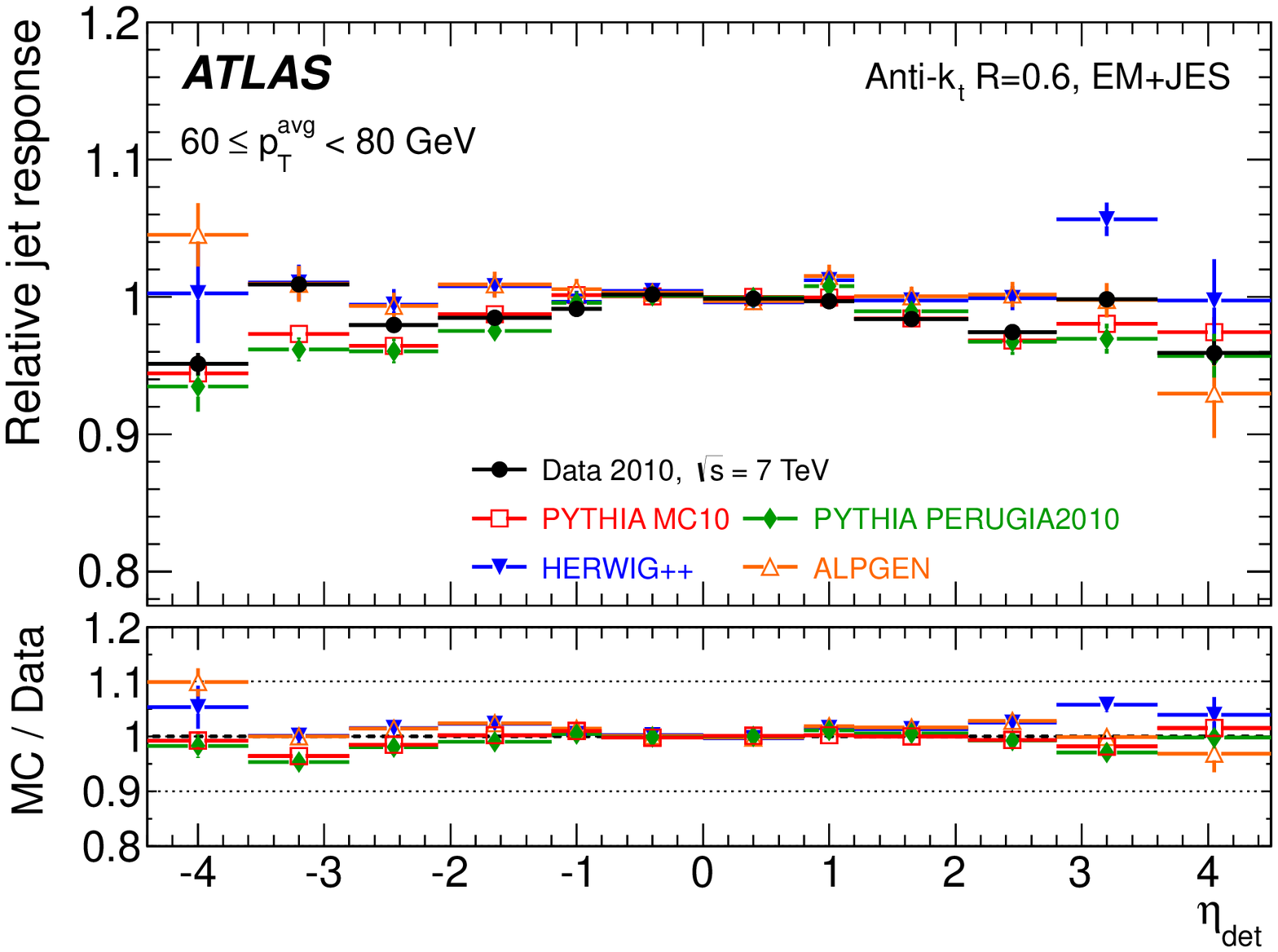}}\quad
  \subfloat[$80 \leq \ptavg < 110$~\GeV]{\includegraphics[width=0.48\textwidth]{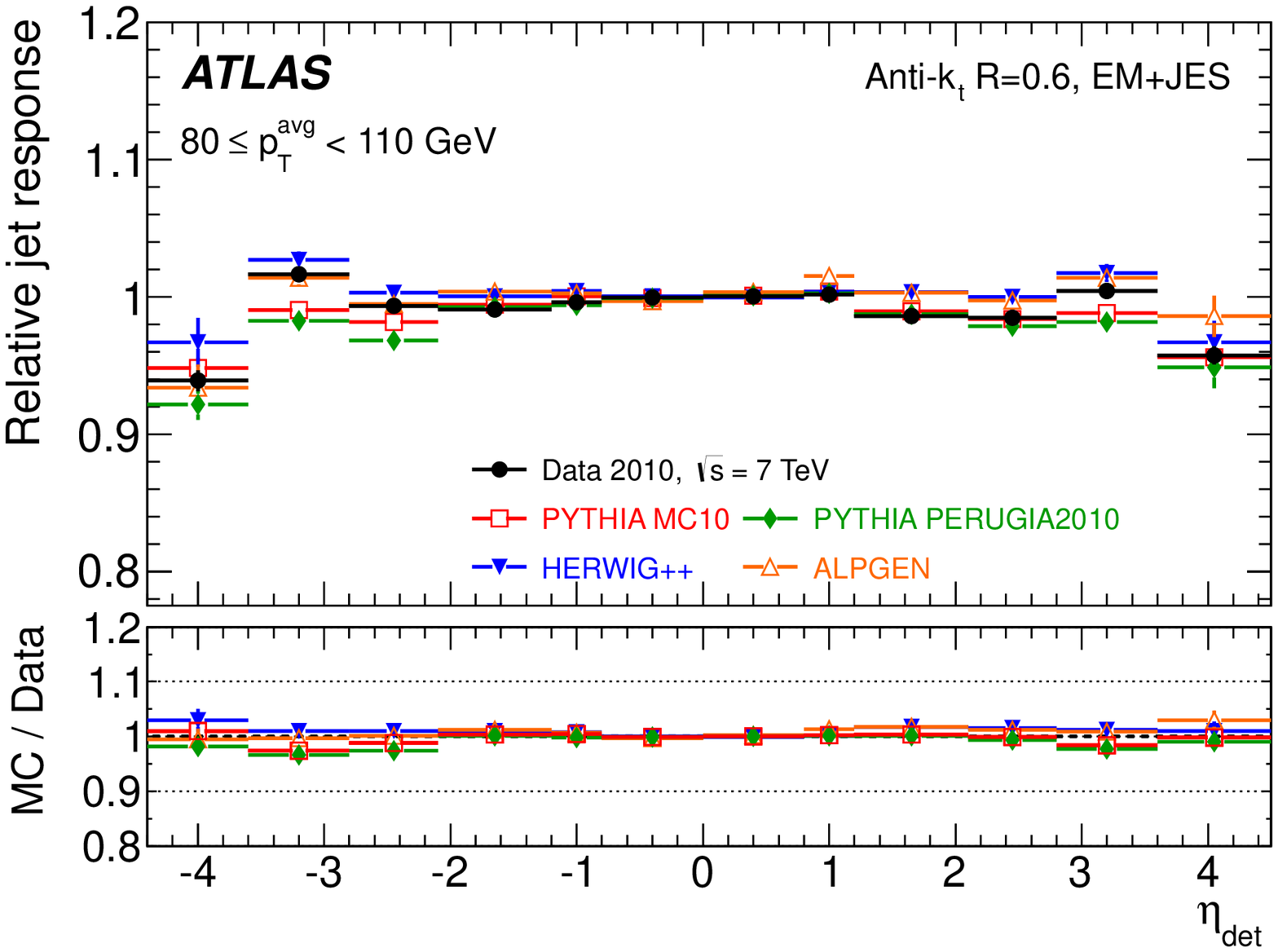}}
  \caption{
    Relative jet response, $1/c$, of \antikt{} jets with $R = 0.6$  as a function of the jet 
    pseudorapidity measured using the matrix $\eta$-intercalibration method in bins of the average \pt{}
    of the two leading jets 
    (a) $20 \leq \ptavg < 30$~\GeV, (b)  $30 \leq \ptavg < 45$~\GeV, 
    (c) $60 \leq \ptavg < 80$~\GeV{} and $80 \leq \ptavg < 110$~\GeV.
    The lower part of each figure shows the ratio of Monte Carlo simulation to data. 
    Only statistical uncertainties are shown.  
    \label{fig:response_dataMCeta}}
\end{figure*}
%
%
\begin{figure*}[htp!]
  \centering
  \subfloat[$1.2 \leq |\eta| < 2.1$]{ \includegraphics[width=.48\textwidth]{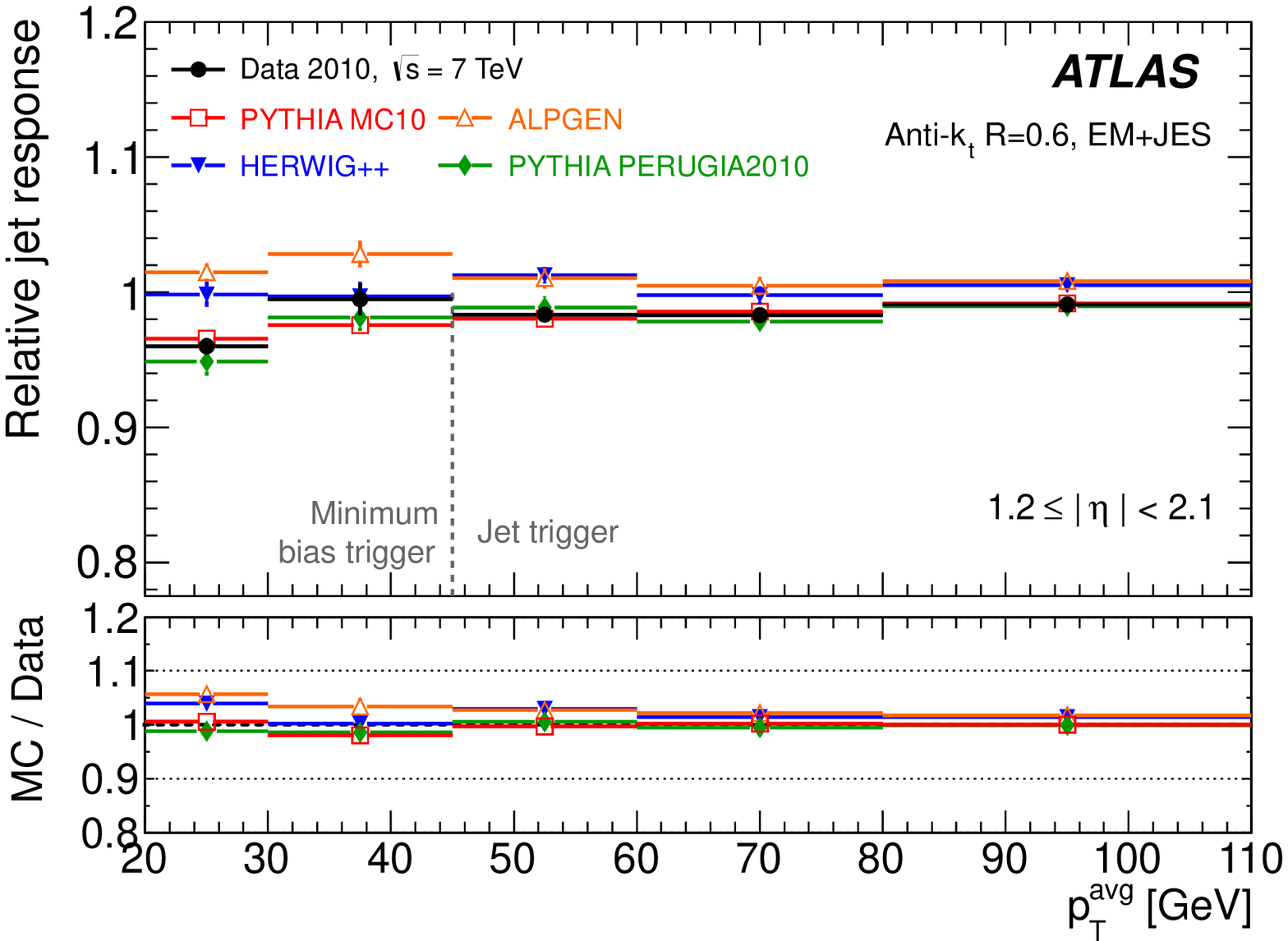}}\quad
  \subfloat[$3.6 \leq |\eta| < 4.5$]{ \includegraphics[width=.48\textwidth]{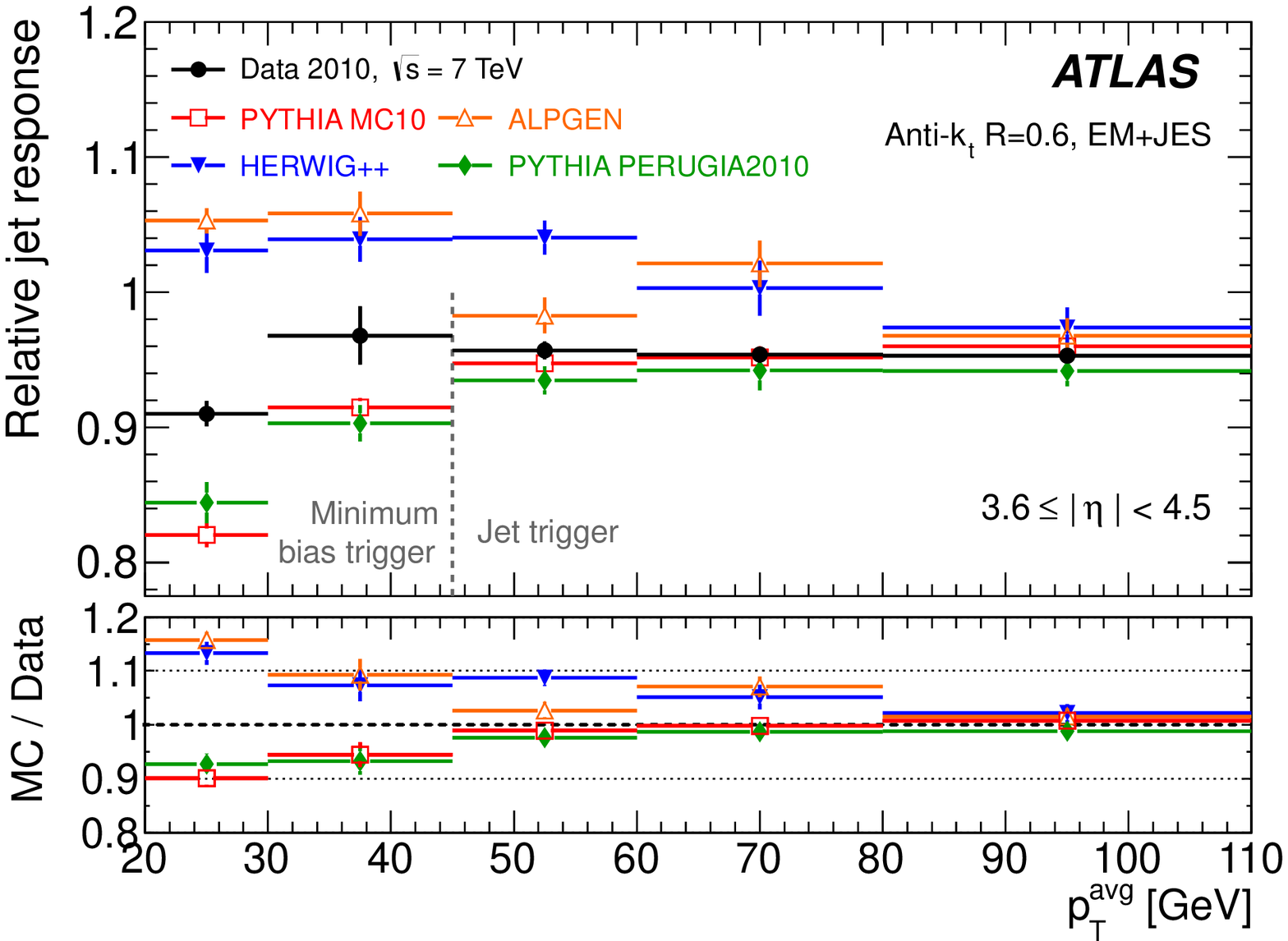}}
  \caption{
    Relative jet response, $1/c$, of \antikt{} jets with $R = 0.6$  
    as a function of \ptavg{} found using the 
    matrix $\eta$-intercalibration  method for (a)  $1.2 \leq |\eta| < 2.1$  and (b) $3.6 \leq |\eta| <4.5$. 
    For $\ptavg{}<45$~\GeV, the data are collected using the minimum bias trigger stream.
    For $\ptavg{}>45$~\GeV, the data are collected using the calorimeter trigger stream.
    The lower part of each figure shows the ratio of Monte Carlo simulation to data.  
    Only statistical uncertainties are shown.  
    \label{fig:response_dataMCpt}}
\end{figure*}


\subsubsection{Comparison of intercalibration methods}
\label{sec:res}
The relative jet response obtained with the matrix
meth\-od is compared to the relative jet response obtained using the
method with a fixed reference region. Figure~\ref{fig:comp}
shows the jet response relative to central jets ($1/c$) 
for two \ptavg{}-bins,  $30 \le \ptavg{} < 40$~\GeV{} and $60 \le \ptavg{} < 80$~\GeV. 
These results are obtained for a reference region  $0.1 \le |\eta|<0.6$ and therefore
not directly comparable to the results discussed below where $0.1 \le |\eta|<0.8$ is used.

The response observed using the fixed reference region meth\-od is compatible with those obtained using the matrix 
meth\-od\footnote{As discussed in Section~\ref{sec:evsel}, even for an ideal detector the asymmetry,
and hence the relative response, is not expected to be exactly flat due to the effects
 of soft QCD radiation and other soft particle activities.}. 
These results are representative of all the phase 
space regions studied in this analysis and the matrix method 
is therefore used to give the final uncertainty on the \insitu{} $\eta$-intercalibration 
due to its higher statistical precision.

\subsubsection{Comparison of data with Monte Carlo simulation}
Figure~\ref{fig:response_dataMCeta} shows the relative response
obtained with the matrix method as a function of the jet pseudorapidity for data 
and Monte Carlo simulations in four \ptavg{} regions.

The response in data is
reasonably well reproduced by the Monte Carlo simulations for
$\ptjet > 60$~\GeV, with the Monte Carlo simulation and data agreeing typically better than
$2 \%$ in the central region ($|\etajet| < 2.8$) and  $5 - 10 \%$ (depending on \ptavg)
in the forward region ($|\etajet|>2.8$). 
At lower values of \pt, the
data do not agree as well with the Monte Carlo simulations and the Monte Carlo
simulations themselves show a large spread around the data. For
$20 \le \ptavg{} < 30$~\GeV, the Monte Carlo simulation deviates from the data by about $10 \%$ for
$|\eta| > 2.8$, with the different Monte Carlo simulations predicting both higher
and lower relative responses than that observed in the data. 

The main differences, due to residual low-$\pt$ jet effects 
(see Section~\ref{sec:evsel}), occur between \pythia{} with 
the MC10 or the \Perugia{} tune on one side and \alpgen/\herwigpp{} on the other. 
The differences therefore apparently reflect a difference in physics modelling  between the 
event generators.

Figure~\ref{fig:response_dataMCpt} shows the relative response as a function of \ptavg{}.
The distributions are shown for jets  in the region $1.2 \le |\etajet| < 2.1$ 
and also for those in the region $3.6 \le |\etajet| < 4.5$. 
Again, the response is reasonably well described by the Monte Carlo simulation 
for all calorimeter regions at high \pt{} and the more central region at low \pt{}.

\subsubsection{Total uncertainties in the forward region}
\label{sec:EtaIntercalibUnc}
The Monte Carlo simulation predictions for the relative jet response diverge at low values of $\ptavg$. 
The data themselves lie between the different predictions.
The uncertainty on the relative jet response must reflect this disagreement 
because there is no {\it a priori} reason to believe one theoretical prediction 
over another. 

The uncertainty on the relative response is taken to be the RMS deviation 
of the Monte Carlo predictions from the data. 
At high $\pt{}$, where the spread of Monte Carlo simulation predictions is small, 
the uncertainty mainly reflects the true difference between the response in data 
and simulation. At low $\pt{}$ and large $|\eta|$, the uncertainty mainly reflects 
the physics modelling uncertainty, although the detector-based differences between 
data and simulation are also accounted for. 
Other uncertainty sources, such as trigger selection or the QCD radiation suppression using the third jet,
are either negligible, or included in the total uncertainty assigned
from the spread of Monte Carlo predictions around the data.
\index{jet calibration $\eta$-intercalibration uncertainty}

\begin{figure*}[ht!]
\centering
\subfloat[Transverse momentum]{ \includegraphics[width=0.48\textwidth]{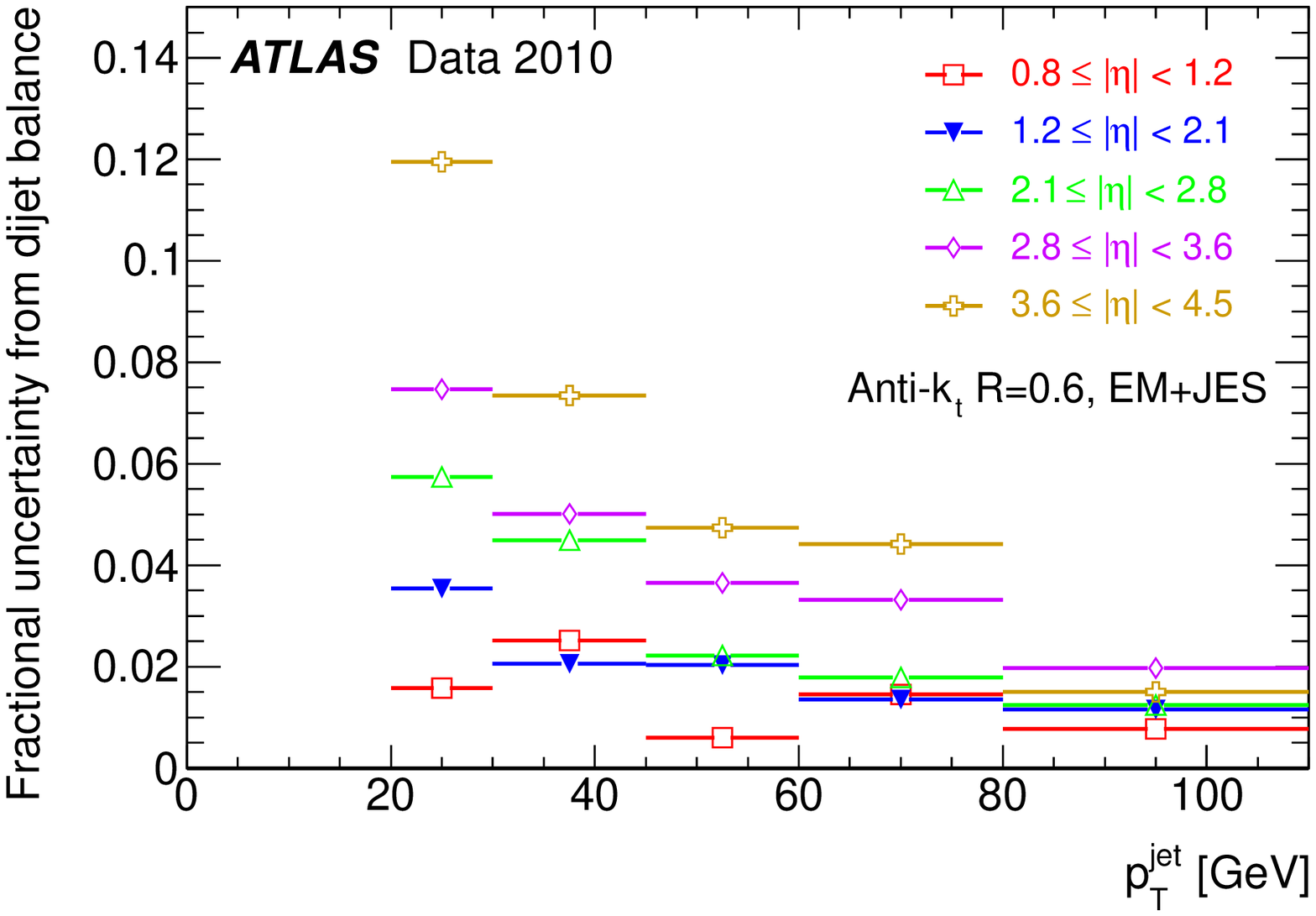}}\quad
\subfloat[Pseudorapidity]     { \includegraphics[width=0.48\textwidth]{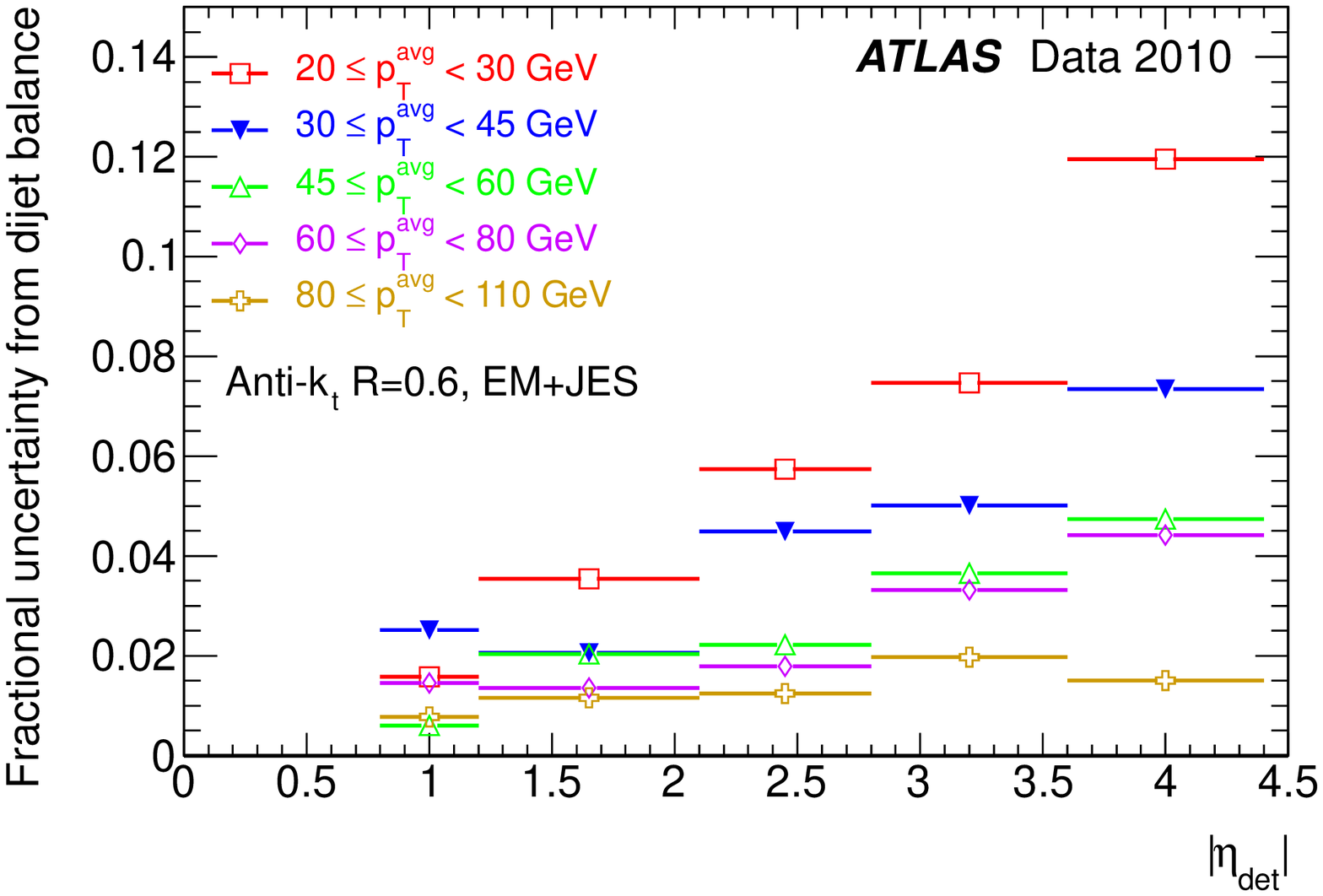}}
\caption{Fractional response
  uncertainty for \antikt{} jets with $R = 0.6$ calibrated with the \EMJES{} scheme
  as obtained from the dijet balance \insitu{} technique as a function of \ptjet for various $|\etajet|$-regions  
  of the calorimeter (a) and as a function of $|\etajet|$ in various \ptjet{} bins (b).
  \label{fig:response_uncert}}
\end{figure*}

Figure \ref{fig:response_uncert} shows the uncertainty in the jet response,
relative to jets in the central region $|\eta|<0.8$, as a function of the jet $\pt{}$ and $|\eta|$.
The JES uncertainty, determined in the central detector region using the
single particle response and systematic variations of the Monte Carlo simulations,
is transferred to the forward regions using the results from the dijet balance.
These uncertainties are included in the final uncertainty as follows:
\begin{enumerate}
\item 
  The total JES uncertainty in the central region $0.3 \leq |\eta|<0.8$ is kept  as a baseline.
\item The uncertainty from the relative intercalibration is taken as the
  RMS deviation of the MC predictions from the data
  and is added in quadrature to the baseline uncertainty.
\end{enumerate}
The measurements are performed for transverse momenta in the range $20\leq\ptavg<110$~\GeV. 
The uncertainty for jets with $\pt>100$~\GeV{} is taken as the uncertainty of the last available $\pt$-bin\footnote{
This is justified by the decrease of the intercalibration uncertainty with $\pt$, but
cannot completely exclude the presence of calorimeter non-linearities for jet energies above those used
for the intercalibration.}.
The uncertainties are evaluated separately for jets reconstructed with distance
parameters $R=0.4$ and $R=0.6$, and are in general found to be
slightly larger for $R=0.4$. 

\begin{figure*}[!ht]
  \centering
  \subfloat[$30 \leq \ptavg < 45$~\GeV]{\includegraphics[width=0.49\textwidth]{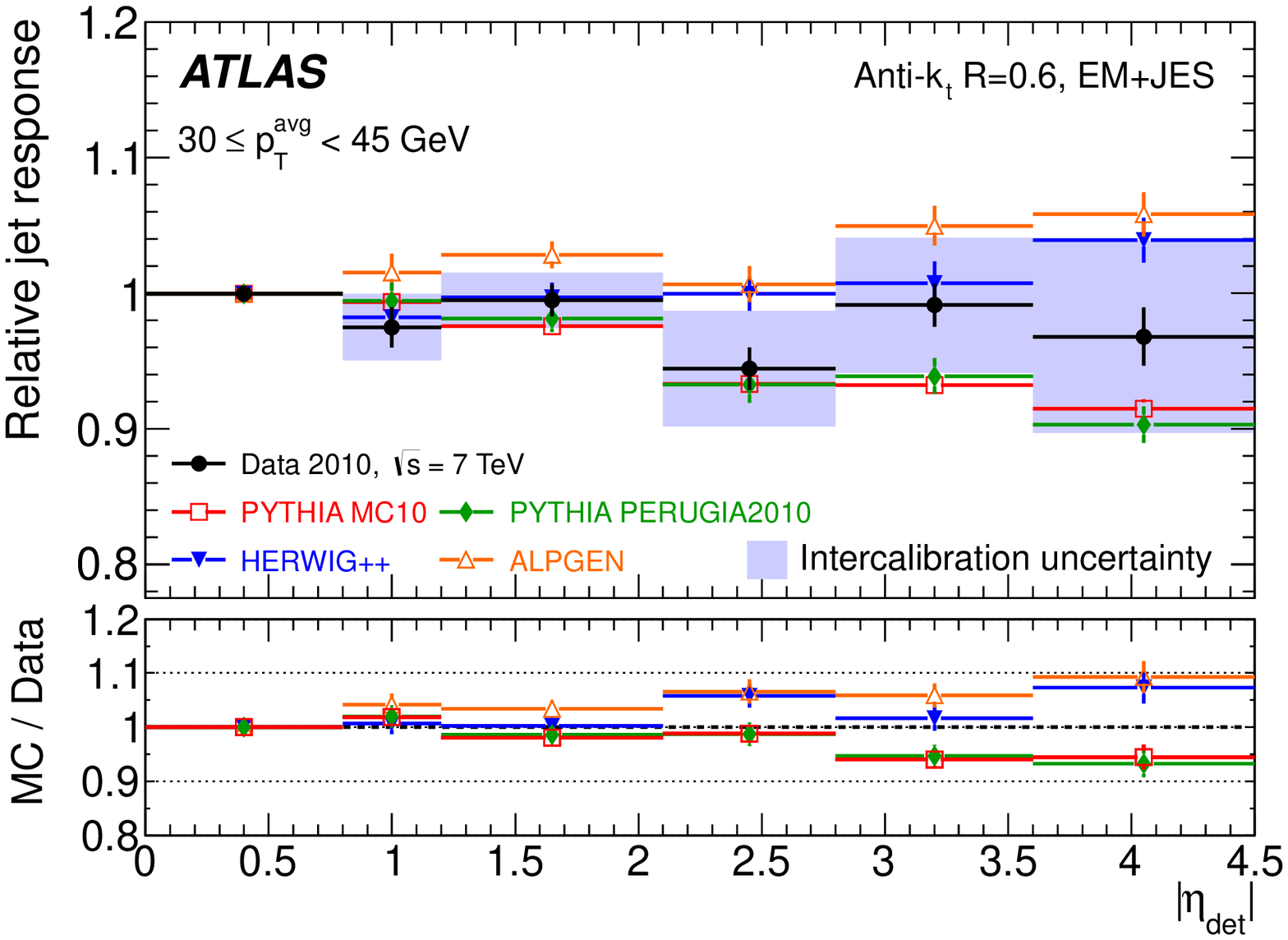}}
  \subfloat[$80 \leq \ptavg < 110$~\GeV]{\includegraphics[width=0.49\textwidth]{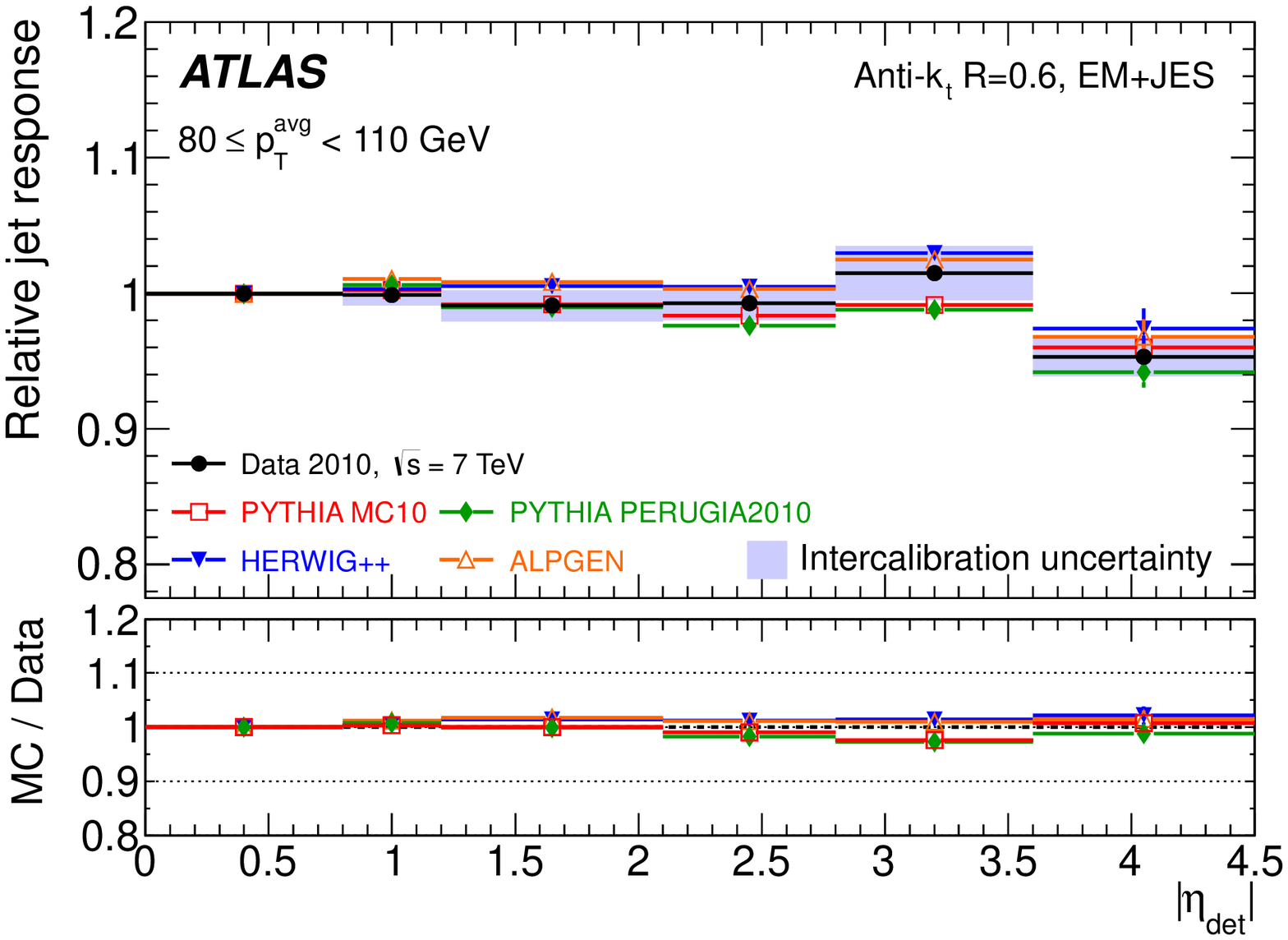}}
  \caption{Average jet response for \antikt{} jets with $R = 0.6$ calibrated with the \EMJES{} scheme
    measured relative to a central reference jet within $|\eta|<0.8$ 
in data and various Monte Carlo generator samples 
    as a function of $|\etajet|$ for \ptavg{} in the
    ranges $30 - 45$~\GeV{} (a) and $80 - 110$~\GeV{} (b). 
    The resulting systematic uncertainty component is shown as a shaded band around the data points.
    The errors bars on the data points only show the statistical uncertainties.
  }
  \label{fig:IntercalibrationUncertainty2128}
\end{figure*}

Figure~\ref{fig:IntercalibrationUncertainty2128} shows the relative
jet response, and the associated intercalibration uncertainty calculated
as detailed above, as a
function of jet $|\eta|$ for two representative $\ptavg$-bins.

%
\begin{figure*}[!!htp]
\begin{center}
\subfloat[Tower jets (tower-based correction)]{ \label{fig:closure:fit:tower:tower}\includegraphics[width=0.45\textwidth]{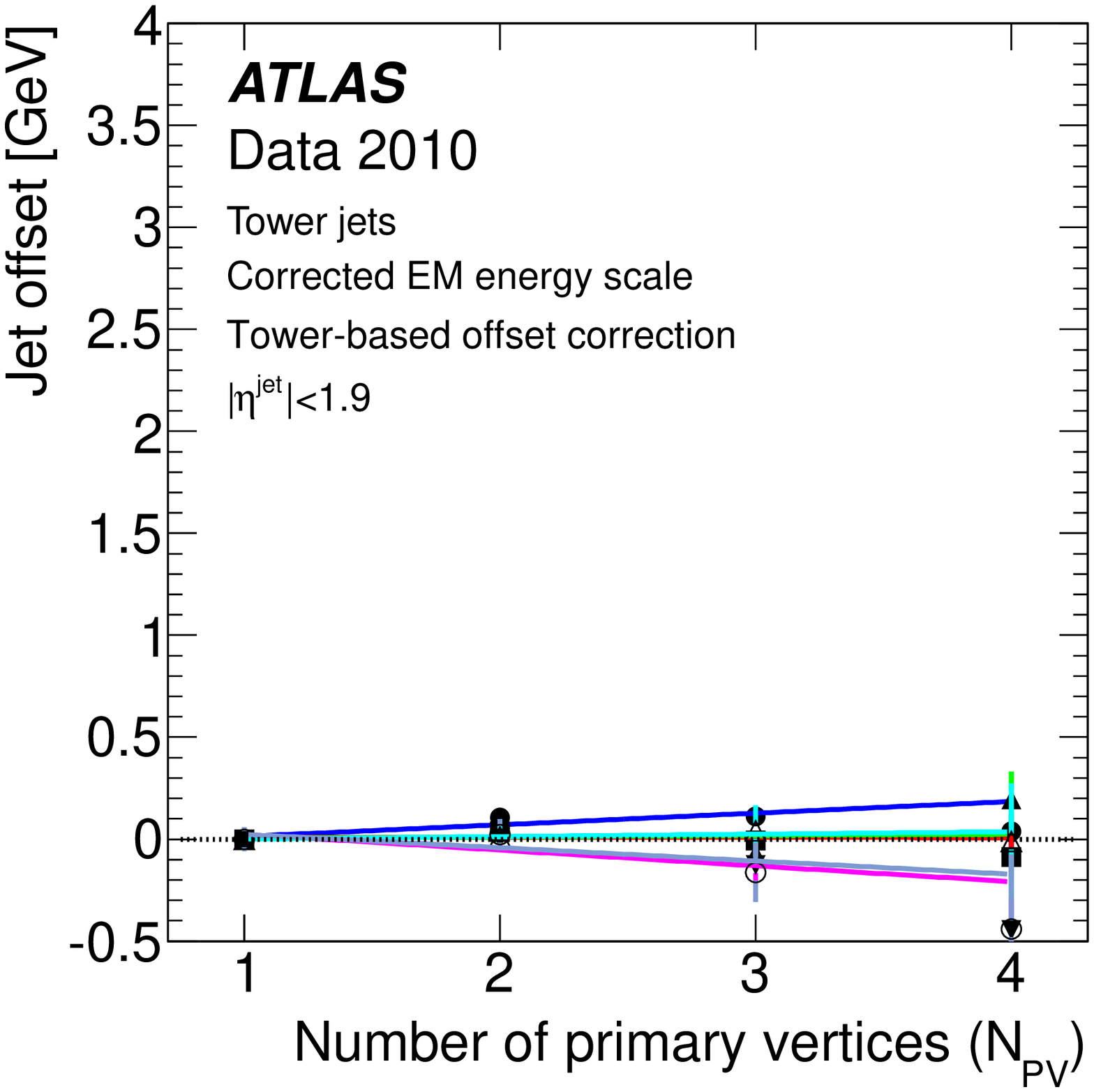}}
\hspace{1.cm}
\subfloat[Tower jets (jet-based correction)]{ \label{fig:closure:fit:tower:jet}\includegraphics[width=0.45\textwidth]{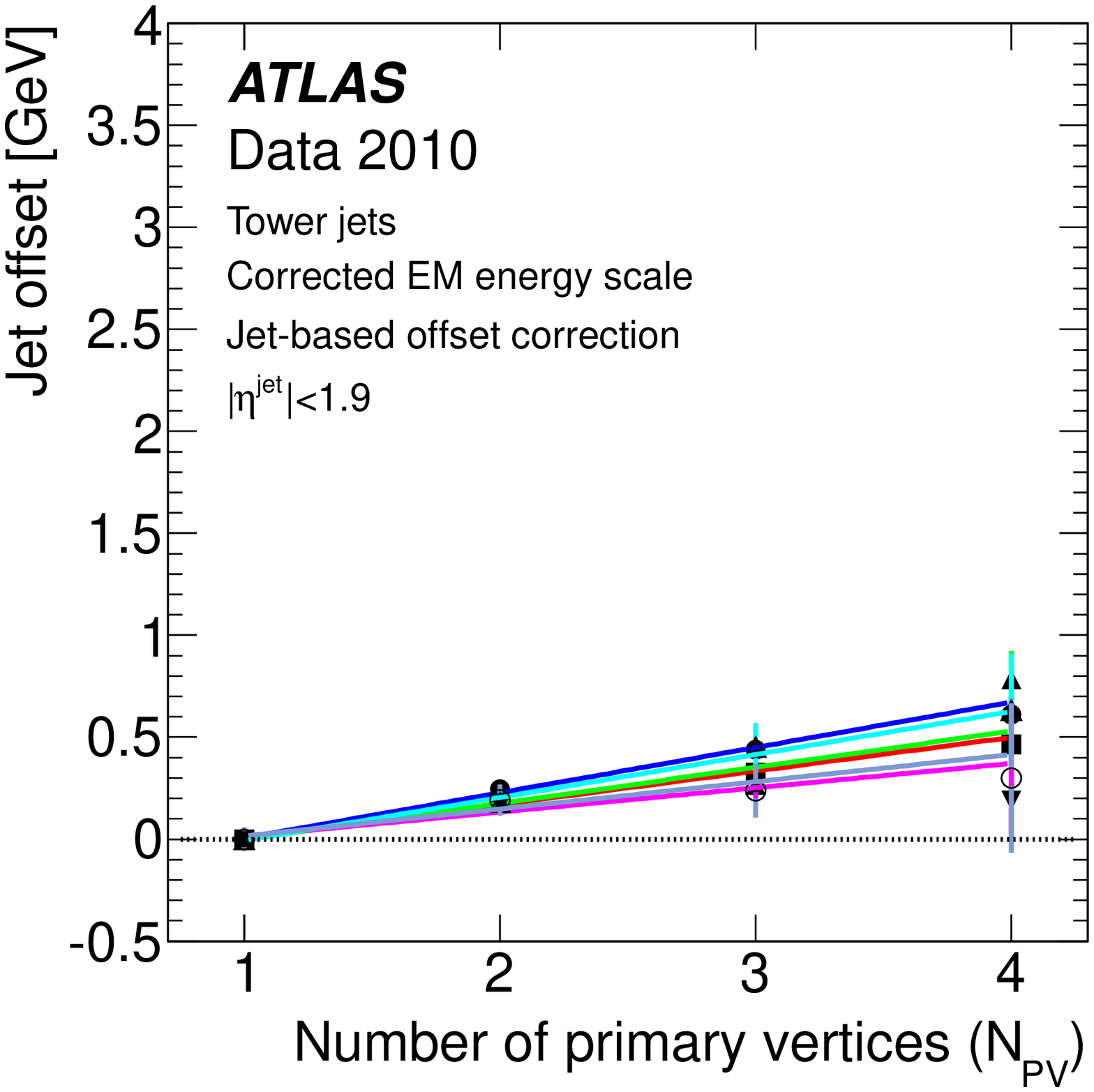}} \\
\subfloat[\Topo{} jets (jet-based correction)]{\label{fig:closure:fit:topo:jet}\includegraphics[width=0.45\textwidth]{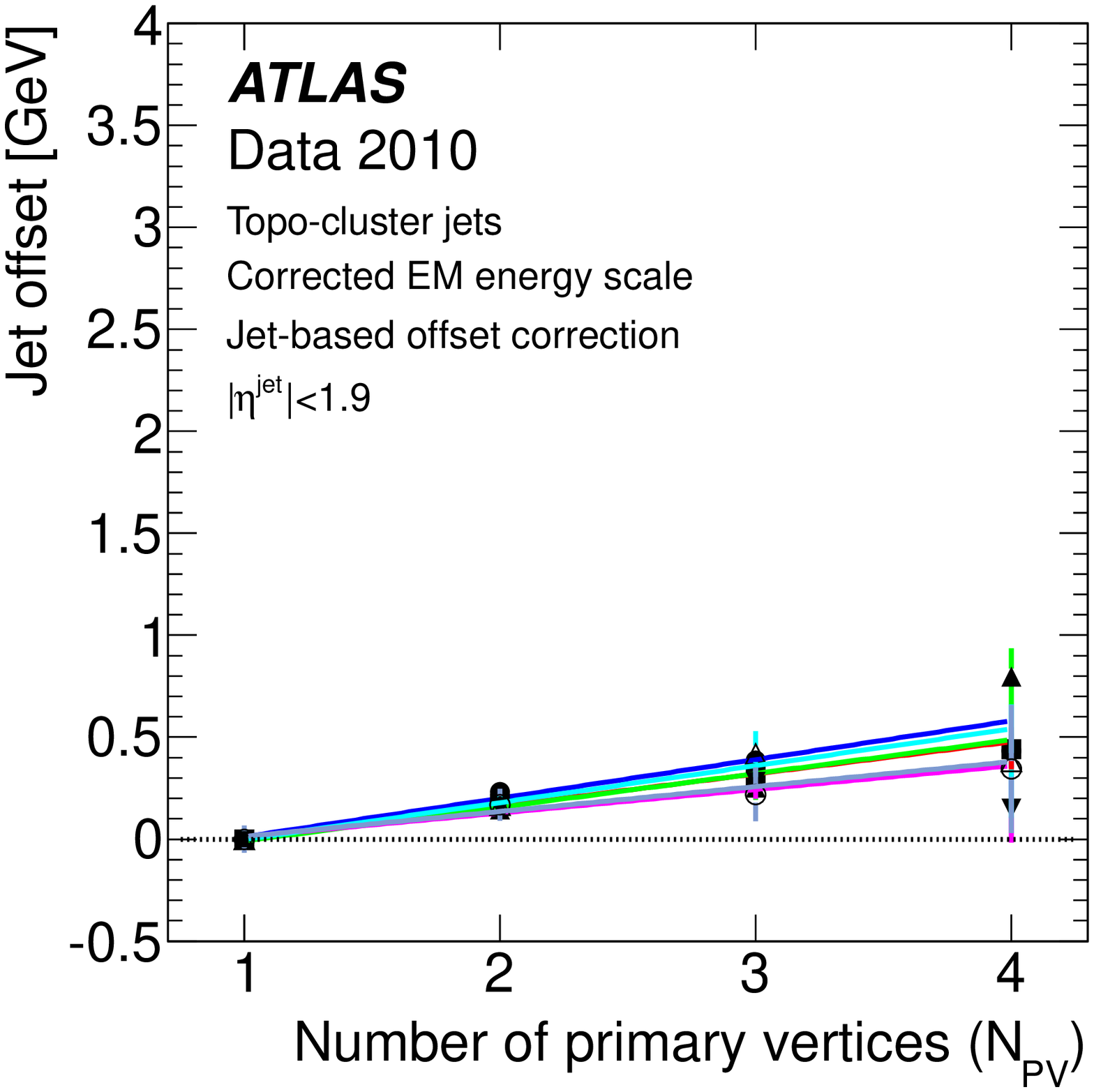}}
\hspace{1.cm}
\includegraphics[width=0.45\textwidth]{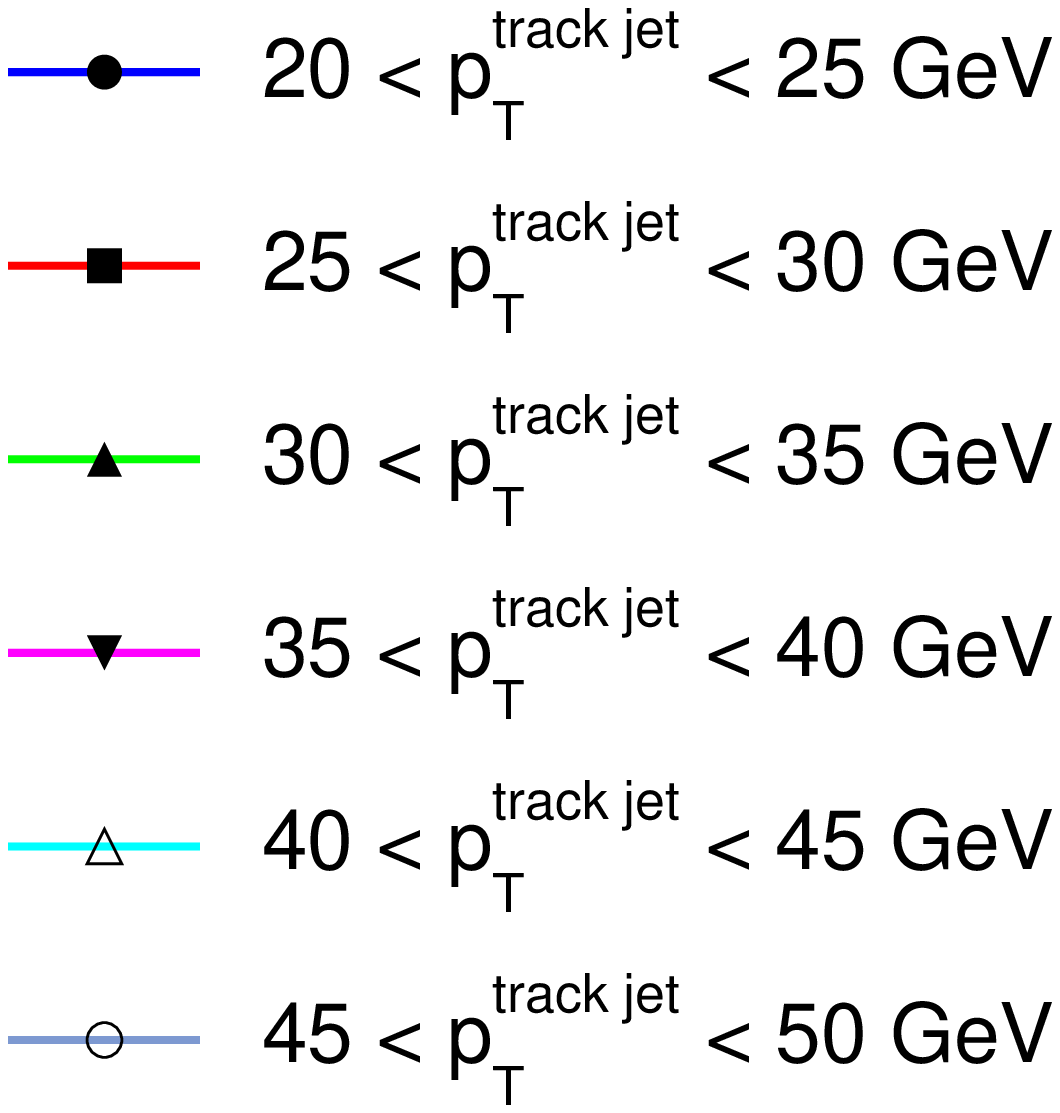} 
\caption{
Jet residual offset measured at the \EM{} scale after pile-up correction
using the most probable value \etjet{} obtained from a fit to a Landau+Gauss distribution 
for various bins in track jet transverse momentum (\pTtrkjet) as a function of the primary vertex multiplicity:
tower jets corrected with tower-based offset correction (using the actual number constituent towers) (a), 
tower jets corrected with the jet-based offset correction (using the average number of constituent towers) (b) and 
\topo{} jets corrected with the jet-based offset correction (using the average number of equivalent constituent towers) (c). 
The axis ranges are identical to Figure~\ref{fig:trkoffset} for ease of comparison.
The jet offset is given for \antikt{} jets at the \EM{} scale with $R = 0.6$. 
Only the statistical uncertainties of the fit results are shown.
\vspace{3.cm}
}
\label{fig:closure:fit}
\end{center}
\end{figure*}

\begin{figure}[ht!]
  \centering
  \includegraphics[width=0.5\textwidth]{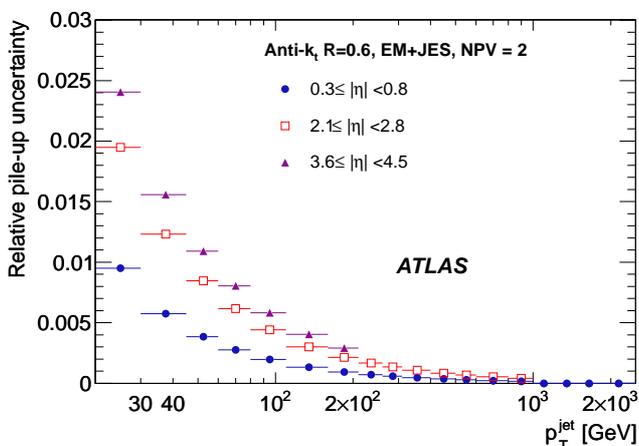}\\[-4mm]
  \caption{Relative \JES{} uncertainty from pile-up for \antikt{} jets with $R = 0.6$
    in the case of two measured primary vertices, $\Npv = 2$, 
  for central (\central{}, full circles), endcap (\ecap{}, open squares) and forward (\forward{}, full triangles) jets 
  as a function of jet \ptjet{}. 
  } %
  \label{fig:Pileup}
\end{figure}

\subsection{Uncertainties due to multiple proton-proton collisions}
\label{sec:pileupSummary}
\begin{table*}[ht!]
        \begin{center}
                \begin{tabular}{l|c|c|p{5cm}}
                        \hline \hline
                        Systematic & Tower-based offset & Jet-based offset & Comments \\
                        \hline
                        Trigger selection & $16 \%$ & $16 \%$ & \MBTS{} vs Jet triggers \\
                        Tower multiplicity variation & -- & $20 \%$ & $\langle N^{\rm jet}_{\rm towers} \rangle$ vs \pTtrkjet{} and \Npv  \\
                        \pTtrkjet{} variation & $21 \%$ & $22 \%$ & Variation of $100$ \MeV/vertex \\
                        \hline
      Total (quadrature sum) & $26 \%$ & $34 \%$ & Assumes uncorrelated errors \\
                        Result from closure test & $2 \%$ & $35 \%$ & Determined from average \\
                        \hline \hline
                \end{tabular}
        \end{center}
\caption{Summary of systematic uncertainties associated with the offset correction for both the tower-based offset applied
jet-by-jet to tower jets and
the jet-level offset applied to \topo{} jets. The uncertainty is expressed as a percentage of the average 
offset correction, shown in Table~\ref{table:slopes}.}
\label{table:systematics}
\end{table*}%

\begin{table*}[ht!]
 \begin{center}
  \begin{tabular}{c|r|r|r|r}
  \hline \hline
   \multirow{2}{*}{Track jet \pT} & \multicolumn{2}{|c|}{Tower jets [\GeV/vertex]} & \multicolumn{2}{|c}{\Topo{} jets [\GeV/vertex]} \\
    & \multicolumn{1}{c}{Before} & \multicolumn{1}{c|}{After} & \multicolumn{1}{c}{Before} & \multicolumn{1}{c}{After} \\
   \hline 
   $20$ - $25$ \GeV & $0.55 \pm 0.02$ & $0.06 \pm 0.02$  & $0.50 \pm 0.02$ & $0.19 \pm 0.02$ \\
   $25$ - $30$ \GeV & $0.47 \pm 0.02$ & $0.00 \pm 0.02$  & $0.47 \pm 0.02$ & $0.16 \pm 0.02$ \\
   $30$ - $35$ \GeV & $0.49 \pm 0.03$ & $0.01 \pm 0.03$  & $0.47 \pm 0.03$ & $0.17 \pm 0.03$ \\
   $35$ - $40$ \GeV & $0.42 \pm 0.03$ & $-0.08 \pm 0.03$ & $0.41 \pm 0.03$ & $0.12 \pm 0.03$ \\
   $40$ - $45$ \GeV & $0.51 \pm 0.05$ & $0.01 \pm 0.05$  & $0.48 \pm 0.05$ & $0.18 \pm 0.05$ \\
   $45$ - $50$ \GeV & $0.42 \pm 0.06$ & $-0.07 \pm 0.06$ & $0.41 \pm 0.06$ & $0.12 \pm 0.06$ \\
   \hline 
   Average     & $0.48 \pm 0.02$ & $-0.01 \pm 0.02$ & $0.46 \pm 0.02$ & $0.16 \pm 0.02$ \\
   \hline \hline
  \end{tabular}
 \end{center}
\caption{Variation of the calorimeter \etjet{} with pile-up for several bins in track jet \pT. 
Slopes are given in \GeV/vertex at the electromagnetic scale for each primary vertex 
from additional proton-proton collisions in the event, and represent the slope of the jet offset 
before and after the tower-based offset correction. 
Tower-based corrections are applied to tower jets and jet-based corrections
are applied to \topo{} jets. 
The reported uncertainties are purely statistical.}

\label{table:slopes}
\end{table*}%

The offset to the jet transverse energy due to pile-up interactions can be measured at the \EM{} scale 
from the average energy in calorimeter towers in minimum bias events.
The uncertainty in the pile-up corrections can be obtained by varying certain analysis choices
and by  studying the jet response with respect to the transverse momentum of track jets as a function
of the number of primary vertices. 

\subsubsection{Tower-based offset closure test using track jets}
The systematic uncertainty in the jet offset correction can be evaluated using track jets.
Figure~\ref{fig:trkoffset} shows the variation of the offset among the various ranges 
of \pTtrkjet{}. The result indicates a systematic uncertainty on the correction of 
approximately $\delta(\mathcal{O}^{\rm EM}_{\rm track \, jet})<100$~\MeV{} per additional vertex at the \EM{} scale 
and $\delta(\mathcal{O}^{\EMJES}_{\rm track \, jet})<200$~\MeV{} per additional vertex at the \EMJES{} scale. 
Since the jet pile-up offset was about $500$~\MeV{} before correction,
even with this conservative estimate the application of the offset correction 
represents an improvement of a factor of five obtained
over the systematic bias associated with pile-up effects 
on the calorimeter jet transverse momentum. 

The full offset correction shows reasonable closure when using the actual constituent tower multiplicity directly
(tower-based)
and a slight under-correction using the average constituent multiplicity in the jet (jet-based). 
Figure~\ref{fig:closure:fit:tower:tower} shows the tower-based correction applied 
to tower jets at the \EM{} scale as a function of the reconstructed vertex multiplicity. 
The tower-based correction exhibits a closure consistent with zero slope in \etjet{} 
as a function of \Npv. 
Figure~\ref{fig:closure:fit:tower:jet} and Figure~\ref{fig:closure:fit:topo:jet} show
the jet-based correction applied to both tower jets and \topo{} jets, respectively. 
The use of the jet-based offset correction slightly 
under-corrects for the effect of pile-up for jets constructed from \textit{both} 
towers \textit{and} \topos. 

The implication of this observation is two-fold:
\begin{enumerate}
  \item There is no significant difference in the sensitivity of \topo{} jets 
        to pile-up as compared to tower jets.
  \item There is a systematic underestimation of the average tower multiplicity in jets 
        due to the effect of pile-up or due to differences in the jet transverse energy
        distribution in the derivation and the validation of the pile-up correction.
\end{enumerate}

\subsubsection{Jet offset correction uncertainties}
The contributions to the jet offset correction uncertainty are estimated from studies that account for:
\begin{enumerate}
 \item The effect of variations of the trigger selection on the measured non-noise-suppressed tower energy distribution 
       that is input to the offset correction.
\item  The variation with \ptjet{} and \Npv{} of the tower multiplicity in jets based on \topos \footnote{
This is determined from the variation in tower multiplicity for $\Npv = 1$ in jets matched to 
track jets with $25 \le \pt < 30$~\GeV{} 
as compared to $\Npv = 4$ in track jets with $35 \le \pt < 40$~\GeV.}.
\item The variation of the offset correction derived from track jets 
as a function of the number of primary vertices for various values of track jet \pt.
\item The non-closure of the tower-based offset correction as evaluated by the dependence of the 
corrected calorimeter jet energy for calorimeter jets matched to track jets as a function of the 
 number of primary vertices.
%
%
%
\end{enumerate}

The \JES{} uncertainty is estimated by add\-ing all uncertainties in
quadrature, including the one from the non-closure of the correction. %
The track jet method can be used only up to $|\eta| = 1.9$, if a full coverage of the jet area
by the tracking acceptance is needed. 
Beyond $|\eta| = 1.9$, the dijet balance method 
detailed in Section~\ref{sec:etaintercalibration} is used. This approach compares the relative jet
response in events with only one reconstructed vertex with the
response measured in events with several reconstructed vertices.
The dijet balance method yields uncertainties similar to those intrinsic to the
method also in the case of $|\eta| < 1.9$.

Each source of systematic uncertainty is summarised in Table~\ref{table:systematics} and 
the resulting effects expressed as a percentage 
of the average offset correction, shown in Table~\ref{table:slopes}. 

For jets based on towers the total systematic uncertainty is significantly larger 
than the validation of the correction using track jets indicate. 
The larger of the two individual uncertainties ($21 \%$) is therefore adopted. This results in
$\delta(\mathcal{O}^{\rm tower-based})=100$~\MeV{} per vertex\footnote{
Using twice the RMS of the variation in the closure test yields a similar value.}. 
The resulting total uncertainty is a factor of five smaller than the bias attributable to 
pile-up ($\approx 500$~\MeV{} per vertex) even with this conservative systematic uncertainty
estimation.

The offset correction for jets based on \topos{} receives an additional uncertainty 
due to the average tower multiplicity approximation. 
This contribution is estimated to introduce a $20\%$ uncertainty in the constituent tower multiplicity 
by comparing jets in events with $\Npv = 1 - 3$ and for the five highest \pTtrkjet-bins. 
This estimation translates directly into a $20 \%$ uncertainty on the jet-based offset. 
The resulting systematic uncertainty on jets corrected by the offset correction 
is estimated to be $\delta(\mathcal{O}^{\rm jet-based})\approx160$~\MeV{} per vertex;
a factor of three smaller than the bias due to pile-up.

Figure~\ref{fig:Pileup} shows the relative uncertainty due to pile-up in the case of two measured
primary vertices. In this case, the uncertainty due to pile-up for central jets with 
$\pt = 20$~\GeV{} and pseudorapidity $|\etajet| \leq 0.8$ is about $1 \%$, while
it amounts to about $2 \%$ for jets with pseudorapidity $2.1 \le |\etajet| < 2.8$ and to less than $2.5 \%$ for all jets 
with $|\eta| \leq 4.5$. In the case of three primary vertices, $\Npv = 3$, 
the pile-up uncertainty is approximately twice
that of $\Npv = 2$, and with four primary vertices the uncertainty for central, endcap and forward jets is 
less than $3 \%$, $6 \%$ and $8 \%$, respectively. The relative uncertainty due to pile-up for events with up to five 
additional collisions becomes less than $1 \%$ for all jets with $\ptjet > 200$~\GeV. 
The pile-up uncertainty needs to be added separately to the estimate of the total jet energy scale 
uncertainty detailed in Section~\ref{sec:JESSummary}.

\subsubsection{Out-of-time pile-up}
The effect of additional proton-proton collisions from previous bunch 
crossings within trains of consecutive bunches (out-of-time pile-up) 
has been studied separately.  
The effect is found to be negligible in the $2010$ data.

\begin{figure}[!ht]
\begin{center}
  \subfloat[Comparison of \Npv = $1$ and \Npv = $2$]{\label{fig:jetShape:1}\includegraphics[width=0.49\textwidth]{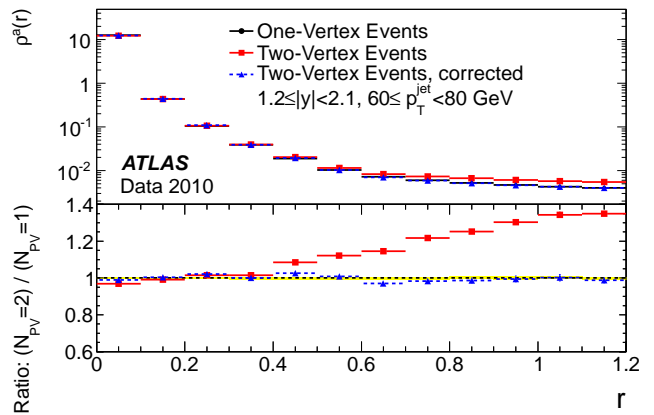}}\\
  \subfloat[Comparison of \Npv = $1$ and \Npv = $3$]{\label{fig:jetShape:2}\includegraphics[width=0.49\textwidth]{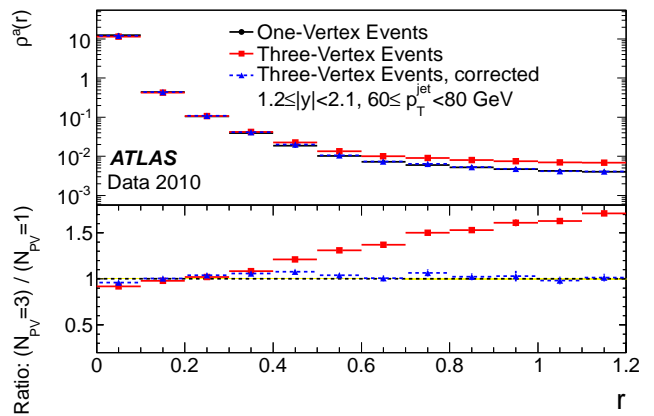}}
        \caption{Measured sum \pt\ in annuli around the jet axis, divided by the total \pt\ around the jet
within $\DeltaR=0.7$ of the jet axis and normalised by the area of each annulus
as a function of the distance of the jet constituent to the jet axis.
The shapes of jets in the rapidity range $1.2 \le |\rapjet| <2.1$ are compared, before and after the offset 
corrections, in events with  one and two reconstructed vertices (a), and  one and three reconstructed vertices (b). 
The corrected distribution is also shown (full triangles). 
Note that the single vertex data (full circles) are partially hidden behind the corrected multi-vertex data.
\Antikt{} jets with $R = 0.6$ reconstructed from calorimeter towers are used and calibrated with the \EMJES{} scheme.
}
        \label{fig:jetShape}
\end{center}
\end{figure}

\subsubsection{Pile-up corrections applied to jet shape measurements}
\label{sec:pileupjetshape}
The measurement of internal jet properties like the energy flow inside jets can be made considerably
more difficult in the presence of additional proton-proton collisions. 
The applicability of the tower-based offset presented in Section~\ref{sec:pileupSummary}
to correct the mean jet energy can also be tested on the internal jet shape measurements.

The offset correction is applied to the measurement of the differential jet shape for $R = 0.6$ tower jets, 
as described in Ref.~\cite{Atlasjetshape}. 

The jet shape variable used, $\rho^a\left( r \right)$, is defined as:
%
\begin{equation}
\rho^a(r) = \frac{1}{ \pi \left[ \left( r+\delta r/2\right)^2 - \left(r-\delta r/2\right)^2 \right] } 
\cdot \left< \frac{ p_T\left( r-\frac{\delta r}{2} , r+\frac{\delta r}{2} \right)} { p_T(0,0.7) } \right>,
\end{equation}
where $r = \sqrt{(d \eta)^2 + (d \phi)^2}$ is the distance 
of the jet constituents to the jet four-momentum vector and
the angled brackets denote an average over all jets, $\pt \left(b, c\right)$ is the sum of the \pt{} 
of all towers with an opening angle $b \le \DeltaR < c $ with respect to the jet axis, 
and $\delta r = 0.1$.  

This definition differs from the canonical jet shape 
variable $\rho\left(r\right)$ in two important ways.  
First, by normalising to area, the variable measures an energy density.  
Therefore, $\rho^a \left ( r \right)$ will approach an asymptotic value far from the jet axis.  
The level of the asymptote is related to the energy density in the calorimeter and is measurably 
higher in events with pile-up.  Second, all towers 
are included in the definition.  This allows an examination of energy outside of the jet cone, 
in some sense measuring ``energy flow'' around the jet axis.

Figure~\ref{fig:jetShape} depicts $\rho^a \left ( r \right )$ with and without a
correction of the tower constituent energy for the mean energy induced by pile-up interactions. 
In events with two (three) reconstructed vertices, differences in this particular jet shape variable 
of up to $35 \%$ ($70 \%$) just outside the jet ($r > 0.6$) and $20 \%$ ($40 \%$) near the nominal jet radius 
($r = 0.6$) are observed. The bulk of the shape ($0.1 \le r < 0.6$) is restored to that observed in events 
with only a single interaction, in both the core ($r < 0.1$) and the periphery ($r > 0.6$) of the jet. 

The results demonstrate that the tower-based offset correction can be applied on a fine scale granularity
and is valid both inside and \textit{near} jets.

\subsection{Summary of jet energy scale systematic uncertainties}
\label{sec:JESSummary}
The total jet energy scale uncertainty is derived by considering all the individual 
contributions described in the previous sections.
In the central region (\AetaRange{0.8}), the estimate proceeds as follows:
\begin{enumerate}
\item For each \ptjet{} and \etajet{} bin, 
the uncertainty due to the calibration procedure is calculated as described 
in Section~\ref{sec:nonclosure} for both jet energy and \pt{} response. 
For each bin, the maximum deviation from unity between the energy and \pt{} response 
is taken as the final non-closure uncertainty. 
\item The calorimeter response uncertainty is estimated as a function of jet \etajet{} and \pt{} 
from the propagation of single particle uncertainties to the jets, 
as detailed in Section~\ref{sec:SingleParticle}.
 \item Sources of uncertainties estimated using Monte Carlo samples with a systematic 
variation are accounted as follows:
\begin{enumerate}
\index{ calorimeter response uncertainty}
\item the response in the test sample $\mathcal{R}_{\mathrm{var}}$ 
and the response in the nominal sample $\mathcal{R}_{\mathrm{nom}}$
is considered as a starting point for the estimate of the \JES{} uncertainty. 
The deviation of this ratio from unity is defined as:
\begin{equation}
\Delta_{\JES}(\pt^{\textrm{jet}},\eta) = \left| 1-\frac{\mathcal{R}_{\mathrm{var}}(\pt^{\textrm{jet}},\eta)}{\mathcal{R}_{\mathrm{nom}}(\pt^{\textrm{jet}},\eta)} \right |.
 \end{equation}
This deviation is calculated from both the energy and \pt{} response,
leading to $\Delta_{\JES}^{\rm E}(\pt^{\textrm{jet}},\eta)$ 
for the deviation in the energy response,
and to $\Delta_{\JES}^{\pt}(\ptjet,\eta)$ 
for the deviation in the transverse momentum response.
\item 
The larger $\Delta_\JES$ in each bin derived from the jet energy 
or transverse momentum response is considered as the contribution 
to the final \JES{} systematic uncertainty due to the specific systematic effect:
\begin{equation}
\Delta_{\JES}(\pt^{\textrm{jet}},|\eta|) = \max(\Delta_{\JES}^{\rm E}(\pt^{\textrm{jet}},\eta), \Delta_{\JES}^{\pt}(\pt^{\textrm{jet}},\eta)).
\end{equation}

\end{enumerate}

\item The estimate of the uncertainty contributions due to additional material 
in the inner detector and overall additional dead material are estimated as described in the previous step. 
These uncertainties are then scaled by the average fraction of particles forming the jet that 
have $p < 20$~\GeV{} (for the inner detector distorted geometry) 
and by the average fraction of particles outside the kinematic range of the single hadron response \insitu{} measurements (for
the overall distorted geometry).
\end{enumerate}

For each (\ptjet{}, $\etajet$)-bin, the uncertainty contributions from the calorimeter, 
the jet calibration non-closure, 
and systematic Monte Carlo simulation variations are added in quadrature.

For pseudorapidities beyond $|\eta|>0.8$, the $\eta$-inter\-cali\-bra\-tion contribution 
is estimated for each pseudorapidity bin in the endcap region as detailed 
in Section \ref{sec:EtaIntercalibUnc}. 
The pseudo\-rapidity inter\-cali\-bra\-tion contribution is added in quadrature to the total \JES{} uncertainty 
determined in the $0.3 \leq |\eta| < 0.8$ region to estimate the \JES{} uncertainty for 
jets with $|\eta|>0.8$, with the exception of the non-closure term that is taken from 
the specific $\eta$-region.
For low \ptjet, this choice leads to partially double counting the 
contribution from the dead material uncertainty, 
but it leads to a conservative estimate in a region where 
it is difficult to estimate the accuracy of the material description.
\index{ $\eta$-intercalibration uncertainty}

The contribution to the uncertainty due to additional proton-proton interactions 
described in Section~\ref{sec:pileupSummary} is added separately, 
depending on the number of primary vertices in the event. 
In the remainder of the section only the uncertainty for 
a single proton-proton interaction is shown in detail. 

\begin{figure}[!ht]
  \centering
  \subfloat[\etaRange{0.3}{0.8} ]{\includegraphics[width=0.45\textwidth]{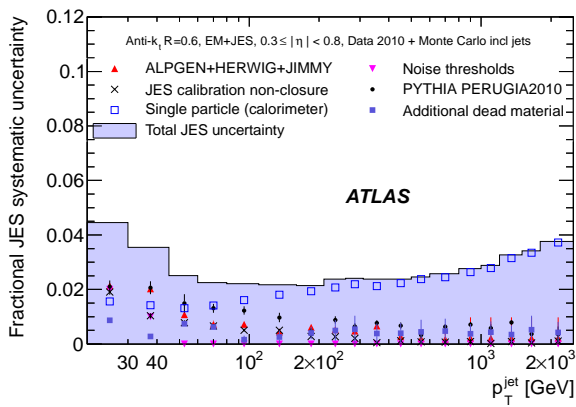}} \\
  \subfloat[\etaRange{2.1}{2.8} ]{\includegraphics[width=0.45\textwidth]{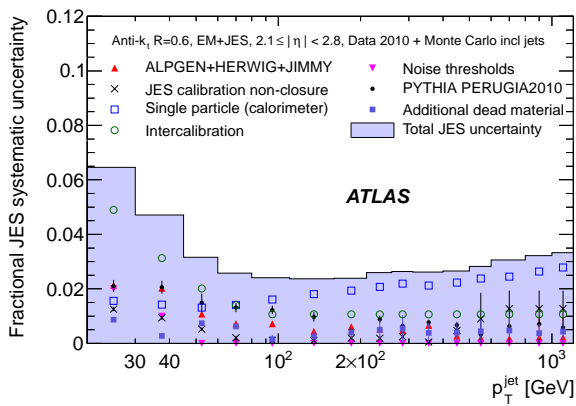}} \\
  \subfloat[\etaRange{3.6}{4.5} ]{\includegraphics[width=0.45\textwidth]{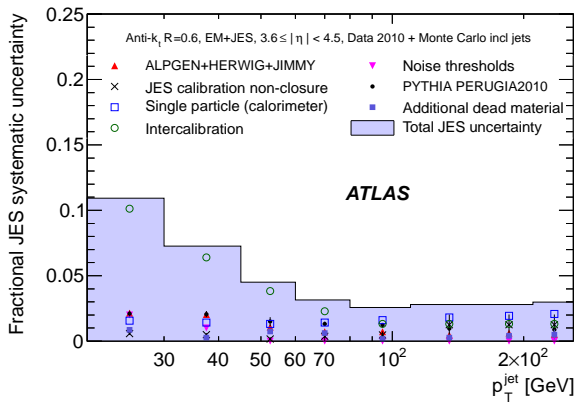}}
  \caption{Fractional jet energy scale systematic uncertainty as a function of \ptjet{} for jets in the 
pseudorapidity region \etaRange{0.3}{0.8} in the calorimeter barrel (a), 
\etaRange{2.1}{2.8} in the calorimeter endcap (b),
and in the forward pseudorapidity region \etaRange{3.6}{4.5}. 
The total uncertainty is shown as the solid light shaded area. 
The individual sources are also shown together with uncertainties from the fitting procedure if applicable.
  \label{fig:FinalJES}
}
\end{figure}

Figure~\ref{fig:FinalJES} shows the final fractional jet energy scale systematic uncertainty 
and its individual contributions as a function of \ptjet{} for three selected \etajet{} regions. 
%
%
The fractional \JES{} uncertainty in the central region amounts to 
$2 \%$ to $4 \%$ for $\ptjet{}< 60$~\GeV, 
and it is between $2 \%$ and $2.5\%$ for \ptRange{60}{800}. 
For jets with $\ptjet{}> 800$~\GeV,
the uncertainty ranges from $2.5 \%$ to $4\%$.
The uncertainty amounts to up to $7\%$ and $3\%$, respectively, for $\ptjet{}< 60$~\GeV{} and 
$\ptjet{}>60$~\GeV{} in the endcap region, where the central uncertainty is 
taken as a baseline and the uncertainty due to the intercalibration 
is added. In the forward region, a $13\%$ uncertainty is assigned for $\ptjet =  20$~\GeV.
The increase in the uncertainty is dominated by the modelling of the soft physics in
the forward region that is accounted for in the $\eta$-intercalibration contribution.  
This uncertainty contribution is estimated conservatively. 

Table~\ref{table:JESSystematics6} 
presents a summary of the maximum uncertainties in the different $\eta$ regions for 
\antikt{} jets with $R=0.6$ and with \ptjet{} of $20$~\GeV, 
$200$~\GeV{} and $1.5$~\TeV{} as examples.

The same study has been repeated for \antikt{} jets with distance 
parameter $R = 0.4$, and the estimate of the \JES{} uncertainty is comparable
to that obtained for \antikt{} jets with $R = 0.6$.
The \JES{} uncertainty for \antikt{} jets with $R = 0.4$ is between 
$\approx 4 \%$ ($8 \%$, $14 \%$) at low \ptjet{} and $\approx$ $2.5 \% - 3 \%$ ($2.5 \% - 3.5 \%$, $5 \%$) 
for jets with $\pt > 60$~\GeV{} in the central (endcap, forward) region,
and is summarised in Table~\ref{table:JESSystematics4}.
%

\begin{table}[!ht]
\begin{center}
\begin{tabular}{c|c|c|c}
\hline
\hline
    $\etajet$ region & \multicolumn{3}{|c}{Maximum fractional \JES{} Uncertainty} \\
  \hline
  \ & \ptjet =$20$~\GeV & $200$~\GeV & $1.5$~\TeV \\
  \hline
 \etaRange{0}  {0.3} & $4.6 $\% & $2.3$\% & $3.1$\% \\
 \etaRange{0.3}{0.8} & $4.5 $\% & $2.2$\% & $3.3$\% \\
 \etaRange{0.8}{1.2} & $4.4 $\% & $2.3$\% & $3.3$\% \\
 \etaRange{1.2}{2.1} & $5.4 $\% & $2.4$\% & $3.4$\% \\
 \etaRange{2.1}{2.8} & $6.5 $\% & $2.5$\% &         \\
 \etaRange{2.8}{3.2} & $7.9 $\% & $3.0$\% &         \\
 \etaRange{3.2}{3.6} & $8.1 $\% & $3.0$\% &         \\
 \etaRange{3.6}{4.5} & $10.9$\% & $2.9$\% &         \\
\hline
\hline
\end{tabular}
\caption{Summary of the maximum \EMJES{} jet energy scale systematic uncertainties for 
different \ptjet{} and \etajet{} regions from Monte Carlo simulation based study for \antikt{} jets with $R=0.6$.
\label{table:JESSystematics6}
}
\vskip10pt
\end{center}

\begin{center}
\begin{tabular}{c|c|c|c}
\hline
\hline
    $\etajet$ region & \multicolumn{3}{|c}{Maximum fractional \JES{} Uncertainty} \\
  \hline
  \ & \ptjet = $20$~\GeV & $200$~\GeV & $1.5$~\TeV \\
  \hline 
 \etaRange{0}  {0.3}  & $4.1$\%  & $2.3$\% & $3.1$\%\\
 \etaRange{0.3}{0.8}  & $4.3$\%  & $2.4$\% & $3.3$\%\\
 \etaRange{0.8}{1.2}  & $4.4$\%  & $2.5$\% & $3.4$\%\\
 \etaRange{1.2}{2.1}  & $5.3$\%  & $2.6$\% & $3.5$\%\\
 \etaRange{2.1}{2.8}  & $7.4$\%  & $2.7$\% &        \\
 \etaRange{2.8}{3.2}  & $9.0$\%  & $3.3$\% &        \\
 \etaRange{3.2}{3.6}  & $9.3$\%  & $3.5$\% &        \\
 \etaRange{3.6}{4.5}  & $13.4$\% & $4.9$\% &        \\
\hline
\hline
\end{tabular}
\end{center}
\caption{Summary of the maximum \EMJES{} jet energy scale systematic uncertainties 
for different \ptjet{} and $\etajet$ regions from Monte Carlo simulation based study for \antikt{} jets with $R=0.4$.
\label{table:JESSystematics4}
}
\end{table}
\index{JES uncertainty table}

\subsection{Discussion of special cases}
\label{sec:JESUncertaintySpecialcase}
\index{Special cases}
The jet energy scale is derived using the simulated sample of inclusive jets
described in Section \ref{sec:NominalSample}, 
with a particular mixture of quark and gluon initiated jets and with a particular
selection of isolated jets. The differences in fragmentation between
quark and gluon initiated jets and the effect of close-by jets give rise to a 
topology and flavour dependence of the energy scale.
Since the event topology and flavour composition (quark and gluon fractions) 
may be different in final states other than the considered inclusive jet sample, 
the dependence of the jet energy response on
jet flavour and topology has to be accounted for in physics analyses.
The flavour dependence is discussed in more detail in Section~\ref{sec:quarkgluon}
and an additional uncertainty specific to jets with heavy quark components
is discussed in Section~\ref{sec:bjet}.

The \JES{} systematic uncertainty is derived for isolated jets\footnote{
This choice is motivated by the minor differences
observed in the average kinematic jet response of isolated and non-isolated jets
in the nominal inclusive jet  Monte Carlo sample and by the need to factorise
the topology dependence of the close-by jet energy scale uncertainty for 
final states other than the inclusive jets considered.}. 
The response of jets as a function of the distance to the closest reconstructed jet
needs to be studied and corrected for separately if the measurement relies on the absolute jet energy scale.
The contribution to the \JES{} uncertainty from close-by jets also needs to be estimated separately, 
since the jet response depends on the angular distance to the closest jet. 
This additional uncertainty can be estimated from the Monte Carlo simulation to data comparison of the \pt-ratio between 
calorimeter jets and matched track jets in inclusive jet events as a function of the isolation radius. 
This is discussed in more detail in Section~\ref{sec:closeby}.

\section{Jet energy scale uncertainties validation with \insitu{} techniques for the EM+JES scheme}
\label{sec:insituvalidation}
The jet energy calibration can be tested \insitu{} using a well-calibrated object as reference
and comparing data to the nominal \pythia{} Monte Carlo simulation. 
The following \insitu{} techniques have been used by \ATLAS:
\begin{enumerate}
\item {\bf Comparison to the momentum carried by tracks associated to a jet:} 
      The mean transverse momentum sum of tracks that are within a cone with size $R$ provides an independent 
      test of the calorimeter energy scale over the entire measured \ptjet{} range within the tracking acceptance.
      The comparison is done in the jet $\etajet$ range \etaRange{0}{2.1}.
\item {\bf Direct \pt{} balance between a photon and a jet:} 
     Events with a photon and one jet at high transverse momentum are used
      to compare the transverse momentum of the jet to that of the photon.
      To account for effects like soft QCD radiation and energy migrating out of
      the jet area the data are compared to the Monte Carlo simulation.
      The comparison is done in the jet $\eta$ range \AetaRange{1.2} and for 
      photon transverse momenta $25 \le \pt^\gamma < 250$~\GeV.
\item {\bf Photon \pt{} balance to hadronic recoil:}
      The photon transverse momentum is balanced against the full hadronic recoil
      using the projection of the missing transverse momentum onto the photon direction (MPF).
      This method does not explicitly involve a jet algorithm. 
      The comparison is done in the same kinematic region as the direct photon balance method. 
\item {\bf Balance between a high-\pt{} jet and low-\pt{} jet system:}
      If jets at low transverse momentum are well-calibrated, 
      jets at high transverse momentum can be balanced against a recoil system of low transverse
      momentum jets. This method can probe the jet energy scale up to the \TeV-regime.
      The $\eta$ range used for the comparison is \AetaRange{2.8}.
\end{enumerate}

All methods are applied to data and Monte Carlo simulation.

The \insitu{} techniques usually rely on assumptions that are only approximately fulfilled.
An example is the assumption that the jet to be calibrated and the reference object are balanced 
in transverse momentum.  This balance can be altered by the presence of additional high-\pt{}
particles.
For the determination of the \JES{} uncertainties the modelling of physics effects 
has to be disentangled from detector effects.
This can be studied by systematically
varying the event selection criteria. The ability of the Monte Carlo simulation
to describe extreme variations of the selection criteria determines the systematic
uncertainty in the \insitu{} methods, since physics effects can be suppressed or amplified by these variations. 

So far the \insitu{} techniques are used to validate the systematic uncertainty
in the jet energy measurement. 
However, they can also be used to obtain jet energy corrections.
This is an interesting possibility when the statistical and systematic uncertainties 
in the samples studied become smaller than the standard \JES{} uncertainty from the single hadron response.
The results of the \insitu{} techniques are discussed in the following sections.

\begin{figure*}[htp!]
\begin{center}
\subfloat[]{\epsfig{file=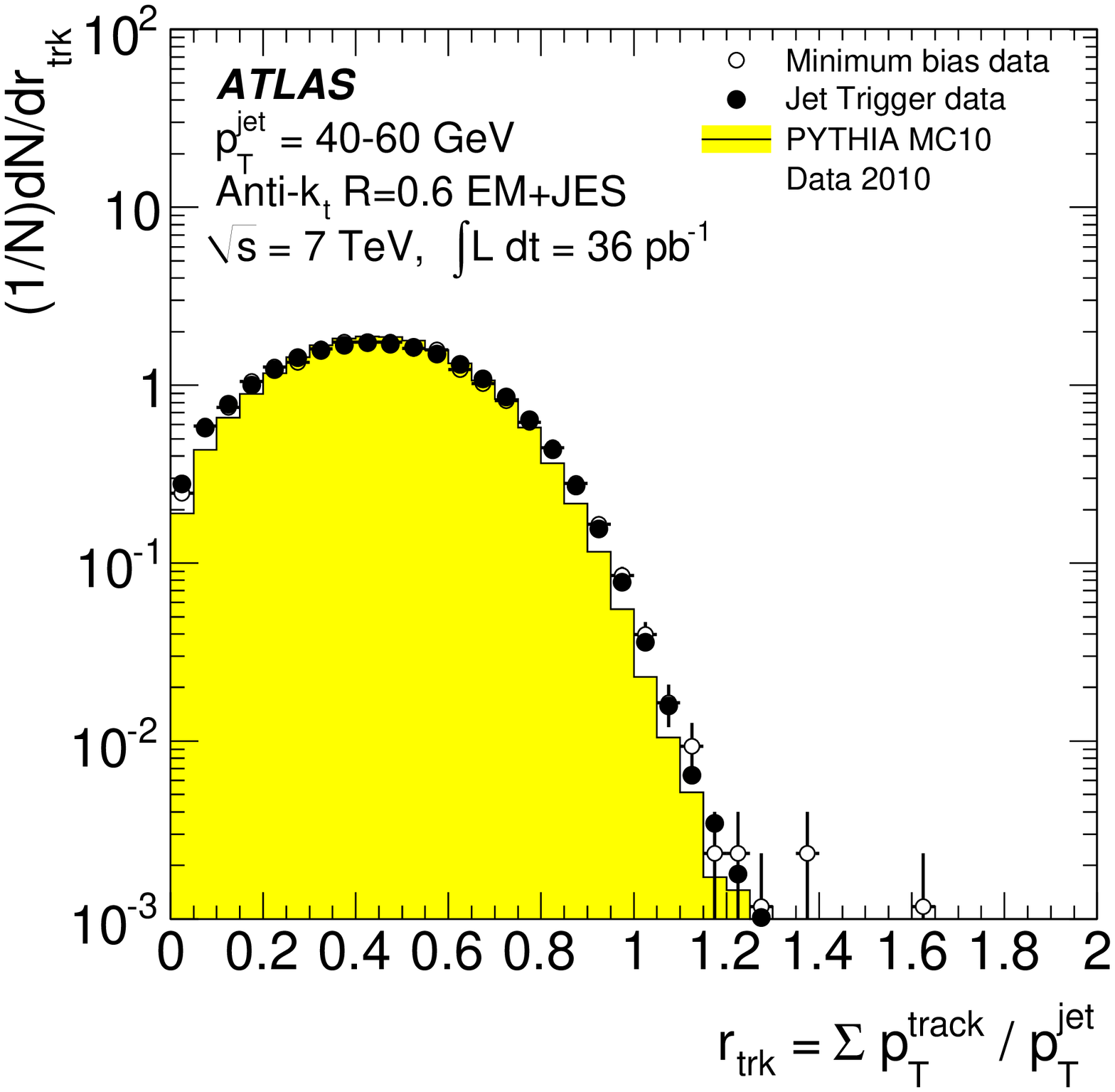,width=0.42\textwidth}}
\hspace{1.cm}
\subfloat[]{\epsfig{file=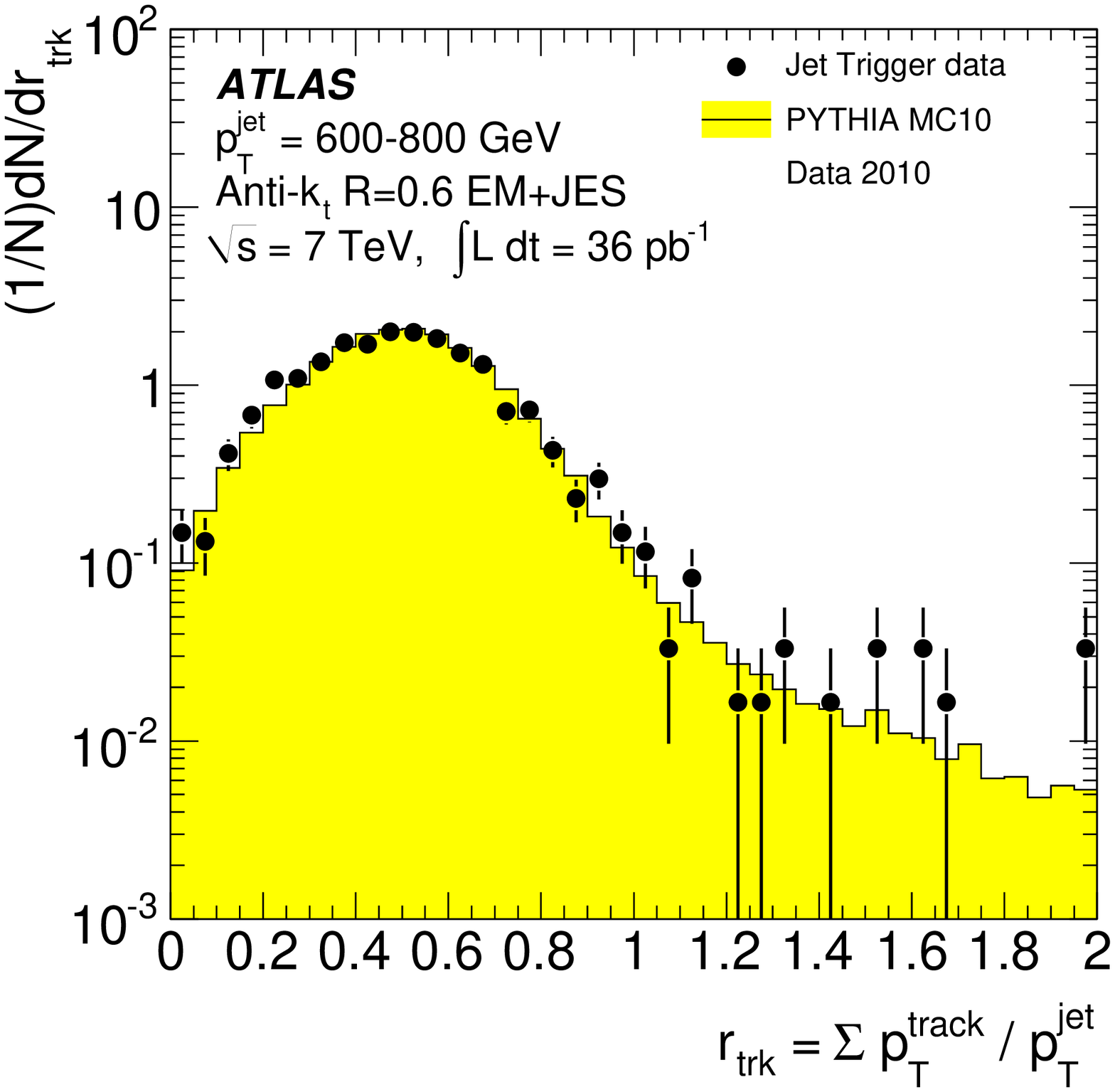,width=0.42\textwidth}} \\
\subfloat[]{\epsfig{file=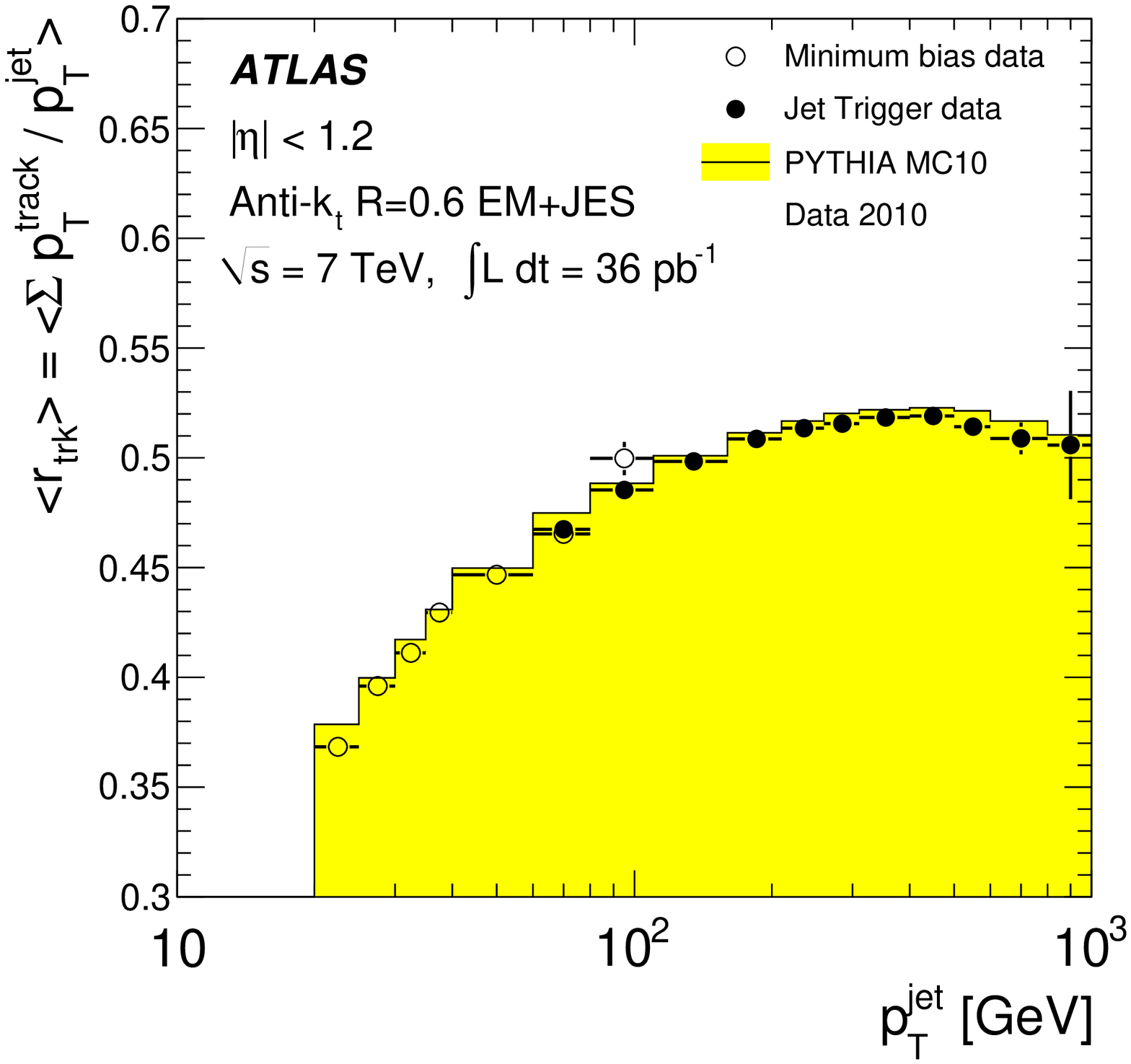,width=0.42\textwidth}}
\hspace{1.cm}
\subfloat[]{\epsfig{file=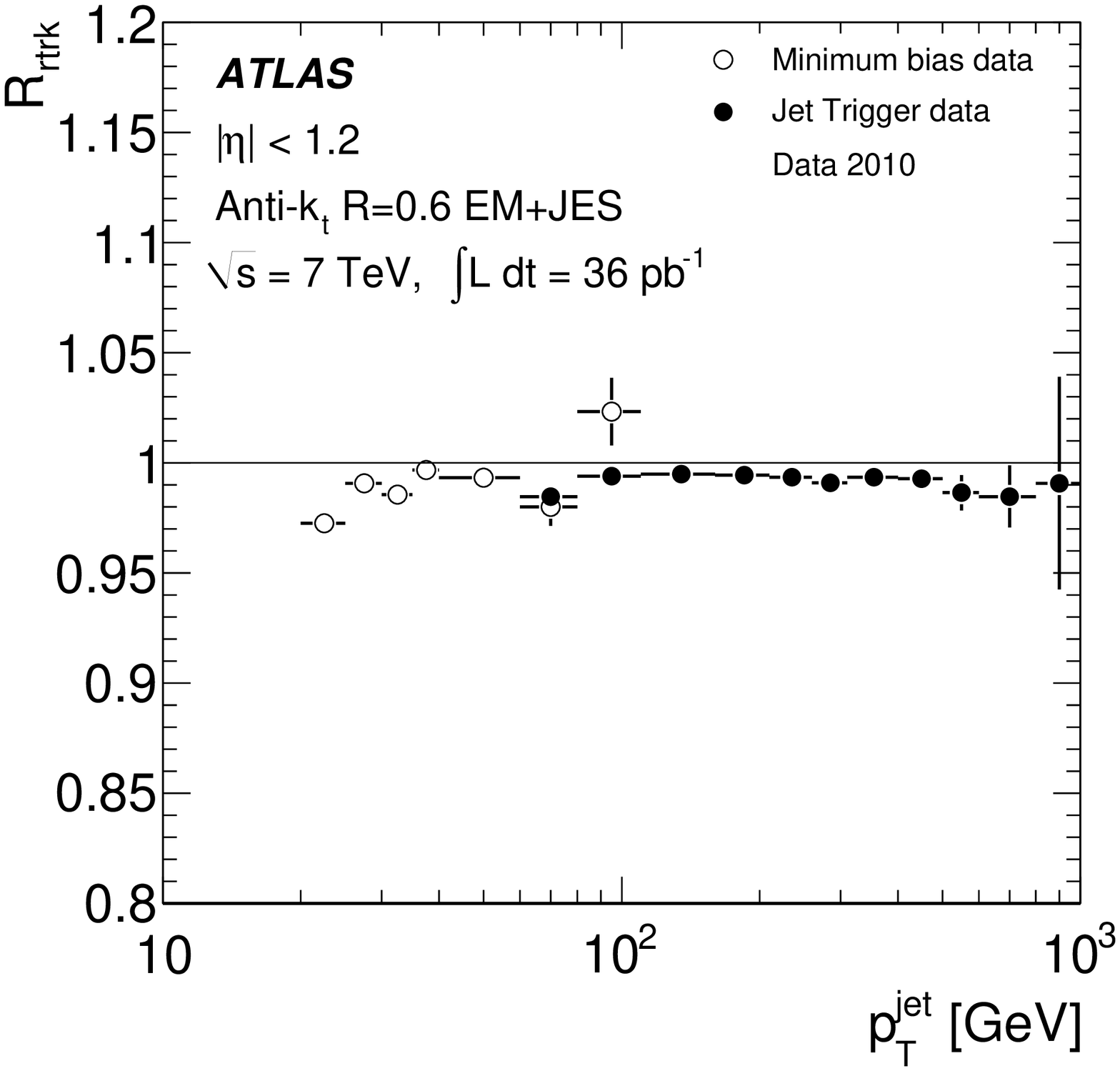,width=0.42\textwidth}}
\end{center}
\caption{
The distribution of the charged-to-total momentum ratio $\rtrk$ for 
$40\ \leq \ptjet <60 \;\GeV$ (a) and for
$600  \leq \ptjet < 800\;\GeV$ (b),
the average charged-to-total momentum ratio 
$\langle \rtrk \rangle$ for data and Monte Carlo simulation as a function of \ptjet{} (c) 
and the ratio of $\langle \rtrk \rangle$ for data and Monte Carlo simulation 
($ R_{\rtrk} $) as a function of \ptjet\ 
for the pseudorapidity range $|\etajet| < 1.2$ (d)
for \antikt\ jets with $R = 0.6$ calibrated using the \EMJES{} scheme. 
The data measured with the jet (minimum bias) trigger are shown as closed (open) circles.
Only statistical uncertainties are shown.
}
\label{fig:rat} 
\end{figure*}

\subsection{Comparison of transverse momentum balance of jets from calorimeter and tracking}
\label{sec:trackjet}
\index{Jet energy scale uncertainty from tracks}
The transverse momentum of each jet can be compared with the total transverse
momentum of tracks associated with the jet by means of a geometrical selection
and the charged-to-total-momentum ratio:
\begin{equation}
\rtrk = \frac{\sum \pt^{\rm \;track}}{\ptjet}
\label{eq:def-rtrk}
\end{equation}
can be used to test the jet calibration. 
\index{\rtrk}
If all produced particles were pions, the symmetry of QCD under isospin transformation
would require that this ratio be $2/3$ once the energy is high enough so that the total particle multiplicity is large
and the initial isospin of the proton-proton system can be ignored. Production of other particles such as
kaons,  $\eta$ mesons, and baryons gives  different fractions, but their contributions can be calculated using
a properly tuned event generator.  

Since the tracking system provides a measurement that is independent of the
calorimeter, the ratio \rtrk\ 
can be used to determine the calorimeter jet energy scale.
The \rtrk\ distribution is broad %
but a meaningful calibration does not require very many events, since the statistical uncertainty 
on the mean scales as $1/\sqrt{N}$.
This calibration can be used for jets confined within the tracking detector coverage.  
Dominant systematic uncertainties result from the knowledge
of the tracking efficiency, variations in the predicted value of \rtrk\
for various generator tunes and loss of tracking efficiency in the dense core
of high-\pt\ jets.

To test the \ptjet\ dependence of the jet energy measurement, the double ratio of charged-to-total momentum observed in
data to that obtained in Monte Carlo simulation is studied:
\begin{equation}
R_{\rtrk} \equiv \frac{{\left [ \langle {\rtrk}  \rangle \right ]}_{\rm Data} }
                    {{\left [ \langle {\rtrk}  \rangle \right ]}_{\rm MC}   }.
\label{eq:def-doublerat}
\end{equation}
\index{$R_{\rtrk}$}
\subsubsection{Jet and track selection}
To ensure that the majority of tracks associated with the jets found in the calorimeter
are within the inner detector fiducial volume, jets are required to have $|\etajet|<2.1$\footnote{
Section~\ref{sec:pileupSummary} discusses ``track jets'' obtained
by running the \antikt\ jet algorithm using tracks as input.  Those
studies are restricted to $|\etajet| <1.9$ to avoid bias in the position
of the centre of the jet due to tracking inefficiencies.  Since the
jets in this section are found using calorimeter information, no such
bias is present and it is therefore possible to extend the pseudorapidity
coverage to $|\etajet|<2.1$.} and  $\ptjet > 20$~\GeV.
To reduce the influence of nearby jets on the measurement,  if two jets are separated
by a distance $\DeltaR < 2 R$ then the softer of these two jets is
rejected from the analysis.

Tracks with $\pttrk > 1$~\GeV{}
are selected using the criteria detailed in Section~\ref{sec:trackjets}.
The  $\pttrk > 1$~\GeV{} requirement  is 
intended to select mainly tracks from fragmentation rather than
those arising from soft and diffuse interactions.

Tracks are associated with jets using a geometric algorithm.  If
the distance $\DeltaR_{\;\rm track,jet}$ between the track and the jet
is less than the distance parameter used in the jet reconstruction ($R=0.4$ or $R=0.6$), 
the track is associated to the jet.  Track parameters
are evaluated at the distance of closest approach to the primary hard-scattering vertex 
and are {\it not} extrapolated to the calorimeter.  
This simple association algorithm facilitates comparison with charged particles from truth jets 
whose parameters correspond to those measured at the origin.

\subsubsection{Comparison of data and Monte Carlo simulation}
\label{sec:comparisons}

The jet response validation using the total momentum measured
in tracks depends on a comparison of the mean value of \rtrk\
observed in the data to that predicted in the Monte Carlo simulation.
It is therefore important to demonstrate that the baseline Monte Carlo generator and 
simulation provide a reasonable description of the data.

ATLAS has measured the charged particle fragmentation function for jets with
$25 \le \ptjet < 500$~\GeV{} and $|\etajet|<1.2$ 
and has compared the measurement with the predictions of several Monte Carlo generators 
and generator tunes~\cite{jetfragmenationpaper}.

The jet fragmentation function and the transverse jet profile
are compared to various Monte Carlo event generators and tunes.
The jet fragmentation function is measured 
using charged particles with momentum fraction $z$ with respect to the jet momentum 
$F(z, \ptjet) = 1/N_{\rm jet} \; dN_{\rm ch}/dz$.

The growth of the mean charged particle multiplicity with \ptjet{} is well
modelled by the Monte Carlo simulation.
The measured jet fragmentation function agrees well with the \pythia{} MC10
and the \Perugia2010 tunes within the measurement uncertainties.
The jet fragmentation function is described by the \pythia{} tunes.
The \herwigpp{} Monte Carlo generator is not consistent with the data.

For observables related to jet properties in the direction transverse
to the jet axis the Monte Carlo generators (\herwig{} and the various
\pythia{} tunes) show reasonable agreement with data, but none
of the generators agrees within the experimental uncertainties
over the full kinematic range. For instance, the \pythia{} MC10 tune shows
an excess of about $10\%$ in the transverse charged particle
distributions close to the jet axis.

These measurements indicate that the \pythia{} MC10 and \Perugia2010 tunes span the range of fragmentation functions
that are consistent with the data.  The studies presented here use the
MC10 tune to obtain the central values of the Monte Carlo predictions.  Systematic
uncertainties are assessed from the difference between the MC10 and \Perugia2010
\pythia{} tunes.

The \rtrk\ distributions used to validate the JES are shown
for data and simulation for two typical bins of jet \pt{} in Figure~\ref{fig:rat}a and Figure~\ref{fig:rat}b. 
Agreement between data and simulation is good, although the data
distribution is somewhat wider than the Monte Carlo simulation.
Figure~\ref{fig:rat}c and Figure~\ref{fig:rat}d show  $\langle \rtrk \rangle$ for data
and simulation and the average double ratio $R_{\rtrk}$, respectively, as a function of \ptjet.  
Figure~\ref{fig:rat}d demonstrates that the
measured JES calibration agrees with that predicted by the Monte Carlo simulation
to better than $2 \%$ for $\ptjet>25\,\GeV$.
Measurements using the minimum bias and jet triggers are consistent for those
\ptjet\ bins where both triggers are accessible.

\begin{figure*}[ht!]
\subfloat[$| \etajet | < 0.3$ ]{
\epsfig{file=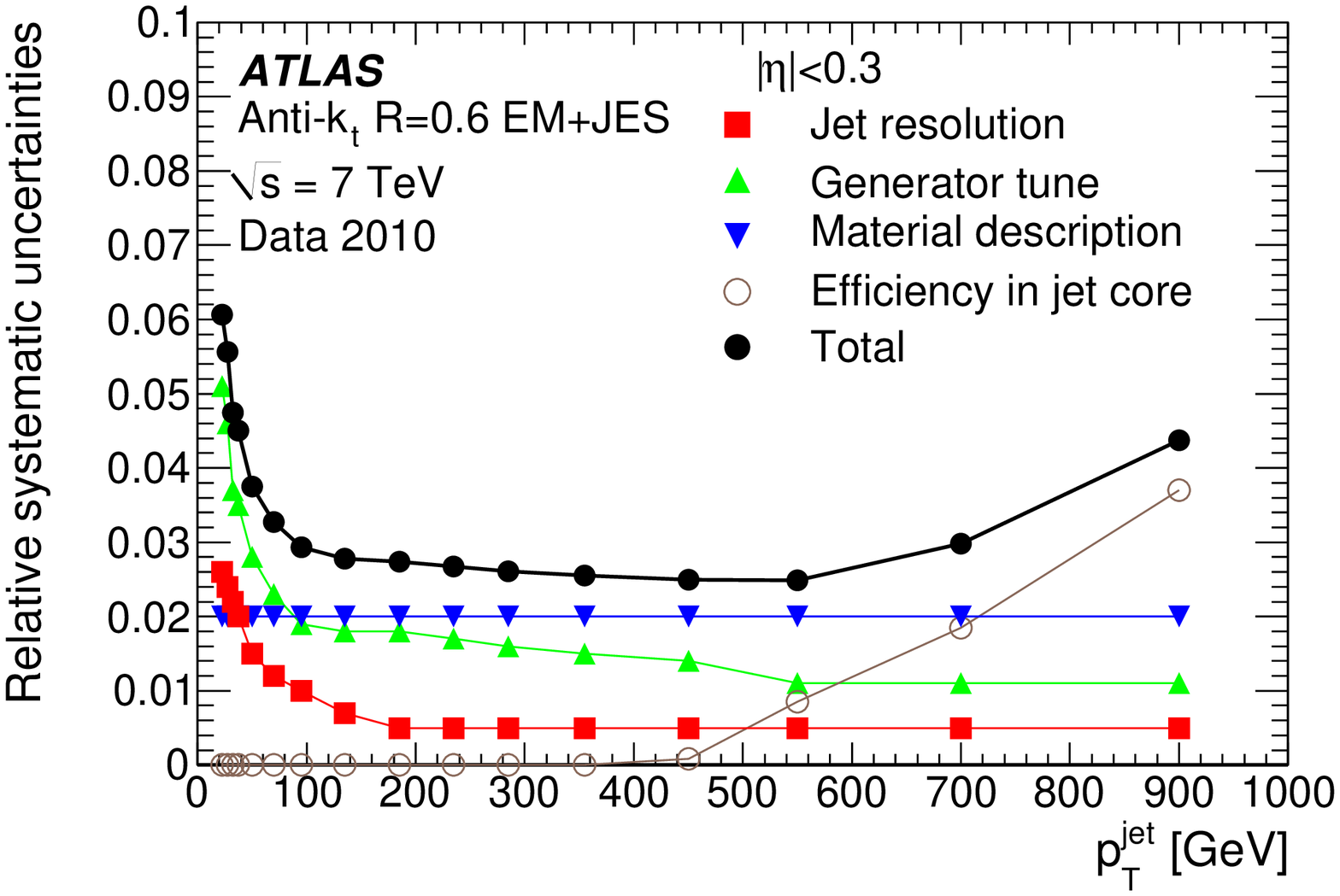,width=0.5\textwidth}
}
\subfloat[\etaRange{0.3}{0.8} ]{
\epsfig{file=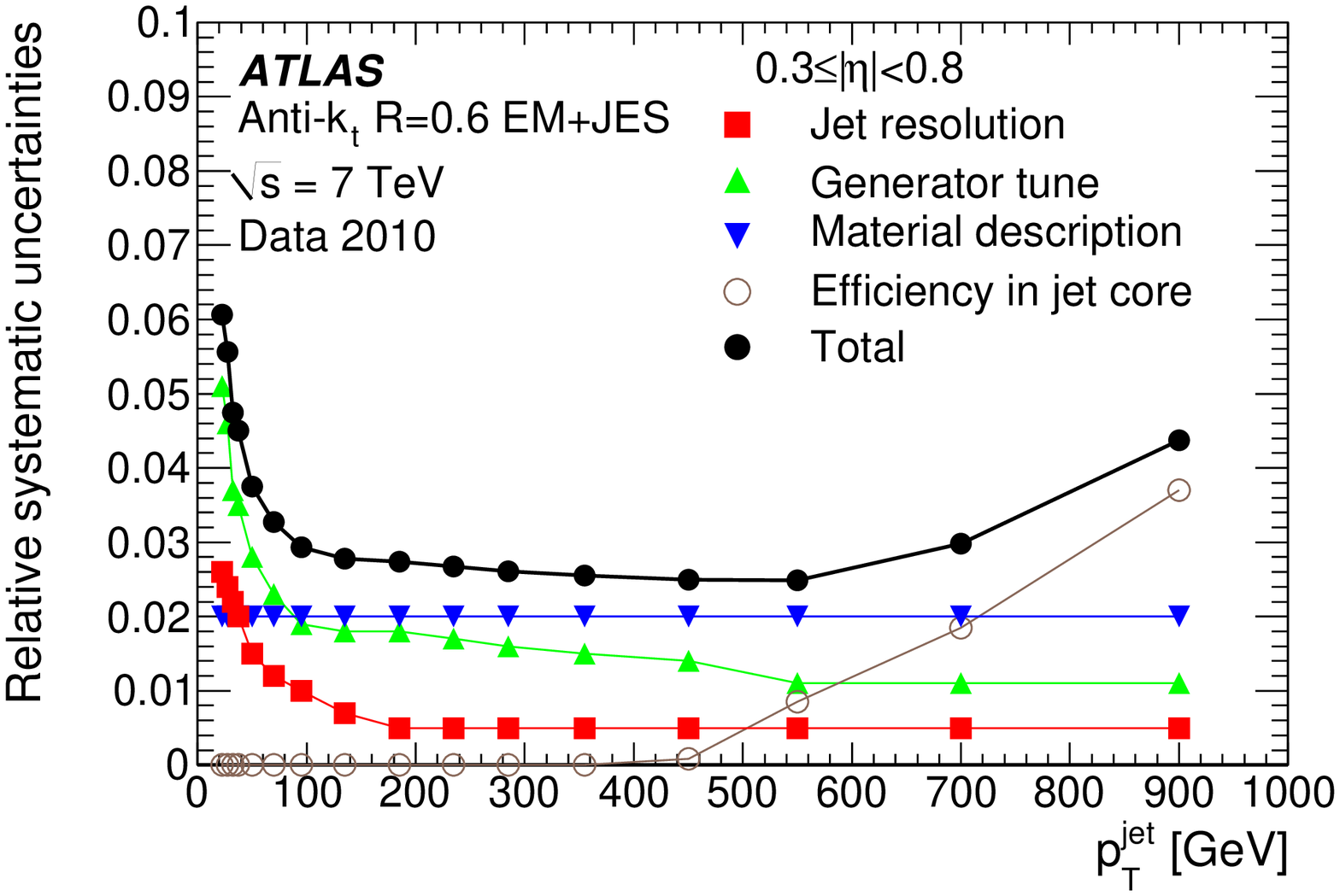,width=0.5\textwidth}
}\\
\subfloat[\etaRange{0.8}{1.2} ]{
\epsfig{file=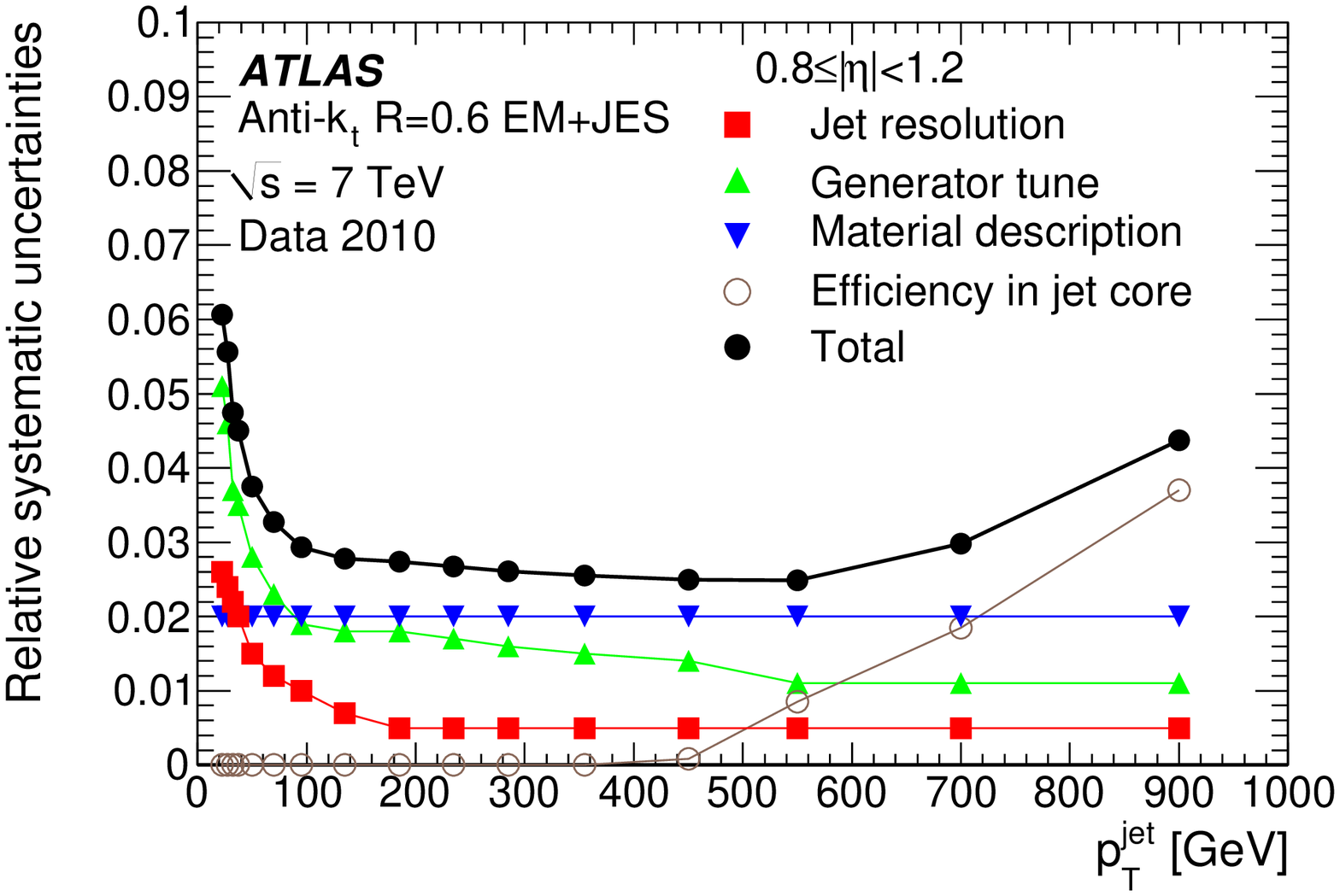,width=0.5\textwidth}
}
\subfloat[\etaRange{1.2}{1.7} ]{
\epsfig{file=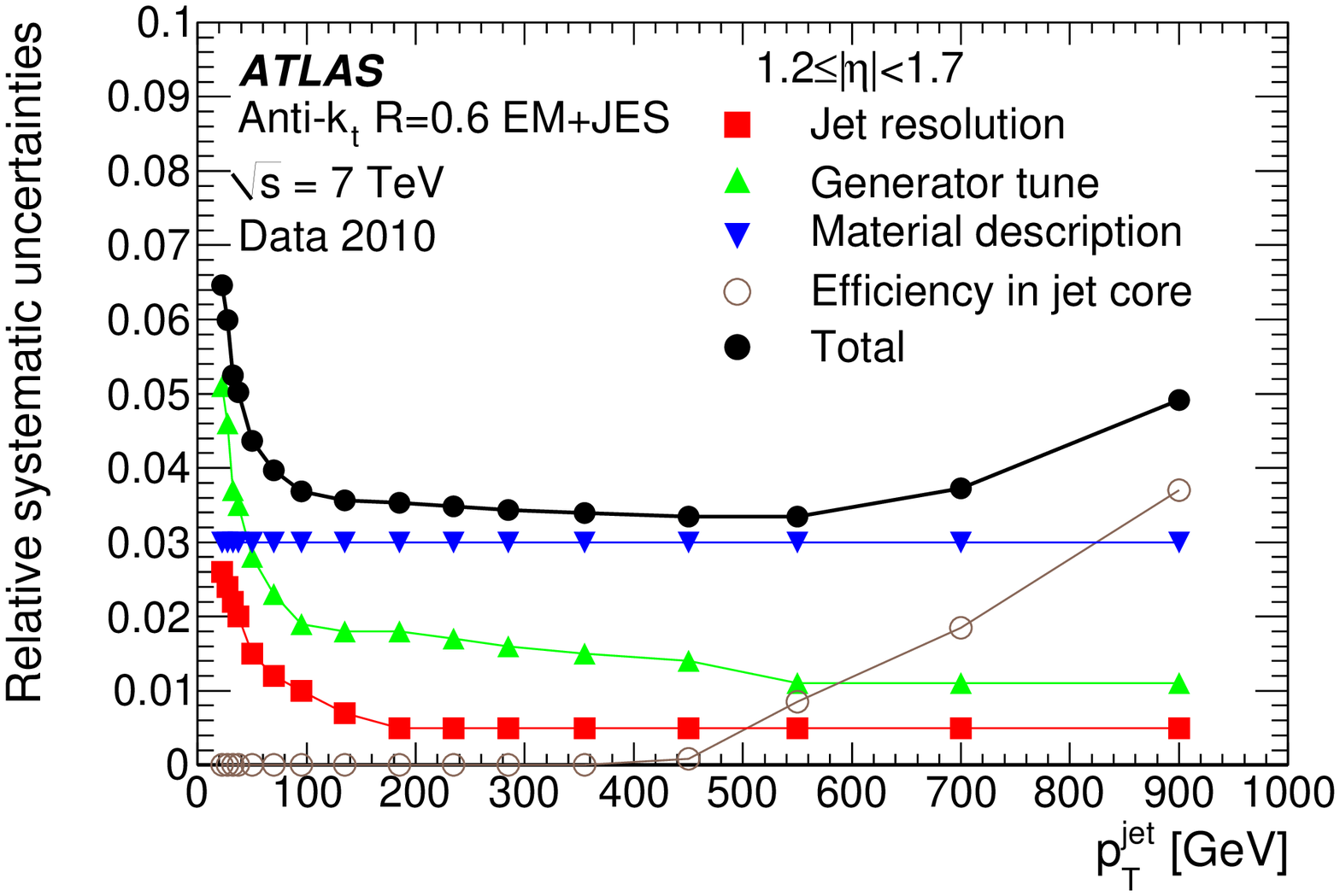,width=0.5\textwidth}
}\\
\subfloat[\etaRange{1.7}{2.1} ]{
\epsfig{file=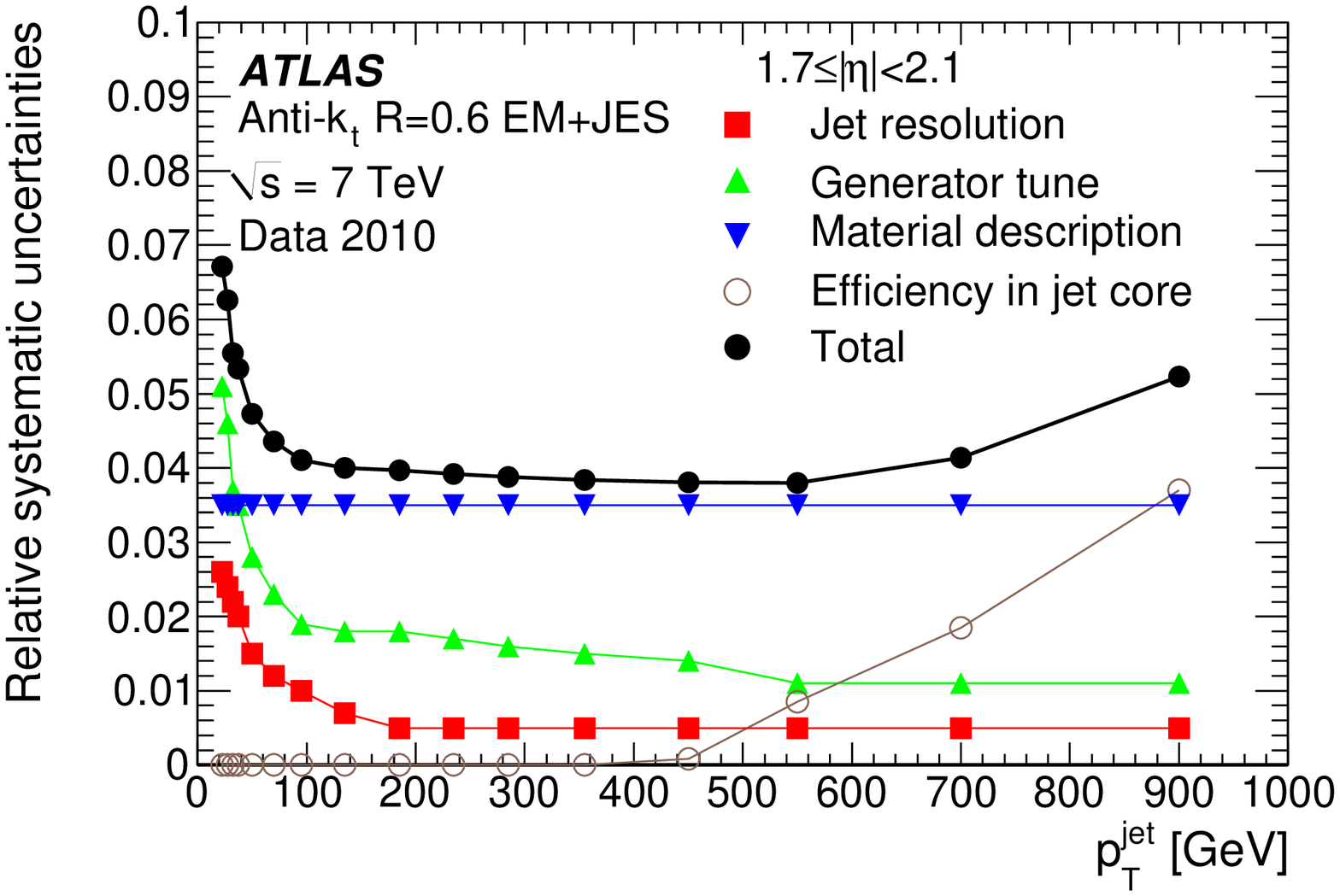,width=0.5\textwidth}
}\\
\caption{Relative systematic uncertainty on the \JES{} obtained by
comparing the total momentum of tracks associated to jets to the calorimeter measurements 
for different \etajet\ regions for \antikt{} jets with $R=  0.6$ calibrated with the \EMJES{} scheme
as a function of \ptjet.
The total and the individual systematic uncertainties, as evaluated from
the inclusive jet Monte Carlo simulation, are shown.
}
\label{fig:totalSys}
\end{figure*}

\subsubsection{Systematic uncertainties}
\label{sec:systematics}
The systematic uncertainties associated with the method using the total track momentum
to test the JES are discussed below.

\paragraph{Generator model dependence} 
While basic isospin arguments constrain the mean fraction of the jet momentum
observed in charged tracks, the prediction for \rtrk\ does depend
on details of the physics model used in the Monte Carlo generator.  
Systematic uncertainties arise from: 
\begin{enumerate}
\item The parametrisation of the fragmentation function and of the underlying event 
      (which mainly affect the fraction of the momentum carried by particles below the $\pt = 1$~\GeV{}
      cut used for this analysis).
\item The model of colour reconnection (which can change the distribution of particles with low momenta). 
\item The probability of producing strange quarks and baryons (which are iso-doublets rather than
      iso-triplets like the pion) and of producing iso-scalars such as the $\eta$.

\end{enumerate}
The size of these uncertainties has been
estimated by studying a wide range of \pythia\ tunes\footnote{Additional information about
the \pythia\ tunes can be found in Ref.~\cite{Perugia2010}.}.
A list of the \pythia\ tunes studied is given in Table~\ref{tab:generators}.

These studies have been done at the generator level and have been cross-checked using
simulated samples when the appropriate tunes were available with full
simulation.  

The data have also been compared to default tunes of
\herwigpp{} and \herwig+\jimmy.
\pythia\ tune 117, and the default \herwigpp\ and \herwig+\jimmy{} tunes are not 
consistent with the measured $f(z)$ distributions.%
Since these generators do not described the fragmentation functions measured by \ATLAS{} \cite{jetfragmenationpaper}
they are excluded from consideration when determining the systematic uncertainty on the JES measurement.

At low \ptjet, the variations between tunes arise mainly from differences in 
the hardness of the jet fragmentation, which affects the fraction of
charged particles falling below the $1$~\GeV{} cut on \pttrk.   In general,
\pythia\ tunes that include the ``colour annealing'' model of colour 
reconnection exhibit harder fragmentation  than similar tunes without colour annealing.  
At high \ptjet, differences among tunes are primarily associated with the
strangeness and baryon content of the truth jets. Versions of \pythia\ tuned
to LEP data (including flavour-dependent fragmentation measurements) using the
tuning software {\sc Professor}~\cite{delphiprofessor} in general 
show a charged fraction about $1 \%$ higher than the other tunes considered here.  
Using a conservative approach, 
the value of systematic uncertainty has been symmetris\-ed around the
baseline tune using the absolute value of the largest deviation of the tunes
considered from the baseline. 

\begin{table*}[ht!]
\begin{center}
\begin{tabular}{l|l|l}
\hline \hline
Tune Name & {\sc pytune} Value & Comments \\
\hline
 MC10 & $-$ & \ATLAS{} default (\pt\ ordered showering)\\
 MC09 & $-$ & \ATLAS{} default for Summer 2010 (\pt\ ordered showering) \\
 RFTA & 100 & Rick Field Tune A  $Q^2$ ordered showering\\
      & 107 & Tune A with ``colour annealing'' colour reconnection \\
      & 110 & Tune A with LEP tune from Professor \\
      & 117 & Tune 110 with ``colour annealing'' colour reconnection\\
      & 129 & Tune of $Q^2$ ordered showering and UE with Professor\\
      & 320 & \Perugia0 (\pt\ ordered showering)\\
\Perugia2010 & 327  & \Perugia0 with updated fragmentation 
and more parton radiation\\
\hline \hline
\end{tabular}
\end{center}
\caption{\pythia\ generator tunes used to study the systematic uncertainty on
the prediction for \rtrk.  Tunes specified by number (e.g. 100) 
refer to the value of the {\sc pytune} parameter~\cite{Perugia2010}. A dash in the table indicates
that the particular tune has no {\sc pytune} value.}
\label{tab:generators}
\end{table*}

\paragraph{Inner detector material description} 
The dominant systematic uncertainty on the reconstruction efficiency for 
isolated tracks is derived from the uncertainty on
the simulation's description of material in the inner detector.  
The systematic uncertainty on the efficiency is independent of \pttrk\
for tracks with $\pttrk > 500$~\MeV{} but is $\eta$-dependent, ranging from
$2 \%$ for  $|\etatrk|<1.3 $ to $7\%$ for $2.3 \le |\etatrk| < 2.5 $~\cite{MinBias2}.
Convolving these uncertainties with the appropriate $\etatrk$ distributions
results in systematic uncertainties on \rtrk\ that range from
$2 \%$ for jet pseudorapidities $|\etajet| < 1.2 $ to $3.5 \%$ for pseudorapidities
$1.7 \le |\etajet| < 2.1$.

Uncertainties in the material distributions also affect the probability 
that photon conversions produce charged particles that can be included 
in the \rtrk{} measurement.  The track selection used here requires at least 
one \Pixel{} hit and most of the material in the \ID{} is at a larger radius than 
the \Pixel{} detector, resulting in a small systematic uncertainties associated with rate of 
conversions.

\paragraph{Tracking efficiency in the jet core} 
There are several effects that change the tracking efficiency and resolution
inside a jet compared to those for isolated tracks:
\begin{enumerate}
\item When two tracks are close together, their hits may
overlap.  While the pattern recognition software allows tracks to share
hits, the resolution is degraded since the calculated position of the hit
is affected by the presence of the other track.  The probability of
not assigning hits to tracks increases.
\item When the hit density becomes high in the core of the jet, failures in
the pattern recognition may result in the creation of tracks by combining
hits that in fact came from several particles.  
Such tracks are called {\it fake tracks}.
\item When two high-\pt\ tracks are close together in space, they
will share hits over many layers.  In this case, one of the two tracks
may be lost.  This effect,  referred to as {\it loss of efficiency}, 
becomes more important as the \ptjet\ increases.
\end{enumerate}
The reliability of the simulation to predict the size of these effects
depends on whether the software properly models merging of \ID{} hits.  
Detailed comparisons of the data and Monte Carlo simulation indicate that 
the simulation accurately reproduces the
degradation of response in the jet core and  models the
degradation in resolution well.  Comparison of the fraction of tracks
with $z>1$ in data and Monte Carlo simulation constrains 
the size of the non-Gaussian tails in the track resolution.
Any residual difference in resolution
between data and simulation is absorbed in the quoted uncertainty due to 
\ID{} alignment.

Fake tracks and loss of efficiency are studied in the simulation using a 
hit-based matching algorithm using truth jets.
These studies indicate that the rate for reconstructing fake tracks remains 
at  $0.1 \%$ for the full \ptjet\ range considered here, but
that there is loss of tracking efficiency near the core of high-\pt\ jets.
This effect has a negligible effect on \rtrk\ for jets 
with $\ptjet < 500\,\GeV$, but increases with \ptjet\ 
such that on average $\sim 7.5 \% $ of the charged track momentum is lost for
jets in the range $800 \le \ptjet < 1000\,\GeV$.
A relative uncertainty of $50 \%$\ is assigned  to the value of the inefficiency 
that is caused by merged hits.
While this effect gives the largest systematic uncertainty 
on the JES for $\ptjet \gtap ~600$~\GeV{} 
($1.9\%$\ for $600 \le \ptjet < 800 \,\GeV$ and $3.7\%$ for 
$800 \le \ptjet < 1000 \,\GeV$), it is still 
smaller than the present statistical uncertainty at these values of  \ptjet.
 
\paragraph{Inner detector alignment} 
For high \pt\ tracks, the momentum resolution achieved in the \ID{} is
worse than that of the simulation.  This degradation in resolution is
attributed to an imperfect alignment of the \ID.  The systematic uncertainty
on \rtrk\ is obtained by degrading the tracking resolution in 
the simulation.  The size of this additional resolution smearing is
determined by studying the width of the measured mass distribution for $Z$-decays
$Z\rightarrow \mu^+\mu^-$.
This procedure results in a systematic uncertainty of less than $0.2\% $ for all \ptjet\  and \etajet. 

\paragraph{Calorimeter jet \pt{} resolution} 
The systematic uncertainty due to jet transverse momentum 
resolution \cite{Resolution2010}
is determined by smearing the jet four-momentum 
(without changing \etajet\ or \phijet ) in Monte Carlo simulation.  
The relative uncertainty on the \ptjet\ resolution is $5\%$\ for $0  \le |\etajet| < 0.8$ and $10\%$\ for
$0.8 \le |\etajet| < 2.1$. %
The effect of this variation is largest for low values of \ptjet\ 
and high values of \etajet;
for $\ptjet < 40 $~\GeV{} and $0.8 < |\etajet| < 2.1 $ the uncertainty on
$R_{\rtrk}$ is $\sim 2\%$.

\paragraph{Combined systematic uncertainty} 
The above uncertainties are assumed to be uncorrelated and are combined in quadrature.  
The resulting total uncertainties are shown in Figure~\ref{fig:totalSys} as a function of \ptjet\ for several regions of \etajet.

\begin{figure*}[ht!!p]
\begin{center}
\vspace{-0.2cm}
\subfloat[$| \etajet | < 0.3$]{
\includegraphics[width=0.39\textwidth]{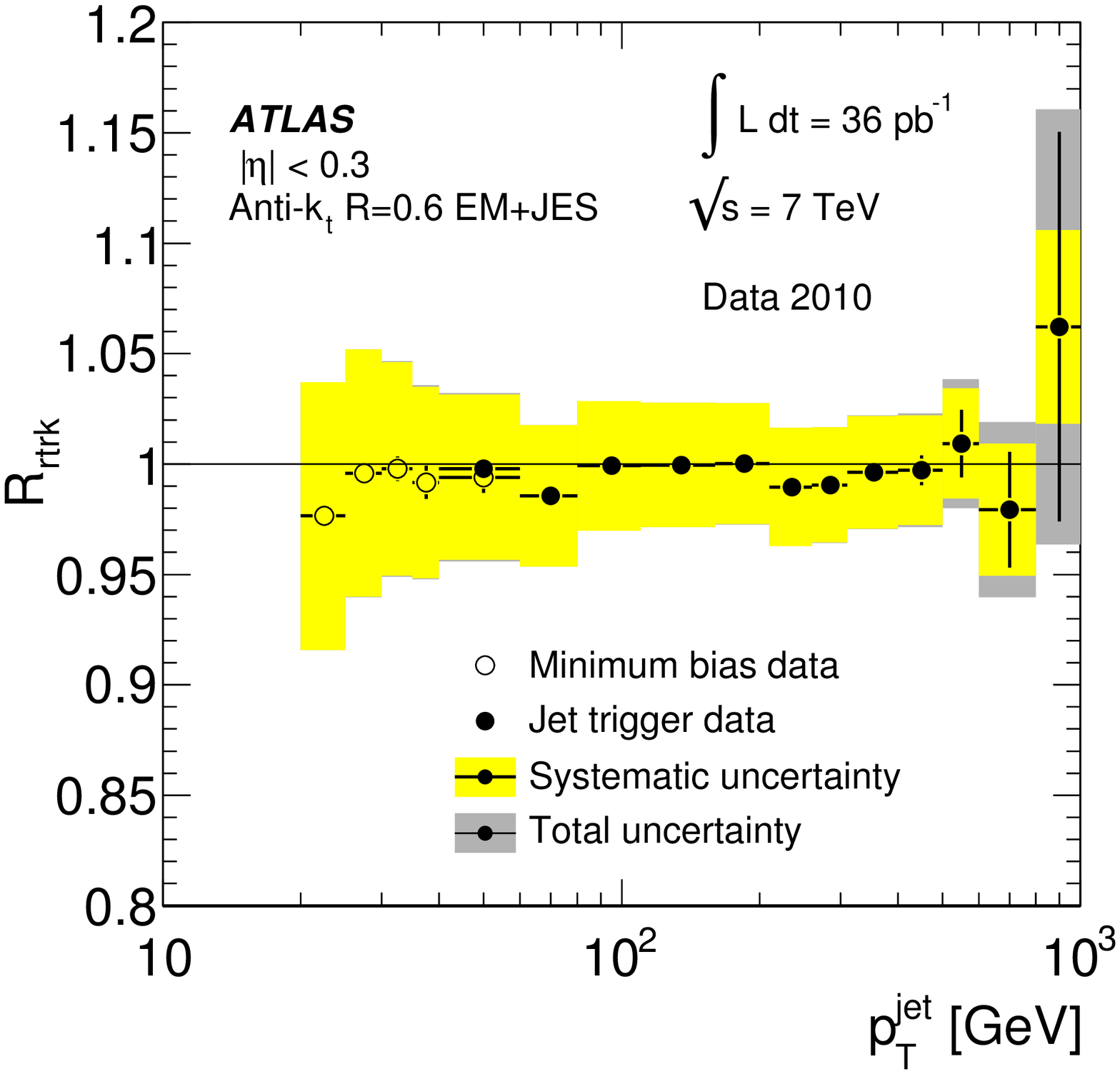}}
\subfloat[\etaRange{0.3}{0.8}]{
\includegraphics[width=0.39\textwidth]{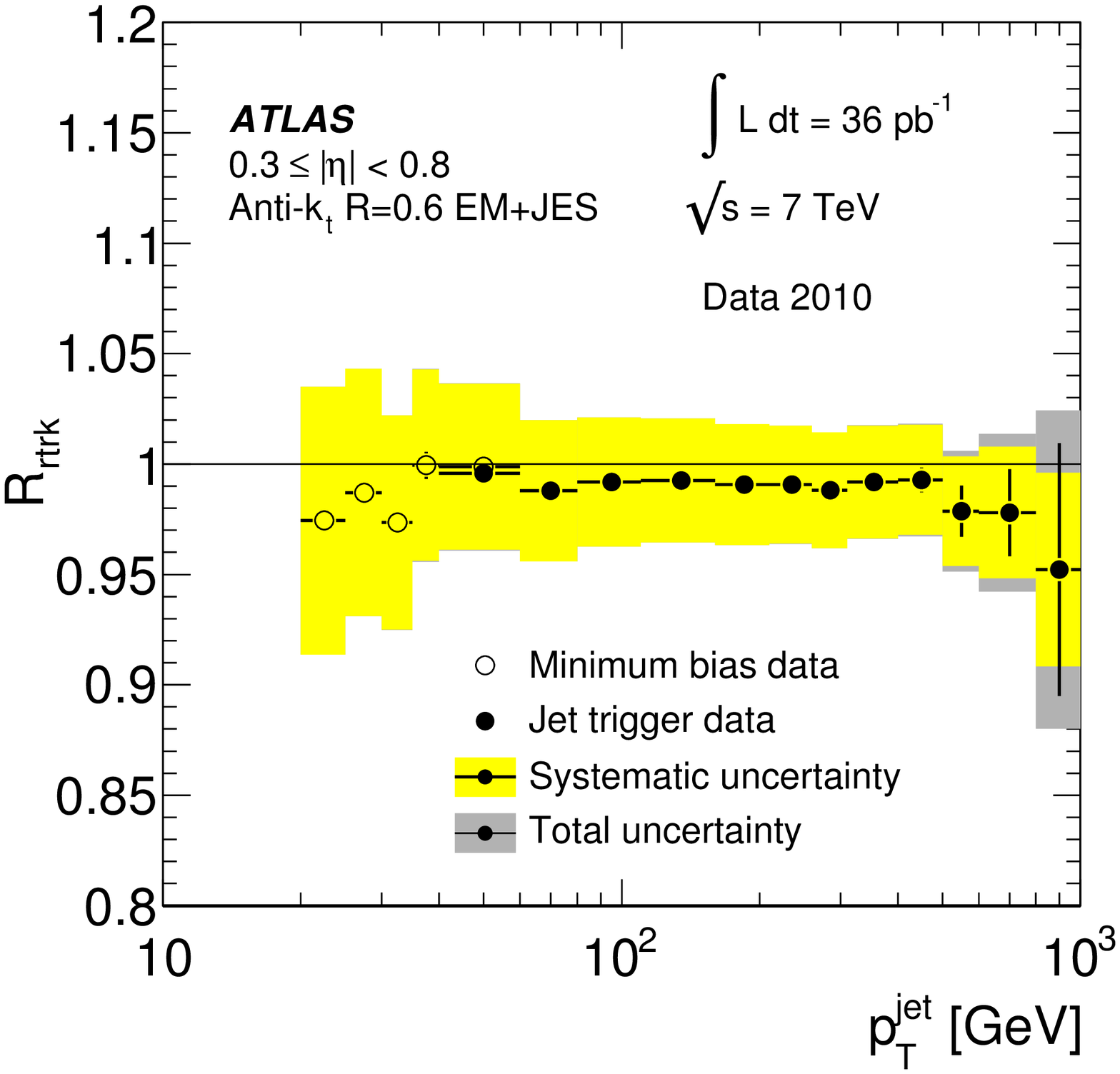}}
\vspace{-0.1cm}
\\
\subfloat[\etaRange{0.8}{1.2}]{
\includegraphics[width=0.39\textwidth]{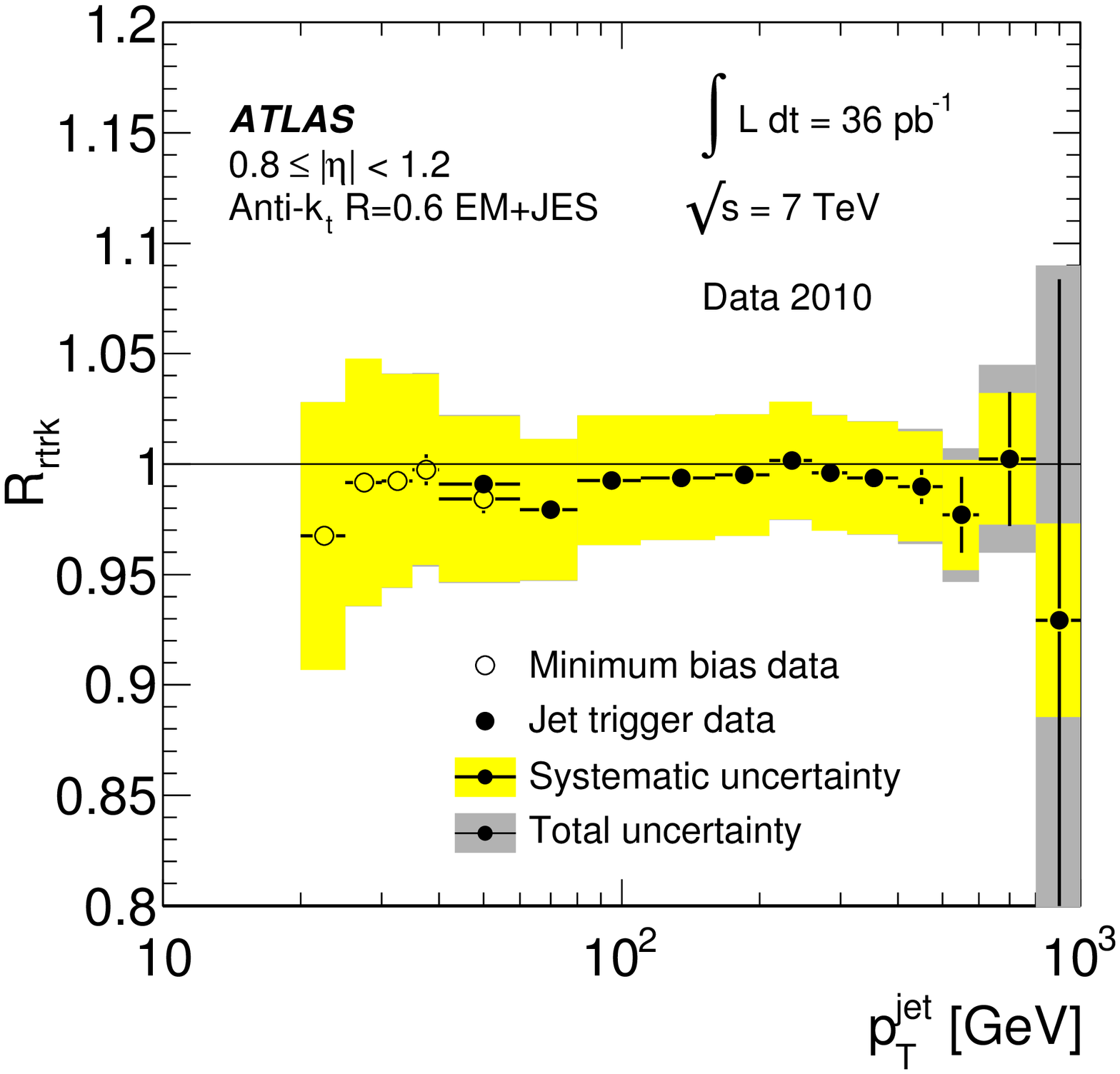}}
\subfloat[\etaRange{1.2}{1.7}]{
\includegraphics[width=0.39\textwidth]{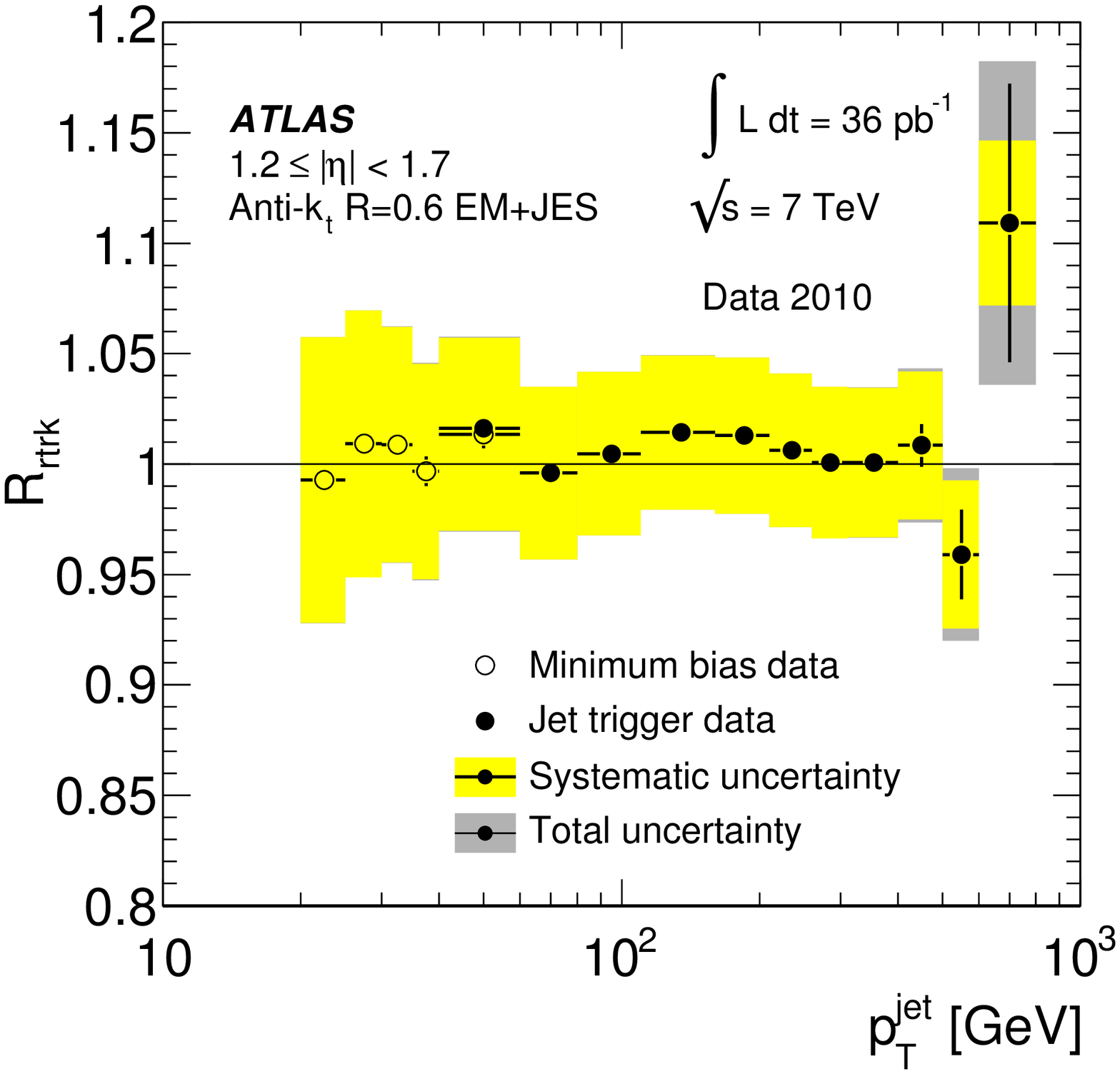}}
\vspace{-0.1cm}
\\
\subfloat[\etaRange{1.7}{2.1}]{
\includegraphics[width=0.39\textwidth]{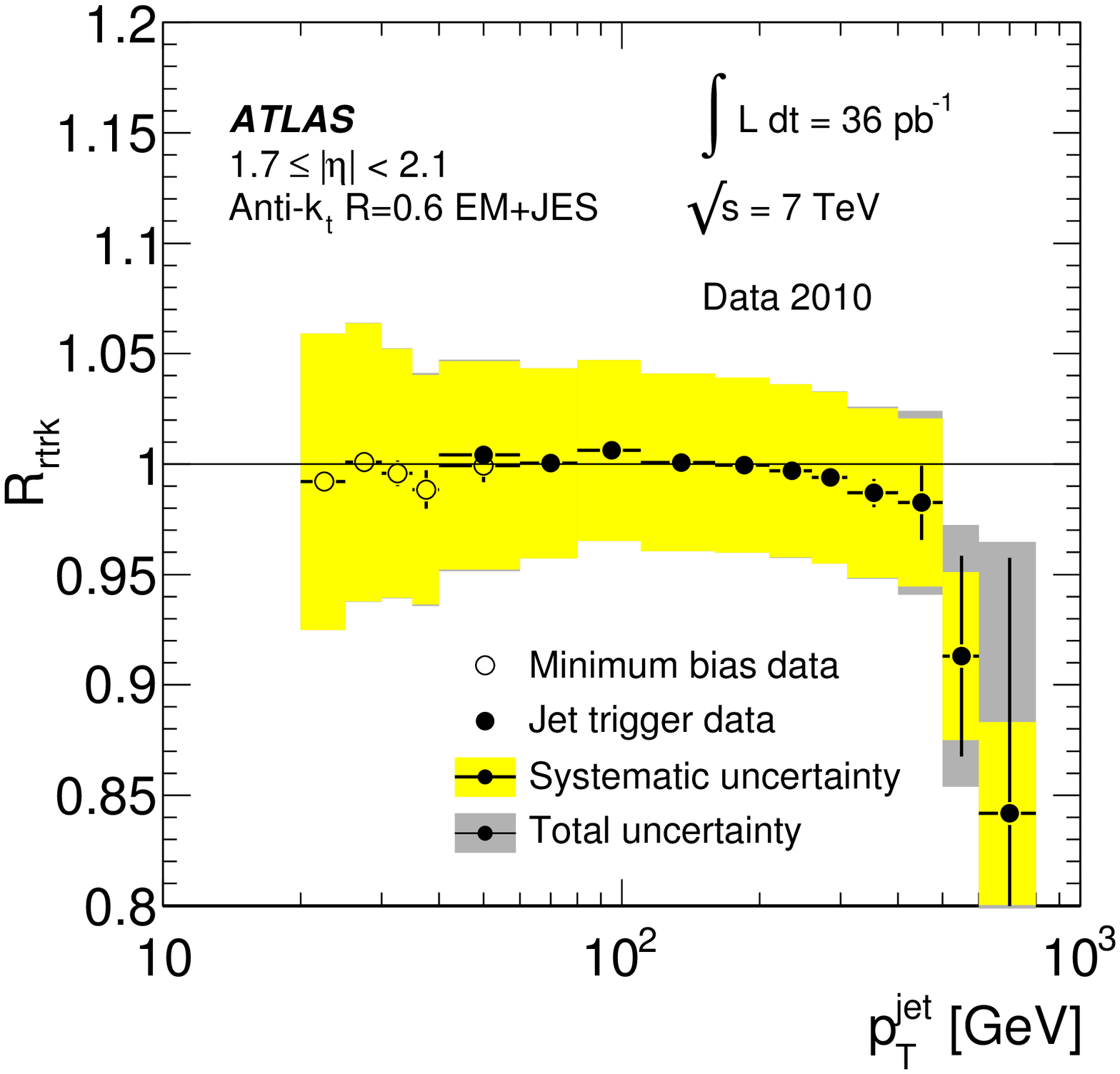}}
\subfloat[\AetaRange{1.2}]{
\includegraphics[width=0.39\textwidth]{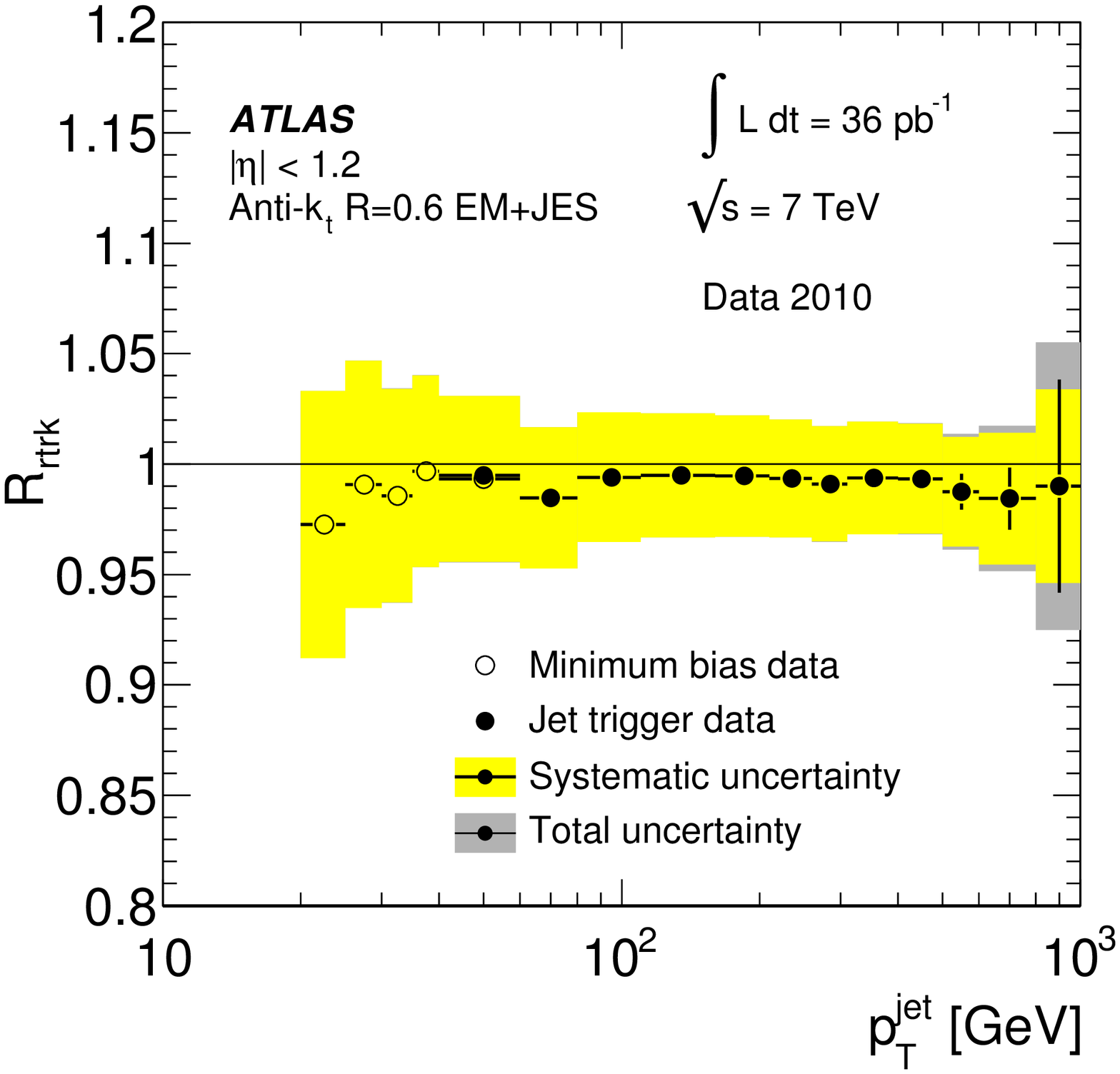}
}\\
\vspace{-0.1cm}
\end{center}
\caption{Double ratio of the mean track to calorimeter response ratio
in data and Monte Carlo simulation 
$R_{\rtrk} = {[\rtrk]}_{\rm Data}/{[\rtrk]}_{\rm MC}$
for \antikt\ jets with $R = 0.6$ calibrated with the \EMJES{} scheme 
as a function of \ptjet{} for various \etajet{} bins.
Systematic (total) uncertainties are shown as a  light (dark) band. 
}
\label{fig:finalAktopo6}
\end{figure*}

\subsubsection{Summary of JES uncertainty from tracks}
\label{sec:finalResultstracks}
Final results for \antikt\ jets with $R = 0.6$ and \EMJES{} corrections are
shown in Figure~\ref{fig:finalAktopo6} for five bins in \etajet{}
with the derived systematic uncertainties.  
To facilitate comparisons at high \ptjet, where the statistical uncertainties
are large, the combined data from the three bins with $|\etajet| < 1.2$ 
are also displayed.
Averaging all data with  $\ptjet > 25\,\GeV$ and $| \etajet | < 1.2$ 
yields a value of \rtrk\ that
agrees  with the simulation to better than $1\%$. This small
discrepancy is well within the quoted systematic uncertainty, which is highly
correlated between bins in \ptjet.  No significant variation of $R_{\rtrk}$
with \ptjet\ is observed.
For $|\etajet|>1.2$, the statistical
uncertainties are large for $\ptjet > 500\,\GeV$.  For $\ptjet < 500\,\GeV$, the
level of agreement between data and simulation is similar to that obtained
at low \etajet.

In summary, $\rtrk$, the ratio of track to calorimeter
transverse momentum, is used to validate the \JES{} for  
\antikt\ jets with $R = 0.4$ and  $R = 0.6$ calibrated with the \EMJES{} calibration scheme.
Systematic uncertainties associated with jet modelling and track reconstruction
are assessed and the me\-thod is shown to provide a \JES{} uncertainty evaluation independent
of the modelling of the calorimeter response.
Systematic uncertainties are below $3\%$ for $0 \le |\etajet| < 0.8$  and rise to $\sim 4\%$
for $1.7 \le |\etajet| < 2.1$ for $40 \le \ptjet < 800$~\GeV.
The results agree within systematic uncertainties with
those predicted using the \ATLAS{} calorimeter simulation and
provide an independent estimate of the overall jet energy scale
and its uncertainty.

\subsection{Photon-jet transverse momentum balance}
\label{sec:gammajet}

\begin{figure}
\begin{center}
        \includegraphics[width=0.49\textwidth]{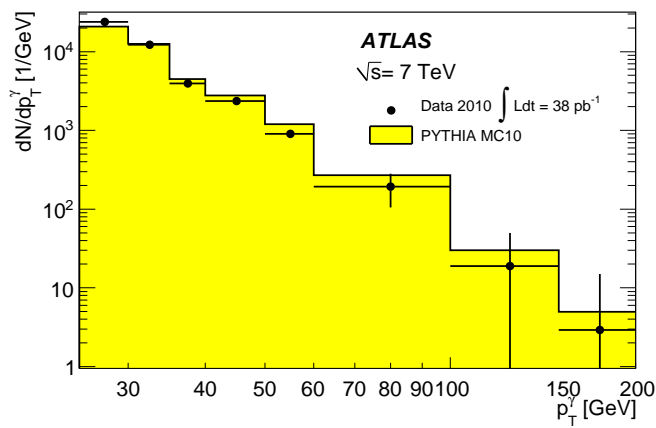}
        \caption{Distribution of the photon transverse momentum for
          events passing the photon selection criteria described in
          Section~\ref{sec:GammaJetEvtSel}.
           A correction is
          made in the first $\pt^{\gamma}$ bin for the pre-scale applied to the
          trigger in this $\pt^{\gamma}$ range. The Monte Carlo simulation is normalised
          to the observed number of events observed in data and corrected for the trigger pre-scale. 
          Uncertainties are statistical only.}
        \label{fig:PhotonPt}
\end{center}
\end{figure}
%
\begin{figure*}[ht]
  \centering
  \subfloat[Direct \pt{} balance technique]{\includegraphics[width=0.42\textwidth]{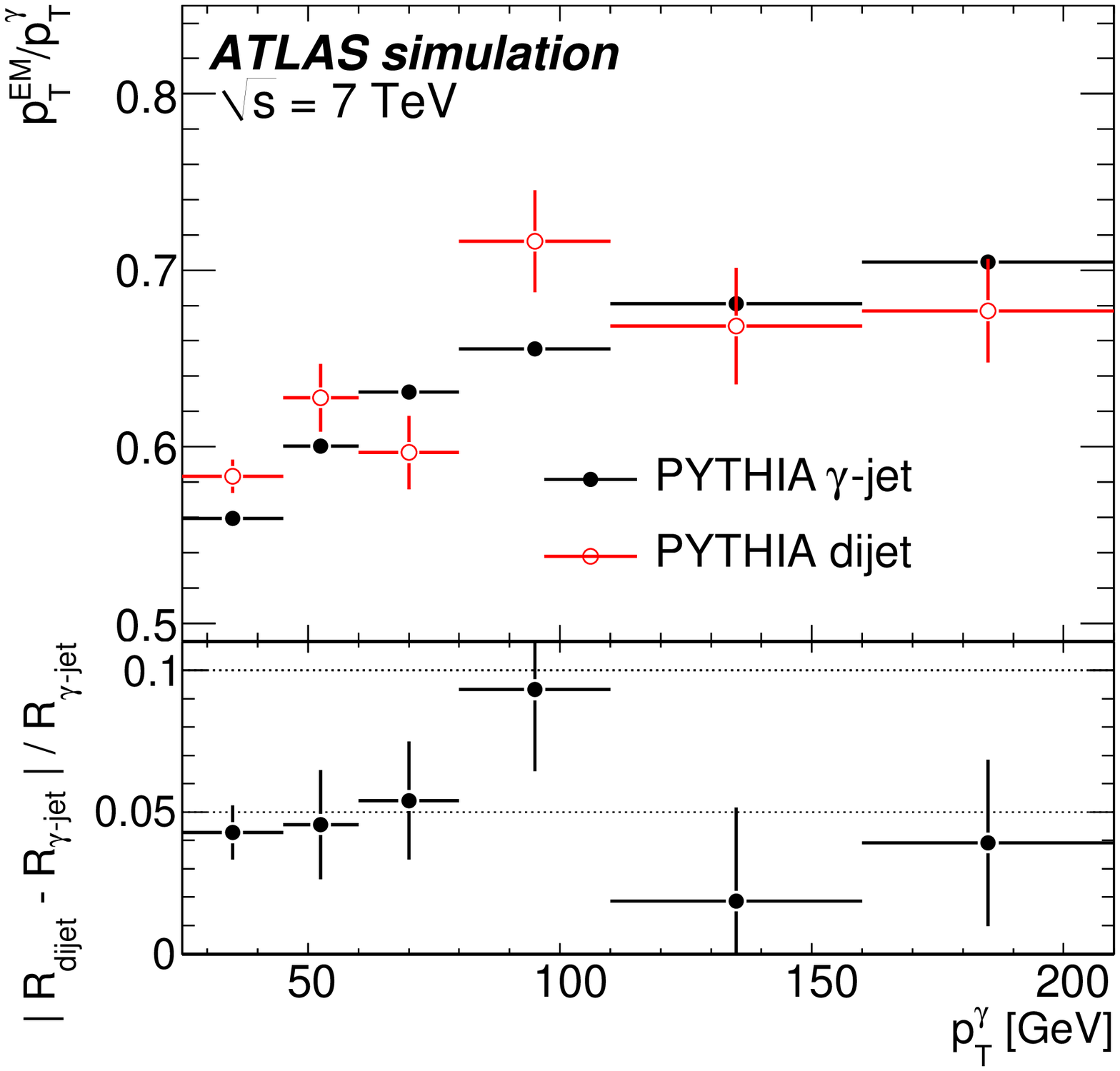}}
\hspace{1.cm}
  \subfloat[MPF technique]       {\includegraphics[width=0.42\textwidth]{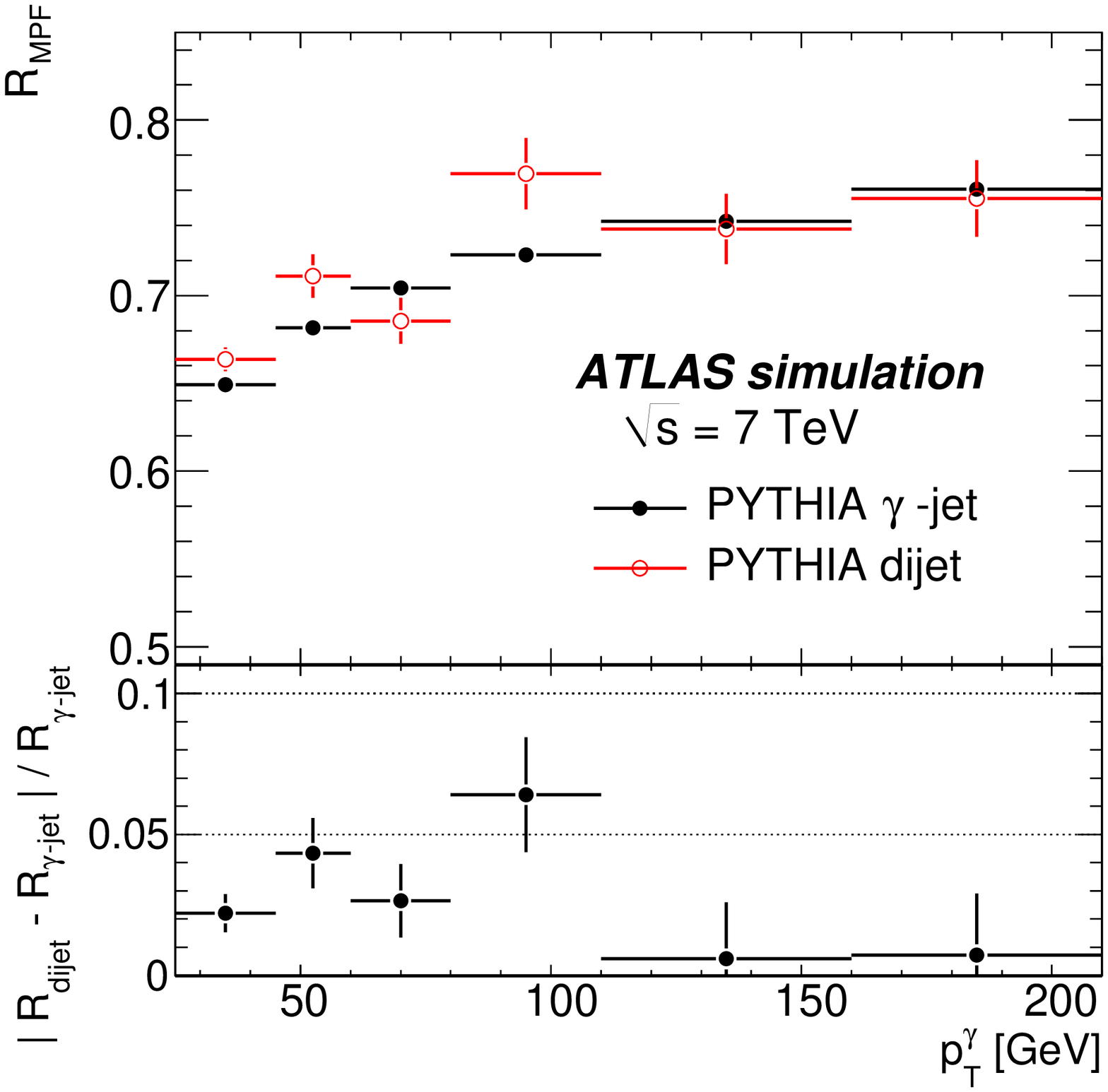}}
  \caption{Average jet response measured at the \EM{} scale 
    as a function of $\pt^{\gamma}$ as determined by the direct \pt{} balance technique for \antikt{} jets with $R = 0.6$
    (a) and by the \MPF{}  technique (b) for \gammajet{} events and dijet events where one jet has been
    reconstructed as a photon, as derived in the Monte Carlo simulation.
    The lower part of the figures shows the absolute response difference between
    the dijet and \gammajet{} events with respect to the response of \gammajet{} events.
    Only statistical uncertainties are shown.
}
  \label{fig:Dijet_Response_Diff}
\end{figure*}

In \gammajet{} events, a jet recoils against a photon at high transverse momentum. 
The photon energy, being accurately measured in the electromagnetic
calorimeter, is used as a reference. Such a topology can be used to
validate the jet energy measurement.
Any discrepancy between data and simulation may be taken
as an uncertainty on the jet energy calibration.

Two methods of balancing the photon and the recoiling jet transverse momentum with different
sensitivities and systematic uncertainties are used: 
the direct  \pt{} balance technique
and
the missing transverse momentum  projection fraction technique. 

\subsubsection{Direct transverse jet momentum balance technique}
\label{sec:directbalance}
The direct \pt{} balance technique exploits the approximate transverse
momentum balance in events with only one photon and one jet with high \pt.
The ratio of the jet \pt{} to the photon \pt{} ($\ptjet/\pt^\gamma$) is used to estimate the jet response.  
Since the photon \pt{} is well-measured and well-described by the simulation,
the quality of the jet \pt{} calibration can be assessed by comparing data and Monte Carlo simulation
using the ratio $\ptjet/\pt^\gamma$.
\index{Direct $\gamma$-jet balance}
This technique was used at the CDF experiment~\cite{cdf06}. 

\subsubsection{Missing transverse momentum projection fraction technique}
\label{sec:MPF}
The missing transverse momentum (\Etmiss) projection fraction (\MPF) technique exploits
the momentum balance, in the transverse plane, of the photon and the
hadronic recoil to derive the detector response to jets. 
This technique has been used in the past for the D0 experiment~\cite{ref:D0_MPF}. 
\index{Missing $E_{\rm T}$ projection fraction technique}

The missing transverse momentum vector (\vecEtmiss) is defined as the opposite of the 
vector sum of the transverse projections of calorimeter energy deposits.
The missing transverse momentum is calculated from the energy deposits in the calorimeter
cells that are included in \topos. 
The calorimeter cell energy is computed using the same calibration as the one used
in the jet calibration scheme to be tested.
The missing transverse momentum is corrected for the photon four-momentum.
The reconstructed jet four-momentum is not directly used in the missing transverse momentum calculation.

The \MPF{} technique is based on the assumption that the only missing transverse momentum
in a \gammajet{} event arises from calorimeter non-compensation, signal losses due to noise
suppression and energy losses in the non-active regions of the
detector  by the hadronic jet.
The transverse momentum balance can be written as:
\begin{equation}
\vec{p}_{\rm T}^{\gamma} + \vec{p}_{\rm T}^{\rm jet} = \vec{0},
\end{equation}
where $\vec{p}_{\rm T}^{\gamma}$ and $\vec{p}_{\rm T}^{\rm jet}$ is the photon and jet 
transverse momentum vector.
The particles  produced by the hard scatter and their interaction in the
calorimeter can be expressed in terms of the observables:
\begin{equation}
\Response^{\gamma} \vec{p}_{\rm T}^{\gamma}  +
\Rcalo \vec{p}_{\rm T}^{\rm jet} = - \vecEtmiss,
\end{equation}
where $\Response_{\gamma}$ is the calorimeter response to photons.
Since the calorimeter is well calibrated for photons, $\Response_{\gamma} = 1$. 
The variable \Rcalo{} denotes the calorimeter response to jets. By using the above two
equations and projecting the \Etmiss~in the direction of
the photon the response can be written as:
\begin{equation}
\RMPF = 1 +
\frac{ \vec{p}_{\rm T}^{\gamma} \cdot \vecEtmiss }
{|p_{\rm T}^{\gamma}|^{2}},
\label{eq:JetResponse-MPF}
\end{equation}
which is defined as the \MPF{} response.

Note that the \MPF{} technique measures the calorimeter response by relying only on the photon and \Etmiss{}
quantities and does not use the jet energy directly. Therefore the \MPF{}
response is independent of the jet algorithm.

\begin{table}
\begin{center}
\begin{tabular}{l|l}
\hline \hline
 Variable & Threshold \\
\hline
$|\etajet| $ & $ < 1.2$ \\
$\pt^{\gamma}$   &   $> 25$ \GeV \\
$|\eta^{\gamma}| $ & $ < 1.37$ \\
$E_{\rm T}^{\gamma \mathrm{Isolation}}$   & $< 3$ \GeV \\
\deltaphijetgamma{}
& $ > \pi - 0.2$ rad \\
$\pt^{\rm jet2} / \pt^{\gamma} $ & $<$ $10 \%$ \\
\hline
\hline
\end{tabular}
\caption{Criteria used to select events with a photon and a jet with high transverse momentum.}
\label{tab:EvtSel}
\end{center}
\end{table}

\begin{figure*}[ht!]
  \centering
  \subfloat[]{\includegraphics[width=0.42\textwidth]{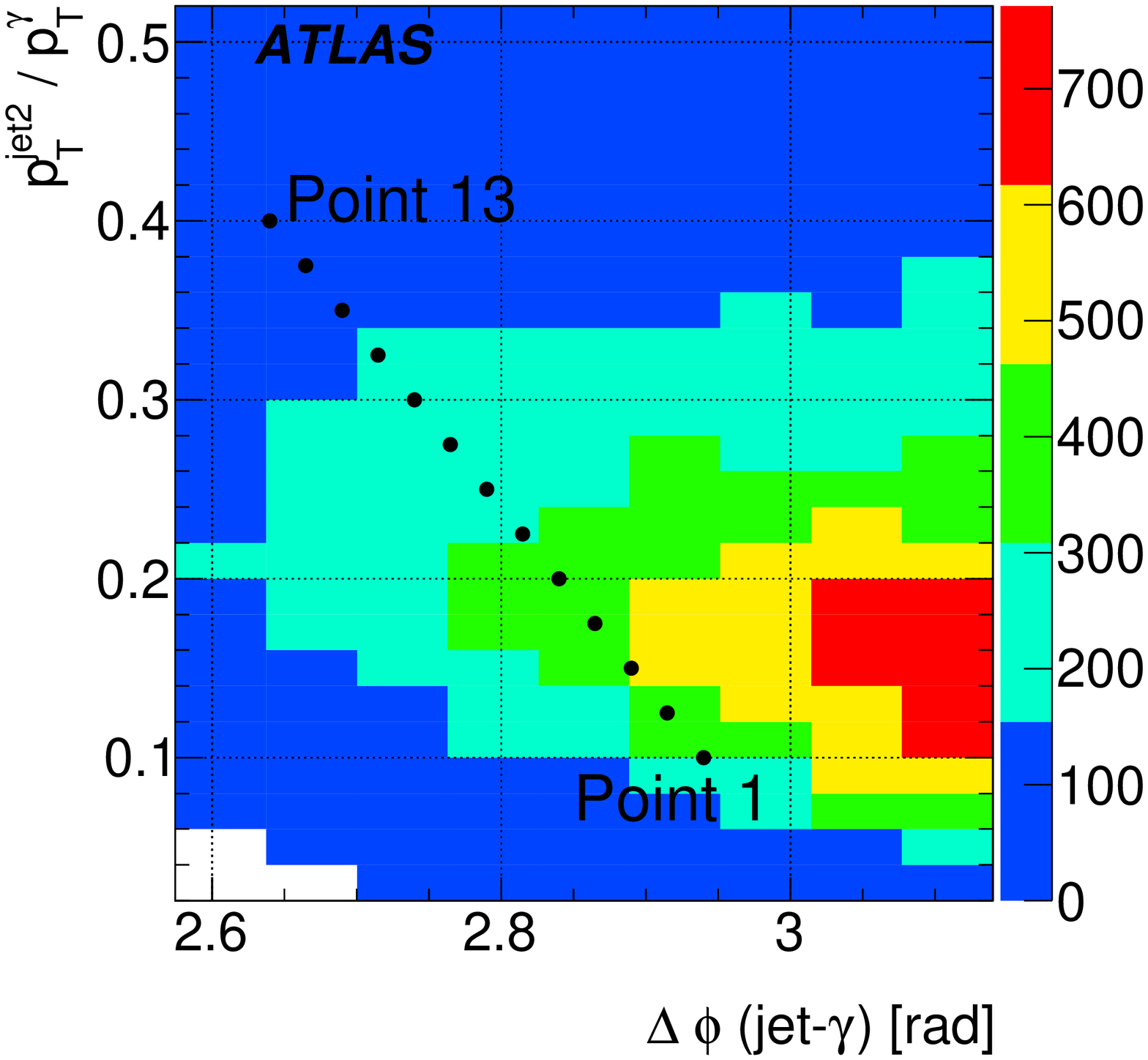}}
\hspace{1.cm}
  \subfloat[]{\includegraphics[width=0.42\textwidth]{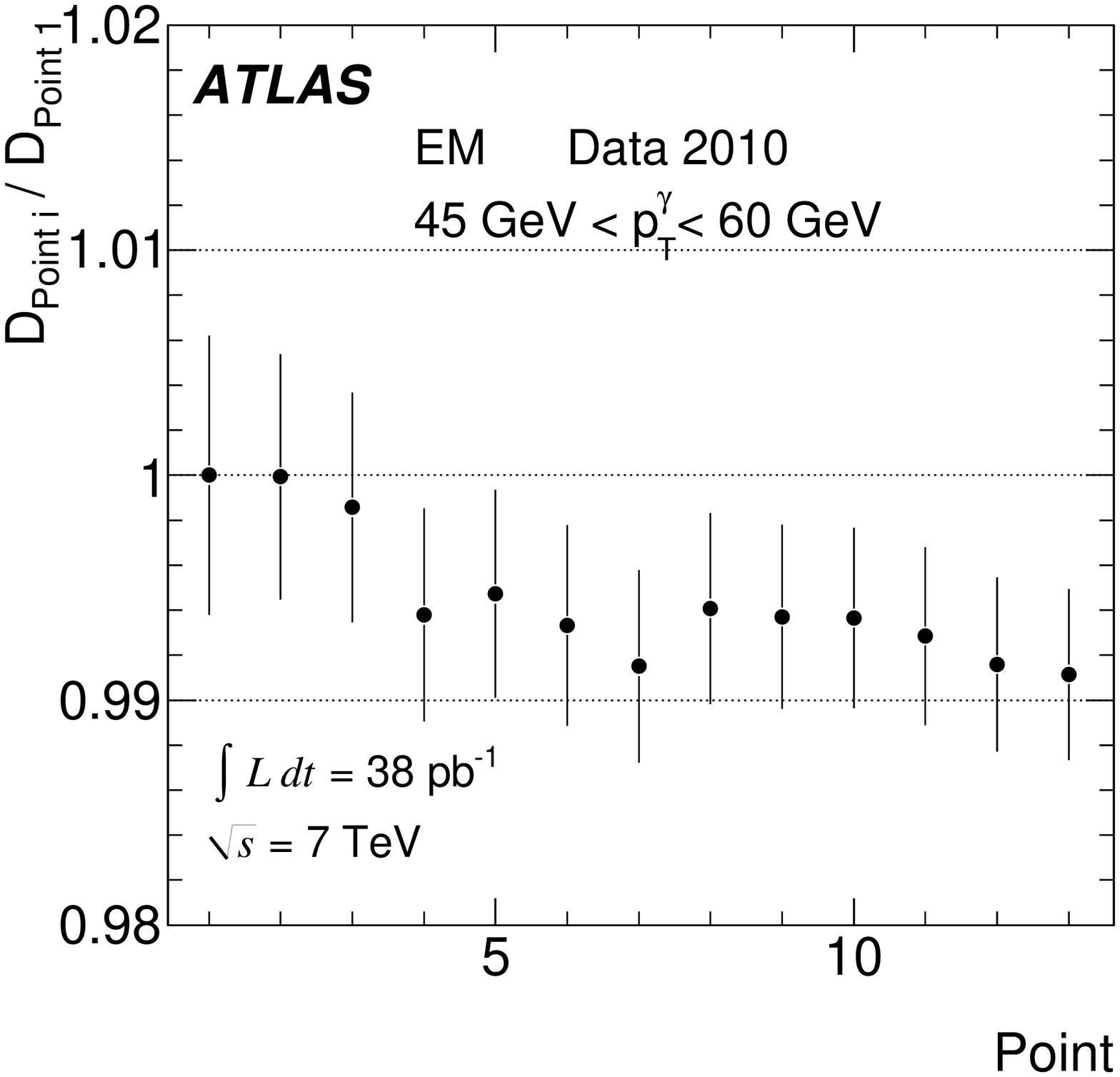}}
 \caption{The values of radiation-suppressing cut thresholds (points)
   used to probe the soft QCD radiation systematic uncertainty, as a
   function of \deltaphijetgamma{} and 
   $\pt^{\rm jet2} / \pt^{\gamma}$ overlaid with the number of events observed in data (a). 
   The nominal selection is the bottom-rightmost point labelled ``Point 1''.  
   Relative change in the \MPF{} response between data and Monte Carlo simulation (b), 
   defined as $D = \left [ \RMPF \right]_{\rm Data}/\left [\RMPF \right]_{\rm MC}$
   from the point given on the $x$-axis to point 1,
   when relaxing the soft QCD radiation suppression as indicated in (a). 
   Only statistical uncertainties are shown.
   }
\label{fig:RadSys_Example}
\end{figure*}

%
\begin{figure}[ht]
  \centering
  \includegraphics[width=0.42\textwidth]{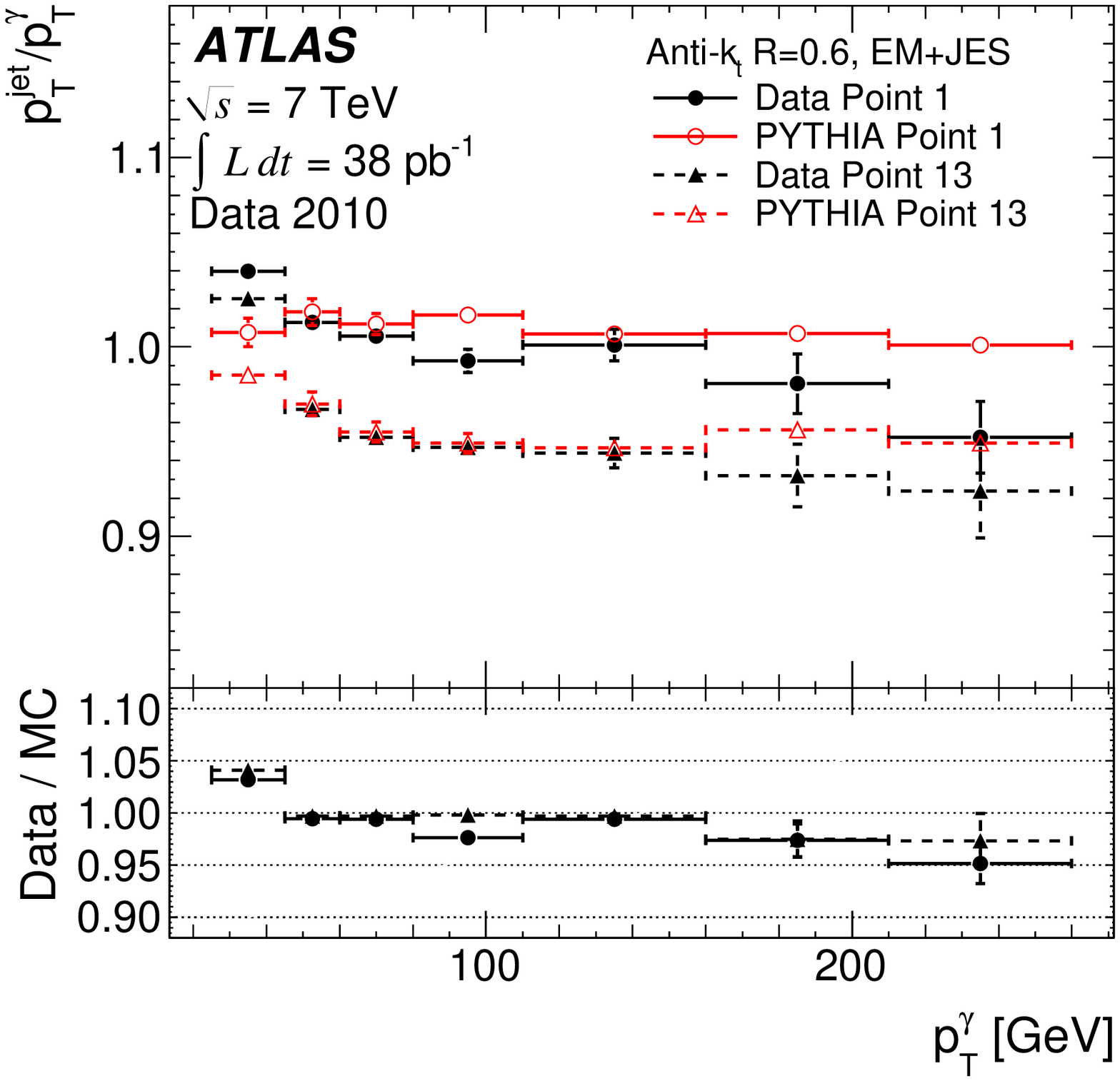}
 \caption{Average jet response as determined by the direct \pt{} balance technique with the nominal selection (Point 1)  
    and with a set of relaxed
    radiation suppression cuts (Point 13), for \antikt{} jets with $R = 0.6$ calibrated with the \EMJES{} scheme 
    as a function of the photon transverse momentum for data and Monte Carlo simulation.
    Only statistical uncertainties are shown.}
\label{fig:RadSys_DirBalanceExample}
\end{figure}

\subsubsection{Photon-jet Monte Carlo simulation sample}
\label{sec:GammaJetMCsample}
The \gammajet{} sample is simulated with the event generator  \pythia{} 
using the \ATLAS{} MC10 tune\cite{MC10}. 

The systematic uncertainty from jets which are identified as photons (fakes) are studied with an
inclusive \pythia{} jet sample using the MC09 tune \cite{MC09}\footnote{Since a large event statistics
is needed for this sample, only a sample with an older tune was available.}. 
To efficiently produce this sample
a generated event is only fully simulated if it contains at least one generated particle jet with $\pt > 17$~\GeV. 
These jets are computed from the sum of the four-momenta of all stable generated particles within a
$0.18 \times 0.18$ region in $\eta \times \phi$.
Events in the dijet sample with prompt photons, e.g. that are produced by radiation are removed.

\subsubsection{Selection of the photon-jet data sample}
\label{sec:GammaJetEvtSel}

The leading photon in each event must have 
$\pt^{\gamma} >  25$~\GeV{} and lie in the pseudorapidity range $|\eta^{\gamma}|<1.37$.
In this range the photon is fully contained within the electromagnetic barrel calorimeter. 
Furthermore, events in which the leading photon is in a calorimeter region
where an accurate energy measurement is not possible are rejected.
In each event only the leading photon is considered.

The leading photon candidate must also satisfy strict photon identification
criteria~\cite{PhotonPaper}, meaning that the pattern of energy deposition in the
calorimeter is consistent with the expected photon showering behaviour. 
The photon candidate must be isolated from other activity in the
calorimeter $(E_{\rm T}^{\gamma \; \mathrm{Isolation}})$ with an
isolation cone of size $R = 0.4$.
If the leading photon does not meet all of these
criteria, the event is rejected.

Only events are retained that fired an online trigger requiring a photon candidate with 
$\pt^{\gamma} >  20$~\GeV{} or $\pt^{\gamma} >  40$~\GeV. At the trigger level
the photon identification requirements are less strict than those of the off-line selection.

The \pt{} distribution of photons in events selected 
with the above criteria is shown in Figure~\ref{fig:PhotonPt}. 
The small discrepancies 
between the $\pt^{\gamma}$ spectrum in data and Monte Carlo simulation do not affect
the comparison of the jet response in data and Monte Carlo simulation.

The leading jet must be in the fiducial region  $|\etajet| < 1.2$.   
Soft QCD radiation can affect the \pt{} balance
between the jet and photon. 
The following two selection cuts are applied to  suppress this effect. 
To select events in which the photon and the leading jet are
back-to-back, 
$\deltaphijetgamma > \pi - 0.2$ radians is required. 
The presence of sub-leading jets is suppressed by requiring that the sub-leading jet 
has \pt{} less than
$10 \%$ of the \pt{} of the leading photon\footnote{This cut is not applied,
if it would be below the jet \pt{} reconstruction threshold of $\ptjet = 7$~\GeV.
If in this case a sub-leading jet with  $\ptjet \ge  7$~\GeV{} is present, the event is rejected.}.  
A summary of the event selection criteria can be found in Table~\ref{tab:EvtSel}.

\subsubsection{Systematic uncertainties of the photon-jet \insitu{} validation technique}
\label{sec:gammajetSystematics}
Uncertainties due to background from jets identified as photons (fakes), 
soft QCD radiation, in-time pile-up, 
non-functional cal\-ori\-me\-ter read-out regions and the photon energy scale are studied.

\begin{figure}[ht]
  \centering
  \includegraphics[width=0.49\textwidth]{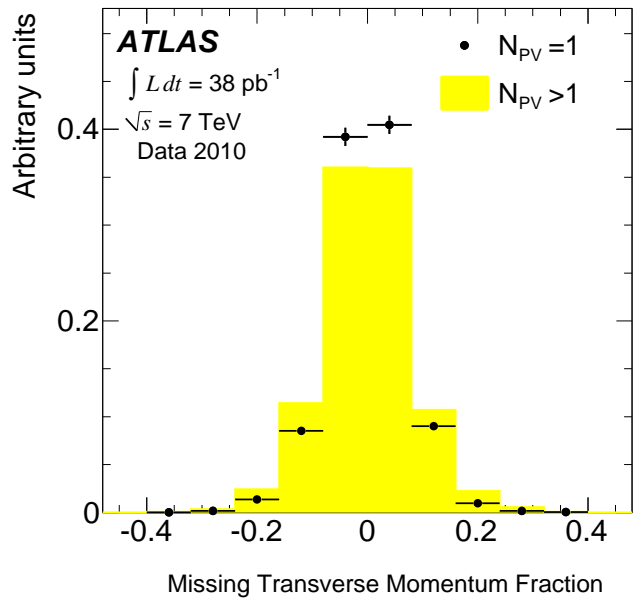}
 \caption{The missing transverse momentum fraction (MTF) distribution for data with exactly one
    reconstructed primary vertex \Npv, and with more than one reconstructed primary vertex.
    Only statistical uncertainties are shown.}
  \label{fig:MTF}
\end{figure}

%
\begin{figure}
  \centering
  \includegraphics[width=0.49\textwidth]{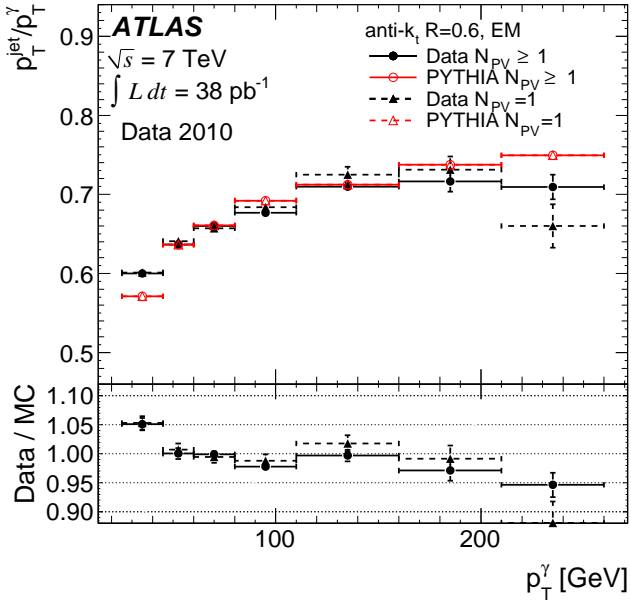}
 \caption{Average jet response for \antikt{} jets with $R = 0.6$ at the \EM{} scale as determined by the direct \pt{}
    balance technique in events with any number of reconstructed primary vertices and in events with exactly one reconstructed vertex
    as a function of the photon transverse momentum for both data and Monte Carlo simulation.
    The lower part of the figure shows the data to Monte Carlo simulation ratio.
    Only statistical uncertainties are shown.}
\label{fig:Pileup_DirBalanceExample}
\end{figure}

\begin{figure*}[ht!p]
\begin{center}
  \subfloat[Direct \pt{} balance technique]{\includegraphics[width=0.44\textwidth]{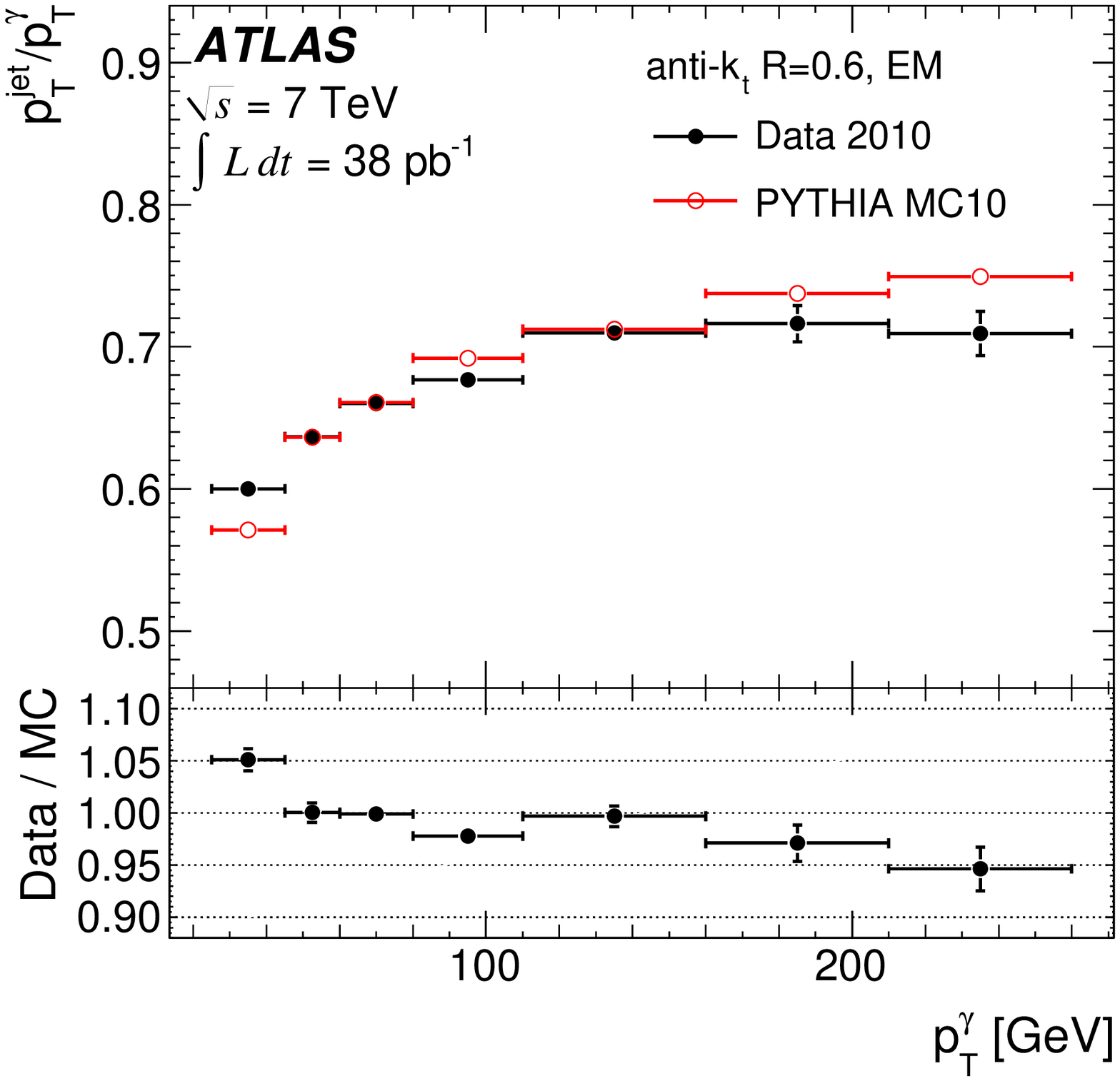}}
\hspace{1.cm}
  \subfloat[MPF technique]{\includegraphics[width=0.44\textwidth]{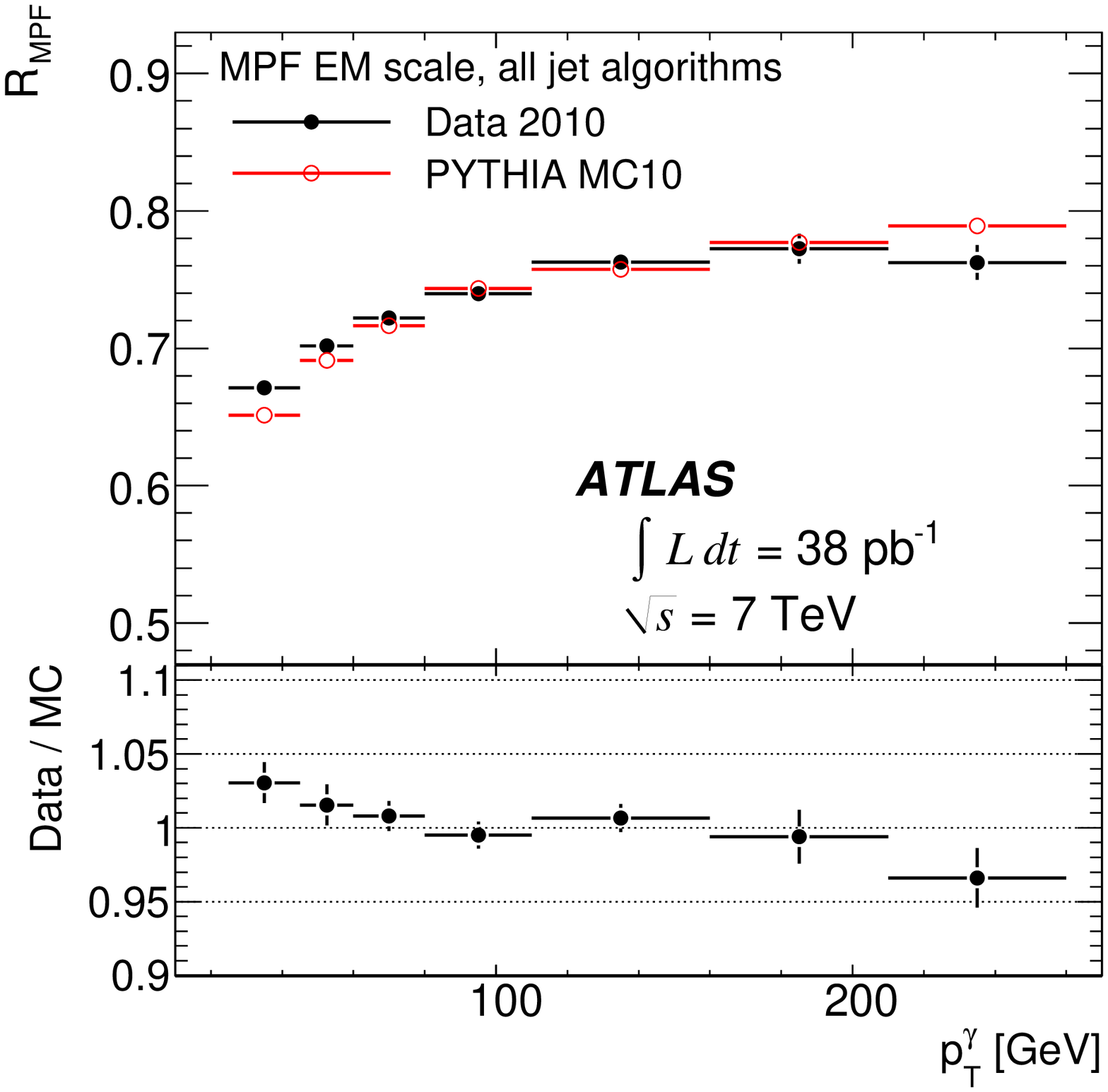}}
 \caption{Average jet response as determined by the direct \pt{} balance for \antikt{} jets with $R = 0.6$ (a) 
   and the \MPF{} technique (b)
    using the \EM{} scale for both data and Monte Carlo simulation
    as a function of the photon transverse momentum.
    The lower part of the figure shows the data to Monte Carlo simulation ratio.
    Only statistical uncertainties are shown.
}
  \label{fig:ResponseEM}
\end{center}
\end{figure*}

\begin{table}[ht!p]
\begin{center}
\begin{tabular}{l|cc|cc}
\hline \hline
  & \multicolumn{2}{|c}{Direct \pt{} balance [\%]} & \multicolumn{2}{|c}{\MPF{} [\%]} \\
\hline
$\pt^{\gamma}$ range [\GeV] & $45 - 60$ & $110 - 160$ & $45 - 60$ & $110 - 160$ \\
\hline
Background        & $\pm 1.0$        & $\pm 0.4$      & $\pm 0.6$       & $\pm 0.1$  \\
Soft QCD radiation    & $\pm 0.8$        & $\pm 0.9$      & $\pm 0.7$       & $\pm 0.4$ \\
In-time pile-up   &  $\pm 0.8$       & $\pm 0.8$      & $\pm 0$         & $\pm 0$ \\
Photon scale      & $^{+0.5}_{-0.3}$  & $^{+0.5}_{-0.3}$& $^{+0.2}_{-0.5}$ & $^{+0.3}_{-0.5}$  \\
\hline
Total systematics  & $^{+1.6}_{-1.5}$ & $^{+1.4}_{-1.3}$ & $^{+0.9}_{-1.0}$ & $^{+0.5}_{-0.6}$ \\
\hline
\hline
\end{tabular}
\caption{Individual systematic uncertainties in the jet energy scale from
  both the direct \pt{} balance and the \MPF{} techniques at two values of $\pt^{\gamma}$. }
\label{tab:SystSummary}
\end{center}
\end{table}

\paragraph{Background in the photon-jet sample}
\index{Photon background}
The systematic uncertainty from jets which are identified as photons (fakes) are studied with the
inclusive jet Monte Carlo simulation sample described in Section~\ref{sec:GammaJetMCsample}. 
Dijet events in which one of the jets is misidentified as a photon contribute to
the data sample but not to Monte Carlo simulation signal sample. 
The rate of dijet events faking photons is sensitive
to the detailed modelling of the jet fragmentation and the detector
simulation, and is therefore subject to large uncertainties.

The systematic uncertainty from this background is determined in two steps. First the difference in
the detector response between the \gammajet{}  ($\Response_{\gammajet }$) and the filtered dijet
sample ($\Response_{\rm dijet}$) is determined in the Monte Carlo simulation
as seen in Figure~\ref{fig:Dijet_Response_Diff}.
Also shown is the absolute response difference $|\Response_{\rm dijet} - \Response_{\gammajet }|$ relative
to the response of the \gammajet{} sample-. 
A response difference of maximally $3 - 5$\% is estimated.

To estimate the contribution from background in the signal region the distribution of 
photon candidates observed in the sidebands of a two-dimensional distribution is used.
The transverse isolation energy, $E_{\rm T}^{\gamma \mathrm{Isolation}}$, and the photon identification
of the photon candidate are used for this estimate.
On the isolation axis, the signal region contains photon candidates
with $E_{\rm T}^{\gamma \mathrm{Isolation}} < 3$~\GeV, while the sideband contains photon candidates 
with $E_{\rm T}^{\gamma \mathrm{Isolation}} > 5$~\GeV.
On the other axis, photon candidates passing the identification criteria belong to the signal region,
while those that fail the tight identification criteria but
pass a background-enriching selection belong to the
photon identification sideband. Further details are found in Ref.~\cite{PhotonPaper}. 

The purity $P$ measured in the signal sample is about $0.6$ at $\pt = 25$~\GeV{} and rises to
about $0.95$ at higher \pt\footnote{This is similar to the purity measured 
in Ref.~\cite{PhotonPaper} and small differences are due to the different data samples.}.
The systematic uncertainty is then calculated as
\begin{equation}
\epsilon = \left ( \frac{\Response_{\rm dijet} - \Response_{\gammajet}}{\Response_{\gammajet}} \right )
\cdot ( 1 - P ).
\label{eq:DijetBackground}
\end{equation}
The systematic uncertainty is below $1 \%$ for the direct balance technique
and below $0.6 \%$ for the \MPF{} technique.
The effect of background contamination in the \gammajet{} sample has
been further validated by relaxing the photon identification criteria.
Both data and Monte Carlo simulation  show a $3 \%$ variation in response 
for the direct \pt{} balance technique, mostly at low \pt.
This is consistent with the systematic uncertainty computed with the purity method
using Equation~\ref{eq:DijetBackground}, e.g. for the lowest \pt{} bin $40 \%$ of the events are expected to be 
dijet background giving a response that is $5 \%$ higher than the response of  \gammajet{} events.

\paragraph{Soft QCD radiation suppression cuts}
\index{Soft QCD radiation suppression cuts}
The stability of the jet response ratio of the data to the Monte Carlo simulation
is explored by varying the radiation suppression cuts.
Figure~\ref{fig:RadSys_Example}a shows the thresholds
for the $\pt^{\rm jet2}/\pt^{\gamma}$ and \deltaphijetgamma{}
cuts for $13$ sets of cuts. 
Figure~\ref{fig:RadSys_Example}b illustrates the change in the ratio of the data
to the Monte Carlo simulation of the \MPF{} response for each of these $13$ sets of cuts, 
for one typical $\pt^{\gamma}$ bin.
The result demonstrates that the ratio of the data response to the Monte Carlo response 
is not sensitive to the exact values of the radiation cuts, within the $1 \%$ level.
The systematic uncertainty
is taken as the difference in the data to Monte Carlo ratio between the nominal
cuts defining the signal sample, and the loosest cuts in all \pt-bins,
labelled as ``Point 13'' in Figure~\ref{fig:RadSys_Example}a.

The \MPF-determined response changes slightly between the data and the Monte Carlo simulation,
the systematic uncertainty is $0.7 \%$ at $\pt^{\gamma} = 50$~\GeV{} and falls to $0.4 \%$
at $\pt^{\gamma} = 135$~\GeV.  
The quoted values are determined from linear fits to the points analogous to those shown in Figure~\ref{fig:RadSys_Example}b.

The stability of the ratio of the data to the Monte Carlo simulation for the response measured with the
direct \pT{} balance technique is shown in
Figure~\ref{fig:RadSys_DirBalanceExample}. The response
measured in either data or in Monte Carlo simulation varies by up to $10 \%$ due to differing
radiation suppression cuts. However, the data to Monte Carlo ratio
with and without the radiation suppression cuts  
is stable within $\sim 1 \%$.

\paragraph {In-time pile-up}
\index{In-time pile-up}
The average number of proton-proton collisions in each bunch crossing 
grew significantly during the data-taking period.
Thus, there is a non-negligible fraction of events containing
in-time pile-up (see Section~\ref{sec:pileup}). 
The additional collisions produce extra particles
which can overlap with the hard interaction of interest in the \ATLAS{} detector.
The increased energy is about $0.5$~\GeV{} per additional reconstructed 
primary vertex (see Section~\ref{sec:track-jet-offset}).

The \MPF{} technique is expected to be insensitive to in-time
pile-up events. Because in-time pile-up is random and symmetric in $\phi$, the 
mean of the quantity $\vec{p}_{\rm T}^{\gamma} \times \vecEtmiss $
should be robust against in-time pile-up. 
The missing transverse fraction (\MTF)  is defined as:
\begin{equation}
\MTF = 
\frac{(\vec{p}_{\rm T}^\gamma \times \vecEtmiss)_z}{\mid \vec{p}_{\rm T}^\gamma \mid^2} =
\frac{\mid \vecEtmiss \mid}{\mid \vec{p}_{\rm T}^\gamma \mid} \  \sin{( \phi_{\vecEtmiss} - \phi_{\vec{p}_{\rm T}^\gamma})},
\end{equation}
where $(\vec{p}_{\rm T}^\gamma \times \vecEtmiss)_z$ is the $z$-component of the vector resulting from the cross product.
The \MTF{} measures the activity in the plane perpendicular to the photon $\vec{p}_{\rm T}$.
The mean of the \MTF{} is zero, if there is no bias due to in-time
pile-up. 

Figure~\ref{fig:MTF} shows the \MTF{} distribution for data with
and without in-time pile-up. For both these distributions the means are
compatible with zero.

From the study of the \MTF{} distribution and other checks, such as the
dependence of the \MPF{} on \Npv, it can be justified that
in-time pile-up can be neglected and no systematic uncertainty is attributed to the \MPF{} method.
In the case of  the direct \pt{} balance technique the impact of in-time pile-up is
explored by comparing the \pt{} balance between events with exactly one identified
primary vertex and events with any number of vertices.
As seen in Figure~\ref{fig:Pileup_DirBalanceExample}
the ratio of the response in data to the response in Monte Carlo simulation
for events with exactly one vertex and for events
with more than one vertex is consistent with a variation of $0.8 \%$.
This is taken as a systematic uncertainty. 

No effect due to the offset correction for in-time pile-up is seen (see Section~\ref{sec:pileup}),
and no systematic uncertainty is attributed to the offset correction
for in-time pile-up.

\begin{figure}[ht!p]
\begin{center}
  \includegraphics[width=0.45\textwidth]{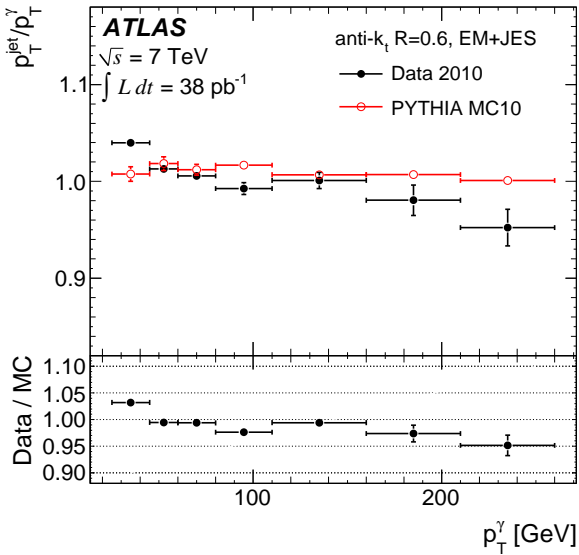}
\caption{Average jet response as determined by the direct \pt{} balance technique for 
\antikt{} jets with $R = 0.6$ calibrated with the \EMJES{} scheme 
 as a function of the photon transverse momentum for both data and Monte Carlo simulation.
The lower part of the figure shows the data to Monte Carlo simulation ratio.
 Only statistical uncertainties are shown.}
\label{fig:DBResponse_EMJES}
\end{center}
\end{figure}

\begin{figure*}[ht!!p]
  \centering
  \subfloat[\EM]   {\includegraphics[width=0.44\textwidth]{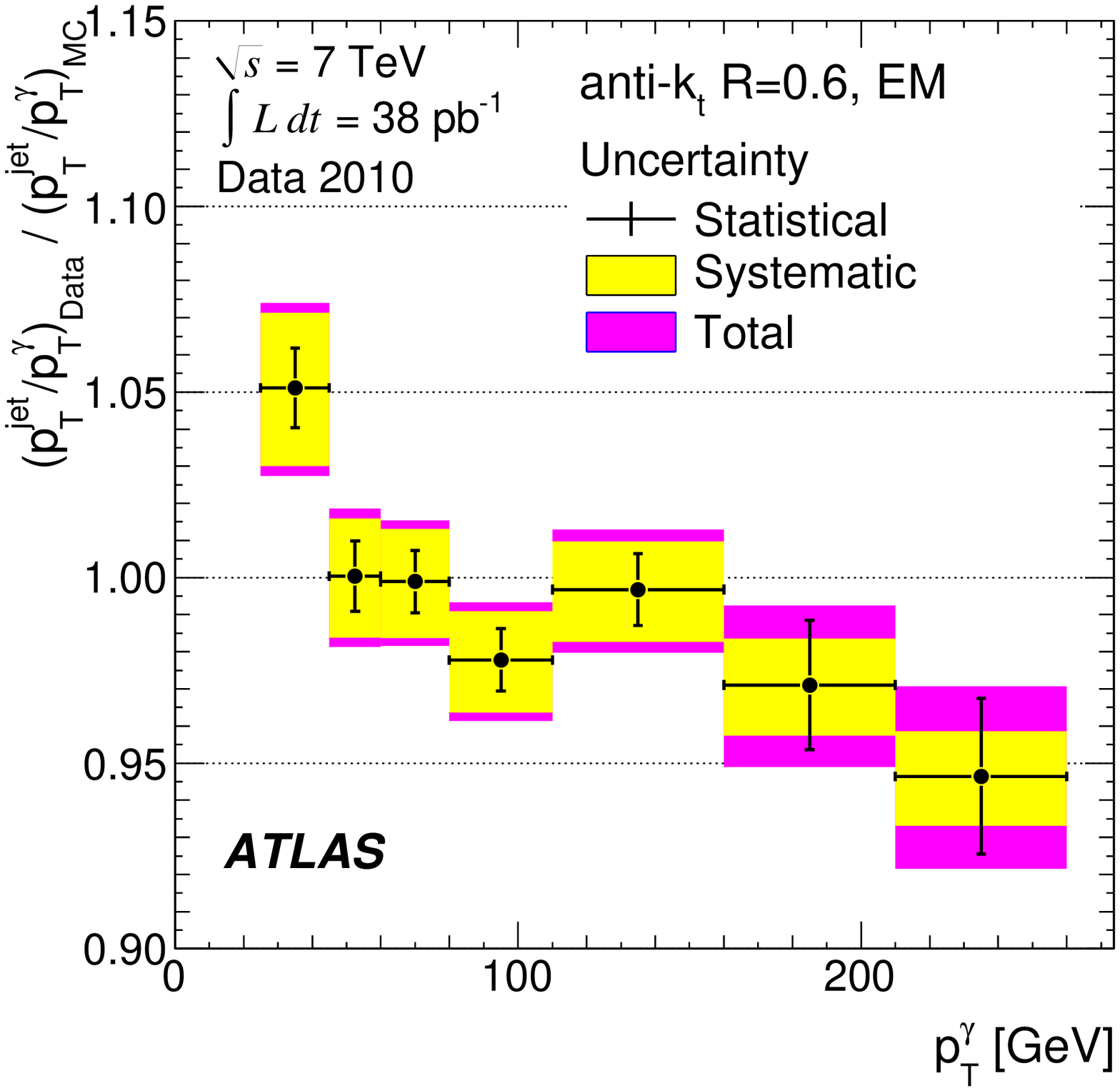}}
\hspace{1.cm}
  \subfloat[\EMJES]{\includegraphics[width=0.44\textwidth]{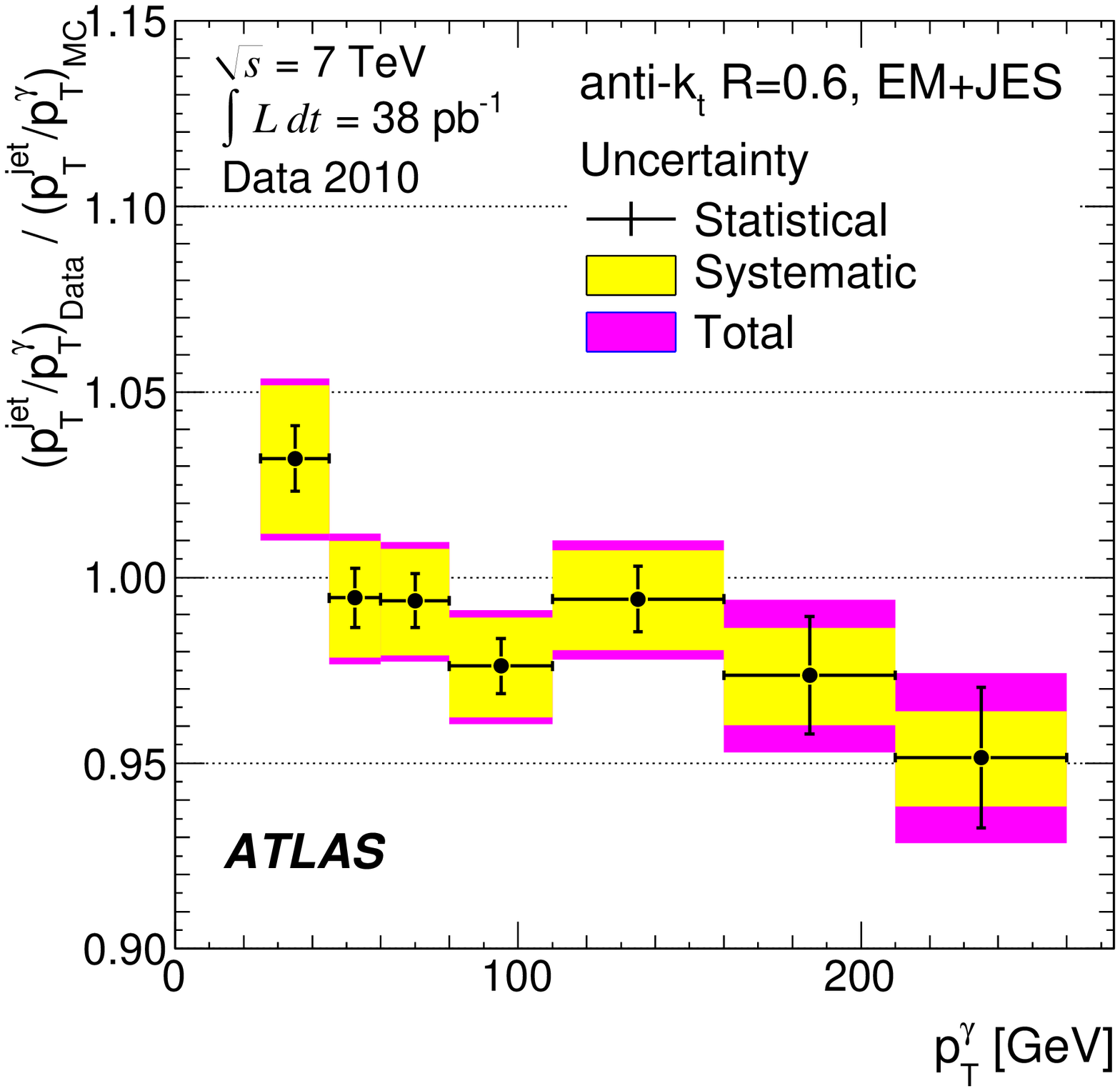}}\\
 \caption{Average jet response ratio of data to Monte Carlo simulation using the
    direct \pt{} balance technique for each input energy scale, \EM{} (a) and \EMJES{} (b),
    as a function of the  photon transverse momentum. Statistical and
    systematic uncertainties (light band) are included with the total uncertainty
    shown as the dark band.}
  \label{fig:TotalErrors_DB_EM}
\end{figure*}

\begin{figure}[ht!!p]
  \includegraphics[width=0.45\textwidth]{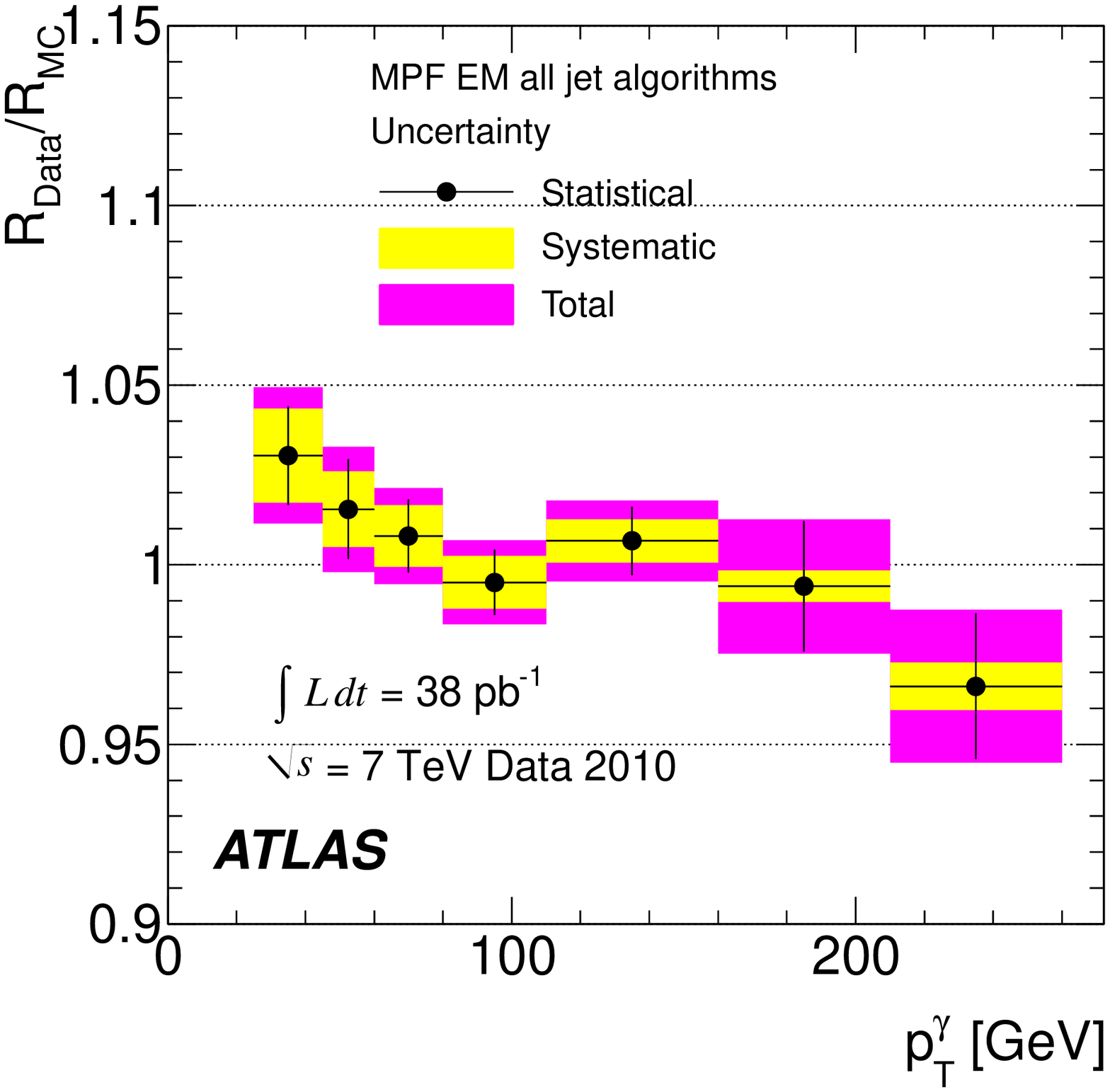}
  \caption{Average jet response ratio of data to Monte Carlo simulation using the
    \MPF{} method at the \EM{} scale as a function of the
    photon transverse momentum. Statistical and systematic uncertainties (light band) 
    are included. The total uncertainty is shown as the dark band.}
  \label{fig:TotalErrors_MPF_EM}
\end{figure}

\paragraph{Impact of missing calorimeter read-out regions}
For a small subset of the calorimeter channels the calorimeter readout
is not functioning properly.
The energy of these calorimeter cells is evaluated using the trigger
tower information, which has larger granularity and less accurate
resolution. While photons reconstructed in or near such a region are
not considered in the analysis, there is no such rejection applied to jets.
A sub-sample of events with no jet containing such a cell has been
used to evaluate a possible systematic uncertainty between data and simulation.
Within the statistical uncertainty, no bias is observed for the \MPF{}
\gammajet{} technique or
the direct \pt{} balance technique, therefore no systematic uncertainty
is  assigned.

\paragraph {Photon energy scale}
\index{Photon energy scale}
Both the direct \pt{} balance and the \MPF{} techniques are sensitive 
to the photon energy scale.  
The absolute electron energy scale has
been measured \insitu{} using the invariant mass constraint in $Z \to e^{+}e^{-}$ for electrons.
The uncertainty on the photon energy scale results in
a systematic uncertainty smaller than $1 \%$, depending on \ptjet{} and $\etajet$. 

The direct \pt{} balance technique and the \MPF{} technique find a systematic uncertainty which is approximately
opposite in sign. This sign difference is caused by the upwards
shift in photon energy leading to an equivalent downwards shift
in \vecEtmiss, and vice versa.

The response measured with both the \MPF{} and the direct
\pt{} balance techniques has been studied for converted and non-converted photons.
The results of both samples agree within the statistical uncertainties. No additional systematic 
uncertainty has been considered for this effect, which is already
accounted for in the photon energy scale and the photon background
systematic uncertainty.

\paragraph{Total systematic uncertainty}
Table~\ref{tab:SystSummary} shows a summary of the systematic
uncertainties studied for the direct \pt{} balance and \MPF{} techniques. 
The total
systematic uncertainties for each method are similar, although each
method is sensitive to different effects. 
Total systematic uncertainties are found on the data to Monte Carlo simulation jet response ratio
of smaller than $1 \%$ for the \MPF{} method and
of smaller than $1.6 \%$ for direct balance method.

\subsubsection{Results from the photon-jet balance}
\label{sec:gammajetDoubleRatios}

The direct \pt{} balance and \MPF{} techniques are used to validate the jet
response \insitu{} by comparing data and Monte Carlo simulation. 
The response in data and Monte Carlo simulation for the \EM{} scale energy 
is shown in Figure~\ref{fig:ResponseEM}.
The jet response in data and Monte Carlo simulation agrees within uncertainties in the range
$\pt^{\gamma} >  45$~\GeV. In the range $25 \le \pt^{\gamma} < 45$~\GeV{}
there is a shift in the data to Monte Carlo ratio of $5 \%$ for
the direct \pt{} balance technique and $3 \%$ for the \MPF{} technique.

Since the \EMJES{} calibration depends only on the \pt{} and $\etajet$ of the
jet, it is possible to validate the \EMJES{}  calibration scheme by
using the \EM{} scale as a function of $\pt^{\gamma}$ and
$\etajet$. 
Figure~\ref{fig:DBResponse_EMJES} shows the jet response measured in both data and Monte Carlo simulation 
using the direct \pt{} balance technique with the \antikt{} jet algorithm 
with $R=0.6$ for the \EMJES{} calibration scheme. 
The data to Monte Carlo simulation agreement is within $\pm 5 \%$.

Figure~\ref{fig:TotalErrors_DB_EM} shows the ratio of
$\ptjet /\pt^\gamma$ between data and Monte Carlo simulation together with the total 
uncertainty on the determination of the data to Monte Carlo simulation ratio, for
\antikt~jets with $R=0.6$. 
Similarly, Figure~\ref{fig:TotalErrors_MPF_EM}  shows the response ratio 
of data to Monte Carlo simulation, as
determined using the \MPF{} technique together with the total uncertainty on the
determination of the data to Monte Carlo simulation ratio.

For $\pt^\gamma > 45$~\GeV, the response in data and Monte Carlo simulation agree to
within $3 \%$ for both \MPF{} and direct balance techniques up to about $210$~\GeV.
In the range $25 \le \pt^{\gamma} < 45$~\GeV{} there is an observed
shift of $5 \%$ for the direct \pt{} balance technique and $3 \%$ for the \MPF{} technique. 
The lower response at the highest $\pt^\gamma$ is further discussed in Section~\ref{sec:insitucombination}. 

The size of these shifts is consistent with the systematic
uncertainty on the \EMJES{} jet energy calibration (see Section~\ref{sec:JESUncertainties}).
At high $\pt^\gamma$ the dominant uncertainty is statistical while the systematic uncertainty 
dominates at low $\pt^\gamma$.

\subsubsection{Summary of the photon-jet balance}
The validation of the \EMJES{} calibration scheme for jets with the \antikt{} jet algorithm reconstructed 
from \topos{}  using \insitu{} methods is presented. 
Agreement between the response in data and Monte Carlo simulation is found to be 
within statistical uncertainties for $45 \le \pt^{\gamma} < 210$~\GeV. 
Both techniques observe a shift in the  data to Monte Carlo simulation ratio for $25 \le \pt^{\gamma} < 45$~\GeV.
The total systematic uncertainties of the \gammajet{} \insitu{} technique 
is estimated to be less than $1.6 \%$ for $45 \le \pt^{\gamma} < 240$~\GeV.

\subsection{Multijet transverse momentum balance}
\label{sec:multijet}
\index{multijet calibration}
The \pt{} reach in the \gammajet{} transverse momentum balance technique
is limited by the available event statistics. 
The multijet balance technique where a recoil system of low-\pt{} jets
balances several jets at lower \ptjet{}
can be used to assess the jet calibration at higher \pt. 
Jet transverse momenta up to  the \TeV{} region can be probed.
The same method can also be used to obtain
correction factors for possible non-linearities at very high \ptjet. 
Here, the method is only used to assess the \JES{} uncertainty.

\begin{figure}[ht!]
  \centering
  \includegraphics[width=0.49\textwidth]{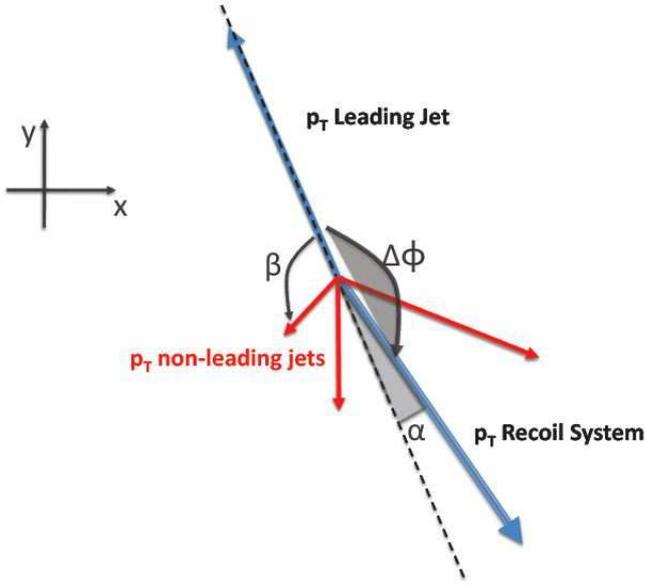}
  \caption{Sketch of the event topology used for the multijet balance technique 
in the $x-y$-plane.}
  \label{fig:MJBSketch}
\end{figure}

\begin{figure*}[ht!]
  \centering
  \subfloat[Before asymmetry cut]{\includegraphics[width=0.49\textwidth]{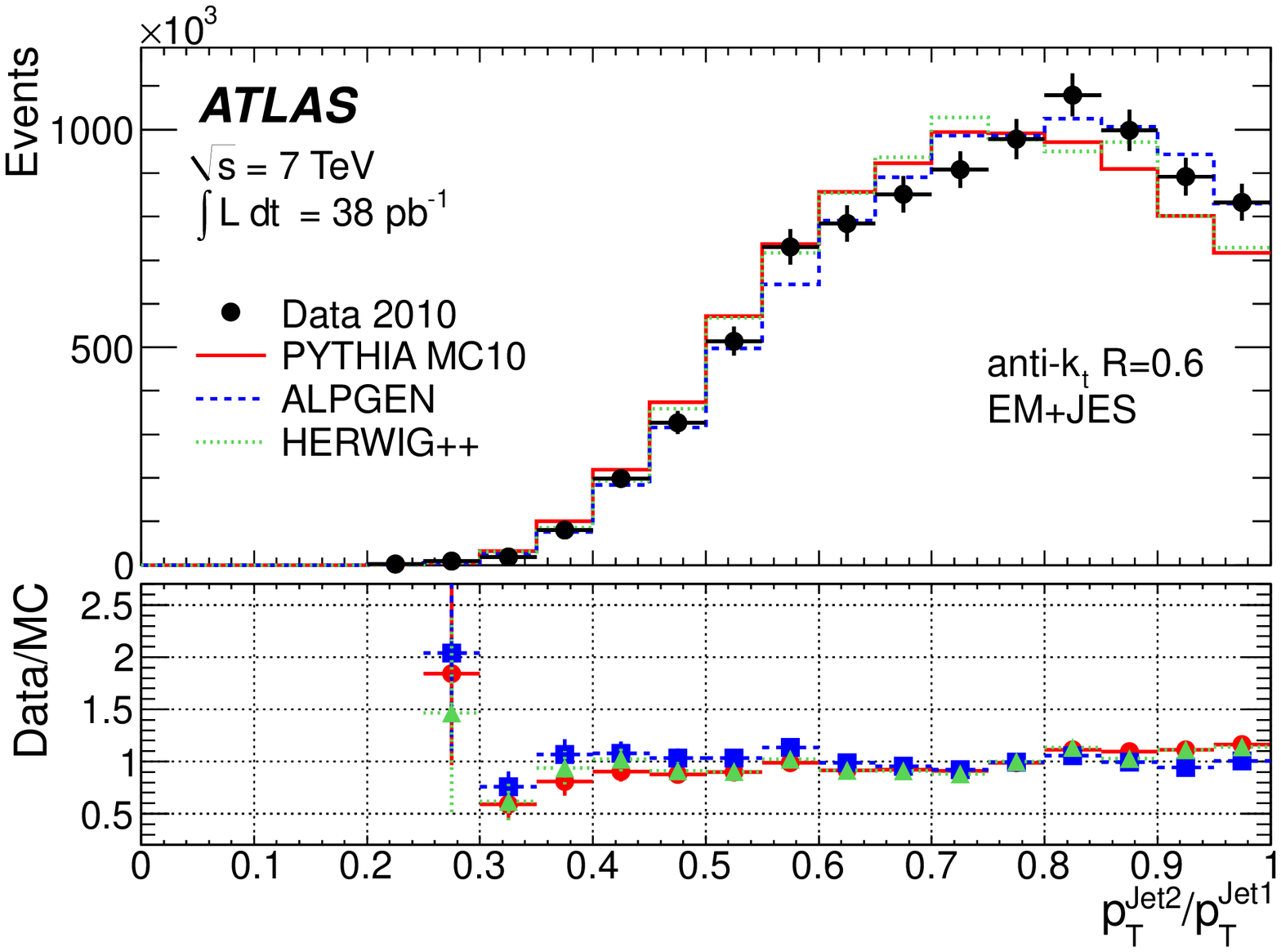}}\hfill
  \subfloat[After asymmetry cut] {\includegraphics[width=0.49\textwidth]{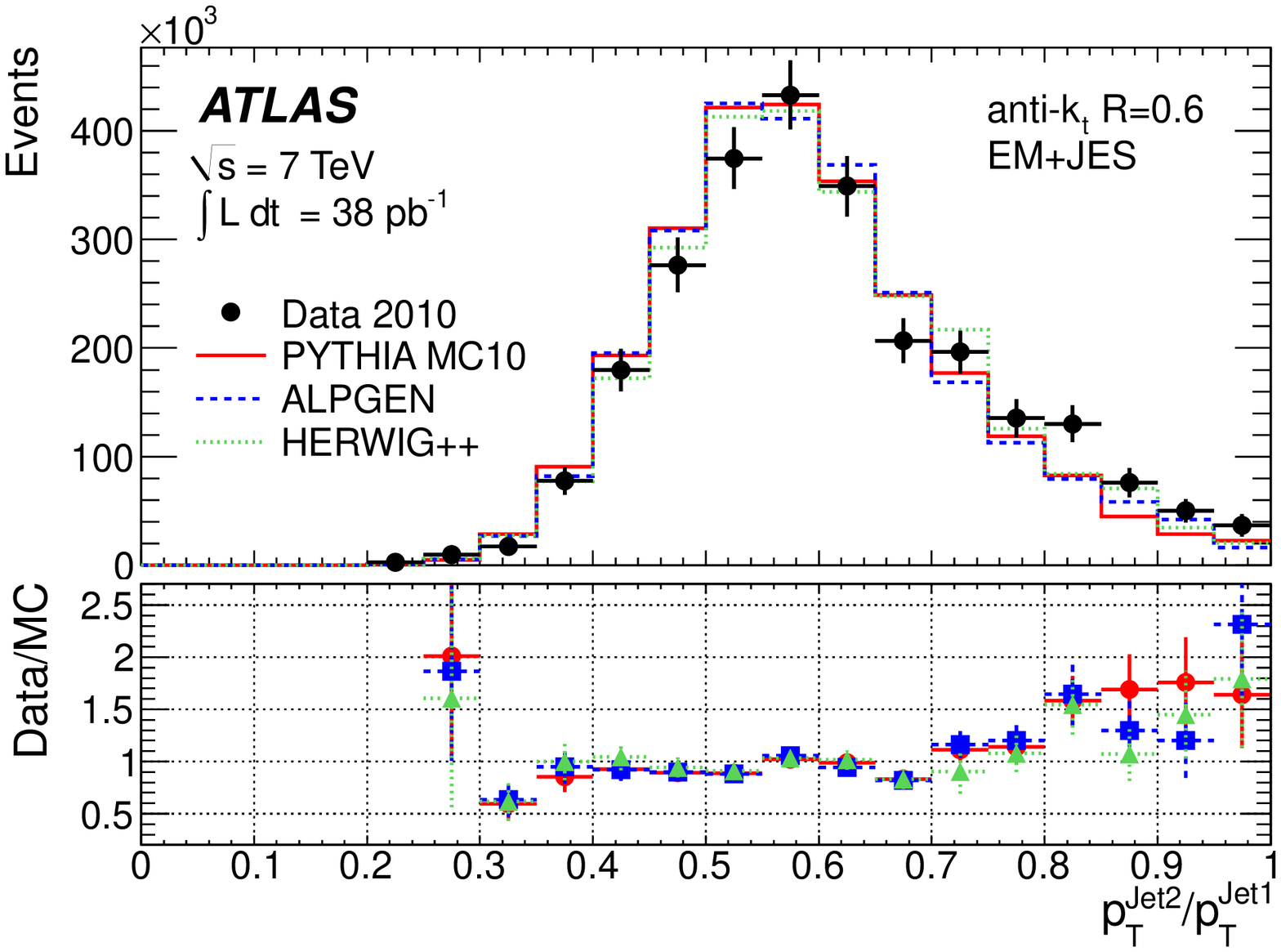}}\\
  \caption{Distribution of the ratio of the sub-leading to the leading jet \pt{} for 
\antikt{} jets with $R = 0.6$ before (a) and after (b) the asymmetry cut, see Equation~\ref{eq:multiasscut}, 
has been applied 
for data (full circles) and for simulation (lines). 
All the distributions in the simulation are normalised to the number of data events. 
Events selected by pre-scaled triggers have entered the histogram weighted by the pre-scale value.
Only statistical uncertainties are shown.} 
  \label{fig:Pt2OverPt1}
\end{figure*}

\subsubsection{The multijet balance technique}
\label{sec:MJBT}
The method exploits the \pt{} balance in events where the highest \pt{} jet (leading jet) is produced
back-to-back in $\phi$ to a multijet system. The leading jet is required to have significantly larger transverse 
momentum than other jets in the event. In this way the leading jet is at a higher \ptjet{} scale compared
to other reconstructed jets, called non-leading jets. 
The ensemble of the non-leading jets passing the selection cuts is referred to as the recoil system. 

\index{\ptRecoil}
The event topology used in this analysis is sketched in Figure~\ref{fig:MJBSketch}. 
The vectorial sum of the transverse momenta of all non-leading jets defines the transverse momentum of
the recoil system (\ptRecoil), which is expected to approximately balance the transverse momentum 
of the leading jet (\ptLeading). 
Thus a correlation between the momentum scale of the leading jet and the scale of the non-leading jets 
can be established. 
If the absolute \JES{} is well-known for all non-leading jets, the \JES{} of the leading
jet can be verified by studying the multijet balance (MJB) that is defined as the ratio:
\begin{equation}
\mathrm{MJB} = \frac{ | \vec{p}_{\rm T}^{\rm Leading} | }{ | \vec{p}_{\rm T}^{\rm Recoil} |  }.
\end{equation}
\index{multijet balance (MJB)}

Moreover, the \ptRecoil{} is a good estimator of the true leading jet \pt, and it is therefore 
interesting to study MJB as a function of \ptRecoil. 
In the ideal case MJB should be equal to one; however, various effects such as the presence of close-by jets, 
soft gluon emission, pile-up or the selection criteria themselves may introduce a bias.

The comparison between the balance measured in the simulation 
(${[{\rm MJB}]}_{\rm MC}$) and the data (${[{\rm MJB}]}_{\rm Data}$) 
can be interpreted as a source of systematic uncertainty and therefore the ratio  
\begin{equation}
r = \left [ {\rm MJB} \right]_{\rm Data}/ \left [{\rm MJB}\right]_{\rm MC}
\end{equation}
 can be used to assess the high \pt{} \JES{} uncertainty.

The jets belonging to the recoil system must be confined to a lower jet energy scale with respect 
to the leading jet in order to ensure that the multijet balance is testing the absolute high \pt{} scale 
and not only the intercalibration between jets. 
There are various analysis methods to constrain the leading jet to a higher \pt{} scale with respect to 
the jets in the recoil system. In this analysis it is done by setting an upper 
limit on the ratio between the transverse momentum of the second highest \pt{} jet ($\pt^{\rm Jet2}$) 
and the \ptRecoil. 
This cut is very efficient in selecting multijet events while minimising the bias on the transverse 
momentum of the leading jet.

\begin{table}
  \begin{center}
    \begin{tabular}{l|c}
      \hline \hline
      Variable     & \multicolumn{1}{c}{Cut value}\\ 
      \hline
      Jet \pt & $>20$~\GeV  \\
      Jet rapidity & $|\rapjet| < 2.8$  \\
      Number of good jets & $\geq 3$ \\ 
      \ptRecoil & $> 80$~\GeV \\
      $\alpha$ & $< 0.3$~radian \\
      $\beta$ & $>1$~radian  \\
      $\pt^{{\rm Jet}2} / \ptRecoil$ & $<0.6$ \\      
      \hline \hline
    \end{tabular}
  \caption{Selection criteria to define the event sample for the multijet balance analysis.}
  \label{tab:AnalysisCuts}
  \end{center}
\end{table}

\subsubsection{Selection of multijet events}
\label{sec:MJBTeventselection}
Two jet trigger selections have been used to cover a wide \pt{} range with large enough statistics. 
The first trigger selection requires at least one jet with $\pt > 15$~\GeV{} at the \EM{} scale 
in the level-1 calorimeter trigger. 
The data collected with this trigger are used to cover the region of $\ptRecoil < 260$~\GeV. 
 The second trigger selection, which requires at least one jet with $\pt > 95$~\GeV{} 
for the level-1 trigger, 
is used to populate the region of $\ptjet \geq 260$~\GeV.  
The two trigger thresholds are fully efficient for jets 
with $\ptRecoil > 80$~\GeV{} and $\ptjet > 250$~\GeV. 
To avoid a trigger bias, the multijet balance 
is studied in events containing a recoil system with transverse momentum larger than $80$~\GeV. 

In order to select events with one jet being produced against a well-defined recoil system,
a selection is applied using two angular variables ($\alpha$ and $\beta$ 
as depicted in Figure~\ref{fig:MJBSketch}):
\begin{enumerate}
\item $\alpha = |\Delta\phi - \pi|$, where $\Delta\phi$ is the azimuthal opening angle between the 
highest \pt{} jet and the recoil system.
\item $\beta$ is the azimuthal opening angle of the non-leading jet that is closest to the leading jet 
in $\phi$, measured with respect to the leading jet. 
\end{enumerate}

Events are selected by requiring:
\begin{enumerate}
\item $\alpha = |\Delta \phi - \pi| < 0.3$ radian.
\item $\beta > 1$ radian, i.e. no jets within $|\Delta \phi| = 1$ radian around the leading jet. 
\end{enumerate}
The cuts applied to $\alpha$ and $\beta$ retain the bulk of the events.

A further selection is applied to ensure that the leading jet is at a higher scale with respect to the 
jets composing the recoil system.
This is done by requiring  that the asymmetry ratio $A$ 
of $\pt^{{\rm Jet}2}$ to the transverse momentum of the recoil system satisfies the following inequality:
\begin{equation}
\label{eq:multiasscut}
A = \frac{\pt^{{\rm Jet}2} }{ \ptRecoil} < 0.6.
\end{equation}
This cut enables the efficient suppression of events with topologies very close to those 
of dijet events. This can be seen from the distributions of the ratio of the $\pt^{{\rm Jet}2}$ to the 
leading jet \pt{} shown in Figure~\ref{fig:Pt2OverPt1} before  
and after the cut is applied. 
Events are weighted according to the pre-scale values applied at the trigger level.

This selection therefore ensures that the leading jet is at a 
higher scale with respect to the jets forming the recoil system. 
At the same time this cut does not bias either the leading jet \pt{} or the recoil system \pt. 
This has been confirmed using Monte Carlo simulation  by checking that the average response 
of the leading jet and recoil system \pt{} is not significantly shifted from one after the 
asymmetry cut is applied. 
A summary of the selection criteria used in the analysis is given in Table~\ref{tab:AnalysisCuts}.

\subsubsection{Measurement of the multijet balance}
\label{sec:Results}
\begin{figure}[ht!]
  \centering
  \includegraphics[width=0.49\textwidth]{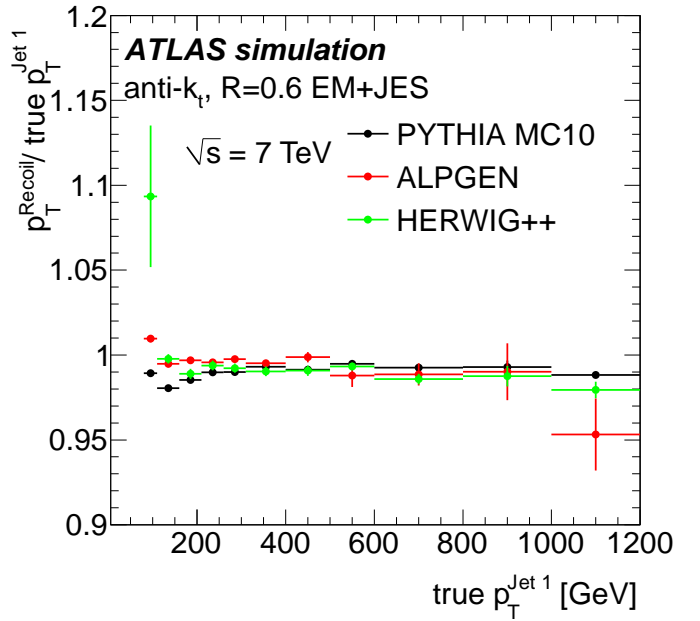}\hfill
  \caption{Ratio of the reconstructed recoil system \pt{} to the true leading jet \pt{} for \antikt{}
   jets with $R = 0.6$ as a function of the true leading jet \pt{} for three samples of Monte Carlo simulations. 
   Only statistical uncertainties are shown.}
  \label{fig:Analysis}
\end{figure}

The multijet balance is studied as a function of the 
transverse momentum of the recoil system, \ptRecoil, which is a good estimator of the true leading 
jet \pt{} as shown in Figure~\ref{fig:Analysis} for various Monte Carlo simulations.
The ratio of reconstructed \ptRecoil{} to the true leading jet  \pt{} as a function 
of the true leading jet \pt{} is, on average, consistent with unity to better than $1 \%$. 

\begin{figure}[ht!]
  \centering
\includegraphics[width=0.49\textwidth]{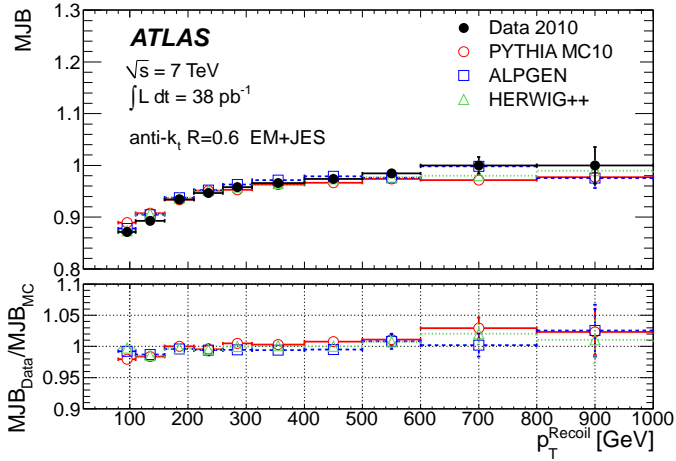}
  \caption{Multijet balance MJB as a function of the recoil system \pt{} 
for data and Monte Carlo simulation for \antikt{} jets with $R = 0.6$. 
Only statistical uncertainties are shown.
}
  \label{fig:PtBalance}
\end{figure}

The multijet balance obtained from the selected events for the \antikt{}  jet algorithm 
with $R = 0.6$ is shown in Figure~\ref{fig:PtBalance} for data and Monte Carlo simulation. 
The transverse momentum of the recoil system ranges from $80$~\GeV{} up to $1.0$~\TeV{} for the 
\antikt{} jets with $R = 0.6$. 

The multijet balance at low \ptRecoil{} values shows a bias towards values lower than one. 
This is a due to effects which broaden the leading jet and the \ptRecoil, and is a direct consequence
of binning in \ptRecoil.
This effect is observed already for truth jets and is, 
after reconstruction,  correctly reproduced by the Monte Carlo simulation. 

The data to Monte Carlo simulation ratio obtained from the multijet balance
distributions are shown in the lower part of Figure~\ref{fig:PtBalance}. 
The average value of the data to Monte Carlo simulation ratio
is within $3 \%$ for transverse jet momenta up to the \TeV-region. 
The data to Monte Carlo simulation 
ratio provides an estimate of the uncertainty on the leading jet \pt{} scale. 

\begin{figure}[ht!]
  \centering
\includegraphics[width=0.49\textwidth]{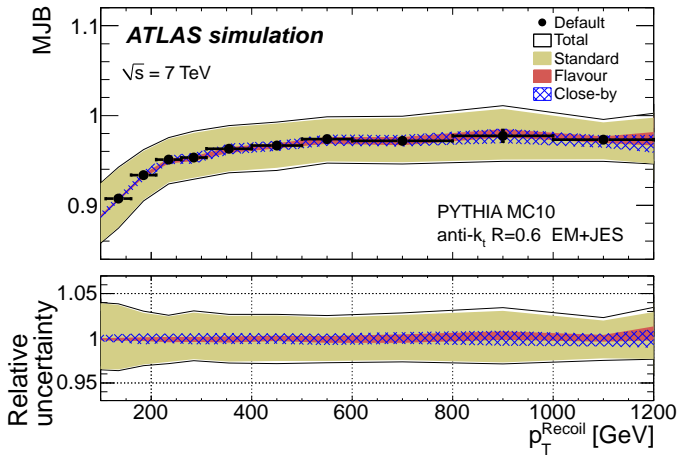}
\caption{The multijet balance MJB as a function of \ptRecoil{} (full dots) 
with statistical uncertainties for \antikt{} jets with $R = 0.6$. 
The three bands are defined by the maximum shift of MJB  
when the jets that compose the recoil system are shifted up and down by the standard \JES{}  uncertainty, 
close-by jet and flavour uncertainties. The black lines show the total uncertainty obtained 
by adding in quadrature the individual uncertainties.
The lower part of the figure shows the relative uncertainty due to the scale uncertainty of the jets 
that compose the recoil system, defined as the maximum relative shift with respect to the nominal value, 
as a function of \ptRecoil. } 
\label{fig:PtRecoilUnc}
\end{figure}

\subsubsection{Estimate of the systematic uncertainty on the multijet balance}
\label{sec:Systematic}
Two main categories of systematic uncertainty have been considered:
\begin{enumerate}  
\item The reference \ptjet{} of the recoil system.
\item The MJB used to probe the leading jet \pt{}, 
due to selection criteria or an imperfect Monte Carlo simulation modelling of the event. 
\end{enumerate}

The standard \JES{} uncertainty has been obtained for isolated 
jets. %
In the case of multijet events the additional uncertainty due to close-by jets 
(see Section~\ref{sec:closeby}) 
and the different flavour composition (see Section~\ref{sec:quarkgluon}) 
should be taken into account.

The systematic uncertainty on the recoil system has been calculated taking into account the following effects:
\begin{enumerate}
\item {\bf JES uncertainty:} The \JES{} uncertainty described in Section~\ref{sec:JESUncertainties}
is applied to each jet composing the recoil system.

\item {\bf Close-by jet:} Jets belonging to the recoil system are often produced 
with another jet nearby in the multijet environment, and the jet response is dependent 
on the angular distance to the closest jet. %
The close-by jet uncertainty has been estimated by studying the \pt{} ratio between the calorimeter jets 
and matched track jets as a function of the jet transverse momentum for different jet isolation cuts. 
This uncertainty is discussed in more detail in Section~\ref{sec:closeby}.

\item{\bf  Flavour composition of the recoil system:}  The \JES{} uncertainty is estimated for the average jet 
composition of the inclusive jet sample. 
A discrepancy in the specific flavour composition between data and Monte Carlo simulation
may result in an additional \JES{} uncertainty.
The procedure described in Section~\ref{sec:quarkgluon} is used to estimate this uncertainty.
It requires as input the average jet response and the flavour composition uncertainty as a function of the jet \pt. 
In the samples used, the uncertainty on the jet \pt{} due to flavour composition is about $1 \%$.
\end{enumerate}

The systematic uncertainty on MJB due to the uncertainty on \ptRecoil{} is estimated by 
calculating the multijet balance after shifting the \pt{} of all jets in the recoil system up and down 
by the systematic uncertainties. 
 The total systematic uncertainty is obtained by summing in quadrature the contribution of each source and
 is shown in Figure~\ref{fig:PtRecoilUnc} for \antikt{} jets 
with $R = 0.6$.  The contributions of each single source are also shown separately. 
The standard \JES{} uncertainty is the dominant source of uncertainty over the entire \ptjet{} range.

The second category of systematic uncertainties includes sour\-ces that affect MJB used to 
probe the jet energy scale at high \ptjet. These are discussed below.

%
\begin{table}[ht!p]
  \begin{center}
    \begin{tabular}{c|c|c}
      \hline \hline
     Variable     & \multicolumn{1}{|c}{Nominal}&\multicolumn{1}{|c}{Range}\\ 
      \hline
      Jet \pt{}                  & $20$~\GeV    & $15$-$35$~\GeV \\
      $\alpha$                   & $0.3$~radian & $0.1$-$0.4$~radian \\
      $\beta$                    & $1.0$~radian & $0.5$-$1.5$~radian \\
      $\pt^{{\rm Jet}2} / \ptRecoil$   & $0.6$        & $0.4$-$0.7$ \\      
      \hline \hline
    \end{tabular}
  \caption{Nominal cut values and the range of variation used to evaluate the systematic uncertainty 
on the selection criteria for the multijet balance technique.
Events below the values are rejected.}
  \label{tab:AnalysisSys}
  \end{center}
\end{table}

\begin{figure*}[ht!p]
  \centering
\subfloat[]{\includegraphics[width=0.49\textwidth]{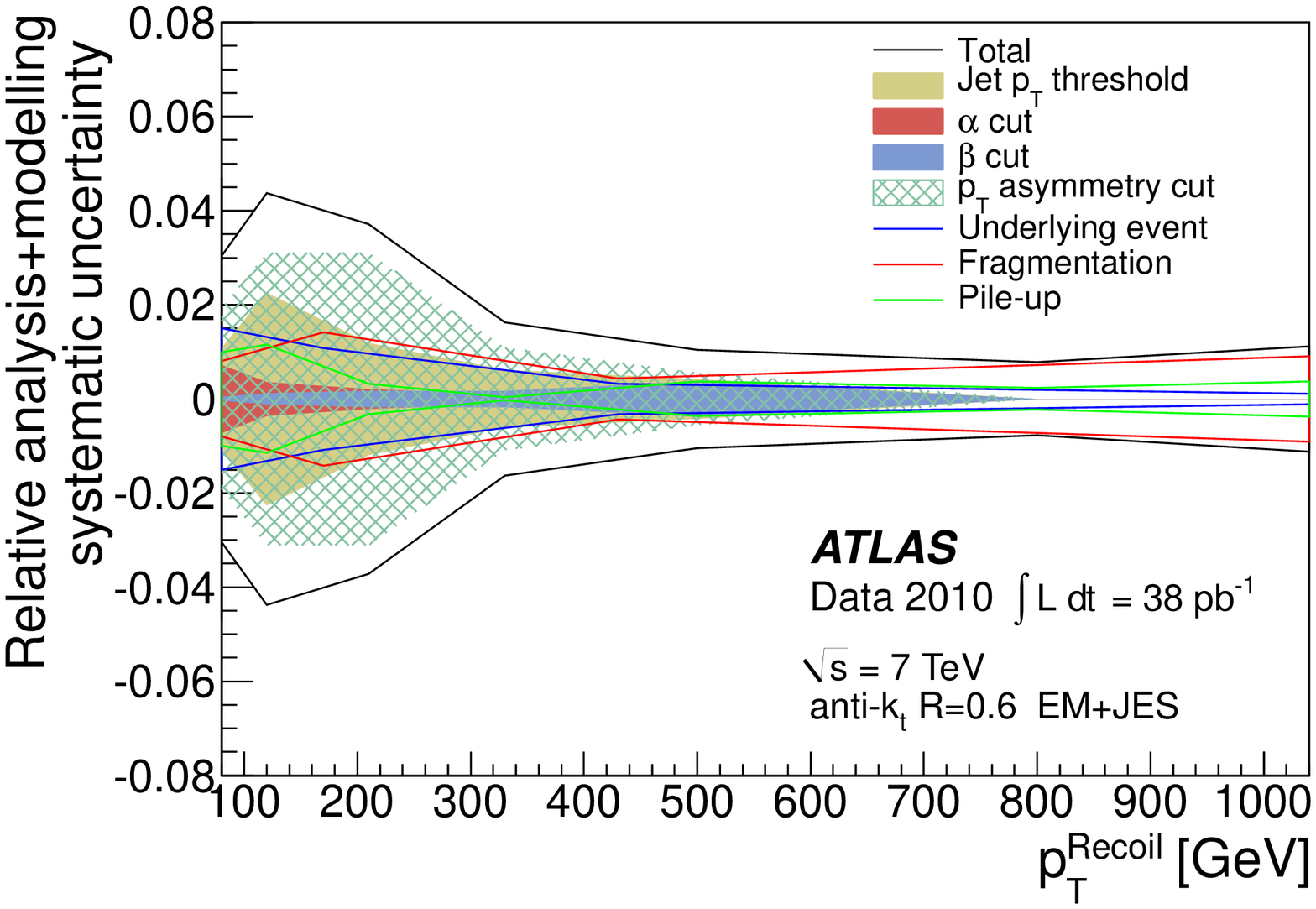}}\hfill
\subfloat[]{\includegraphics[width=0.49\textwidth]{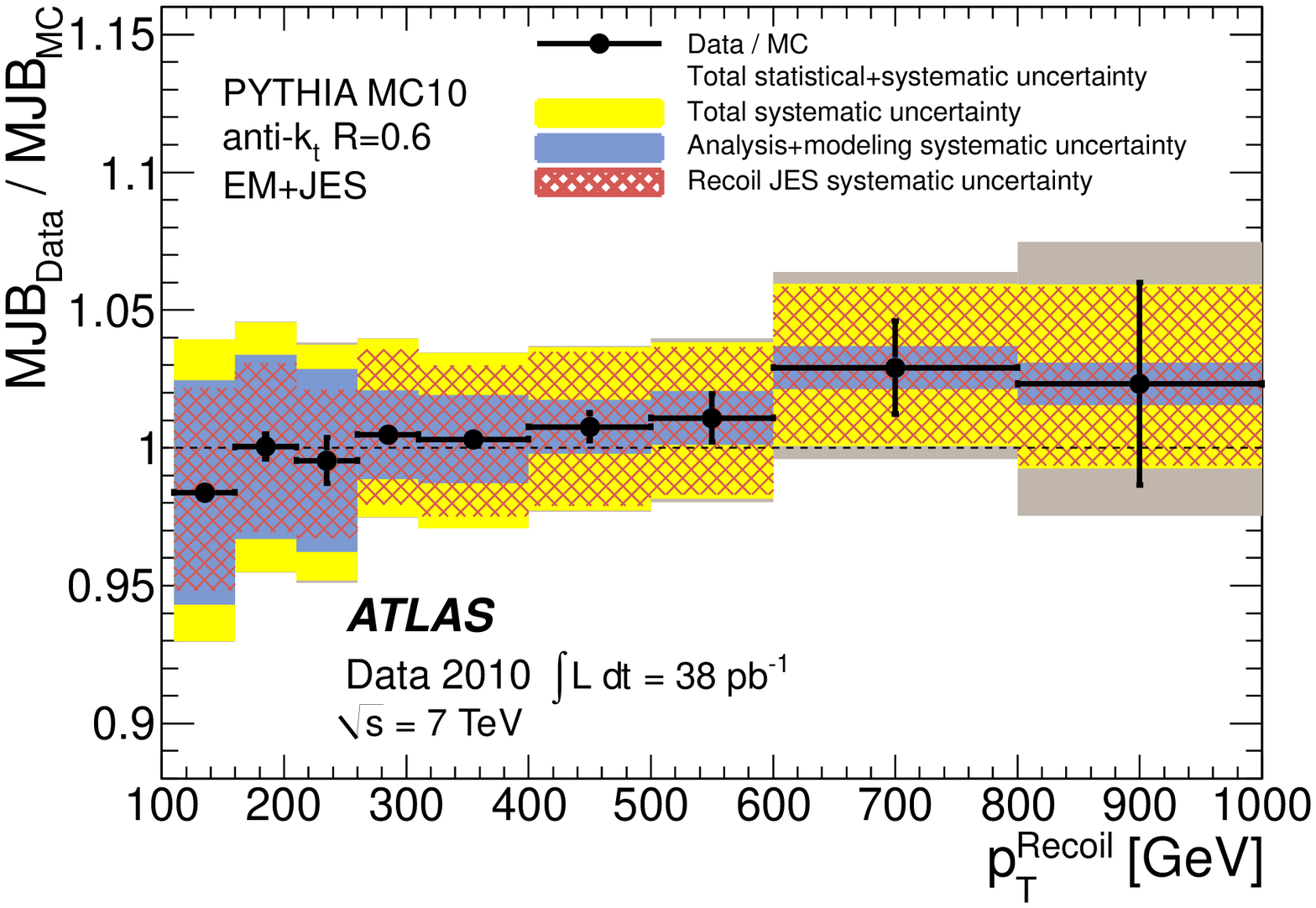}}\hfill
  \caption{
a) Single contributions as a function of \ptRecoil{} to the relative uncertainty on MJB due to 
the sources considered in the selection criteria and event modelling for \antikt{} jets with $R = 0.6$
(various lines) 
and the total uncertainty (full line) obtained as the squared sum of  all  uncertainties.
b) Ratio of data to Monte Carlo simulation for the multijet balance (MJB) as a function of the recoil system \pt{} for \antikt{} jets 
with $R = 0.6$. The various shaded regions show the total uncertainty (dark band) 
obtained as the squared sum of all total systematic uncertainties (light band) 
and of the statistical uncertainty (error bars). 
Also displayed are the contributions to the systematic uncertainty due to multijet analysis cuts 
and event modelling (darkest band) and to the jet energy scale for jets in the recoil system (hatched band).
} 
 \label{fig:AnaModelUnc}
\end{figure*}

\begin{table}[ht!p]
  \begin{center}
    \begin{tabular}{l|c}
      \hline \hline
      Source     & \multicolumn{1}{|c}{uncertainty}\\ 
      \hline
      Jet energy scale of the recoil system  & $4 \%$  \\
      Flavour composition & $\simeq 1 \%$  \\      
      Close-by jets & $2 \%$ \\
       \hline
      Jet \pt{} threshold & $<2 \%$ \\
      $\alpha$ cut & $< 1 \%$ \\
      $\beta$ cut & $<1 \%$ \\
      $p_T^{{\rm Jet}2}/\ptRecoil$ cut & $3 \%$ \\ 
      Underlying event modelling & $2 \%$ \\
      Fragmentation modelling & $1.5 \%$ \\
      Pile-up & $<1 \%$ \\
      \hline \hline
    \end{tabular}
  \caption{Maximum values of the systematic uncertainties in the whole \ptRecoil{} range
 for \antikt{} jets with $R = 0.4$ or $R = 0.6$, 
 for all effects considered in the multijet balance analysis.}
  \label{tab:SystSumm}
  \end{center}
\end{table}

In the following the various sources considered are discussed:
\begin{enumerate}
\item {\bf Selection criteria:} The imperfect description given by the Monte Carlo simulation 
for the variables used to select the events might induce a systematic uncertainty on the multijet balance. 
In order to evaluate this systematic uncertainty, all relevant selection criteria are varied in a range 
where the corresponding kinematic variables are not strongly biased and can be examined 
with small enough statistical fluctuations. 
The nominal values and the range of variations of the selection criteria are listed in Table~\ref{tab:AnalysisSys}.

The systematic uncertainty on MJB originating from these sources is evaluated by calculating the 
multijet balance after varying the cut for each variable in the range mentioned above.
For each value of the selection criteria the ratio ($r$) between the MJB values 
calculated from data and Monte Carlo simulation
is evaluated as a function of the recoil system \pt.
The maximum deviation of the $r$ with varied cuts ($r_{\rm varied}$) with respect to the nominal 
ratio ($r_{\rm nominal}$), being expressed in the double ratio
\begin{equation}
r_{\rm varied}/r_{\rm nominal}
\end{equation}
is assumed to represent the systematic uncertainty for the source. A quadratic sum of the systematic
uncertainties for all sources is taken as the total systematic uncertainty.\\

\item {\bf Jet rapidity acceptance:} The analysis uses only jets with $|\rapjet|<2.8$ 
to have a smaller jet energy scale uncertainty on the recoil system. 
This selection, however, could cause an additional systematic uncertainty,
if the fraction of jets produced outside the rapidity range differs in the data and Monte Carlo simulation.
This effect is evaluated by studying MJB (calculated as usual from only jets with $|\rapjet|<2.8$) 
for events with $\ptRecoil >80$~\GeV, as a function of the total transverse energy ($\sum E_{\rm T}$) summed over all 
jets with $|\rapjet|<4.5$, in the data and Monte Carlo simulation. 
The agreement between the data and Monte Carlo simulation is satisfactory, and MJB is stable 
over the entire $\sum E_{\rm T}$ range with the largest deviations up to  $3 \%$
with the largest deviations at relatively high $\sum E_{\rm T}$.
Since the majority of events have a very small $\sum E_{\rm T}$, this effect is 
considered to be negligible. \\

\item {\bf Soft physics modelling:}  
Imperfect modelling of multiple parton
interactions, of fragmentation and of parton shower radiation 
may affect the multijet balance in two ways. 
Firstly the selection criteria may act differently on samples with different 
modelling of the event topology. 
Secondly  MJB itself can be directly affected, since the modelling variation acts differently on the leading jet 
and the recoil system.

The systematic uncertainty for each of these sources is estimated by evaluating the ratio between the 
MJB measured using the nominal Monte Carlo simulation and an alternative Monte Carlo simulation sample 
where the particular source of uncertainty is varied.
As alternative Monte Carlo simulation samples \herwigpp{} and \pythia{} with the \Perugia{} tune
are used. 

In addition, the parameter controlling the centre-of-mass energy dependence of the
cut-off parameter determining whether an event is produced via a matrix element or 
by the underlying event model ({\rm PARP(90)})
is lowered from \pythia{} ${\rm PARP}(90)$$=0.25$ to  \pythia{} ${\rm PARP}(90)$$=0.16$. 
This change increases the energy in the forward region.
The systematic uncertainty introduced by these variations is at most $2 \%$. \\

\item {\bf Pile-up:}  Imperfect description of the pile-up %
may introduce a systematic uncertainty.
This effect is estimated by evaluating the ratio 
\begin{equation}
{\rm MJB}^{\rm pile-up}/{\rm MJB}^{\rm nominal},
\end{equation}
where the nominal sample is simulated without pile-up collisions. 
The systematic uncertainty due to pile-up is smaller than $1 \%$ for the whole \ptjet{} range considered.
\end{enumerate}

All systematic uncertainties due to the selection criteria, event modelling and pile-up, 
and the total uncertainty obtained by summing them in quadrature are shown as a function of \ptRecoil{} 
in Figure~\ref{fig:AnaModelUnc} for \antikt{} jets with $R = 0.6$. 

The final systematic uncertainty resulting from the uncertainties of the recoil reference
system and from the multijet balance variable added in quadrature
is presented in Figure~\ref{fig:AnaModelUnc}b for \antikt{} jets with $R = 0.6$. 
The total systematic uncertainty amounts to about $4 \%$ for jets of $\ptjet = 1$~\TeV. 
At high transverse momentum the main contribution to the systematic uncertainty is due to the standard 
\JES{} uncertainty of the \EMJES{} scheme. 
The maximum values of the uncertainties in the \ptjet{} range considered for each source are summarised 
in Table~\ref{tab:SystSumm}.

\subsubsection{Summary of the multijet balance results}
The data sample collected in $2010$ allows the validation of the high-\pt{} 
jet energy scale to within $5 \%$ up to $1$~\TeV{} for  \antikt{} jets with $R = 0.6$ and up to $800$~\GeV{} 
for jets with $R = 0.4$ calibrated with the \EMJES{} scheme. 
In this range the statistical uncertainty is roughly equivalent to, or smaller than, the systematic uncertainty.

\subsection{Summary of JES validation using \insitu{} techniques }
\label{sec:insituvalidationsummary}
The jet energy calibration can be tested \insitu{} using a well-calibrated object as reference
and comparing data to the \pythia{} Monte Carlo simulation tuned to \ATLAS{} data~\cite{mc10chargedparticles}. 
The \insitu{} techniques have been discussed in the previous sections,
i.e. 
the comparison of jet calorimeter energy to the momentum carried by tracks 
associated to a jet (Section~\ref{sec:trackjet}), the
direct transverse momentum balance between a jet and a photon
and the photon balance using the missing transverse momentum projection technique
(Section~\ref{sec:gammajet})  as well as
\pt{} balance between a high-\pt{} jet recoiling against a system of lower \pt{} jets 
(Section~\ref{sec:multijet})

\begin{figure}[!htb]
  \centering
  \includegraphics[width=0.49\textwidth]{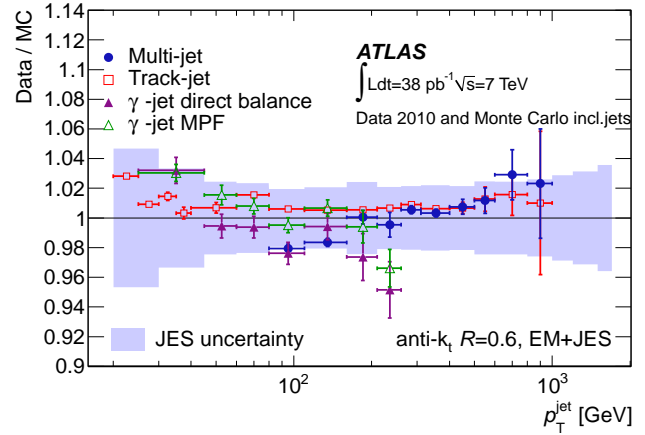}
  \caption{Ratio of \ptjet{} over reference \pt{} in
    data and Monte Carlo simulation   
    for several \insitu{} techniques for \AetaRange{1.2}.
    Only statistical uncertainties are shown.
    Superimposed is the jet energy scale uncertainty obtained from single hadron response measurements
    and systematic Monte Carlo simulation variations
    as a function of \ptjet{} (light band)
    for \AetaRange{0.3}.
  }
\label{fig:insitusummary}
\end{figure}

The comparison of data to Monte Carlo simulation for all \insitu{} techniques
for the pseudorapidity range \AetaRange{1.2} 
is shown in Figure~\ref{fig:insitusummary} together with the JES uncertainty 
region as estimated from the single hadron response measurements and systematic variations of the 
Monte Carlo simulations.
The results of the \insitu{} techniques support the estimate of the JES uncertainty 
obtained using the independent method described in Section~\ref{sec:JESUncertainties}.
\index{\Insitu{} validation summary}

%
\subsection{JES uncertainty from combination of in situ techniques}
\label{sec:JesUncertaintyinsitu}
The \JES{} uncertainty can also be obtained by combining the results of the
\insitu{} techniques described in the previous sections. 
In this combination the ability of the Monte Carlo simulation to describe the data,
the individual uncertainties of the \insitu{} techniques
and their compatibility are considered.

\subsubsection{Combination technique}
\label{sec:hvptools}
The requirements for combining the uncertainties from the individual 
\insitu{} techniques are:
\begin{enumerate}
\item Propagate all uncertainties of the individual \insitu{} techniques to the final uncertainty.
\item Minimise biases on the shapes of the measured distributions, i.e. on the
      \pt{} dependence of the data to Monte Carlo simulation ratio.
\item Optimise the uncertainties on the average while respecting the two previous requirements.
      This is equivalent to minimise the $\chi^2$ between the average and the individual measurements.
\end{enumerate}
The combination proceeds in the following steps:
\begin{enumerate}
\item {\bf Toy Monte Carlo method:}
Monte Carlo pseu\-do-ex\-per\-i\-ments are created that 
represent the ensemble of measurements and contain 
the full data treatment chain including interpolation and averaging (as described in the following steps).
These pseudo-experiments are used to consistently propagate all uncertainties into the
evaluation of the average.
The pseudo-experiments represent the full list of available measurements and take into account all known 
correlations. 

\item {\bf Interpolation method:}
A linear interpolation is used to obtain the nominal values\footnote{A second order polynomial 
interpolation provides in principle a better shape description.
However, due to the smooth variations in the results of each \insitu{} measurement,
the differences between the results obtained with the two interpolation procedures are found to be negligible.}.
The final interpolation function per measurement, within the \pt{} range, is discretised into
small ($1$~\GeV) bins for the purpose of averaging.

\item {\bf Averaging:}
 The data are averaged taking into account all known correlations
to minimise the spread in the average measured from the Monte Carlo pseudo-experiments.
\end{enumerate}

The combination of the \insitu{} calibration data is performed using the 
software package HVPTools \cite{Davier:2009zi}.
 The systematic uncertainties are introduced in HVPTools for each component as an algebraic function of \pt{} or as a numerical
 value for each data point. The systematic uncertainties belonging
 to the same source are taken to be fully correlated throughout all measurements affected.

The HVPTools package transforms the \insitu{} data and associated statistical and systematic covariance matrices into 
fine-grained \pt{} bins, taking into account the best knowledge of the correlations between the points within 
each \insitu{} measurement.
Statistical and systematic
correlations between the measurements could also be included, but as the different measurements use independent events,
these correlations are neglected\footnote{Care was taken to avoid an overlap of the multijet balance 
and $\gamma$-jet result. Allowing for an overlap
would have required taking into account the (strong) correlations, without a potential gain in precision.}.
The covariance matrices are obtained by assuming systematic uncertainties corresponding to the same source are fully correlated.  
Statistical uncertainties, taken as independent between the data points, are added in quadrature to these matrices.

The interpolated measurements from different \insitu{} methods contributing 
to a given momentum bin 
are averaged taking correlations between measurement points into account.
The measurements are performed at different \ptjet{} values
and use different binning (point densities)\footnote{The method avoids replacing 
missing information in case of a lower point density (wider binning)
by extrapolating information from the polynomial interpolation.}.

To derive proper {\em averaging weights} for each \insitu{} method, wider
{\em averaging regions}\footnote{
   For example, when averaging two measurements with unequal point spacing, a
   useful averaging region would be defined by the measurement of the \insitu{} method with the larger point spacing,
   and the points of the other measurement would be statistically merged before computing
   the averaging weights.
}  
are defined. These regions are constructed such that
all \insitu{} method covering the corresponding \pt{} range have
at least one measurement inside.
The averaging regions are used to compute weights for the \insitu{} methods,
which are later applied in the bin-wise average in fine $1$~\GeV{} bins.

The averaging weights for each \insitu{} method are computed as follows:
   \begin{enumerate}
   \item The generation of pseudo-experiments fluctuates the data points
         around the original measurements taking into account all known correlations.
         The polynomial interpolation is redone for each pseudo-experiment
         for each \insitu{} method.

   \item For each \insitu{} measurement and each Monte Carlo pseudo-experiment
         the new bin content for each wider region is calculated from
         the integral of the interpolating polynomials.
   \item 
         The contents of the wide bins are treated as new measurements
         and are again interpolated with polynomials.
         The interpolation function is used to obtain new measurements in
         small~($1$ \GeV) bins for each \insitu{} method in the \ptjet{} range 
         covered by it.
  \item In each small bin a covariance matrix~(diagonal here) between the measurements 
        of each \insitu{} method is computed. Using this matrix
        the averaging weights are obtained by $\chi^2$ minimisation.
   \end{enumerate}

For the averaging weights the procedure using the large averaging regions as an intermediate 
step is important in order to perform a meaningful comparison of the precision of the different \insitu{} methods. 
The average is computed avoiding shape biases which would come from the use of large bins.
Therefore at this next step the fine $1$~\GeV{} bins are obtained directly from the
interpolation of the original bins.

 The bin-wise average between measurements is computed as follows:
   \begin{enumerate}
   \item The generation of Monte Carlo pseudo-experiments fluctuates the data points
         around the original measurements taking into account all known correlations. 
         The polynomial interpolation is redone for each generated Monte Carlo pseudo-experiment
         for each \insitu{} method.
   \item For each generated pseudo-experiment, small~($1$~\GeV) bins are filled for each measurement
         in the momentum intervals covered by that \insitu{} method, using the polynomial interpolation.
   \item The average and its uncertainty are computed in each small bin using the weights
         previously obtained. 
         This will be displayed as a band with the central value given by the average
         while the total uncertainty on the average is represent by the band width. 
   \item The covariance matrix among the measurements is computed in each small bin.
   \item $\chi^2$ rescaling corrections are computed for each bin as follows:
        if the $\chi^2$ value of a bin-wise average exceeds the number of degrees of freedom ($n_{\rm dof}$), 
        the uncertainty on the average
        is rescaled by $\sqrt{\chi^2/n_{\rm dof}}$ to account 
        for inconsistencies\footnote{Such (small) inconsistencies are seen 
        in the comparison of the $\gamma$-jet and track jet results in one \ptjet{} bin.}. 
   \end{enumerate}

The final systematic uncertainty for a given jet momentum is (conservatively) estimated by the maximum
deviation between the average band and unity.
The central value (measured bias) and the uncertainty on the average measurement are hence taken into account.
If a correction for the measured bias were performed, only the relative uncertainty on the average would affect
the final JES calibration.

A smoothing procedure, using a variable-size sliding interval with a Gaussian kernel, is applied to the systematic uncertainty. 
It removes spikes due to statistical fluctuations in the measurements, as well as discontinuities at the first and/or 
last point in a given measurement.

%
\begin{figure}[ht!]
\begin{center}
\includegraphics[width=0.49\textwidth]{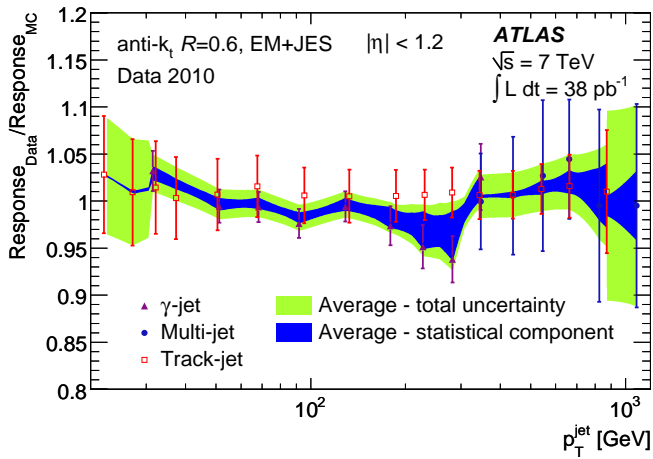}
\caption{Average jet response ratio of the data to the Monte Carlo simulation 
for jets with $|\eta|<1.2$ as a function of the transverse jet momentum \pt{}
for three \insitu{} techniques.
The error displays the statistical and systematic
uncertainties added in quadrature.
Shown are the results for \antikt{} jets with $R = 0.6$
calibrated with the \EMJES{} scheme.
The light band indicates the total uncertainty
from the combination of the \insitu{} techniques.
The inner dark band indicates the statistical component.
\label{fig:responseratioinsitu}}
\end{center}
\end{figure}
\begin{figure}[ht!]
\begin{center}
\includegraphics[width=0.49\textwidth]{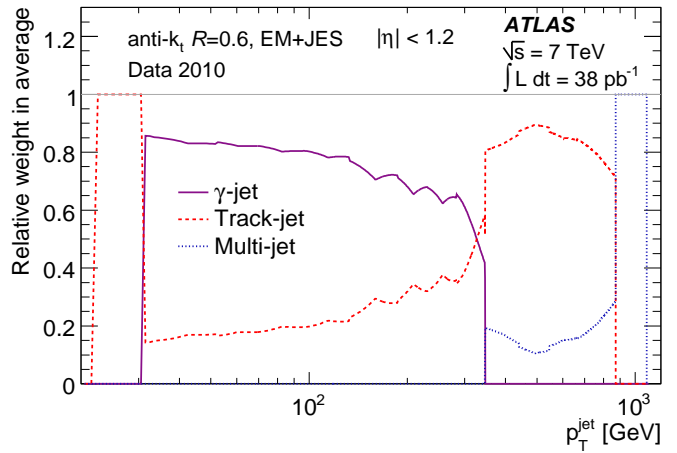}
\caption{%
Weight carried by each \insitu{} technique in the combination to
derive the jet energy scale uncertainty 
as a function of the jet transverse momentum \ptjet{}
for \antikt{} jets with $R = 0.6$ calibrated with the \EMJES{} scheme.
\label{fig:weightinsitu}}
\end{center}
\end{figure}

\begin{figure}[ht!]
\begin{center}
\includegraphics[width=0.49\textwidth]{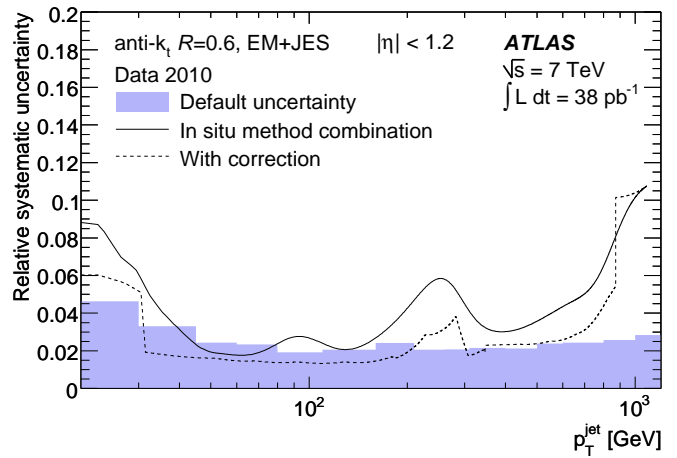}
\caption{Jet energy scale uncertainty from the combination of \insitu{} techniques (solid line) 
as a function of the jet transverse momentum \pt{}
for  \antikt{} jets with $R = 0.6$ calibrated with the \EMJES{} scheme for $|\eta|<1.2$.
The dashed line shows the \JES{} uncertainty that could have been achieved,
if \insitu{} techniques had been used to recalibrate the jets.
For comparison, the shaded band indicates the JES uncertainties as derived from
the single hadron response measurements and systematic Monte Carlo variations 
for $|\etajet|<0.3$.
\label{fig:uncertaintyinsitu}}
\end{center}
\end{figure}

\subsubsection{Combination results}
\label{sec:insitucombination}
Following the method described in the previous section the \JES{} uncertainty
for jets with $|\etajet|<1.2$ can be obtained.
The multijet balance analysis is repeated for jets with $| \etajet | < 1.2$.
and the uncertainty for low-\pt{} jets is taken from the \gammajet{} analysis.
The resulting uncertainty is larger than the one in Section~\ref{sec:multijet}.

Figure~\ref{fig:responseratioinsitu} shows the ratio of the jet response
in data and Monte Carlo simulation as a function of the transverse jet momentum
for the three \insitu{} techniques using as reference objects photons (\gammajet), 
a system of low-energetic jets (multijet)
or the transverse momentum of all tracks associated to jets (track jet). 
The errors shown for each \insitu{} technique are the statistical and systematic
uncertainties added in quadrature.

The results from the track jets cover the widest \ptjet{} range from the lowest
to the highest \ptjet{} values. 
Compared to the \gammajet{} results they have a relatively large systematic uncertainty. 
The \gammajet{} results cover a \ptjet{} range up to about $300$~\GeV. 
From this point onwards the multijet balance method helps to constrain the \JES{} uncertainty.

Figure~\ref{fig:weightinsitu} shows the contribution of each \insitu{} technique to the total
\JES{} uncertainty in form of their weight.
In the region $30 \lesssim \ptjet \lesssim 300$~\GeV{} the \gammajet{} results make the highest
contribution to the overall \JES{} uncertainty determination. The contribution is about
$80 \%$ at $\ptjet = 30$~\GeV{} and decreases to about $60  \%$ at 
 $\ptjet = 300$~\GeV. 
At the lowest \ptjet{} the method based on tracks determines the \JES{} uncertainty. 
At about $\ptjet = 300$~\GeV{} the \gammajet{} results and the ones based
on tracks have an about equal contribution. Above  $\ptjet = 300$~\GeV{} 
the results based on tracks have the highest contribution to the \JES{}
uncertainty. In this region the multijet balance contributes to the \JES{}
uncertainty to about $20 \%$. For the highest \ptjet{} only
the multijet balance is used to determine the \JES{} uncertainty.

The final \JES{} uncertainty obtained from the combination of the \insitu{}
techniques is shown in Figure~\ref{fig:uncertaintyinsitu}.
The \JES{} uncertainty is about $9 \%$ at $\ptjet = 20$~\GeV{} and decreases
to about $2 - 3 \%$ for $50 \leq \ptjet < 200$~\GeV. 
At the lowest \ptjet{} the systematic uncertainty is determined by the \insitu{} method
based on tracks, for which the data have a higher central value than the Monte Carlo simulation.
At $250$~\GeV, the uncertainty increases because the \gammajet{} results are $5 \%$ below unity and therefore pull the central
value of the average down as shown in Figure~\ref{fig:responseratioinsitu}. 
Moreover, the \gammajet{}  and the track methods give different results.
While for all other \ptjet{} values the $\chi^2/n_{\rm dof}$ is within 
$0.2 \le \chi^2/n_{\rm dof} < 0.8$, 
it rises to $\chi^2/n_{\rm dof} = 2$ at $250$~\GeV.

For $\ptjet > 350$~\GeV{} the multijet balance 
contributes to the uncertainty and the resulting uncertainty is about $4 - 5 \%$ up to $700$~\GeV. 
At the highest reachable \ptjet{} the \JES{} uncertainty 
increases to $10 \%$. %

Figure~\ref{fig:uncertaintyinsitu}  also compares the \JES{} uncertainty
obtained from a combination of \insitu{} techniques to the
one derived from the single hadron response measurements and the
systematic Monte Carlo simulation variations (see Section~\ref{sec:JESUncertainties}).
The \insitu{} \JES{} uncertainty is larger than the standard \JES{} uncertainty
in most \ptjet{} regions. 
It is similar in the region $30 \lesssim \ptjet \lesssim 150$~\GeV.
Figure~\ref{fig:uncertaintyinsitu} also shows the \JES{} uncertainty, 
that could have been ach\-iev\-ed, if the \insitu{} techniques had been used 
to correct the jet energy scale.
In this case the \JES{} uncertainty obtained from a combination of \insitu{} techniques
would be slightly smaller than the standard \JES{} uncertainty over a wide \ptjet{} range 
of $30 - 700$~\GeV. 
\section{Jet energy calibration based on global jet properties} 
\label{sec:calibTechnique}
\index{Global sequential jet calibration}
\subsection{Global sequential technique}
\label{sec:GStechnique}
The global sequential calibration (\GS) technique is a multi-variate extension of the \EMJES{} calibration. 
Any variable $x$ that is correlated with the detector response to the jet can be used. 
A multiplicative correction to the jet energy measurement is derived by inverting the 
calibrated jet response \Response{} as a function of this variable: 
\begin{equation}
C(x) = \Response^{-1}(x) / \langle  \Response^{-1}(x) \rangle, 
\end{equation}
where $\langle  \Response^{-1}(x) \rangle$ denotes the average inverse jet response.
After this correction, the remaining dependence of the response on the variable $x$ is removed without 
changing the average energy, resulting in a reduction of the spread of the reconstructed jet energy and, thus, 
an improvement in resolution. 

Several variables can be used sequentially to achieve the optimal resolution. 
This procedure requires that the correction for a given variable $x_{i}$ ($C^i$) is calculated using jets 
to which the correction for the previous variable $x_{i-1}$ ($C^{i-1}$) has already been applied. 
The jet transverse momentum after correction number $i$ is given by :
\begin{equation}
\pt^{i} = C^i(x_i) \times \pt^{i-1} = C^i(x_i) \times C^{i-1}(x_{i-1}) \times \pt^{i-2} = ...
\end{equation}

\subsection{Properties derived from the internal jet structure}
\label{sec:props}
The jet properties used in the \GS{} calibration characterise the longitudinal and transverse topology of the 
energy deposited by the jet. 
A large energy deposit in the hadronic layers indicates, for example, a larger hadronic component of the jet implying an on average lower detector response in the non-compensating \ATLAS{} calorimeter. 
Close to a crack region, the transverse extent of the jet is correlated to how many particles of the jet hit 
the poorly instrumented transition region.

Each of these jet properties may be sensitive to several effects: energy deposited in the dead material, 
non-compensation of the calorimeter, or unmeasured energy due to the noise suppression. 
In the \GS{} calibration, no attempt is made to separate these effects. 
The jet properties help to significantly improve the jet energy resolution, 
and implicitly correct on average for these effects. 

The longitudinal structure of the jet\footnote{Here, longitudinal refers to the direction along the jet axis.} 
is characterised by the fractional energy deposited in the different  
layers of the calorimeters before any jet calibration is applied (``layer fractions'') : 
\begin{equation}
f_{\rm layer} = \frac{E_\EM^{\rm layer}}{E_\EM^{\rm jet}},
\end{equation}
%
where $E_\EM^{\rm jet}$ is the jet energy at the \EM{} scale and $E_\EM^{\rm layer}$ the energy deposited 
in the layer of interest, also defined at the \EM{} scale. 
The transverse jet structure can be characterised by the jet width defined as:
\begin{equation}
\width = \frac{\displaystyle \sum_{i} p^i_{\rm T} \; \Delta R_{i,{\rm jet}}} {\displaystyle \sum_{i} p^i_{\rm T}},
\label{eq:jetwidth}
\end{equation}
%
where the sums are over the jet constituents ($i$) and \pt{} is the transverse constituent momentum.
$\Delta R_{i,{\rm jet}}$ is the distance in $\eta \times \phi$-space between the jet constituents and the jet axis.
In the following study \topos{} are used as jet constituents.

\subsection{Derivation of the global sequential correction}
The \GS{} corrections are determined in jet $|\etajet|$ bins of width $0.1$ from $|\etajet| = 0$ to $|\eta| = 4.5$. 
In each bin, the jet properties that provide the largest improvement in 
jet energy resolution have been selected in an empirical way. 
The chosen jet properties and the order in which they are applied are summarised in Table~\ref{tab:properties}. 
The improvement in resolution obtained is found to be independent of which property is used first to derive a correction. 
\begin{table}[ht!]
  \centering
  \begin{tabular}{c|c|c|c|c}
    \hline    \hline
	$|\etajet|$ region & Corr 1 & Corr 2 & Corr 3 & Corr 4 \\
	    \hline
	$|\etajet| < 1.2$     & \ftile & \fem  & \fpres & \width \\
	\etaRange{1.2}{1.4}   & \ftile &       &        & \width \\
	\etaRange{1.4}{1.7}   & \ftile & \fhec &        & \width \\
	\etaRange{1.7}{3.0}   &        & \fhec &        & \width \\
	\etaRange{3.0}{3.2}   &        & \fem  &        & \width \\
	\etaRange{3.2}{3.4}   &        & \fem  &        &       \\
	\etaRange{3.4}{3.5}   &        & \fem  &        & \width \\
	\etaRange{3.5}{3.8}   & \ffcal &       &        & \width \\
	\etaRange{3.8}{4.5}   & \ffcal &       &        &       \\
    \hline    \hline
  \end{tabular}
  \caption{Sequence of corrections in the \GS{} calibration scheme in each $|\etajet|$ region.}
  \label{tab:properties}
\end{table}

In the following section, ``\GSL'' refers to the calibration applied up to the third correction 
(containing only the calorimeter layer fraction corrections) 
and ``\GS'' to the calibration applied up to the last correction (including the \width{} correction).

%

\begin{figure*}[!ht]
\centering
\subfloat[\fpres]{\includegraphics[width=0.4\textwidth]{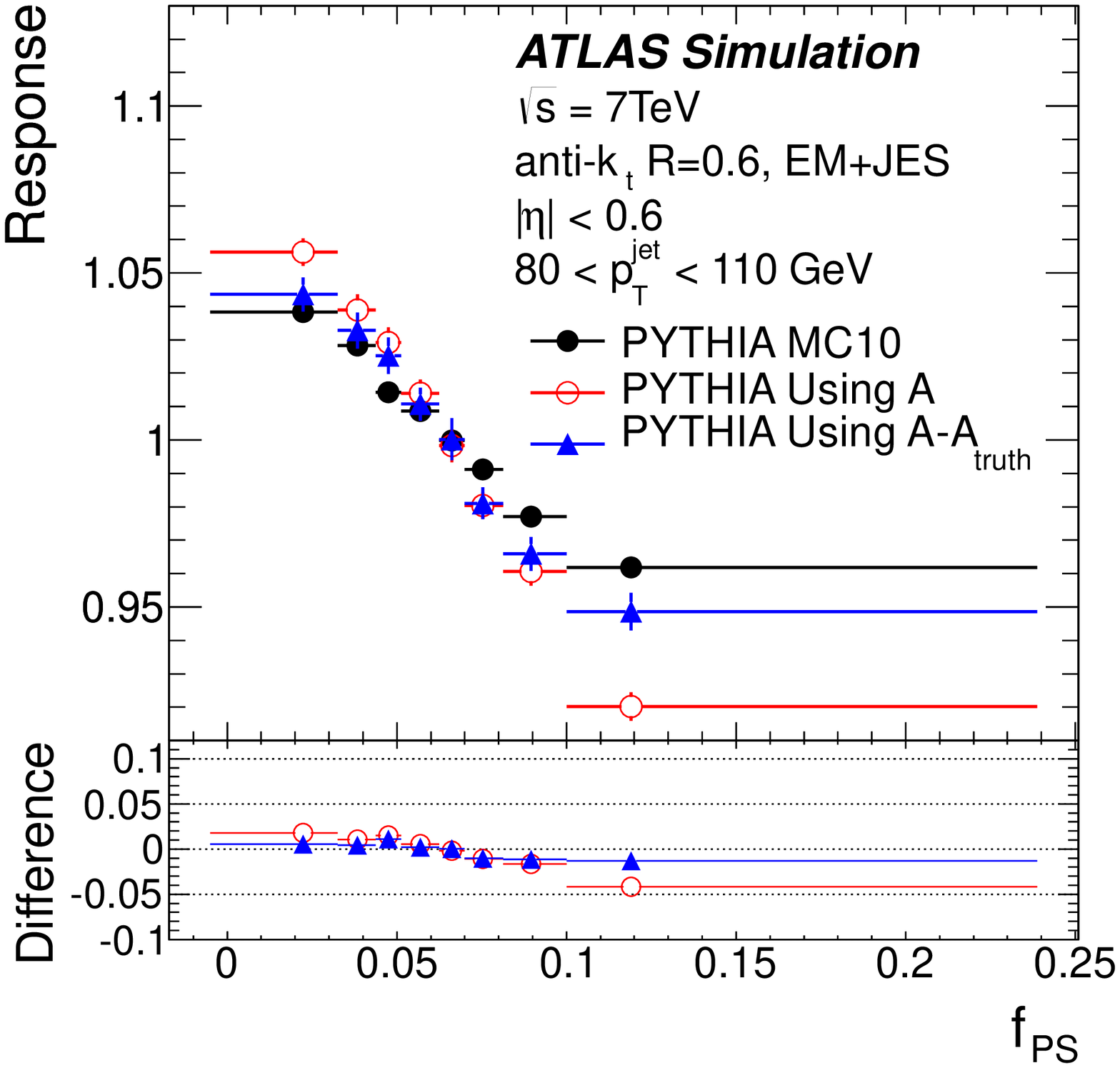}}
\hspace{1.cm}
\subfloat[\fem]  {\includegraphics[width=0.4\textwidth]{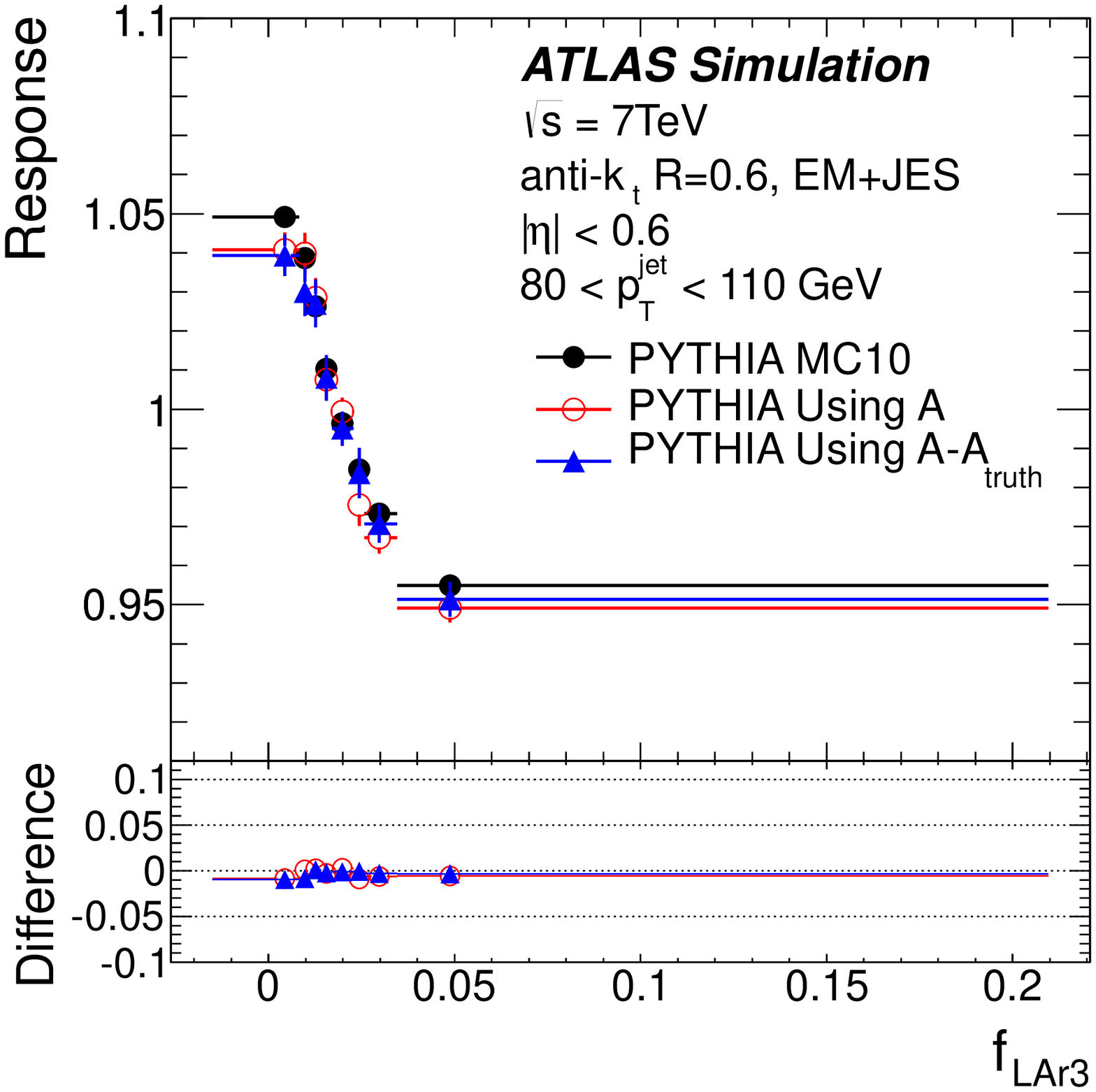}}\\
\subfloat[\ftile]{\includegraphics[width=0.4\textwidth]{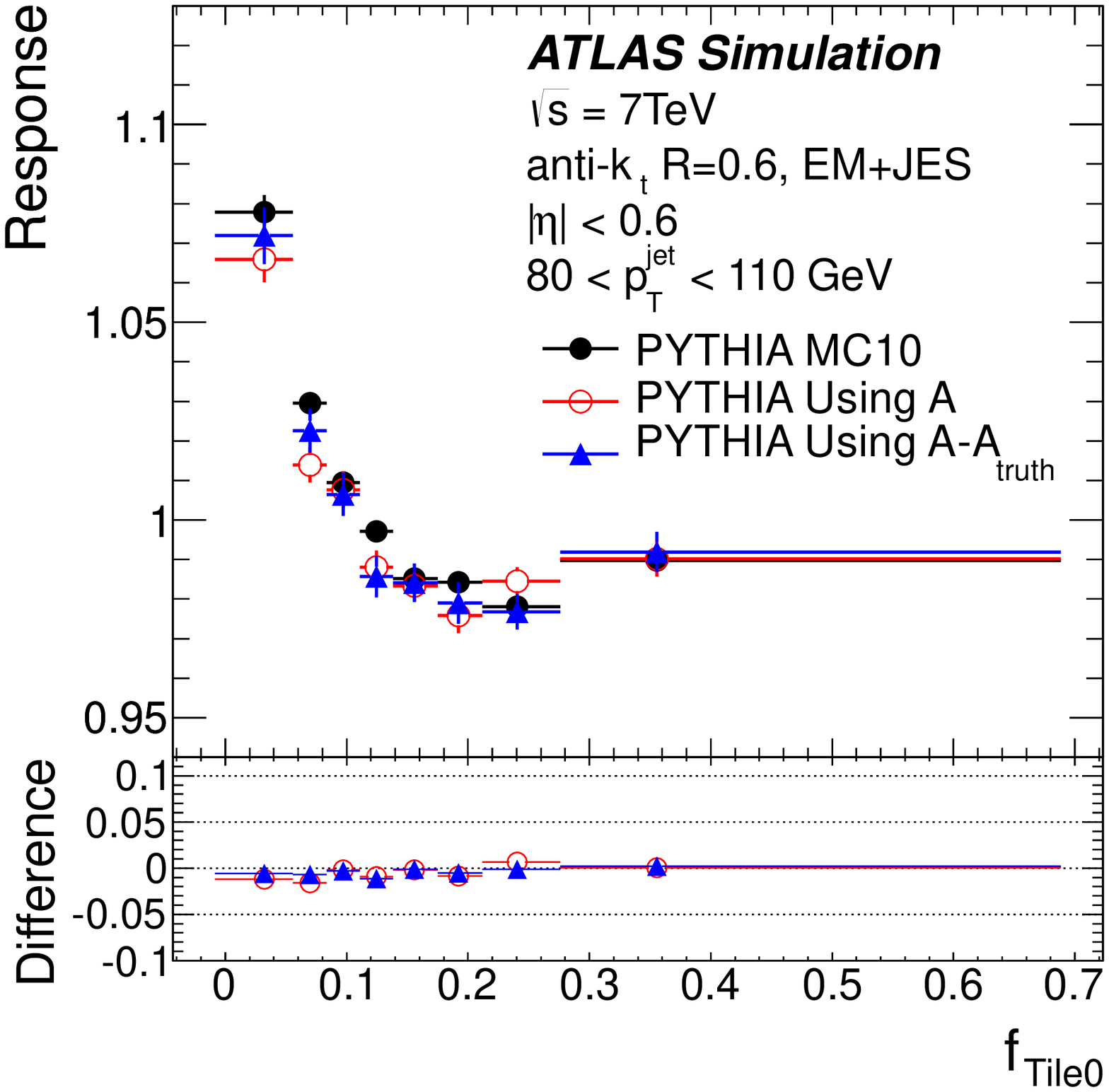}}
\hspace{1.cm}
\subfloat[jet \width]{\includegraphics[width=0.4\textwidth]{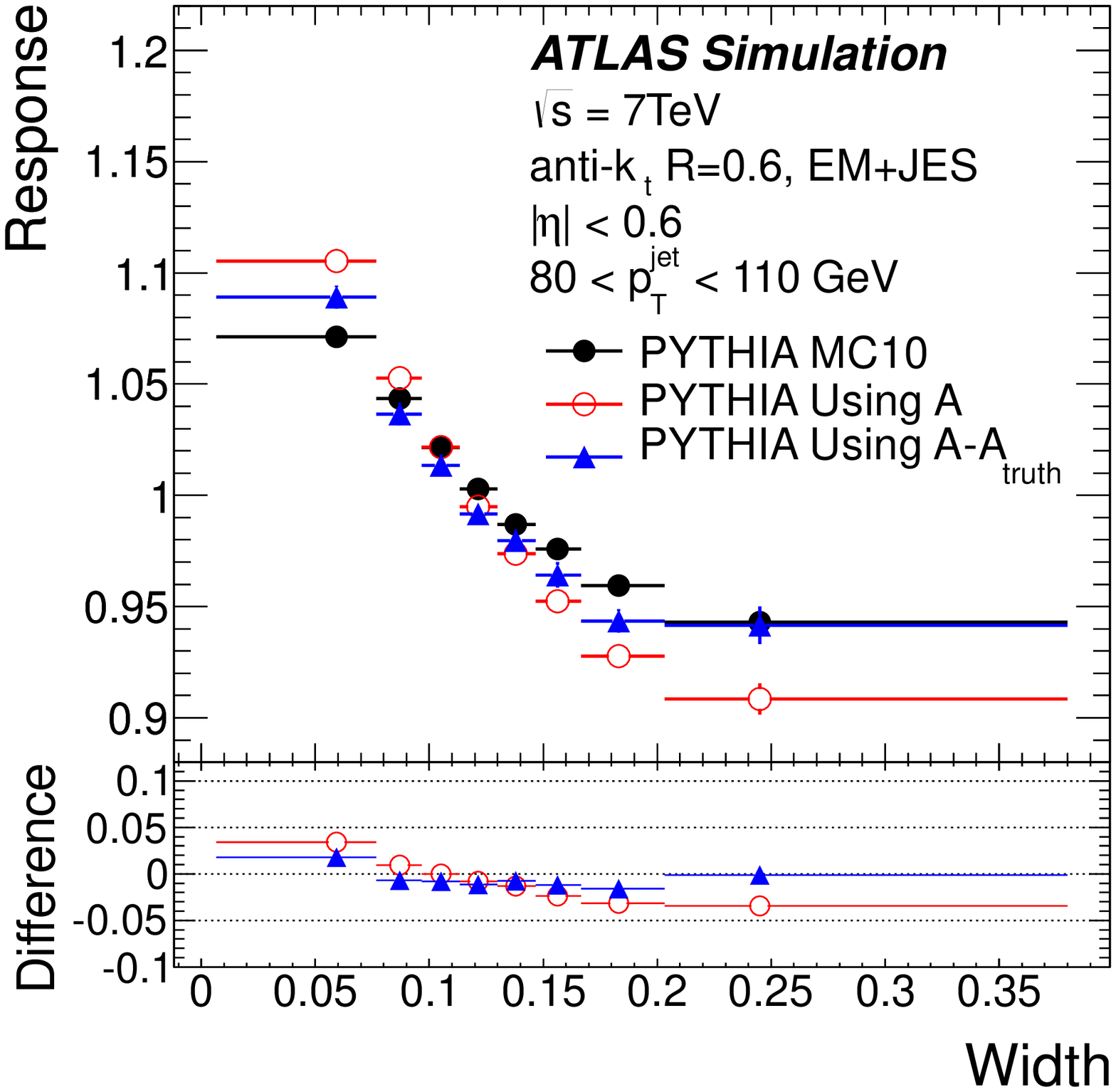}}
\caption{Average jet response calculated using truth jets 
(full circles), using the reconstructed asymmetry
$A$ (open circles), and using $A - A_{\rm true}$ (triangles) as a function of the calorimeter layer energy fraction
\fpres{} (a), \fem{} (b), \ftile{} (c) and the lateral jet \width{} (d) in the \pythia{} MC10 sample. 
The lower part of each figure shows the differences between the response calculated using the truth jet
and the one calculated with the dijet balance method without $A_{\rm true}$ (full triangles) and with $A_{\rm true}$ (open circles). 
\Antikt{} jets with $R=0.6$ calibrated with the \EMJES{} scheme are used and
have $80 \le \ptjet < 110$~\GeV{} and $|\etajet| < 0.6$.}
\label{fig:DijetMethodMC}
\end{figure*}

\begin{figure*}[ht!]
\centering
\subfloat[\fpres]{\includegraphics[width=0.4\textwidth]{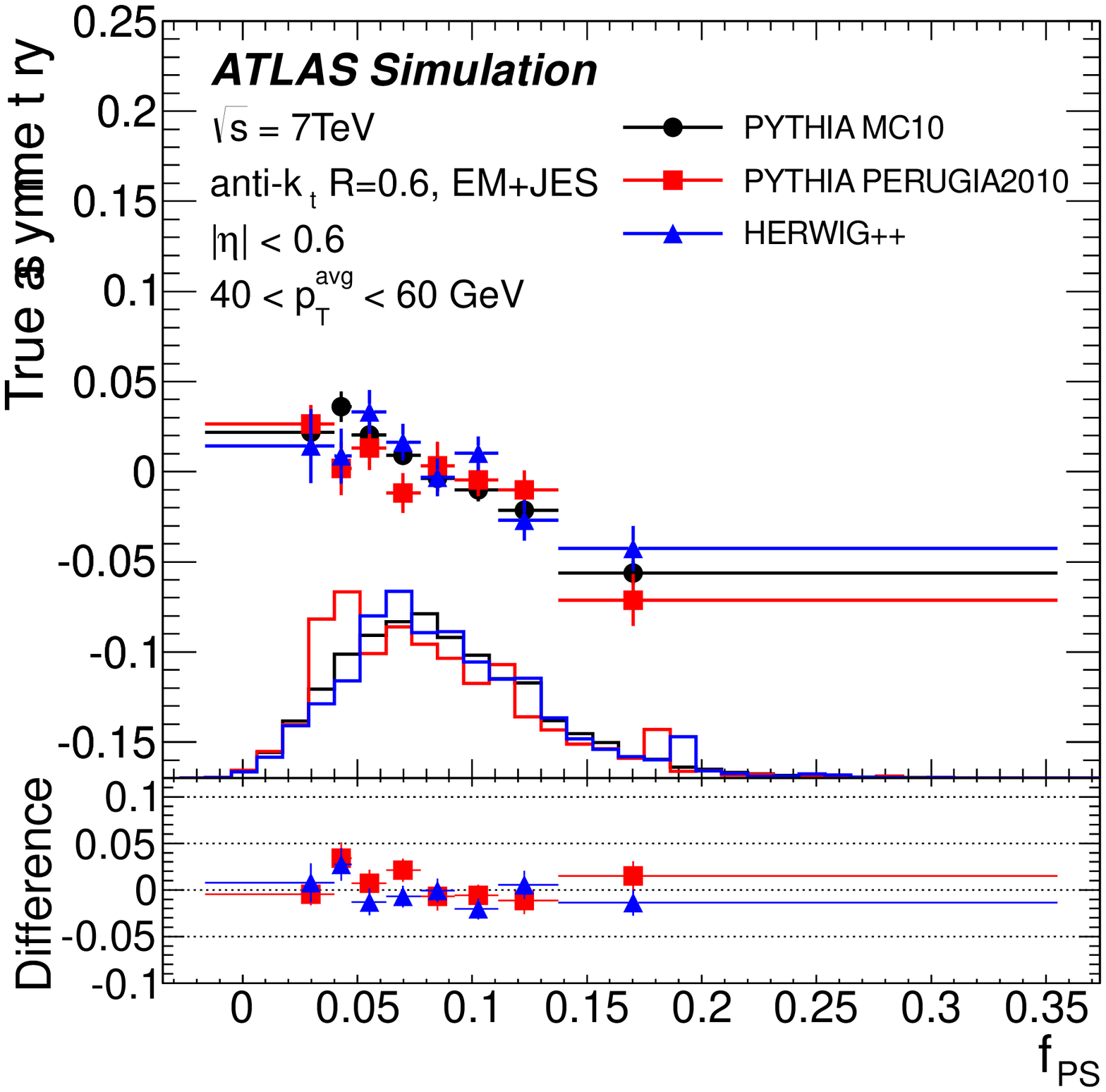}}
\hspace{1.cm}
\subfloat[\fem]{\includegraphics[width=0.4\textwidth]{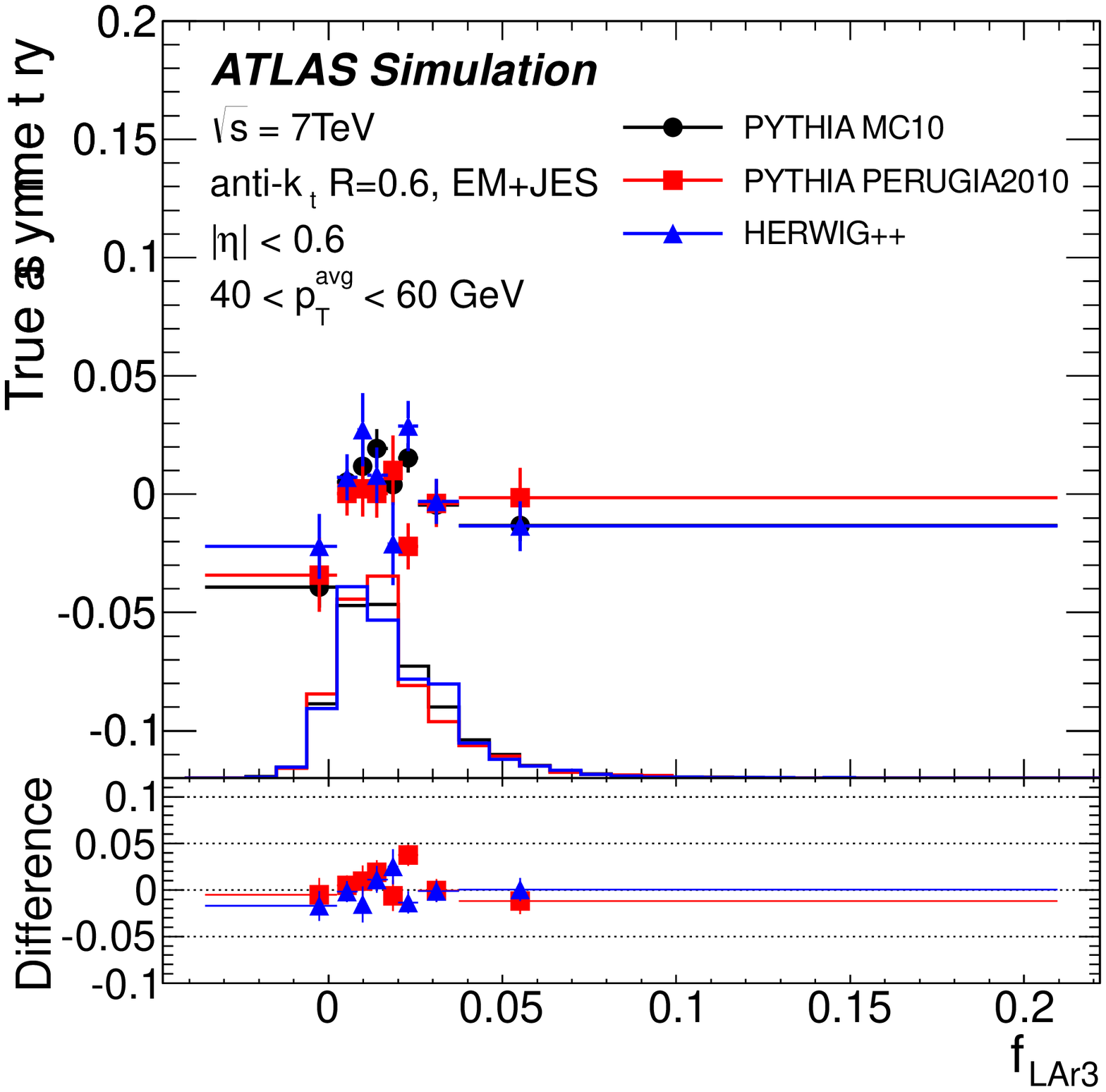}} \\
\subfloat[\ftile]{\includegraphics[width=0.4\textwidth]{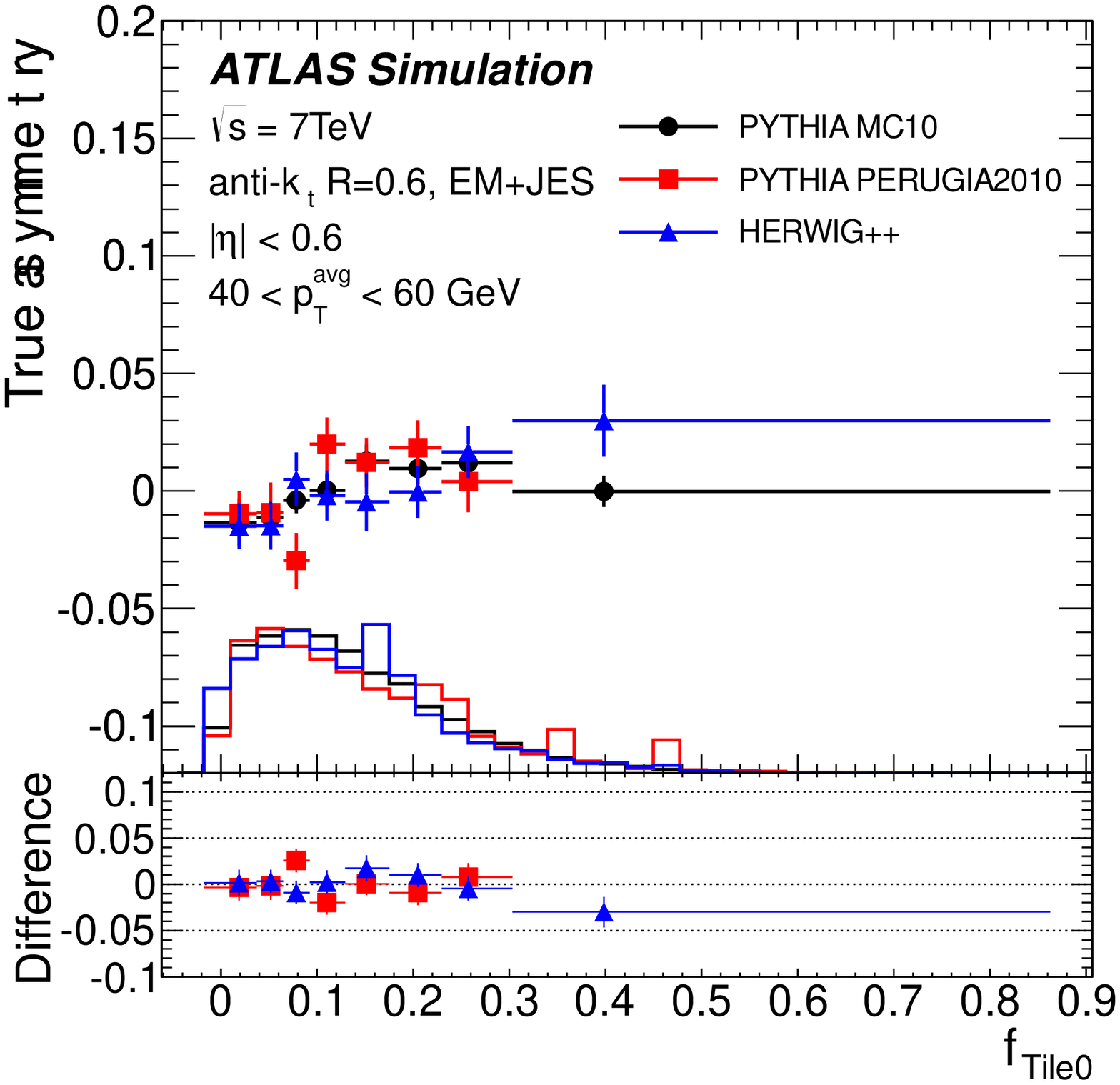}}
\hspace{1.cm}
\subfloat[Jet \width]{\includegraphics[width=0.4\textwidth]{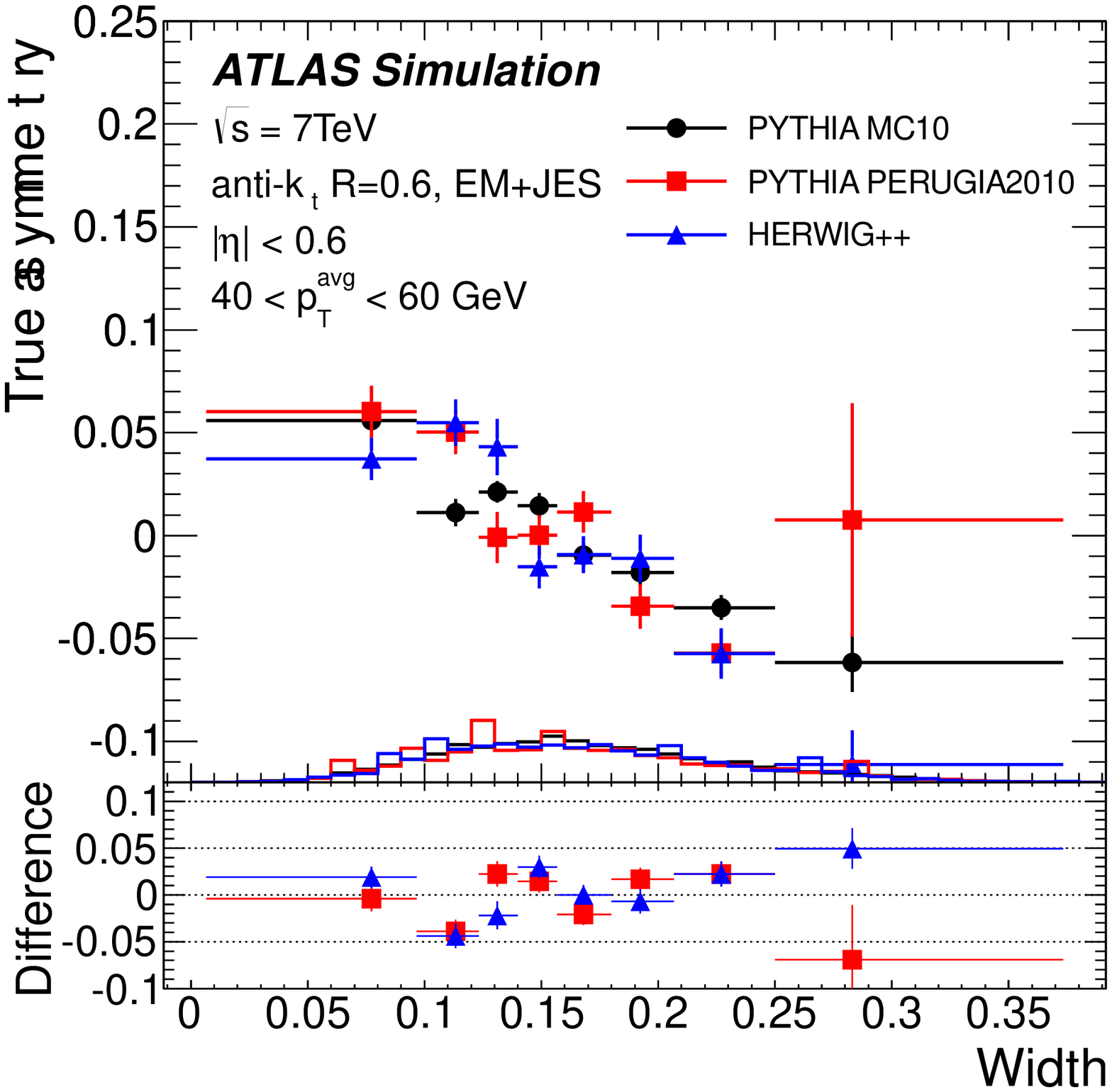}}
\caption{Average asymmetry for truth jets obtained from various Monte Carlo event generators and tunes 
(\pythia{} with the MC10 and the \Perugia 2010 tune and \herwigpp) as a function of the
calorimeter layer fraction $\fpres$ (a),  $\fem$ (b), $\ftile$ (c) and the lateral jet \width{} (d) of the probe jet. 
\Antikt{} jets with $R=0.6$ calibrated with the \EMJES{} scheme are used and 
have $40 \le \ptavg < 60$~\GeV{} and $|\etajet| < 0.6$. 
The distributions of the jet properties are superimposed on each figure.
The lower part of each figure shows the differences between \pythia{} MC10 and the other Monte Carlo generators.
}
\label{fig:AtruthAllSamples}
\end{figure*}

\begin{figure*}[!htb]
\centering
\subfloat[\fpres]{\includegraphics[width=0.4\textwidth]{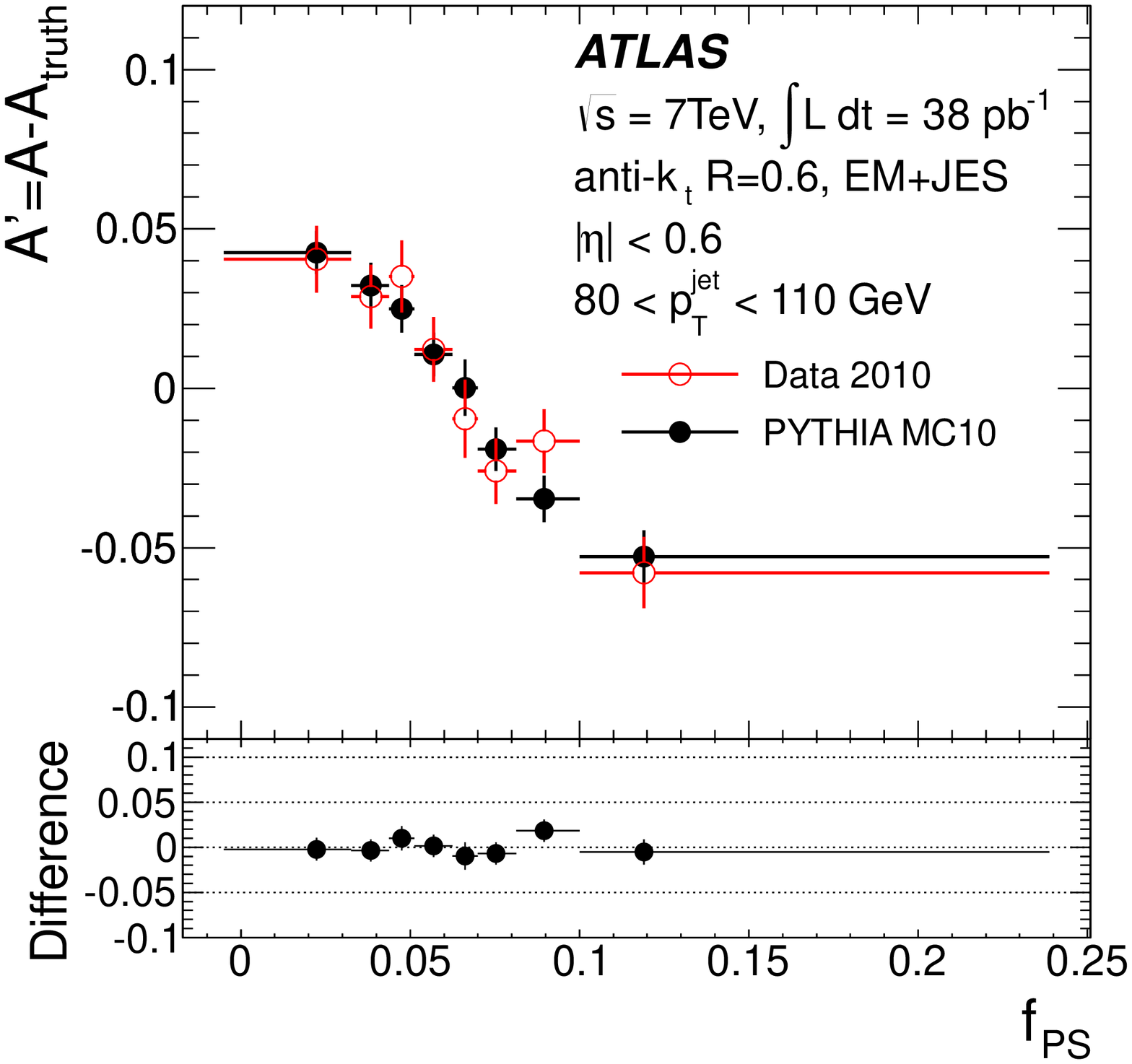}}
\hspace{1.cm}
\subfloat[\fem]  {\includegraphics[width=0.4\textwidth]{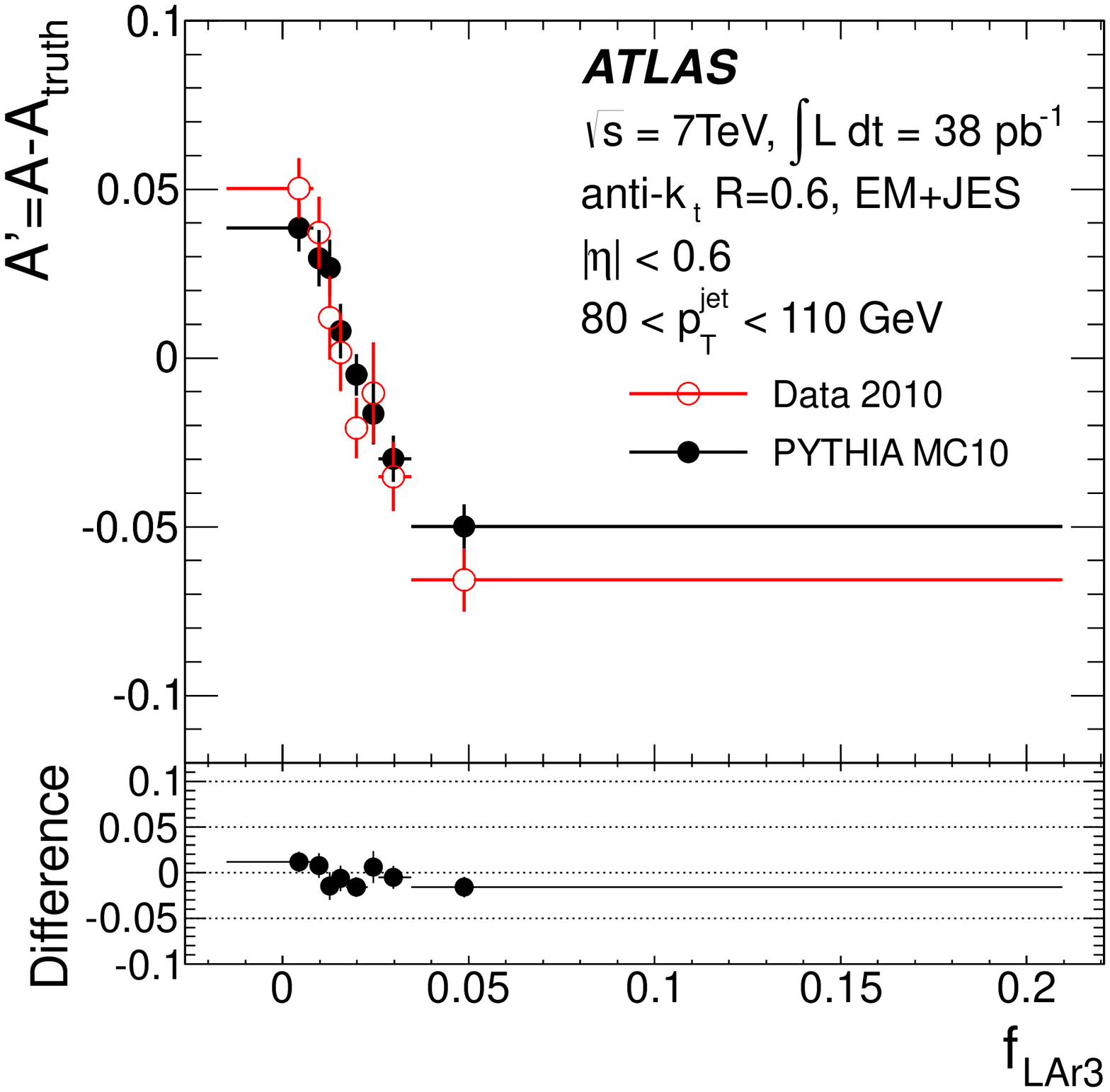}}\\
\subfloat[\ftile]{\includegraphics[width=0.4\textwidth]{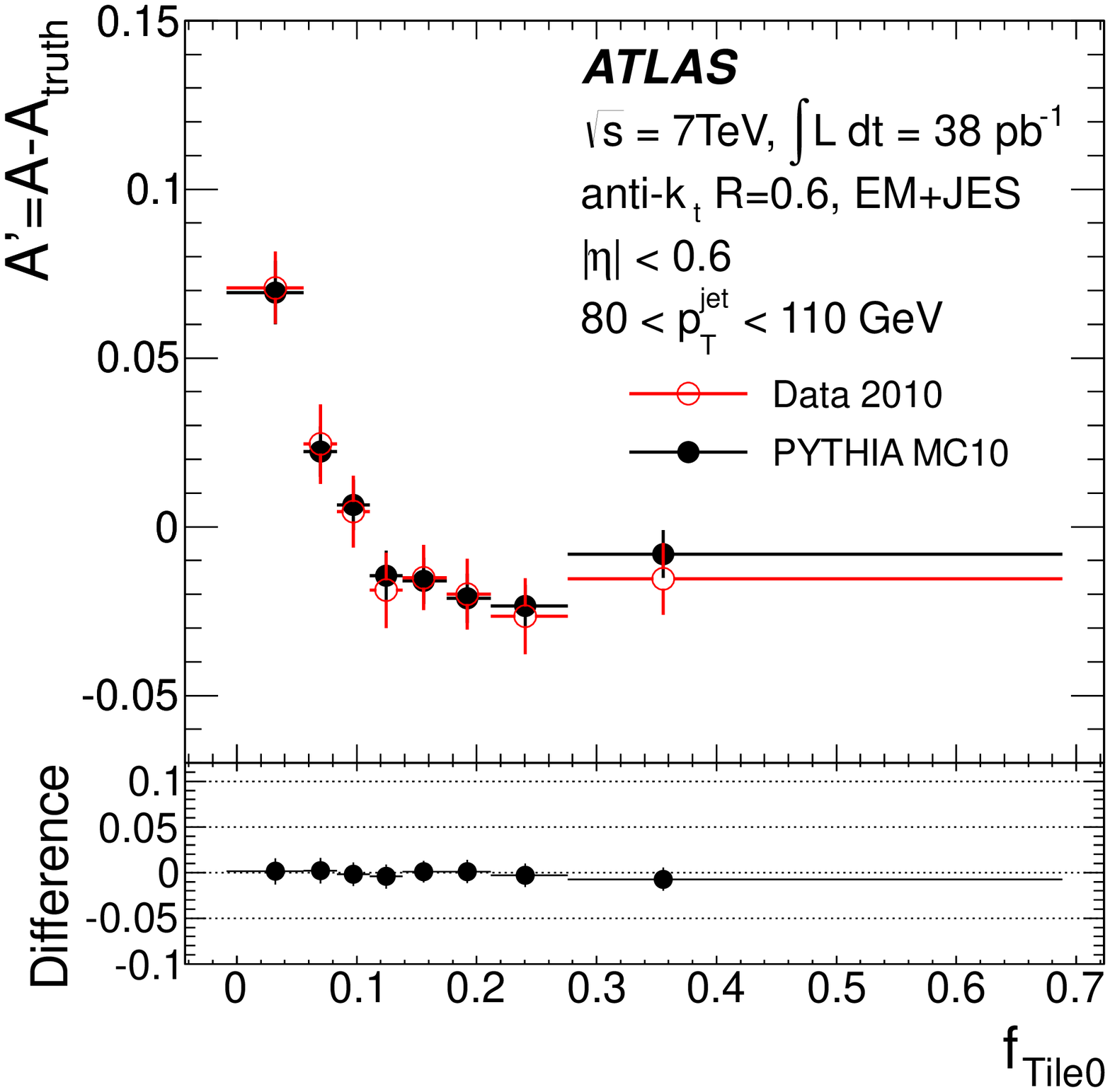}}
\hspace{1.cm}
\subfloat[Jet \width]{\includegraphics[width=0.4\textwidth]{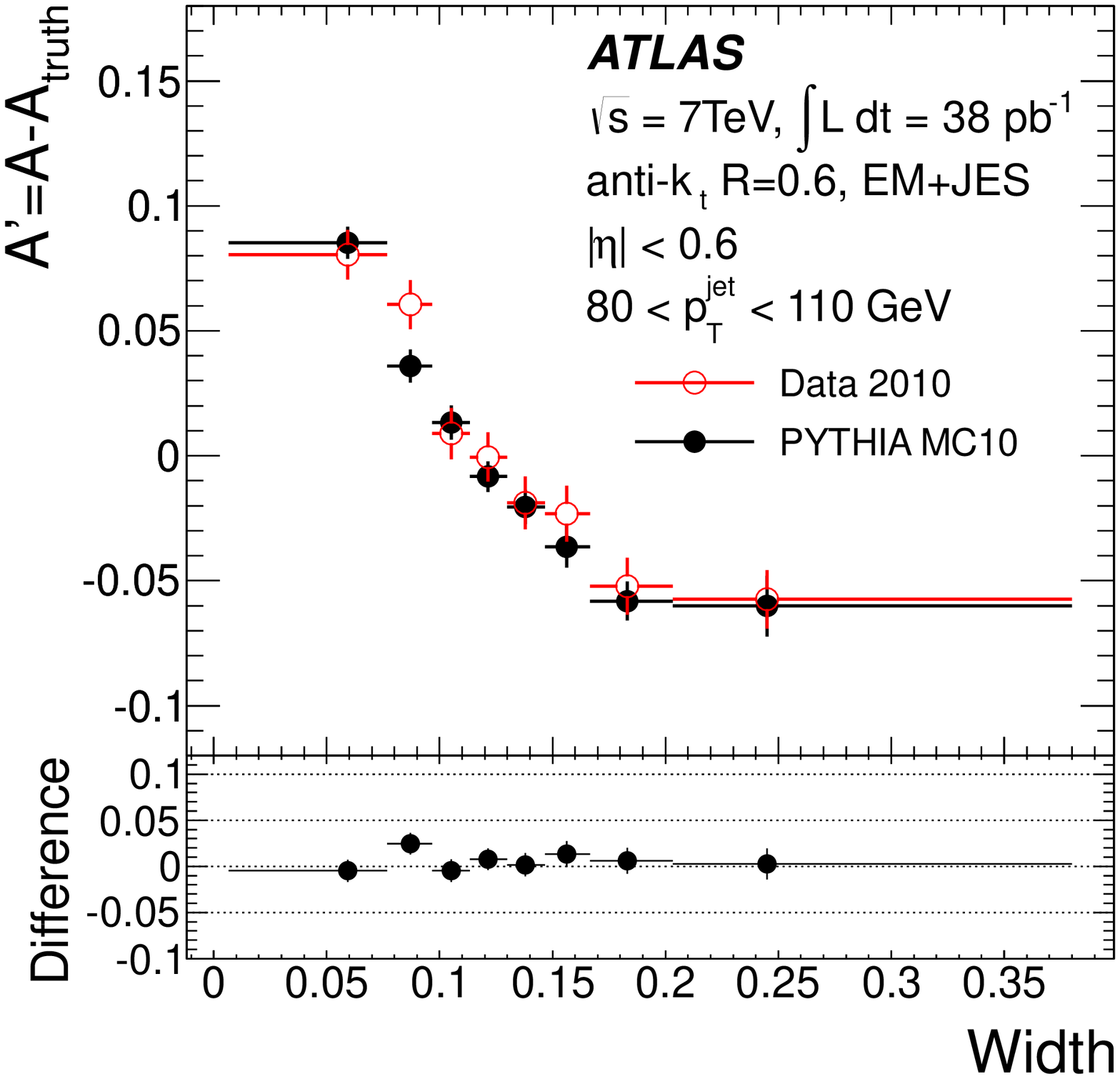}}
\caption{Difference between the average reconstructed asymmetry and the average true asymmetry in data (open circles) 
and in the reference \pythia{} MC10 sample (full circles) as a function of the calorimeter layer fractions
$\fpres$ (a), $\fem$ (b), $\ftile$ (c) and the lateral jet \width{} (d). 
The lower part of each figure shows the differences between data and Monte Carlo simulation. 
\Antikt{} jets with $R=0.6$ calibrated with the \EMJES{} scheme are used and 
have $80 \le \ptjet < 110$~\GeV{} and $|\etajet| < 0.6$.}
\label{fig:DijetMethodDATA}
\end{figure*}
%

%
\begin{figure*}[ht!]
\centering
\subfloat[\fpres]{\includegraphics[width=0.4\textwidth]{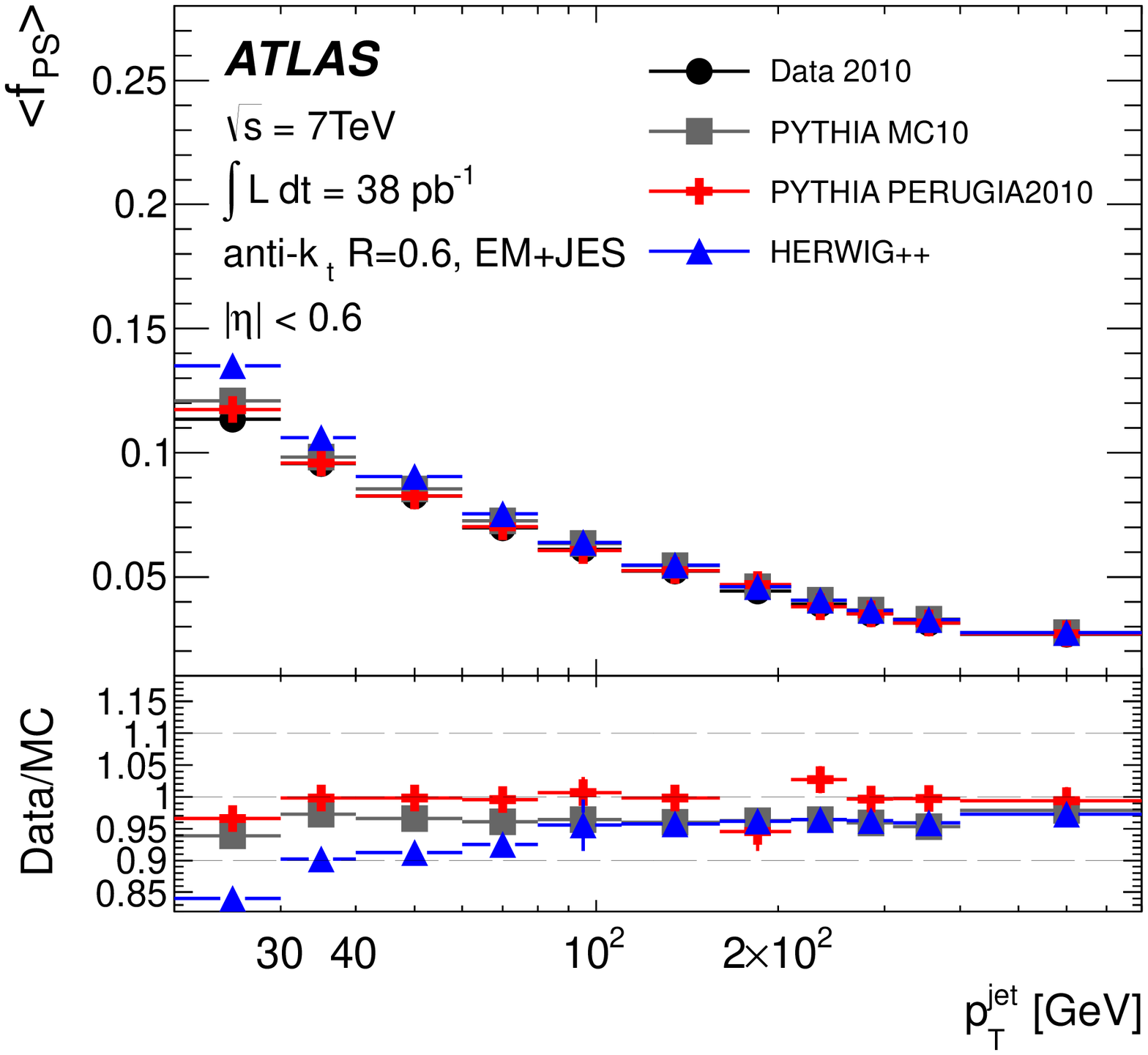}}
\hspace{1.cm}
\subfloat[\fem]  {\includegraphics[width=0.4\textwidth]{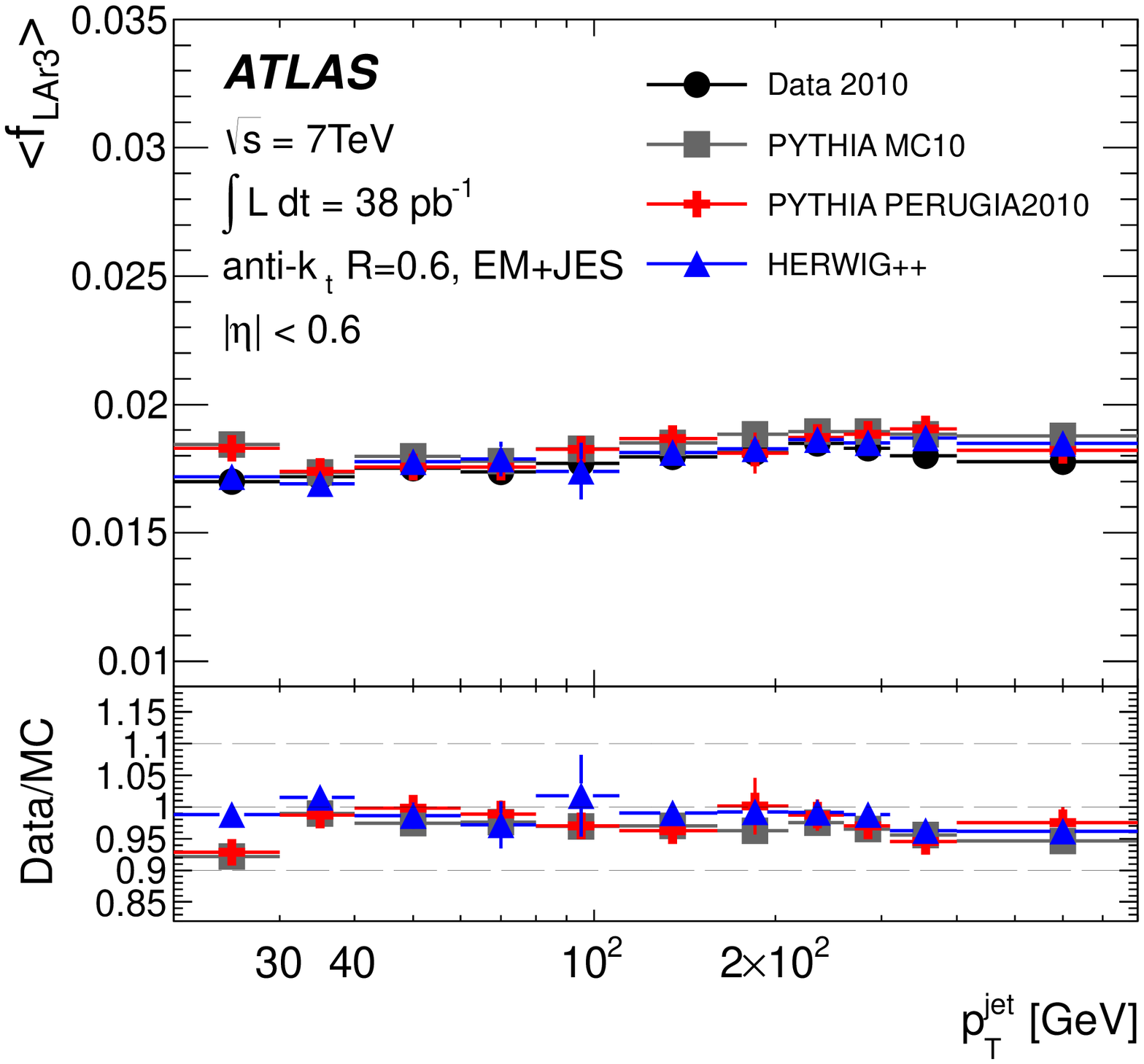}}\\
\subfloat[\ftile]{\includegraphics[width=0.4\textwidth]{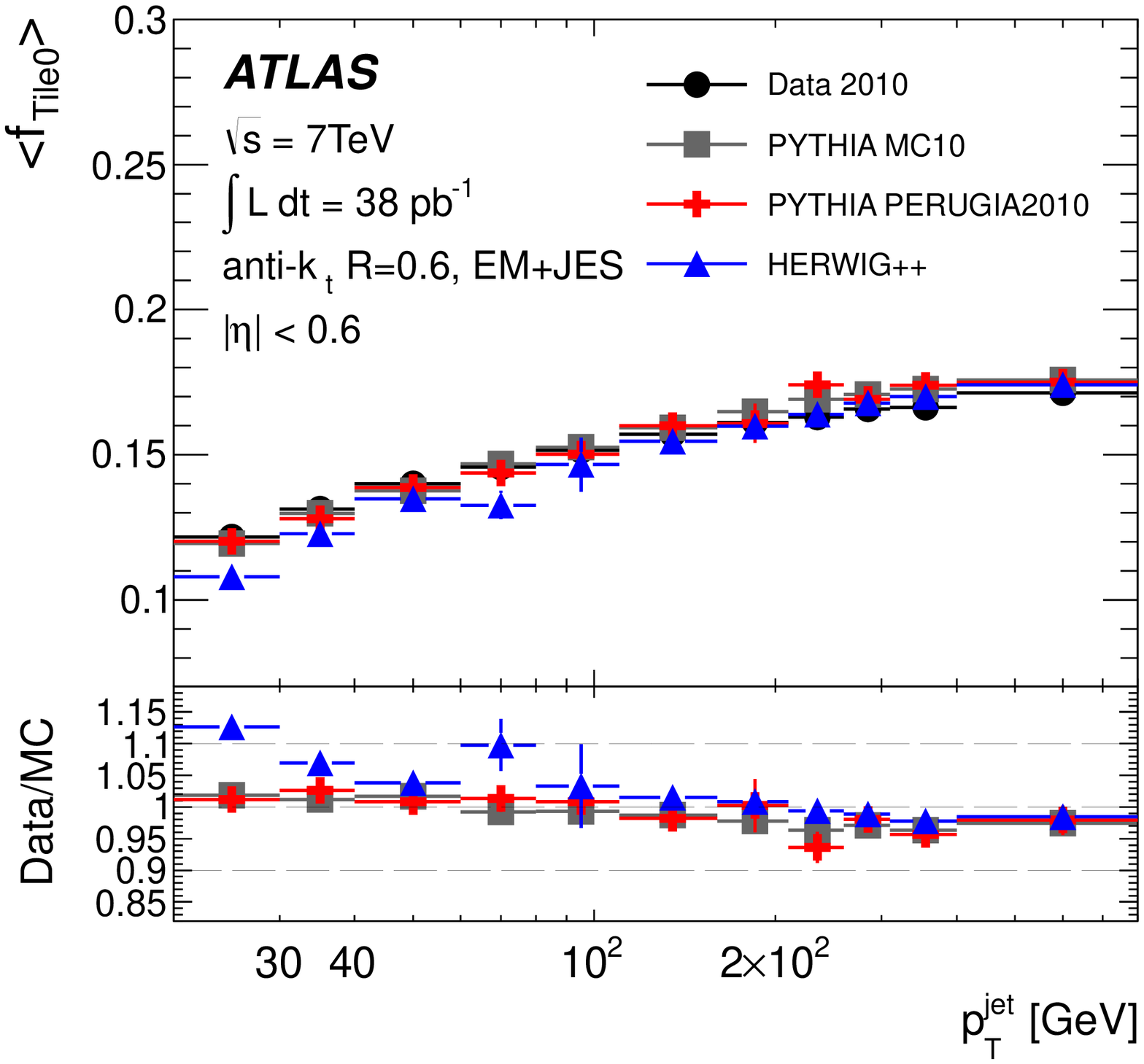}}
\hspace{1.cm}
\subfloat[Jet \width]{\includegraphics[width=0.4\textwidth]{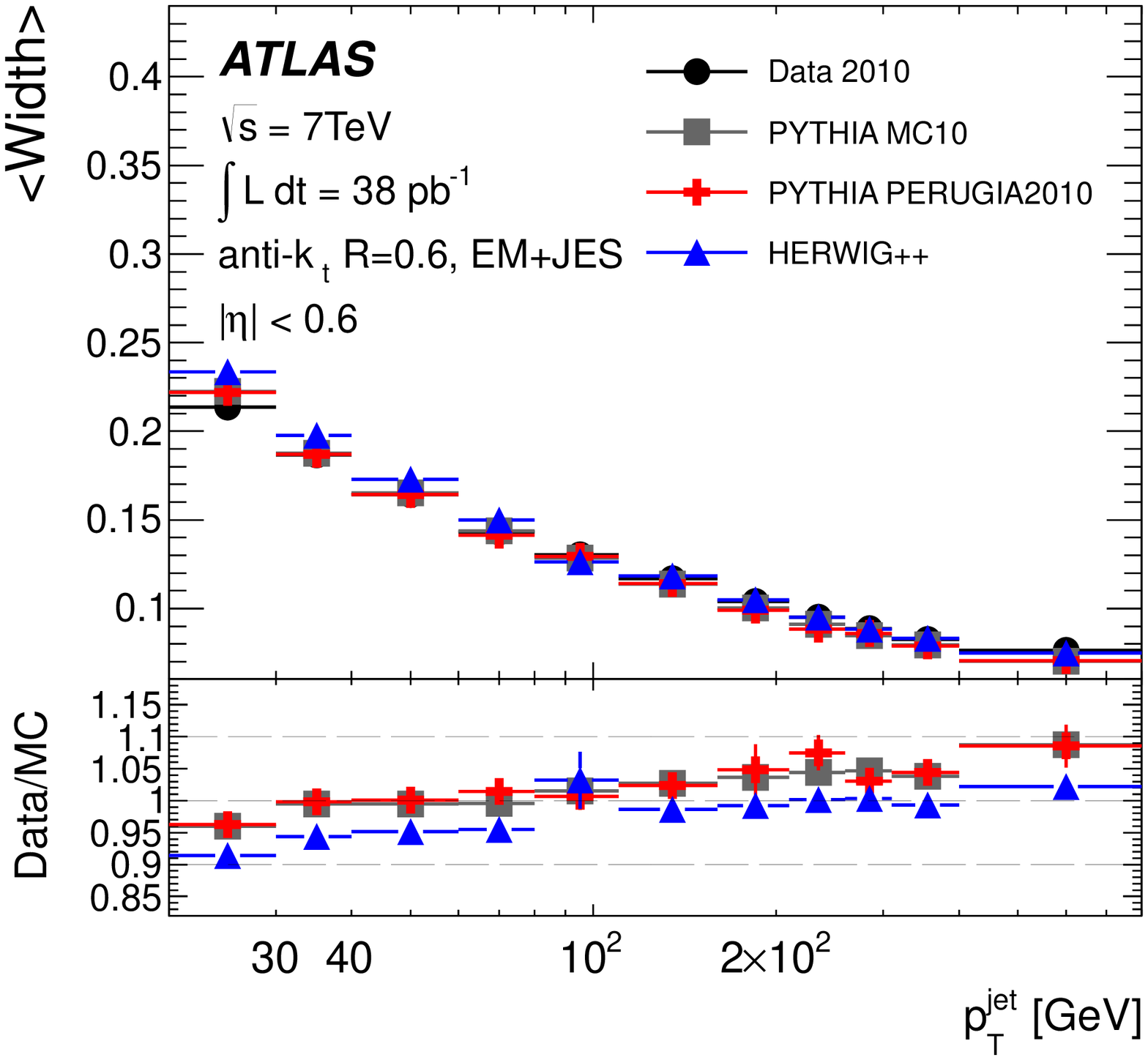}}
\caption{Mean value of the jet calorimeter layer fractions $\fpres$ (a),  $\fem$ (b), $\ftile$ (c) and 
the jet \width{} (d) as a function of \ptjet{} 
for $|\etajet| < 0.6$ for data and various Monte Carlo simulations. 
\Antikt{} jets with $R=0.6$ calibrated with the \EMJES{} scheme are used.
The ratio of data to Monte Carlo simulation is shown in the lower part of each figure.
}
\label{fig:jetprop_vs_pT}
\end{figure*}

%
\begin{figure*}[ht!]
\centering
\subfloat[\GSL{} jet \pt{} comparison]{
\includegraphics[width=0.4\textwidth]{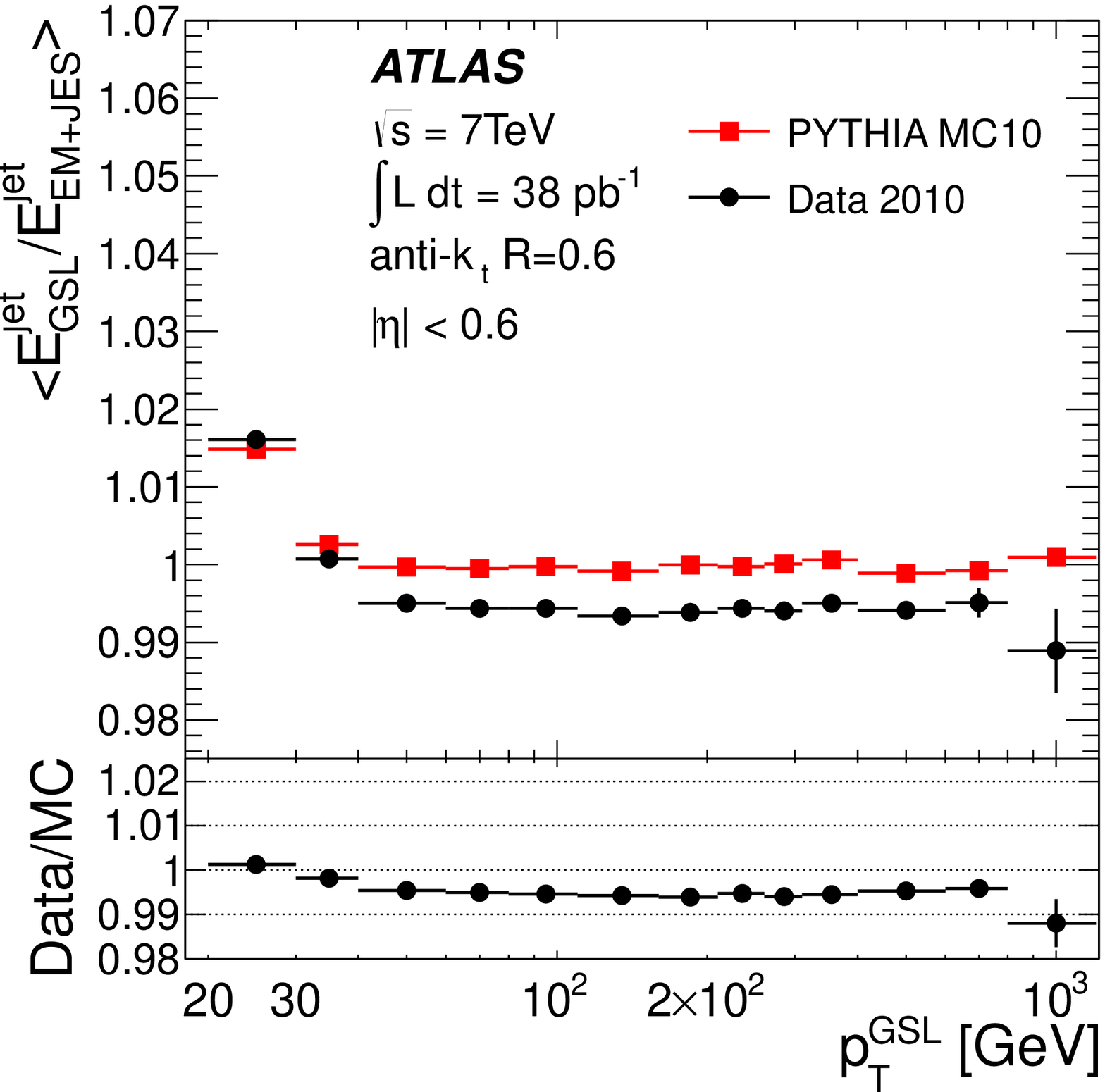}
}
\hspace{1.cm}
\subfloat[\GS{} jet \pt{} comparison]{
\includegraphics[width=0.4\textwidth]{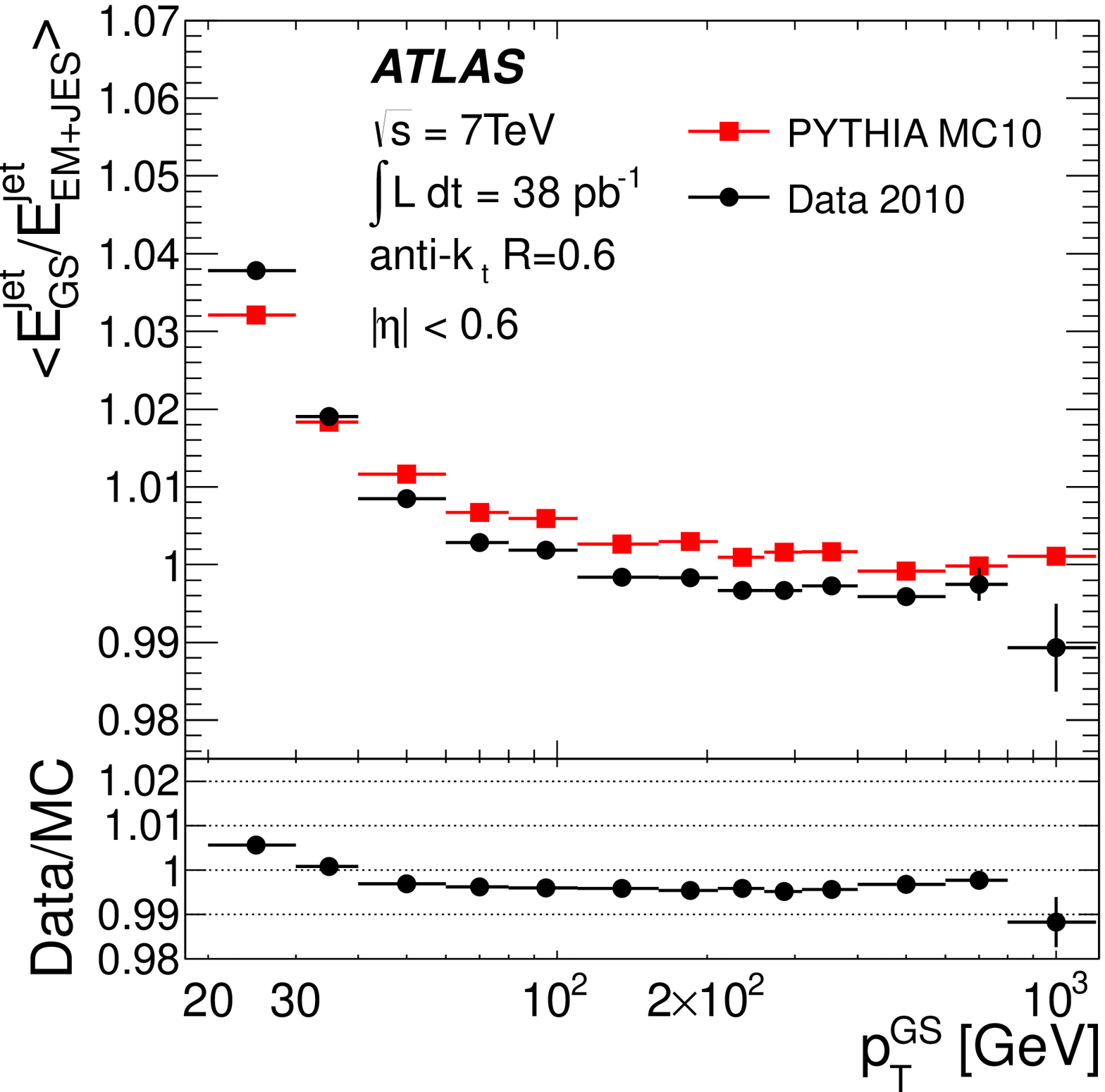}
}\\
\subfloat[\GSL{} jet $\eta$ comparison]{
\includegraphics[width=0.4\textwidth]{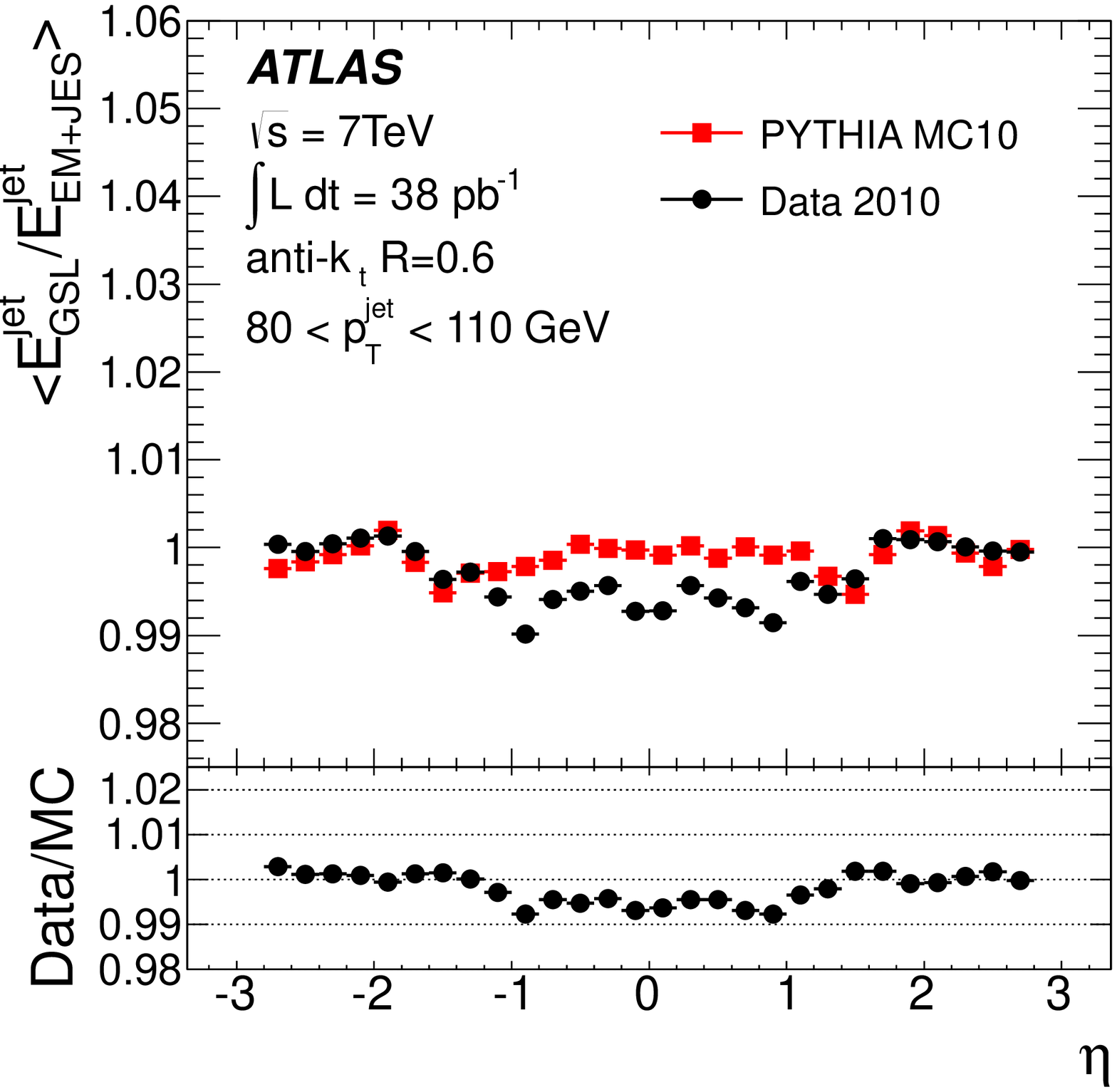}}
\hspace{1.cm}
\subfloat[\GS{} jet $\eta$ comparison]{
\includegraphics[width=0.4\textwidth]{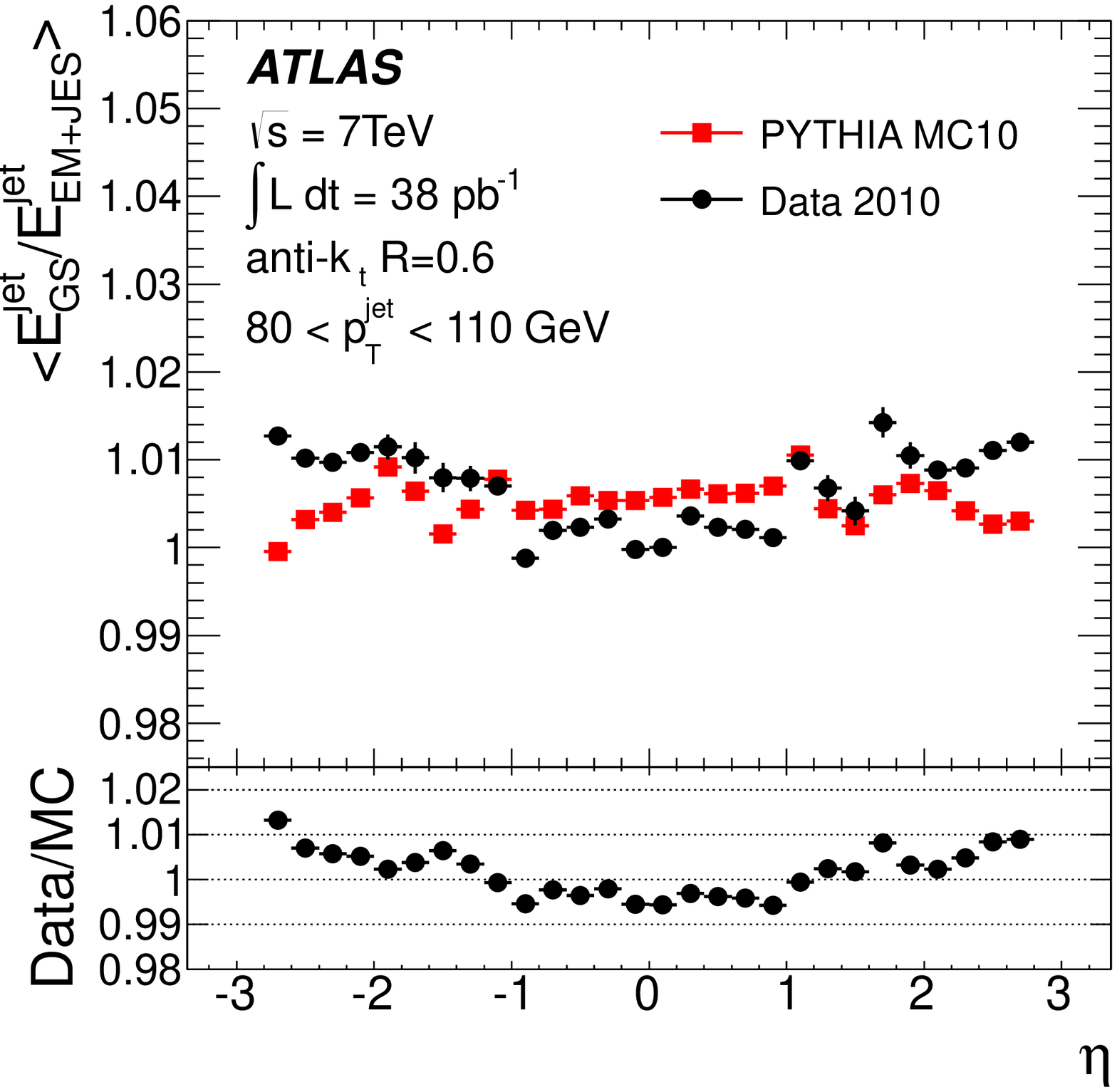}
}
\caption{Average jet energy after \GSL{} (a,c) and \GS{} (b,d) corrections divided by the average jet energy after 
the \EMJES{} calibration as a function of \ptjet{} (a,b) in the calorimeter barrel and 
as a function of \etajet{} for $80 \le \ptjet < 100$~\GeV{} (c,d) 
in data and the Monte Carlo simulation. 
\Antikt{} jets with $R=0.6$ are used.
The double ratio $[E_{\GS{} (\GSL)}/E_{\EMJES}]_{\rm Data}/[E_{\GS{} (\GSL)}/E_{\EMJES}]_{\rm MC}$ is shown in the
lower part of each figure.}
\label{fig:EcalEem}
\end{figure*}

\section{Jet energy scale uncertainties for jet calibrations based on global jet properties}
\label{sec:JESCalibGSC}
The \JES{} uncertainties in the global sequential jet calibration sche\-me are evaluated
using the transverse momentum balance in events with only two jets
at high transverse momentum. 

By construction the \GS{} calibration scheme preserves the energy scale 
of the \EMJES{} calibration scheme for the event sample
from which the corrections have been derived. 
Possible changes of the \JES{} in event samples with different topologies
or jet flavours 
are studied in Section~\ref{sec:gsc-quark-gluon}.

\subsection{Validation of the global sequential calibration using dijet events}
\label{sec:dijetBalance}
%
\subsubsection{Dijet balance method}
\label{sec:DijetBalanceMethod}
The \GS{} corrections can be derived from dijet events using the dijet balance method. 
This method is a tag-and-probe technique exploiting the \pt{} imbalance between two back-to-back jets. 
In contrast to the method presented in Section~\ref{sec:etaintercalibration},
a correction for a truth jet imbalance is applied.  

Dijet events are selected by requiring that the two highest \pt{} jets are back-to-back ($\Delta \phi > 2.8$ radian). 
The two jets are required to be in the same pseudorapidity region. 

The jet whose response dependence on the layer fractions or \width{} is studied, is referred to as the probe jet, 
while the other is referred to as the reference jet. 
The average transverse momentum of the probe and the reference jet is defined as 
\begin{equation}
\pt^{\rm avg} = (\pt^{\rm probe}+ \pt^{\rm ref})/2. 
\end{equation}
Since the choice of  the reference jet and the probe jet is arbitrary, events are always used twice, 
inverting the roles of reference and probe.

The \GS{} corrections are measured through the asymmetry variable defined as:
\begin{equation}
\label{eq:AsymReco}
A(x) = \frac{\pt^{\rm probe}(x) - \pt^{\rm ref}}{\pt^{\rm avg}(x)},
\end{equation}
\noindent where $x$ is any of the properties used in the \GS{} calibration (see Table~\ref{tab:properties}). 
Both $\pt^{\rm probe}$ and $\pt^{\rm ref}$ depend on $x$, but the dependence is explicitly written only for the probe jet, 
because the jet property used to build the correction belongs to the probe jet. 

The probe and the reference jet transverse momenta are defined with the same calibration. 
When computing correction factor $i$, they are both corrected up to the $(i-1)^{\rm th}$ correction 
(see Section~\ref{sec:GStechnique}). 
The mean response as a function of $x$ is given by:
\begin{equation}
\langle R(x) \rangle= \frac{1 + \langle A(x) \rangle /2}{1 -\langle A(x) \rangle /2}.
\label{eq:RespVsAsym}
\end{equation}

The measurement of the response through the asymmetry defined in Equation~\ref{eq:AsymReco} assumes that the 
asymmetry is zero. This is true on average, but not when computed in bins of $x$. 
The measured asymmetry $A(x)$ is therefore a mixture of detector effects and imbalance at the level
of the generated particles. 
In order to remove the effect of imbalance at the level of generated particles, a new asymmetry is defined: 
\begin{equation}
A'(x) = A(x) - A_{\rm true}(x),
\end{equation}
\indent where $A(x)$ is given by Equation~\ref{eq:AsymReco} and $A_{\rm true}(x)$ is: 
\begin{equation}
A_{\rm true}(x) = \frac{p_{{\rm T},{\rm true}}^{\rm probe}(x)-p_{{\rm T},{\rm true}}^{\rm ref}}{p_{{\rm T},{\rm true}}^{\rm avg}(x)},
\end{equation}
\noindent where $p_{{\rm T},{\rm true}}^{\rm avg}(x) = (p_{{\rm T},{\rm true}}^{\rm probe}(x)+p_{{\rm T},{\rm true}}^{\rm ref})/2$. 
The variable $A_{\rm true}$ denotes the asymmetry for truth jets (or true asymmetry) and 
is calculated by matching reconstructed jets to truth jets. 
The asymmetry $A_{\rm true}$ is determined in the Monte Carlo simulation. 
When using $A'(x)$ instead of $A(x)$ in Equation~\ref{eq:RespVsAsym}, the effects of imbalance at the level 
of generated particles are removed
and the resulting response depends only on detector effects. 
Accounting for the truth jet imbalance is particularly important for the corrections 
that depend on the energy in the presampler and the jet \width. 

\subsubsection{Validation of the dijet balance method in the Monte Carlo simulation}
\label{sec-ValidationDiJetMC}
The dijet balance method can be checked in two different ways.

The first uses the default \pythia{} event sample with the MC10 tune and compares 
the response calculated using Equation~\ref{eq:RespVsAsym} to the response calculated using 
the truth jets. 
Figure~\ref{fig:DijetMethodMC} shows this comparison for jets after the \EMJES{} calibration 
for  $80 \le \ptjet < 110$~\GeV{} and $|\etajet| < 0.6$. 
The results obtained using the asymmetry defined as in Equation~\ref{eq:AsymReco} and when
incorporating the true asymmetry are shown. 
If the true asymmetry were ignored, the calculated response would be different
from the the true jet response by up to $4 \%$ for high values 
of the jet \width{} and the presampler fraction in this particular \ptjet{} bin. 
This difference increases with decreasing \ptjet{} reaching $8 \%$ for jets of $\ptjet \approx 20 \GeV$. 
These differences %
are reduced to less than $2  \%$ 
when a correction for $A_{\rm true}$ is used. 
Similar results are found in the other \ptjet{} and $|\etajet|$ bins.

The second test compares the true asymmetry between different simulated samples. 
Figure~\ref{fig:AtruthAllSamples} shows the true asymmetry as a function of 
\fpres{}, \fem{}, \ftile{} and the jet \width{} in the central region for $40 \le \ptjet < 60$~\GeV{} 
for various event samples:
the reference \pythia{} sample with the MC10 tune, 
the \pythia{} sample with the \Perugia 2010 tune and the \herwigpp{} sample.
The last two samples test the sensitivity to the description of
soft physics or the specifics of the hadronisation process 
that could cause differences in the truth jet imbalance. The true asymmetry differs by no more than $5 \%$ 
in this particular \ptjet{} and \etajet{} bin. 
For $\ptjet > 60$~\GeV{} and other $|\etajet|$ bins, the true asymmetries differ by less 
than $2 \%$. At low \ptjet{} (below $40$~\GeV{} in the barrel), the $\Delta \phi$ cut, in particular 
combined with the small \Perugia 2010 and \herwigpp{} samples yield 
statistical uncertainties of the order of $5 \%$.

In summary, the dijet balance method allows the determination of the response as a function of the layer fractions and 
the jet \width{} over the entire transverse jet momentum and pseudorapidity ranges. 
This method can therefore be applied to data to validate the corrections derived in the Monte Carlo simulation.

\subsubsection{Differences between data based and Monte Carlo based corrections}
\label{sec:MCBased_vs_DataBased}
%
Figure~\ref{fig:DijetMethodDATA} shows the difference between the reconstructed asymmetry 
and the true asymmetry for the \pythia{} MC10 sample
as a function of $\fpres$, $\fem$, $\ftile$ and \width{} for jets 
with $80 \le \ptjet < 110$~\GeV{} and $|\etajet| < 0.6$. 
The reconstructed asymmetries in
data and the \pythia{} MC10 sample are compatible within statistical uncertainties. 
Similar agreement is found in the other \etajet{} and \ptjet{} regions. 

The asymmetries as shown in Figure~\ref{fig:DijetMethodDATA} are used to
derive data based corrections.
The difference between data and Monte Carlo simulation
provides a quantitative measure of the additional jet energy scale uncertainty
introduced by the \GS{} calibration.
After the first two corrections in Table~\ref{tab:properties}
the response changes by less than $1 \%$ for data based and Monte Carlo based corrections. 
The response changes by an additional $1 \%$ to $2 \%$ after the third (\Presampler) and the fourth (\width) corrections 
are applied in the barrel. The agreement in the endcap is within $2 \%$ ($4 \%$) for $\pt^{\rm truth} > 60$~\GeV{} ($ < 60$~\GeV).

Data based corrections are also derived with true asymmetries coming from the \Perugia 2010 and \herwigpp{} samples. 
These corrections are then applied to the reference \pythia{} MC10 sample and 
the response yielded is compared to the response obtained after applying the reference data based corrections using 
the true asymmetry from the reference \pythia{} MC10 sample. The difference in response is found to be 
lower than $0.5 \%$ in all the \ptjet{} and $|\etajet|$ bins where the statistical uncertainty is small enough.

\label{sec:jet-properties-syst}
As a further cross-check the same \GS{} corrections (here the Monte Carlo based ones) 
are applied to both data and Monte Carlo simulation samples. The difference between data and simulation 
reflects differences in the jet properties used as input to the \GS{} calibration in the inclusive samples. 

Figure~\ref{fig:jetprop_vs_pT} shows the mean value of $\fpres$, $\fem$, $\ftile$ and 
\width{} as a function of \ptjet{} 
in the barrel for data and various Monte Carlo simulation samples: 
the nominal \pythia{} MC10, \pythia{} \Perugia 2010 and \herwig++. 
The agreement for $\ftile$ and $\fpres$ between data and \pythia{} with the MC10 tune is within $5 \%$
over the entire \ptjet{} range. For $\fem$, this agreement is also within $5 \%$ except for $20 \le \ptjet < 30$~\GeV{} 
where a disagreement of $7.5 \%$ is observed. A larger disagreement is found for the jet width. 
Jets are $5 \%$ ($10 \%$) wider in data than in Monte Carlo simulation at $200$~\GeV{} ($600$~\GeV).

The standard deviations of the \fem{}  and the \fpres{} distributions
show also agreement within $5 \%$  between data and \pythia{} MC10 simulation for $\fem$ and $\fpres$  
over the entire \ptjet{} range. 
For $\ftile$ and \width, disagreements of $10 \%$ are observed in some \ptjet{} bins. 
Similar results are found in the other $|\etajet|$ bins for the calorimeter layer fractions
and the jet \width, except for \etaRange{2.1}{2.8}, where the agreement for the \width{} is slightly
worse than in the other eta ranges.

Figure~\ref{fig:jetprop_vs_pT} 
shows that \pythia{} with the MC10 and the \pythia{}
\Perugia 2010 tunes agree to within a few per cent. The agreement of the \herwig{} sample with data is  
as good as for the other samples for $\fem$ and $\ftile$, except for $20 \le \ptjet < 30$~\GeV. 
For $\fpres$ and the \width, disagreements of $5 - 10 \%$ are observed between \herwigpp{} 
and the other samples for $\ptjet < 60$~\GeV. For $\ptjet > 160$~\GeV, \herwigpp{} is found to describe 
the \width{} observed in data better than the other samples.

The systematic uncertainty can be quantitatively estimated by comparing how the correction coefficients 
$E_{\GS}^{\rm jet}/E_{\EMJES}^{\rm jet}$ differ between data and Monte Carlo simulation. 
The correction coefficient as a function of \ptjet{} in the barrel calorimeter in data 
and in the \pythia{} MC10 sample after \GSL{} and \GS{} corrections
are shown in Figure~\ref{fig:EcalEem}a and Figure~\ref{fig:EcalEem}b. 
The ratios of data to Monte Carlo simulation  are shown 
in the lower part of each figure. 
Figure~\ref{fig:EcalEem}c and Figure~\ref{fig:EcalEem}d show the same quantity, 
but as a function of $\etajet$ for $80 \le \ptjet < 110$~\GeV. 

Deviations from unity in the ratios between data and Monte Carlo simulation as shown in Figure~\ref{fig:EcalEem}
represent the systematic uncertainty associated to the \GS{} corrections. 
This uncertainty is added in quadrature to the \EMJES{} uncertainty.
The results for all the \ptjet{} and $\etajet$ ranges are the following:

For $20 \le \ptjet < 30$~\GeV{} and \etaRange{0}{2.1}, 
the data to Monte Carlo ratio 
varies from $0.5 \%$ to $0.7 \%$ depending on the $|\etajet|$ region. 
For $\ptjet > 30$~\GeV{} and \etaRange{0}{2.1}, 
the uncertainty is lower than $0.5 \%$. 
For \etaRange{2.1}{2.8}, the the data to Monte Carlo ratio 
varies from $0.4 \%$ to $1 \%$ depending on the \ptjet{} bin. 
For a given \ptjet, the uncertainty is higher for \etaRange{2.1}{2.8} than for \etaRange{0}{2.1},
because of the poorer description of the jet width.
For \etaRange{2.1}{2.8} the \GSL{} scheme shows slightly larger difference than the \GS{} scheme.
In general, the uncertainty on the data to Monte Carlo ratio 
is lower than $1 \%$ for $20 \le \ptjet < 800$~\GeV{} and \etaRange{0}{2.8}.

The uncertainty coming from the imperfect description of the jet properties 
and the differences between data based and Monte Carlo simulation based corrections presented 
in Section~\ref{sec:dijetBalance} are not independent. 
The average jet response after the \GS{} calibration in each \ptjet{} and \etajet{} bin, 
which depends on both the distribution of the properties and the \GS{} corrections, is close 
to the response after the \EMJES{} calibration.

A change in the distribution of a jet property therefore translates into a change in the \GS{} correction 
as a function of this property such that the average jet response stays the same in the sample used to derive the correction. 
The differences described in Section~\ref{sec:dijetBalance} are therefore partly caused by differences in the jet properties.

\subsection{Sensitivity of the global sequential calibration to pile-up}
\label{sec:gsc-pileup}
An important feature of the \GS{} calibration is its robustness when applied in the presence of pile-up interactions,
which translates into small variations in the size of each of the corrections 
and the distributions of the jet properties. 
The corrections derived in the sample without pile-up are directly applicable to the sample with pile-up 
with only a small additional effect on the jet energy scale. 

The difference between the response after each \GS{} correction 
and the response after the \EMJES{} calibration in the Monte Carlo simulation samples, 
after the offset correction as described in Section~\ref{sec:pileup} is applied,
changed by less than $1 \%$ for  $\pt^{\rm truth} > 30$~\GeV{}  
after each of the \GS{} corrections, and by  $2 \%$ for lower $\pt^{\rm truth}$,
when samples with and without pile-up are compared.
These variations are smaller than the uncertainty on the jet energy in the absence of 
pile-up over the entire \pt{} range, thus demonstrating the robustness of the additional corrections with respect to pile-up.

\subsection{Summary on the JES uncertainty for the global sequential calibration}
\label{sec:gsc-conclusions}
The systematic uncertainty on the global sequential calibration in the inclusive jet sample has been evaluated.
It is found to be lower than $1 \%$ for \etaRange{0}{2.8} and $20 \le \ptjet < 800$~\GeV. 
This uncertainty is added in quadrature to the \JES{} based on the \EMJES{} calibration scheme.

\section{Jet calibration schemes based on cell energy weighting}
\label{sec:JetCalibSchemes}
\index{jet calibration schemes}
Besides the simple \EMJES{} calibration scheme,
\ATLAS{} has developed several calibration schemes \cite{cscbook} with different levels 
of complexity and different sensitivity to systematic effects.
The \EMJES{} calibration facilitates the evaluation of systematic uncertainties for the early analyses, 
but the energy resolution is rather poor and it exhibits a rather high sensitivity of the
jet response to the flavour of the parton inducing the jet.
These aspects can be improved using more sophisticated calibrations.

The \ATLAS{} calorimeters are non-compensating and give
a lower response to hadrons than to electrons or photons. 
Furthermore reconstruction inefficiencies and energy deposits outside the
calorimeters lower the response to both electromagnetic and hadronic
particles, but in different ways.  
The main motivation for calibration schemes based on cell energy density
is to improve the jet energy resolution by weighting differently
energy deposits from electromagnetic and hadronic showers.
The calorimeter cell energy density is a good indicator, since the radiation length $X_0$ 
is much smaller than the hadronic interaction length $\lambda_{\mathrm{I}}$.

Two calibration schemes implementing this idea have been developed:
\begin{enumerate}
\item For the global calorimeter cell energy density calibration (\GCW) the weights depend on the cell energy density and
are obtained from Monte Carlo simulation by optimising the reconstructed jet
energy resolution with respect to the true jet energy. 
This calibration is called ``global'' because 
the jet is calibrated as a whole and, furthermore, the weights 
that depend on the calorimeter cell energy density
are derived such that fluctuations in the measurement of the jet energy are minimised 
and this minimisation corrects for all effects at once. 
\item For the local cluster calibration (\LCW{}) multiple variables 
  at the calorimeter cell and the \topo{} levels are considered in a modular approach treating
  the various effects of non-compensation, dead material deposits and
  out-of-cluster deposits independently. The corrections are obtained
  from simulations of charged and neutral particles. The \topos{} in
  the calorimeter are calibrated ``locally'', without considering the
  jet context, and jets are then reconstructed directly from calibrated
  \topos{}.
\end{enumerate}
\index{global cell weighting \GCW}
\index{local cluster calibration \LCW}

Final jet energy scale corrections also need to be applied to the \GCW{} and \LCW{} calibrated jets, but
they are numerically smaller than the ones for the \EMJES{} calibration scheme.
These corrections are derived with the same procedure as described in Section~\ref{sec:JetCalib}.
The resulting jets are referred to as calibrated with \GCWJES{} and \LCWJES{} schemes.

\subsection{Global cell energy density weighting calibration} 
\label{sec:JetCalibSchemeGCW} 
\index{Global cell energy density weighting calibration} 
\begin{table}
 \centering
 \begin{tabular}{c|c|c} 
\hline \hline
 \multirow{2}{*} {Calorimeter Layer}  & {Nb. E/V} & {Poly. Degree} \\
& {bins}  & {on E/V} \\ 
\hline
\Presampler B                          & $1$  & $1$ \\
\Presampler E                          & $1$  & $1$ \\
\EMB 1                                 & $1$  & $1$ \\
\EME 1                                 & $1$  & $1$ \\
\EMB 2 and \EMB 3 with $|\eta|<0.8$    & $16$ & $4$ \\
\EMB 2 and \EMB 3 with $|\eta|\ge 0.8$ & $16$ & $4$ \\
\EME 2 and \EME 3 with $|\eta|< 2.5$   & $16$ & $4$ \\
\EME 2 and \EME 3 with $|\eta|\ge 2.5$ & $16$ & $4$ \\
\TileBar 0, \TileBar 1 and \TileBar 2  & $16$ & $4$ \\
\TileExt 0, \TileExt 1 and \TileExt 2  & $16$ & $4$ \\
\HEC 0-3 with $|\eta| < 2.5 $          & $16$ & $4$ \\
\HEC 0-3 with $|\eta| \ge 2.5 $        & $16$ & $4$ \\
\FCAL 0                                & $16$ & $3$ \\
\FCAL 1 and \FCAL 2                    & $16$ & $3$ \\
\Cryo{} term                           & $1$  & $1$ \\
\Gap{}                                 & $1$  & $1$ \\
\Scint{}                               & $1$  & $1$ \\ 
\hline \hline
\end{tabular}
\caption{ \label{tab:GCWlayers} Number of energy density bins per calorimeter layer used in the \GCW{}
jet calibration scheme and the degree of the polynomial function used in the weight parametrisation.}
\end{table}
%
This calibration scheme (\GCW) attempts to assign a larger cell-level weight to hadronic energy depositions 
in order to compensate for the different calorimeter response to hadronic and electromagnetic energy depositions. 
The weights also compensate for energy losses in the dead material.

In this scheme,
jets are first found from  \topos{} or calorimeter towers at the \EM{} scale. 
Secondly the energies of the calorimeter cells
forming jets are weighted according to their energy density. 
Finally, a JES correction is derived from the sum of the weighted energy in the calorimeter cells 
associated to the jet as a function of the jet \pt{} and pseudorapidity.

The weights are derived using Monte Carlo simulation information. 
A reconstructed jet is first matched to the nearest truth jet requiring $\Rmin < 0.3$. 
No second truth jet should be within a distance of $\DeltaR = 1$. The nearest truth jet should 
have a transverse energy $\ET > 20$~\GeV. 
The transverse energy of the reconstructed jet should be $\ET^{\EM} > 5$~\GeV, 
where $\ET^\EM$ is the transverse energy of the reconstructed jet measured at the electromagnetic scale. 

For each jet, calorimeter cells are identified  with an integer number $i$ denoting a calorimeter layer 
or a group of layers in the \ATLAS{} calorimeters. Afterwards, each cell is classified according 
to its energy density which is defined as the calorimeter cell energy measured at the electromagnetic scale 
divided by the geometrical cell volume ($E/V$).

A weight $w_{ij}$ is introduced for each calorimeter cell 
within a layer $i$ at a certain energy density bin $j$. 
The cells are classified in up to $16$ $E/V$ bins according to the following formula:
\begin{equation}
j = \frac{ \ln \frac{E/\GeV}{V/{\rm mm^{3}}} }{\ln 2} + 26,
\end{equation}
\noindent
where $j$ is an integer number between $0$ and $15$.
Calorimeter cells in the presampler, the first layer of the electromagnetic calorimeter, 
the gap and crack scintillators (\Gap, \Scint)
are excluded from this classification. 
A constant weight is applied to these cells independent of their $E/V$. 
The cryostat (\Cryo) term is computed as the geometrical average of the energy
deposited in the last layer of the electromagnetic barrel \LAr{} calorimeter and the first layer of the 
\Tile{} calorimeter.
This gives a good estimate of the energy loss in the material between the \LAr{} and the \Tile{}
calorimeters.

In the case of the seven layers without energy density segmentation 
the weights are denoted by $v_{i}$. 
Table~\ref{tab:GCWlayers} shows the number of energy density bins for each calorimeter layer.

The jet energy is then calculated as: 
\begin{equation}
\label{eq:h1_coarse_2}
E^{\rm jet}_\GCW  = \sum_{i = 1}^{10}\sum_{j = 1}^{16} w_{ij} \; E_{ij} + \sum_{i = 1}^{7} v_{i} \; E_{i},
\end{equation}
where $w_{ij}$ ($v_{i}$) are the \GCW{} calibration constants. In order to reduce the number of degrees of freedom, for a given layer $i$, the energy density dependence of each element $w_{ij}$ is parameterised by a common polynomial 
function of third and fourth degree depending on the layer (see Table~\ref{tab:GCWlayers}). 
In this way the number of free parameters used to calibrate any jet is reduced from $167$ to $45$.

The weights are computed by minimising the following function:
\begin{equation}
\chi^{2} = \frac{1}{N_{\rm jet}}\sum_{{\rm jet}=1}^{N_{\rm jet}}\left( \frac{E^{\rm jet}_\GCW}{E^{\rm jet}_{\rm truth}} - 1 \right)^{2},
\end{equation}
where $N_{\rm jet}$ is the total number of jets in the Monte Carlo sample used. 
This procedure provides weights that minimise the jet energy resolution. 
The mathematical bias on the mean jet energy that is introduced in particular at low jet energies
(see Ref.~\cite{Lincoln:1993yi}) is corrected by an additional jet energy calibration following the method 
described in Section~\ref{sec:JetCalib} and discussed in Section~\ref{sec:GCWLCWJES}.

\subsection{Local cluster weighting calibration} 
\index{Local cluster weighting calibration} This calibration
scheme~\cite{Barillari:2009zza,EndcapTBelectronPion2002} corrects locally the
\topos{} in the calorimeters independent of any jet context.  The
calibration starts by classifying \topos{} as mainly electromagnetic
or hadronic %
depending on cluster shape variables~\cite{TopoClusters}. The cluster
shape variables characterise the topology of the energy deposits of electromagnetic
or hadronic showers and are defined as observables derived from 
calorimeter cells with positive energy 
in the cluster and the cluster energy. All
weights depend on this classification and both hadronic and
electromagnetic weights are applied to each cluster.

\subsubsection{Barycentre of the longitudinal cluster depth} 
\label{eq:barycentrelambda}
The barycentre of the longitudinal depth of the \topo{} ($\lambda_{\rm centre}$) is 
defined as the distance along the shower axis from the front of the calorimeter to the shower centre. 
\index{longitudinal cluster depth $\lambda_{\rm centre}$} 
The shower centre has coordinates:
\begin{equation} 
\langle i \rangle = \frac{\sum_{k|E_k > 0} E_k \; i_k}{\sum_{k|E_k > 0} \; E_k},
\end{equation}
\noindent with $i$ taking values of the spatial coordinates $x,y,z$
and $E_k$ denoting the energy in the calorimeter cell $k$. 
Only calorimeter cells with positive energy are used.

The shower axis is determined from the spatial correlation matrix of all cells in the \topo{} 
\index{cluster axis} 
with positive energies:
\begin{equation}
C_{ij}=\frac{\sum_{k|E_k>0} E^2_k(i_k-\langle i\rangle)(j_k-\langle j\rangle)}{\sum_{k|E_{k>0}}E^2_k},
\label{eq:corrMatrix}
\end{equation}
with $i,j=x,y,z$. The shower axis is the eigenvector of this matrix closest to the direction joining
the interaction point and the shower centre. 

\subsubsection{Cluster isolation} 
\label{eq:clusterisolation}
The cluster isolation is defined as the ratio of the number of unclustered calorimeter cells\footnote{
Unclustered calorimeter cells that are not contained in any \topo.} that are neighbours of
a given \topo{} to the number of all neighbouring cells.
The neighbourhood relation is defined in two dimensions,
i.e. within the individual calorimeter layer\footnote{In general,
\topos{} are formed in a three dimensional space defined by $\eta$, $\phi$ and the calorimeter depth.}. 

After calculating the cluster isolation for each individual calorimeter layer,
the final cluster isolation variable is obtained by
weighting the individual layer cell ratios by the energy fractions
of the \topo{} in these layers. This assures that the isolation is
evaluated where the \topo{} has most of its energy.

The cluster isolation is zero for \topos{} where all neighbouring calorimeter cells
in each layer are inside other \topos{} and one for \topos{} with no
neighbouring cell inside any other \topo.

\begin{figure*}
 \centering
  \subfloat[\GCW]{\includegraphics[width=0.49\textwidth]{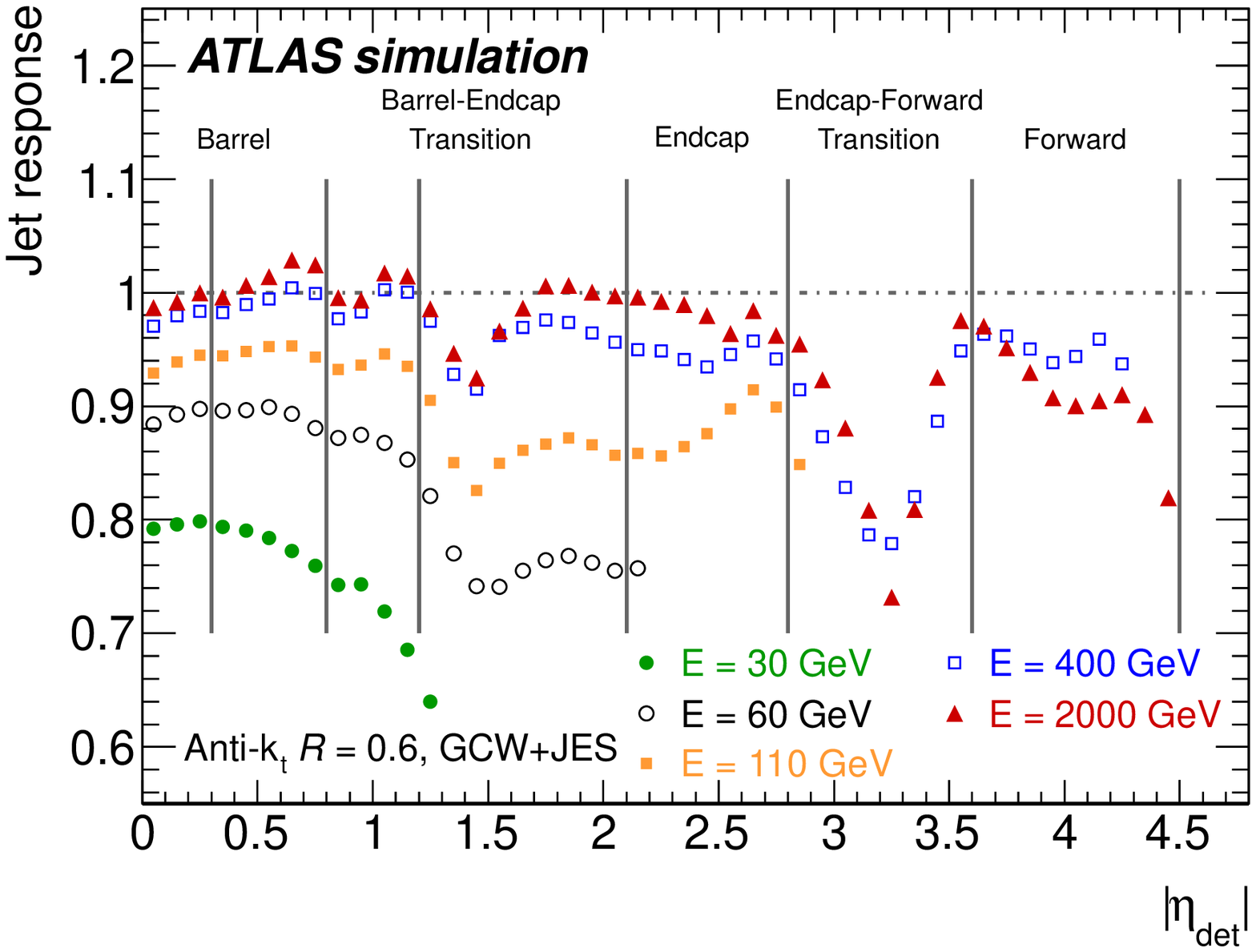}}
  \subfloat[\LCW]{\includegraphics[width=0.49\textwidth]{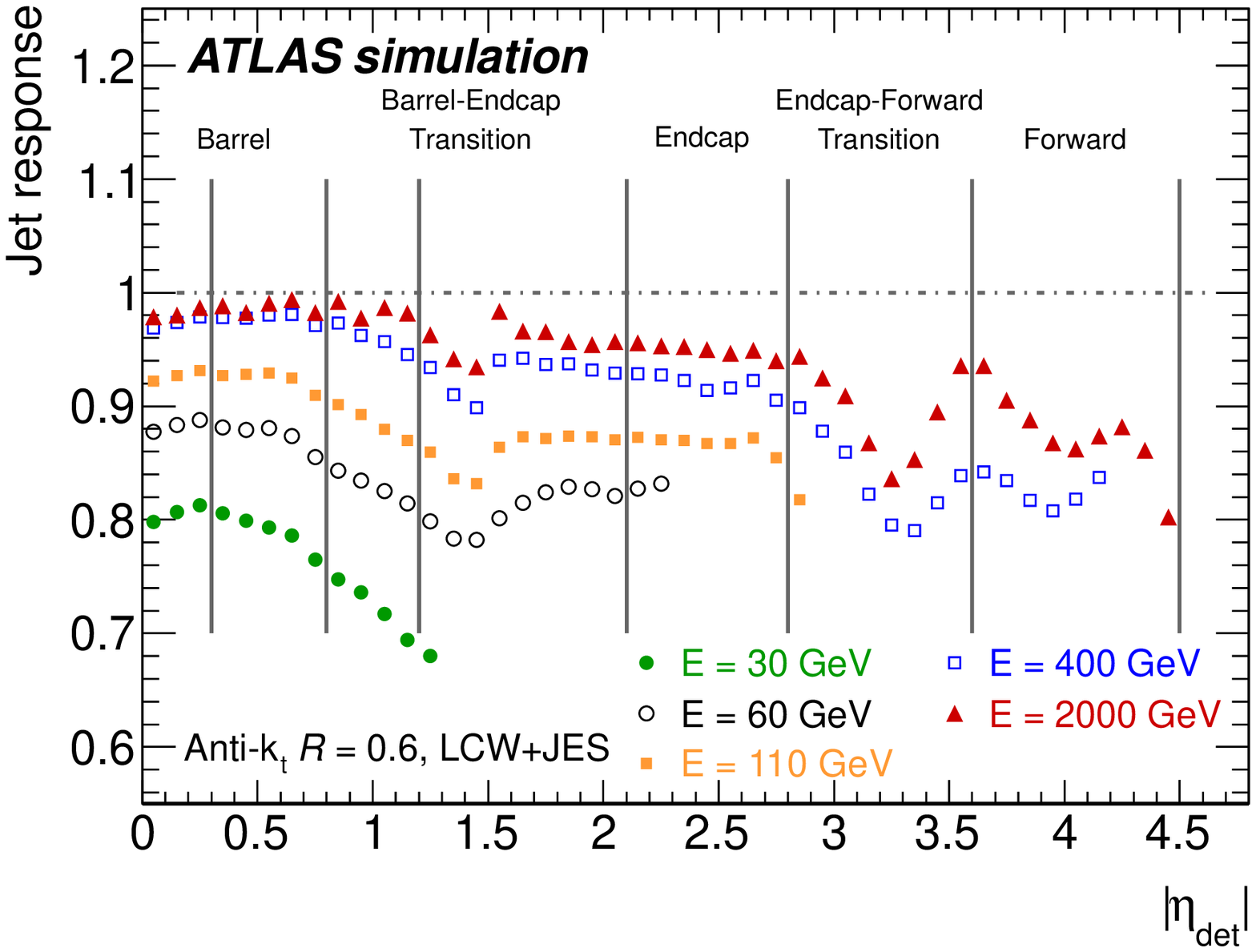}}
 \caption{
   Average simulated jet energy response at the \GCW{} (a) and the \LCW{} (b) scale in bins of the
   \GCWJES{} and \LCWJES{} calibrated jet energy and as a function of the detector pseudorapidity
   $|\etaDet|$. 
 }
 \label{fig:GCWLCWJES_vs_eta}
\end{figure*}

\subsubsection{Cluster energy correction} 
All corrections are derived from the Monte Carlo simulations for single
charged and neutral pions. The hadronic shower simulation model used is QGSP\_BERT.
The detector geometry and \topo{} reconstruction is the same as in the nominal
Monte Carlo simulation sample.
A flat distribution in the logarithm of pion energies from $200$~\MeV{} to $2$~\TeV{} is used.

The corrections are derived with respect to the true deposited energy
in the active and inactive detector region (``calibration hits'').

True energy depositions are classified in three types by the \ATLAS{} software:
\begin{enumerate}
\item The {\bf visible energy}, like the energy deposited by ionisation.
\item The {\bf invisible energy}, like energy absorbed in nuclear reactions.
\item The {\bf escaped energy}, like the energy carried away by neutrinos\footnote{
The escaped energy is recorded at the place where the
particle that escapes the detector volume (``world volume'' in \geant{} terminology) is produced.}.
\end{enumerate}
\index{calibration hits} 

The local cluster calibration proceeds in the following steps:
\begin{enumerate}
 \item {\bf Cluster classification}: \index{\LCW{} \Topo{} classification} 
The expected 
   population in logarithmic bins of the \topo{} energy, the cluster
   depth in the calorimeter, and the average cell energy density are used
   to calculate classification weights. The weights are calculated for
   small $|\eta|$ regions by mixing neutral and charged pions with a
   ratio of $1:2$.  This assumes that $2/3$ of the pions should be charged.  
   Clusters are classified as mostly
   electromagnetic or mostly hadronic. The calculated weight denotes
   the probability $p$ for a cluster to stem from a hadronic
   interaction.
\item {\bf Hadronic weighting}: \index{\LCW{} Hadronic weighting}
\Topos{} receive calorimeter cell correction weights derived from detailed
Monte Carlo simulations of charg\-ed pions. 
Calorimeter cells in \topos{} are weighted according to the \topo{} energy and the 
calorimeter cell energy density. 
The hadronic energy correction weights are calculated from the true energy
deposits as given by the Monte Carlo simulation ($w_{\rm HAD}$) multiplied by a weight to 
take into account the different nature of hadronic and electromagnetic showers.

The applied weight is 
\begin{equation}
w_{\rm HAD} \cdot p + w_\EM \cdot (1 - p), 
\end{equation}
where $w_\EM = 1$ and $p$ is the probability of the \topo{} to be hadronic as determined 
by the classification step.
Dedicated correction weight tables for each calorimeter layer
in $0.2$-wide $|\eta|$-bins are used.
The correction weight tables are binned logarithmically in \topo{} energy and cell energy density ($E/V$).

\item {\bf Out-of-cluster (OOC) corrections}: \index{\LCW{} Out-of-cluster (OOC) corrections}
A correction for isolated energy deposits inside the calorimeter, but
outside \topos{} is applied. These are energy depositions not passing
the noise thresholds applied during the clustering.  These corrections
depend on $|\eta|$, the energy measured around the \topo{} and the
cluster barycentre $\lambda_{\rm centre}$. There are two sets of constants for hadronic
and electromagnetic showers and both are used for each cluster with
the respective weights of $p$ and $1 - p$.
The OOC correction is finally multiplied  with the cluster isolation value
discussed in Section~\ref{eq:clusterisolation} in order to avoid double counting.

\item {\bf Dead material (DM) corrections}: 
\index{\LCW{} Dead material (DM) corrections} 
Energy deposits in materials outside the calorimeters are corrected.
For energy deposits in upstream material like the inner wall of the cryostat, 
the presampler signals are highly correlated to the lost energy.
The corrections are derived from the sum of true energy depositions
in the material in front and behind the calorimeter and from the presampler signal.

The correction for energy deposited in the outer cryostat wall between
the electromagnetic and the hadronic barrel cal\-ori\-meters is based on the
geometrical mean of the energies in the layers just before and just
beyond the cryostat wall. 
Corrections for other energy deposits without clear
correlations to \topo{} observables are obtained from look-up tables
binned in \topo{} energy, the pseudorapidity $|\eta|$, and the shower depth.
Two sets of DM weights for hadronic and electromagnetic showers are used.
The weights are applied according to the classification probability $p$ defined above.

\end{enumerate}

All corrections are defined with respect to the electromagnetic scale energy of the \topo.
Since only calorimetric information is used,
the \LCW{} calibration does not account for low-energy
particles which do not create a \topo{} in the calorimeter.
This is, for instance, the case when the energy is absorbed entirely in inactive detector material 
or particles are bent outside of the calorimeter acceptance.

\subsection{Jet energy calibration for jets with calibrated constituents} 
\label{sec:GCWLCWJES}
\index{calibrated \GCW{} and \LCW{} constituents} 
The simulated response to jets at the \GCW{} and \LCW{} energy scales,
i.e. after applying weights to the calorimeter cells in jets or
after the energy corrections to the \topos, are shown 
in Figure~\ref{fig:GCWLCWJES_vs_eta} as a function of  $\etaDet$ for various
jet energy bins. The inverse of the response shown in each bin is equal to the
average jet energy scale correction. 
The final jet energy correction needed to restore the reconstructed jet energy to
the true jet energy is much smaller than in the case of the \EMJES{} calibration
shown in Figure~\ref{fig:EMJES_vs_eta}.

\begin{figure*}[!ht]
\centering
\subfloat[Barrel \Presampler]{
\label{fig:cellDens_Barr:3}
\includegraphics[width=0.3\textwidth]{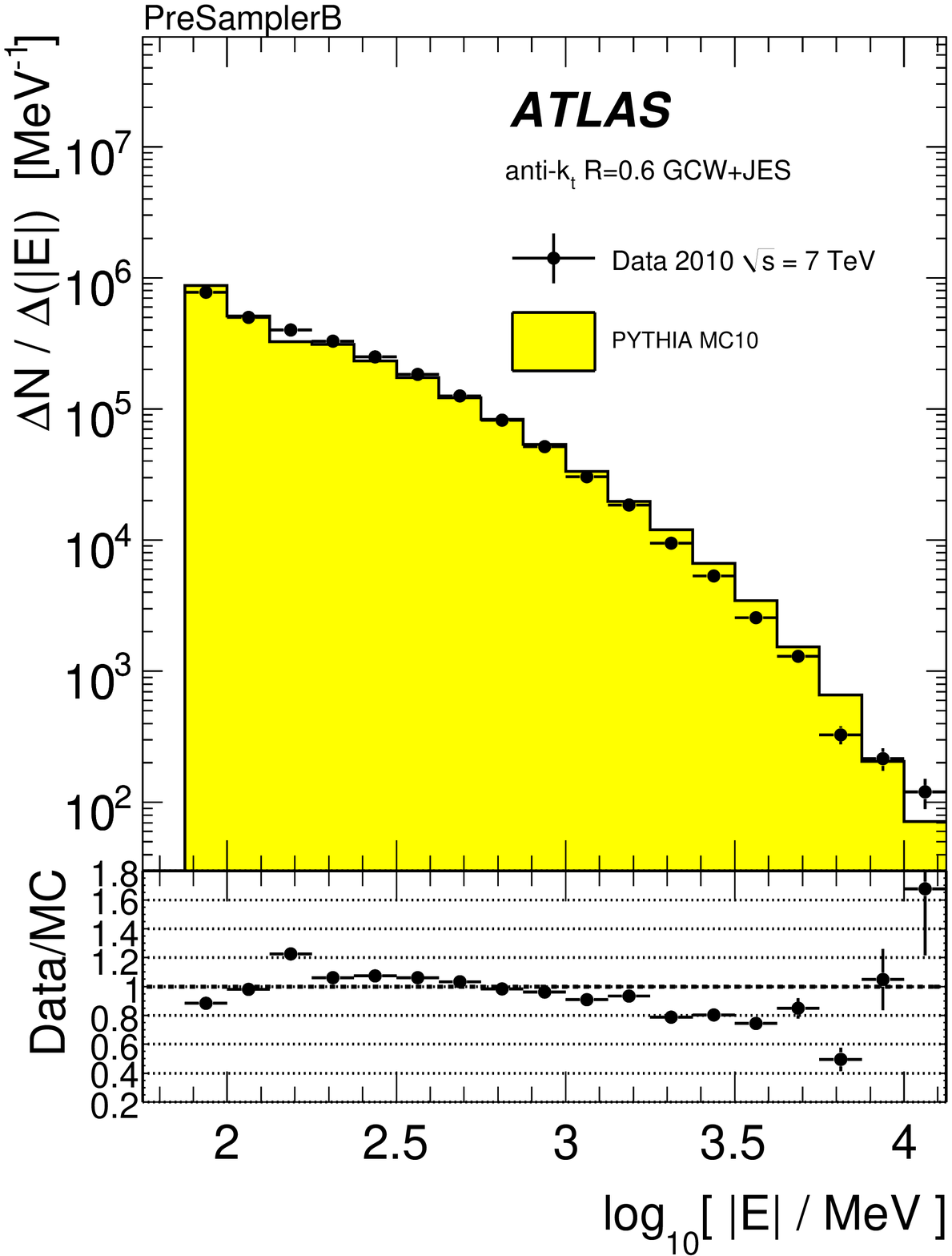}}
\subfloat[Second layer \EMB]{
  \label{fig:cellDens_Barr:1}
 \includegraphics[width=0.3\textwidth]{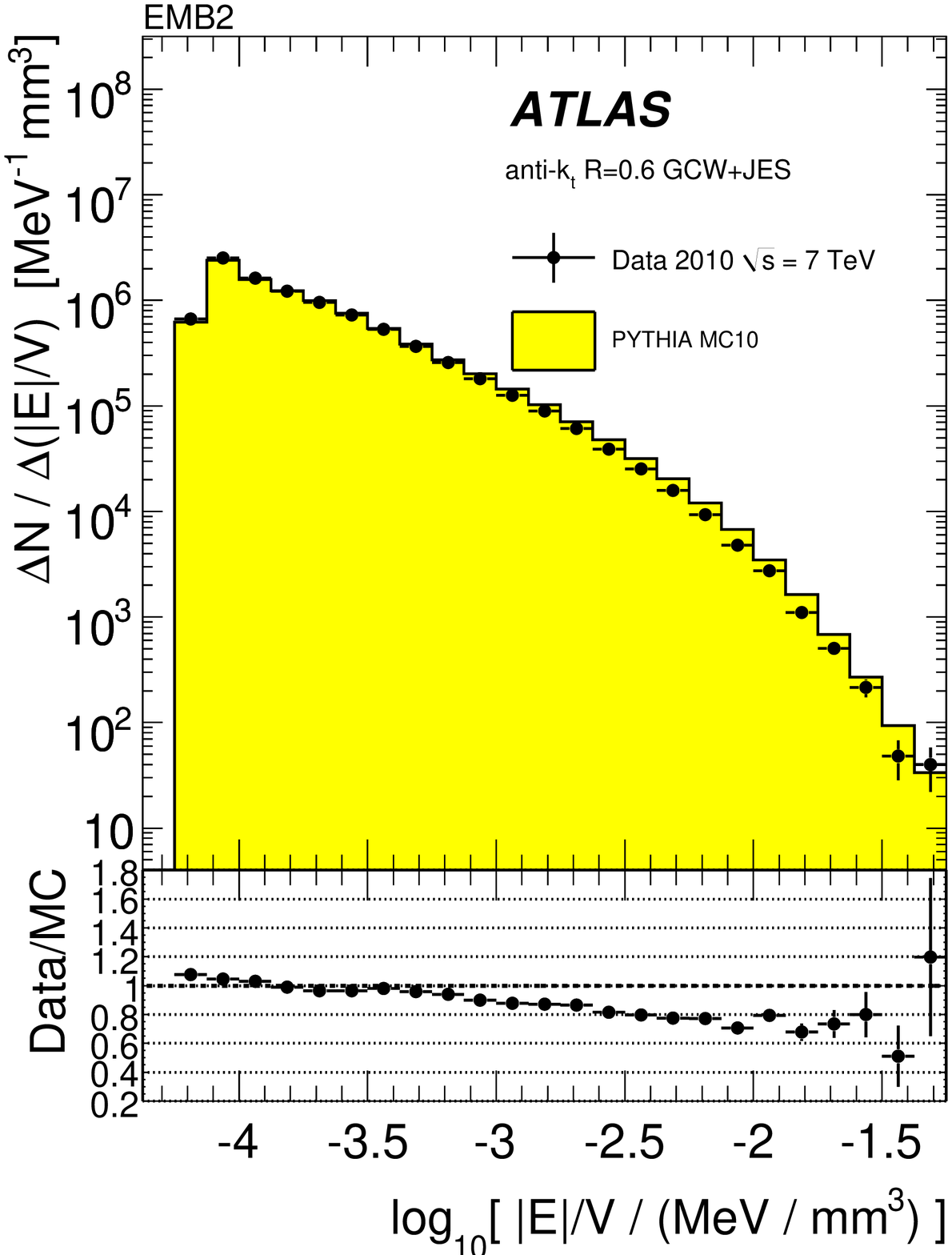}} 
\subfloat[Second layer \Tile{} in barrel]{
  \label{fig:cellDens_Barr:2}
\includegraphics[width=0.3\textwidth]{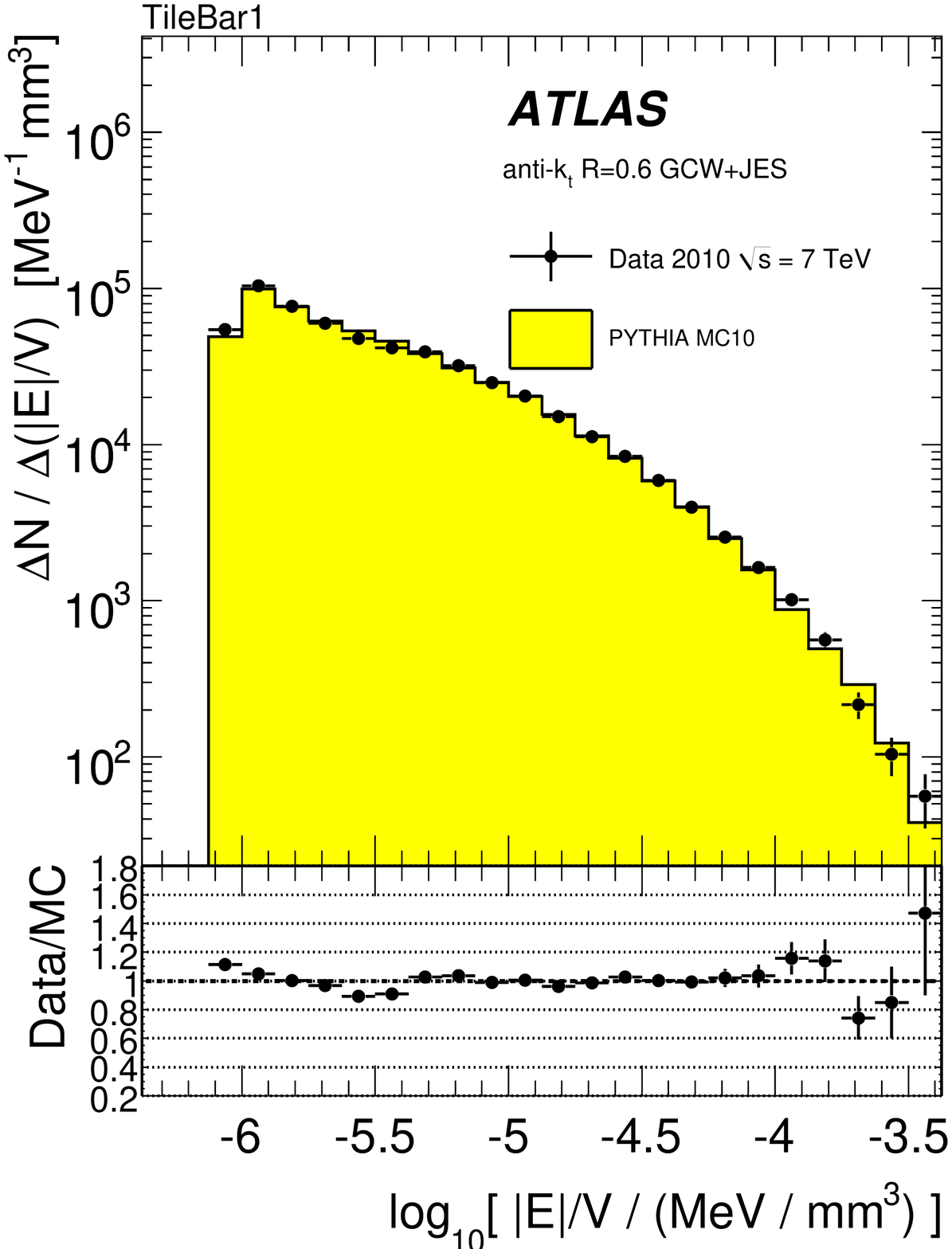}}\\ 
\subfloat[Second layer of \EMEC]{
  \label{fig:cellDens_Endc:1}
 \includegraphics[width=0.3\textwidth]{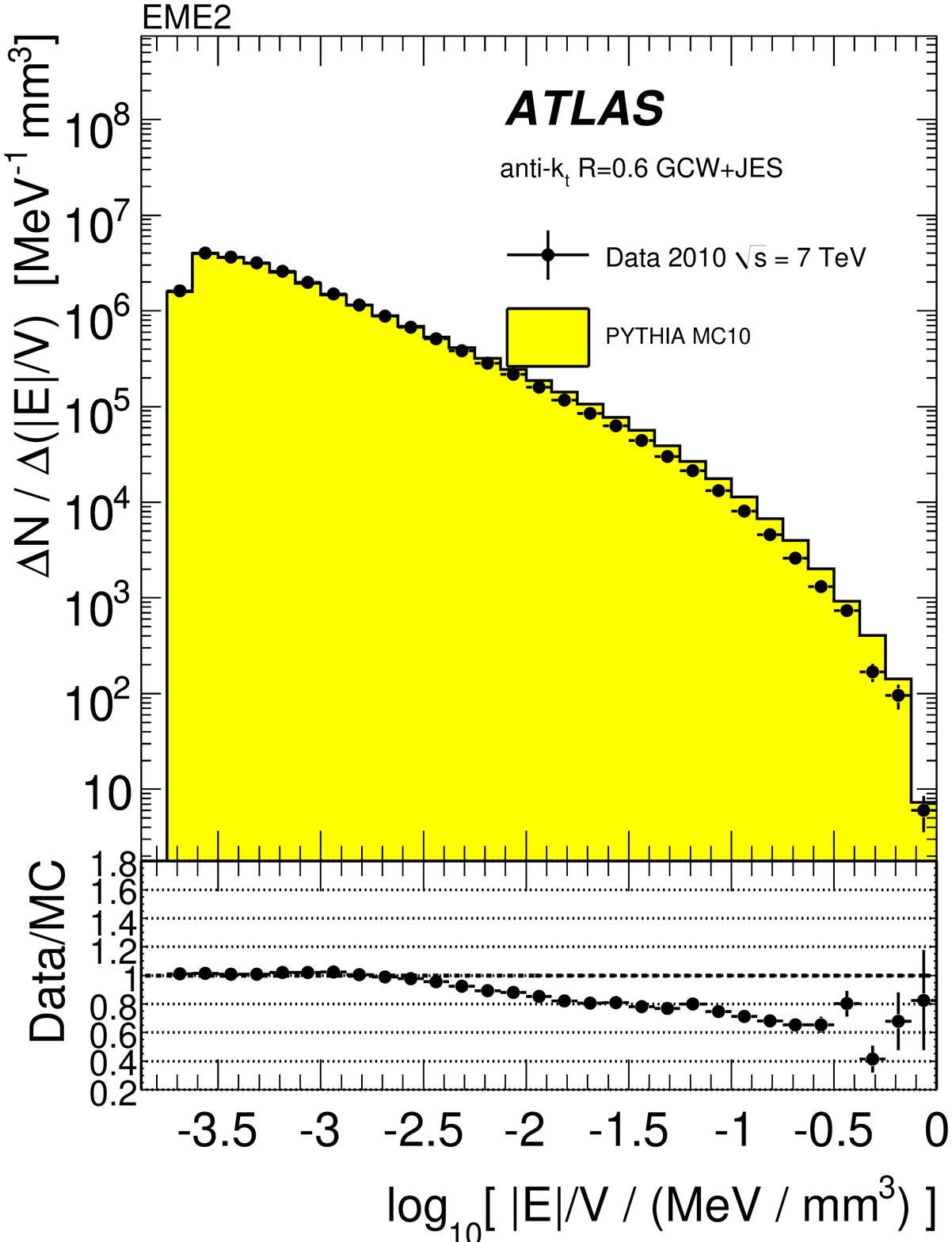}}
\subfloat[First layer of \HEC]{
  \label{fig:cellDens_Endc:2}
\includegraphics[width=0.3\textwidth]{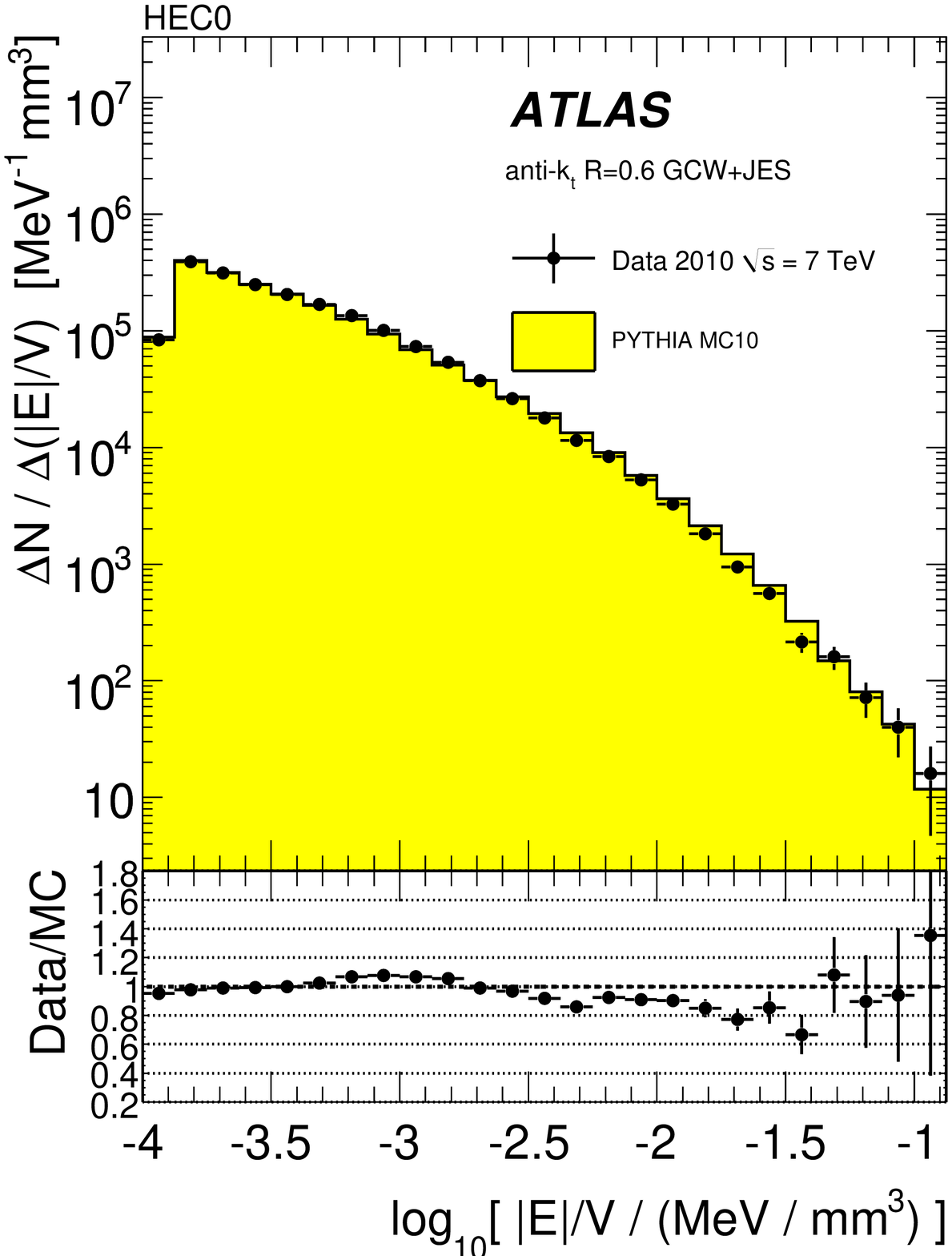}}
\subfloat[First layer of \FCAL]{
  \label{fig:cellDens_Endc:3}
\includegraphics[width=0.3\textwidth]{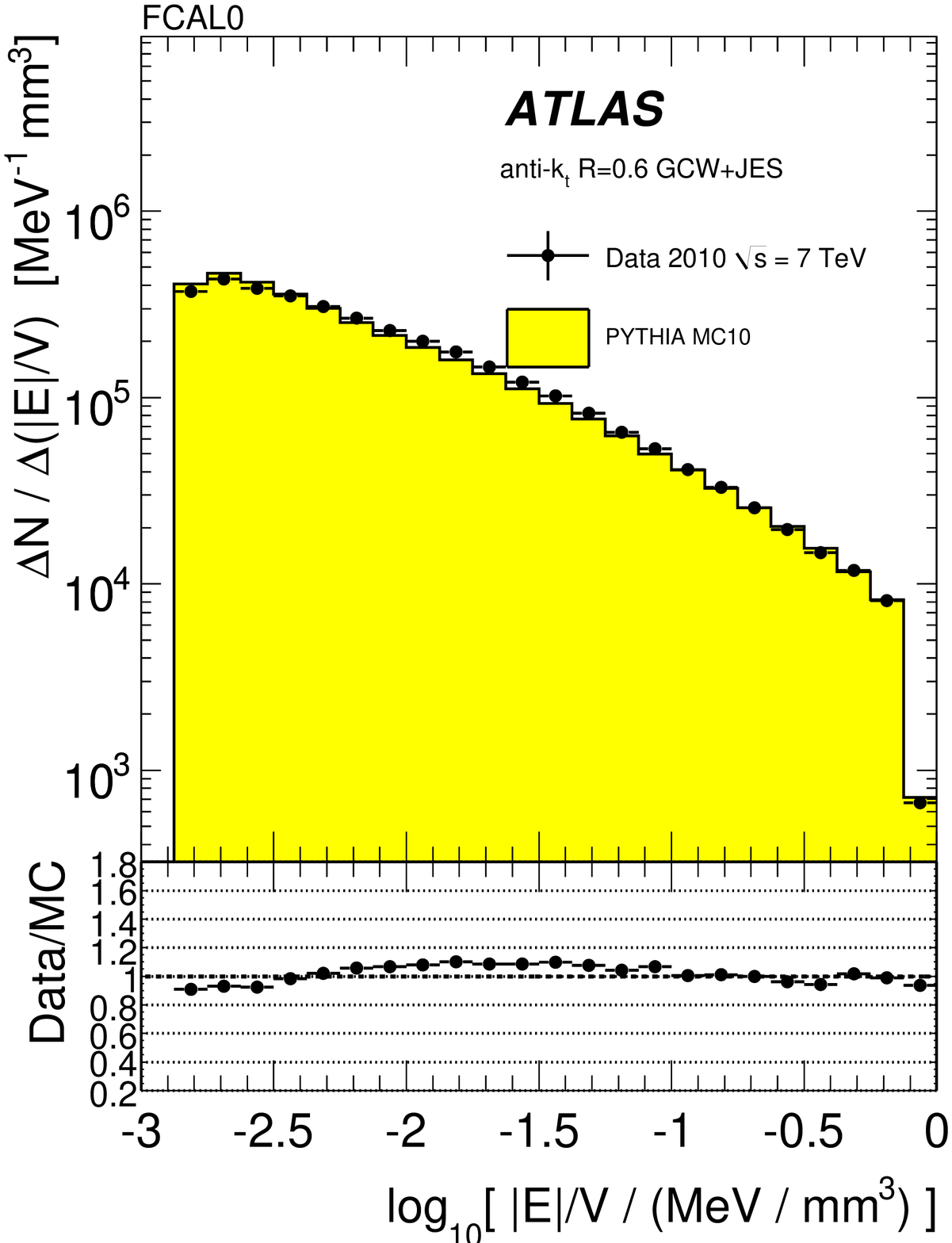}}
\caption{Calorimeter cell energy density distributions used in the \GCW{} jet calibration scheme
        in data (points) and Monte Carlo simulation (shaded area) for calorimeter cells in 
the barrel presampler (a), 
the second layer of the barrel electromagnetic calorimeter (b), 
the second  layer of the barrel hadronic Tile calorimeter  (c),
the second layer of the endcap electromagnetic calorimeter (d), 
the first layer of the endcap hadronic calorimeter (e) 
and the first layer of the forward calorimeter (f). 
\Antikt{} jets with $R=0.6$ requiring $\ptjet>20$~\GeV and $y<2.8$ calibrated with the \GCWJES{} scheme are used.
Monte Carlo simulation distributions are normalised to the number of cells in data distributions. 
The ratio of data to Monte Carlo simulation  is shown in the lower part of each figure. 
Only statistical uncertainties are shown.
}
 \label{fig:cellDens}
\end{figure*}

\section{Jet energy scale uncertainties for jet calibrations based on cell weighting}
\label{sec:JESUncertaintiescellweighting}
The jet energy scale uncertainty for jets based on cell weighting is obtained
using the same \insitu{} techniques as described in Section~\ref{sec:insituvalidation}.
The results for each \insitu{} technique together with the combination of
all \insitu{} techniques are discussed in Section~\ref{sec:insituvalidationcellweighting}.

In order to build up confidence in the Monte Carlo simulation 
the description of the variables used as inputs to the cell weighting 
by the Monte Carlo simulation 
is discussed in
Section~\ref{sec:inputsglobalcellweighting} 
for the global cell weighting scheme 
and in Section~\ref{sec:calibClustersProp}
for the local cluster weighting scheme. 

\subsection{Energy density as input to the global cell weighting calibration}
\label{sec:inputsglobalcellweighting}
The global cell energy density weighting calibration scheme 
(see Section~\ref{sec:JetCalibSchemeGCW}) applies weights to the energy 
deposited in each calorimeter cell according to the calorimeter cell energy density ($E/V$, 
where $V$ is the calorimeter cell volume defined before). 
This attempts to compensate for the different calorimeter response to hadronic and electromagnetic 
showers, but it also compensates for energy losses in the dead material. 
The description of the calorimeter cell energy density in the Monte Carlo simulation
is therefore studied to validate this calibration scheme.
\index{Cell energy densities as input to \GCW}

Only calorimeter cells inside jets with $\ptjet > 20$~\GeV{} and $|y|<2.8$
built of \topos{} and with a cell energy of at 
least two standard deviations above the noise thresholds are considered for this comparison. 
Similar results have been obtained using cells inside jets built from calorimeter towers. 
The Monte Carlo simulation reproduces the generic features of the data over
many orders of magnitude. 
However, the following paragraphs discusses those differences, 
all of which are on the order of a few percent.

Figure~\ref{fig:cellDens} shows the calorimeter cell energy density distributions in data and 
Monte Carlo simulation for cells in representative longitudinal segments of the barrel and forward
 ca\-lor\-ime\-ters.
Fewer cells with high energy density are observed in data than predicted 
by Monte Carlo simulation in the barrel presampler (a) and in the second layer 
of the barrel electromagnetic calorimeter (b). 
This behaviour is observed for other segments of the barrel electromagnetic calorimeter,
but not for the second layer of the \Tile{} barrel calorimeter (c). Here,
a good agreement between data and Monte Carlo simulation is found 
over the full energy density spectrum. Only for the lowest energy densities  are slight differences found. 
Good agreement is also present in the first layer of the \Tile{} extended barrel calorimeter, 
while the energy density is on average smaller for the second and third layer in the data 
than in the Monte Carlo simulation. 
Such a deficit of high energy density cells in data is also observed
for the second and third layer of the scintillators placed in the gap between the 
\Tile{} barrel and extended barrel modules. Better agreement is found between data 
and Monte Carlo simulation for the first layer of the scintillators. 

The second layer of the endcap electromagnetic calorimeter (d) shows a similar behaviour 
to that observed in the barrel: 
fewer cells are found at high energy density in the data than in the Monte Carlo simulation. 
This effect is present in all three layers of the endcap electromagnetic calorimeter, 
yet it  becomes more pronounced with increasing calorimeter depth. 
A similar effect, but of even larger magnitude has been observed 
for cells belonging to the endcap presampler. 
The first layer of the endcap hadronic calorimeter (e) shows a better agreement 
between data and Monte Carlo simulation. This agreement is also present for other 
layers of the \HEC. In the first layer of the forward calorimeter 
more cells with energy densities 
in the middle part of the spectrum are found in data than in
Monte Carlo simulation (f). 
This effect has been observed in other \FCAL{} layers, 
and it becomes slightly more pronounced with increasing \FCAL{} depth.

\begin{figure*}[ht!!p]
\centering
   \subfloat[Electromagnetic \topos]{
    \label{fig:clustersInJetsISO:1}
    \includegraphics[width=0.45\textwidth]{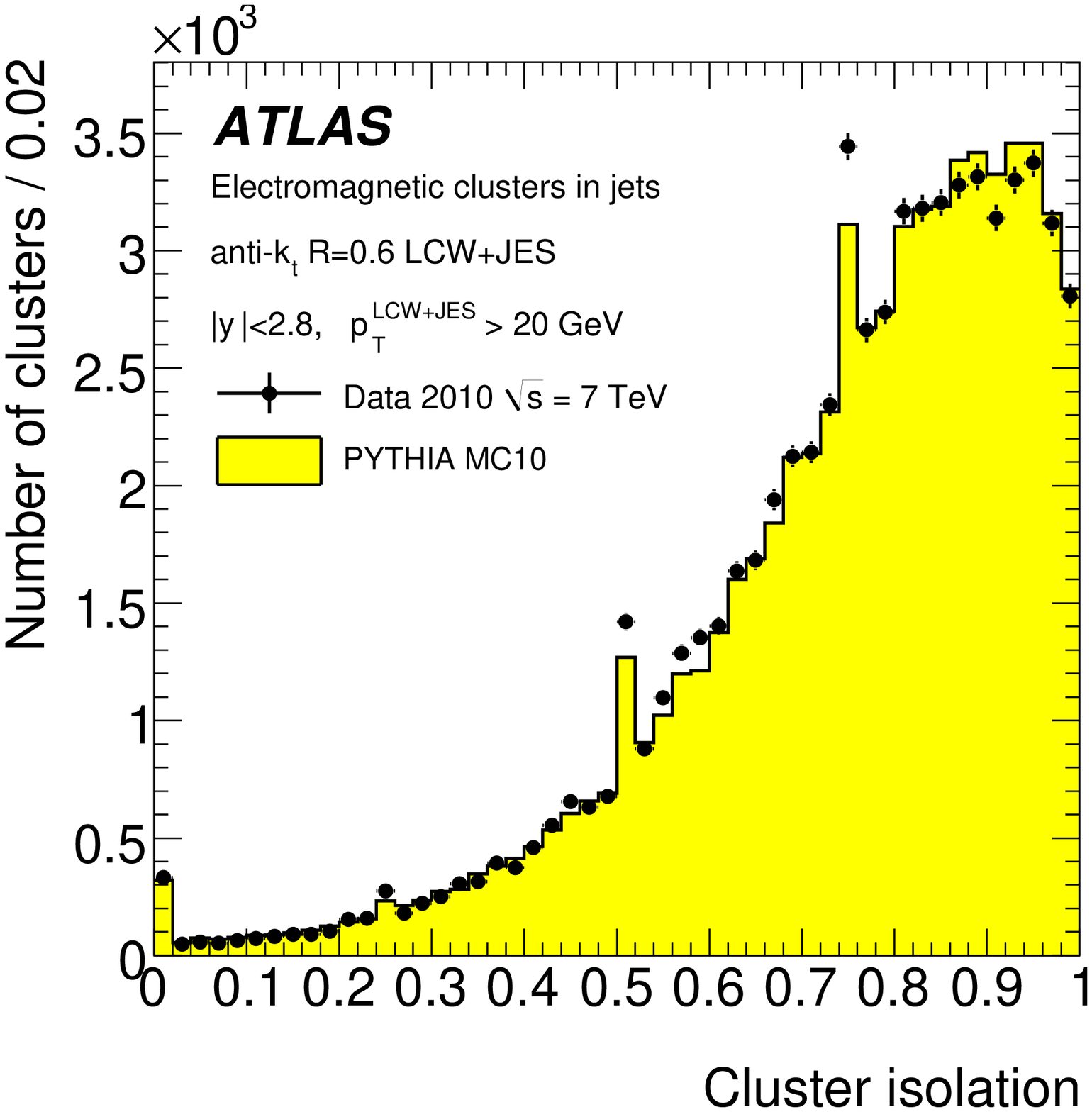}
   }
   \subfloat[Hadronic \topos]{
    \label{fig:clustersInJetsISO:2}
    \includegraphics[width=0.45\textwidth]{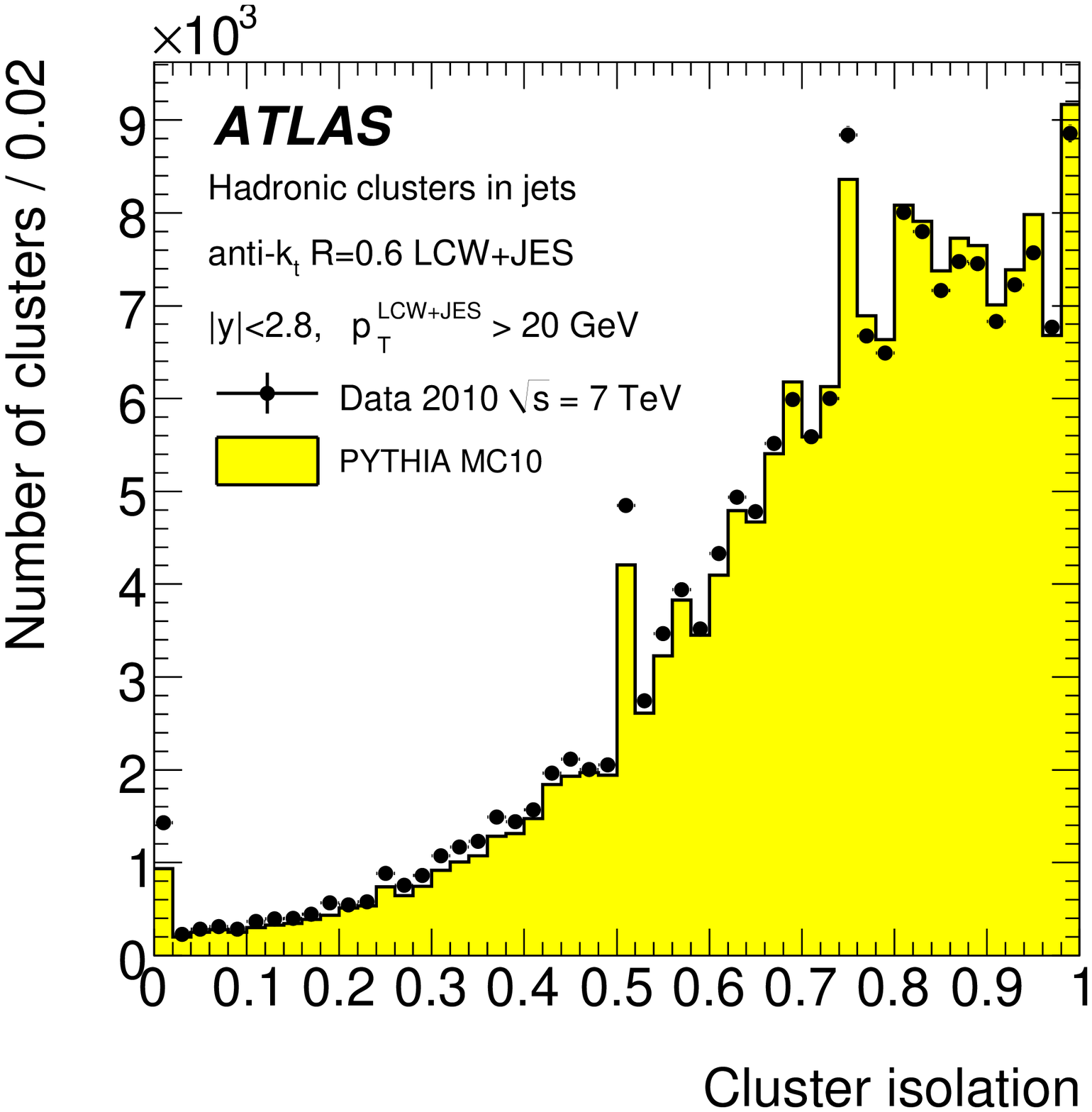}
    }
\caption{Distributions of the isolation variable for \topos{} classified as electromagnetic (a) and as hadronic (b) 
in data (points) and Monte Carlo simulation (shaded area). 
\Topos{} associated to \antikt{} jets with $R=0.6$ with $\ptjet > 20$~\GeV{} and $| \rapjet |< 2.8$ calibrated with the
\LCWJES{} scheme are used. 
}
\label{fig:clustersInJetsISO}
\end{figure*}

\begin{figure*}[ht!!p]
\centering
  \subfloat[Electromagnetic \topos]{
    \label{fig:clustersInJetsPtISO:1}
    \includegraphics[width=0.45\textwidth]{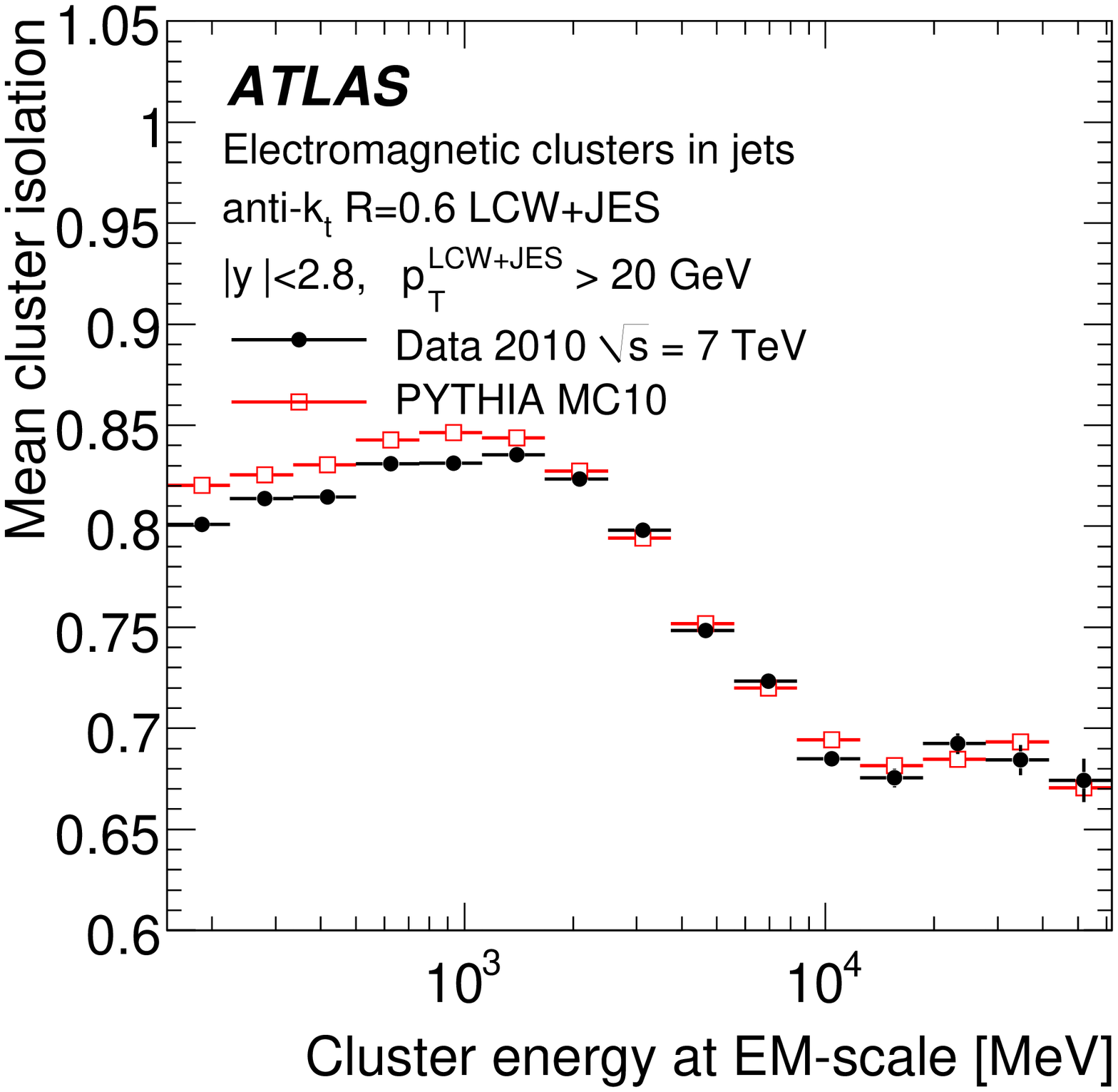}
    }
  \subfloat[Hadronic \topos]{
    \label{fig:clustersInJetsPtISO:2}
    \includegraphics[width=0.45\textwidth]{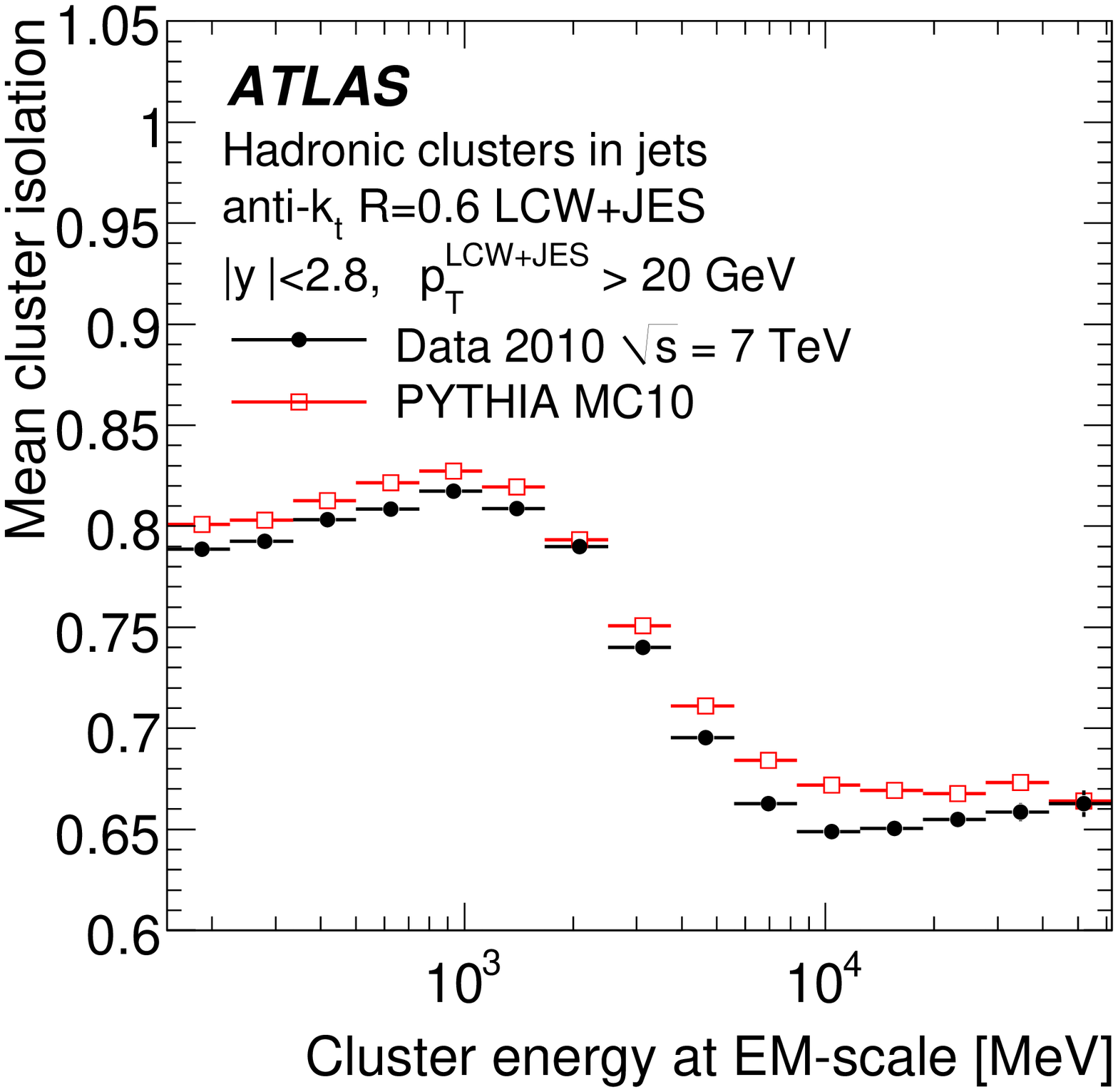}
    }
\caption{Mean value of the cluster isolation variable for \topos{} classified as
electromagnetic (a) and  as hadronic (b) as a function of the \topo{} energy measured at the \EM{} scale,
in data (closed circles) and Monte Carlo simulation (open squares). 
\Topos{} associated to \antikt{} jets with $R=0.6$ with $\ptjet > 20$~\GeV{} and $| \rapjet |< 2.8$ calibrated with the
\LCWJES{} scheme are used. 
}
\label{fig:clustersInJetsPtISO}
\end{figure*}

%
\begin{figure*}[ht!p]
\centering
  \subfloat[Electromagnetic \topos]{
    \label{fig:clustersInJetsLAM:1}
    \includegraphics[width=0.45\textwidth]{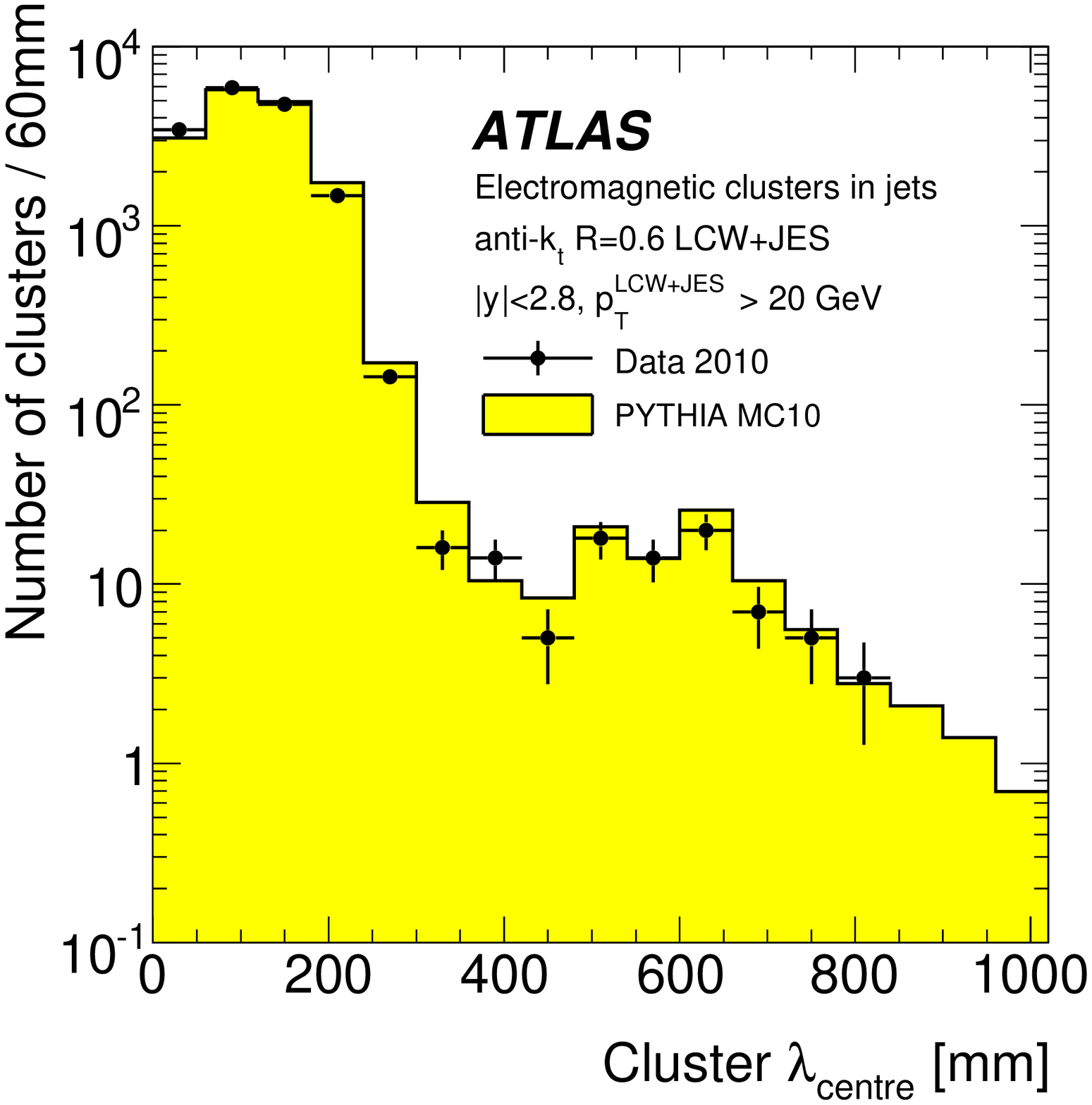}
}
  \subfloat[Hadronic \topos]{
    \label{fig:clustersInJetsLAM:2}
   \includegraphics[width=0.45\textwidth]{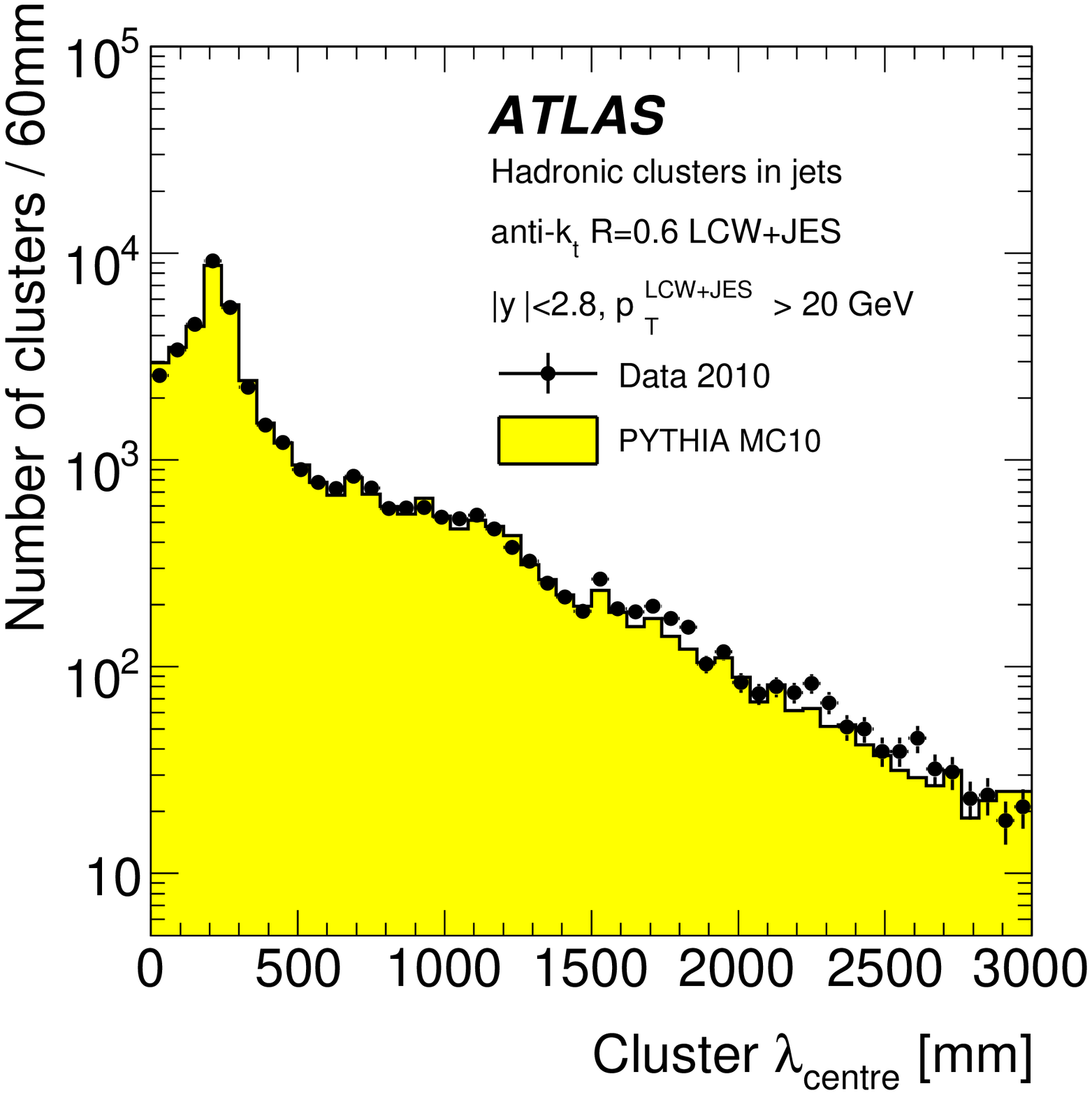}
}
\caption{Distributions of the longitudinal cluster barycentre $\lambda_{\rm centre}$ for \topos{} 
classified as electromagnetic (a) and as hadronic (b) in data (points) and Monte Carlo simulation (shaded area). 
\Topos{} associated to \antikt{} jets with $R=0.6$ with $\ptjet > 20$~\GeV{} and $| \rapjet |< 2.8$ calibrated with the
\LCWJES{} scheme are used. 
}
\label{fig:clustersInJetsLAM}
\end{figure*}
\begin{figure*}[!htp]
\centering
  \subfloat[Electromagnetic \topos]{
    \label{fig:clustersInJetsPtEM:1}
   \includegraphics[width=0.45\textwidth]{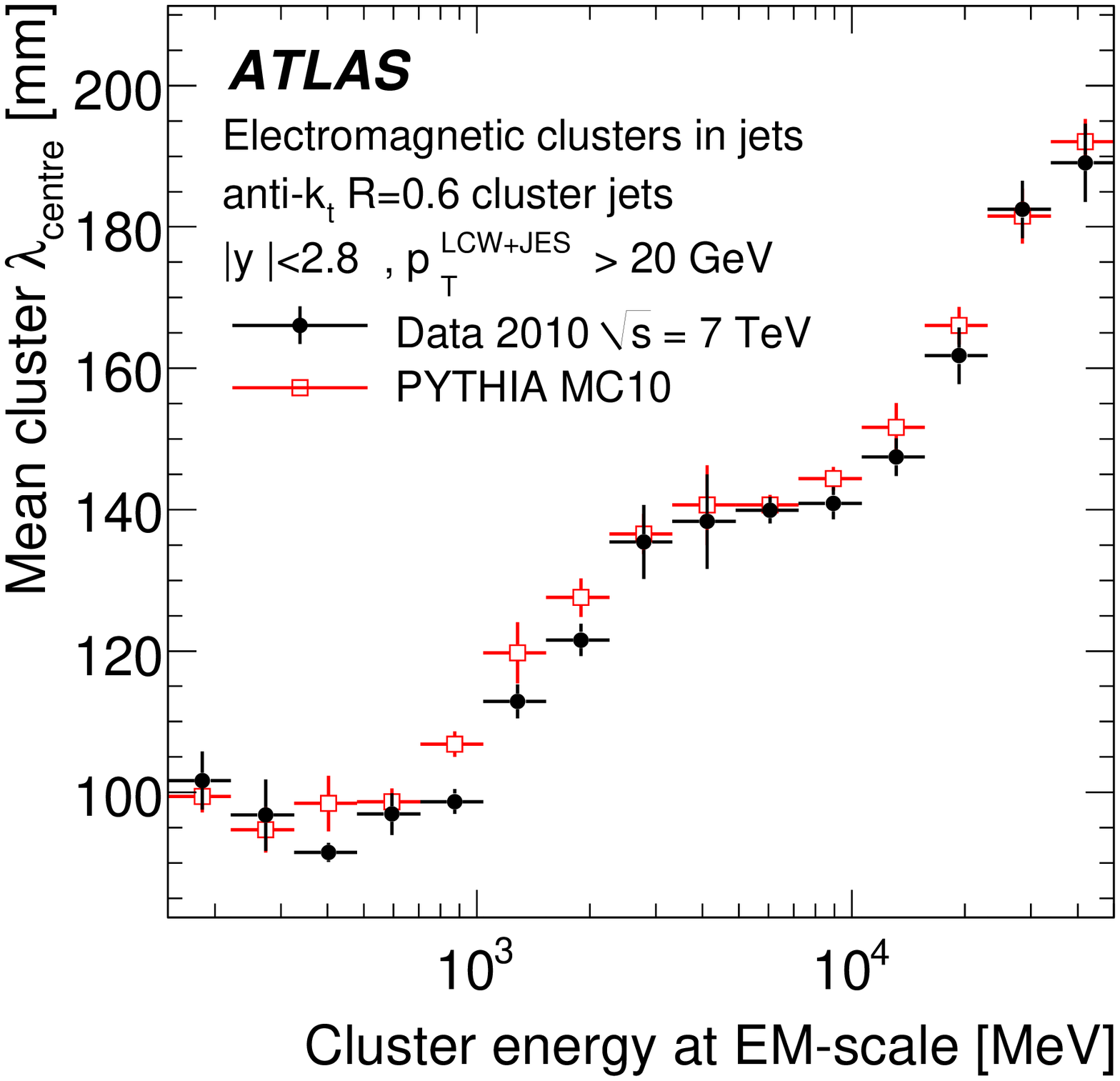}
}
  \subfloat[Hadronic \topos]{
    \label{fig:clustersInJetsPtHAD:1}
   \includegraphics[width=0.45\textwidth]{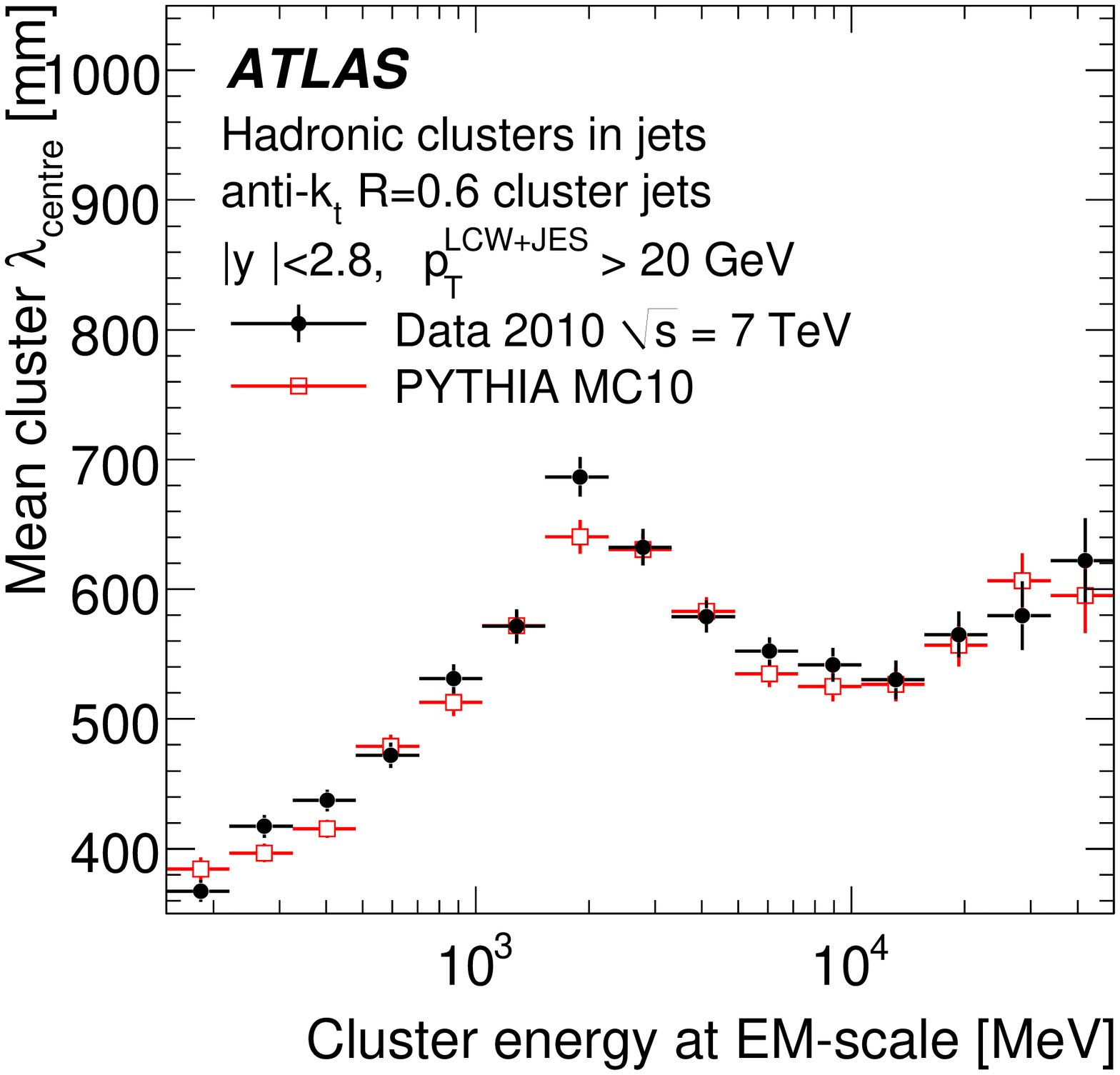}
}
\caption{Mean value of the longitudinal cluster barycentre $\lambda_{\rm centre}$ as a function of the \topo{}
 energy measured at the \EM{} scale
for \topos{} classified as electromagnetic (a) and as hadronic in data (b) in data (closed circles)
and Monte Carlo simulation (open squares). 
\Topos{} associated to \antikt{} jets with $R=0.6$ with $\ptjet > 20$~\GeV{} and $| \rapjet |< 2.8$ calibrated 
with the \LCWJES{} scheme are used. 
}
\label{fig:clustersInJetsPtLAM}
\end{figure*}
\begin{figure*}[!htp]
\centering
\vspace{-0.2cm}
\subfloat[Hadronic response weights]{
  \label{fig:calibClusters:1}
  \includegraphics[width=0.39\textwidth]{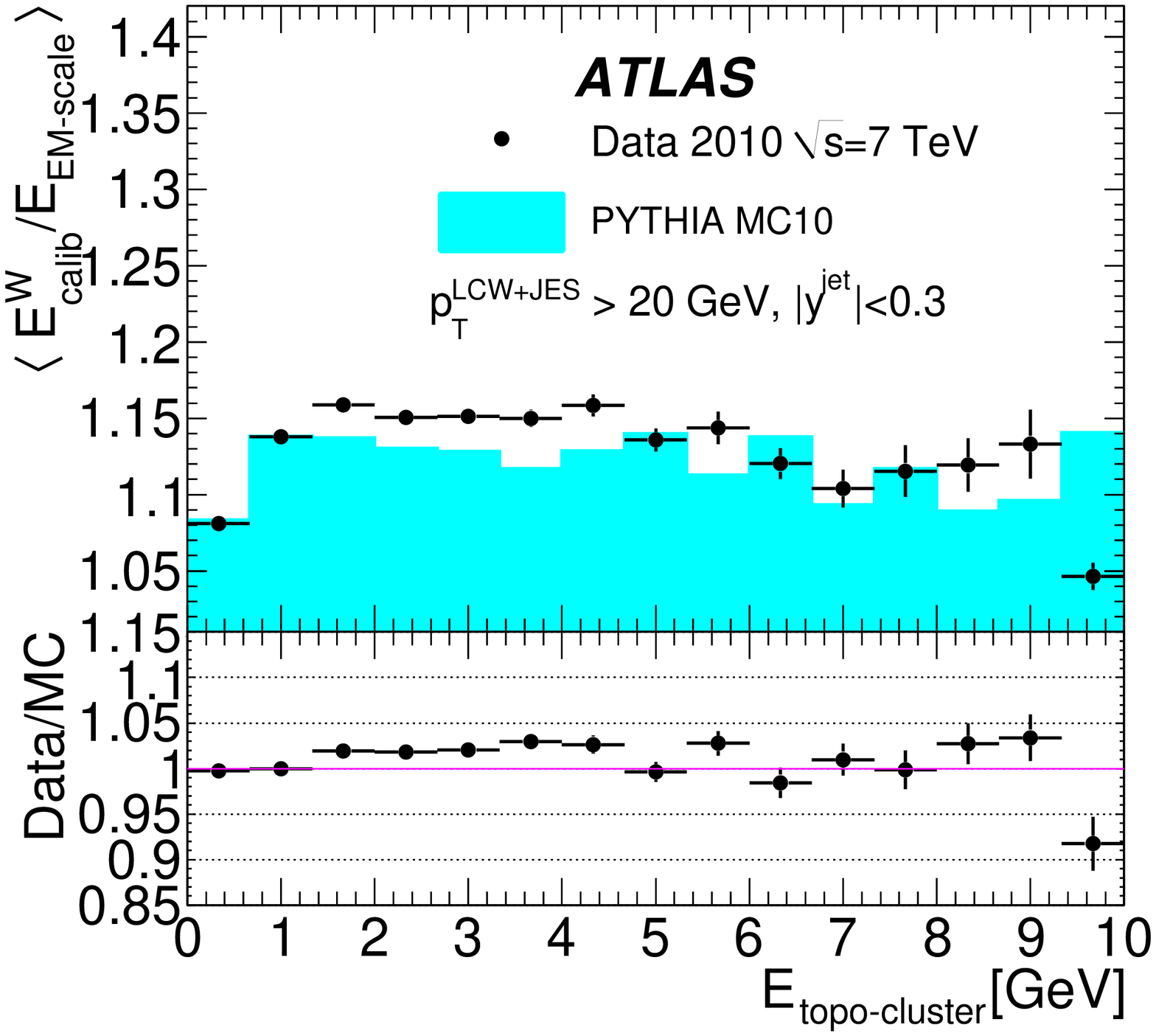}
  \includegraphics[width=0.39\textwidth]{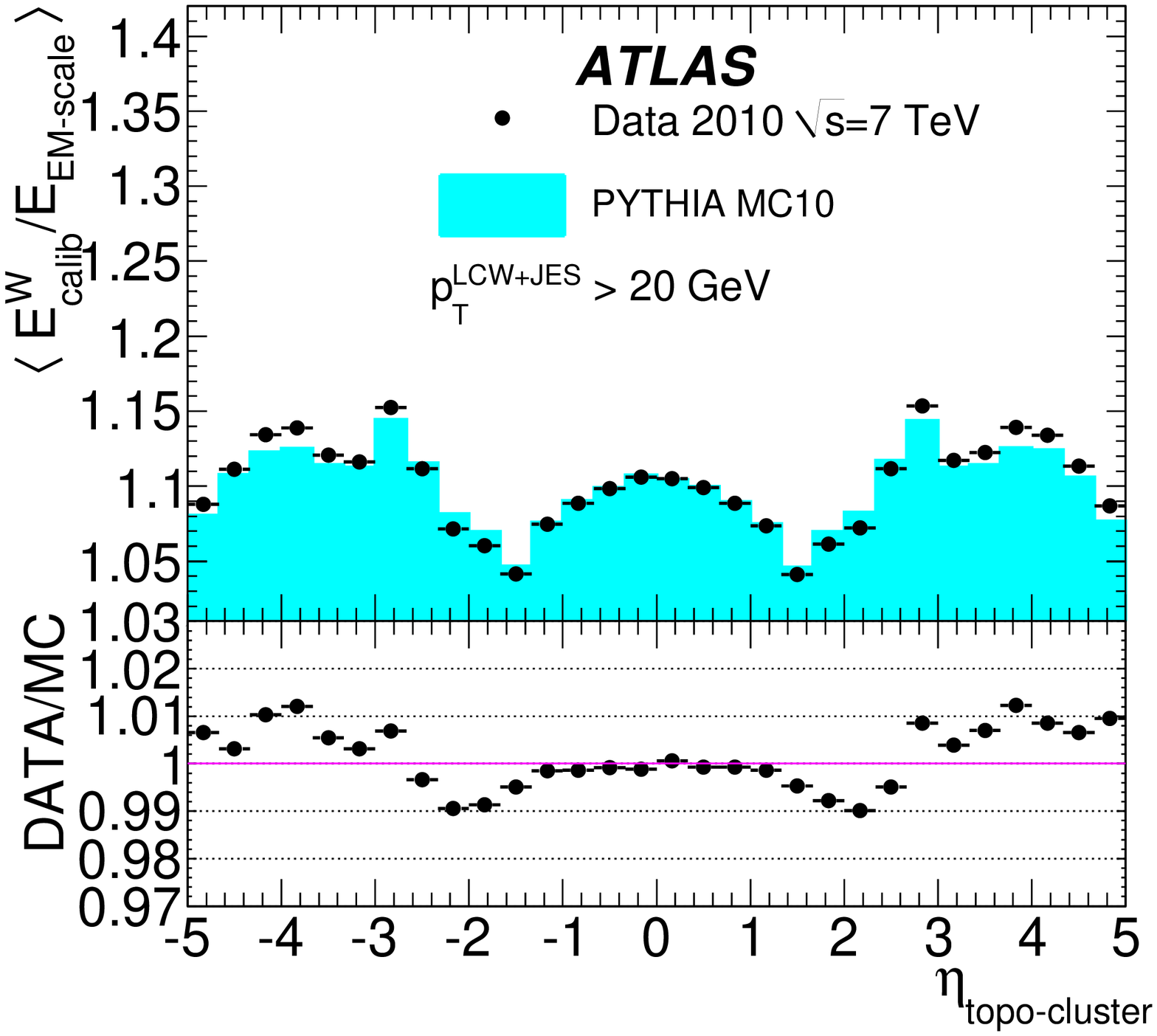}
}\\
\vspace{-0.2cm}
\subfloat[Hadronic response and out-of-cluster weights]{
  \label{fig:calibClusters:2}
  \includegraphics[width=0.39\textwidth]{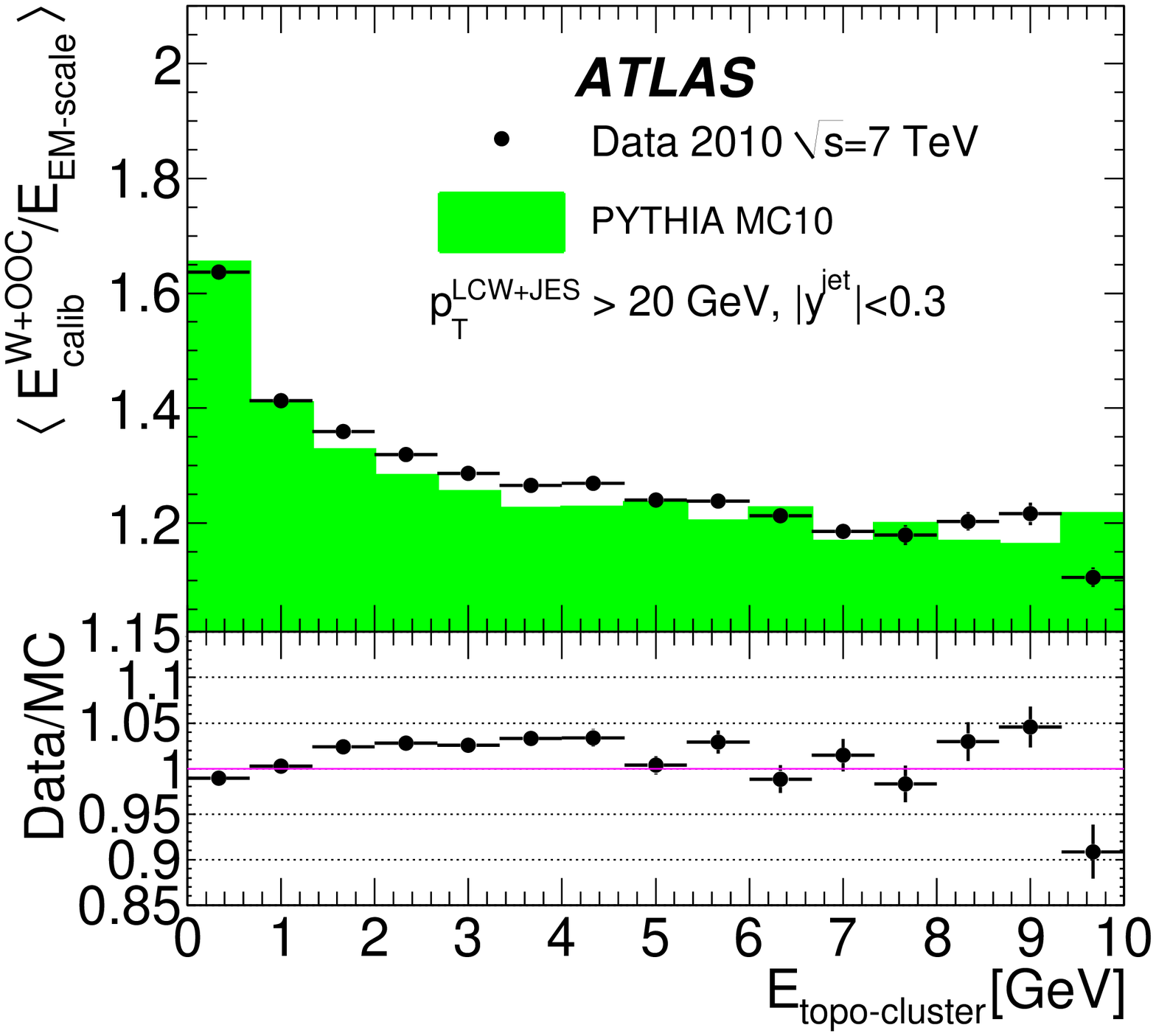}
  \includegraphics[width=0.39\textwidth]{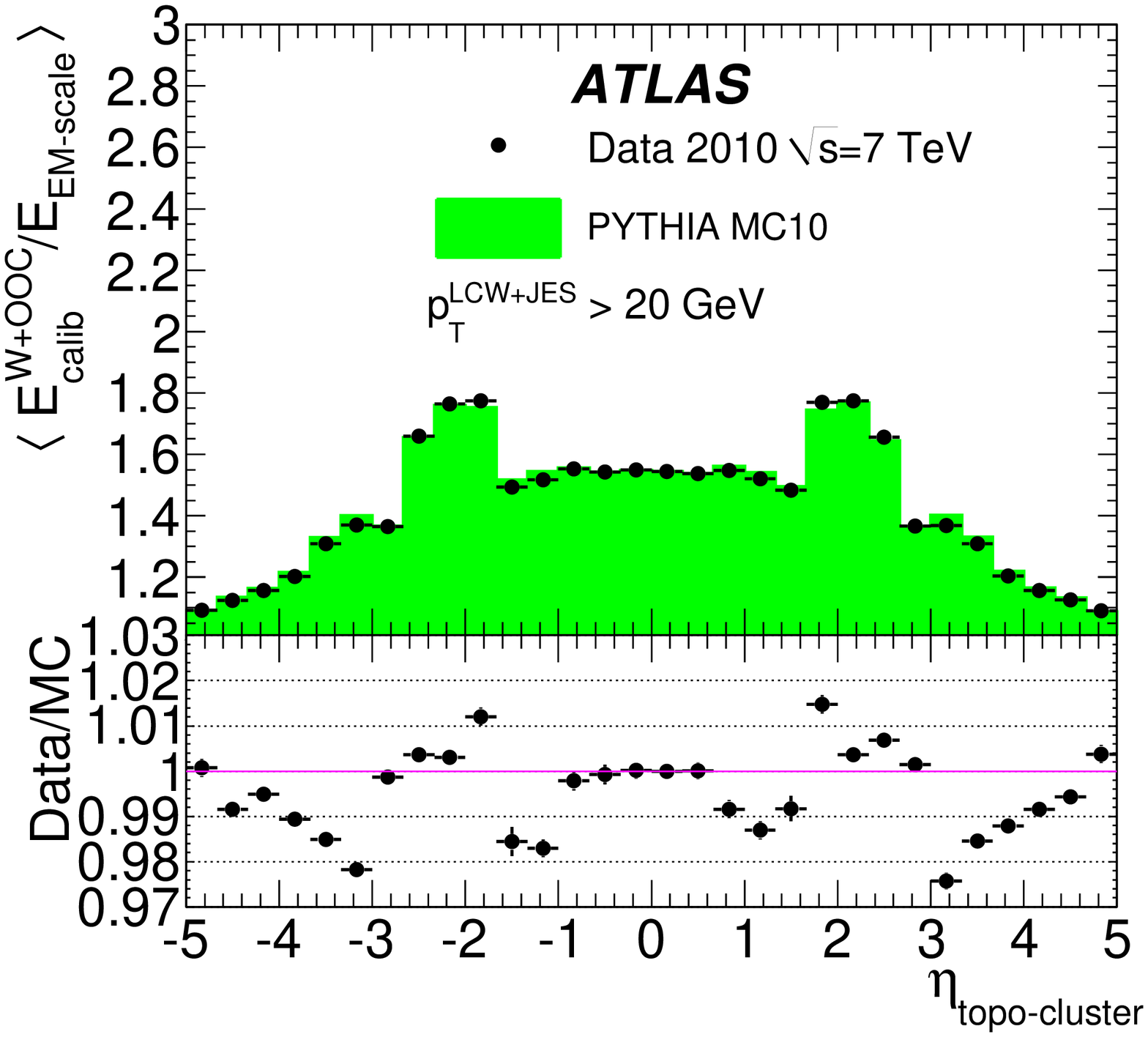}
 }\\
\vspace{-0.2cm}
\subfloat[Hadronic response, out-of-cluster and dead material weights]{
  \label{fig:calibClusters:3}
  \includegraphics[width=0.39\textwidth]{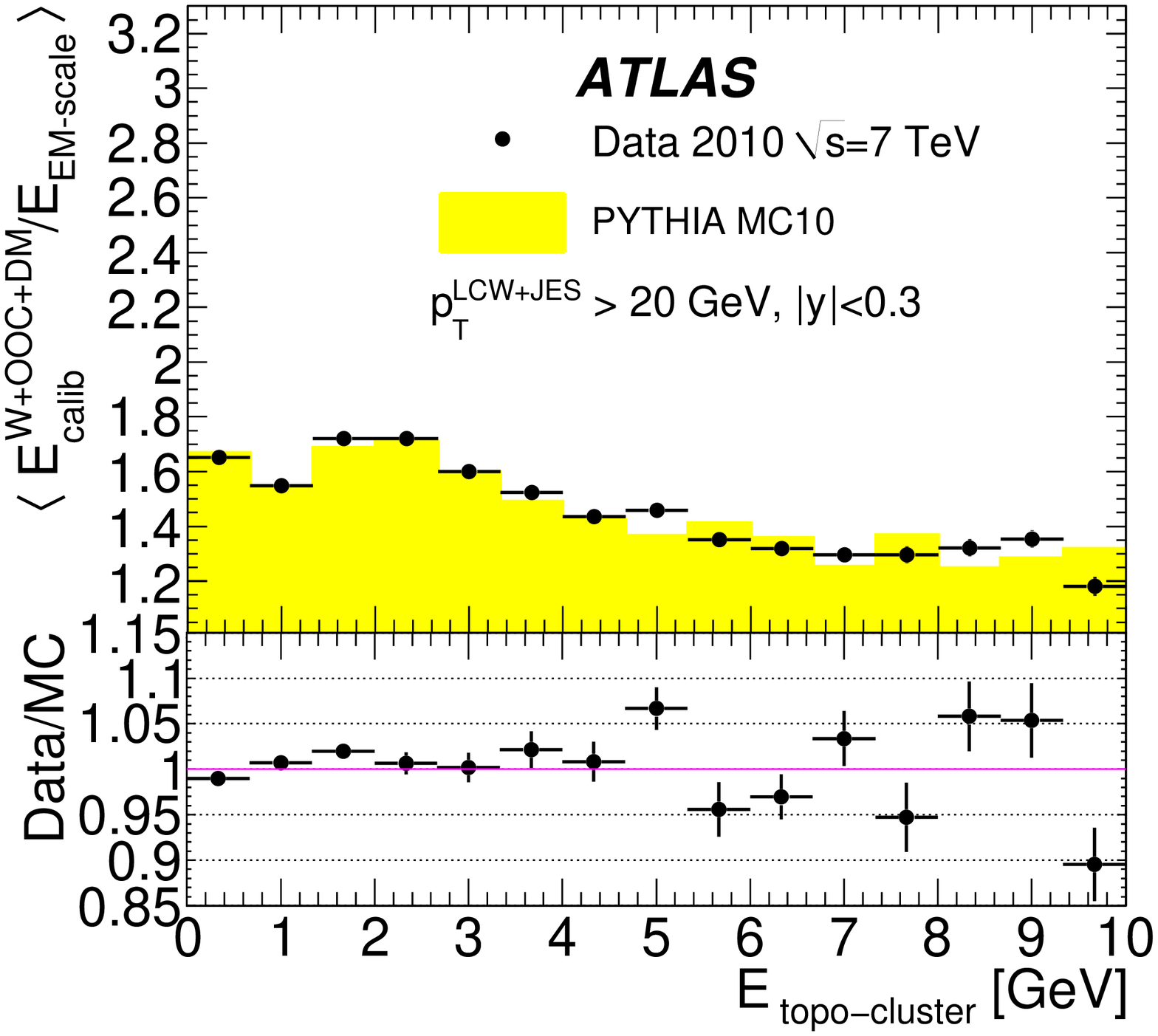}
  \includegraphics[width=0.39\textwidth]{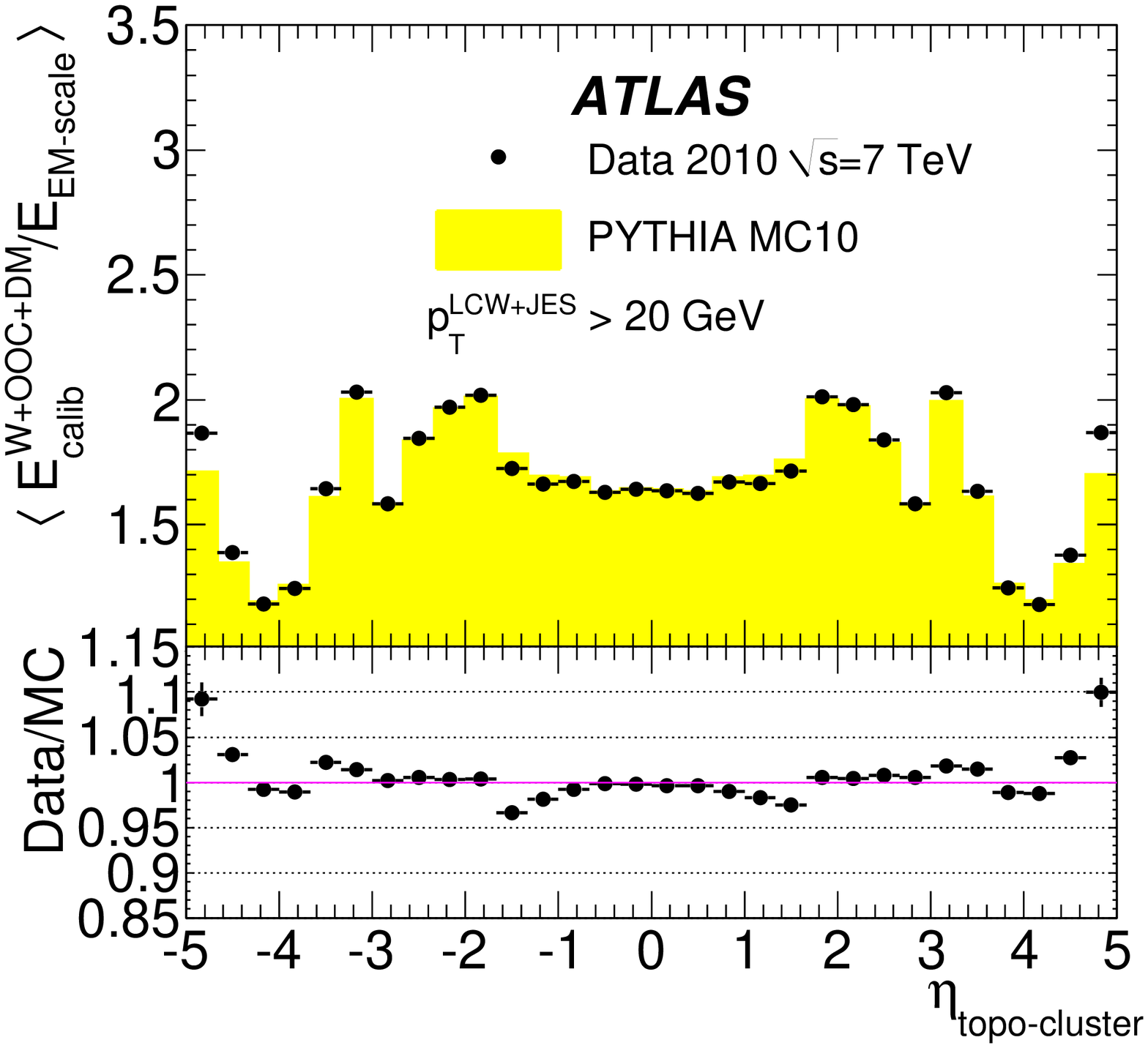}
 }
\vspace{-0.2cm}
\caption{Mean calibrated \topo{} energy divided by the uncalibrated \topo{} energy in data (points) and
 Monte Carlo simulation (shaded area) as a function of the uncalibrated \topo{} energy (left)
 and pseudorapidity (right) after hadronic response weighting (a), adding out-of-cluster corrections (b), 
 and adding dead material corrections (c) applied to \topos{} in jets. 
The corrections are  sequentially applied.
 \Antikt{} jets with $R=0.6$ in the \LCWJES{} scheme are required to have $\ptjet > 20$~\GeV. 
In addition, for the results as a function of the \topo{} energy (left) the jets have been 
restricted to $|\rapjet| < 0.3$.
}
\label{fig:calibClusters}
\end{figure*}

%
\subsection{Cluster properties inside jets as input to the local cluster weighting calibration}
\label{sec:calibClustersProp}
The \LCW{} weights are defined with respect to the electromagnetic scale energy of the \topos{} 
and can therefore be applied in any arbitrary order.
This allows systematic checks of the order in which the corrections are applied.
There are four cluster properties used in the \LCW{} calibration scheme:
\begin{enumerate}
\item The energy density in cells in \topos.
\item The cluster energy fraction deposited in different calo\-ri\-me\-ter layers.
\item The isolation variable characterising the energy around the cluster.
\item The depth of the cluster barycentre in the calorimeter. 
\end{enumerate}

In addition, the cluster energy after each correction step
and the cluster location can be compared in data and Monte Carlo simulation.

\subsubsection{Cluster isolation}
\label{sec:clusterisolation}
Figure~\ref{fig:clustersInJetsISO} shows the distributions of the cluster isolation variable for all \topos{}
in calibrated jets with $\ptjet > 20$~\GeV{} and $| \rapjet | < 2.8$ 
for \topos{} classified as electromagnetic (a) and hadronic (b).

The cluster isolation variable is bounded between $0$ and $1$, with higher values corresponding to higher isolation
(see Section~\ref{eq:clusterisolation}). 
Most of the \topos{} in lower energetic jets have a high degree of isolation.
The peaks at $0.25$, $0.5$ and $0.75$ are due to the \topos{} 
in boundary regions which are geometrically difficult to model 
or regions with a small number of calorimeter cells. 
Such \topos{} contain predominantly gap scintillator cells or are located
at the boundary of the \HEC{} and the \FCAL{} calorimeters.

The features observed are similar for \topos{} classified as mostly electromagnetic and those classified 
as mostly hadronic. A reasonable agreement between data and Monte Carlo simulation 
(see Fig.~\ref{fig:clustersInJetsISO}) is found.
The agreement in the peaks corresponding to the transition region between calorimeters
is not as good as in the rest of the distribution.

Figure~\ref{fig:clustersInJetsPtISO} shows the mean value of the \topo{} isolation variable as a function of 
the \topo{} energy for all \topos{} in jets with $\ptjet > 20$~\GeV{} 
and $| \rapjet |< 2.8$ for \topos{} classified as electromagnetic (a) or as hadronic (b).
The Monte Carlo simulation consistently predicts more isolated \topos{} 
than observed in the data, particularly at \topo{} energies $E < 2$~\GeV{} 
and for both hadronic and electromagnetic cluster classifications. 
This feature is present in all rapidity regions, except for very low energy \topos{} classified as 
mostly electromagnetic in  very central jets. 

\subsubsection{Longitudinal cluster barycentre}
\label{sec:clusterlambdacentre}
Figure~\ref{fig:clustersInJetsLAM} shows the 
cluster barycentre $\lambda_{\rm centre}$ distributions
for all \topos{} in \LCW{} calibrated jets with 
$\ptjet > 20$~\GeV{} and $|\rapjet| < 2.8$ and for both cluster classifications.
Most \topos{} classified as electromagnetic have their centre in the electromagnetic calorimeter, as expected.
Those \topos{} classified as mostly hadronic are very often in the electromagnetic calorimeter, 
since these low $\pt$ jets do not penetrate far into the hadronic calorimeter.
However, a structure is observed, related to the position of the different longitudinal layers in the 
hadronic calorimeter. This structure is more prominent when looking at individual rapidity regions, 
being smeared where the geometry is not changing
in this inclusive distribution.
Good agreement is observed between data and Monte Carlo simulation.

Figure~\ref{fig:clustersInJetsPtLAM} shows the mean value of distributions of $\lambda_{\rm centre}$
as a function of the cluster energy for all \topos{} in jets with $\ptjet > 20$~\GeV{} and $|\rapjet| < 2.8$, 
again for both types of \topos.
In this case, \topos{} classified as mostly electromagnetic have their barycentre deeper in the calorimeter on average 
as the cluster energy increases. A different behaviour is observed for clusters tagged as hadronic,
for which the mean depth in the calorimeter increases until approximately $2$~\GeV, at which point the 
mean depth decreases again. The shape of the mean depth as a function of energy is different for different
jet rapidities due to the changing calorimeter geometry. However, the qualitative features are similar, 
with a monotonic increase up to some \topo{} energy, and a decrease thereafter. This is likely due to an
increased probability of a hadronic shower to be split into two or more clusters with increased cluster energy. 
A good agreement is observed between data and Monte Carlo simulation. 

\subsubsection{Cluster energy after LCW corrections}
\label{sec:calibClusters}
%
In this section the size of each of the three corrections of the \topo{} calibration is studied in data and 
Monte Carlo simulation. This provides a good measure of how the differences between data and Monte Carlo simulation 
observed in previous sections impact the size of the corrections applied. 

Figure~\ref{fig:calibClusters} shows the mean value of the ratio of the calibrated \topo{} energy to the
uncalibrated \topo{} energy after each calibration step 
as a function the \topo{} energy and pseudorapidity.
Only \topos{} in \LCW{} calibrated jets with $\ptjet > 20$~\GeV{} are considered. 
For the results shown as a function of 
\topo{} energy the pseudorapidity of the jets is, in addition, restricted to $|\rapjet| < 0.3$.

The agreement between data and Monte Carlo simulation is within $5 \%$
for the full pseudorapidity range 
and is generally better for lower \topo{} energies where the correction for the 
out-of-cluster energy dominates.
 As the \topo{} energy increases the largest corrections become the hadronic response 
and the dead material corrections. 
An agreement to about $1 \%$ is observed in a wide region in most of the barrel region after each correction. 
The agreement between data and Monte Carlo simulation is within $2 \%$ for all 
\topo{} pseudorapidities after the hadronic and the out-of-cluster corrections. 
Larger differences are observed between data and Monte Carlo simulation 
in the transition region between the barrel and the endcap and in the forward region once
 the dead material correction is applied.

%
\begin{figure*}[htp!!]
\begin{center}
\vspace{-0.2cm}
\subfloat[$| \etajet | < 0.3$]{
\epsfig{file=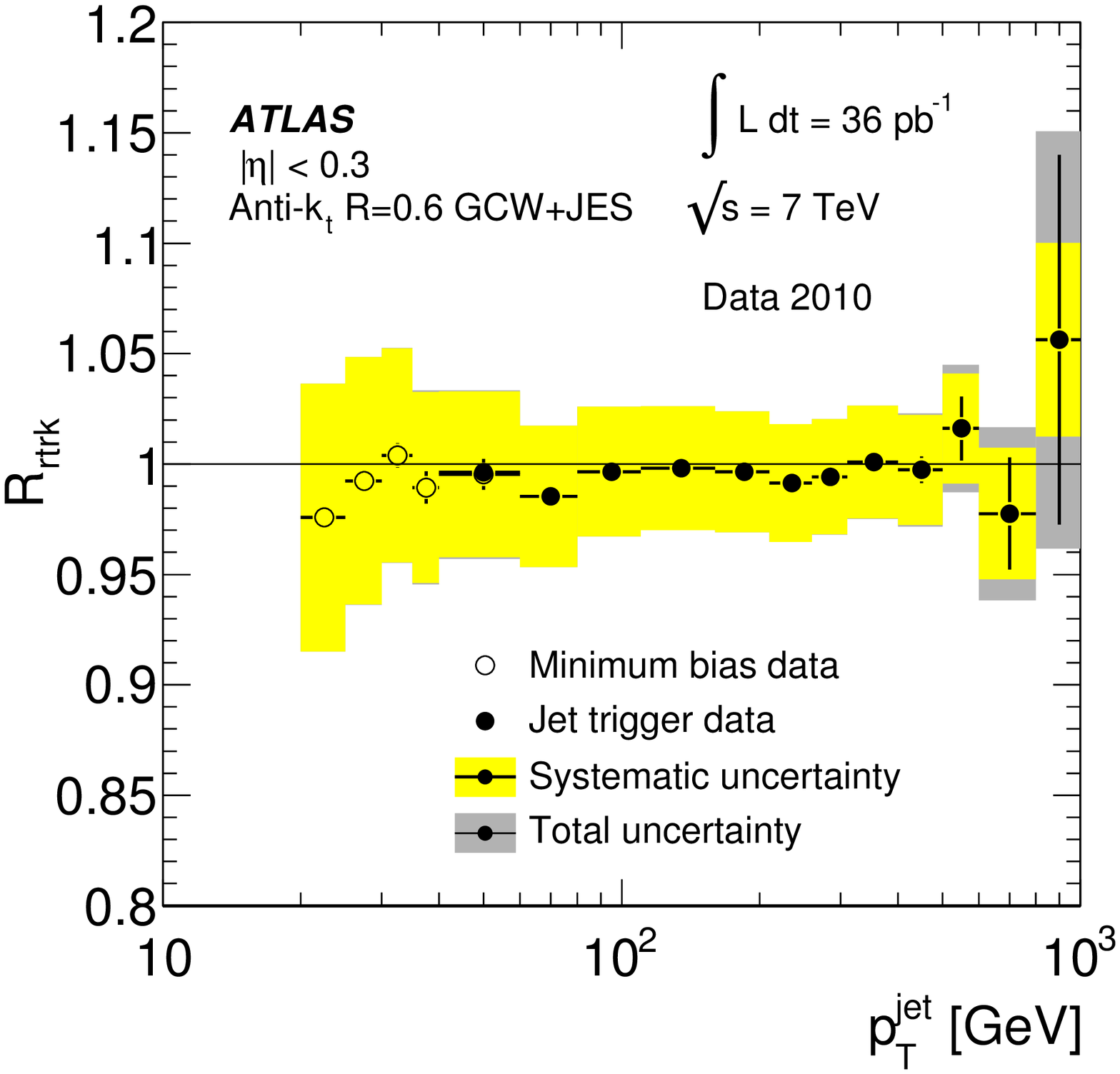,width=0.39\textwidth}
}
\subfloat[\etaRange{0.3}{0.8}]{
\epsfig{file=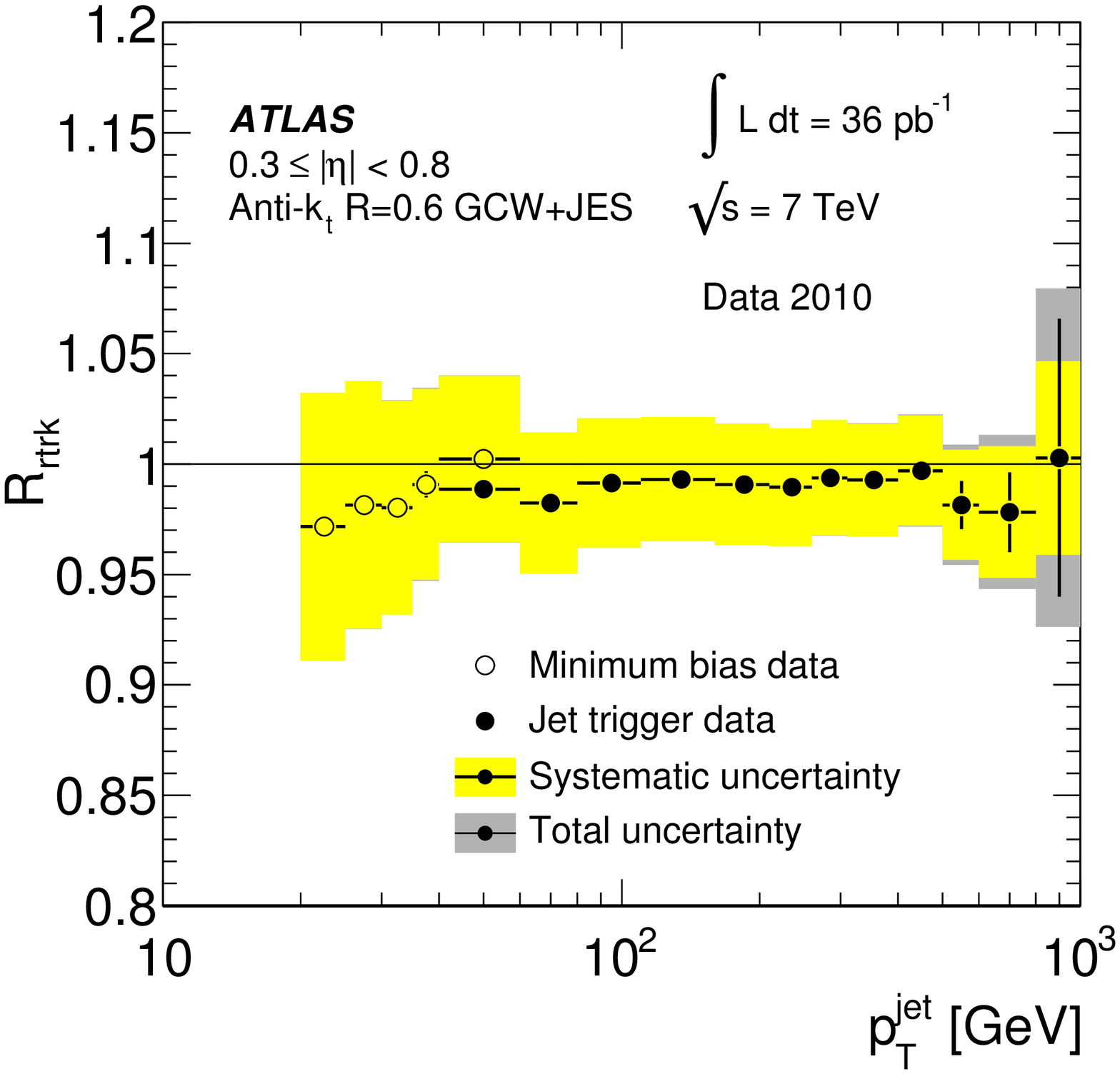,width=0.39\textwidth}
}
\vspace{-0.1cm}
\\
\subfloat[\etaRange{0.8}{1.2}]{
\epsfig{file=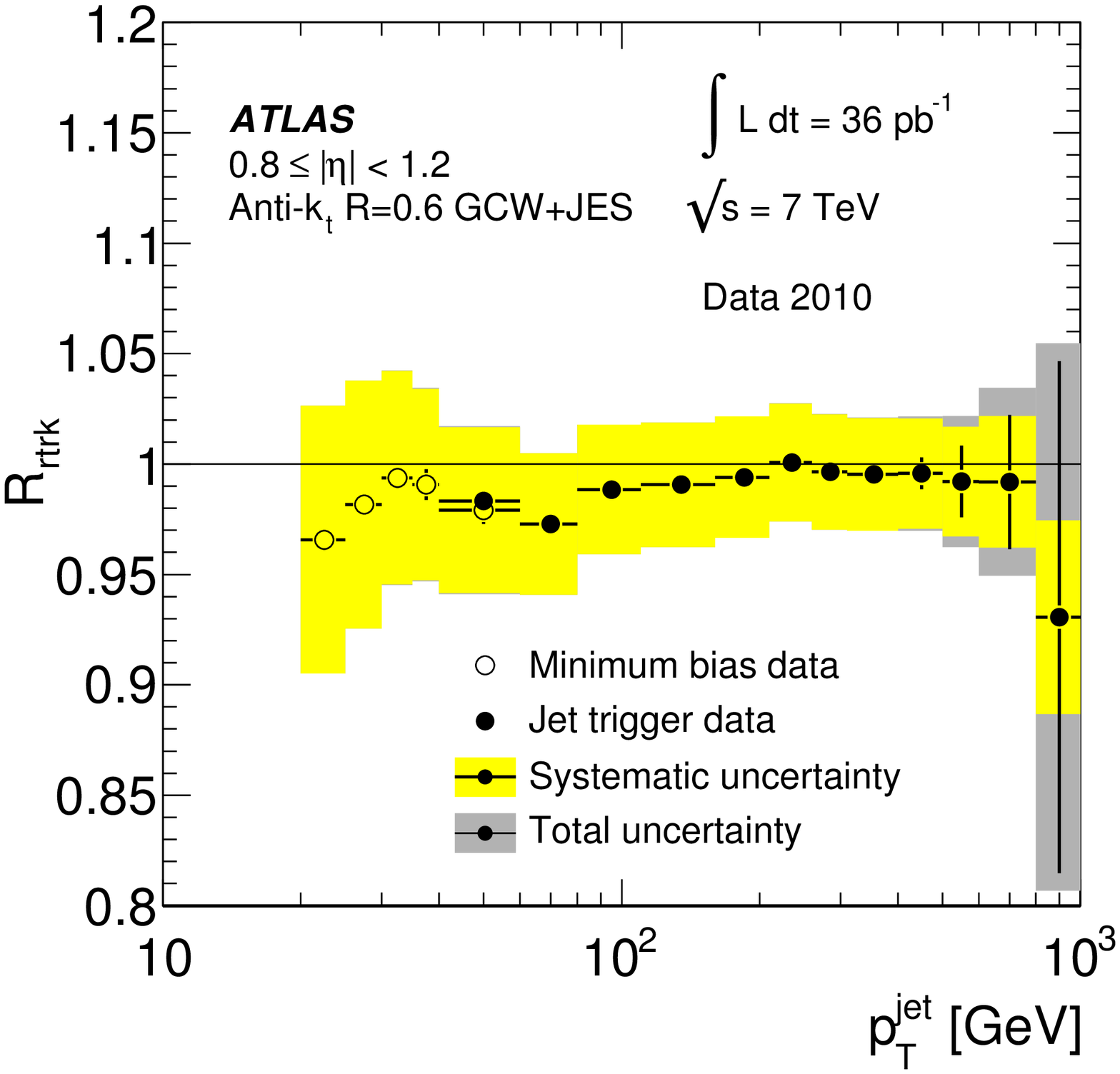,width=0.39\textwidth}
}
\subfloat[\etaRange{1.2}{1.7} ]{
\epsfig{file=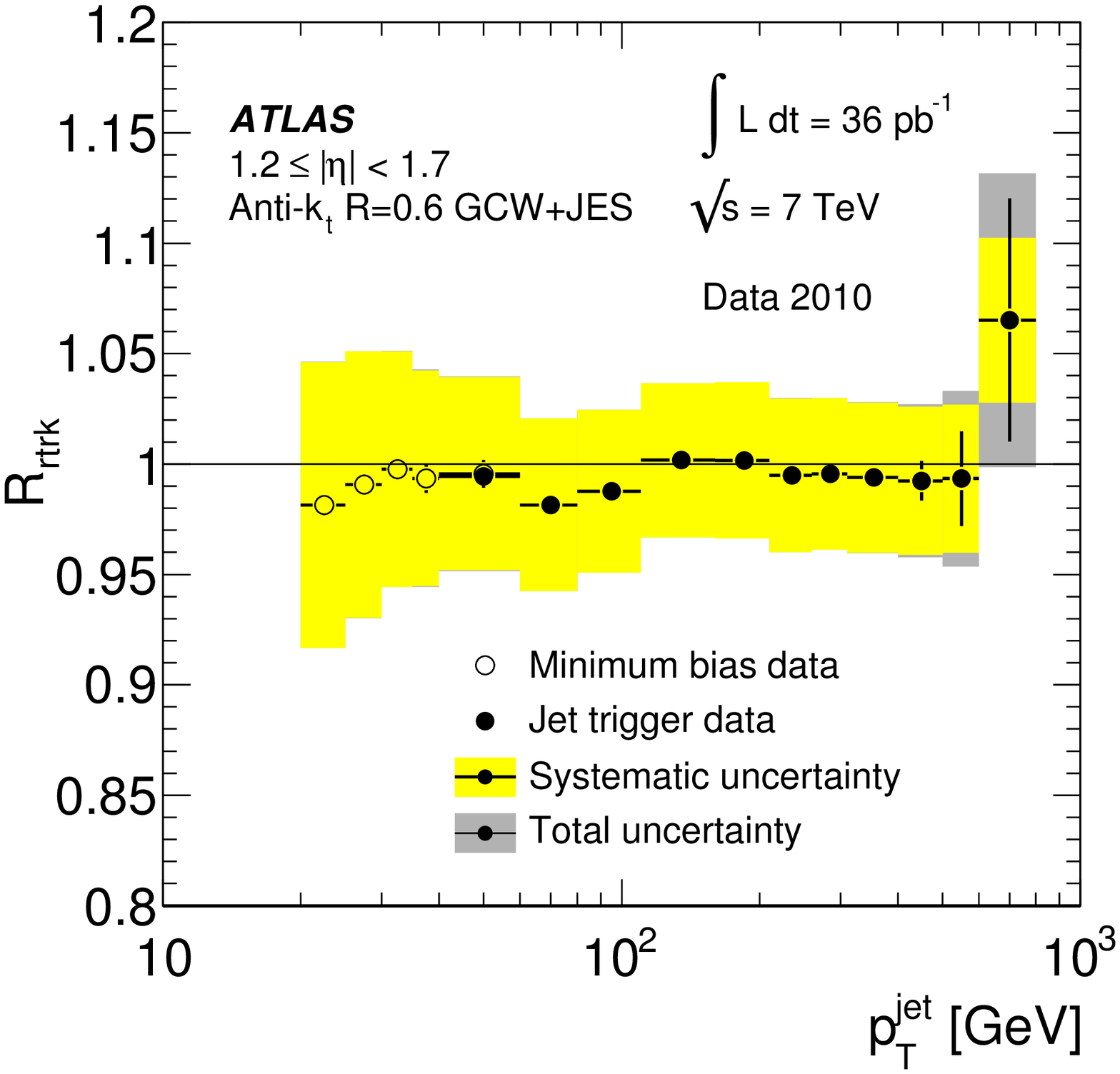,width=0.39\textwidth}
}
\vspace{-0.1cm}
\\
\subfloat[\etaRange{1.7}{2.1}]{
\epsfig{file=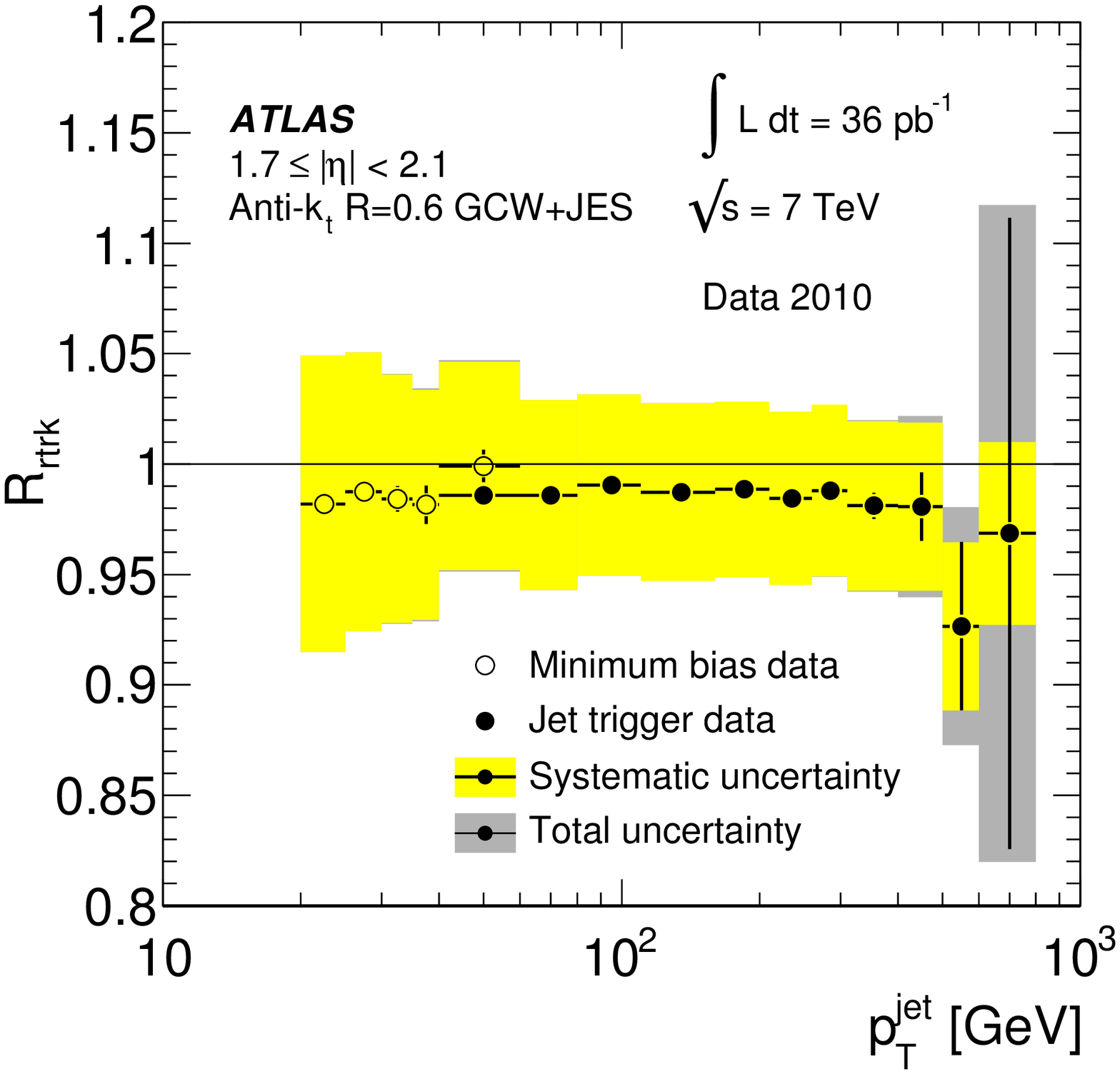,width=0.39\textwidth}
}
\subfloat[\AetaRange{1.2}]{
\epsfig{file=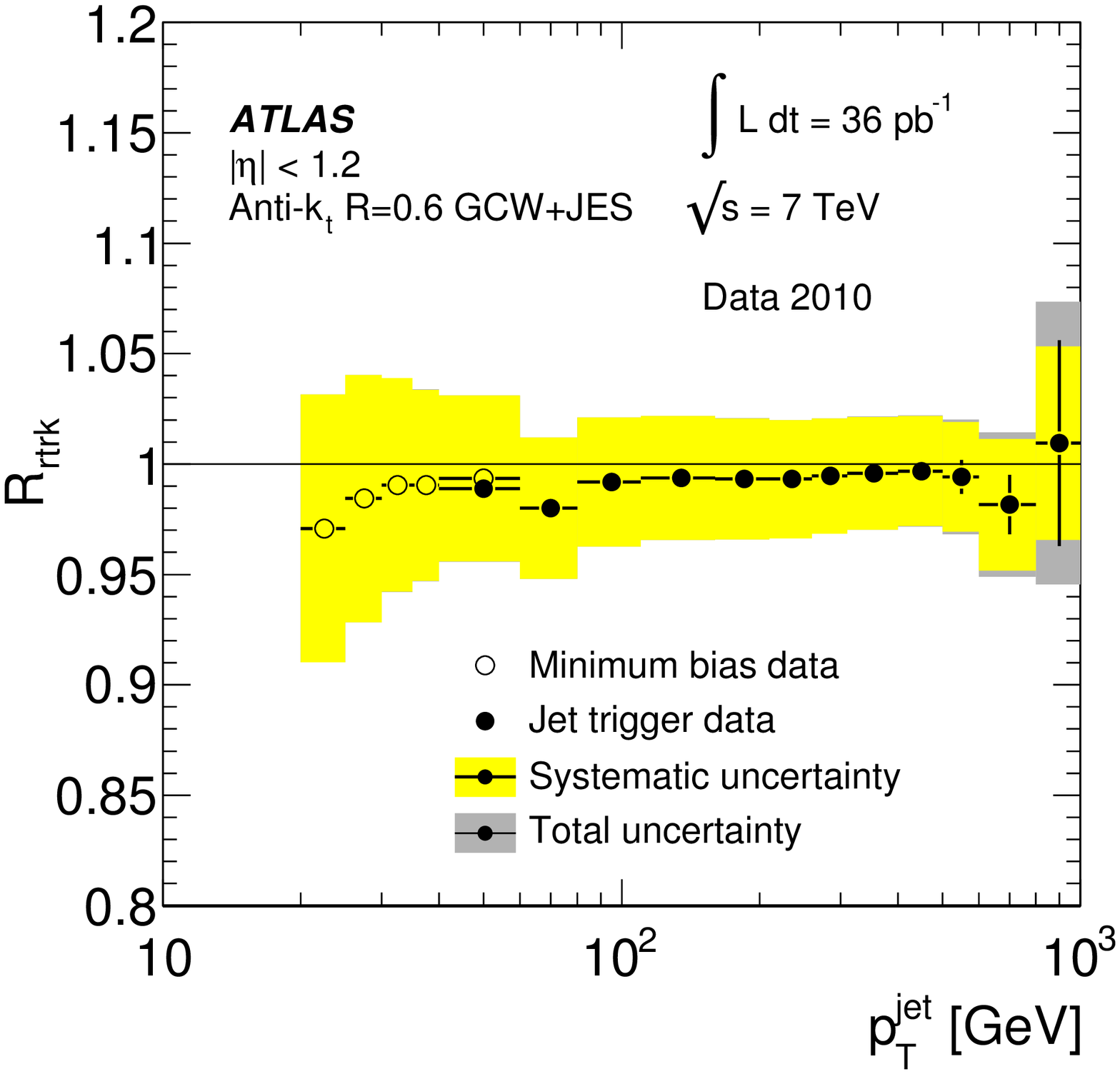,width=0.39\textwidth}
}\\
\end{center}
\vspace{-0.1cm}
\caption{Double ratio of the track to calorimeter response comparison
in data and Monte Carlo simulation,  
$R_{\rtrk} = [<\rtrk>]_{\rm Data} / [<\rtrk>]_{\rm MC}$,
for \antikt\ jets with $R = 0.6$ using the \GCWJES{} calibration scheme 
as a function of \ptjet{} for various $\etajet$ bins.
Systematic (total) uncertainties are shown as a  light (dark) band. 
}
\label{fig:finalAktopo6gcw}
\end{figure*}

\begin{figure*}[ht!!p]
\begin{center}
\vspace{-0.2cm}
\subfloat[$| \etajet | < 0.3$]{
\epsfig{file=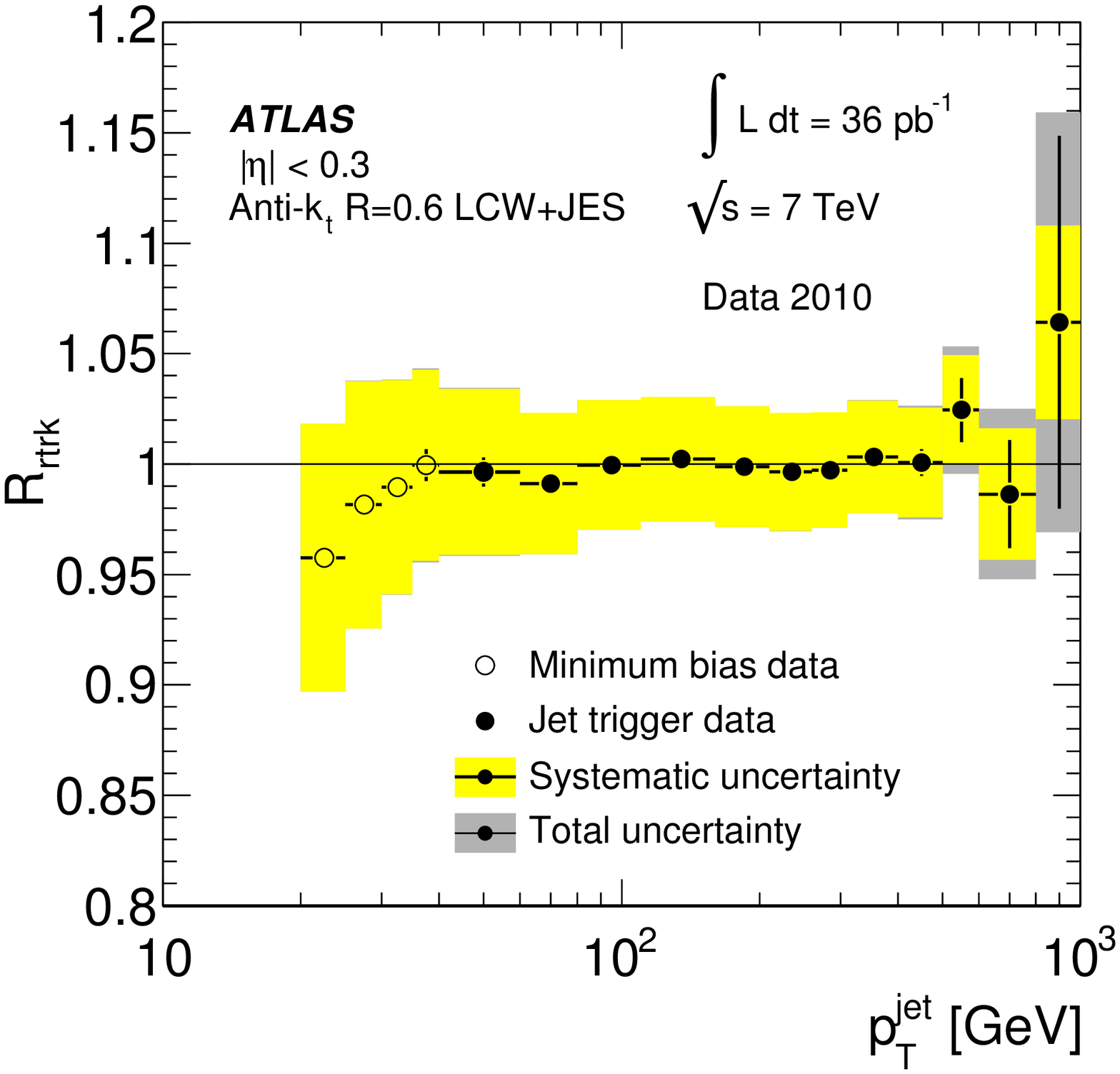,width=0.39\textwidth}
}
\subfloat[\etaRange{0.3}{0.8}]{
\epsfig{file=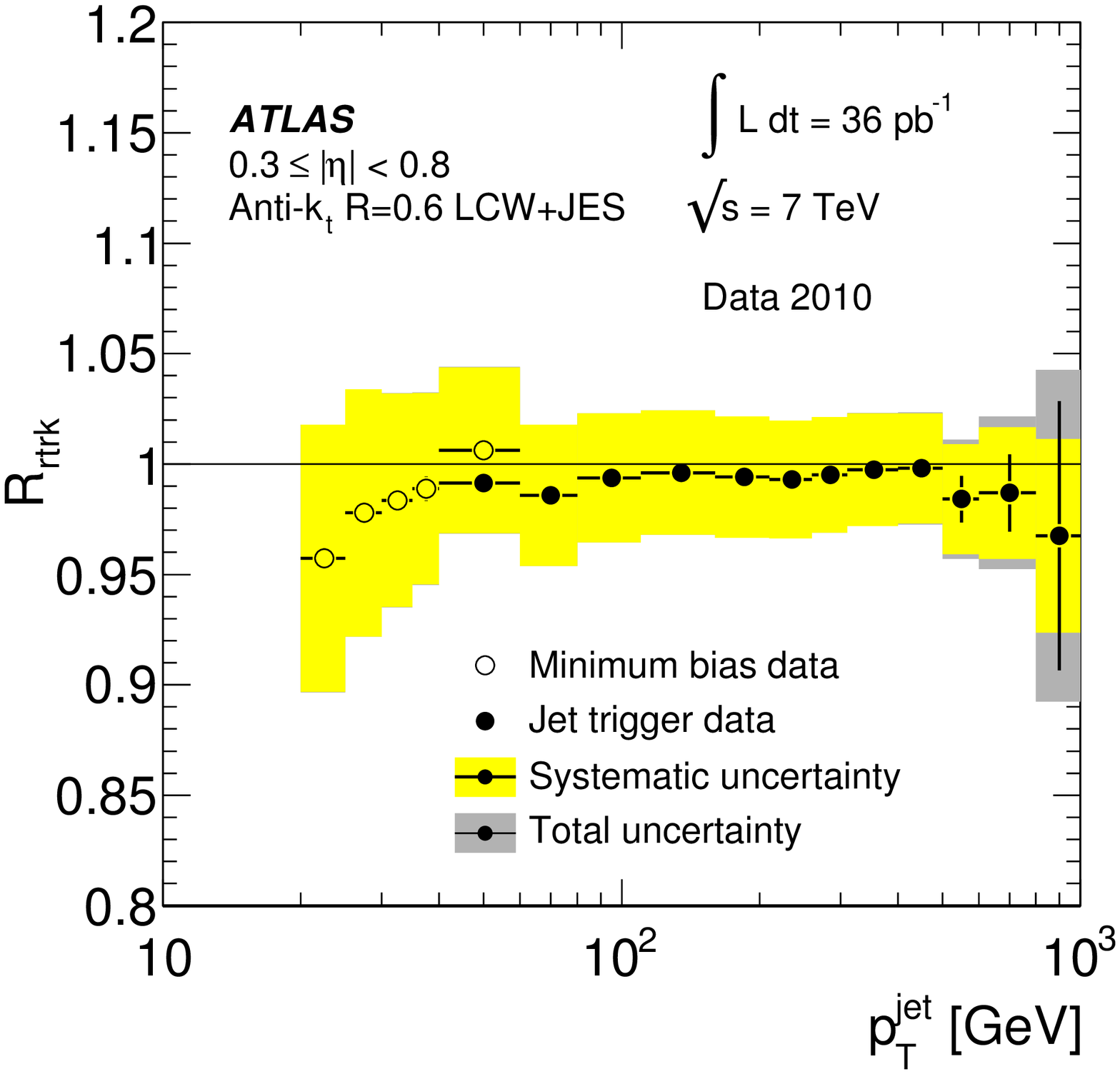,width=0.39\textwidth}
}
\vspace{-0.1cm}
\\
\subfloat[\etaRange{0.8}{1.2}]{
\epsfig{file=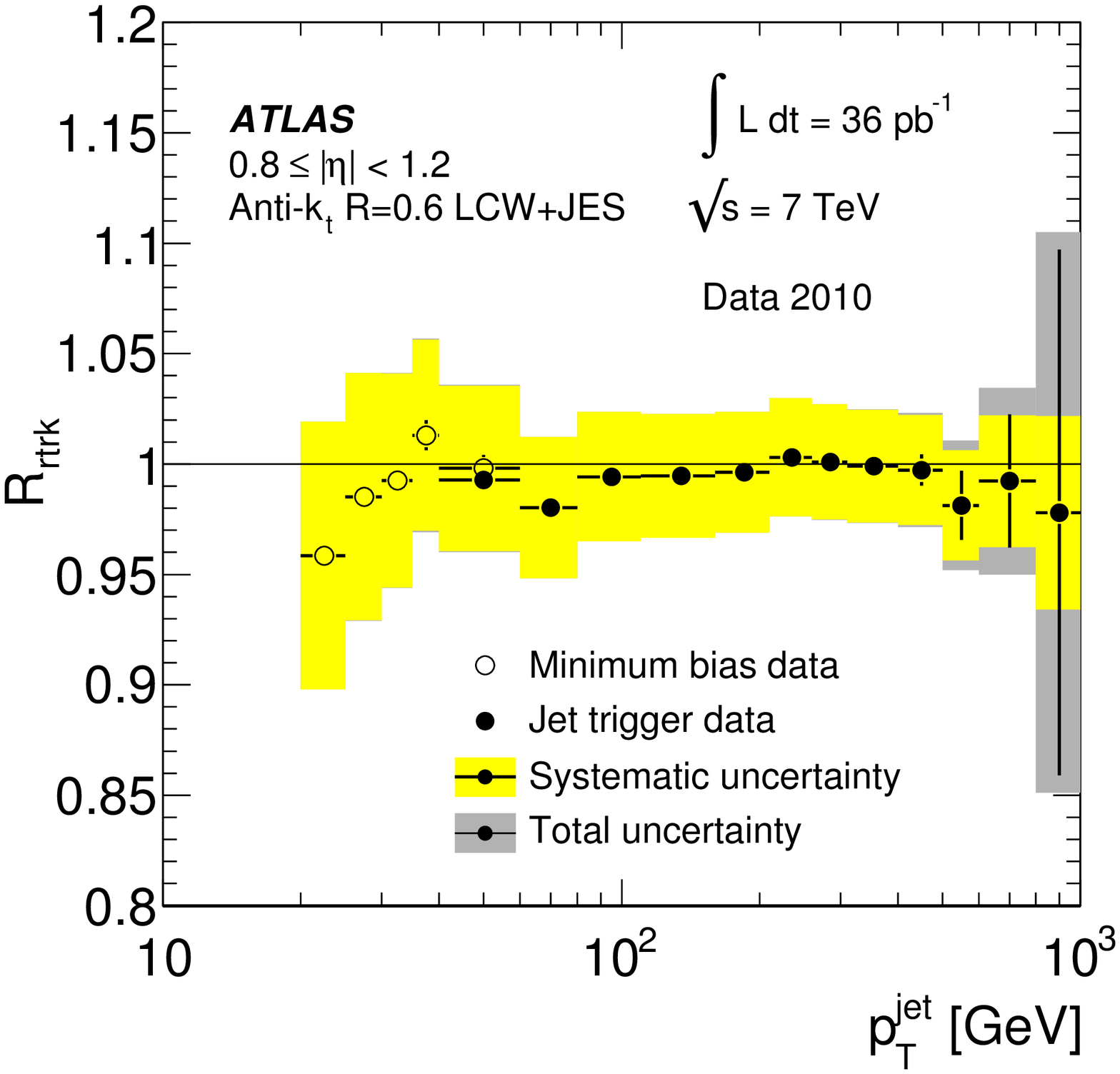,width=0.39\textwidth}
}
\subfloat[\etaRange{1.2}{1.7}]{
\epsfig{file=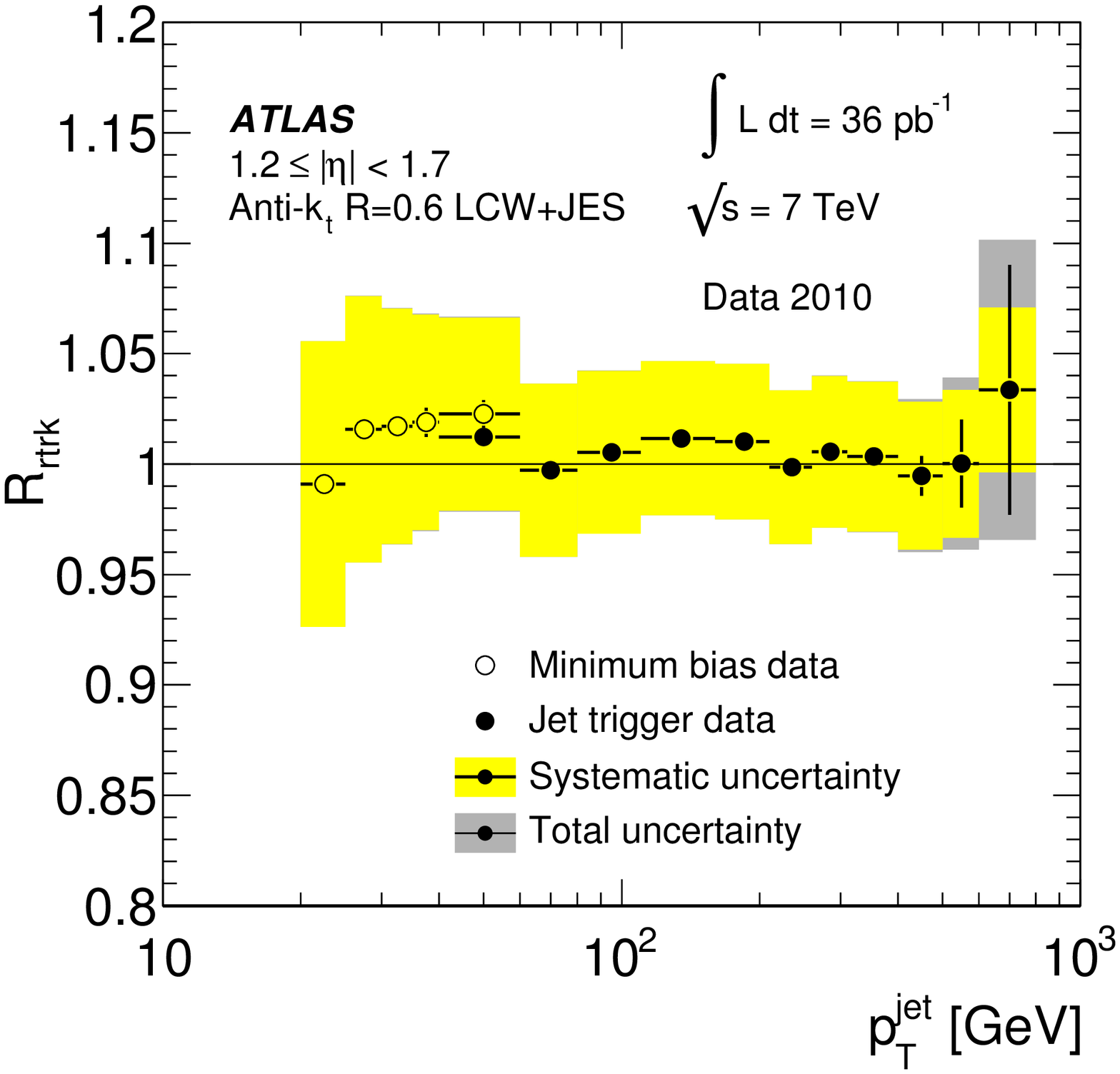,width=0.39\textwidth}
}
\vspace{-0.1cm}
\\
\subfloat[\etaRange{1.7}{2.1}]{
\epsfig{file=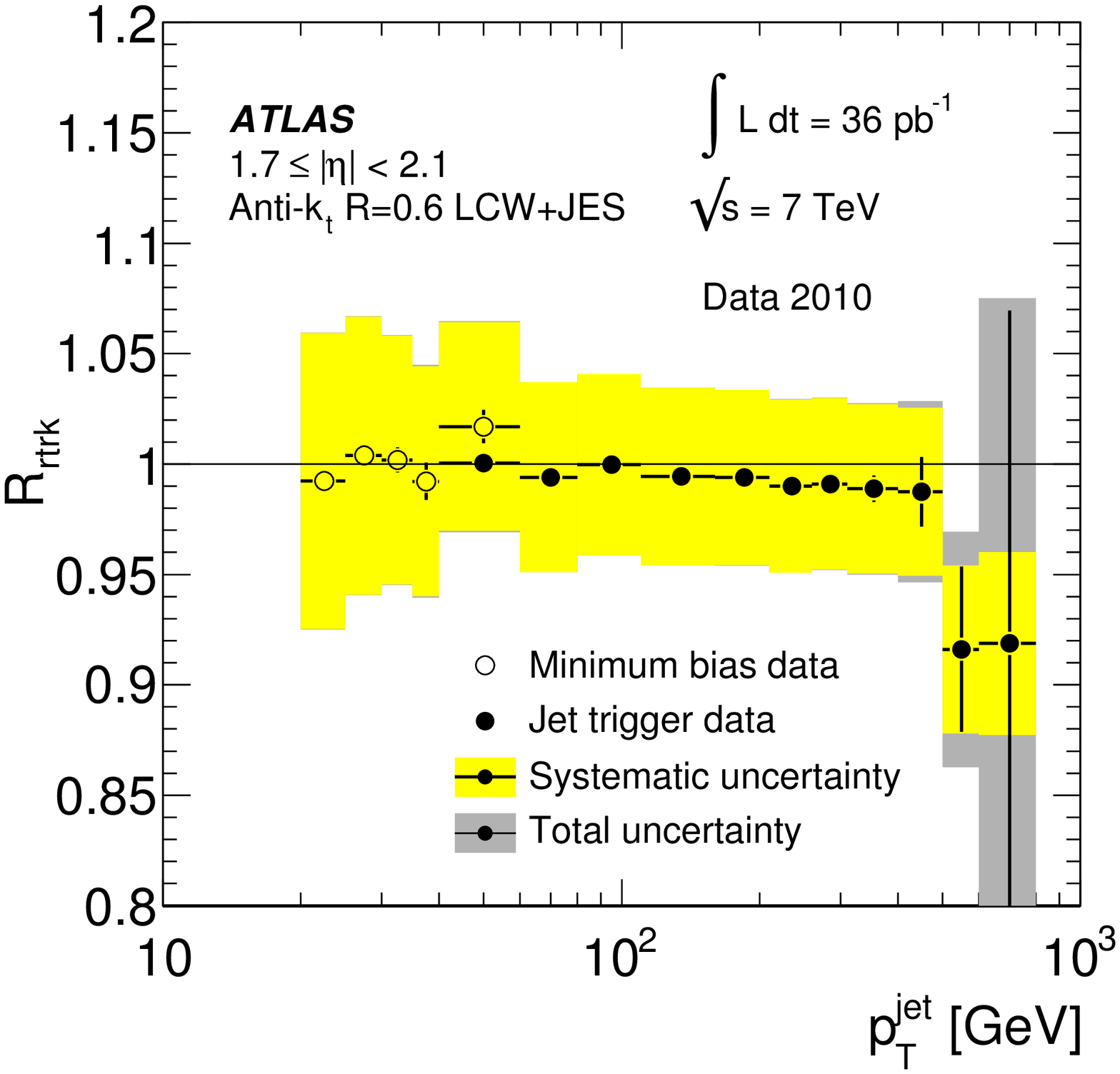,width=0.39\textwidth}
}
\subfloat[\AetaRange{1.2}]{
\epsfig{file=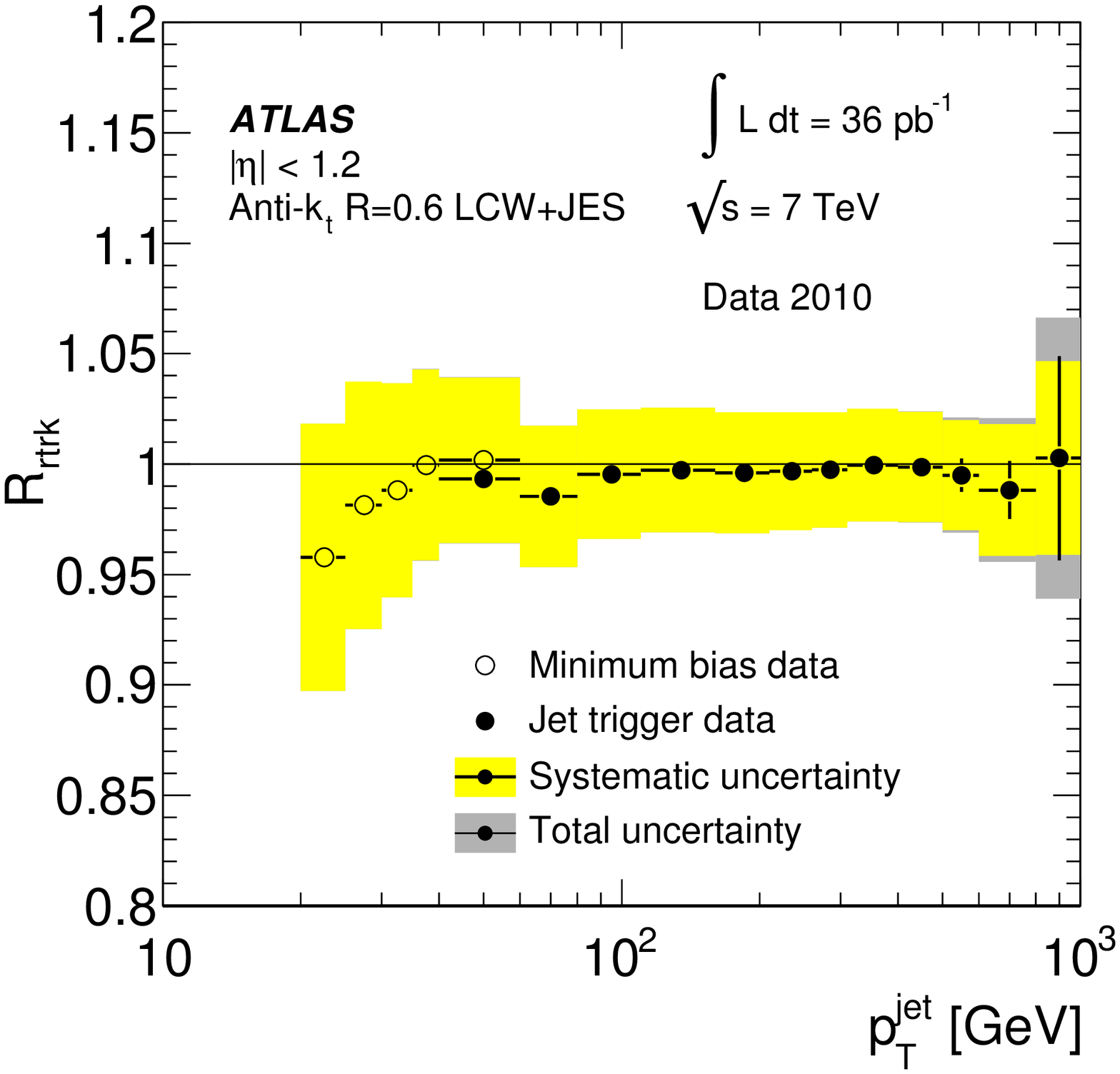,width=0.39\textwidth}
}
\vspace{-0.1cm}
\\
\end{center}
\vspace{-0.1cm}
\caption{Double ratio of the track to calorimeter response comparison
in data and Monte Carlo simulation,  
$R_{\rtrk} = [<\rtrk>]_{\rm Data} / [<\rtrk>]_{\rm MC}$,
for \antikt\ jets with $R = 0.6$ using the \LCWJES{} calibration scheme 
as a function of \ptjet{}
for various $\etajet$ bins.
Systematic (total) uncertainties are shown as a  light (dark) band. 
}
\label{fig:finalAktopol6lcw}
\end{figure*}

\begin{figure*}[ht!!p]
\begin{center}
 \subfloat[\GCWJES]{\includegraphics[width=0.4\textwidth]{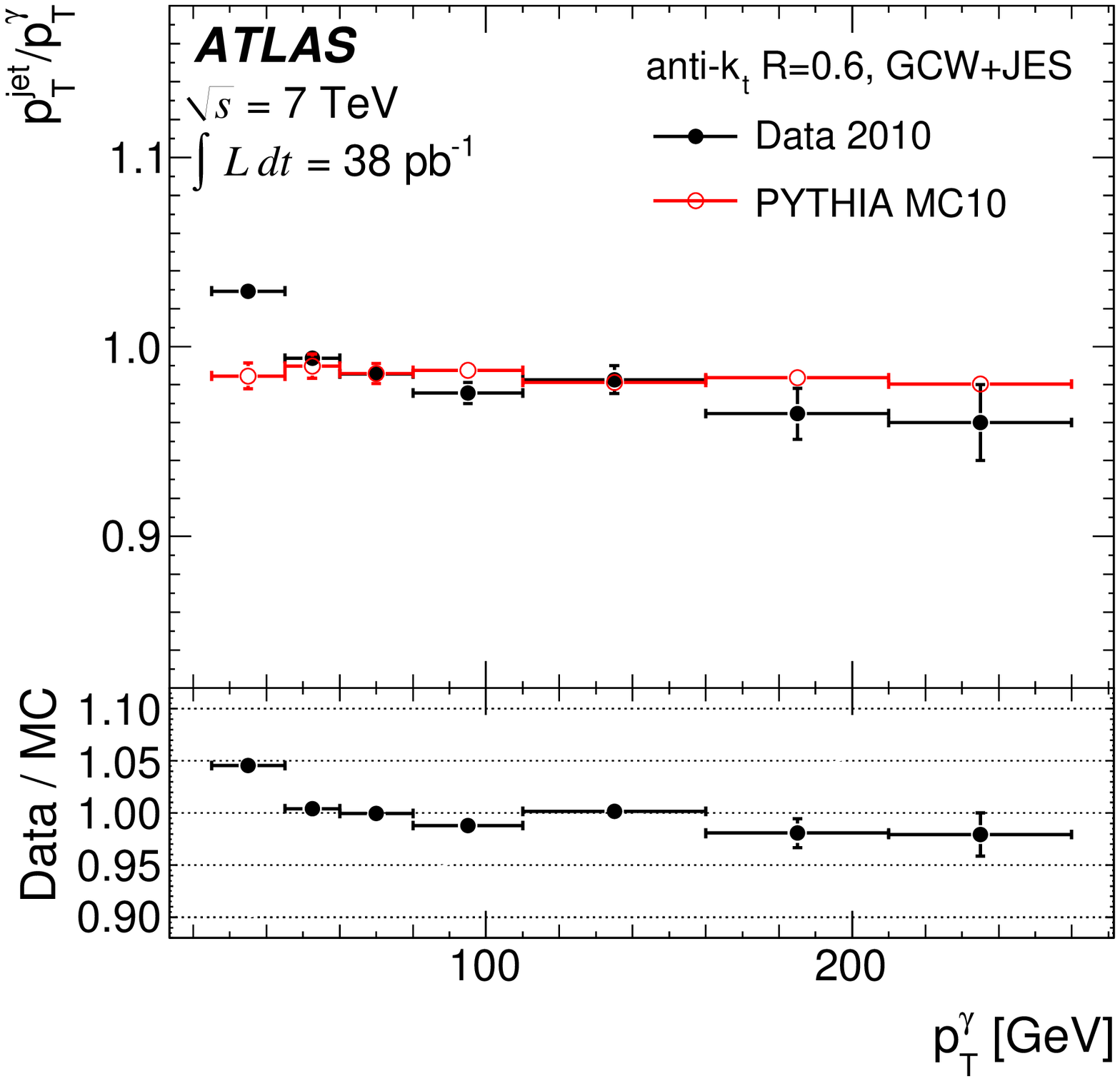}}
\hspace{1.cm}
 \subfloat[\LCWJES]{\includegraphics[width=0.4\textwidth]{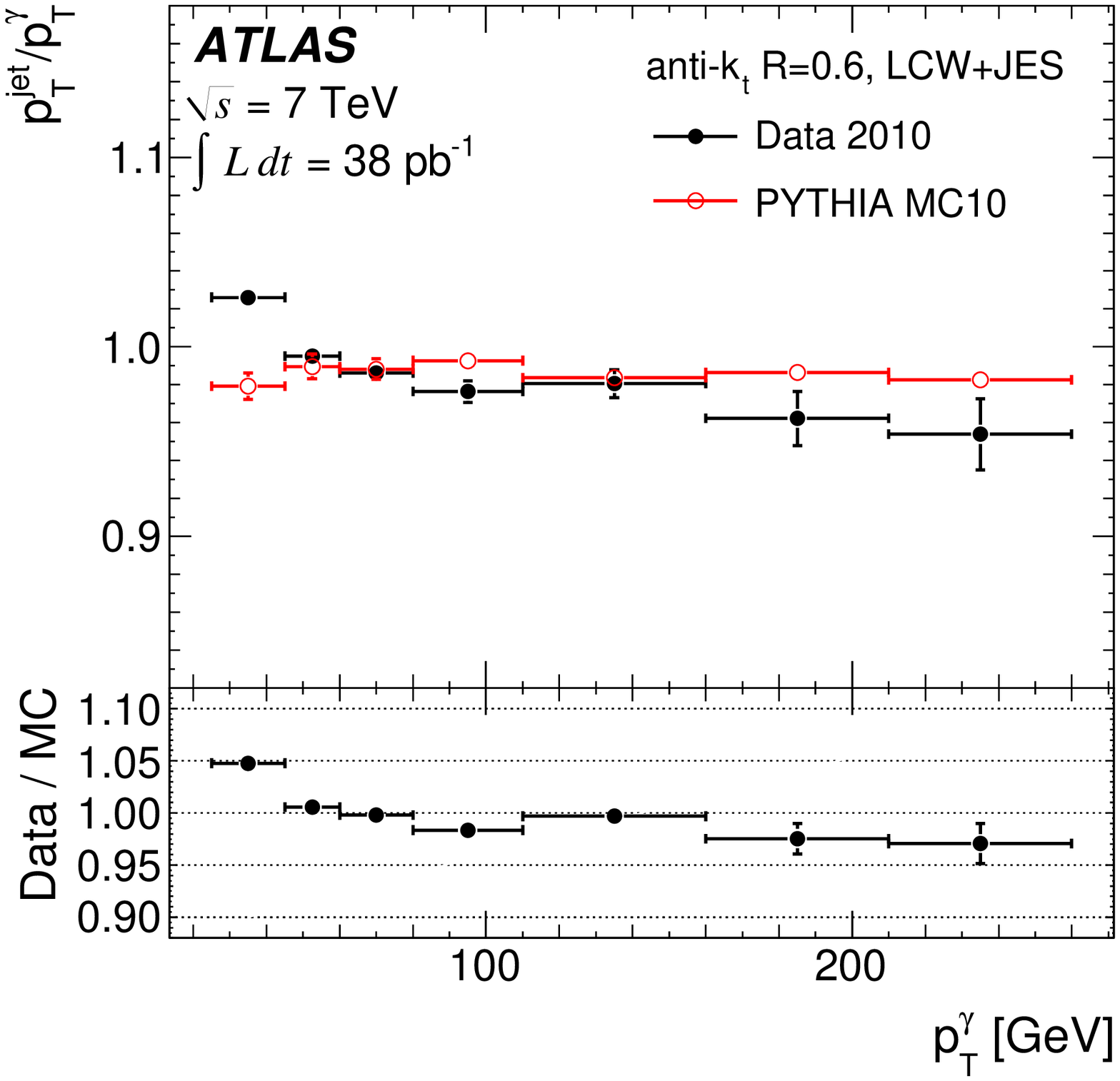}}
        \caption{Average jet response as determined by the direct \pt{}
          balance technique for \antikt{} jets with $R=0.6$ calibrated with the \GCWJES{} (a) and \LCWJES{} (b) 
          scheme as a function of photon transverse momentum for both data and Monte Carlo simulation.
The lower part of each figure shows the data to Monte Carlo simulation ratio.
Only statistical uncertainties are shown.}
        \label{fig:DBResponse}
\end{center}
\end{figure*}

\begin{figure*}[ht!!p]
\begin{center}
  \subfloat[\GCW{}]{\includegraphics[width=0.4\textwidth]{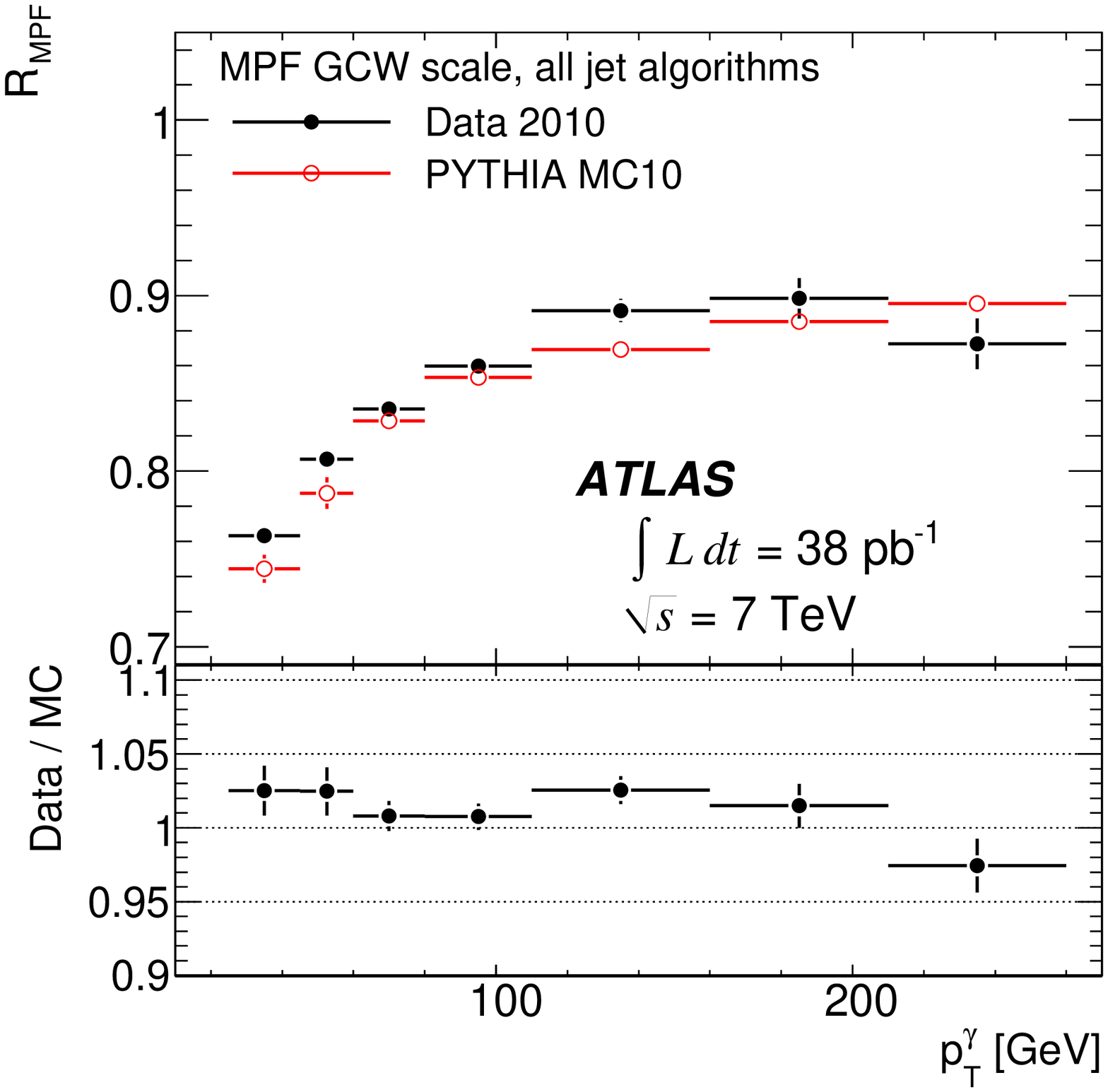}}
\hspace{1.cm}
  \subfloat[\LCW{}]{\includegraphics[width=0.4\textwidth]{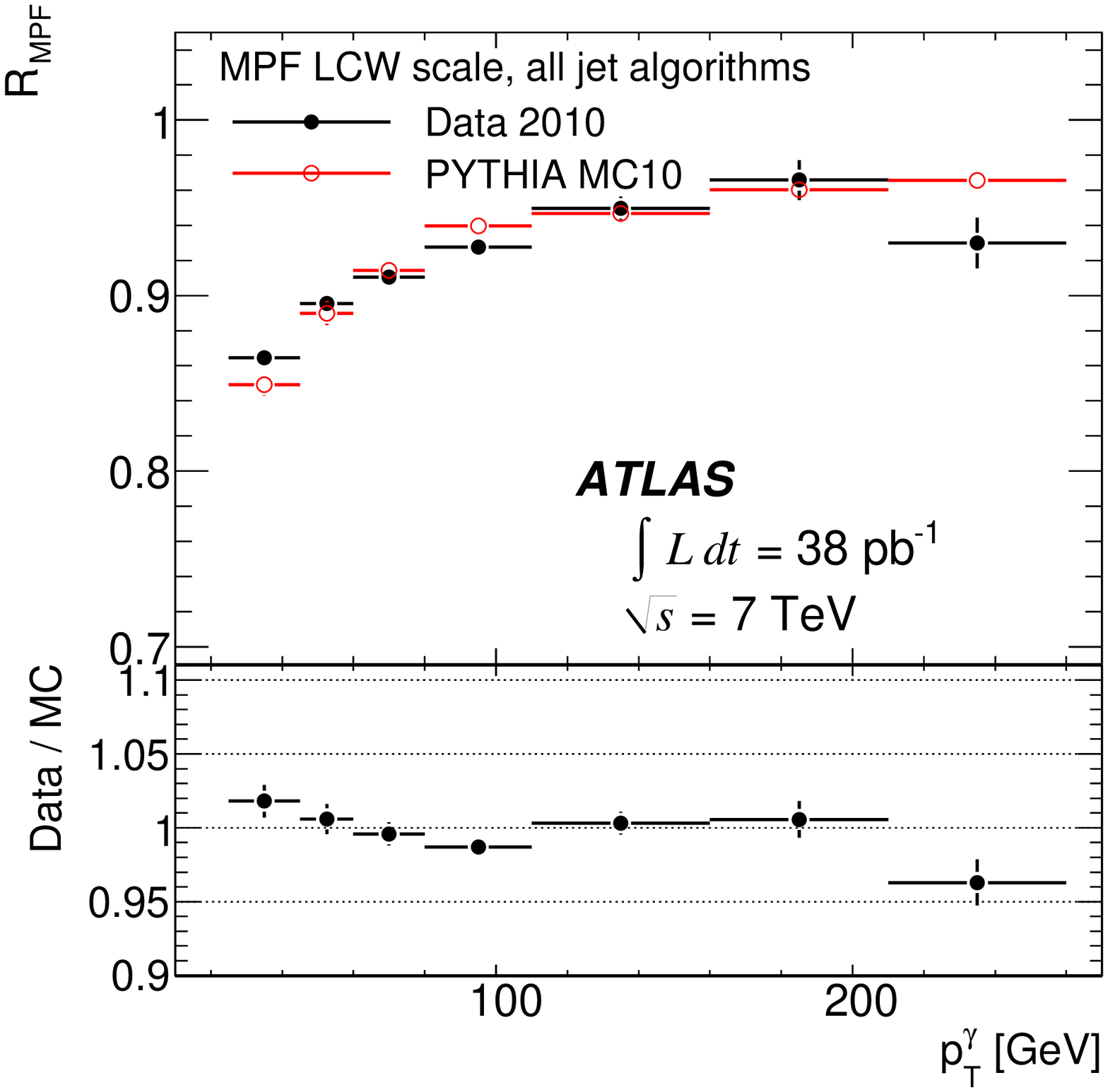}}
        \caption{Average calorimeter response as determined by the \MPF{} technique
          for the \GCW{} (a) and \LCW{} (b) calibration scheme as a function of
          photon transverse momentum for both data and Monte Carlo simulation.
The lower part of each figure shows the data to Monte Carlo simulation ratio.
Only statistical uncertainties are shown.}
        \label{fig:MPFResponse_GCW}
\end{center}
\end{figure*}

\begin{figure*}[ht!!p]
  \centering
  \subfloat[\GCWJES]{\includegraphics[width=0.4\textwidth]{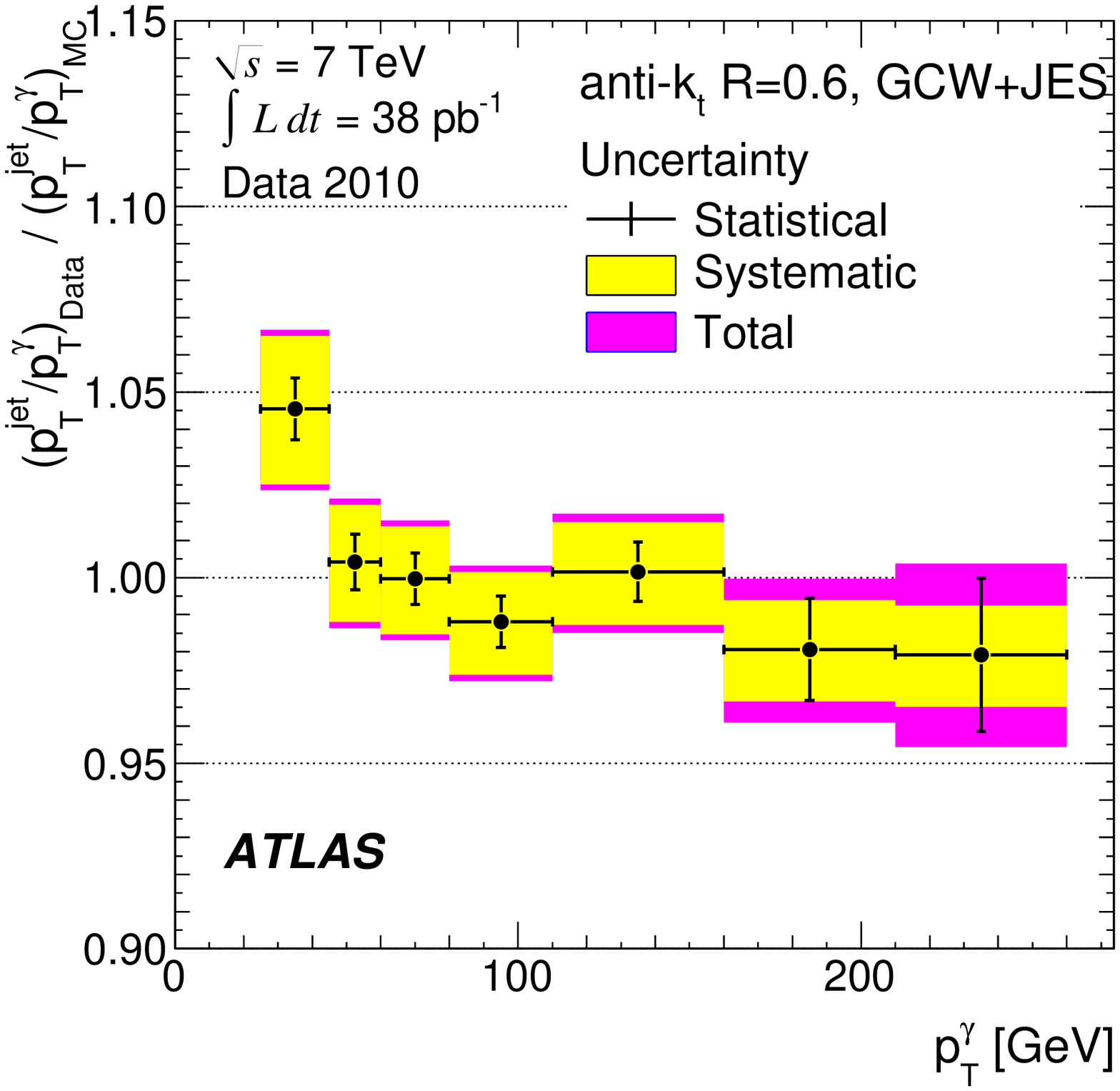}}
\hspace{1.cm}
  \subfloat[\LCWJES]{\includegraphics[width=0.4\textwidth]{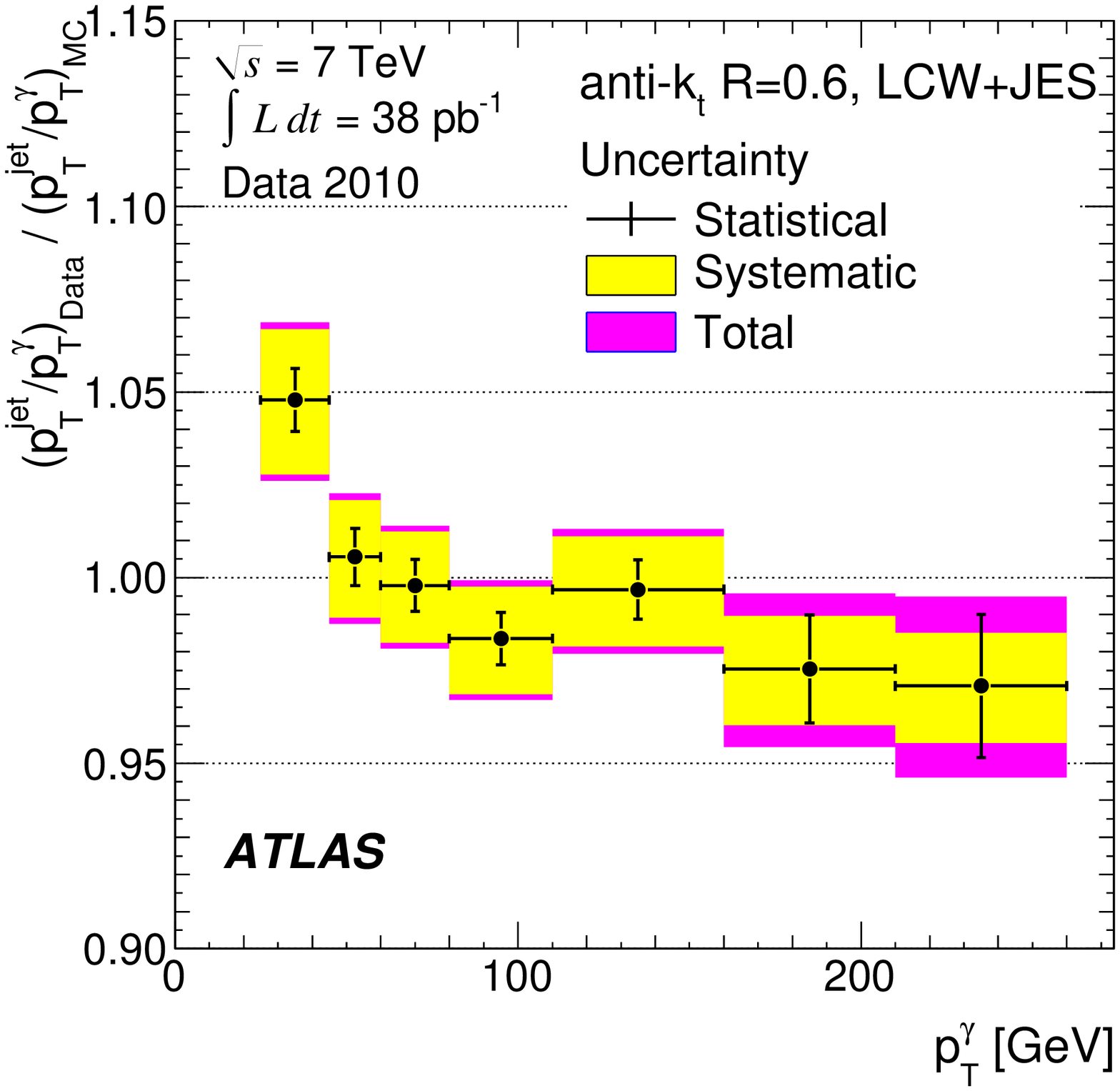}}
 \caption{Average jet response in data to the response in Monte Carlo simulation using the
    direct \pT{} balance technique of \antikt{} jets with $R  = 0.6$ calibrated with the
    \GCWJES{} (a) and \LCWJES{} (b) scheme as a function of
    photon transverse momentum. Statistical and systematic uncertainties (light band) are included with the total uncertainty
    shown as the dark band.}
  \label{fig:TotalErrors_DB}
\end{figure*}

\begin{figure*}[ht!!p]
  \centering
  \subfloat[\GCW]{\includegraphics[width=0.4\textwidth]{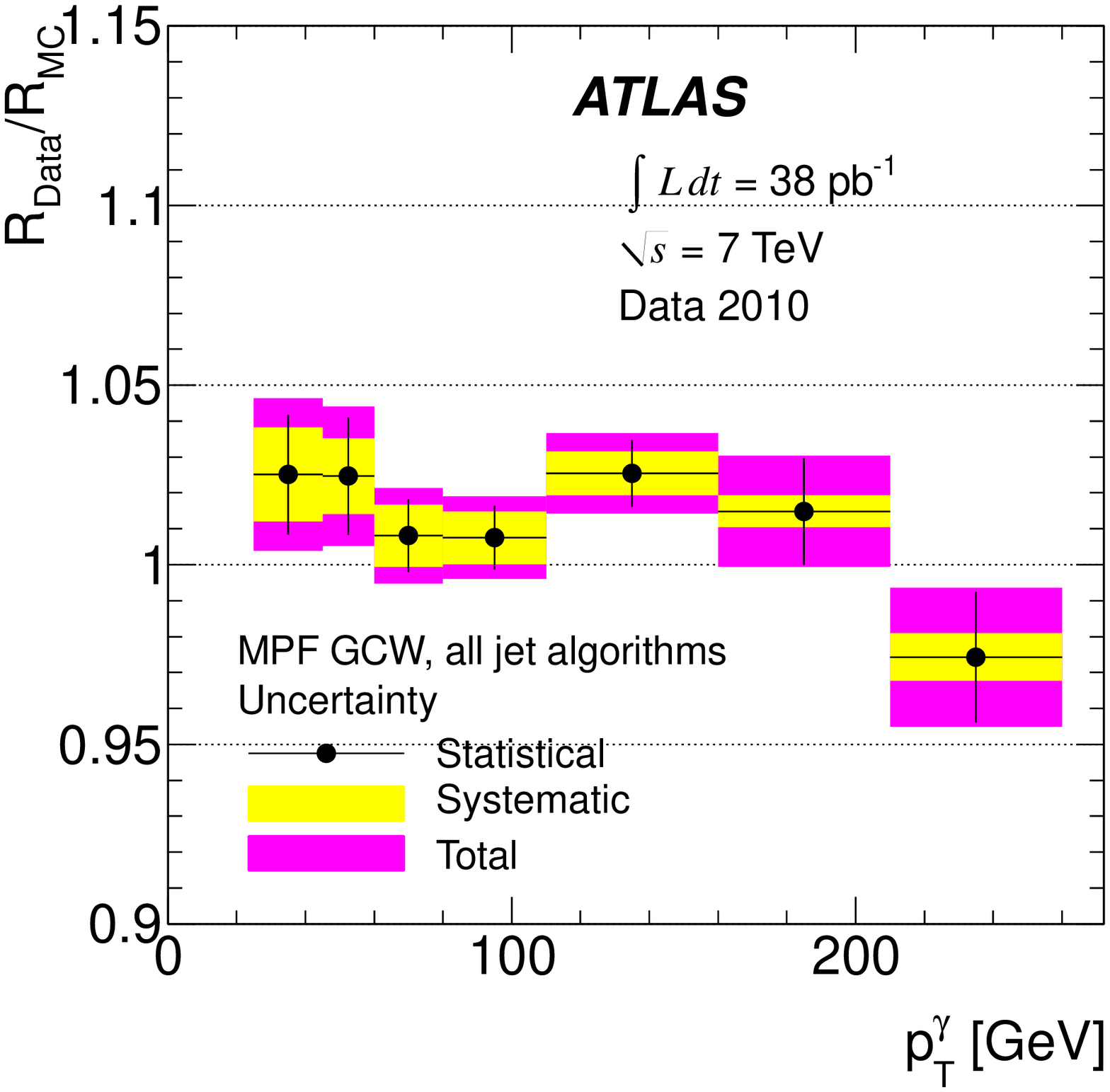}}
\hspace{1.cm}
  \subfloat[\LCW]{\includegraphics[width=0.4\textwidth]{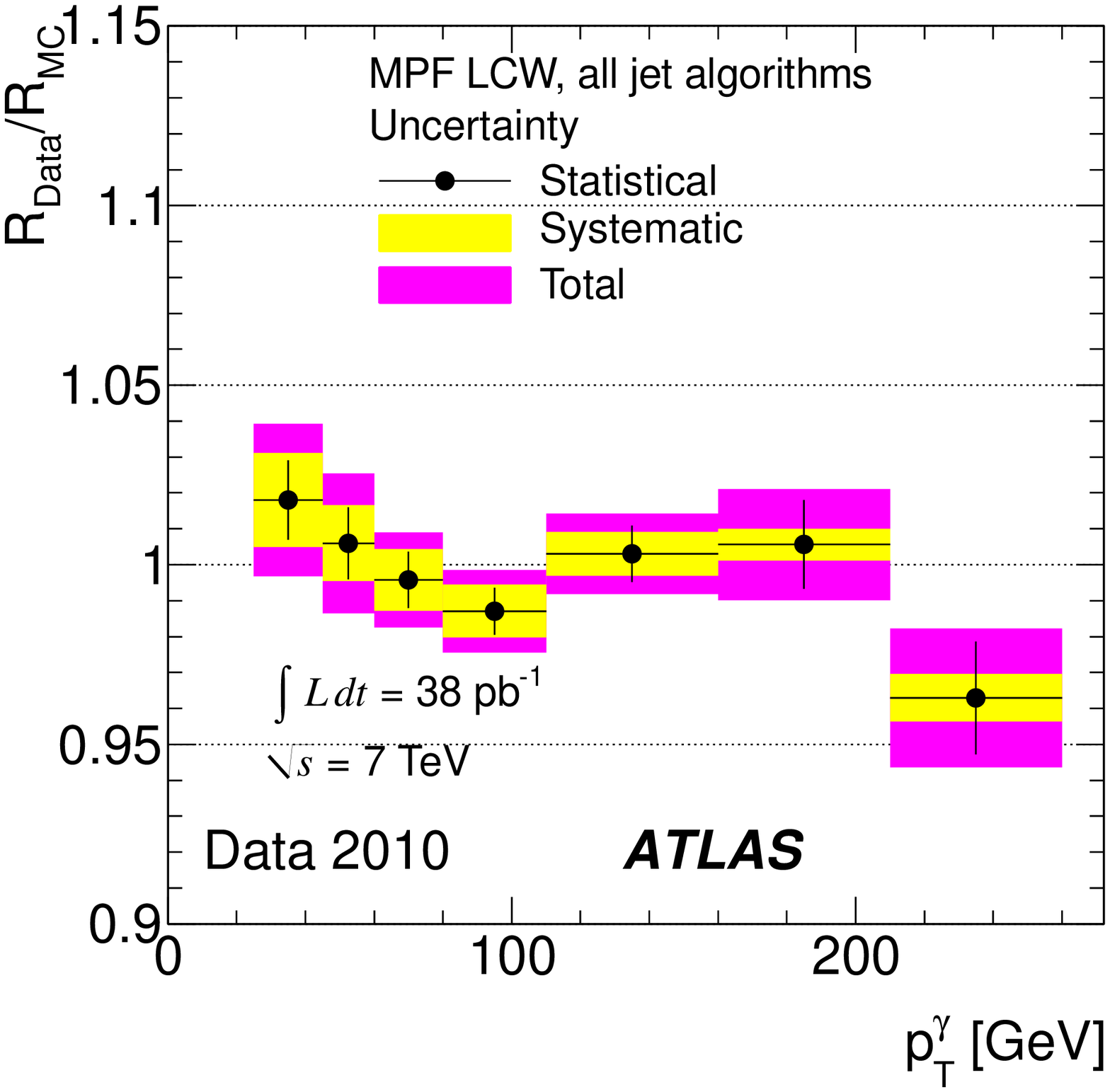}}
  \caption{The ratios of the \MPF{} calorimeter response in data to the response in  Monte Carlo simulation using the
    \MPF{} method for each input energy scale
    \GCW{} (a), and \LCW{} (b) as a function of the photon transverse momentum. 
    Statistical and systematic uncertainties (light band) are included. The total uncertainty is shown
    as the dark band.}
  \label{fig:TotalErrors_MPF}
\end{figure*}

\begin{figure*}[ht!!p]
\centering
\subfloat[\GCWJES]{\includegraphics[width=0.49\textwidth]{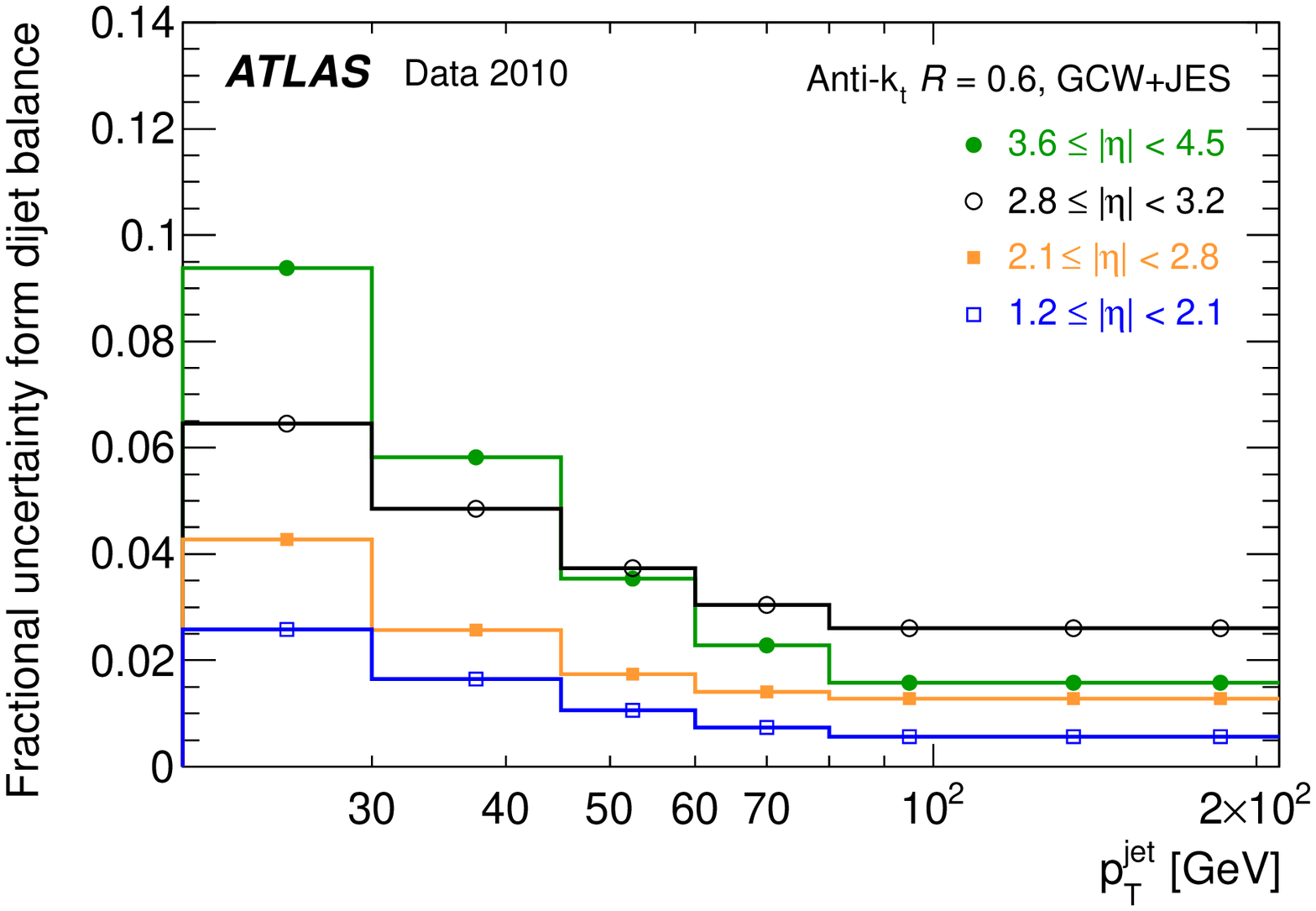}}
\subfloat[\LCWJES]{\includegraphics[width=0.49\textwidth]{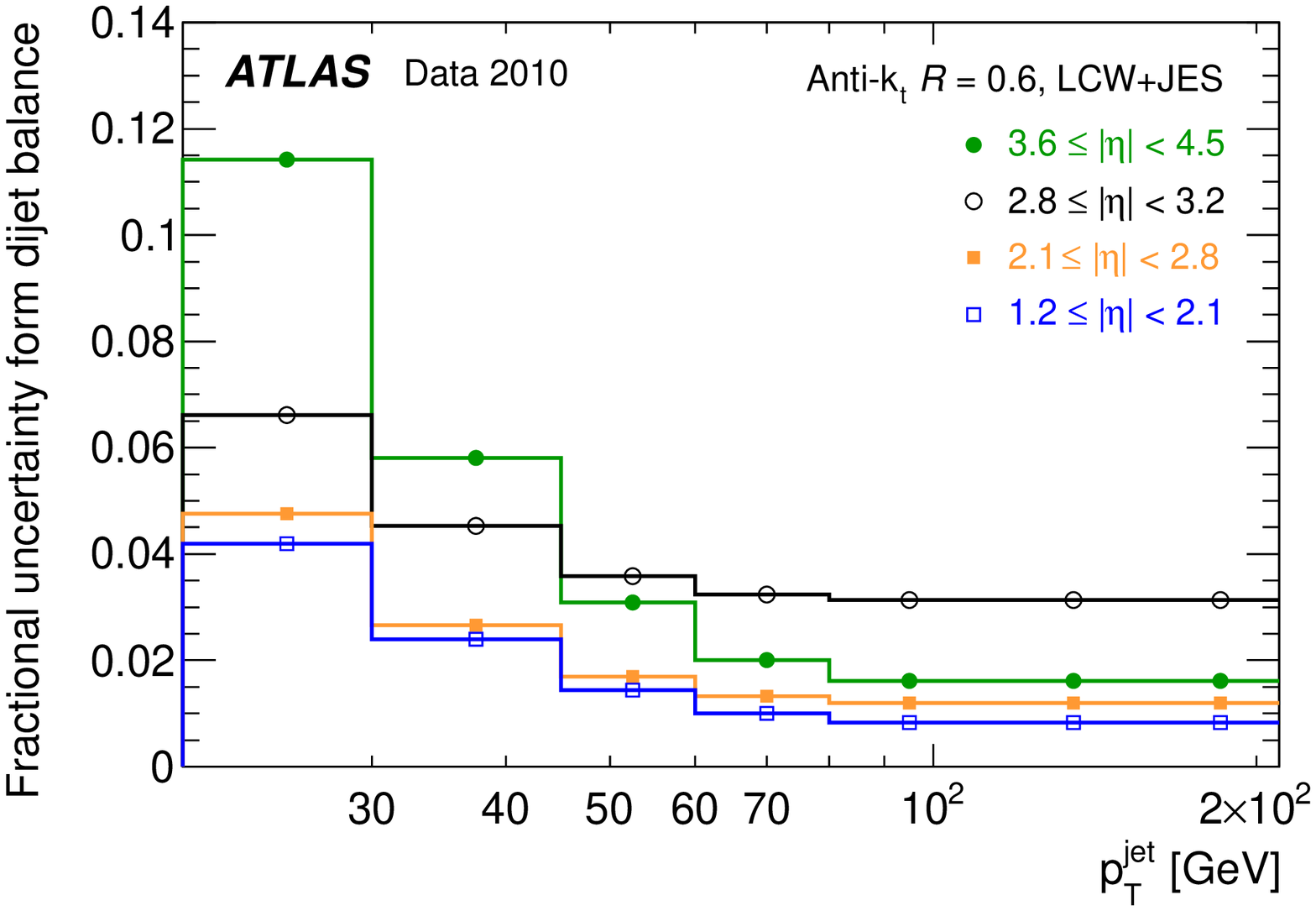}}
\caption{Uncertainty in the jet response obtained from the dijet $\eta$-intercalibration technique
for \antikt{} jets with $R = 0.6$ as a function of the jet \pt{} for various 
$|\eta|$-regions of the calorimeter.
The jets are calibrated with the \GCWJES{} (a) and the \LCWJES{} (b) calibration schemes.
Only statistical uncertainties are shown.}
\label{fig:etainterCellWeight}
\end{figure*}

%
\subsection{Jet energy scale uncertainty from \insitu{} techniques for jets based on cell weighting}
\label{sec:insituvalidationcellweighting}

For the jet calibration schemes based on cell weighting the \JES{} uncertainty
is evaluated using \insitu{} techniques. The same techniques as 
described in Section~\ref{sec:insituvalidation} are employed.
The final \JES{} uncertainty is obtained from a combination of all
\insitu{} techniques following the prescription in Section~\ref{sec:JesUncertaintyinsitu}.

\subsubsection{Comparison of transverse momentum balance from calorimeter and tracking}
\label{sec:trackjetcellweighting}
The result of the JES validation using the total transverse momentum of the tracks
associated to jets (see Section \ref{sec:trackjet}) is shown 
in Figure~\ref{fig:finalAktopo6gcw} for jets calibrated with the \GCWJES{} scheme 
and 
in Figure~\ref{fig:finalAktopol6lcw} for jets calibrated with the \LCWJES{} scheme
in various jet pseudorapidity regions within \AetaRange{2.1}.
The bin \AetaRange{1.2} is obtained by combining the \AetaRange{0.3}, \etaRange{0.3}{0.8} and \etaRange{0.8}{1.2} bins.

Similar results as for the \EMJES{} scheme are obtained.
In both cases, the agreement between data and simulation is excellent and within the uncertainties
of the \insitu{} method.  
The calibration schemes agree to within a few per cent, except for the bins with very low 
numbers of events.

\subsubsection{Photon-jet transverse momentum balance}
\label{sec:gammajetcellweighting}

The response measured by the direct \pt{} balance technique 
(see Section~\ref{sec:directbalance}) for the 
\GCWJES{} and \LCWJES{} calibrations is shown in Figure~\ref{fig:DBResponse}.
The agreement of the Monte Carlo simulation with data is similar
for both calibration schemes. The data to Monte Carlo agreement is $3$ to $5 \%$.  

Figure~\ref{fig:MPFResponse_GCW} shows the comparison of the response determined by the MPF technique
(see Section \ref{sec:MPF}), measured in data and Monte Carlo simulation at the \GCW{} and \LCW{} 
jet energy scales. 
To calculate the response using the \MPF{} technique at these energy scales the \Etmiss{} 
is calculated using \GCW{} or \LCW{} calibrated \topos{} as an input\footnote{
For the \GCW{} calibration scheme the cell energies in the \topos{} are multiplied by the
cell energy weights described in Section~\ref{sec:JetCalibSchemeGCW}.}. 
All the JES calibrations are found to be consistent
between data and Monte Carlo simulation to within $3$ to $4 \%$. 

The ratios of jet response in data to the response in  Monte Carlo simulation using the
direct \pT{} balance technique for the \GCWJES{}  and \LCWJES{}  jet calibration schemes 
as a function of the photon transverse momentum
are shown in Figure \ref{fig:TotalErrors_DB}. The agreement of data and Monte Carlo simulation
is within $5 \%$ and is compatible with unity within the statistical and systematic uncertainties.
A similar result for the \MPF{} technique is shown in Figure~\ref{fig:TotalErrors_MPF}.
Good agreement between data and Monte Carlo simulation is found.

%
\begin{figure*}[htp!]
  \centering
\subfloat[\GCWJES]{\includegraphics[width=0.49\textwidth]{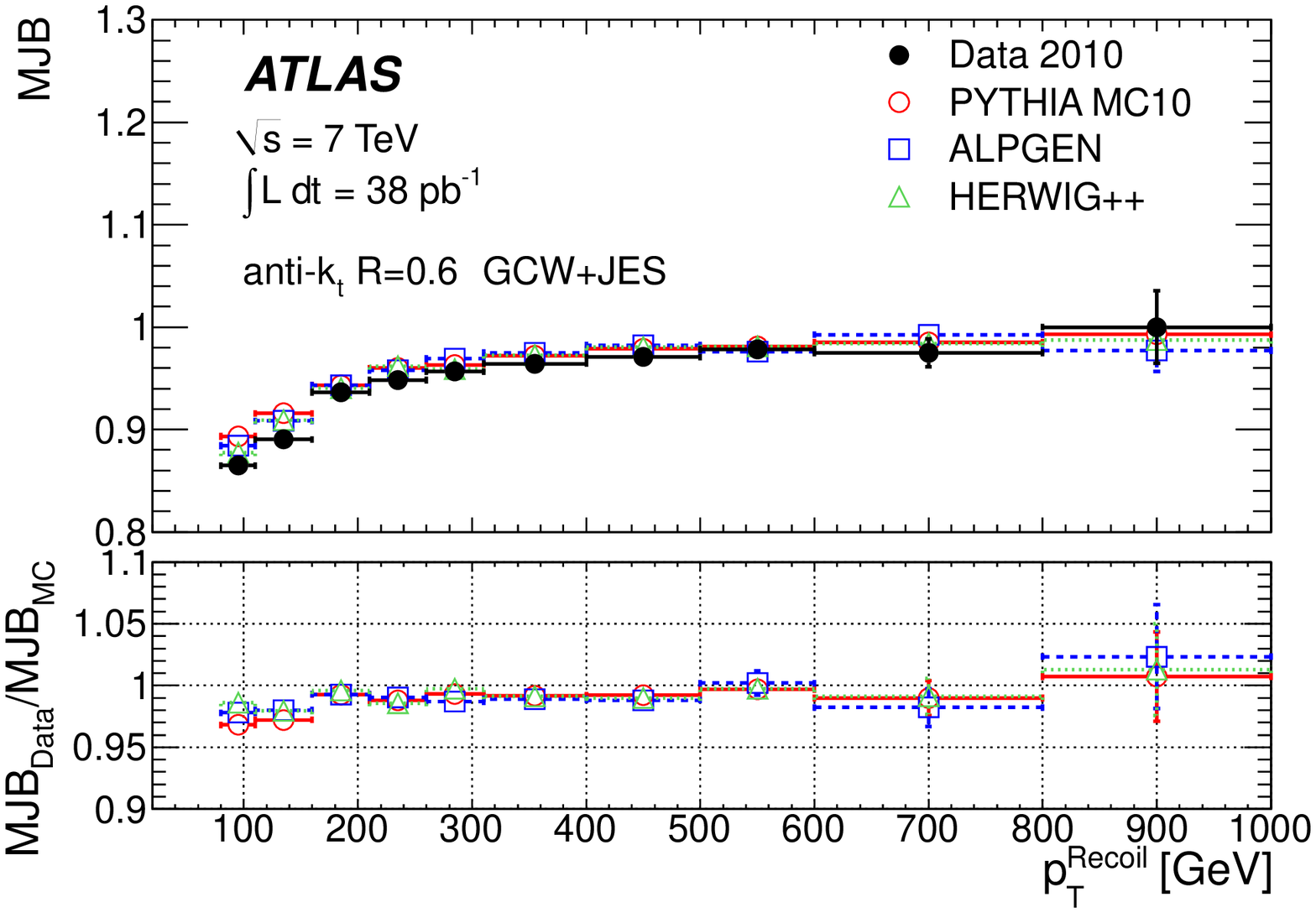}}\hfill
\subfloat[\LCWJES]{\includegraphics[width=0.49\textwidth]{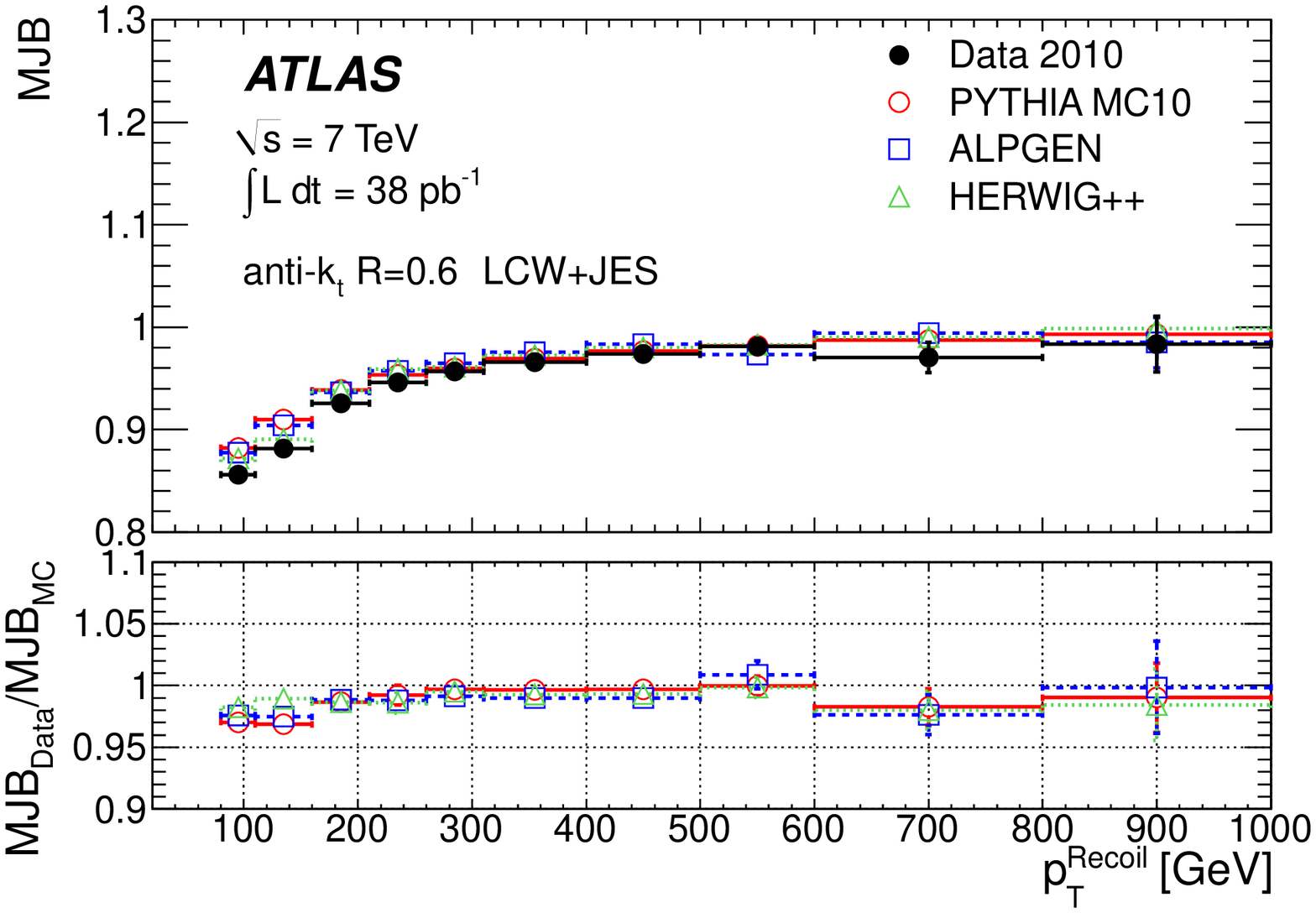} }\\
  \caption{Multijet balance MJB as a function of the recoil system \pt{} 
for data and Monte Carlo simulation for the  \antikt{} algorithm with $R = 0.6$ using the \GCWJES{} (a) and 
\LCWJES{} (b) calibration scheme.
Only statistical uncertainties are shown.}
  \label{fig:PtBalanceCellWeighting}
\end{figure*}

\begin{figure*}[htp!]
\centering
 \subfloat[\GCWJES]{\includegraphics[width=0.49\textwidth]{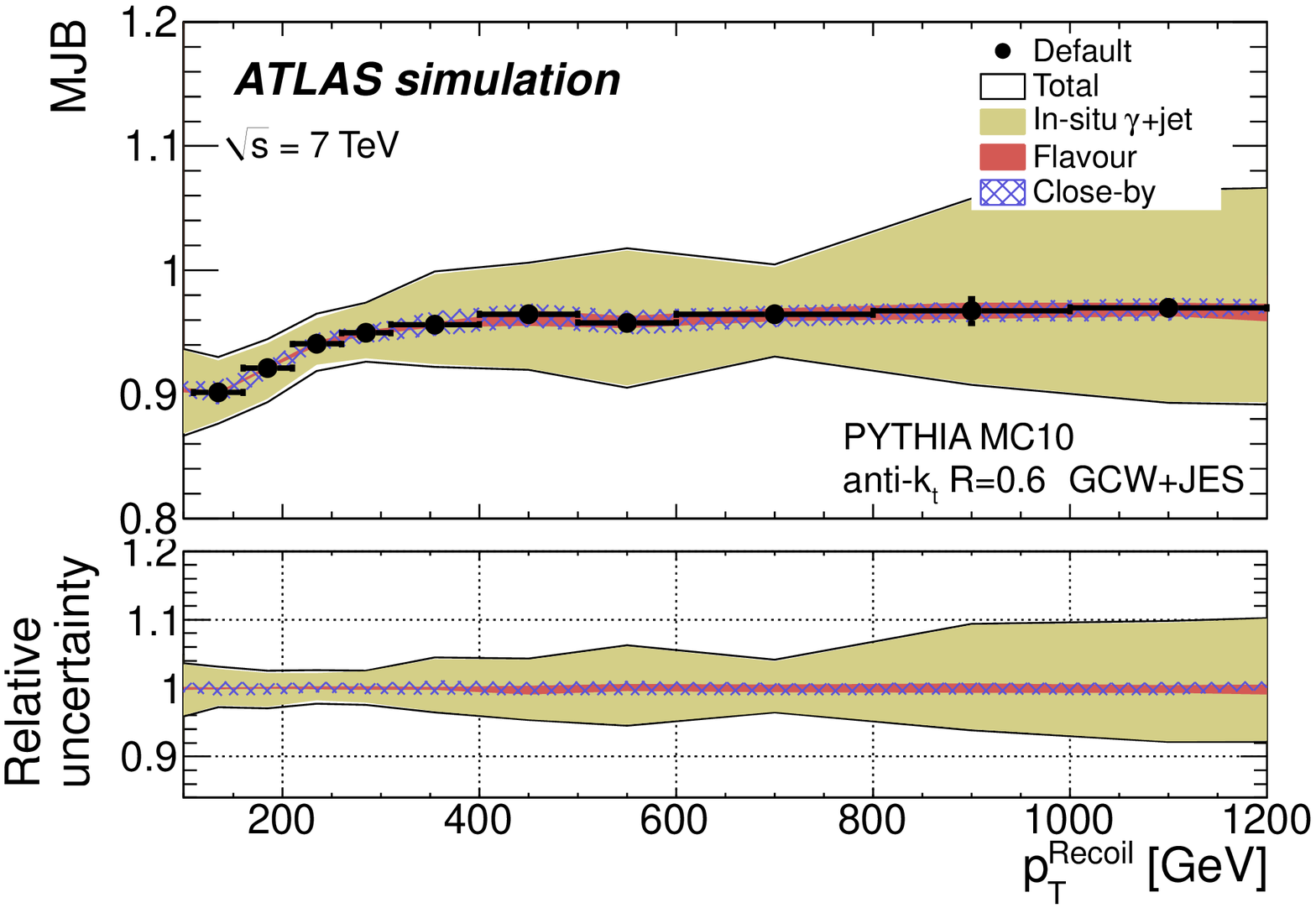}}\hfill
 \subfloat[\LCWJES]{\includegraphics[width=0.49\textwidth]{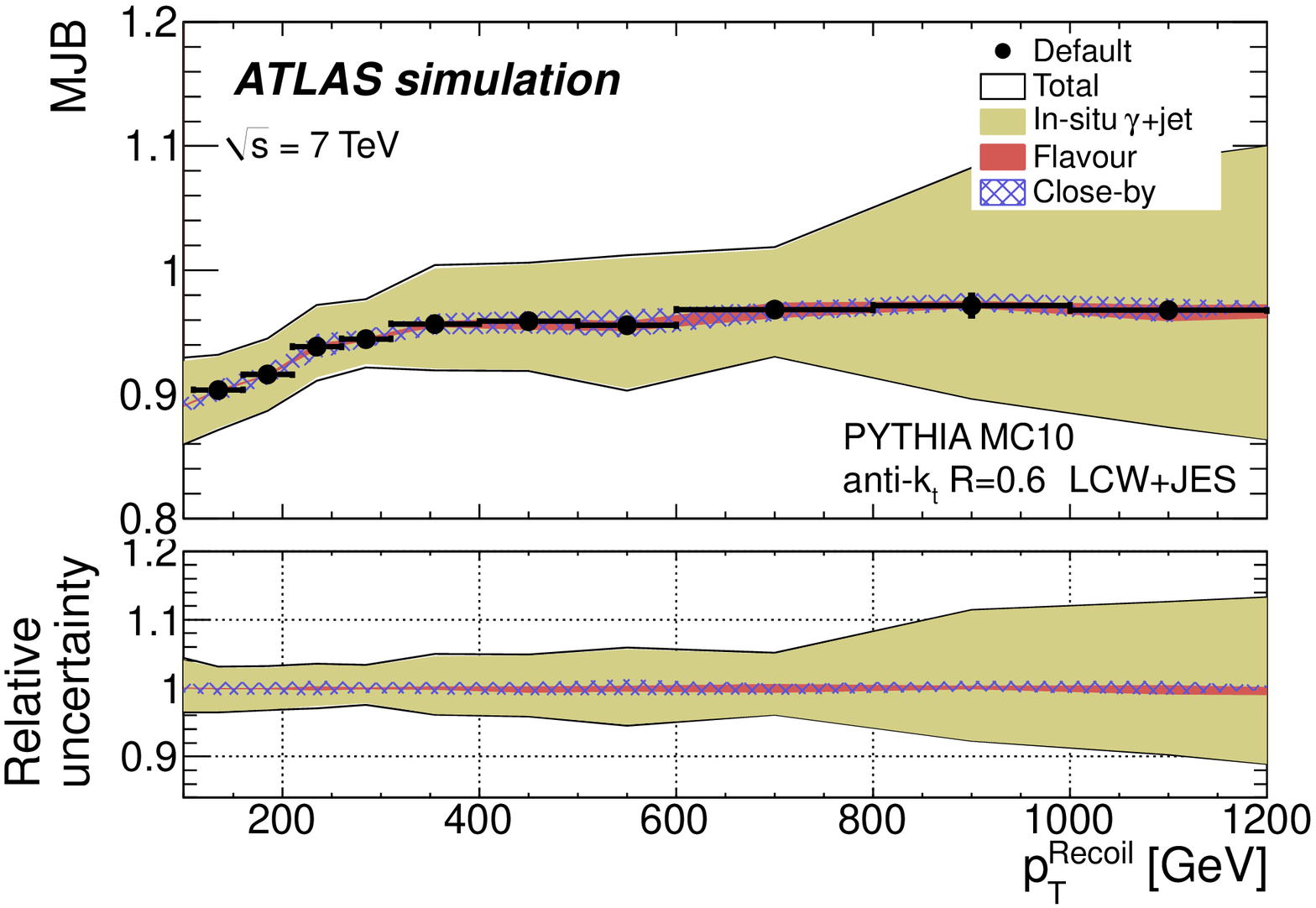}}\\
\caption{The multijet balance ${\rm MJB}_{\rm MC}$ as a function of the recoil system \pt{}  (full dots) 
for \antikt{} jets with $R = 0.6$ using the \GCWJES{} (a) and \LCWJES{} (b) calibration schemes. 
The three bands are defined by the maximum shift of MJB  
when the jets that compose the recoil system are shifted up and down by the \JES{} uncertainty 
determined from the \gammajet{} balance, 
close-by jet and flavour uncertainties. The black lines show the total uncertainty obtained 
by adding in quadrature the individual uncertainties. 
The lower part of the figure shows the relative uncertainty due to the scale uncertainty of the jets 
that compose the recoil system, defined as the maximum relative shift with respect to the nominal value, 
as a function of \ptRecoil. } 
\label{fig:PtRecoilUncCellWeighting}
\end{figure*}

\begin{figure*}[htp!]
  \centering
\subfloat[\GCWJES]{\includegraphics[width=0.49\textwidth]{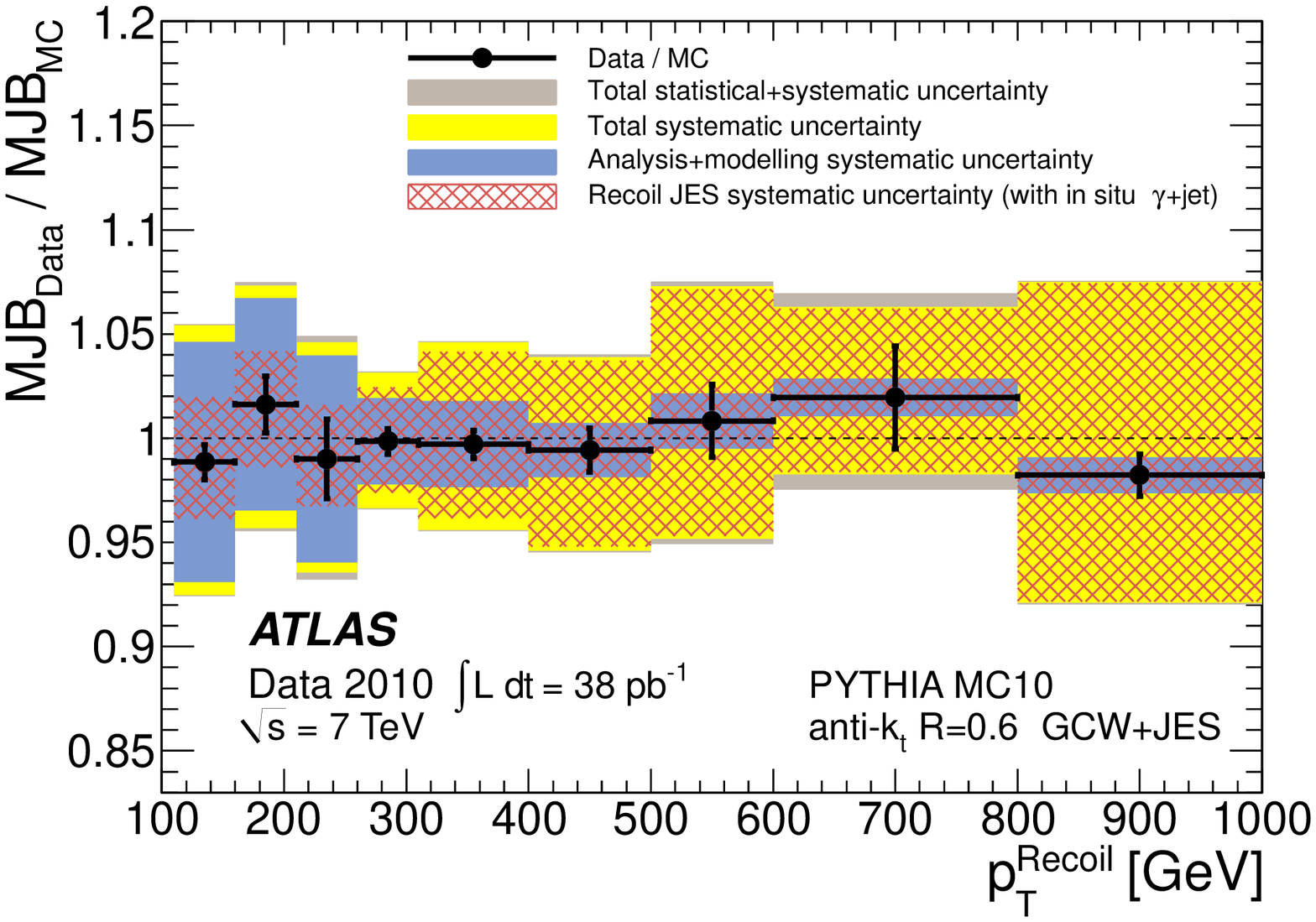}}\hfill
\subfloat[\LCWJES]{\includegraphics[width=0.49\textwidth]{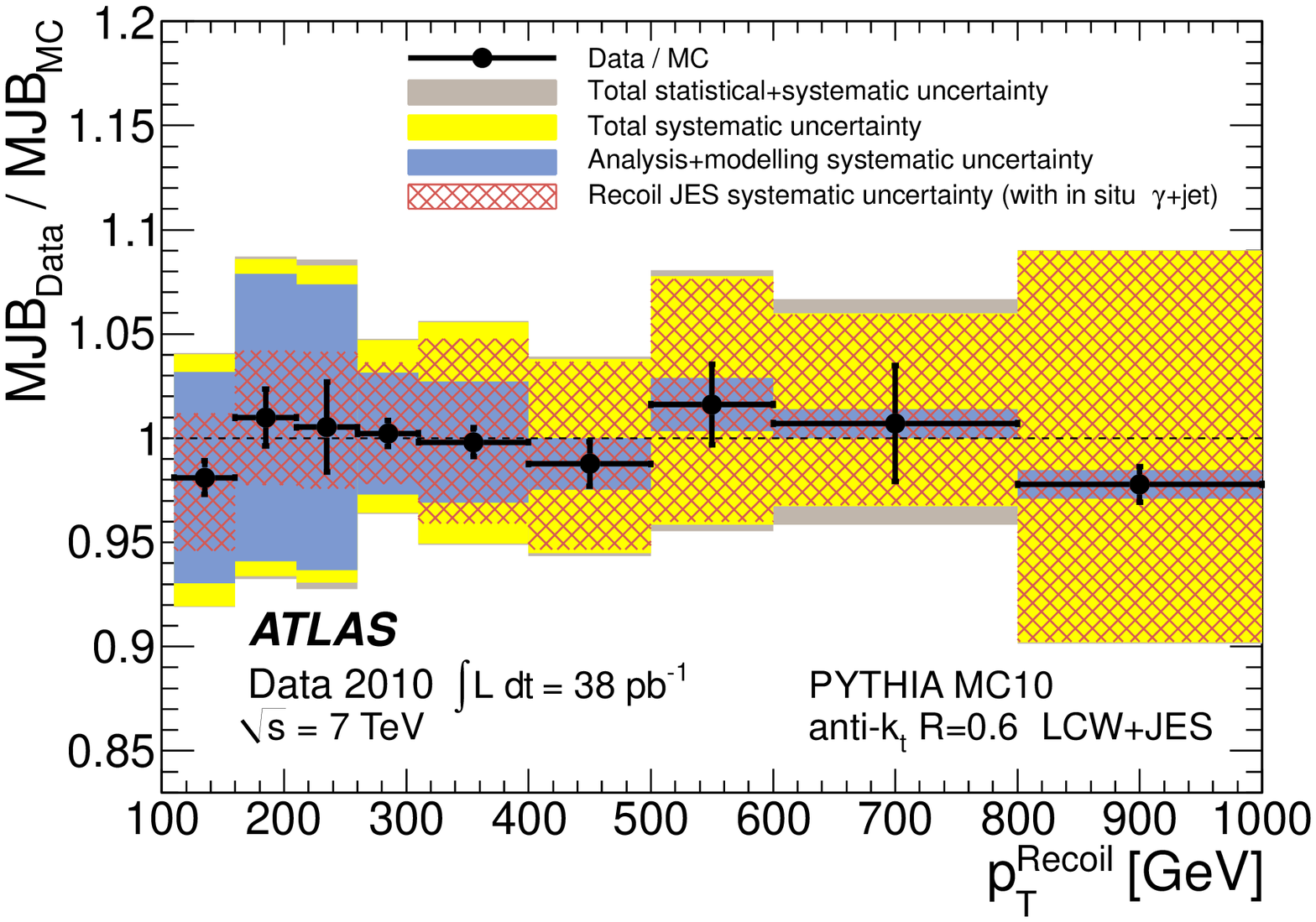}}\\
  \caption{Ratio of the data to MC for the multijet balance as a function of the recoil system \pt{} for \antikt{} jets 
with $R = 0.6$ using the \GCWJES{} (a) and \LCWJES{} (b) calibration schemes. 
The various shaded regions show the total uncertainty (dark band) 
obtained as the squared sum of the total systematic uncertainty (light band) 
and of the statistical uncertainty (error bars). 
Also displayed are the contributions to the systematic uncertainty due to analysis cuts 
and event modelling (darkest band) and to the jet energy scale for jets in the recoil system (hatched band).} 
    \label{fig:FinalResultCellWeighting}
\end{figure*}

\subsubsection{Intercalibration of forward jets using events with dijet topologies}
\label{sec:dijetcellweighting}
The transverse momentum balance in events with only two jets at high transverse
energy can be used to determine the \JES{} uncertainty for jets in the 
forward detector region.
The matrix method, described in Section~\ref{sec:etaintercalibration},
is used in order to test the performance of the \GCWJES{} and \LCWJES{} calibrations for jets with $|\eta|>1.2$
and to determine the \JES{} uncertainty in the forward region 
based on the well calibrated jet in the central reference region.

The same selection and method as for the test of the \EMJES{} calibration is applied, 
with two exceptions: 
the reference region is defined by $|\etaDet| < 1.2$ instead of $|\etaDet| < 0.8$, 
and a fit is applied to smooth out statistical fluctuations. 
The \JES{} uncertainty in the reference regions is obtained from the $\gamma$-jet results
(see Section~\ref{sec:gammajetcellweighting})
and using the sum of track momenta (Section~\ref{sec:trackjetcellweighting}).

Figure~\ref{fig:etainterCellWeight} shows the resulting uncertainties as a function 
of \ptjet{} for various $\eta$-bins. The uncertainty is taken as the RMS 
spread of the relative response from the Monte Carlo predictions around the relative response measured 
in data (see Section~\ref{sec:EtaIntercalibUnc}).
The \JES{} uncertainty introduced by the dijet balance is largest at lower \ptjet{} and smallest at higher \ptjet.
For $\ptjet > 100$~\GeV{} the \JES{} uncertainty for the \GCWJES{} scheme
is less than $1 \%$ for \etaRange{1.2}{2.1} and
about $2.5 \%$ for \etaRange{2.8}{3.2}.
For $\ptjet = 20$~\GeV{} the \JES{} uncertainty 
is about $2 \%$ for \etaRange{1.2}{2.1} and
about $9.5 \%$ for \etaRange{3.6}{4.5}.

The \JES{} uncertainties for the \LCWJES{} calibration sche\-me are slightly larger than those
for \GCWJES{} scheme.

%

%

\begin{figure*}[ht!p]
\begin{center}
\subfloat[\GCWJES]{\includegraphics[width=0.49\textwidth]{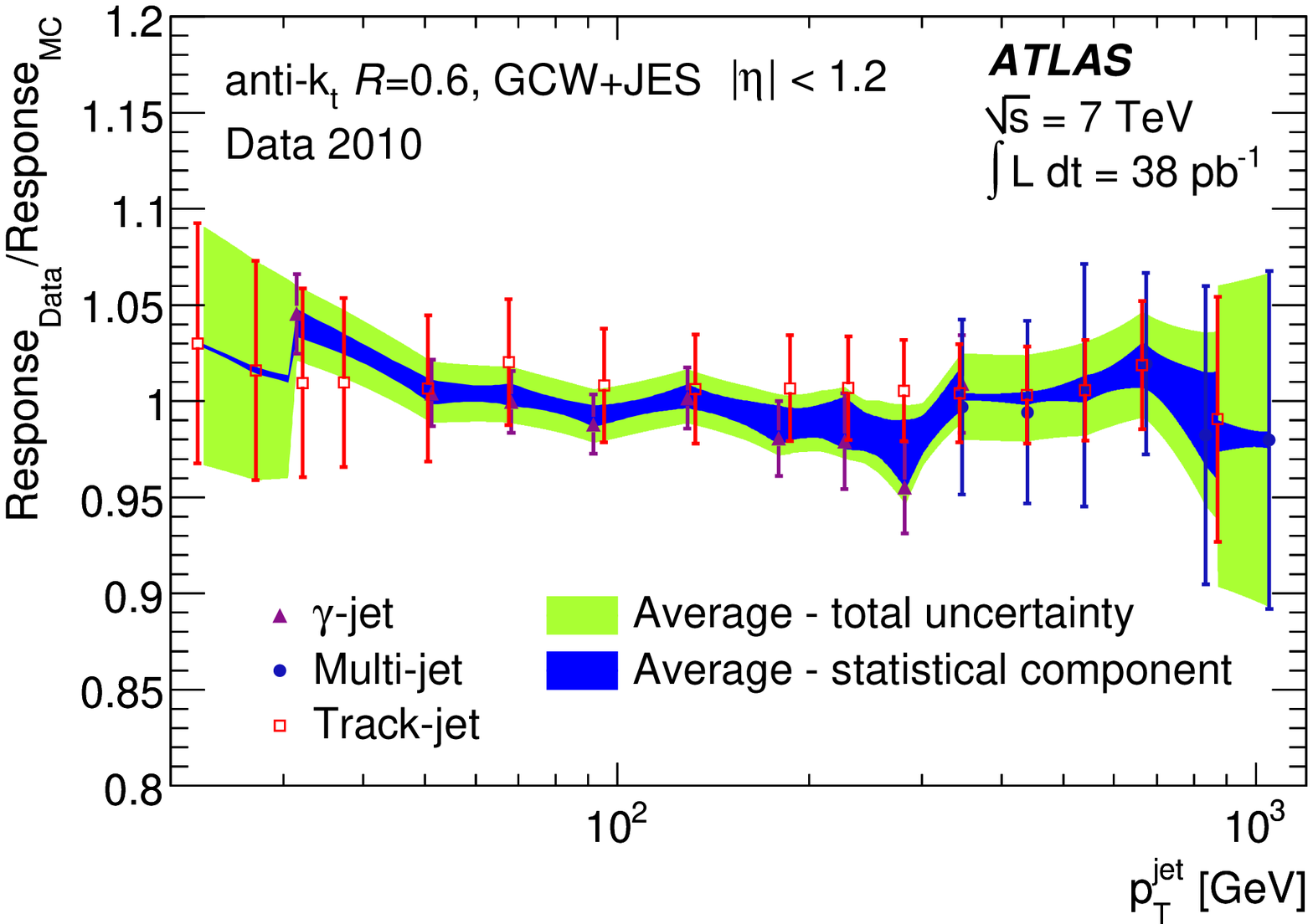}}
\subfloat[\LCWJES]{\includegraphics[width=0.49\textwidth]{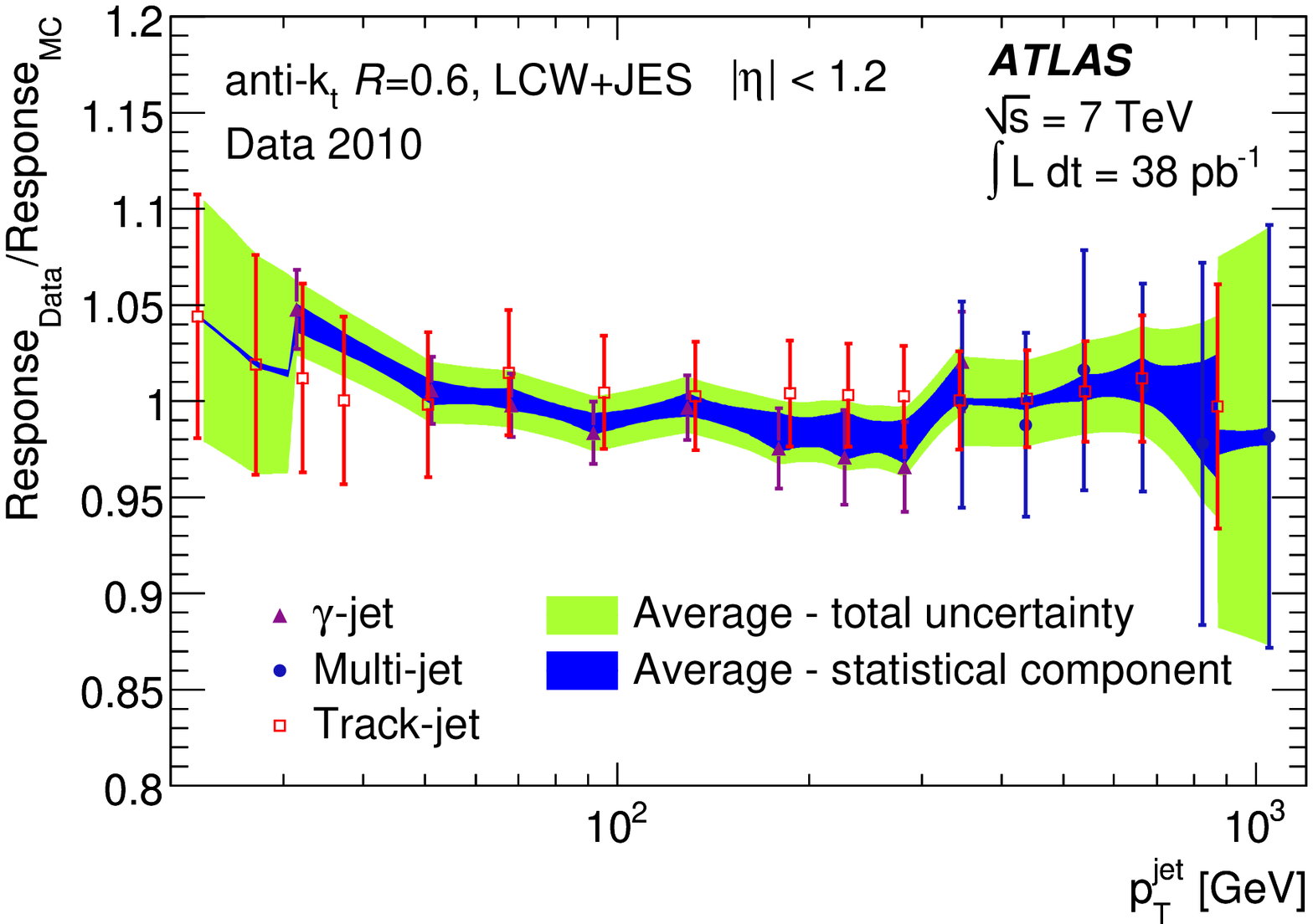}}
\caption{Jet response ratio of the data to the Monte Carlo simulation  as a function of \ptjet{}
for three \insitu{} techniques using as reference objects:
photons (\gammajet), a system of low energetic jets (multijet)
or the transverse momentum of all tracks associated to jets
(tracks in jets). The error bar displays the statistical and systematic
uncertainties added in quadrature.
Shown are the results for \antikt{} jets with $R = 0.6$
calibrated with the \GCWJES{} (a) and \LCWJES{} (b) calibration schemes.
The light band indicates the combination of the \insitu{} techniques.
The inner dark band shows the fraction due to the statistical uncertainty. 
\label{fig:responseratioinsitucellweighting}}
\end{center}
\end{figure*}

\begin{figure*}[ht!p]
\begin{center}
\subfloat[\GCWJES]{\includegraphics[width=0.49\textwidth]{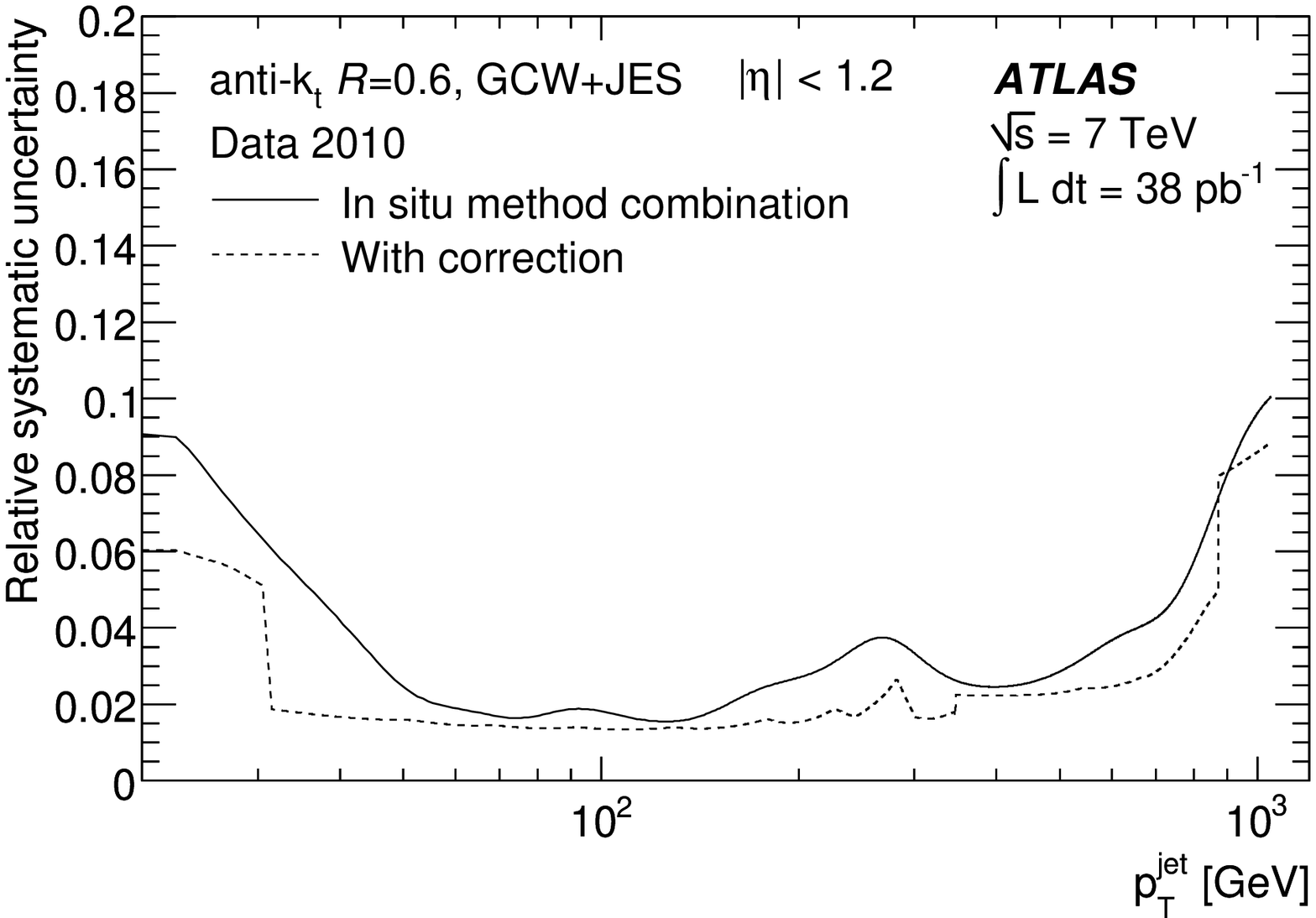}}
\subfloat[\LCWJES]{\includegraphics[width=0.49\textwidth]{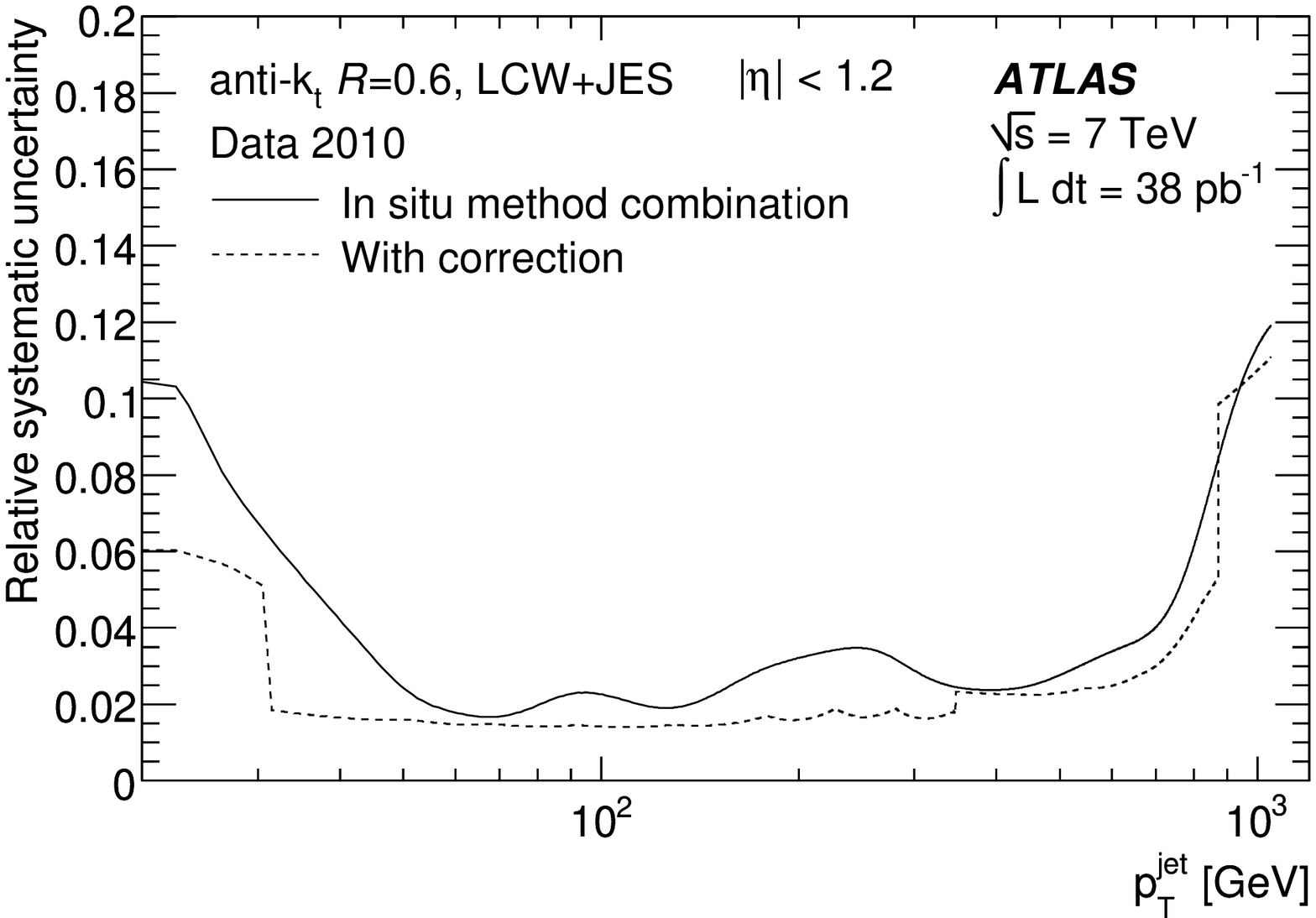}}
\caption{Jet energy scale uncertainty (solid line) as a function of \ptjet{}
for \antikt{} jets with $R = 0.6$ for $|\etajet| < 1.2$ calibrated with the
\GCWJES{} (a) and the \LCWJES{} (b) calibration scheme.
The dashed line shows the JES uncertainty that could have been achieved,
if \insitu{} techniques had been used to recalibrate the jets.
\label{fig:uncertaintyinsitucellweighting}}
\end{center}
\end{figure*}

\begin{figure*}[ht!!!p]
\begin{center}
\vspace{-0.4cm}
\subfloat[$|\etajet| < 0.3$]  {\includegraphics[width=0.39\textwidth]{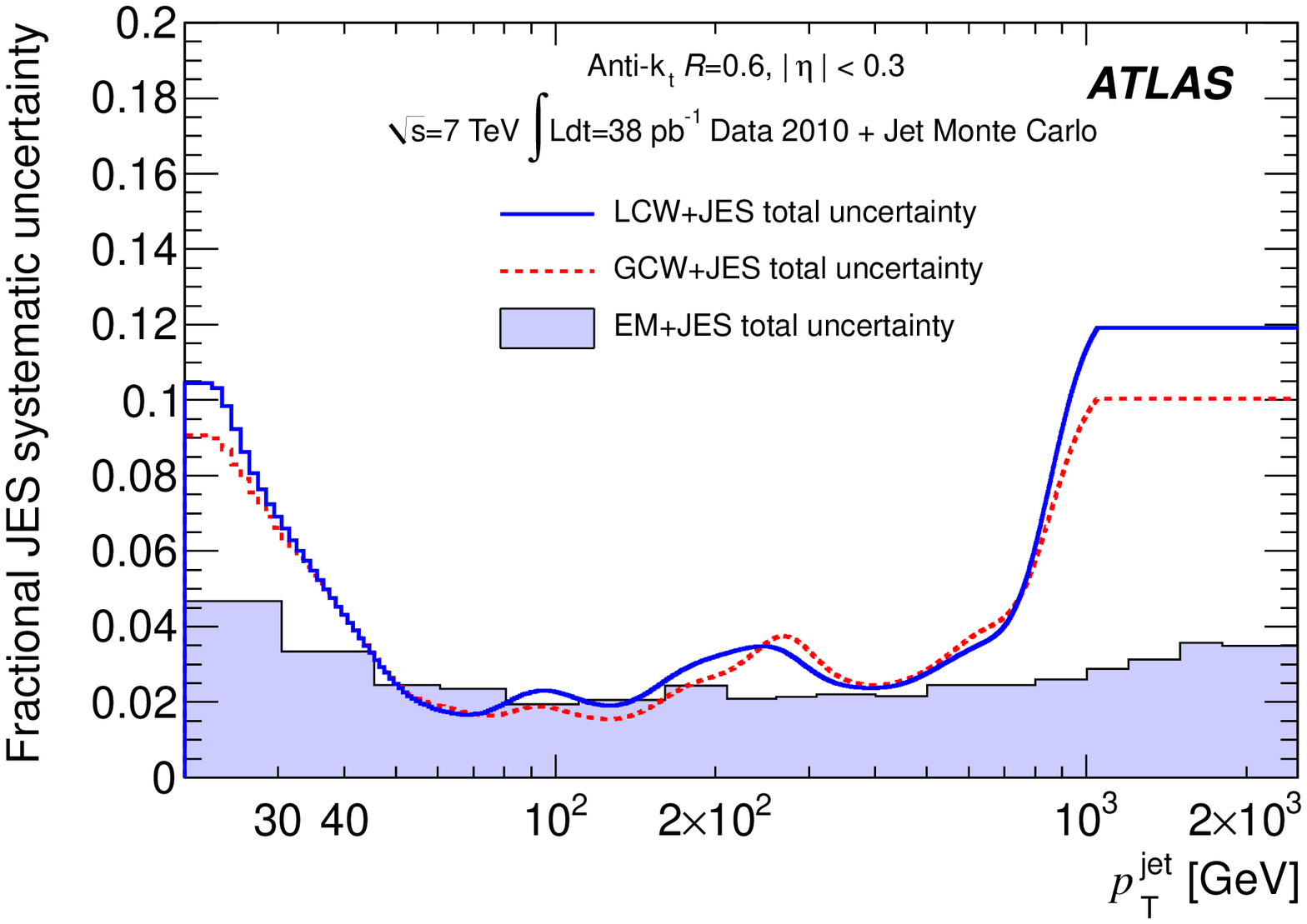}}
\subfloat[\etaRange{0.3}{0.8}]{\includegraphics[width=0.39\textwidth]{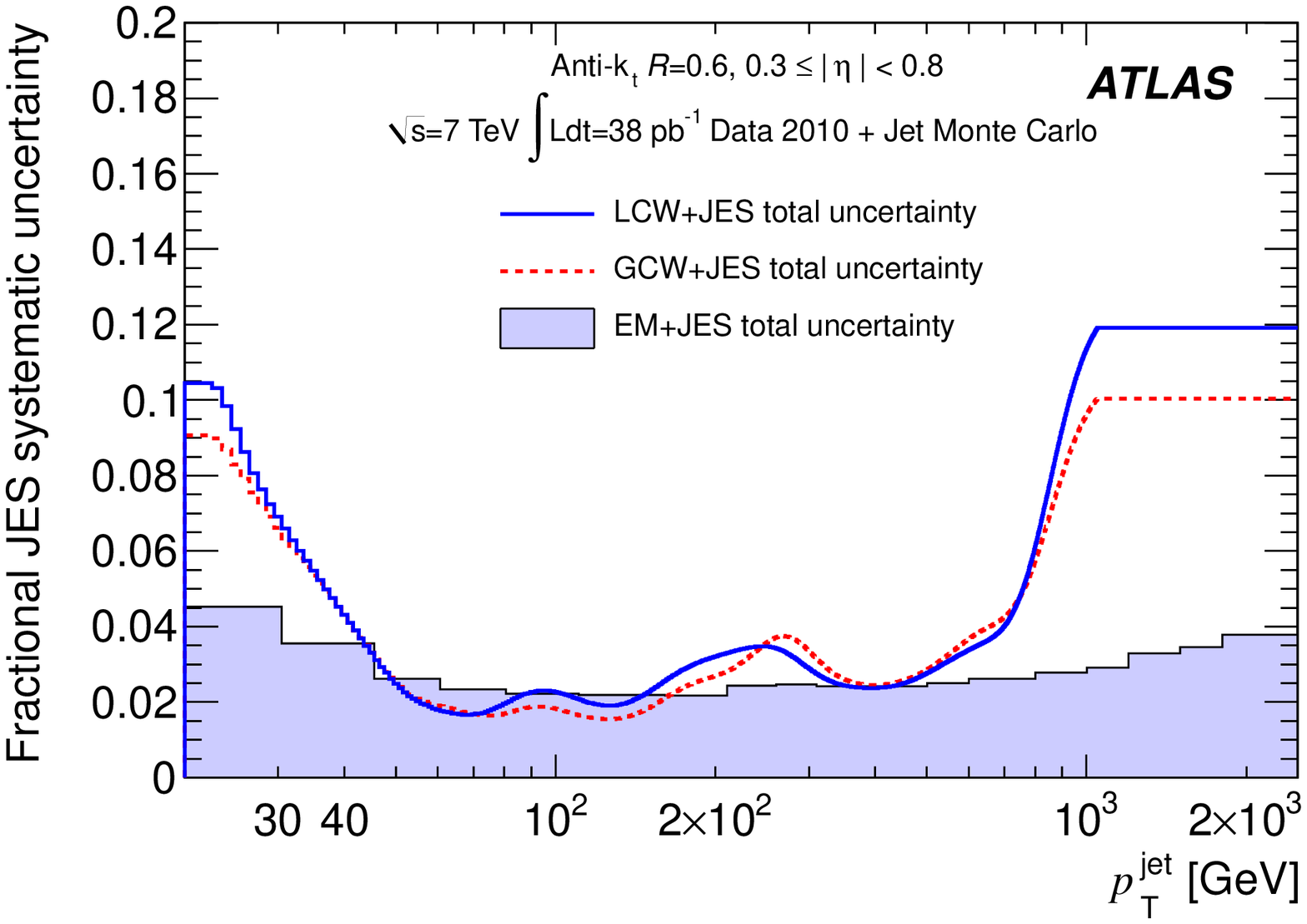}}
\vspace{-0.2cm}
\\
\subfloat[\etaRange{0.8}{1.2}]{\includegraphics[width=0.39\textwidth]{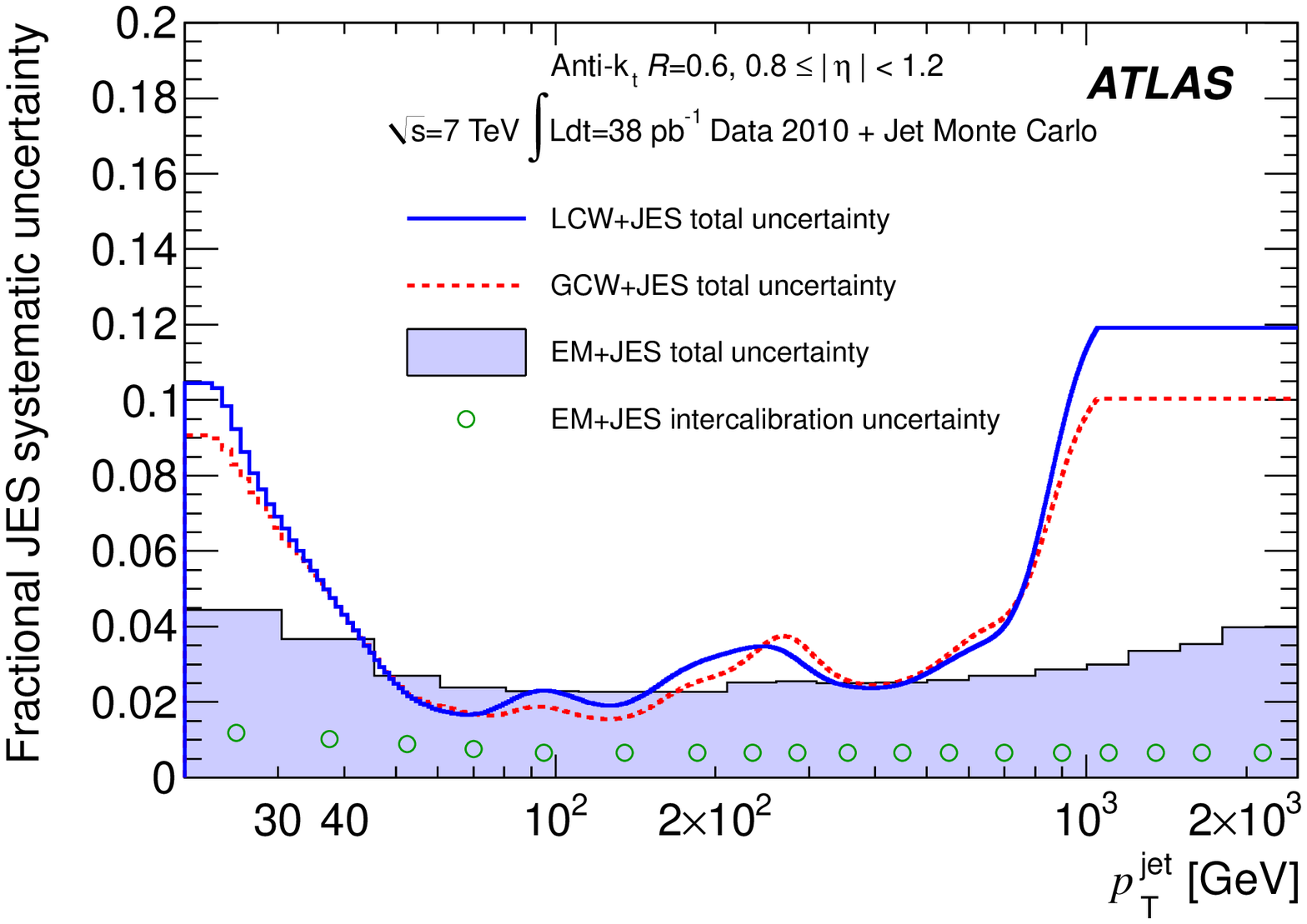}}
\subfloat[\etaRange{1.2}{2.1}]{\includegraphics[width=0.39\textwidth]{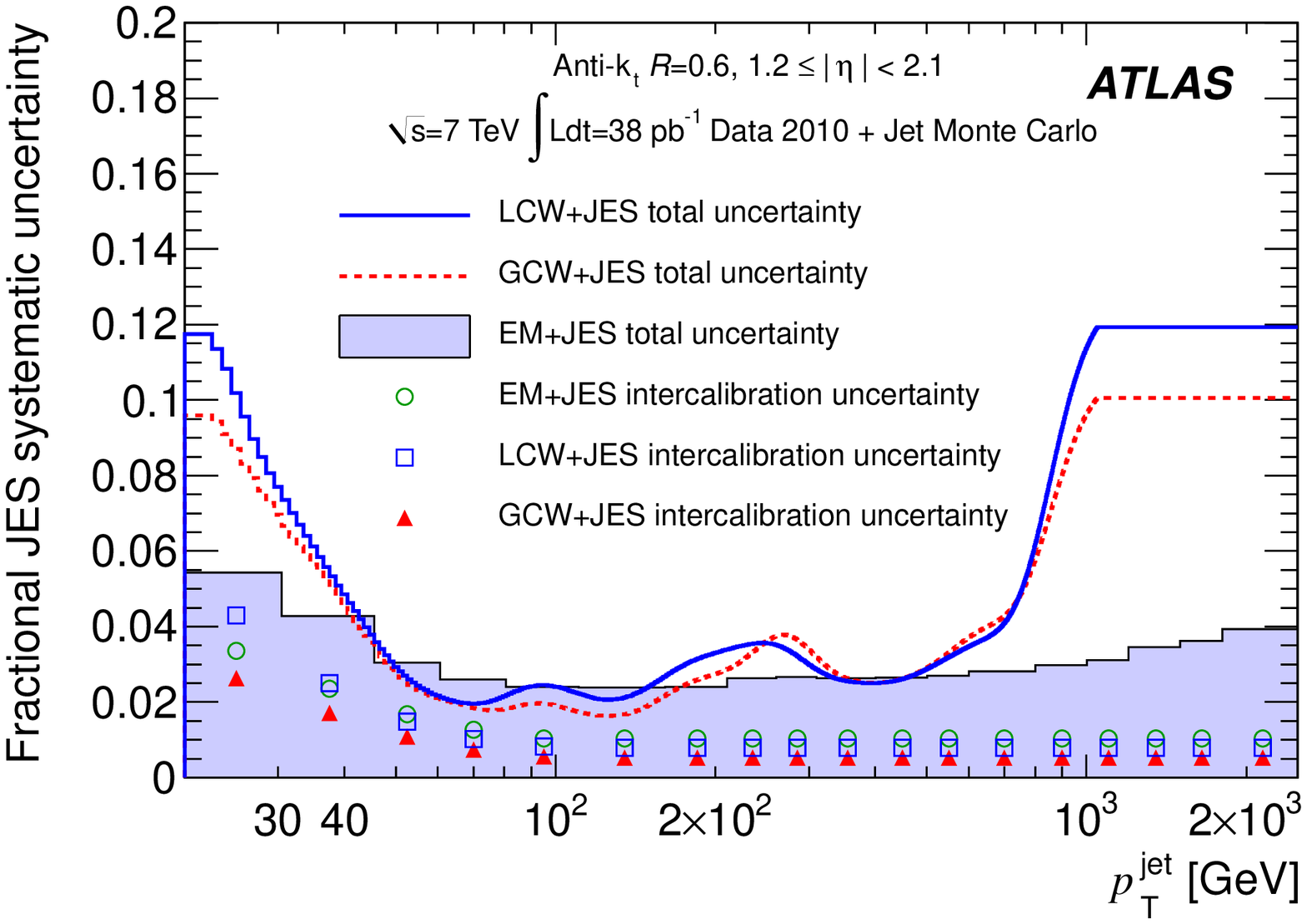}}
\vspace{-0.2cm}
\\
\subfloat[\etaRange{2.1}{2.8}]{\includegraphics[width=0.39\textwidth]{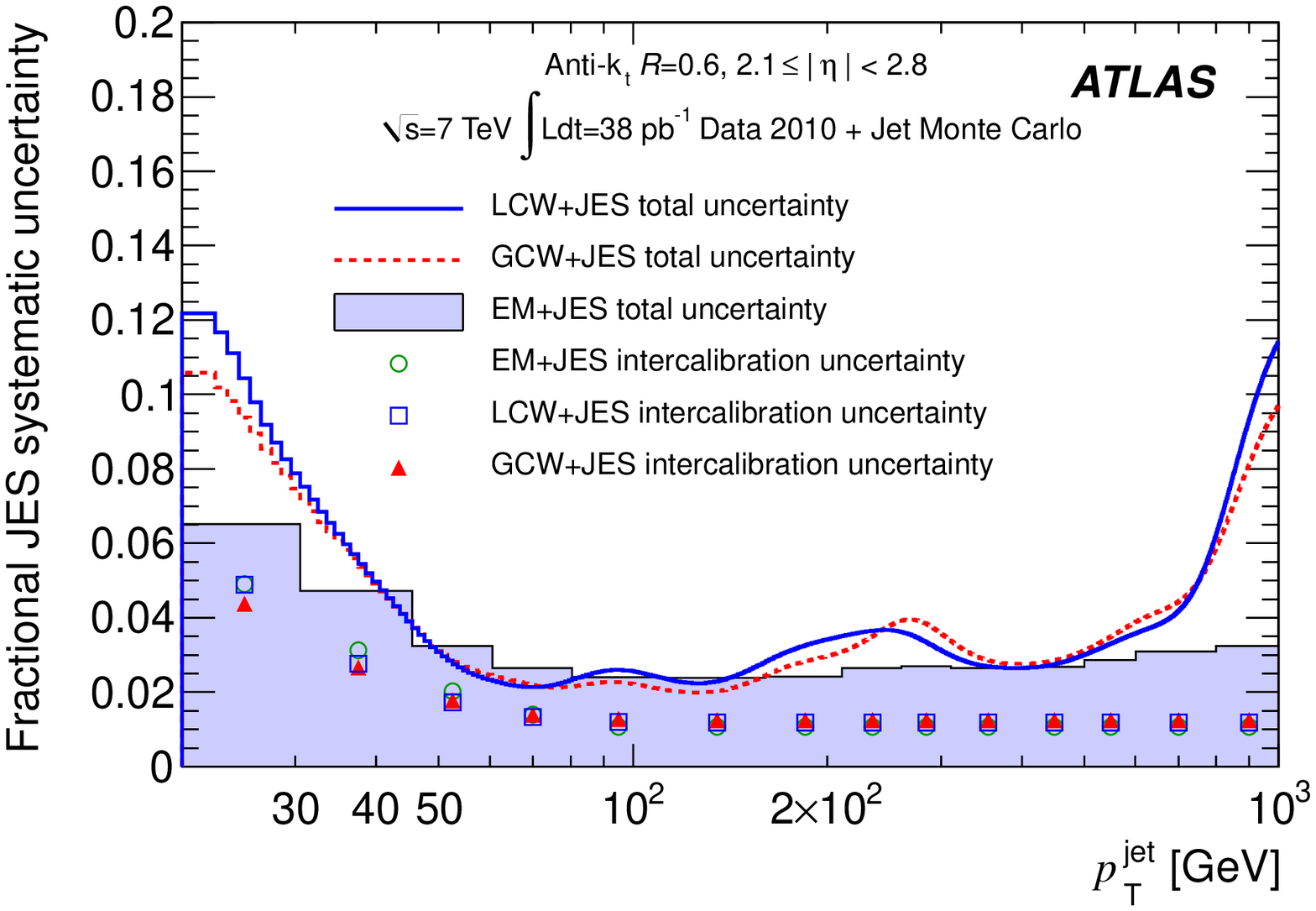}}
\subfloat[\etaRange{2.8}{3.2}]{\includegraphics[width=0.39\textwidth]{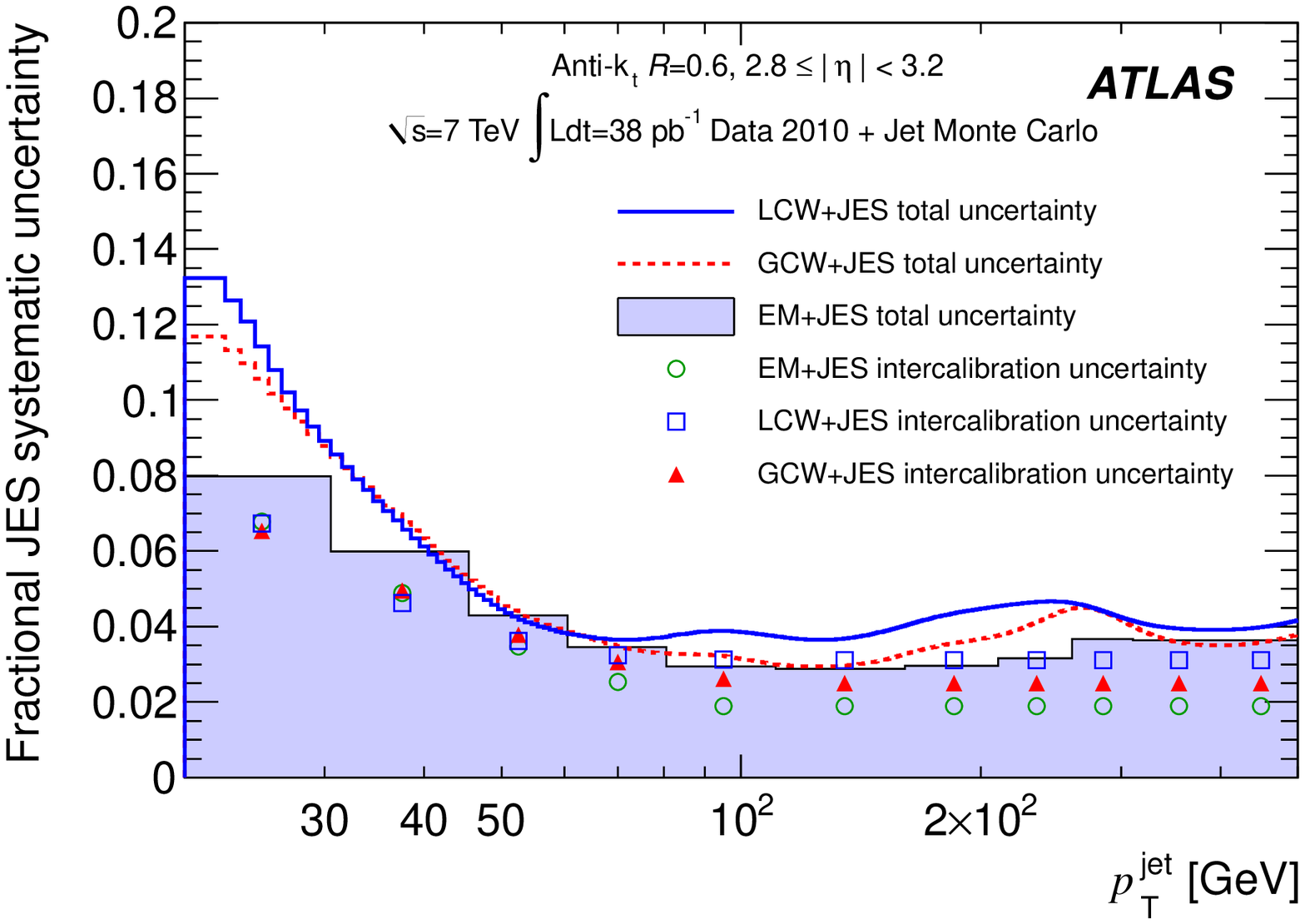}}
\vspace{-0.2cm}
\\
\subfloat[\etaRange{3.2}{3.6}]{\includegraphics[width=0.39\textwidth]{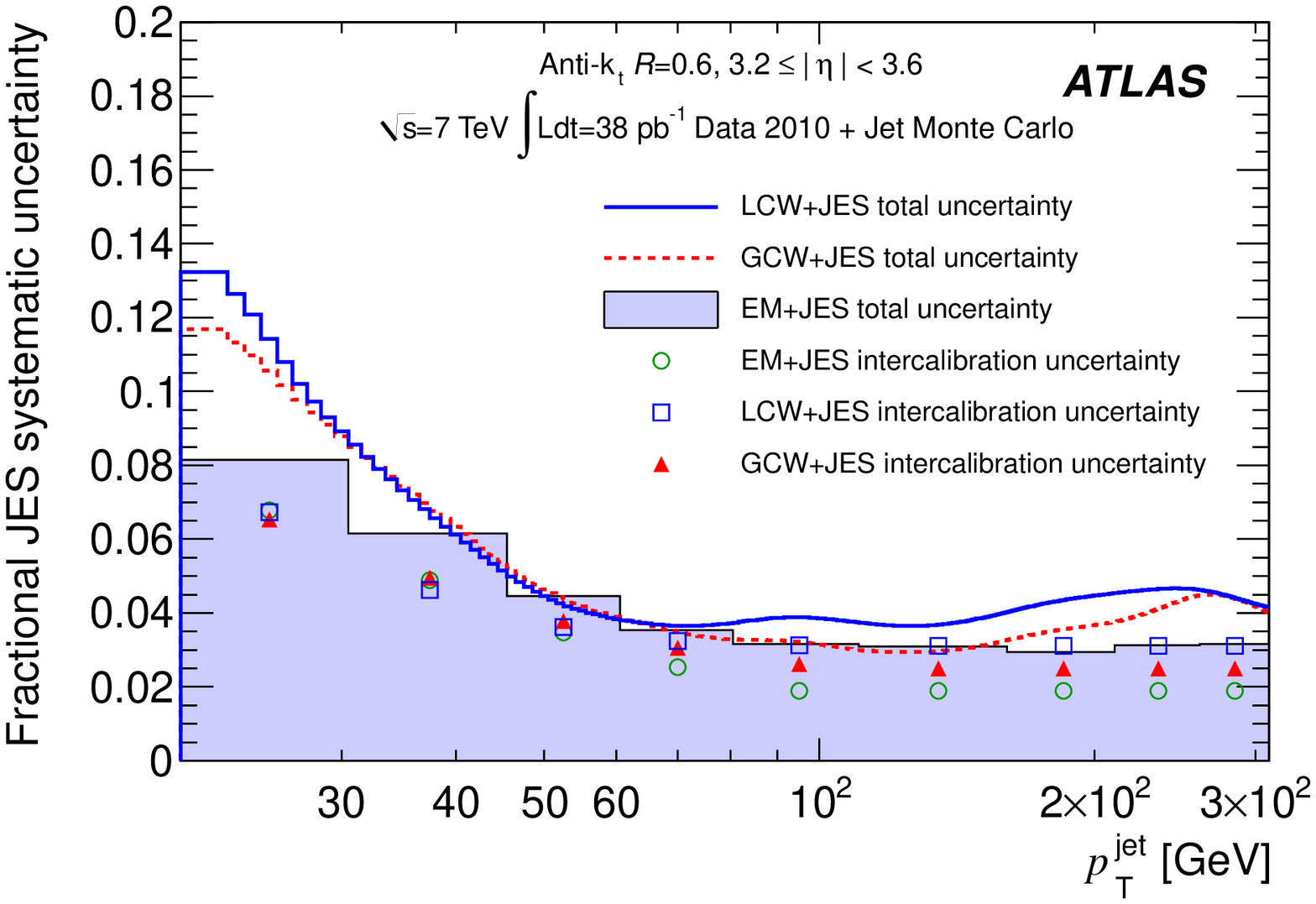}}
\subfloat[\etaRange{3.6}{4.5}]{\includegraphics[width=0.39\textwidth]{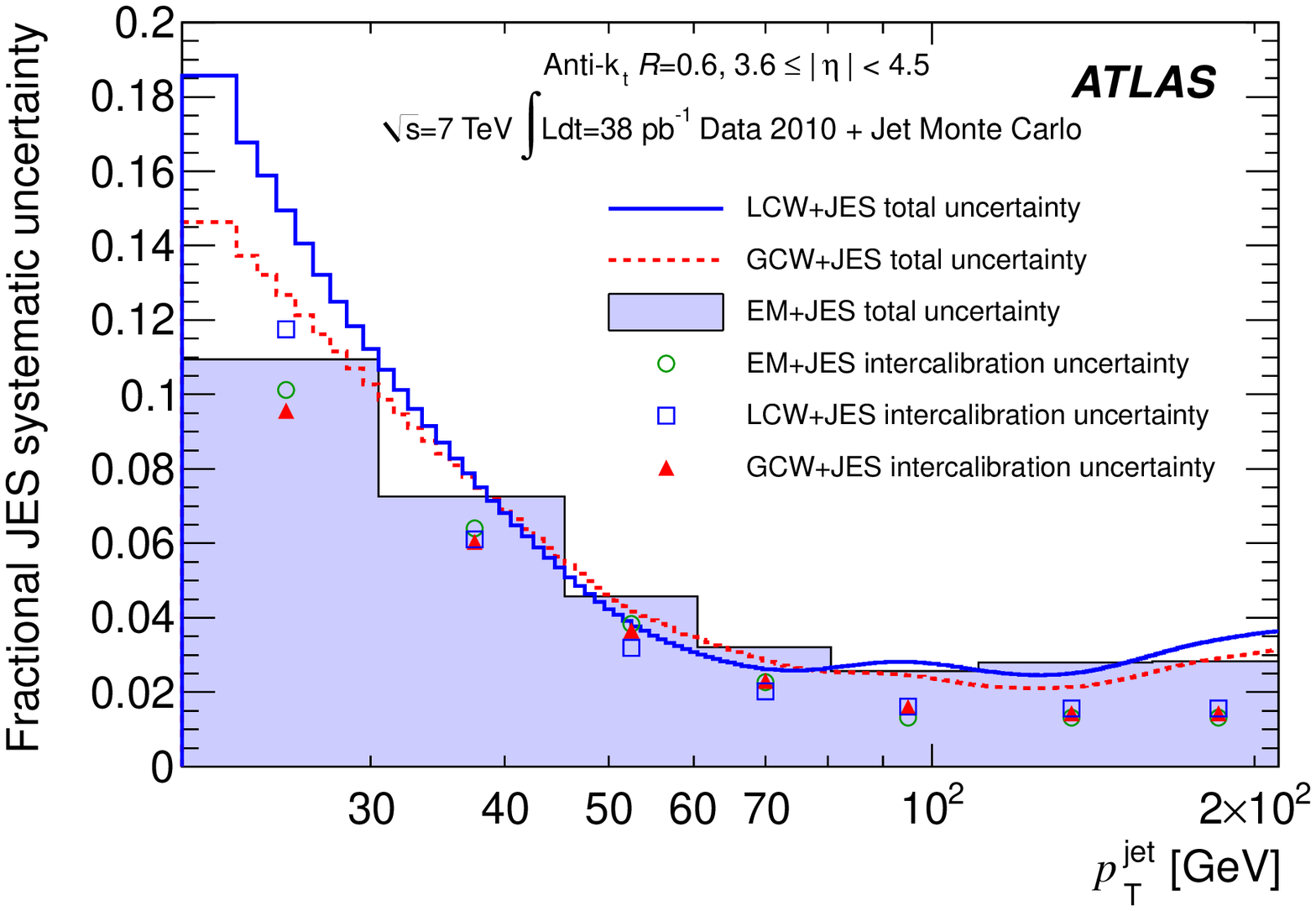}}
\vspace{-0.2cm}
\caption{
Fractional \JES{} uncertainties as a function of \ptjet{}
for \antikt{} jets with $R = 0.6$ for the various $\etajet$ regions
for the \LCWJES{} (full line) and the \GCWJES{} (dashed line) schemes. 
These are derived from a  combination of the \insitu{} techniques which are limited
in the number of available events at large \ptjet. 
The fractional \JES{} uncertainty for \EMJES{} derived from single hadron response measurements
and systematic Monte Carlo simulation variations is overlaid as shaded area for comparison.
The $\eta$-intercalibration uncertainty is shown as open symbols 
for $|\etajet| > 0.8$ for the \EMJES{} and for $|\etajet| > 1.2$ for the \LCWJES{} and \GCWJES{} schemes.
\label{fig:insitucomparison}}
\end{center}
\end{figure*}

\subsubsection{Multijet transverse momentum balance}
\label{sec:multijetcellweighting}
The multijet balance (MJB) technique, described in Section~\ref{sec:multijet}, is used to evaluate the
\JES{} uncertainty in the high transverse momentum region for the \GCWJES{} and \LCWJES{} calibration schemes. 
The method and selection cuts used are the same as those for the \EMJES{} calibrated jets.

Figure~\ref{fig:PtBalanceCellWeighting} shows the MJB for \antikt{}  jets with $R = 0.6$ obtained
using the \GCWJES{} and \LCWJES{} calibrations in the data and Monte Carlo simulation
as a function of the recoil \pt. 
The agreement between the data and MC simulations, 
evaluated as the data to Monte Carlo simulation ratio, are very similar to those for the \EMJES{} calibration. 

The systematic uncertainties on the MJB for these cell energy weighting calibration schemes are evaluated in 
the same way as the \EMJES{} calibration, described in Section~\ref{sec:Systematic}, except for the component of
the standard \JES{} uncertainty on the recoil system. The \JES{} uncertainty for jets in the recoil system is obtained
from the \insitu{} $\gamma$-jet balance discussed in Section~\ref{sec:gammajetcellweighting}. 
In this case, the systematic uncertainty
on the MJB due to the recoil system \JES{} uncertainty is then calculated by shifting the \pt{} of recoil jets
up and down by the \gammajet{} \JES{} uncertainty. In order to apply the \gammajet{} \JES{} uncertainty to the 
recoil system, the MJB analysis is performed with jets selected within the range $|\etajet| < 1.2$, where the 
\JES{} uncertainty based on \gammajet{} events has been derived.
The close-by jet and flavour composition systematic uncertainties are also re-evaluated 
for the \GCWJES{} and \LCWJES{} jets
using the same method (see Section~\ref{sec:closeby}).

Figure~\ref{fig:PtRecoilUncCellWeighting} shows the total and individual \JES{} systematic uncertainties 
on the recoil system for \antikt{} jets with $R=0.6$ 
calibrated by the \GCWJES{} and \LCWJES{} schemes. The increase of the
\JES{} uncertainty at high \ptjet{} above $800$~\GeV{} is caused by a large 
\JES{} systematic uncertainty due to limited \gammajet{} event statistics at high \pt.

The systematic uncertainties associated with the analysis method and event modelling are re-evaluated in the 
same way as for the \EMJES{} calibration scheme and then added to the recoil system \JES{} systematic uncertainties.
The summary of all systematic uncertainties and the total uncertainty obtained by adding the statistical and 
systematic uncertainties in quadrature is shown in Figure~\ref{fig:FinalResultCellWeighting} for \antikt{} 
jets with $R = 0.6$.

\subsubsection{Cell weighting JES uncertainty from combination of \insitu{} techniques}
\label{sec:JesUncertaintyinsitucellweighting}
Figure~\ref{fig:responseratioinsitucellweighting} shows the jet response ratio
of data to Monte Carlo simulation for the various \insitu{} techniques
as a function of the jet transverse momentum for the \GCWJES{} (a) and the \LCWJES{} (b)
calibration schemes. Statistical and systematic uncertainties are displayed.
The average from the combination of all \insitu{} techniques is overlaid.

The weight of each \insitu{} technique contributing to the average 
is similar to the one for the \EMJES{} calibration scheme shown in Figure~\ref{fig:weightinsitu}. 
The contributions are also similar for the \LCWJES{}  and the \GCWJES{} calibration schemes.

Figure~\ref{fig:uncertaintyinsitucellweighting} shows the final \JES{} uncertainty
for the \GCWJES{} (a) and the \LCWJES{} (b)  calibration schemes  for $|\etajet| < 1.2$.
At the lowest \pt{} the JES uncertainty is about $9 \%$ to $10 \%$ 
and decreases
for increasing \ptjet.  For $\ptjet > 50$~\GeV{} it is about $2 \%$
and at  $\ptjet = 250$~\GeV{} it is about $3 $ to  $4 \%$. For jets in the 
\TeV-regime the \JES{} uncertainty is $10$ to $12 \%$.

Figure~\ref{fig:uncertaintyinsitucellweighting} also shows 
the \JES{} uncertainty attainable, if the \insitu{} techniques had been used to correct the jet energy. 
Using the \insitu{} techniques for jet calibration would have resulted
in an improved \JES{} uncertainty for both jet calibration schemes
based on cell energy weighting.

The \JES{} uncertainty obtained in the central reference region
($|\etajet| < 1.2$) is used to derive the \JES{} uncertainty in the
forward region using the dijet balance technique. The central
region \JES{} uncertainty is combined with the uncertainties from
the dijet balance shown in Figure~\ref{fig:etainterCellWeight}.

\section{Summary of jet energy scale uncertainties of various calibration schemes}
\label{sec:JesUncertaintycomparison}
The \EMJES{} uncertainties are derived from single hadron response measurements
and from systematic variations of the Monte Carlo simulation (see Section~\ref{sec:JESUncertainties}).

The JES uncertainty for the \GS{} jet calibration scheme is given by the sum in 
quadrature of the \EMJES{} uncertainty and the uncertainty associated to the \GS{} corrections. 
The latter, derived in Section~\ref{sec:JESCalibGSC}, is conservatively taken to be 
$0.5\%$ for $30 < \pt < 800$~\GeV{} and $|\etajet| < 2.1$ and $1 \%$ for $\pt < 30$~\GeV{} and $2.1 < |\eta| < 2.8$.  
These uncertainties are also supported by \insitu{} techniques.

The JES uncertainties in the \LCWJES{} and \GCWJES{} jet calibration schemes
are derived from a combination of several \insitu{} techniques.

Figure~\ref{fig:insitucomparison} shows a comparison of the JES uncertainties
for the  \EMJES{}, the \LCWJES{} and the \GCWJES{} jet calibration schemes for various \etajet-regions.
The uncertainties in the \LCWJES{} and \GCWJES{} schemes derived
in Section~\ref{sec:JESUncertaintiescellweighting} are similar, but
the uncertainty for the \GCWJES{} calibration scheme is a bit smaller for very
low  and very large \ptjet.

Over a wide kinematic range, $40 \lesssim \ptjet \lesssim 600$~\GeV, all calibration schemes
show a similar JES uncertainty. At $\ptjet \approx 250$~\GeV{} the uncertainties
based on the \insitu{} techniques are about $2 \%$ larger compared to the uncertainty results 
from the \EMJES{} calibration scheme. 

For $\ptjet < 40$~\GeV{}  and $\ptjet >  600$~\GeV{}  the \EMJES{} calibration
scheme has a considerably smaller uncertainty. For the high \pt{} regions
the JES calibration based on \insitu{} suffers from the limited number of events in the data samples.
At low \pt{} the systematic uncertainty on the 
\insitu{} methods leads to a larger JES uncertainty.

 \begin{figure*}[ht!p]
   \begin{center}
      \subfloat[Efficiency from truth jets]        {\includegraphics[width=0.48\textwidth]{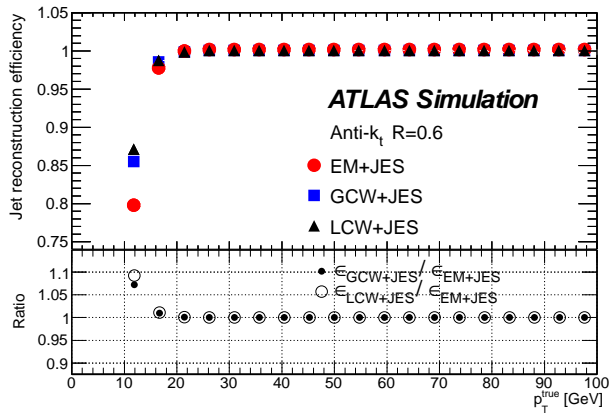}}
      \subfloat[Efficiency from track jets for \EMJES] {\includegraphics[width=0.48\textwidth]{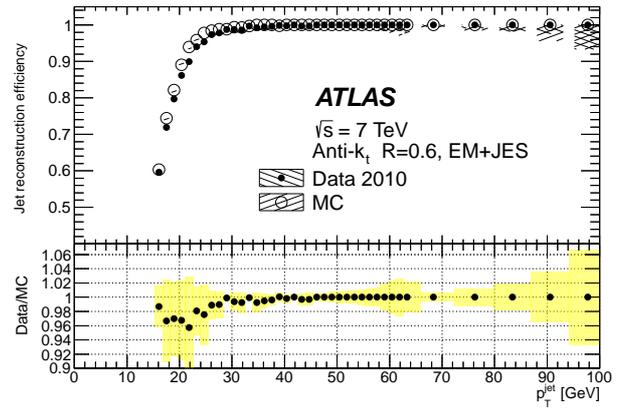}}\\
      \subfloat[Efficiency from track jets for \GCWJES]{\includegraphics[width=0.48\textwidth]{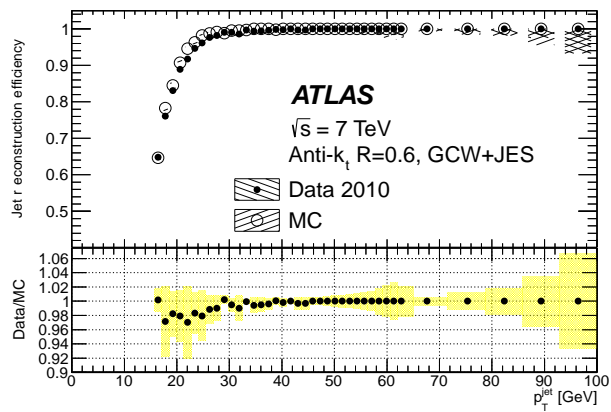}}
      \subfloat[Efficiency from track jets for \LCWJES]{\includegraphics[width=0.48\textwidth]{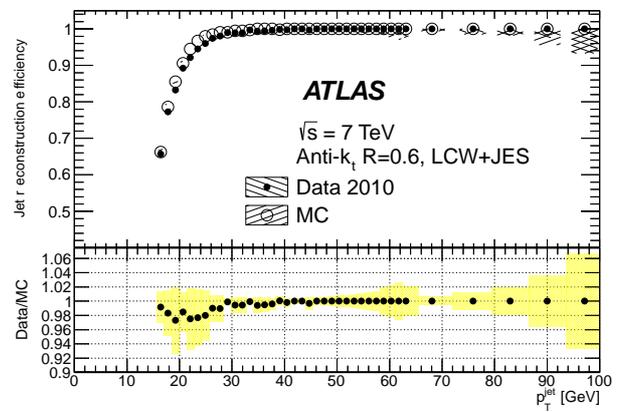}}
 \end{center}
 \caption{Calorimeter jet reconstruction efficiency 
with respect to truth jets (a) and track jets (b,c,d) 
as a function of the truth jet (a) or the calorimeter jet (b,c,d) \pt{} for the three calibration schemes: 
\EMJES{} (b), \GCWJES{} (c) and \LCWJES{} (d). 
The lower part of the figure (a) shows ratio of the efficiency of the \LCWJES{} and the \GCWJES{}  calibration schemes 
to that of the \EMJES{} calibration scheme. 
The ratio of data to Monte Carlo simulation is also shown in the lower part of the figure for (b), (c) and (d).
The hatched area correspond to the systematic uncertainty obtained by variations in the \insitu{} method.}
 \label{fig:plot_jetRecoEfficiency}
 \end{figure*} 

%
\begin{figure*}
  \centering
  \subfloat[Calorimeter jet]{\includegraphics[width=0.45\textwidth]{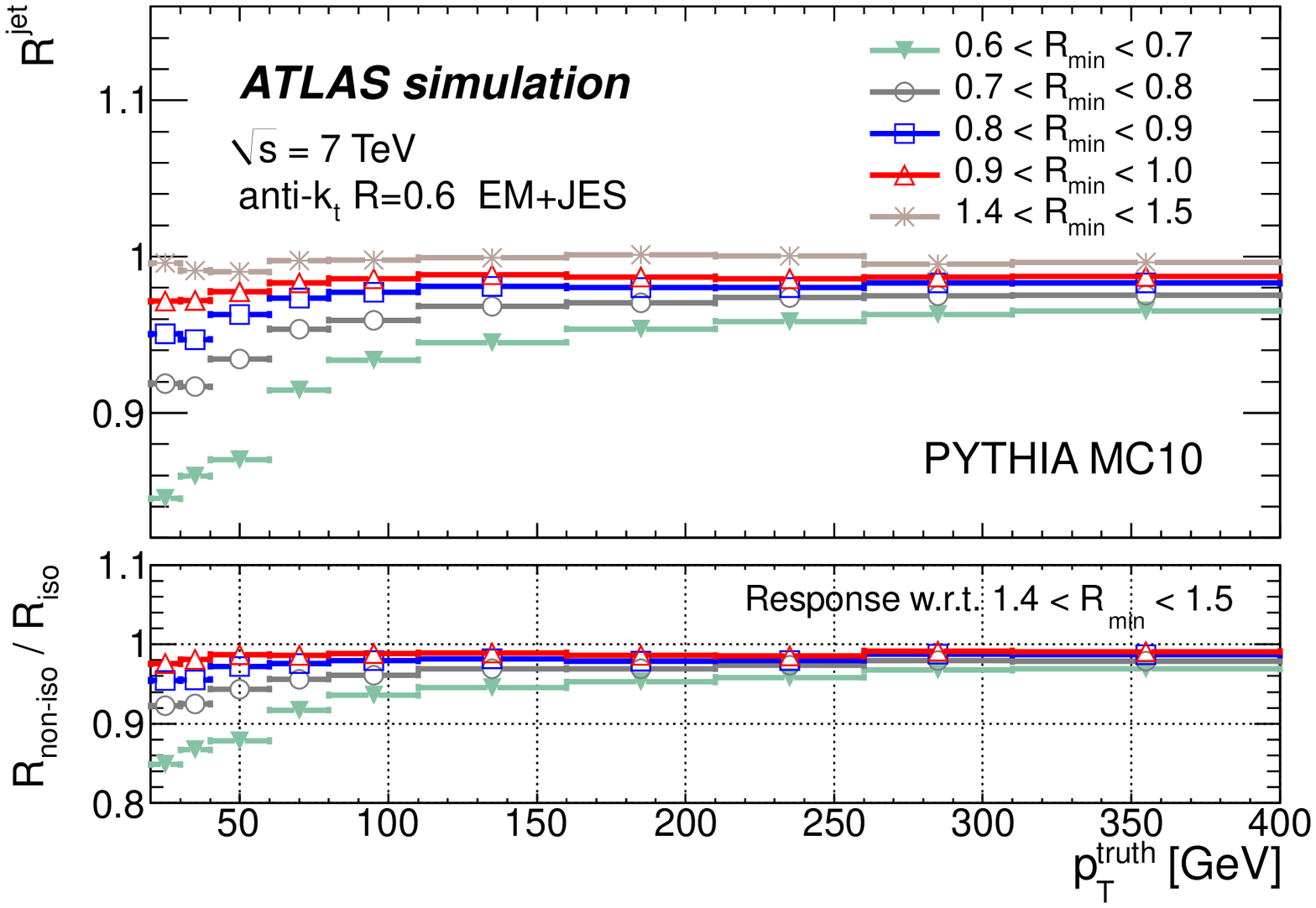}}\hfill
  \subfloat[Track jet]      {\includegraphics[width=0.45\textwidth]{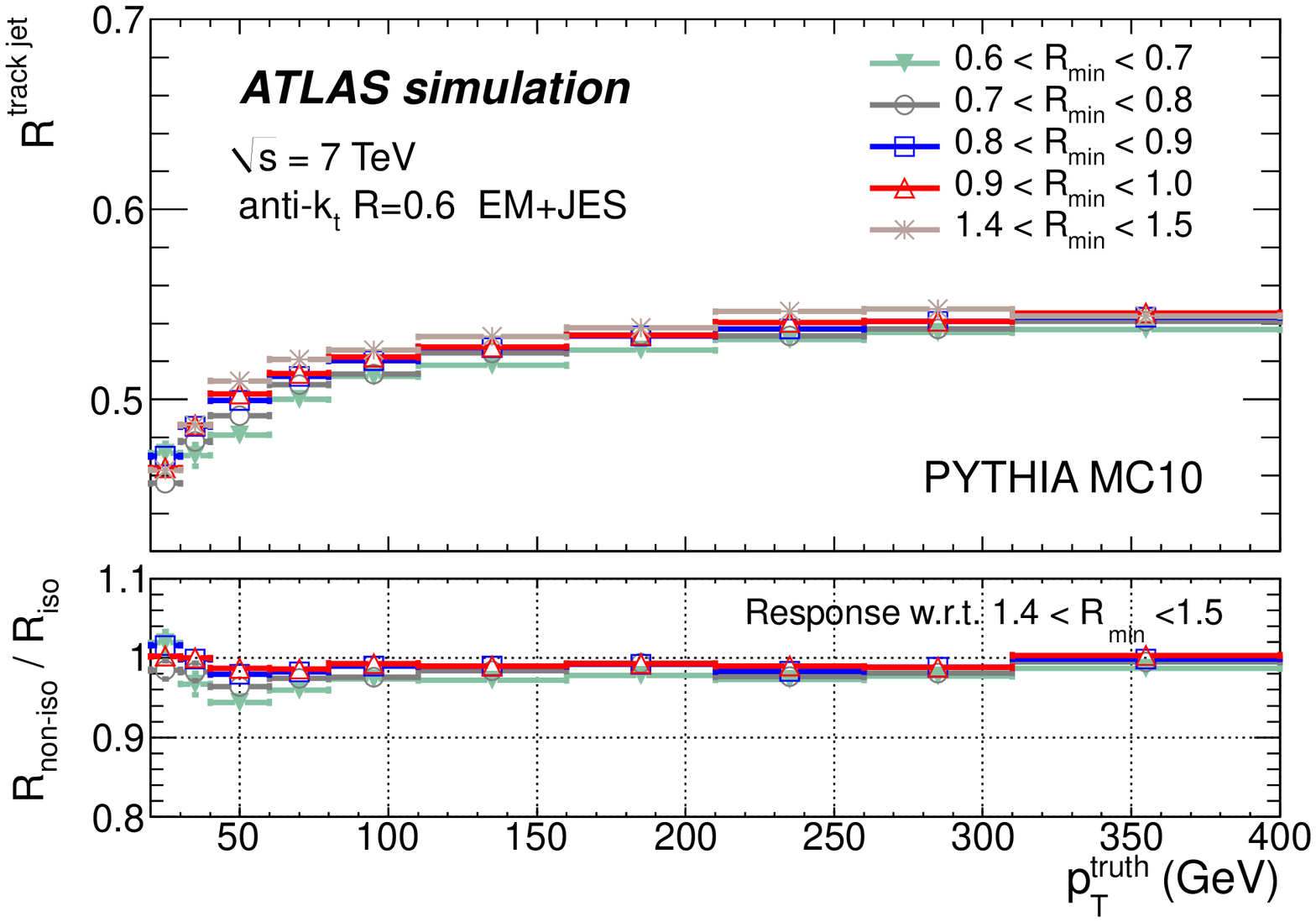}}
  \caption{Average ratio of calorimeter jet (a) and the track jet (b) $\pt$ to the matched truth jet \pt{} as a function of 
truth jet \pt{} for \antikt{} jets with $R = 0.6$, for different  \Rmin{} values. 
The bottom part of the figure shows the relative response of non-isolated 
jets with respect to that of isolated jets, obtained as the calorimeter or track jet response 
for $ \Rmin < 1.0$ divided by the jet response for $1.4 \le \Rmin <1.5$.} 
  \label{fig:JetResponseMC06}
\end{figure*}

\begin{figure*}
  \centering
  \subfloat[Data]{\includegraphics[width=0.45\textwidth]{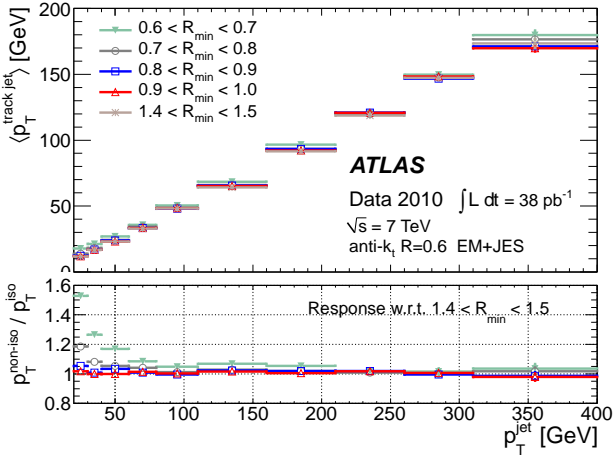}}\hfill
  \subfloat[MC]{\includegraphics[width=0.45\textwidth]{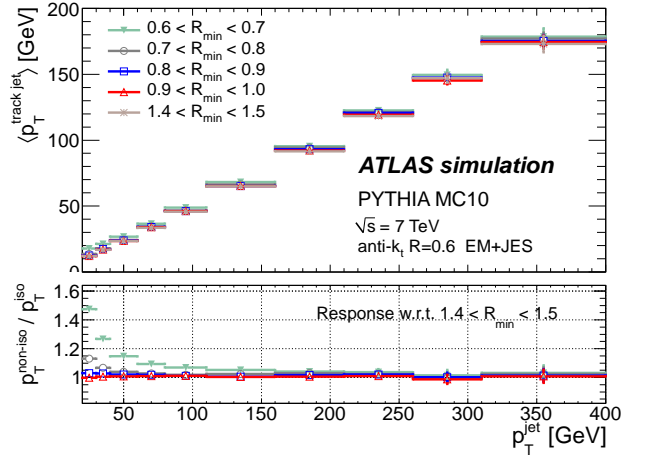}}
  \caption{Average track jet $\pt$ as a function of calorimeter jet $\pt$ for \antikt{} jets with $R = 0.6$ 
in data (a) and MC simulations (b) for different \Rmin{} values. 
The lower part shows the relative response of non-isolated jets with respect to that of isolated jets, 
obtained as the track jet $\pt$ for $\Rmin < 1.0$ divided by that for $1.4 \le \Rmin < 1.5$.} 
  \label{fig:TrackJetPt06}
\end{figure*}

\begin{figure}
  \centering
\includegraphics[width=0.45\textwidth]{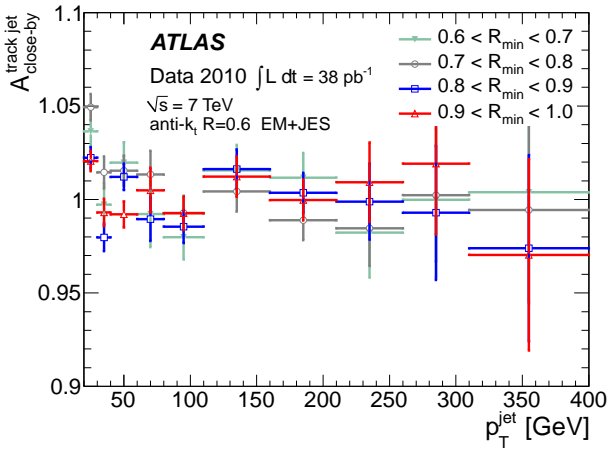} 
  \caption{Ratio of data to Monte Carlo simulation of the track jet \pTtrkjet{} for non-isolated jets divided by the 
track jet \pTtrkjet{}  for isolated jets as a function of the jet \ptjet. 
Only statistical uncertainties are shown.
} 
  \label{fig:DataMCTrackJetPt}
\end{figure}

\begin{figure*}
  \centering
  \subfloat[Data]{\includegraphics[width=0.45\textwidth]{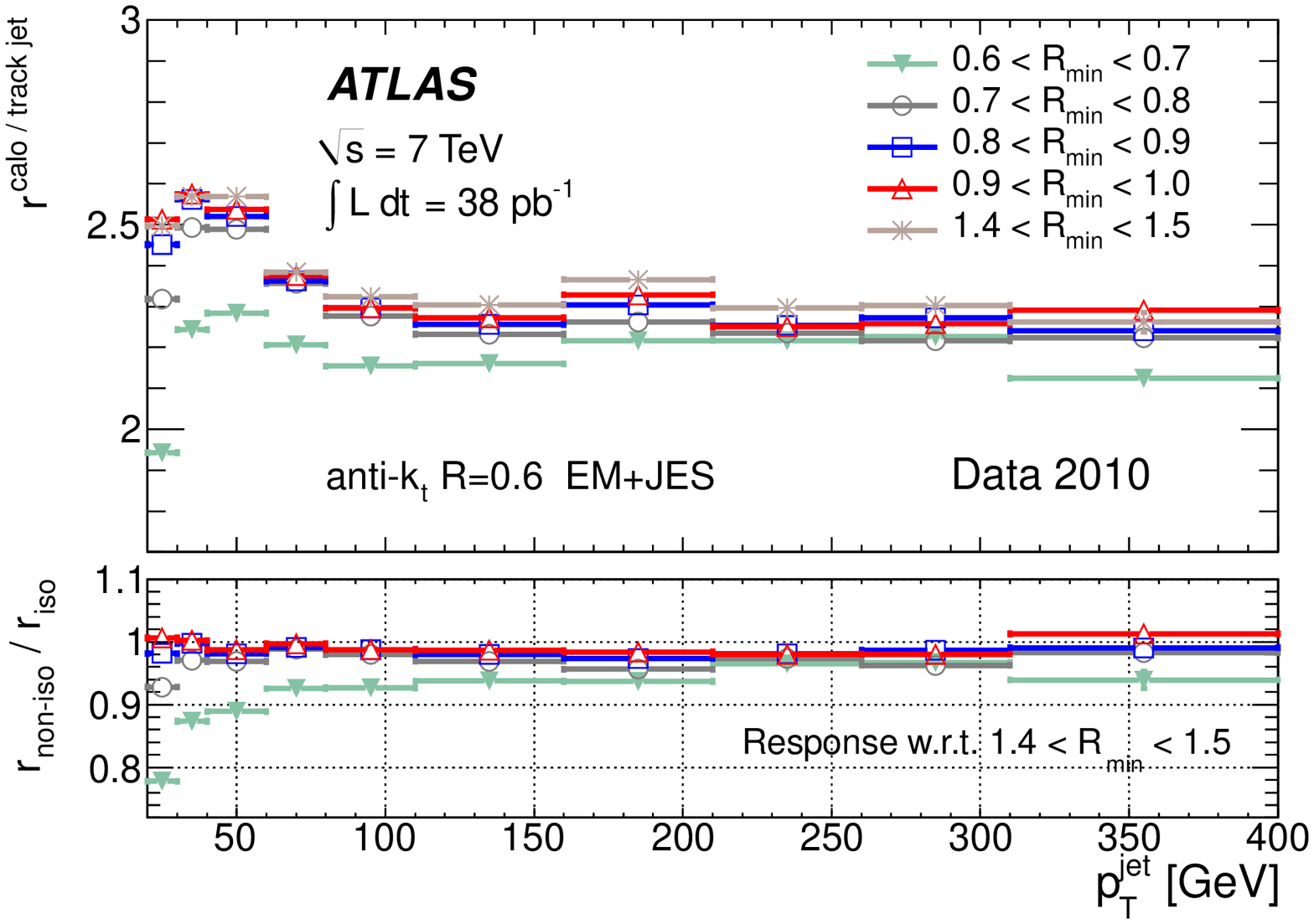}}\hfill
  \subfloat[Monte Carlo simulation]{\includegraphics[width=0.45\textwidth]{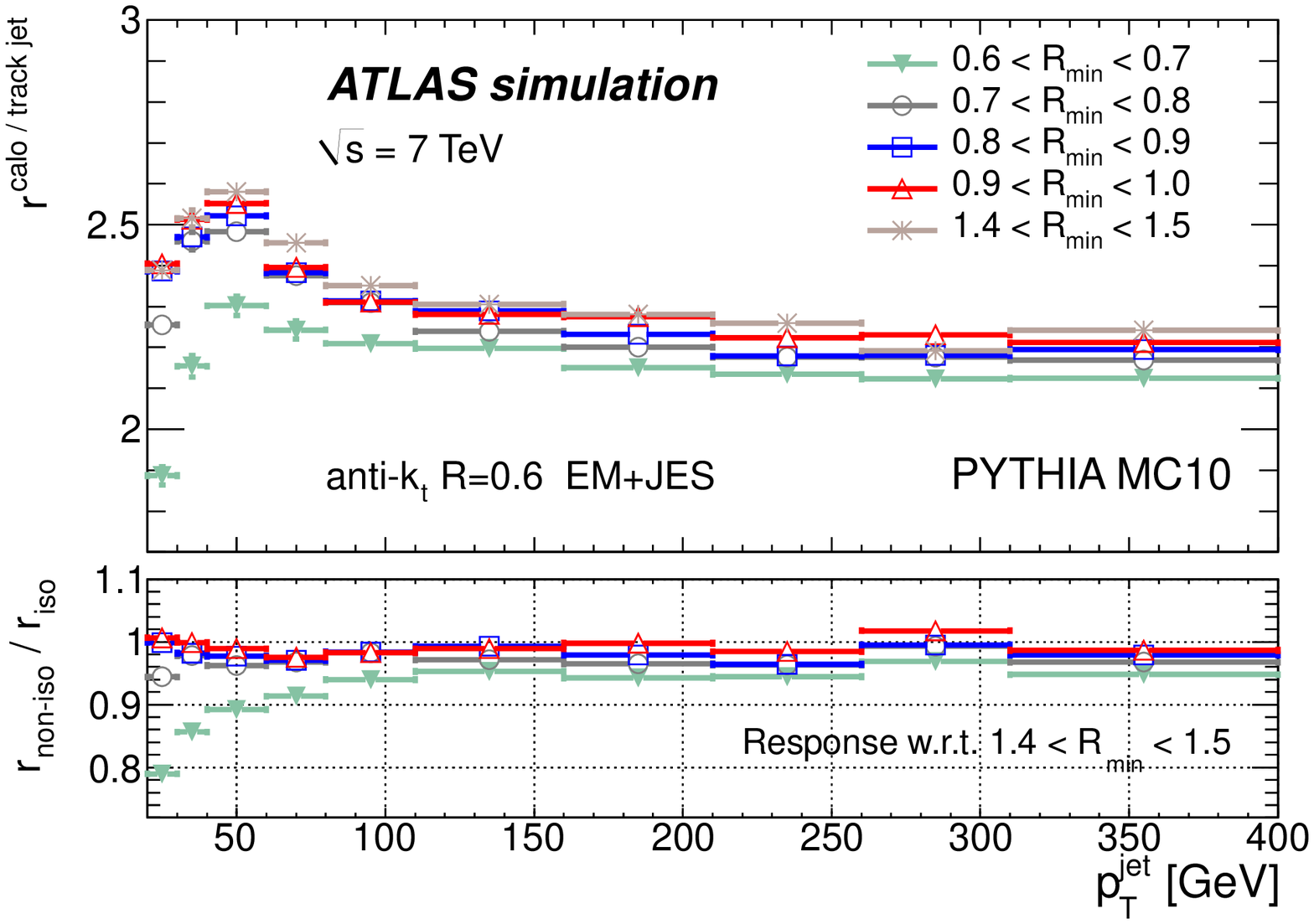}}
  \caption{Ratio of calorimeter jet $\ptjet$ to the matched track jet \pt{} as a function of 
calorimeter jet $\ptjet$ for \antikt{} jets with $R = 0.6$ in data (a) and Monte Carlo simulations (b)
for different \Rmin{} values. The lower part shows the relative response of non-isolated jets with 
respect to that of isolated jets, obtained as the jet response for $\Rmin < 1.0$ divided by the 
jet response for $1.4 \le \Rmin < 1.5$.} 
  \label{fig:CalTrackRatio06}
\end{figure*}

\begin{figure}
  \centering
\includegraphics[width=0.45\textwidth]{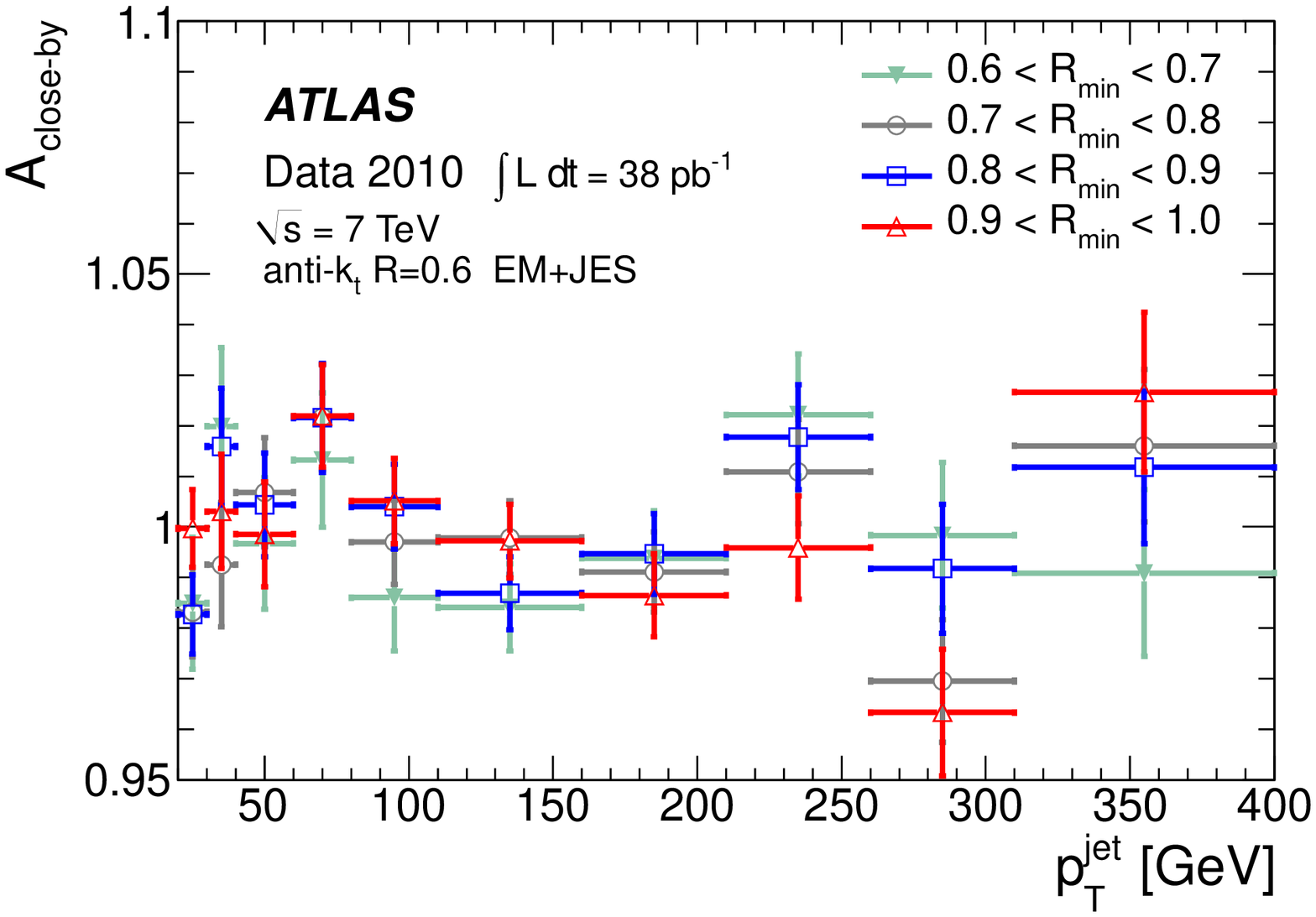}
  \caption{Data to Monte Carlo simulation ratio of the relative response of non-isolated jets with respect to that of 
isolated jets for \antikt{} jets with $R = 0.6$ calibrated with the \EMJES{} scheme.
Only statistical uncertainties are shown. 
} 
  \label{fig:DataMCCalTrack}
\end{figure}

\section{Jet reconstruction efficiency}
\label{sec:jetrecoeff}
A tag-and-probe method is implemented to measure \insitu{} the jet reconstruction efficiency relative to track jets. 
Because track jets (see Section~\ref{sec:trackjets})
and calorimeter jets (see Section~\ref{sec:calorimeterjets})
are reconstructed by independent \ATLAS{} sub-detectors, 
a good agreement between data and Monte Carlo simulation for this matching efficiency means 
that the absolute jet reconstruction efficiency can be determined from the simulation. 
\subsection{Efficiency in the Monte Carlo simulation}
The jet reconstruction efficiency is determined in the Monte Carlo simulation by
counting in how many cases a calorimeter jet can be matched to a truth jet.
Reconstructed jets are matched to truth jets, if their jet axes are within $\DeltaR < 0.4$.

Figure~\ref{fig:plot_jetRecoEfficiency}a shows the jet reconstruction efficiency 
for \antikt{} jets with $R = 0.6$ calibrated with the \EMJES, \GCWJES, and \LCWJES{} calibration schemes
as a function of the transverse momentum of the truth jet. 
The efficiency reaches its maximum value for a truth jet transverse momentum of $20$~\GeV. 
The lower part of the figure shows the ratio of the efficiency in the \GCWJES{} and \LCWJES{} calibration schemes
to that obtained from the \EMJES{} scheme.
Similar performance is found for all calibration schemes.

The small differences at low \ptjet{} might be caused by the slightly better jet energy resolution obtained
with the \GCWJES{} and the \LCWJES{} calibration schemes. 
Moreover, jets based on  the \LCWJES{} scheme are built from calibrated
\topos{} while the jets calibrated with the \EMJES{} and the \GCWJES{} calibration schemes 
use \topos{} at the electromagnetic
scale.

\subsection{Efficiency \insitu{} validation}
The ability of the Monte Carlo simulation to correctly reproduce the jet reconstruction in the data
is tested using track jets that provide an independent reference.

A tag-and-probe  technique is used as described in the following steps: 
\begin{enumerate}
\item Only track jets with $\pt > 5$~\GeV{} and $|\etajet| < 1.9$ are considered.
\item The track jet with the highest \pt{} in the event is defined as the reference object.
\item The reference object is required to have $\pt > 15$~\GeV\footnote{
Reference track jets with $\pt < 15$~\GeV{} are not used, since they would result in a sample of biased probe track jets. 
In this case, mostly events where the probe track jet has fluctuated up in energy (such that it passes the $5$ \GeV{} threshold) 
would be kept.
The $15$~\GeV{} cut has been determined by measuring the jet reconstruction efficiency relative to track jets 
as a function of the reference track jet \pt. The measured efficiency for low probe track jet \pt{} was found to be dependent 
on the reference track jet \pt{} when the latter is smaller than $15$~\GeV. 
The jet reconstruction efficiency is stable for a reference track jet \pt{} greater than $15$~\GeV.}.
\item The reference track jet is match\-ed to a calorimeter jet with $\ptjet> 7$~\GeV{}, 
if $\DeltaR({\rm tag}, {\rm calo jet}) < 0.6$\footnote{The less restrictive matching criterion with
respect to previous sections is motivated by the lower \pt.}.
\item The probe track jet must be back-to-back to the reference jet in $\phi$ with $|\Delta \phi| \geq 2.8$ radian. 
\item Events with additional track jets within $|\Delta \phi|\geq 2.8$ radian are rejected.
\item The calorimeter reconstruction efficiency with respect to track jets is then defined as the fraction of 
probe jets matched to a calorimeter jet using $\Delta R({\rm probe}, {\rm calo jet}) < R$ 
(with $R = 0.4$ or $R = 0.6$) with respect 
to all probe jets.
\end{enumerate} 

The jet reconstruction efficiency is measured in a sample of minimum bias events 
and is compared to a minimum bias Monte Carlo simulation.
Due to the restriction of $|\eta| < 1.9$ on track jets, the measurement is only valid for calorimeter jets with
$|\etajet| < 1.9 + R$, where $R =0.4$ or $R =0.6$.

Figures~\ref{fig:plot_jetRecoEfficiency}b-d show the measured
calorimeter reconstruction efficiency with respect to track jets as 
a function of the calorimeter jet transverse momentum for \antikt{} jets with $R = 0.6$ 
calibrated with the \EMJES, \GCWJES, and \LCWJES{} calibration
schemes\footnote{Technically, the efficiency is first measured as a function of the track jet \pt{}. Using the known relation
between the average track jet and the average calorimeter jet \pt, the track jet \pt{} is 
then converted to the calorimeter jet \ptjet.}. 
The reconstruction efficiency reaches a plateau close to $100$\% at a jet transverse calorimeter momentum of about $25$~\GeV . 
The matching efficiency in data ($\epsilon_{\rm Data}$) and in Monte Carlo simulation ($\epsilon_{\rm MC}$) shows a good overall agreement 
except at low \ptjet{} where the efficiency in data is slightly lower than in the Monte Carlo simulation. 
Similar performance is found for all calibration schemes.

The systematic uncertainties on the jet reconstruction efficiency measured \insitu{}
are obtained by varying the following event selection requirements for both data and 
Monte Carlo simulation: the opening angle
$|\Delta \phi|$ between the reference and the probe track jets,
the $\DeltaR$ requirement between the tag track jet and the calorimeter jet
and the probe track jet and the calorimeter jet.

The sensitivity in both data and Monte Carlo simulation  to the azimuthal opening angle  
as well as to the $\Delta R({\rm tag}, {\rm calo jet})$
variation is small. However, the efficiency shows a sensitivity with respect to 
the $\DeltaR({\rm probe} \; {\rm jet}, {\rm calo} \; {\rm jet})$. 
The variation of $\epsilon_{\rm Data} /\epsilon_{\rm MC} $ for these different parameters is shown
in Figure~\ref{fig:plot_jetRecoEfficiency}. 
At high \ptjet{} the statistical uncertainties after the cut variations lead to an enlarged uncertainty band.

The systematic uncertainty of the \insitu{} determination is larger than the observed shift between
data and Monte Carlo simulation.
For $\ptjet < 30$~\GeV{} a systematic uncertainty of $2 \%$ for jets is assigned. 

\subsection{Summary of jet reconstruction efficiency}
The jet reconstruction efficiency is derived using the nominal inclusive jet Monte Carlo simulation sample.
The systematic uncertainty is evaluated using a tag-and-probe technique using track jets
in both data and Monte Carlo simulation.

The jet reconstruction efficiency is well described by the Monte Carlo simulation
and is within the systematic uncertainty of the \insitu{} method.
A systematic uncertainty of $2 \%$ for jets
with $\ptjet < 30$~\GeV{} is assigned and negligible for higher \ptjet.

\section{Response uncertainty of non-isolated jets}
\label{sec:closeby}
The standard \ATLAS{} jet calibration and associated \JES{} uncertainty 
is obtained using only isolated jets (see Section~\ref{sec:EMJES}).
Jets are, however, often produced with nearby jets in a busy environment 
such as found in multijet topologies or in events where 
top-quark pairs are produced. Therefore a separate study is needed to determine the additional \JES{} uncertainty for jets 
with nearby jet activity.

Jets with $\ptjet > 20$~\GeV{} and $|\rapjet|<2.8$ calibrated with the \EMJES{} scheme
are used. The close-by \JES{} uncertainty is evaluated within $|\rapjet|<2.0$.

\subsection{Evaluation of close-by jet effects}
\label{sec:CloseBy}
The effect due to close-by jets is evaluated in the Monte Carlo simulation
by using truth  jets as a reference.
Similarly, track jets are used as a reference in both data and Monte Carlo simulation 
(see Sections \ref{sec:trackjets} and \ref{sec:truthjets} for comparison). 
The calorimeter jet response relative to these reference jets is examined for 
different values of $R_{\rm min}$, the distance from the calorimeter jet to the closest
jet in $\eta$-$\phi$ space.

The relative calorimeter jet response to the truth jets provides an absolute $\pt$ scale 
for the calorimeter jets, 
while the relative response to the track jets allows \insitu{} validation of the 
calorimeter jet response and the evaluation of the systematic uncertainty. 
For this purpose, the track jet response in data needs to be established for the non-isolated case 
and the associated systematic uncertainty has to be understood.

In the relative response measurement in the Monte Carlo simulation, 
the truth jet is matched to the calorimeter jet or track jet 
in $\eta$-$\phi$ space by requiring $\DeltaR <0.3$. Similarly, the track jet
is matched to the calorimeter jet within $\DeltaR < 0.3$ when the relative response to the track jet 
is examined. If two or more jets are matched within the \DeltaR{} range, 
the closest matched jet is taken. 

The calorimeter response to the matched track jet is defined as the ratio of the 
calorimeter jet \ptjet{} to the track jet transverse momentum (\pTtrkjet)
\begin{equation}
\rtrackjet = \ptjet/\pTtrkjet.
\end{equation}
This response is examined as a function of the jet transverse momentum \ptjet{} and
for different \Rmin{} values measured relative to the closest calorimeter jet with 
$\pt > 7$~\GeV{} at the \EM{} energy scale\footnote{Unless otherwise stated, 
calorimeter jets (selected as listed below) and nearby jets 
(selected with $\pt > 7$~\GeV{} at the \EM{} scale) are both used in the jet response measurement, 
if a matched track jet is found.}.
The ratio of the calorimeter jet response for non-isolated 
(i.e. small \Rmin) to the response of isolated (large \Rmin) jets, 
is given by 
\begin{equation}
\rtrackjetisoratio = \rtrackjetnoniso / \rtrackjetiso.
\end{equation}
This ratio is compared between data and Monte Carlo simulations.
\begin{equation}
  A_{\rm close-by} = \left [ \rtrackjetisoratio \right ]_{\rm Data}\:/\: \left [\rtrackjetisoratio \right ]_{\rm MC}.
  \label{eq:A_closeby}
\end{equation}
The deviation of $A_{\rm close-by}$ is assumed to represent the component of calorimeter \JES{} uncertainty due to close-by jets. 
This uncertainty, 
convolved with the systematic uncertainty in the track jet response due to a nearby jet,
provides the total \JES{} systematic uncertainty due to the close-by jet effect.

\begin{table}
  \begin{center}
    \begin{tabular}{c|cc|cc}\hline\hline
                            & \multicolumn{2}{c|}{$R = 0.6$} & \multicolumn{2}{c}{$R = 0.4$} \\\hline
       \ptjet{} [\GeV]         & $20 - 30$& $> 30$    & $20 - 30$ & $> 30$   \\\hline
      $0.4 \le \Rmin < 0.5$ & -        & -        & $2.7 \%$  & $2.8 \%$ \\
      $0.5 \le \Rmin < 0.6$ & -        & -        & $1.7 \%$  & $2.3 \%$ \\
      $0.6 \le \Rmin < 0.7$ & $3.9 \%$ & $1.9 \%$ & $2.5 \%$  & $2.7 \%$ \\
      $0.7 \le \Rmin < 0.8$ & $5.1 \%$ & $1.6 \%$ & -         & \\
      $0.8 \le \Rmin < 0.9$ & $2.5 \%$ & $1.9 \%$ & -         & \\\hline \hline
    \end{tabular}
  \caption{Summary of jet energy scale systematic uncertainty assigned for non-isolated jets accompanied 
by a close-by jet within the denoted \Rmin{} ranges. 
The second row in the table indicates the \pt{} range of the non-isolated jets.
\Antikt{} jets with $R=0.6$ and $R=0.4$ are used.}
  \label{tab:SystUncertVsRmin}
  \end{center}
\end{table}

\subsection{Non-isolated jet response}
\label{sec:closebyAnalysis}
Events that contain at least two jets with $\ptjet > 20$~\GeV{} 
and absolute rapidity $|\rapjet|<2.8$ are selected.
The response of non-isolated jets is studied in the Monte Carlo simulation
using the calorimeter jet response $\RcaloCALIB = \ptjet/\pttruth$.

Figure~\ref{fig:JetResponseMC06}a shows the calorimeter jet response 
as a function of \pttruth{} for \antikt{} jets with $R = 0.6$. 
The jet response was measured for nearby jets 
in bins of \Rmin{} values. The lower part of the figure shows the ratio of the 
non-isolated jet response for $\Rmin < 1.0$ to the isolated response
$1.4 \le \Rmin < 1.5$, 
\begin{equation}
\RcaloCALIB_{\rm non-iso}/\RcaloCALIB_{\rm iso}. 
\end{equation}

The observed behaviour at small \Rmin{} values 
indicates that the non-isolated jet response is lower by up to $15 \%$ 
relative to the isolated jet response for $\ptjet > 20$~\GeV{}, if 
the two jets are within $\Rmin < R + 0.3$.
The magnitude of this effect depends on \ptjet{} and is largest at low \ptjet.

The track jet response relative to the matched truth jet is defined as 
\begin{equation}
\Rtrack = \pTtrkjet/\pttrue.
\end{equation} 
Figure~\ref{fig:JetResponseMC06}b shows \Rtrack{}
as a function of \pttrue{} for \antikt{} jets with $R = 0.6$.
The track jet response is more stable against the presence of close-by jets and has a much weaker \Rmin{}
dependence than the calorimeter jet response.
This results from the smaller ambiguity in the matching between the truth and track jets 
that are both measured from the primary interaction point. Moreover, track jets are less
influenced by magnetic field effects than calorimeter jets.

\subsection{Non-isolated jet energy scale uncertainty}
\label{sec:CloseByJES}
Figure~\ref{fig:TrackJetPt06} shows the average track jet transverse momentum as a function of \ptjet{}
for \antikt{} jets with $R = 0.6$ in both data and Monte Carlo simulations for various \Rmin{} values. 
The lower part of the figure shows the \pt{} ratio of non-isolated to isolated track jets defined as
\begin{equation}
r^{\rm track \; jet}_{\rm non-iso/iso} = {p^{\rm  track \; jet}_{\rm T, non-iso}/p^{\rm  track \; jet}_{\rm T, iso}}.
\end{equation}

The data to MC ratio defined as
\begin{equation}
  A_{\rm close-by}^{\rm track \; jet} = \left [r^{\rm track \; jet}_{\rm non-iso/iso} \right ]_{\rm Data}\:/\: \left [ r^{\rm track \; jet}_{\rm non-iso/iso} \right]_{\rm MC}
  \label{eq:A_closeby_Track}
\end{equation}
is compared between data and Monte Carlo simulations in Figure~\ref{fig:DataMCTrackJetPt}.
This ratio can be used 
to assess the potential of track jets to test close-by effects in the small \Rmin{} range. 
The agreement between data and Monte Carlo simulation is quite satisfactory: 
within $2$ to $3 \%$ for $\ptjet > 30$~\GeV{} and slightly worse for $20 \le \ptjet < 30$~\GeV.

Therefore, the track jet response systematic uncertainty is assigned separately for the 
two $\ptjet$ regions: $|1 - A_{\rm close-by}^{\rm track \; jet}|$ is used as the uncertainty 
for $20 \le \ptjet < 30$~\GeV, while for $\ptjet > 30$~\GeV{} a standard deviation of 
the $A_{\rm close-by}^{\rm track \; jet}$ is calculated and assigned as the uncertainty. 
These uncertainties are typically $1.5\%$ ($2.0\%$) for \antikt{} jets with $R = 0.6$ ($0.4$).

The calorimeter jet \ptjet{} relative to the matched track jet \pTtrkjet{} (\rtrackjet)  
is shown in Figure~\ref{fig:CalTrackRatio06} 
as a function of \ptjet{} for \antikt{} jets with $R = 0.6$ in data and Monte Carlo simulations. 
The non-isolated jet response relative to the isolated jet response, \rtrackjetisoratio,
shown in the bottom part of Figure~\ref{fig:CalTrackRatio06} reproduces
within a few per cent the behaviour in
the ratio $\RcaloCALIB_{\rm non-iso}/\RcaloCALIB_{\rm iso}$ for the Monte Carlo simulation response 
of calori\-me\-ter to truth jet \pt{} in Figure~\ref{fig:JetResponseMC06}.

The \rtrackjetisoratio{} data to Monte Carlo ratio $A_{\rm close-by}$  (see Equation~\ref{eq:A_closeby})  
is shown in Figure~\ref{fig:DataMCCalTrack}. 
The \Rmin{} dependence of the 
non-isolated jet response in the data is well described by the Monte Carlo simulation.

Within the statistical uncertainty, $A_{\rm close-by}$ differs from unity by at most $\sim 3 \%$ 
depending on the \Rmin{} value in the range of $R \leq \Rmin < R + 0.3$.
No significant \ptjet{} dependence is found over the measured \pt{} range of $20 \leq \ptjet < 400$~\GeV.

The overall \JES{} uncertainty due to nearby jets is taken as the track jet response
systematic uncertainty added in quadrature with the deviation from one of the weighted
average of $A_{\rm close-by}$ over the entire \pt{} range, but added separately for each \Rmin{} range.
The final uncertainties are
summarised in Table~\ref{tab:SystUncertVsRmin} for the two jet distance parameters. 

The $A_{\rm close-by}$ ratio has been examined 
for each of the two close-by jets either with the lower or the higher \ptjet, and
no apparent difference is observed with respect to the inclusive case shown in 
Figure~\ref{fig:DataMCCalTrack}. Therefore, both calorimeter jets which are close to each other are
subject to this uncertainty.

\subsection{Summary of close-by jet uncertainty}
The uncertainty is estimated by comparing in data and Monte Carlo simulation
the track jet response. They are both examined as function of 
the distance \Rmin{} between the jet and the closest jet in the calorimeter.

The close-by jet systematic uncertainty on the jet energy scale is
$2.5 - 5.1 \%$ ($1.7 - 2.7 \%$) and $1.6 - 1.9 \%$ ($2.3 - 2.8 \%$) for $R = 0.6$ ($R = 0.4$) 
jets with $20 \le \ptjet < 30$~\GeV{} and $\ptjet > 30$~\GeV, respectively, 
in the range of $R \le \Rmin < R + 0.3$ and jet rapidity $|\rapjet|<2.0$.
When the two jets are separated in distance by $R + 0.3$ or more, 
the jet response becomes similar to that for the isolated jets and hence 
no additional systematic uncertainty is required. No significant jet $\pt$ dependence 
is observed at $\ptjet > 30$~\GeV{} for the close-by jet systematic uncertainty.

%
%
\begin{figure*}[ht!!]
\begin{center}
\subfloat[$|\etajet|<0.8$]  {\includegraphics[width=0.45\textwidth]{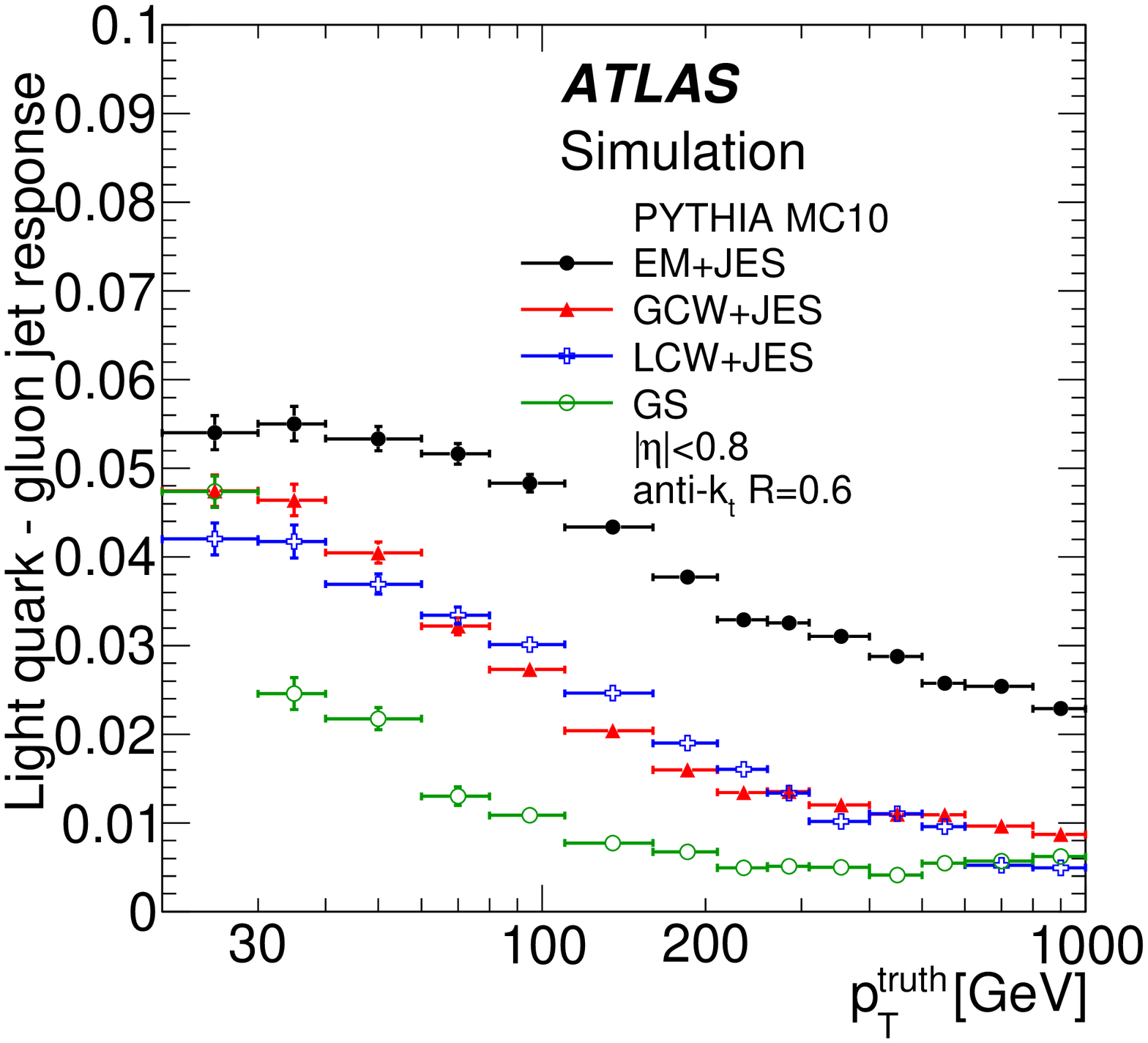} }
\hspace{1.cm}
\subfloat[\etaRange{2.1}{2.8}]{\includegraphics[width=0.45\textwidth]{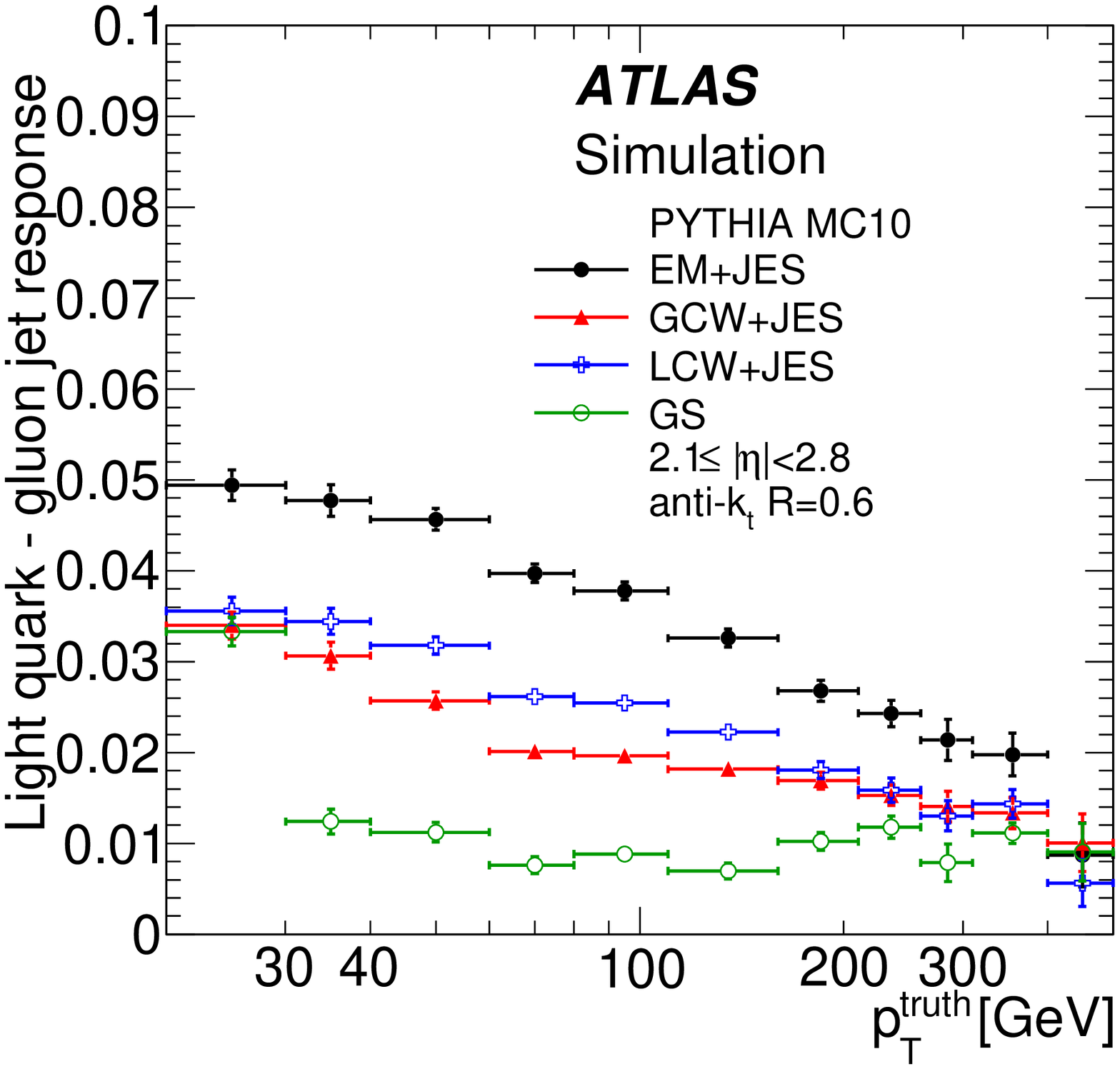} }
\caption{ Difference in average response of gluon and light quark jets as a function of the truth jet \pt{} 
for \antikt{} jets with $R = 0.6$ in the barrel (a) and the endcap (b) calorimeters as determined in Monte Carlo simulation.
Various calibration schemes are shown. The data sample used contains at least two jets with
$\ptjet>60$ \GeV{} and $|\eta|<2.8$. Only statistical uncertainties are shown.}
\label{fig:respLightGluCalibs}
\end{center}
\end{figure*}

\begin{figure*}[ht!p]
\begin{center}
\subfloat[Number of tracks $n_{\rm trk}$]{\includegraphics[width=0.45\textwidth]{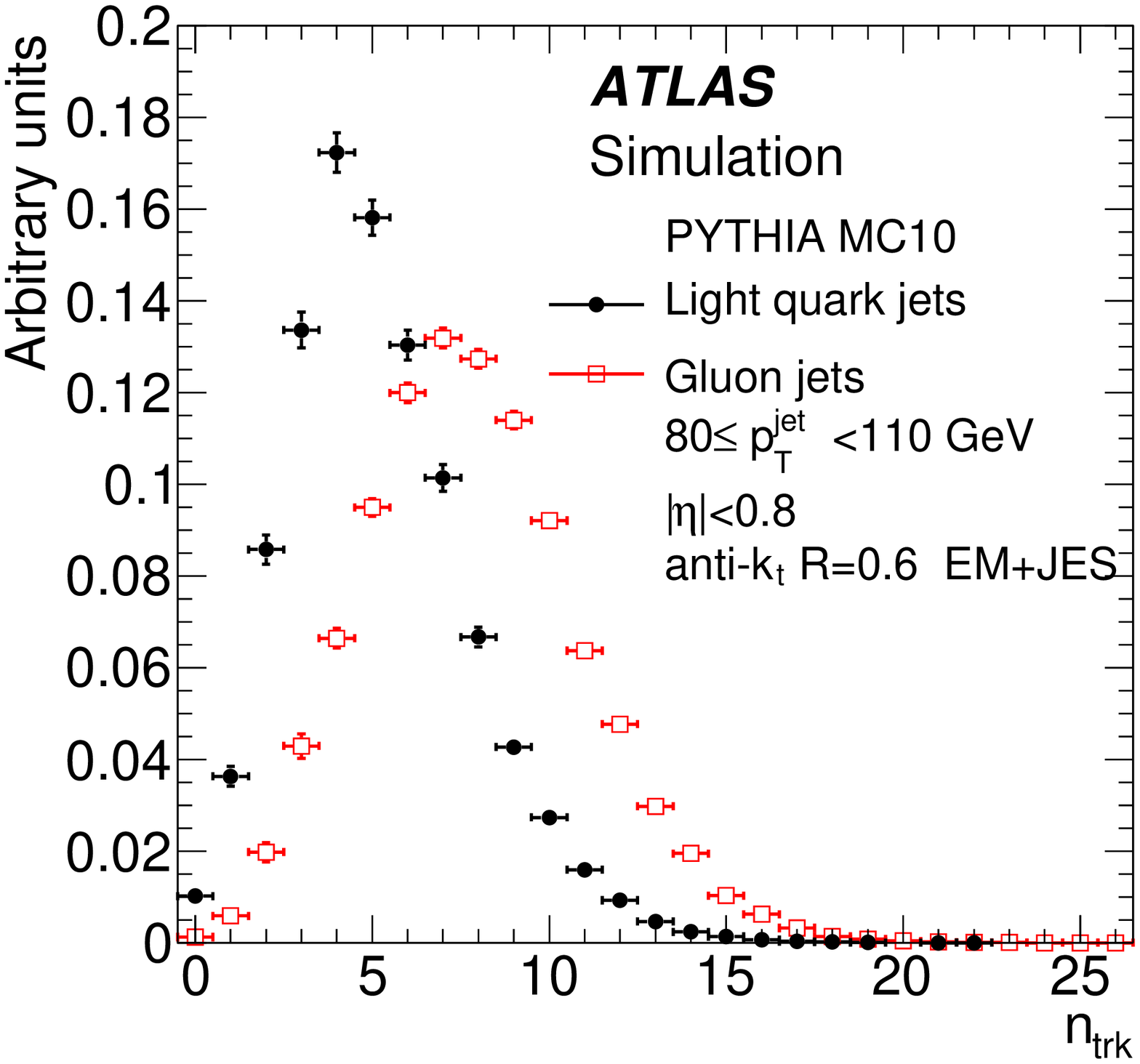} }
\hspace{1.cm}
\subfloat[Jet \width{}]                {\includegraphics[width=0.45\textwidth]{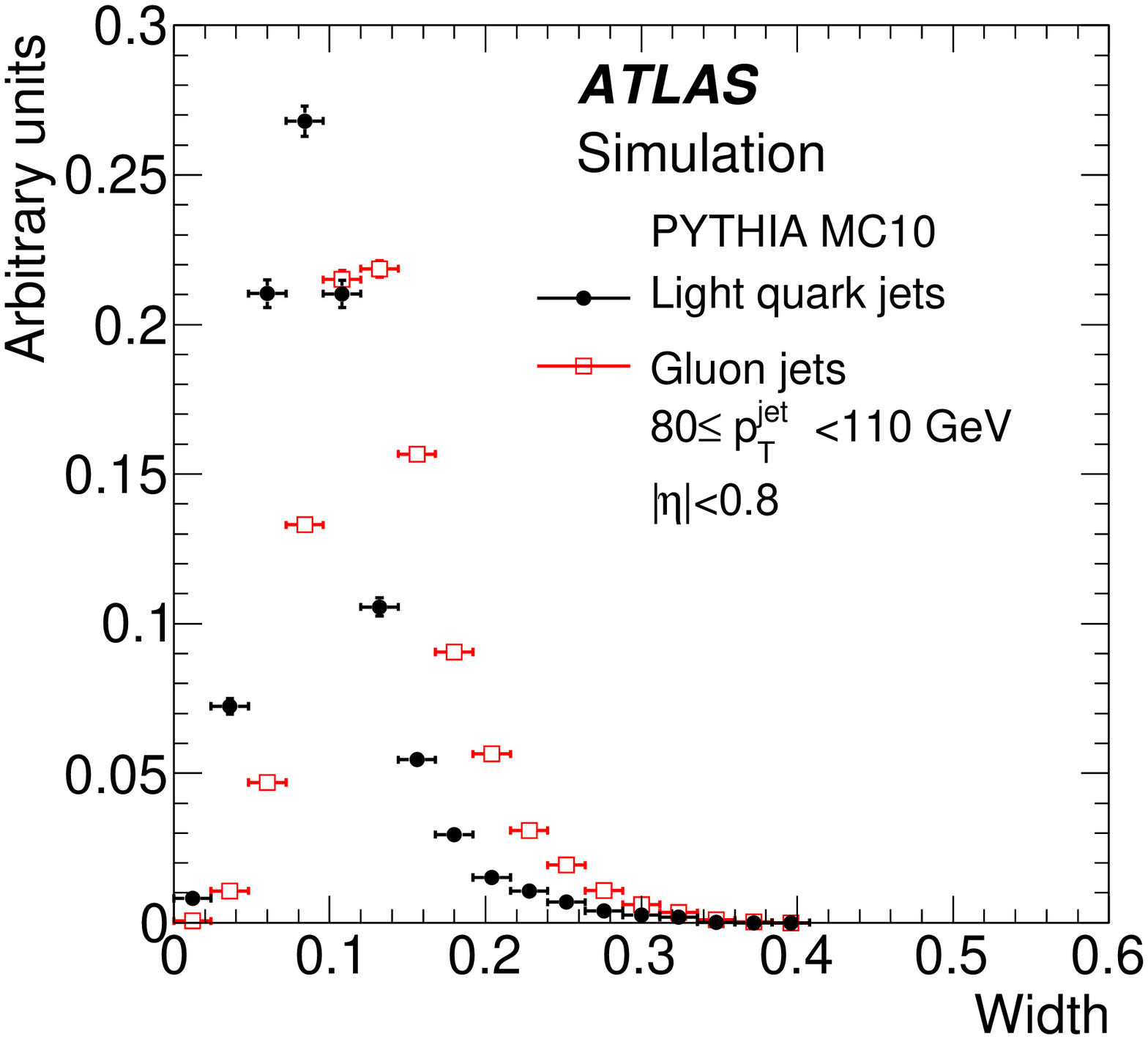} }
\caption{Distribution of the number of tracks associated to the jet $n_{\rm trk}$ (a) 
and the jet \width{}  (b) for isolated \antikt{} jets with $R = 0.6$ classified as light quark jets (solid circles) 
and gluon jets (open squares) in the Monte Carlo simulation. 
Jets with $| \etajet | < 0.8$ and $80 \leq \ptjet < 110$~\GeV\ are shown. 
The distributions are normalised to unit area. Uncertainties are statistical only.
\vspace{1.cm}
}
\label{fig:properties}
\end{center}
\end{figure*}

\begin{figure*}[hb!p]
\begin{center}
\subfloat[Number of tracks $n_{\rm trk}$]{\includegraphics[width=0.45\textwidth]{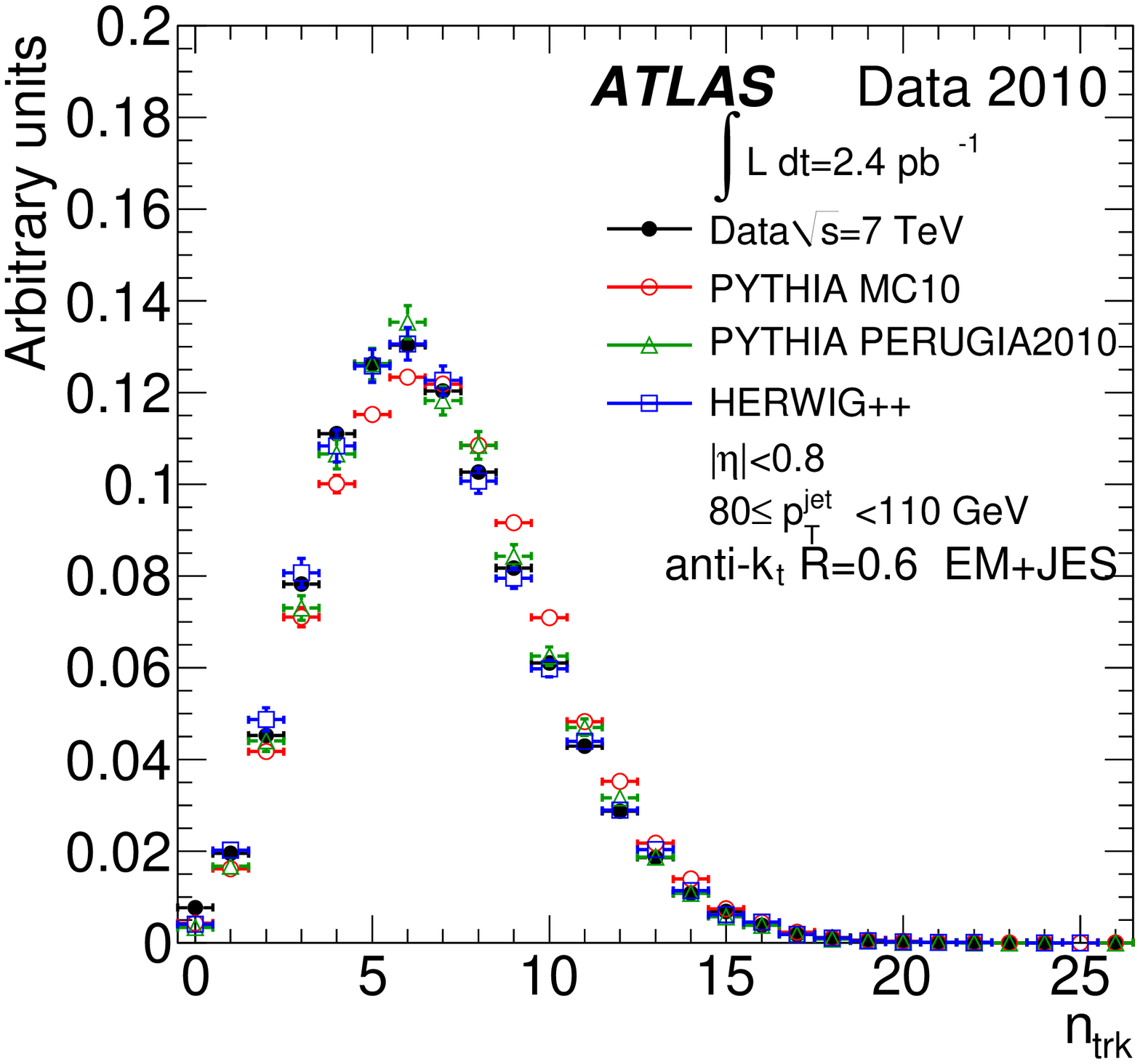} }
\hspace{1.cm}
\subfloat[Jet \width{}]                {\includegraphics[width=0.45\textwidth]{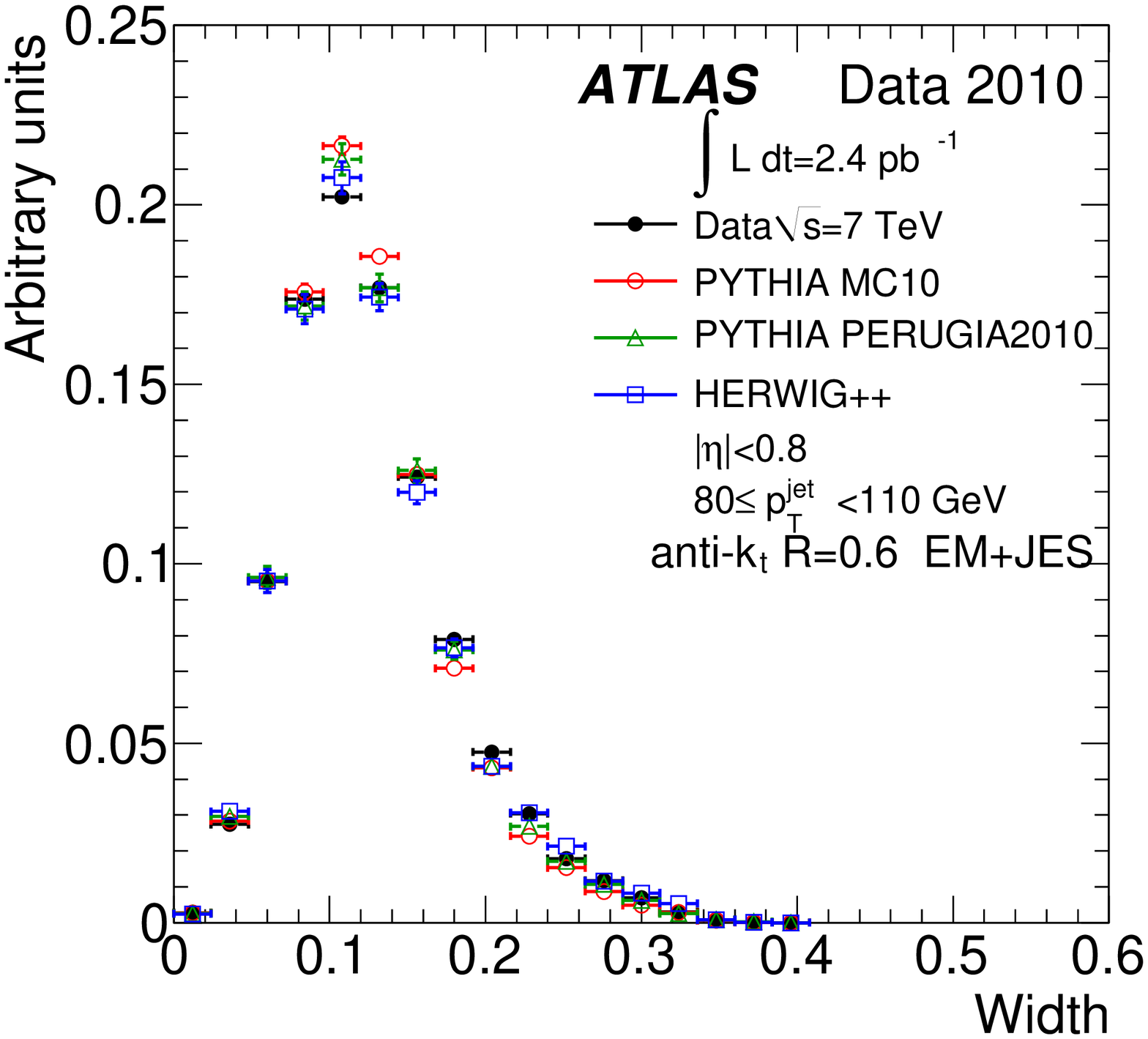} }
\caption{
Distribution of the number of tracks associated to the jet, $n_{\rm trk}$ (a) and the jet \width{}  (b) 
for isolated \antikt{} jets with $R = 0.6$ in data (solid circles) and Monte Carlo simulation.
The \pythia{} MC10 tune (open circles) and \Perugia 2010 tune (open triangles), 
and \herwigpp{} (open squares) distributions 
are shown for jets with $|\eta| < 0.8$ and $80 \leq \ptjet < 110$~\GeV. 
The distributions are all normalised to unity. Uncertainties are statistical only.
}
\label{fig:propertiesDataMC}
\end{center}
\end{figure*}

\begin{figure*}[ht!p]
\begin{center}
\subfloat[Number of tracks $n_{\rm trk}$]{\includegraphics[width=0.45\textwidth]{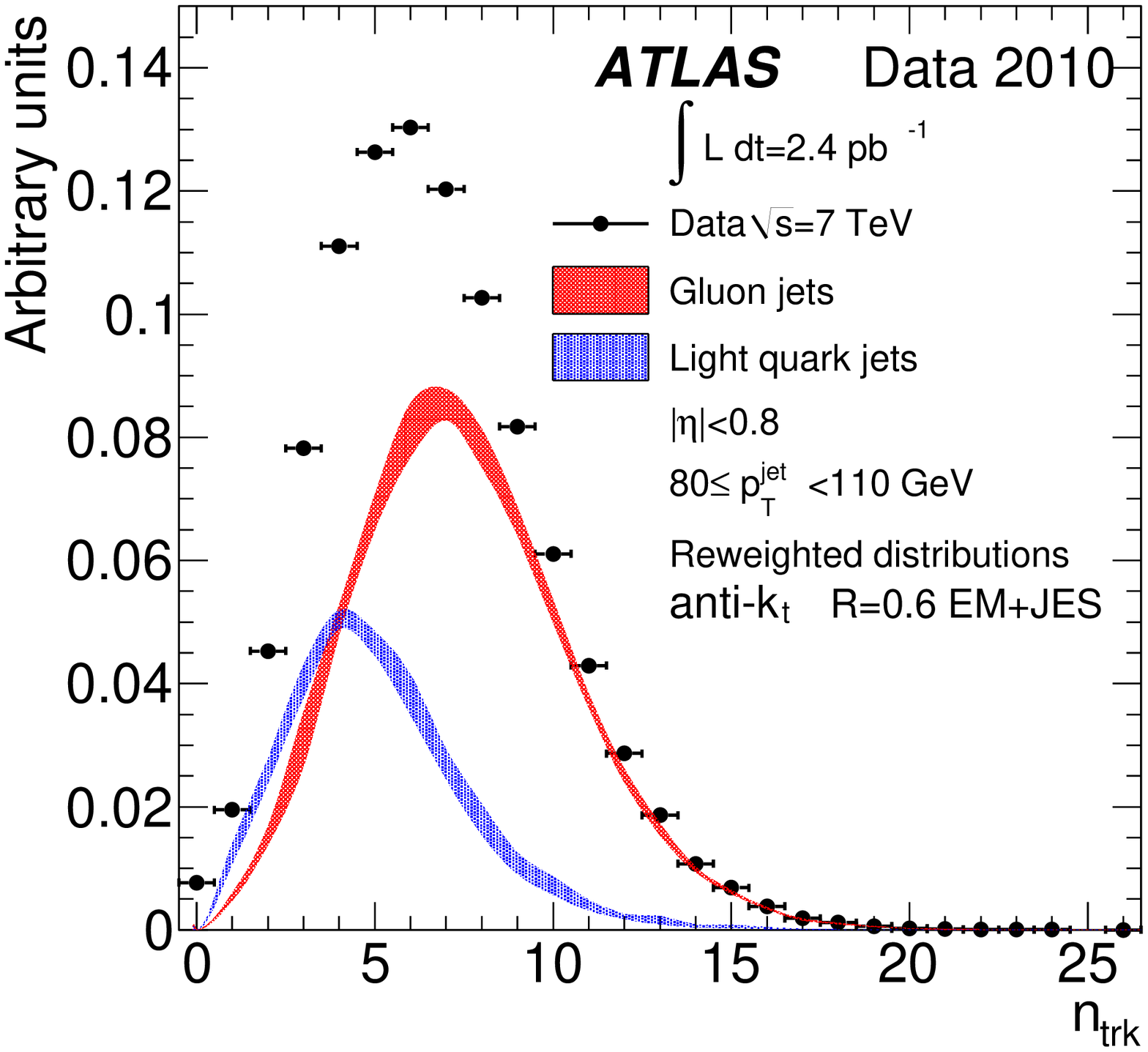} }
\hspace{1.cm}
\subfloat[Jet \width{} ]               {\includegraphics[width=0.45\textwidth]{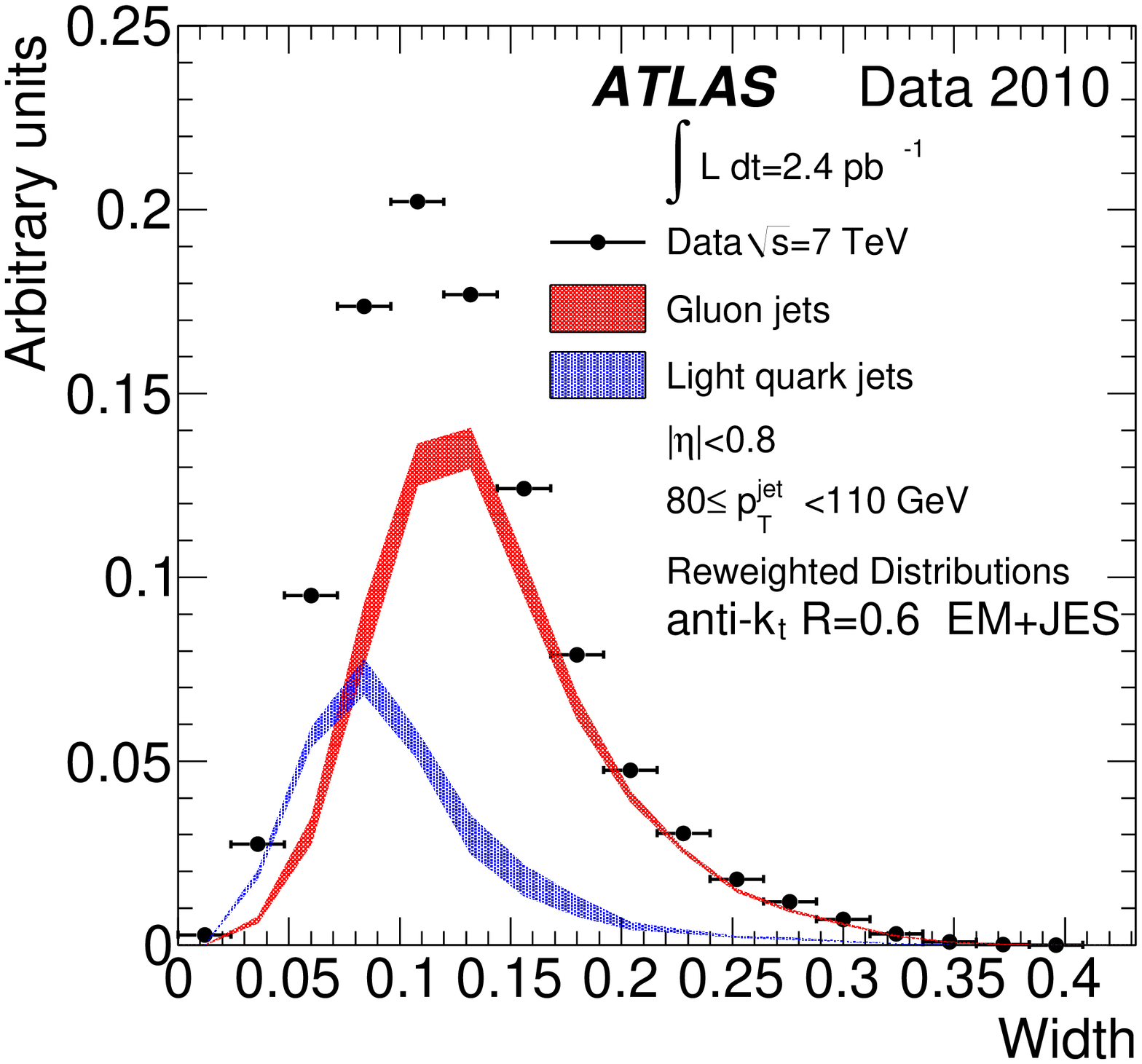} }
\caption{Distribution of the number of tracks associated to the jet, $n_{\rm trk}$ (a) and the jet \width{}  (b) 
for isolated \antikt{} jets with $R = 0.6$ in data (closed circles) and Monte Carlo simulation (bands).
Light quark jets are shown as a dark band, gluon jets are shown as a light band. The width of the band represents the maximum 
variation among the \pythia{} MC10 and \Perugia 2010 tunes and the \herwigpp{} Monte Carlo simulation samples.
Jets with $|\eta| < 0.8$ and $80 \leq\ \ptjet < 110$~\GeV\ are included.
The inclusive distributions are all normalised to unity.
The inclusive Monte Carlo distributions, including the heavy quark jet contributions (not shown), 
are reweighted to match the inclusive distribution of the data. Uncertainties are statistical only.
}
\label{fig:propertiesVal}
\end{center}
\end{figure*}

%
\section{Light quark and gluon jet response and sample characterisation}
\label{sec:quarkgluon}
\begin{figure}[ht!]
\begin{center}
\includegraphics[width=0.49\textwidth]{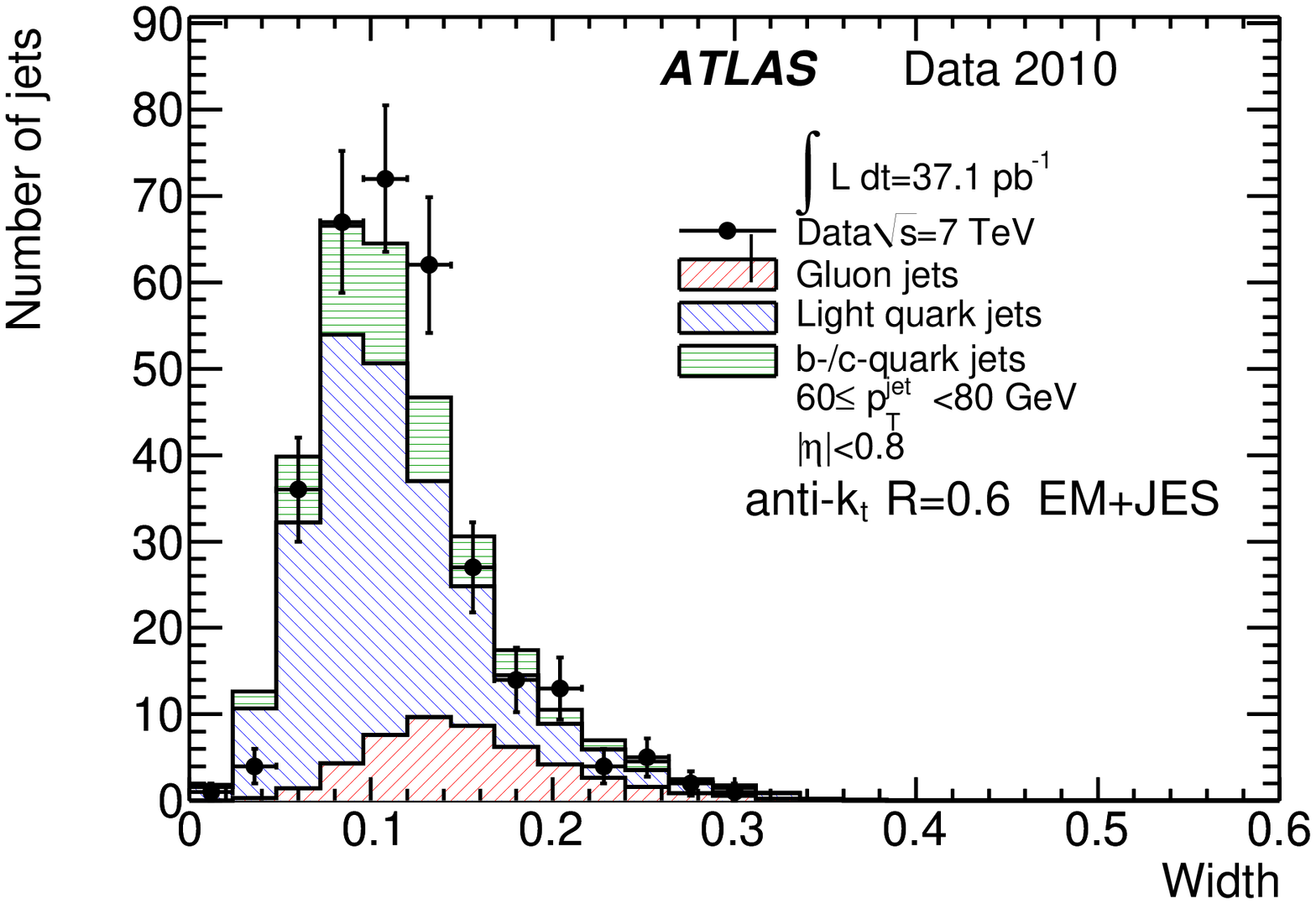} 
\caption{ The jet \width{} template fit in a \gammajet{} data sample using templates 
derived from the inclusive jet Monte Carlo simulation sample created using the \pythia{} MC10 tune. 
Jets with $|\eta| < 0.8$ and $60 \leq \ptjet <80$~\GeV\ are shown. 
The fraction of heavy quark jets is taken directly from the MC simulation. 
}
\label{fig:fitGammaJetWidth}
\end{center}
\end{figure}

\begin{figure*}[ht!]
\begin{center}
\subfloat[Number of tracks $n_{\rm trk}$ template fits ]{\includegraphics[width=0.45\textwidth]{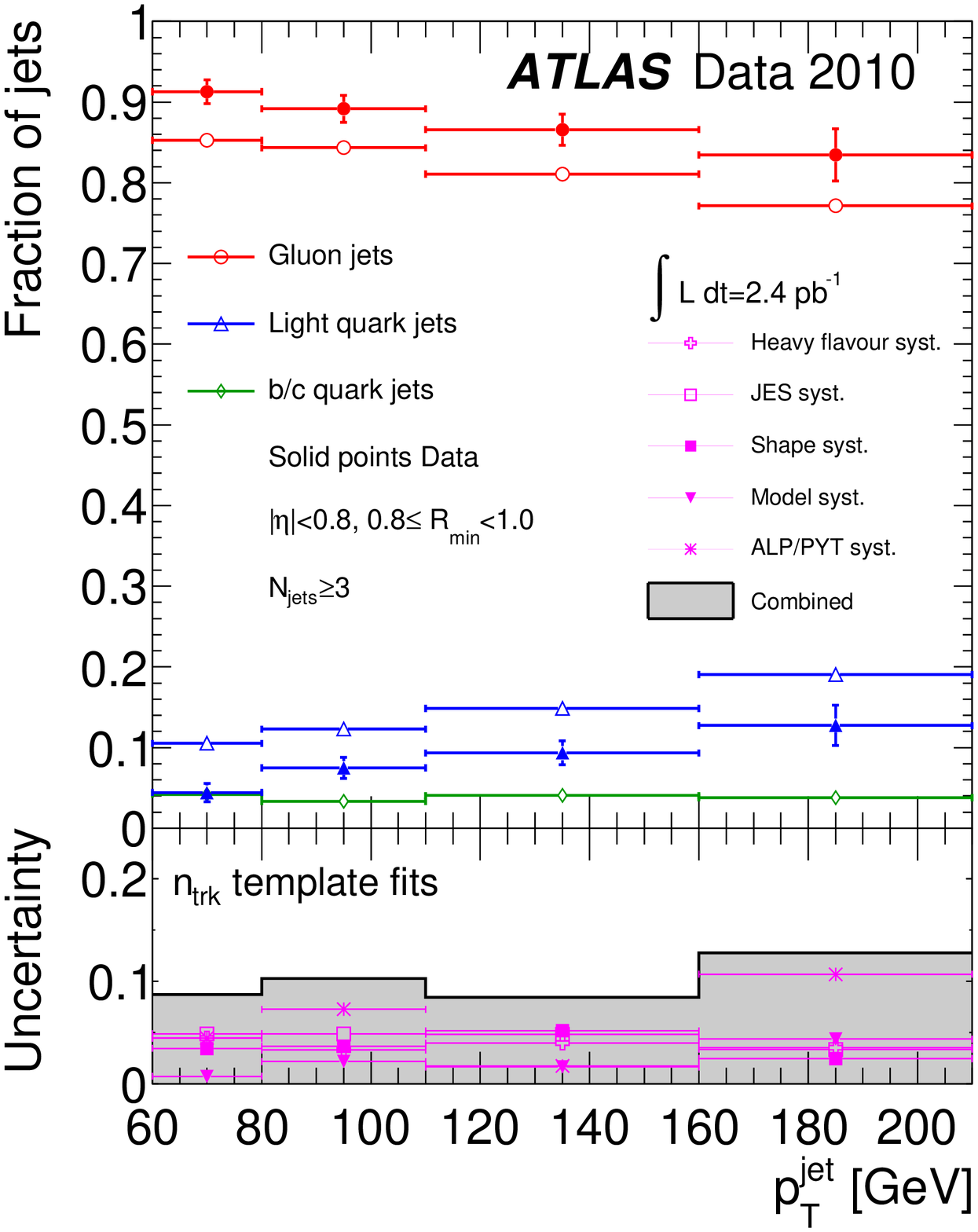}}
\hspace{1.cm}
\subfloat[Jet \width{}  template fits]                  {\includegraphics[width=0.45\textwidth]{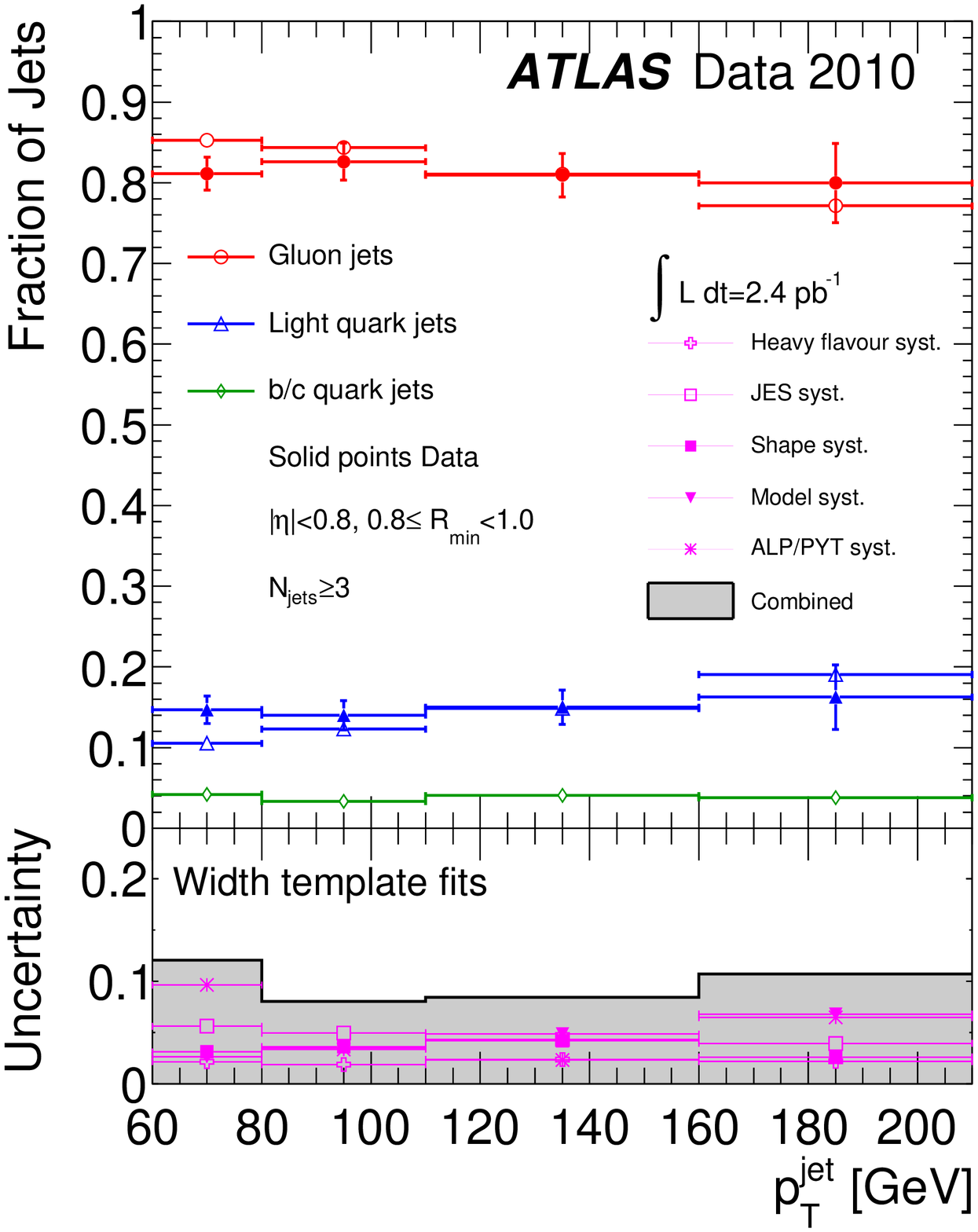}}
\caption{ 
Fitted values of the average light quark and gluon jet fraction in events with three or more jets as a function of \ptjet{}
calculated using the number of tracks $n_{\rm trk}$ templates (a) and the jet \width{}  templates (b). 
Non-isolated \antikt{} jets ($0.8 \leq \Rmin < 1.0$) with $R=0.6$ and with $|\etajet| < 0.8$ 
calibrated with the \EMJES{} scheme are shown. 
The fraction of heavy quark jets is fixed to that of the Monte Carlo simulation.
The flavour fractions obtained in data are shown with closed markers, 
while the values obtained from the Monte Carlo simulation are shown with open markers. 
The error bars indicate the statistical uncertainty of the fit.
Below each figure the impact of the different systematic effects is shown with markers
and the combined systematic uncertainty is indicated by a shaded band. }
\label{fig:flavCompWidth}
\end{center}
\end{figure*}

\begin{figure*}[ht!]
\begin{center}
\subfloat[Number of tracks $n_{\rm trk}$ template fits]  {\includegraphics[width=0.45\textwidth]{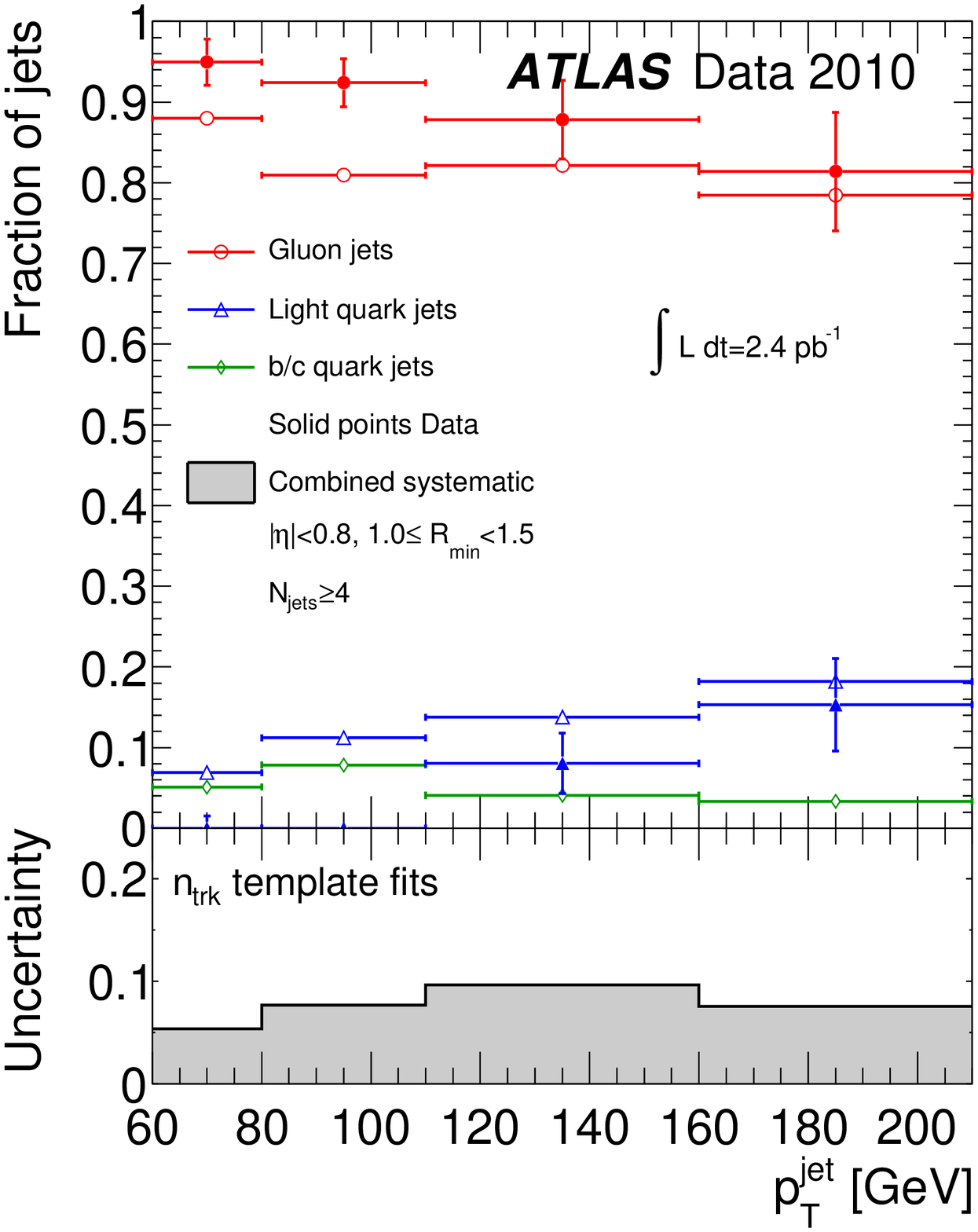} }
\hspace{1.cm}
\subfloat[Jet \width{} template fits]                 {\includegraphics[width=0.45\textwidth]{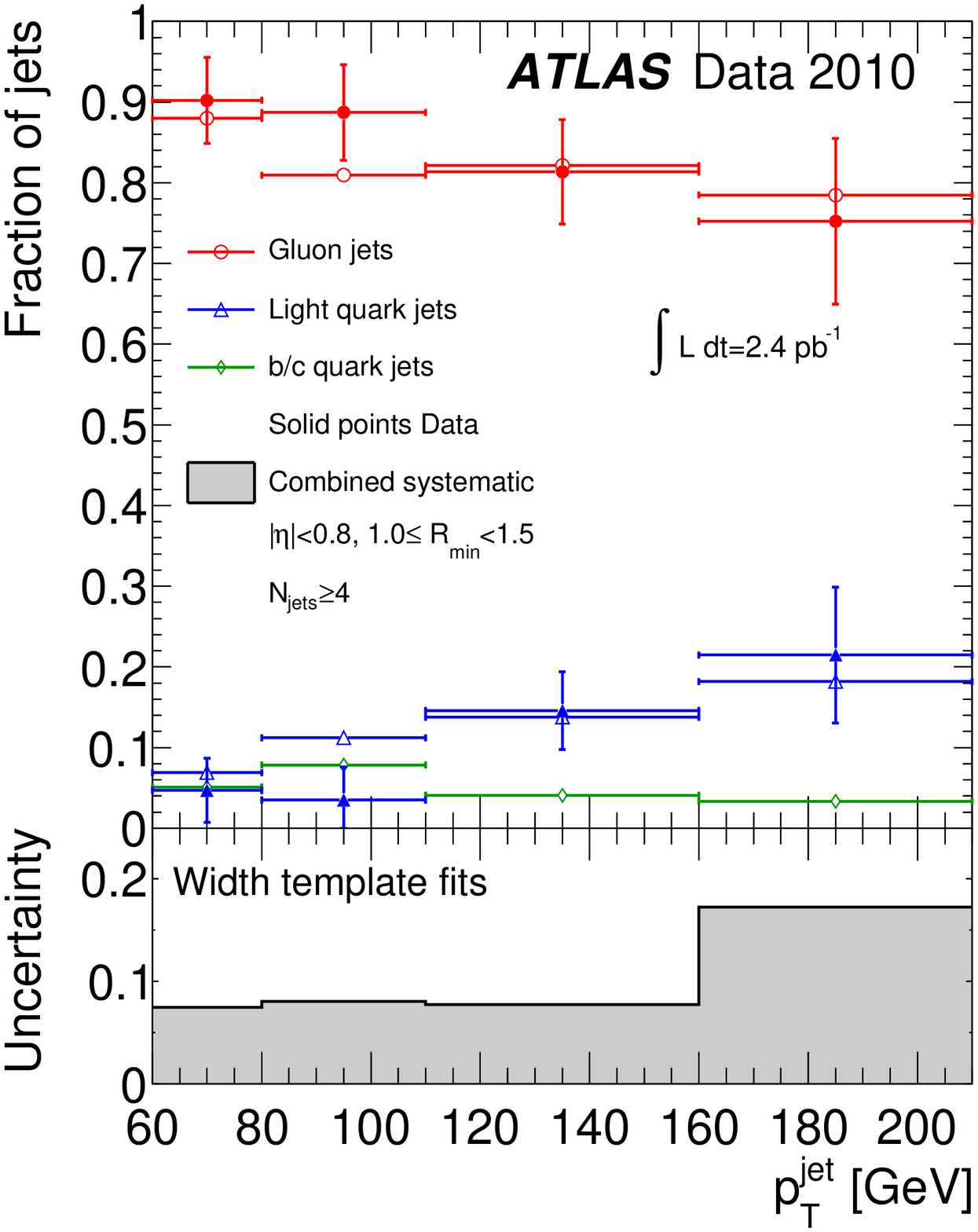} }
\caption{ 
Fitted values of the average light quark and gluon jet fraction in events with four or more jets 
as a function of \ptjet{} for isolated \antikt{} jets with $R = 0.6$ and with $|\etajet| < 0.8$ calibrated with the \EMJES{} scheme. 
The fraction of heavy quark jets is fixed from the Monte Carlo simulation. 
The number of tracks $n_{\rm trk}$ (a) and the jet \width{}  (b) template distributions are used in the fits. 
The flavour fractions obtained in data are shown with closed markers, 
while the values obtained from the Monte Carlo simulation are shown with open markers. 
The error bars indicate the statistical uncertainty of the fit.
Below each figure the systematic uncertainty is shown as a shaded band. }
\label{fig:flavCompNtrk}
\end{center}
\end{figure*}

\begin{figure*}[ht!]
\begin{center}
\subfloat[Number of track $n_{\rm trk}$  template fits] {\includegraphics[width=0.45\textwidth]{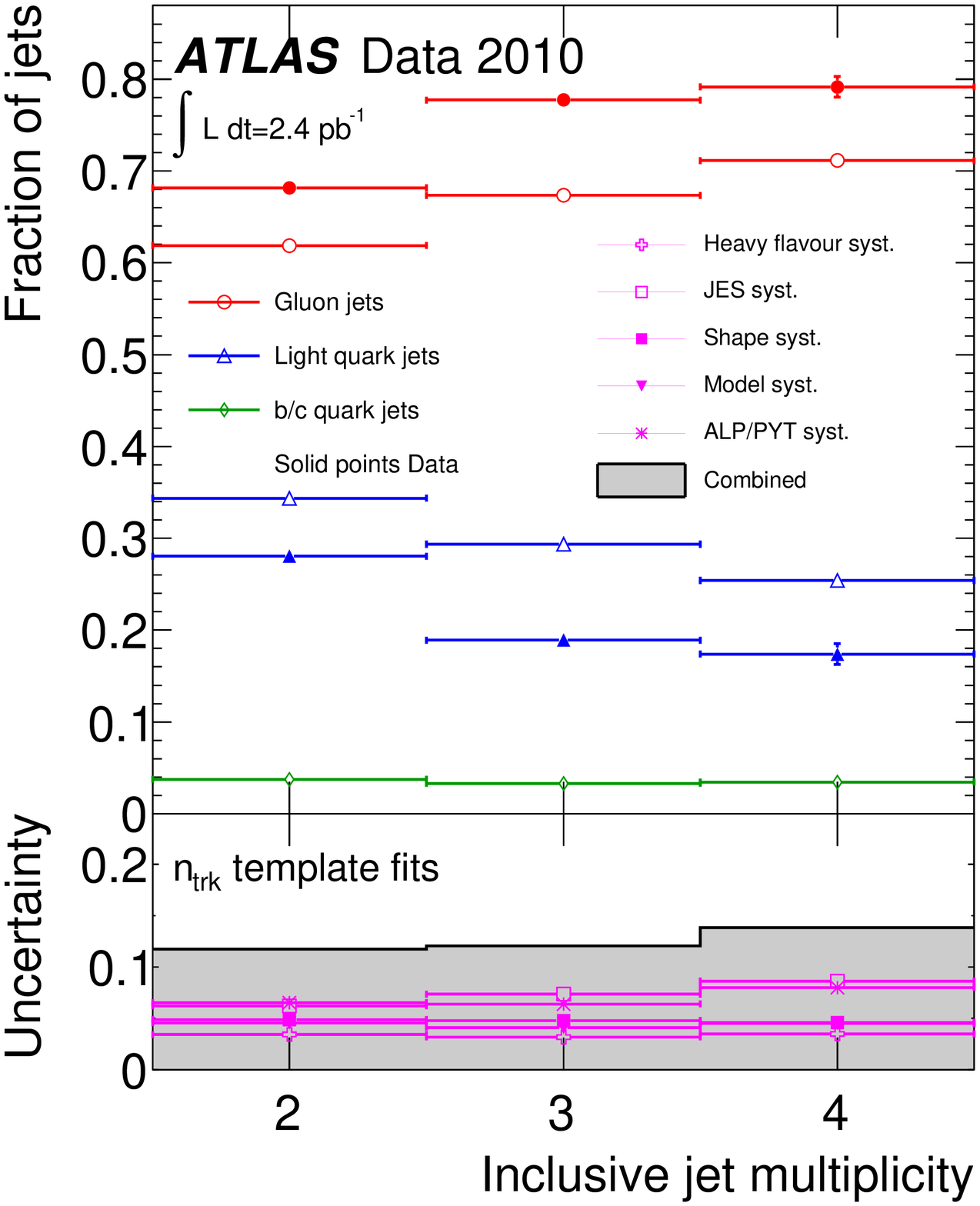} }
\hspace{1.cm}
\subfloat[Jet \width{}  template fits]               {\includegraphics[width=0.45\textwidth]{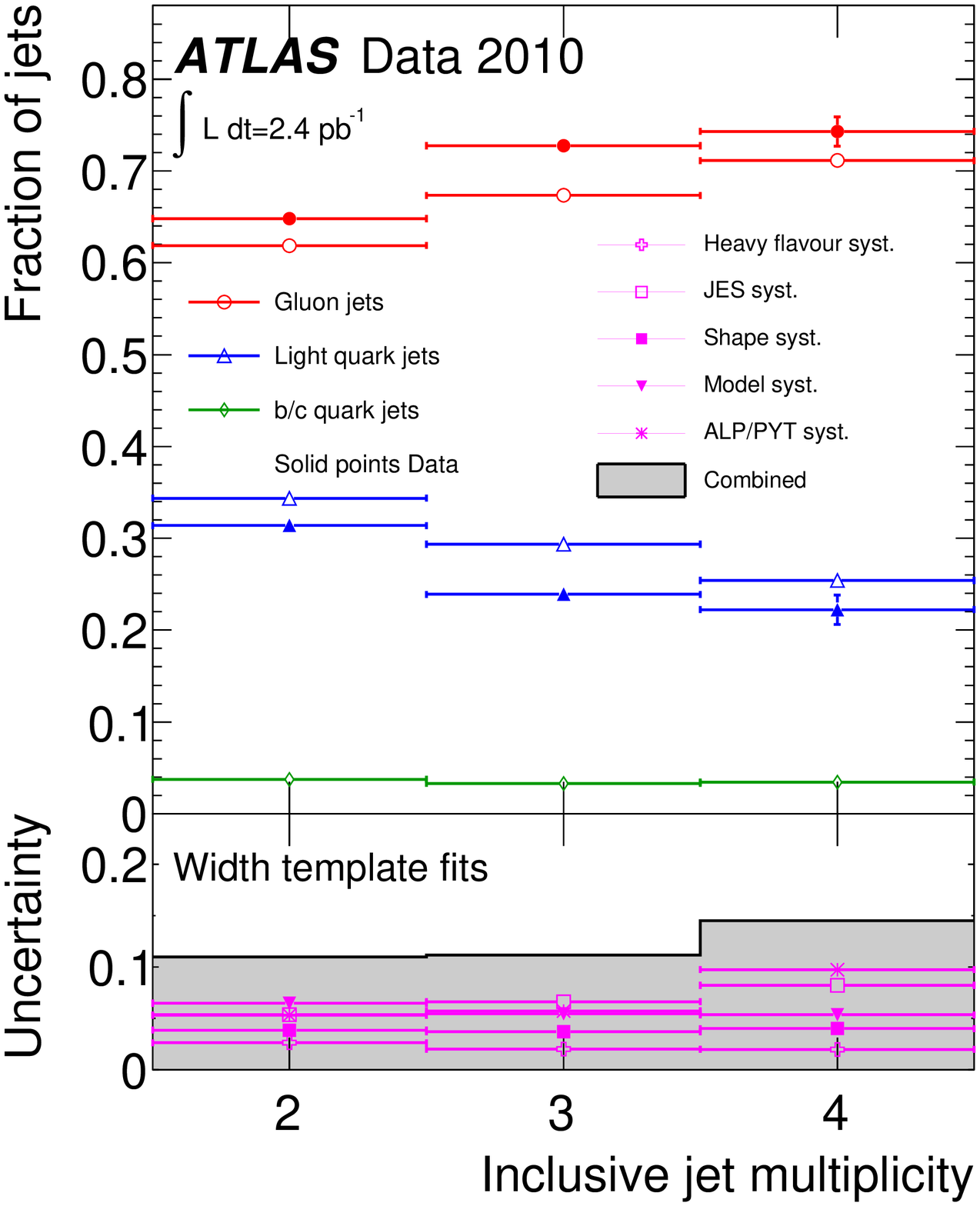} }
\caption{ 
Fitted values of the average light quark and gluon jet fraction as a function of inclusive jet multiplicity
with total uncertainties on the fit as obtained using the number of tracks $n_{\rm trk}$ (a) 
and the jet \width{}  (b) distributions. 
The fraction of heavy quark jets is fixed from the Monte Carlo simulation. 
The flavour fractions obtained in data are shown with closed markers, while the values obtained from the Monte Carlo simulation 
are shown with open markers. 
\Antikt{} jets with $R = 0.6$ calibrated with the \EMJES{} scheme are used.
The error bars indicate the statistical uncertainty of the fit.
Below each figure the impact of the different systematic effects is indicated by markers. 
and the combined systematic uncertainty is shown at the bottom of the figure as a shaded band. }
\label{fig:flavCompMult}
\end{center}
\end{figure*}

In the previous sections the \JES{} uncertainty for inclusive jets was determined.
However, details of the jet fragmentation and showering properties
can influence the jet response measurement. 
In this section the \JES{} uncertainties due to jet fragmentation
which is correlated to the flavour of the parton initiating that jet (e.g. see Ref.~\cite{QGopal})
are investigated.
An additional term in the \JES{} uncertainty is derived for event samples
that have a different flavour content than the nominal Monte Carlo simulation sample.

The jet energy scale systematic uncertainty due to the difference in response 
between gluon and light quark initiated jets (henceforth gluon jets and light quark jets) 
can be reduced by measuring the flavour composition of a sample of jets using template fits to certain jet properties 
that are sensitive to changes in fragmentation.
Although these jet properties may not have sufficient discrimination power to determine the partonic origin of a specific jet, 
it is possible to determine the average flavour composition of a sufficiently large sample of jets.
The average flavour compositions can be determined using
jet property templates built in the Monte Carlo simulation for pure samples.

Templates are constructed in dijet events, which are expected to comprise mostly 
gluon jets at low transverse momentum  and central rapidities.
They are then applied to events with a high-\pt{} photon balancing a high-\pt{} jet (\gammajet{} events), 
which are expected to comprise mostly light quark jets balancing the photon. 
The application of this technique is further demonstrated 
with a sample of multijet events, wherein the jets are initiated mostly by gluons from radiation.


\subsection{Data samples for flavour dependence studies }
\label{sec:datasampleflavour}
Two data samples in addition to the inclusive jet sample discussed before
are used for the studies of the flavour dependence
of the jet response.
\begin{enumerate}
\item {\bf  $\mathrm{\gamma}$-jet sample}
Photons with $\pt > 45$~\GeV{} are selected in the barrel calorimeter 
(with pseudorapidity $|\eta|<1.37$) and a jet  back-to-back ($\Delta \phi > \pi - 0.2$ radians)
to the photon is required.
The second-leading jet in the event is required to
have a \ptjet{} below $10 \%$ of the \ptjet{} of the leading jet.
\Antikt{} jets with $R=0.6$ are used.\Antikt{} jets with $R=0.6$ are used.
\item {\bf Multijet sample}
Jets with $\ptjet{} > 60$~\GeV\ and $|\etajet|<2.8$ are selected and the number
of selected jets defines the sample of at least two, three or four jets.
\end{enumerate}

\subsection{Flavour dependence of the calorimeter response}
\label{sec:calorResponse}
Jets identified in the Monte Carlo  simulation as light quark jets have significantly 
different response from those identified as gluon jets (see Section~\ref{sec:truthjets}).

The flavour-dependence of the jet response is in part
a result of the differences in particle level properties of the two types of jets.
For a given jet \pt{}
jets identified as gluon jets tend to have more particles, and those particles tend to be softer 
than in the case of light quark jets.
Additionally, the gluon jets tend to be wider (i.e. with lower energy density in the core of the jet) 
before interacting with the detector.
The magnetic field in the inner detector amplifies the broadness of gluon jets, since their low-\pt{} 
charged particles tend to bend more than the higher \pt{} particles in light quark jets.
The harder particles in light quark jets additionally tend to penetrate further into the calorimeter.

The difference in calorimeter response between gluon jets and light quark jets in the Monte Carlo simulation
is shown in Figure~\ref{fig:respLightGluCalibs}. 
Jets in the barrel ($|\eta| < 0.8$) and in the endcap ($2.1 \leq |\eta| < 2.8$) calorimeters
are shown separately.
For jets calibrated with the \EMJES{} scheme light quark jets have a $5 - 6 \%$ higher response
than gluon jets at low \ptjet. This difference decreases to about $2 \%$ at high \ptjet.

Since response differences are correlated with differences in the jet properties, 
more complex jet calibration schemes that are able to account for jet shower properties variations
can partially compensate for the flavour dependence.
At low \ptjet{} the difference in response between light quark jets and gluon jets is reduced 
to $4 - 5 \%$ for the \LCWJES{} and \GCWJES{} schemes and about $3 \%$ for the \GS{} scheme. 
For $\ptjet > 300$~\GeV{} the flavour dependence of the jet response is below $1\%$
for the \LCWJES{} and \GCWJES{} and the \GS{} schemes. 

The closer two jets are to one another, the more ambiguous the flavour assignment becomes.
The flavour assignment can become particularly problematic when one truth jet is matched to two reconstructed calorimeter jets 
(``splitting'') or two truth jets are matched to one reconstructed calorimeter jet (``merging'').
Several different classes of close-by jets are examined for changes in the flavour dependence of the jet response.
No significant deviation from the one of isolated jets is found.
Therefore, the cases can be treated separately.
The jet energy scale uncertainty specific to close-by jets is examined further in Section~\ref{sec:closeby}.

\subsection{Systematic uncertainties due to flavour dependence}
\label{sec:flavoursystematics}
Each jet energy calibration schemes restore the average jet energy to better than $2\%$
with small uncertainties in a sample of inclusive jets.
However, subsamples of jets are not perfectly calibrated, as in the case of light quark jets and gluon jets.
The divergence from unity is flavour dependent and may be different in Monte Carlo simulation and data, 
particularly if the flavour content in the data sample is not well-described by the Monte Carlo simulation.
This results in an additional term in the systematic uncertainty for any study using an event or jet selection 
different from that of the sample in which the jet energy scale was derived.

\subsubsection{Systematic uncertainty from MC variations}
In order to test the response uncertainties of exclusive samples of either gluon or light quark jets, 
a large number of systematic variations in the Monte Carlo simulation are investigated
(see Ref.~\cite{atlasjet2010} for details on the variations).
The response difference of quark and gluon jets to that of the inclusive jets is found
to be very similar for each of the systematic Monte Carlo variations.
Therefore the additional uncertainty on the response of gluon jets is neglected.

These conclusions are in good agreement with the studies which derive the calorimeter jet response 
using the single hadron response in Refs.~\cite{singleparticle,atlassingleparticle2011}, where
the uncertainties of the quark and gluon response are similar within  $0.5 \%$.

The results are found to be stable under variations of the Monte Carlo simulation samples
including soft physics effects like colour reconnections.
With more data, a variety of final states may be tested to investigate more details
of the light quark and gluon jet response.

\subsubsection{Systematic uncertainty from average flavour content}
The flavour dependent uncertainty term depends on both the average flavour content of the sample 
and on how well the flavour content is known, e.g. the uncertainty for 
a generic new physics search with an unknown jet flavour composition is different from the uncertainty 
on a new physics model in which only light quark jets are produced.
The response for any sample of jets, $\Response_s$, can be written as\footnote{The following equations
are strictly speaking only valid for a given bin in \pt{} and $\eta$  or in other variables that influence the flavour composition.}:
\begin{eqnarray}
\Response_s = f_g \times \Response_g + f_q \times \Response_q 
+ f_b \times \Response_b + f_c \times \Response_c =  \nonumber \\ 
1 + f_g \times \left( \Response_g - 1 \right) + f_q \times \left( \Response_q - 1 \right) \nonumber \\ 
  + f_b \times \left( \Response_b - 1 \right) + f_c \times \left( \Response_c - 1 \right),
\end{eqnarray}
\noindent where $R_x$ is the detector response to jets and $f_x$ is the fraction of jets for 
$x=$ $g$ (gluon jets), $q$ (light quark jets), $b$ (\bquark{} jets), and $c$ ($c$-quark jets) and $f_g+f_q+f_b+f_c=1$. 
For simplicity, the fraction of heavy quark jets is taken to be known.
This approximation will be dealt with in the systematic uncertainty analysis 
for heavy quarks in Section~\ref{sec:templateFits}.

Since variations in the flavour fractions and the jet flavour response translate
into variations of the jet response for a given sample,
the uncertainty on the jet response can be approximately expressed as:
\begin{eqnarray}
\Delta \Response_s = \Delta f_g \times \left( \Response_g - 1 \right)  +
     \Delta f_q \times \left( \Response_q - 1 \right) +  \nonumber \\ 
      f_g \times \Delta \Response_g + f_q \times \Delta \Response_q + 
      f_b \times \Delta \Response_b + f_c \times \Delta \Response_c ,
\end{eqnarray}
\noindent where $\Delta$ denotes the uncertainty on the individual variables.
Since $f_b$ and $f_c$ are fixed here (i.e. without uncertainty), $\Delta f_g = - \Delta f_q$.
Also, the uncertainties on the response for the exclusive flavour samples (light quark, gluon, $b$, and $c$ quarks) 
are approximately the same as the inclusive jet response uncertainty ($\Delta \Response_j$).

The expression can therefore be simplified:
\begin{eqnarray}
\Delta \Response_s \approx - \Delta f_q \times \left( \Response_g - 1   \right) + 
      \Delta f_q \times \left( \Response_q - 1 \right) + \nonumber \\ 
      f_g \times \Delta \Response_j + f_q \times \Delta \Response_j 
     + f_b \times \Delta \Response_j + f_c \times \Delta \Response_j = \nonumber \\ 
     \Delta f_q \times \left( \Response_q - \Response_g \right) + 
     \left( f_g + f_q + f_b + f_c \right) \times \Delta \Response_j \nonumber \\
     \approx   \Delta f_q \times \left( \Response_q - \Response_g \right) + \Delta \Response_j.
\label{eq:drs}
\end{eqnarray}
The second term is the inclusive jet energy scale systematic uncertainty, 
and the first term is the additional flavour dependent contribution.

Dropping the inclusive jet energy scale systematic uncertainty 
and rewriting Equation~\ref{eq:drs} as a fractional uncertainty,
the flavour dependent contribution becomes:
\begin{equation}
\frac{\Delta \Response_s}{\Response_s} = 
\Delta f_q \times \left( \frac{\Response_q - \Response_g}{\Response_s} \right).
\end{equation}
The uncertainty on the flavour content ($\Delta f_q$) and the inclusive response of the 
sample ($\Response_s$)  depends on the specific analysis.
The difference in response between light quark and gluon jets depends only on the calibration used, 
as discussed in Section~\ref{sec:calorResponse}.

\subsection{Average jet flavour determination}
\label{sec:templateFits}
One way of investigating the flavour composition of a sample is to use different MC generators 
that cover a reasonable range of flavour compositions.
However, these different samples may suffer from under- or overcoverage of the uncertainty or 
from changes in other sample characteristics, e.g. jet \pt{} spectra, 
which may result in a poor estimate of the true uncertainty.
Another approach, pursued in this section, is to estimate the flavour composition of the samples 
by using experimental observables that are sensitive to different jet flavours. 

As described in Section~\ref{sec:calorResponse}, gluon jets tend to have a wider transverse profile 
and have more particles than light quark jets with the same \pt. 
The jet \width{}, as defined in Equation~\ref{eq:jetwidth},
and the number of tracks associated to the jet ($n_{\rm trk}$) are thus expected to be sensitive 
to the difference between light quark jets and gluon jets. 
The jet \width{} may have contributions from pile-up interactions.
In the following discussion only events with exactly one reconstructed primary vertex 
enter the jet \width{}  distributions\footnote{
Techniques to correct for these additional interactions are being developed 
and are discussed in Section~\ref{sec:pileupjetshape}.
}.

The number of tracks associated to a jet is defined by counting the tracks with $\pt > 1$~\GeV\ coming 
from the primary hard scattering vertex with an opening angle between the jet
and the track momentum direction $\Delta R < 0.6$. 
Figure~\ref{fig:properties} shows the jet \width{}  and $n_{\rm trk}$ distributions for 
isolated light quark and gluon jets with $|\eta| < 0.8$ and $80 \le \pt < 110$~\GeV\ in the 
inclusive jet  Monte Carlo simulation sample. 
The gluon jets are broader and have more tracks than light quark jets. 
For this study \antikt{} jets with $R = 0.6$ calibrated with the \EMJES{} scheme are used.

Templates are built from the inclusive jet Monte Carlo sample for the jet \width{}  and $n_{\rm trk}$ 
of light quark and gluon jets separately\footnote{
The $n_{\rm trk}$ and jet \width{}  templates are dealt with independently, 
and the results of their estimates of flavour fraction are not combined.}, 
using the flavour tagging algorithm of Section~\ref{sec:truthjets}.
The templates are constructed in bins of \ptjet, $\etajet$, 
and isolation ($\Delta R$ to the nearest jet, \Rmin).
Fits to the data are performed with these templates to extract the flavour composition. 

Comparisons of the inclusive jet \width{}  and $n_{\rm trk}$ distributions in Monte Carlo simulation 
and data are shown in Figure~\ref{fig:propertiesDataMC} for isolated jets with $R = 0.6$. 
The jet \width{} in Monte Carlo simulation is narrower than in the data for the \pythia{} samples, 
in agreement with other \ATLAS{} analyses~\cite{Atlasjetshape}.

The inclusive $n_{\rm trk}$ and jet \width{}  Monte Carlo simulation distributions are reweighted bin-by-bin 
according to the data distribution. This accounts for the differences observed between the data and Monte Carlo 
simulation.
The same reweighting is applied to the light quark jet and gluon jet distributions.
The reweighted $n_{\rm trk}$ and jet \width{}  distributions for the various Monte Carlo simulation 
samples are shown in Figure~\ref{fig:propertiesVal}.
Since the reweighting is applied to all flavours equally 
the average flavour content of the sample does not change.

After reweighting, the flavour composition of the dijet sample extracted from the data is consistent 
with that of the Monte Carlo simulation.
The extracted values for two representative jet bins are shown in Table~\ref{table:fitResults}.
This result is an important closure test and provides some validation of the templates.

\subsection{Systematic uncertainties of average flavour composition}
Uncertainties on the MC-based templates used in fits to the data result in a systematic uncertainty 
on the extracted flavour composition. 
Systematic effects from the Monte Carlo modelling of the jet fragmentation, 
the jet energy scale and resolution as well as the flavour composition of the sample used
to extract the templates
are discussed in the following. 
Since there is no single dominant uncertainty,
each is individually considered for the extraction of the flavour composition of a sample of jets.
\subsubsection{Monte Carlo modelling of jet width and $n_{\rm trk}$ distributions}
Monte Carlo simulation samples generated with \pythia{} with the MC10 and the \Perugia 2010 tunes 
and \herwigpp{} 
all show reasonable agreement with data (see Figure~\ref{fig:propertiesDataMC}).
Therefore, two separate fits with templates obtained from the latter two alternative Monte Carlo simulation samples are performed.
Reweighting of these alternate samples is performed in the same manner as for the 
nominal \pythia{} MC10 sample.
The largest of the differences in the flavour fractions with respect to the nominal fits is taken 
as the uncertainty due to Monte Carlo modelling.
This estimate should cover physics effects that may impact light quark and gluon jets differently.

\subsubsection{The jet energy scale uncertainty and finite detector resolution} 
The uncertainties in the jet measurement combined with the rapidly falling jet \pt{} spectrum, 
lead to \pt{} bin migrations that affect the templates.
Therefore, the templates are rebuilt with all jet momenta scaled up and down according 
to the inclusive jet energy scale systematic uncertainty.
The difference in the flavour content estimated with the modified templates is taken as a systematic uncertainty.
\subsubsection{Flavour composition of the MC simulation}
The fraction of heavy quark jets in the data is assumed to be the same as that predicted by the \pythia{} MC10
Monte Carlo simulation in the template fits. 
The uncertainty associated with this assumption is estimated by increasing and 
decreasing this Monte Carlo simulation based fraction of heavy quark jets in the template fits 
by a factor of two and repeating the fits with the light quark and gluon jet templates.
The factor of two is taken in order to be conservative in the \gammajet{} and multijet samples, 
due to the lack of knowledge of gluon splitting fraction to $b \bar{b}$.

The \pythia{} Monte Carlo simulation was produced using the modified LO parton distribution functions, 
which may not accurately reproduce the true flavour composition.
Particularly in the more forward pseudorapidity bins, this could produce some inherent biases in the fits.
In order to estimate this uncertainty, the light quark  and gluon jet templates from the standard MC sample 
are combined according to the flavour content of a jet sample generated using \alpgen.
This Monte Carlo generator also uses a leading order PDF, but produces more hard
partons via multiparton matrix elements.
This new combination is then reweighted to match the inclusive distribution in data, 
and the reweighted templates are used to extract the flavour composition of the samples.
The difference between the flavour composition derived in this manner and the flavour composition derived using 
the nominal \pythia{}  Monte Carlo simulation is taken as a systematic uncertainty.
%

\begin{table*}[ht!]
\begin{center}
\begin{tabular}{l|l|l|l}
\hline \hline
       &     & \multicolumn{2}{|c}{Gluon / light / heavy quark jet fraction} \\
Sample &  Selection  &    Data        &                  MC           \\
\hline
\noalign{\smallskip}
Dijet    & $80 \leq \pt < 110$~\GeV, $|\eta| < 0.8$,& $73$ / $22$ / $5\%$                              & $72$ / $23$ / $5\%$ \\ 
         & $1.0\leq \Rmin<1.5$         & $\pm2({\rm stat.})\pm9({\rm syst.})$\%   &               \\ 
\hline
Dijet    & $80\leq p_T<110$~\GeV, $2.1\leq |\eta|<2.8$,        & $45$ / $52$ / $3\%$                    & $39$ / $58$ / $3\%$ \\ 
          & $1.0\leq \Rmin<1.5$                   & $\pm3({\rm stat.})\pm 12({\rm syst.})$\%  &               \\ 
\hline
\gammajet\    & $60\leq \pt <80$~\GeV, $|\eta|<0.8$,          & $16$ / $65$ / $19\%$                             & $6$ / $74$ / $19\%$ \\ 
          & Isolated                                            & $\pm10({\rm stat.})\pm19({\rm syst.})$\% &               \\ 
\hline
Multijet & 3-jet, $80\leq \pt <110$~\GeV, $|\eta|<0.8$,         & $83$ / $13$ / $4\%$                              & $84$ / $12$ / $4\%$ \\ 
          & $0.8\leq \Rmin<1.0$                   & $\pm2({\rm stat.})\pm7({\rm syst.})$\%   &               \\ 
\hline
Multijet & 4-jet, $80\leq \pt<110$~\GeV, $|\eta|<0.8$,         & $89$ / $3$ / $8\%$                               & $81$ / $11$ / $8\%$ \\ 
          & $1.0\leq \Delta R_{\rm min}<1.5$                   & $\pm6({\rm stat.})\pm8({\rm syst.})$\%   &               \\ 
\noalign{\smallskip}\hline \hline
\end{tabular}
\end{center}
\caption{
The results of flavour fits using jet \width{} templates in three data samples: 
dijet events, \gammajet{} events, and multijet events.
The Monte Carlo simulation flavour predictions are taken from \alpgen{} for the dijet and multijet samples 
and \pythia{} for the \gammajet{} sample.
The first uncertainty listed is statistical and the second uncertainty is systematic, 
and both apply to the measured gluon  and light quark jet fractions.
The heavy quark jet fractions in the data are constrained to be the same as those in the MC simulation.
\label{table:fitResults}}
\end{table*}

\subsection{Flavour composition in a photon-jet sample}
\label{sec:flavourdependencephotonjetsample}
The validity of the MC-based templates and fitting method is tested by applying
the method to the \gammajet{} data sample 
and comparing the extracted flavour compositions with the \gammajet{} Monte Carlo simulation predictions. 
This sample should contain a considerably higher fraction of light quark jets than the inclusive dijet sample.
Figure~\ref{fig:fitGammaJetWidth} shows the fit to the jet \width{}  in the \gammajet{} data for jets with 
$|\etajet| < 0.8$ and $60 \le \ptjet <80$~\GeV. 
The heavy quark jet fractions are fixed to those obtained from the \gammajet{} Monte Carlo simulation. 
The extracted light quark and gluon jet fractions are consistent with the true fractions in Monte Carlo simulation, 
though with large uncertainties, as shown in Table~\ref{table:fitResults}.

\subsection{Flavour composition in a multijet sample}
\label{sec:flavourdependencemultijetsample}
The template fit method is also useful for fits to multijet events for various jet multiplicities. 
These events contain additional jets that mainly result from gluon radiation 
and hence include a larger fraction of 
gluon jets than does the \gammajet{} sample.

For this particular analysis, the templates built from the inclusive jet sample are 
used to determine the flavour content of the $n$-jet bin.
However, the \pt{} spectrum of the sub-leading jets is more steeply falling than the leading jet \pt.
An additional systematic uncertainty is estimated to account 
for the difference in \pt{} spectra.
This uncertainty is determined by rederiving templates built with a flat \pt{} distribution 
and a significantly steeper \pt{} distribution than that of the dijet sample.
The slope of the steeply falling distribution is taken from the \pt{} 
of the sixth leading jet in Monte Carlo events with six jets, generated using \alpgen.
The fits are repeated with these modified templates, and the largest difference 
is assigned as a \ptjet{} spectrum shape systematic uncertainty.

Figure~\ref{fig:flavCompWidth} compares the fractions of light quark and gluon jets obtained with a fit of the jet \width{}  
and $n_{\rm trk}$ distributions in events with three or more jets
in data and Monte Carlo simulation as a function of \ptjet{} for non-isolated 
($0.8 \le \Rmin < 1.0$) jets with $|\etajet| < 0.8$. 
The higher gluon jet fractions predicted by the Monte Carlo simulation are reproduced by the fit, 
and the data and the Monte Carlo simulation are consistent.
The total systematic uncertainty on the measurement is below $10 \%$ over the measured \ptjet{} range.

The average flavour fractions obtained from fitting the jet \width{}  and $n_{\rm trk}$ distributions in 
events with four or more jets are shown in Figure~\ref{fig:flavCompNtrk}. 
In both cases, the extracted fractions are consistent with the Monte Carlo predictions
within the systematic uncertainties, 
and the total systematic uncertainty is similar to the one for the three-jet bin.

The extracted light quark and gluon jet fractions, with the total systematic uncertainty 
from the \width{}  and $n_{\rm trk}$ fits, are summarised in Figure~\ref{fig:flavCompMult} 
as a function of inclusive jet multiplicity.
The fractions differ by $10\%$ between the data and the Monte Carlo simulation, 
but are consistent within uncertainties. 
The total systematic uncertainty is around $10 \%$ for each multiplicity bin.
Thus, for the four-jet bin, the flavour dependent jet energy scale
systematic uncertainty can be reduced by a factor of $\sim 10$, from about $6 \%$ obtained assuming a $100 \%$
flavour composition uncertainty to less than $1\%$ after having determined the flavour composition with 
a $10 \%$ accuracy.
A summary of the flavour fit results using the jet \width{}  templates for the different samples 
is provided in Table~\ref{table:fitResults}.

\subsection{Summary of jet response flavour dependence}
The flavour dependence of the jet response has been studied, 
and an additional term to the jet energy scale systematic uncertainty has been derived.

A generic template fit method has been developed to reduce this uncertainty significantly for any given sample of events.
Templates derived in dijet events were applied to both \gammajet{} and multijet events, 
demonstrating the potential of the method to reduce the systematic uncertainty.
The flavour dependent jet energy scale systematic uncertainty can be 
reduced from $\sim 6 \%$ to below $ 1 \%$.

\section{Global sequential calibrated jet response for a quark sample}
\label{sec:gsc-quark-gluon}
%
%
In this section, the performance of the \GS{} calibration (see Section~\ref{sec:calibTechnique})
is tested for a \gammajet{} sample.
The jet energy scale after each \GS{} correction can  be verified using the \insitu{} techniques such as
the direct \pt{} balance technique in \gammajet{} events (see Section~\ref{sec:gammajet}),
where mainly quark induced jets are tested.
The flavour dependence of the \GS{} calibration is tested for jets with $|\eta| < 1.2$.

The measurement is first made with jets calibrated with the \EMJES{} calibration and 
is repeated after the application of each 
of the corrections that form the \GS{} calibration. 
To maximise the available statistics one pseudorapidity bin is used $|\etajet| < 1.2$. 
The Monte Carlo based \GS{}  corrections are applied to both data and Monte Carlo simulation. 
The systematic uncertainty associated with the \GS{} calibration is evaluated by computing the data to Monte Carlo 
simulation ratio of the response after the \GS{} calibration relative to that for the \EMJES{} calibration. 

For $25 \le \ptjet < 45$~\GeV, the agreement between the response in data and Monte Carlo simulation is $3.2 \%$ 
after \EMJES{} and $4.2 \%$ after \GS{} calibration. For $210 \le \ptjet <260$~\GeV, the agreement is $5 \%$ after \EMJES{} 
and $2.5 \%$ after \GS{} calibration. %
Therefore systematic uncertainties derived from the agreement of data and Monte Carlo simulation
vary from $1 \%$ at $\ptjet = 25$~\GeV{} to $2.5 \%$ for $\ptjet = 260$~\GeV. 
These results are compatible within the statistical uncertainty with the uncertainty evaluated using inclusive 
jet events (see Section~\ref{sec:MCBased_vs_DataBased}).

The obtained results indicate that the uncertainty in a sample with a high fraction of light quark jets
is about the same as in the inclusive jet sample.
%

%
\begin{figure*}[ht!p]
\centering
\subfloat[MC generator and detector geometry]{\includegraphics[width=0.45\textwidth]{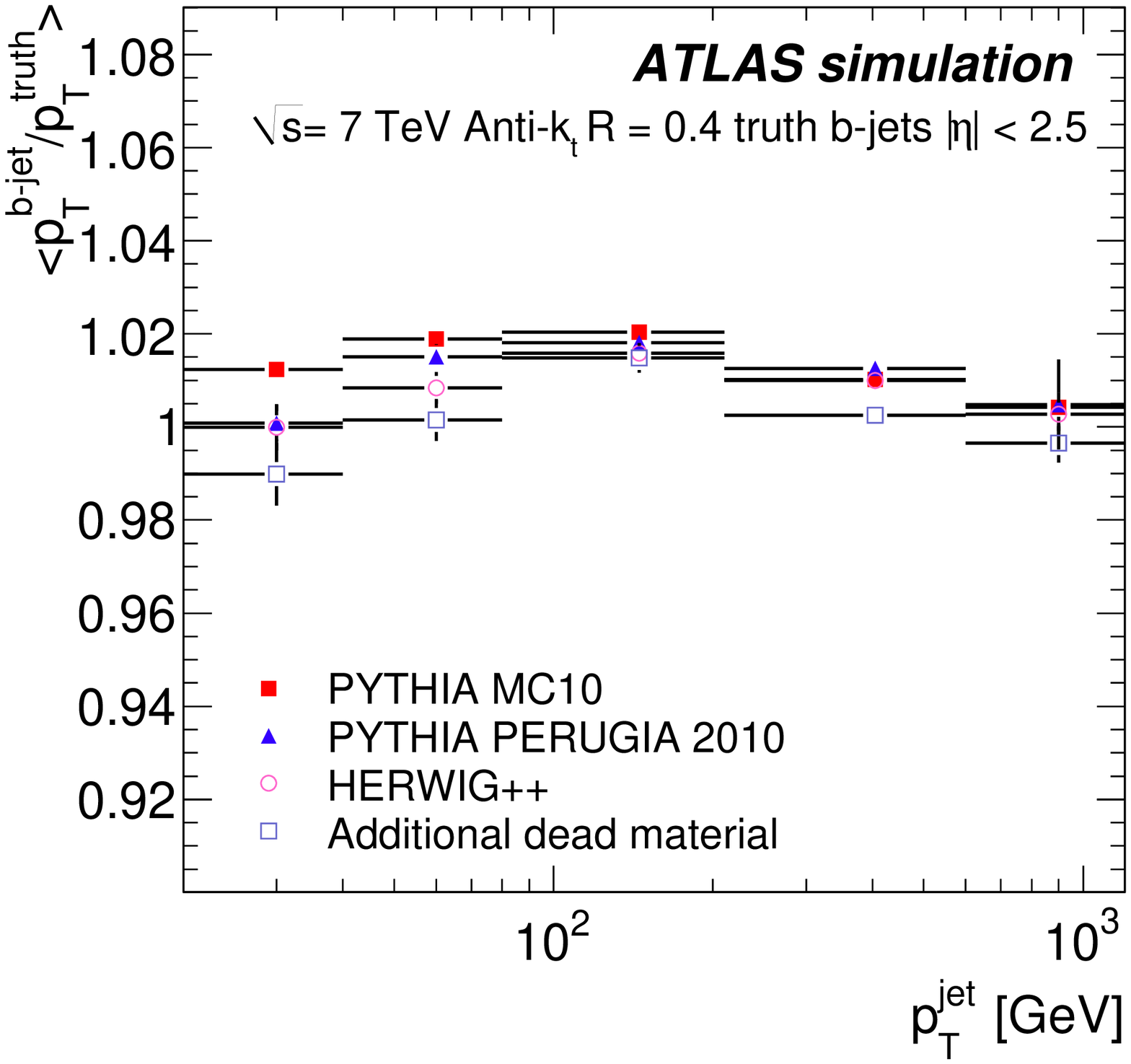}}
\subfloat[\bquark{} fragmentation]{\includegraphics[width=0.45\textwidth]{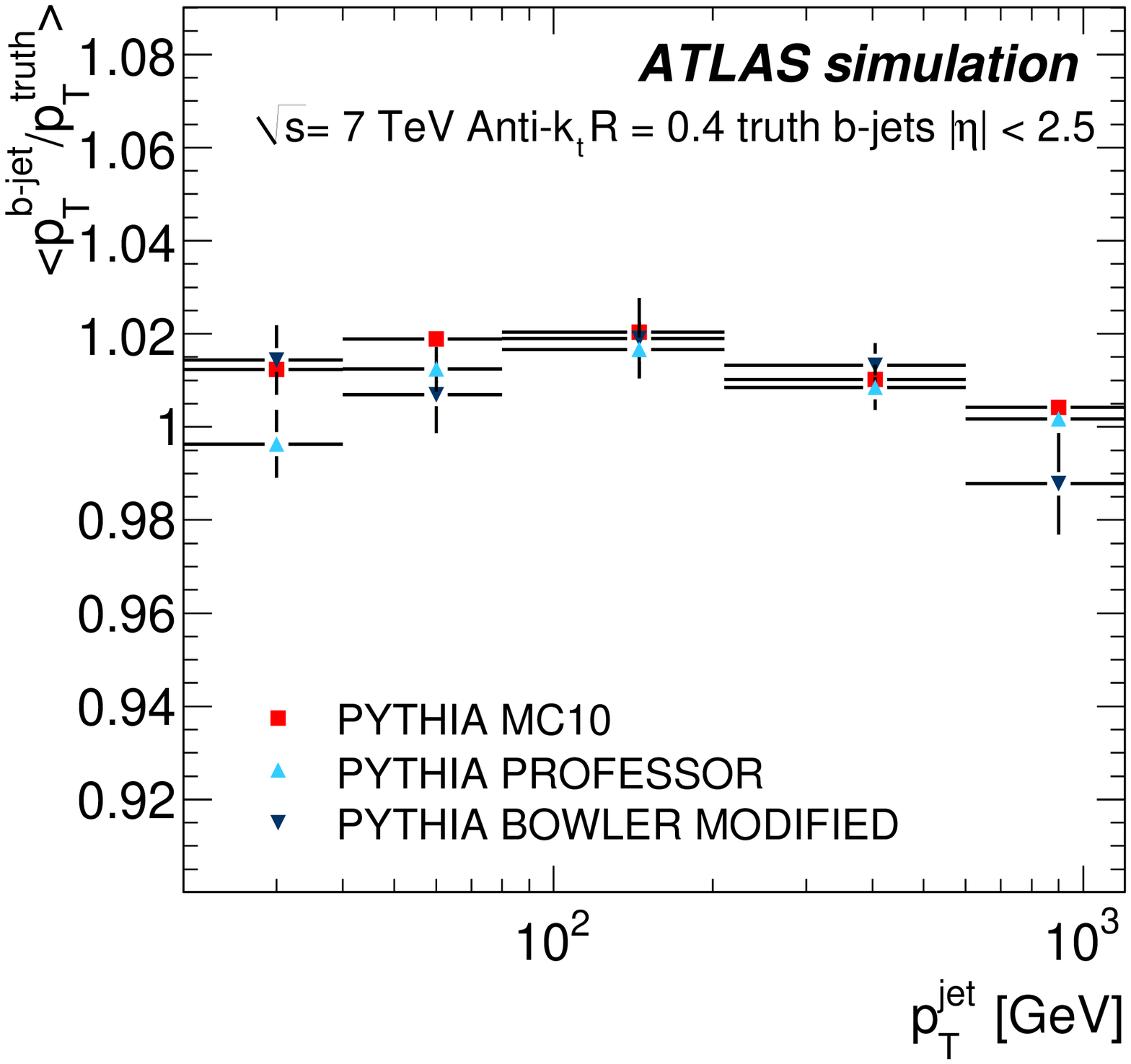}}
\caption{Average response for \bjets{} as a function of \ptjet{}
obtained with the Monte Carlo event generators
\pythia{} with the MC10 and \Perugia 2010 tunes and  \herwigpp{} (a)
and \pythia{} simulations with additional dead detector material.  
Average response  for \bjets{} using 
the \pythia{} Professor  tune and the \pythia{} modified Bowler-Lund fragmentation function
evaluated with respect to the nominal \pythia{} inclusive jet sample (b).
Only statistical uncertainties are shown.
}
\label{fig:bjet1}
\end{figure*}

\section{JES uncertainties for jets with identified heavy quark components}
\label{sec:bjet}
Heavy flavour jets such as jets induced by bottom ($b$) quarks (\bjets)
play an important role in many physics analyses.

The calorimeter jet response uncertainties for \bjets{} is
evaluated using single hadron response measurements in samples of inclusive dijet and $b\bar b$ dijet 
events. 
The \JES{} uncertainty arising from the modelling of the \bquark{} production mechanism and 
the \bquark{} fragmentation can be determined from systematics variations of the Monte Carlo simulation.

Finally, the calorimeter \ptjet{} measurement can be compared to the one from tracks associated to the jets
for inclusive jets and identified \bjets. From the comparison of
data to Monte Carlo simulation the \bjet{} energy scale uncertainty 
relative to the inclusive jet sample is estimated. 

\subsection{Selection of identified heavy quark jets}
Jets are reconstructed using the \antikt{} jet algorithm with $R = 0.4$
and calibrated with the \EMJES{} scheme. Jets with $\ptjet > 20$~\GeV{} and
$|\etajet| < 2.5$ are selected. %

A representative sample of identified \bjets\ is selected by a track-based \btag\ algorithm, 
called the SV0 tagger \cite{cscbook,sv0}. 
This algorithm iteratively reconstructs a secondary vertex in jets and calculates the decay length with respect 
to the primary vertex. The decay length significance is assigned to each jet as a 
tagging weight. 
A jet is identified as a \bjet{} if this weight exceeds a threshold of $5.85$ as explained in Ref.~\cite{sv0}. 
To adjust the Monte Carlo simulation to the $b$-tagging performance in data, a dedicated $b$-tagging calibration 
consisting of ``scale factors'' \cite{sv0} is applied to the simulation and systematic uncertainties 
for the calibration are evaluated. 
For Monte Carlo studies, a sample of \bjets{} is selected using a geometrical matching of
the jet ($\DeltaR < 0.4$) to a \textit{true} $B$-hadron.

\subsection{Calorimeter response uncertainty}
\label{ssec:calob}
The uncertainty of the calorimeter response to identified \bjets{} has been evaluated
using single hadron response measurements \insitu{} and in test-beams~\cite{atlassingleparticle2011}. 
The same method as described in Section~\ref{sec:SingleParticle} is used
to estimate the \bjet{} response uncertainty in events
with top-quark pairs with respect to the one of inclusive jets.

For jets within $|\etajet| < 0.8$ and $20 \leq \ptjet < 250$~\GeV{}
the expected difference in the calorimeter response uncertainty of identified \bjets{}
with respect to the one of inclusive jets is less than $0.5\%$.
It is assumed that this uncertainty extends up to $|\etajet| < 2.5$.

\begin{table}[ht!]
\begin{center}
\begin{tabular}{c|c|c|c}
\hline \hline
Parameter & Nominal & Professor & Bowler-Lund \\ \hline
MSTJ(11) & $4$      &     $5$   &  $4$    \\ 
MSTJ(22) & $2$      &     $2$   &  $2$    \\ 
PARJ(41) & $0.3$    &   $0.49$  &  $0.85$ \\ 
PARJ(42) & $0.58$   &   $1.2$   &  $1.03$ \\ 
PARJ(46) & $0.75$   &   $1.0$   &  $0.85$ \\ 
PARJ(54) & $-0.07$  &           &         \\ 
PARJ(55) & $-0.006$ &           &         \\ 
\hline \hline
\end{tabular}
\caption{\pythia{} steering parameters for the considered variations of the \bquark{} fragmentation functions.
}
\label{fragtable}
\end{center}
\end{table}

\subsection{Uncertainties due to Monte Carlo modelling}
\label{ssec:mcb}
The following uncertainties for \bjets{}  are studied
using systematic variations of the Monte Carlo simulation:
\begin{enumerate}
\item Fragmentation and hadronisation modelling uncertainty obtained by comparing the Monte Carlo generators \herwig{} vs \pythia.
\item Soft physics modelling uncertainty obtained by comparing the \pythia{} MC10 to the 
\pythia{} \Perugia 2010 tune.
\item Modelling uncertainty of the detector material in front and in between the calorimeters.
\item Modelling uncertainty of the fragmentation of \bquarks.
\end{enumerate}

The event generators
\pythia{} and \herwigpp{} are used to evaluate the influence of different hadronisation models,
different parton showers, as well as differences in the underlying event model
(see Section~\ref{sec:MC}). 
Variations in proton parton density functions are also included.

The influence of the soft physics modelling is estimated by replacing the standard 
\pythia{} MC10 tune by the \pythia{} \Perugia 2010 tune.
The impact of additional dead material is tested following the prescription detailed in Section~\ref{sec:JESUncertainties}.

The fragmentation function is used to estimate the momentum carried by the $B$-hadron 
with respect to that of the \bquark{} after quark fragmentation.
The contribution of the \bquark{} fragmentation to the \JES{} uncertainty
is estimated using Monte Carlo samples generated with different sets of tuning parameters 
of two fragmentation functions (see Table \ref{fragtable}).

The fragmentation function included as default in \pythia{} originates from a detailed study of the 
\bquark{} fragmentation function in comparison with OPAL \cite{Abbiendi:2002vt} and SLD \cite{Abe:2002iq} data.  
The data are better described using the symmetric Bowler fragmentation function 
with $r_{Q} = 0.75$ (\pythia{} PARJ(46)), assuming the same modification for $b$- and $c$-quarks.	
The $a$ (\pythia{} PARJ(41)) and $b$ (\pythia{} PARJ(42)) 
parameters of the symmetric Lund function were left with the values 
shown in Table~\ref{fragtable}. A more detailed discussion of uncertainties in the \bquark{} fragmentation function
can be found in Refs.~\cite{Carli:2003cx,Cacciari:1998it,Banfi:2007gu,Ball:2001pq,Cacciari:2002pa}.

The choice of the fragmentation function for this study is based on comparisons to LEP experimental data,
mostly from ALEPH \cite{Heister:2001b} and OPAL \cite{Abbiendi:2002vt},  
as well as from the SLD experiment \cite{Abe:2002iq} included in
a phenomenological study of the \bquark{} fragmentation in top-quark decay \cite{CorcellaMescia:2010}.

To assess the impact of the \bquark{} fragmentation,
the nominal parameters of the \pythia{} fragmentation function
are replaced by the values from a recent tune using the Professor framework \cite{Buckley:2009bj}.
In addition, the nominal fragmentation function is replaced by the
modified Bowler-Lund fragmentation function \cite{Bowler:1981sb}.

For each effect listed above the \bjet{} response uncertainty is
evaluated from the ratio between the response of \bjets{} in the Monte Carlo samples 
with systematic variations to the nominal \pythia{} MC10 \bjet{} sample. 
The deviation from unity of this ratio is taken as uncertainty:
\begin{equation}
{\rm Uncertainty} = 1 - 
\left ( \frac{\Response^{\bjet  }_{\rm variation}}{\Response^{\bjet  }_{\rm nominal}} \right ).
\label{bjetsTotUncertainty}
\end{equation}

The \bjet{} response obtained with \pythia{} for the MC10 and the \Perugia 2010 tunes, the \herwigpp{}
Monte Carlo event generator and using a simulation with additional dead material
is shown in Figure~\ref{fig:bjet1}a. 
Figure~\ref{fig:bjet1}b shows the variation with various fragmentation functions, i.e.
the standard one in the nominal \pythia{} sample versus the ones in the \pythia{} Professor tune sample and the
\pythia{} modified Bowler-Lund fragmentation function sample. 
The response variations are well within about $2 \%$.

\begin{figure}[ht!]
\centering
\includegraphics[width=0.45\textwidth]{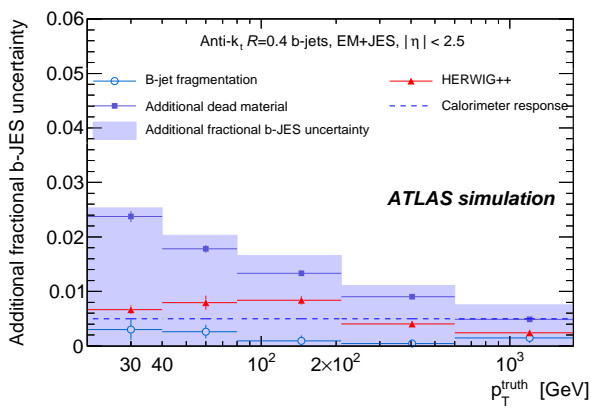}
\caption{Additional fractional \bjet{} \JES{} uncertainty as a function of the truth jet transverse momentum
for \antikt{} jets with $R = 0.4$ calibrated with the \EMJES{} scheme for $| \etajet | < 2.5$.
Shown are systematic Monte Carlo variations using different modelling of the \bquark{} fragmentation and 
physics effects as well as variations in the detector geometry and the uncertainty in the 
calorimeter response to \bjets{} as evaluated from single hadron response measurements.
Uncertainties on the individual points are statistical only.
}
\label{fig:TotalBJESUncertaintyHadron}
\end{figure}

\subsection{Final bottom quark JES uncertainty }
The \bjet{} \JES{} uncertainty is obtained adding the 
calorimeter response uncertainty (see Section~\ref{ssec:calob} for generator details)
and the uncertainties from the systematic Monte Carlo variations (see Section~\ref{ssec:mcb})
in quadrature.

To avoid double counting when combining the \bjet{} uncertainty with the \JES{} uncertainty
of inclusive jets the following effects need to be considered: 
\begin{enumerate}
\item  The uncertainty component due to the \Perugia 2010 tune is not
added, since the effect on \bjets{} is similar to the one on inclusive jets
where it is already accounted for.
\item  The average uncertainty for inclusive jets due to additional dead detector material
is subtracted from the corresponding \bjet{} uncertainty component.
The \JES{} uncertainty due to dead material is smaller for inclusive jets, since
\insitu{} measurements are used.
\end{enumerate}

The resulting additional \JES{} uncertainty for \bjets{} 
is shown in Figure~\ref{fig:TotalBJESUncertaintyHadron}.
It is about $2 \%$ up to $\ptjet \approx 100$~\GeV{} and below $1 \%$ for higher \ptjet.
To obtain the overall \bjet{} uncertainty this uncertainty needs to be added in quadrature to the
\JES{} uncertainty for inclusive jets described in Section~\ref{sec:JESUncertainties}.

 \begin{figure*}[htp!!!]
   \begin{center}
     \subfloat[$|\rapjet\,| < 1.2$]{
       \includegraphics[width=0.5\textwidth]{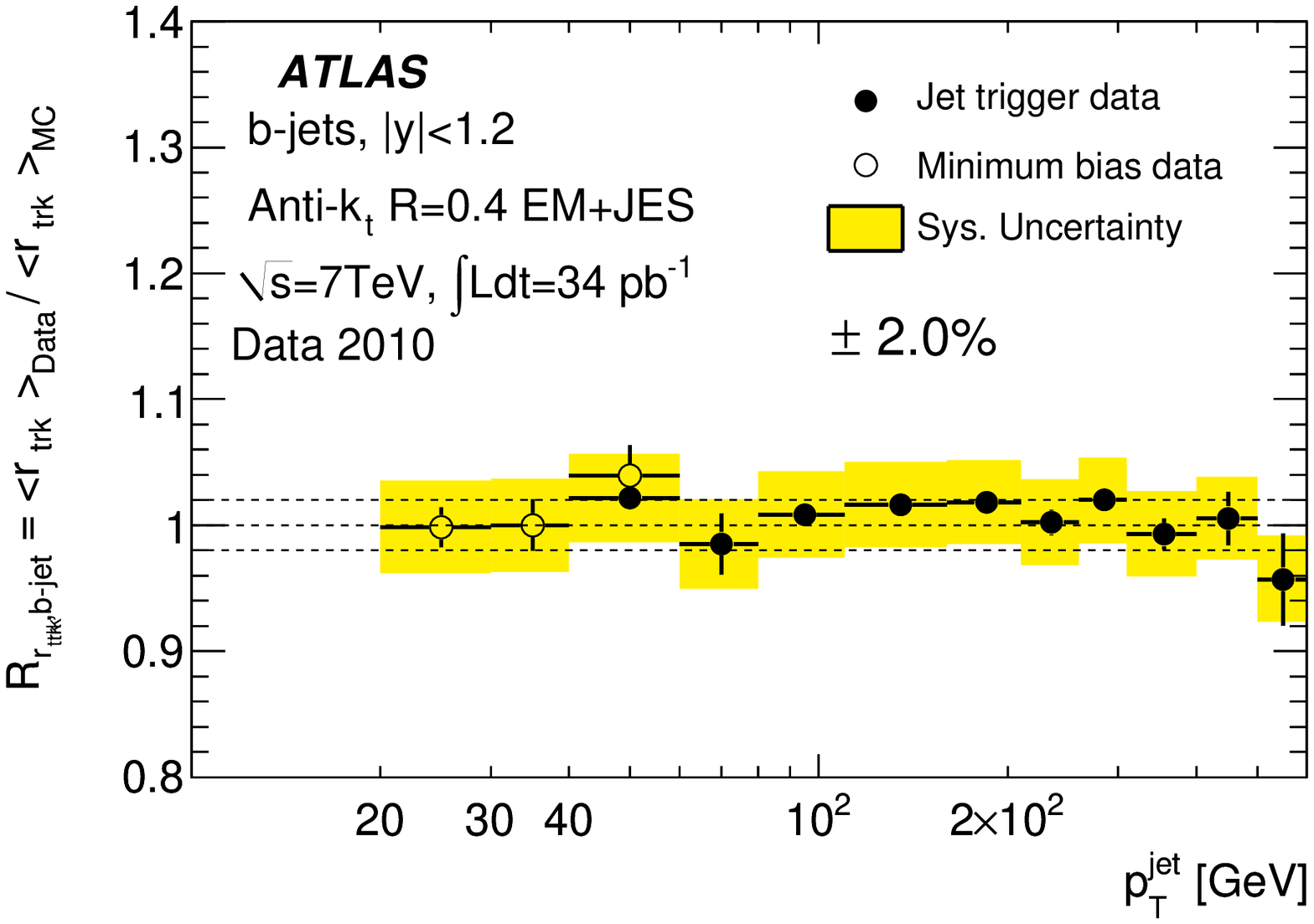}
       \includegraphics[width=0.5\textwidth]{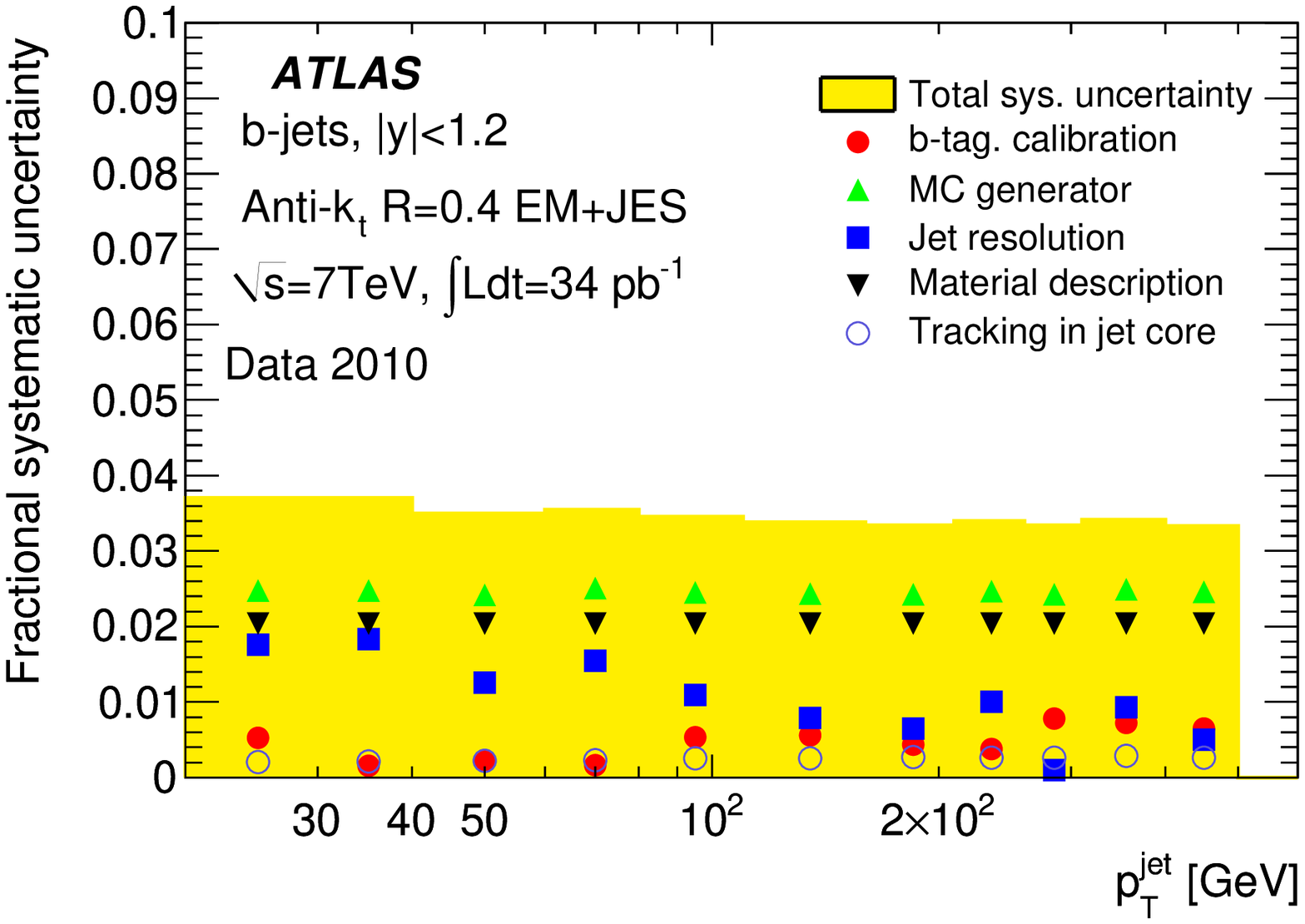}}\\
     \subfloat[$1.2 \leq |\rapjet\,| < 2.1$]{
       \includegraphics[width=0.5\textwidth]{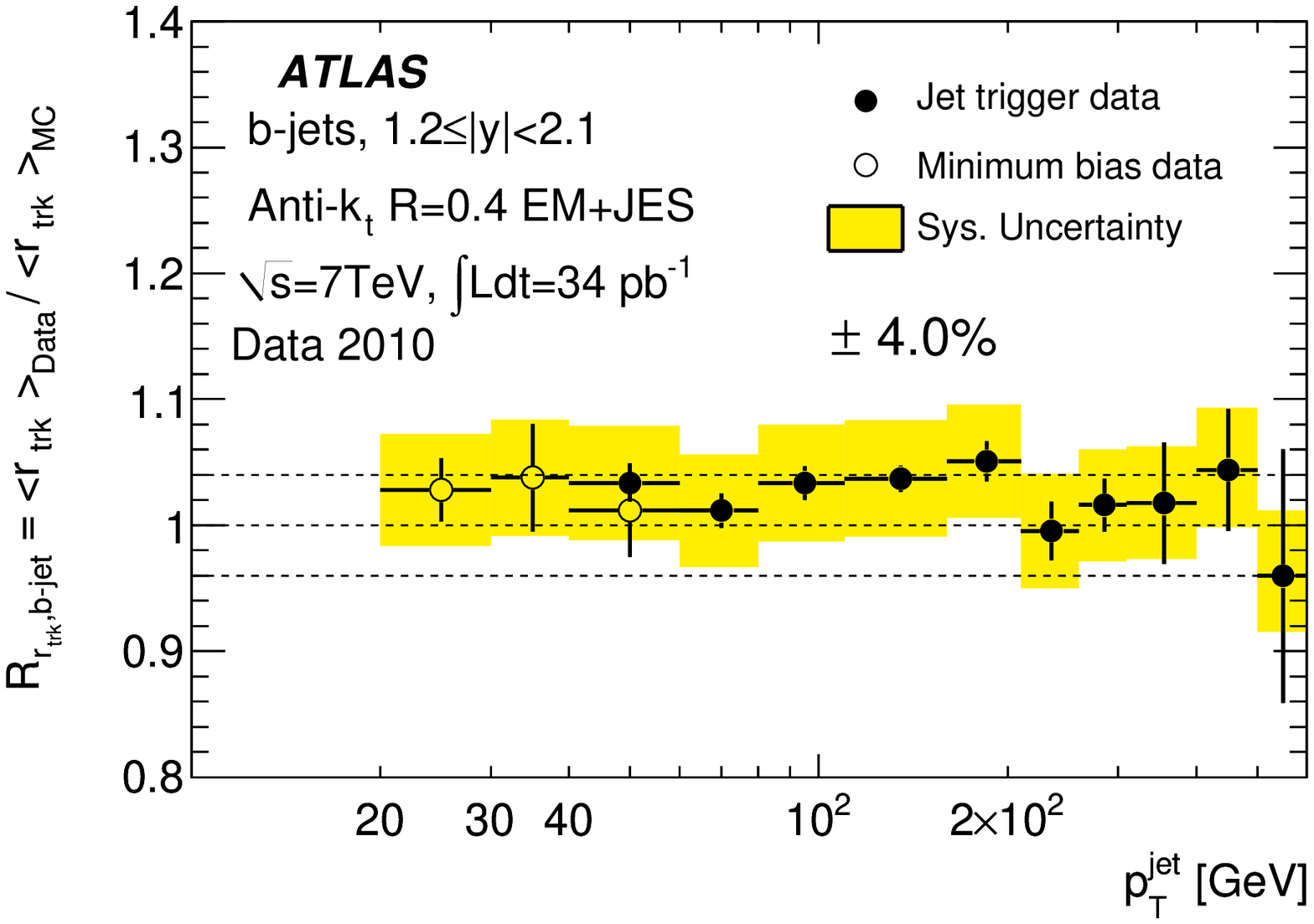}
       \includegraphics[width=0.5\textwidth]{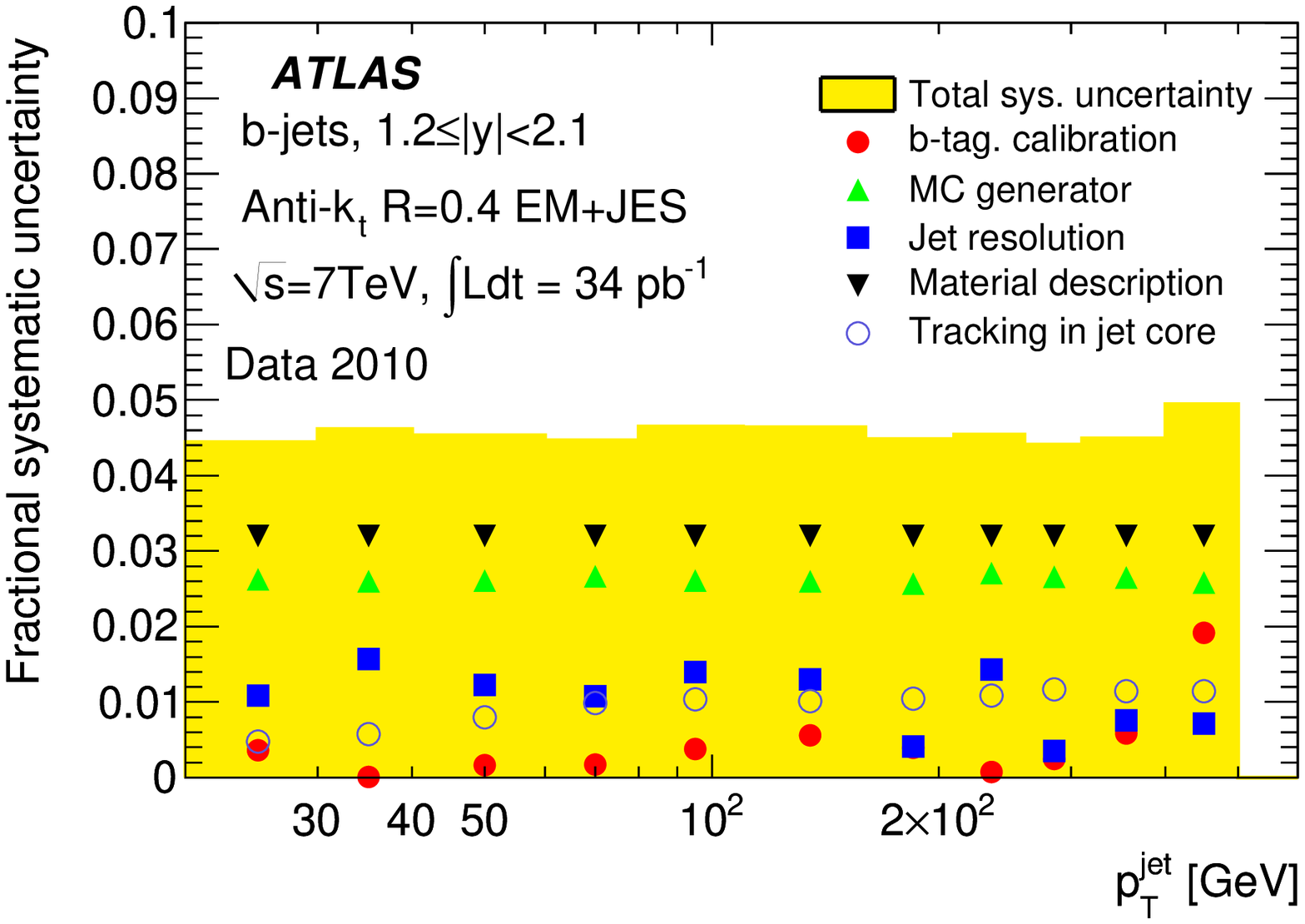}}\\
     \subfloat[$2.1 \leq |\rapjet\,| < 2.5$]{
       \includegraphics[width=0.5\textwidth]{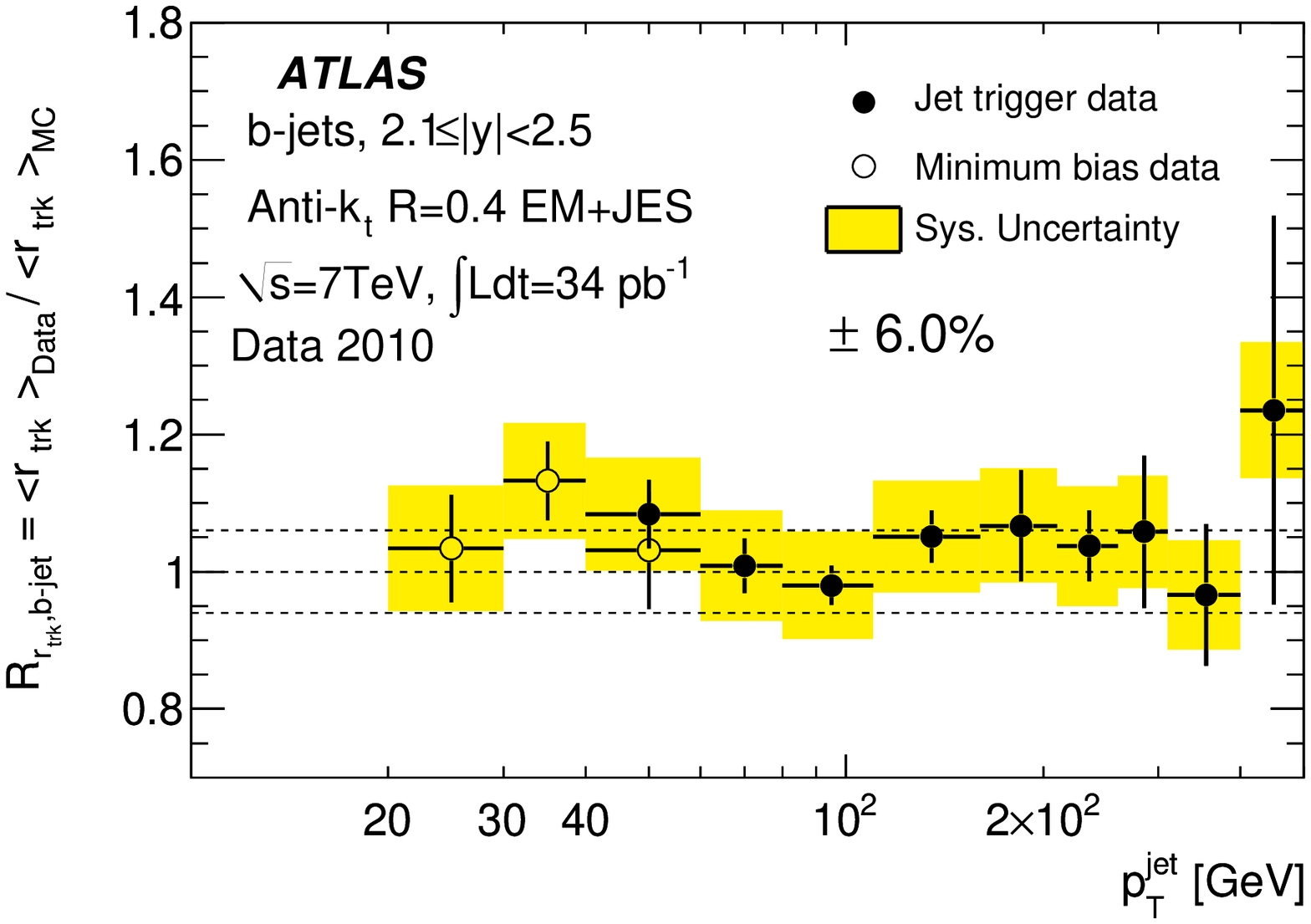}
       \includegraphics[width=0.5\textwidth]{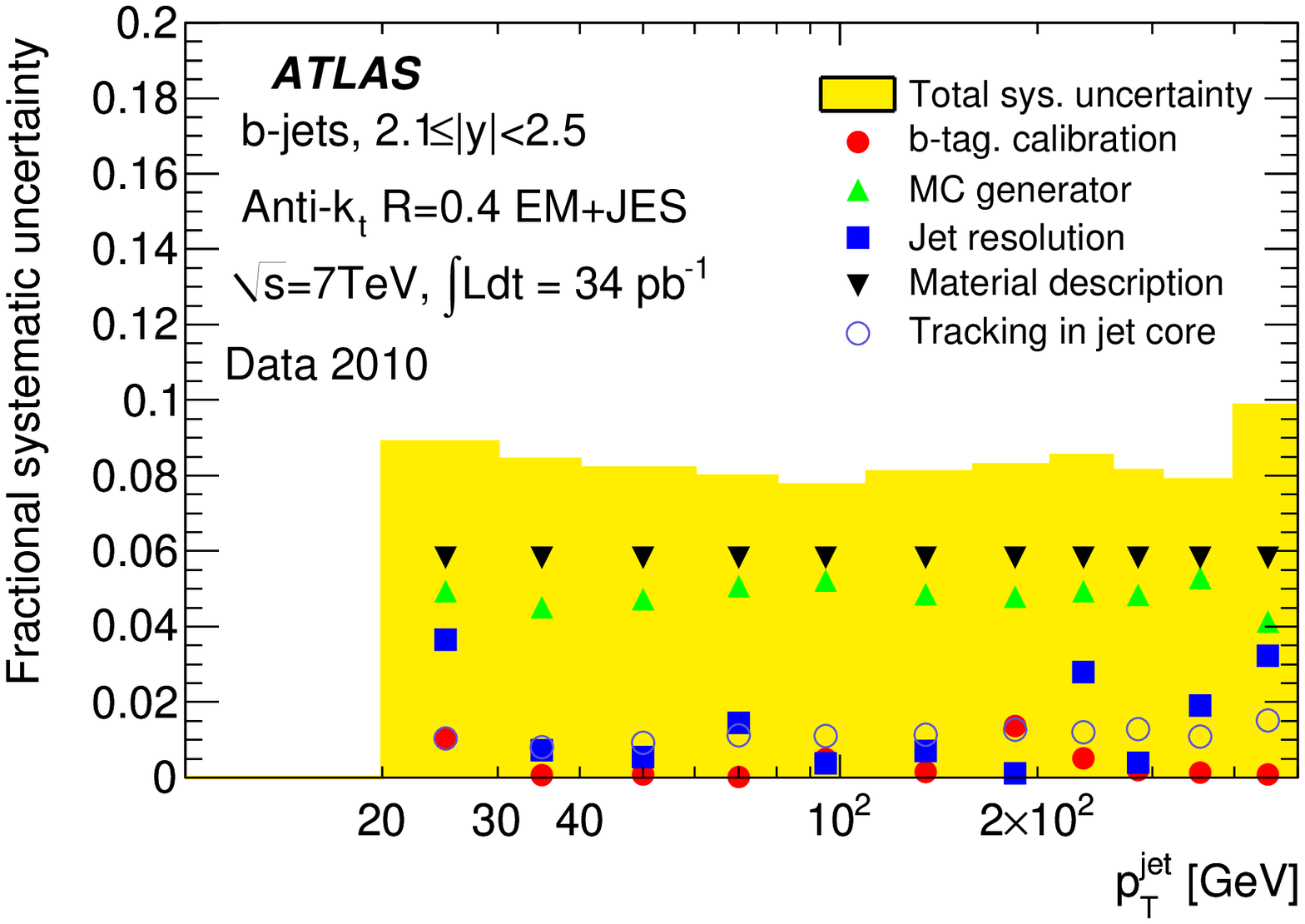}}
     \caption{The ratio of the mean value of \rtrk{} in data and Monte Carlo (left) 
and the fractional systematic uncertainty (right) as a function of \ptjet{} for $|\rapjet\,|<1.2$ (a), $1.2 \leq |\rapjet\,| < 2.1$ (b) and $2.1 \leq |\rapjet\,| < 2.5$ (c).
\Antikt{} jets with $R=0.4$ calibrated with the \EMJES{} scheme are used.
The dashed lines indicate the estimated uncertainty from the data and Monte Carlo simulation agreement. 
Note the changed axis ranges in (c). 
Only statistical uncertainties are shown on the data points.
\label{fig:bjetrtrkratio}}
   \end{center}
 \end{figure*}

 \begin{figure*}[htp!!!]
   \begin{center}
     \subfloat[$|\rapjet\,| < 1.2$]{
       \includegraphics[width=0.5\textwidth]{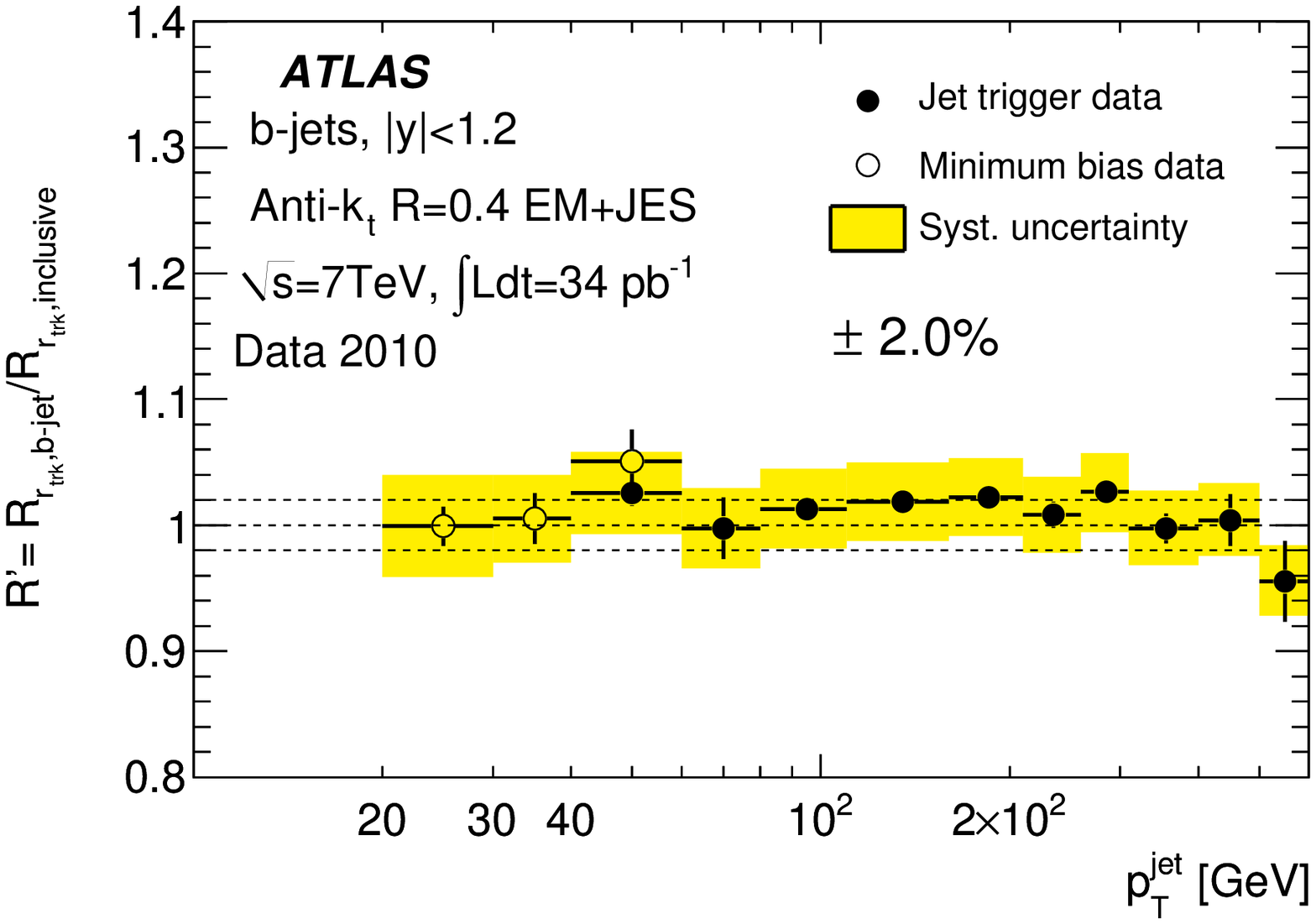}
       \includegraphics[width=0.5\textwidth]{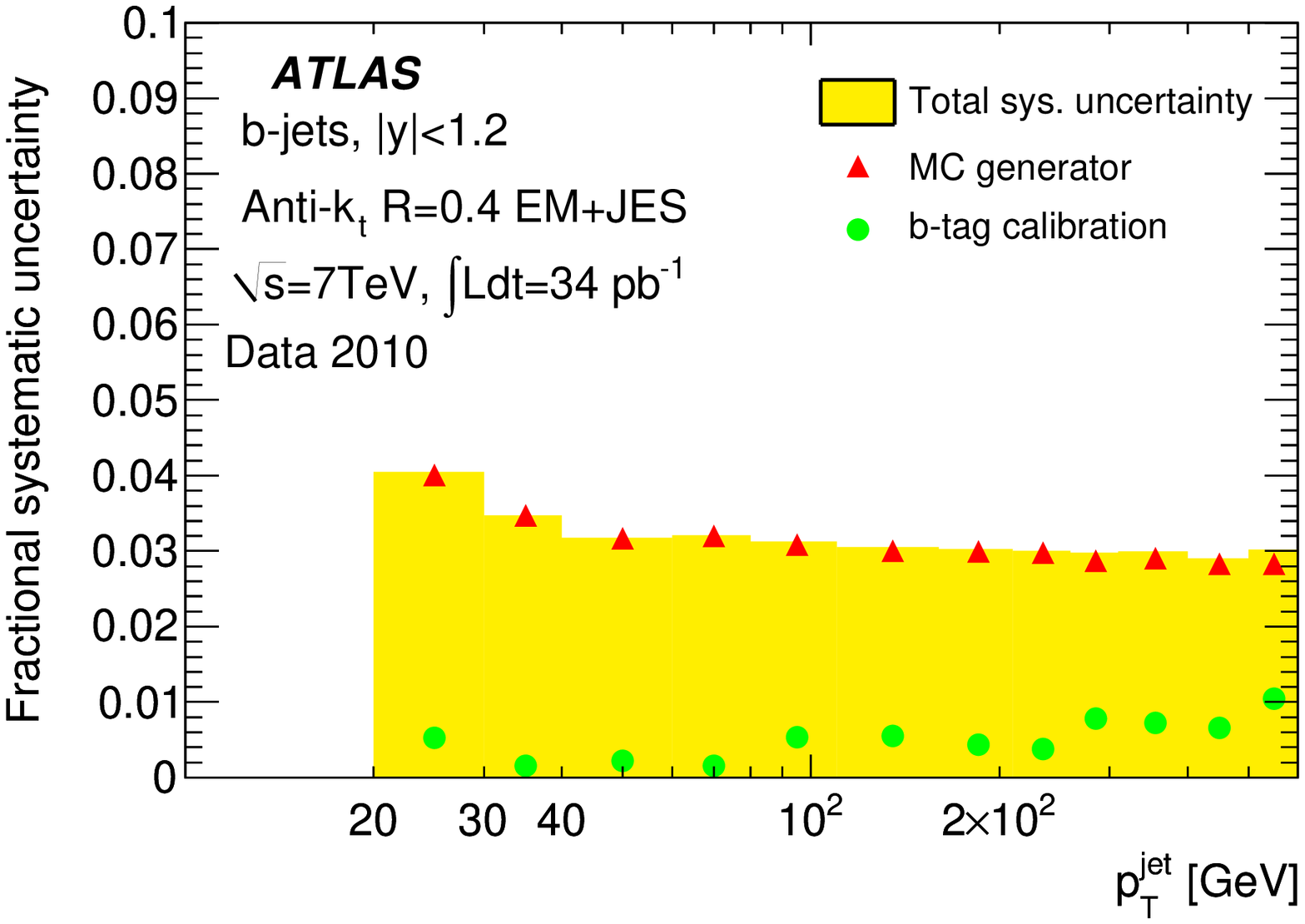}} \\
     \subfloat[$1.2 \leq |\rapjet\,| < 2.1$]{
       \includegraphics[width=0.5\textwidth]{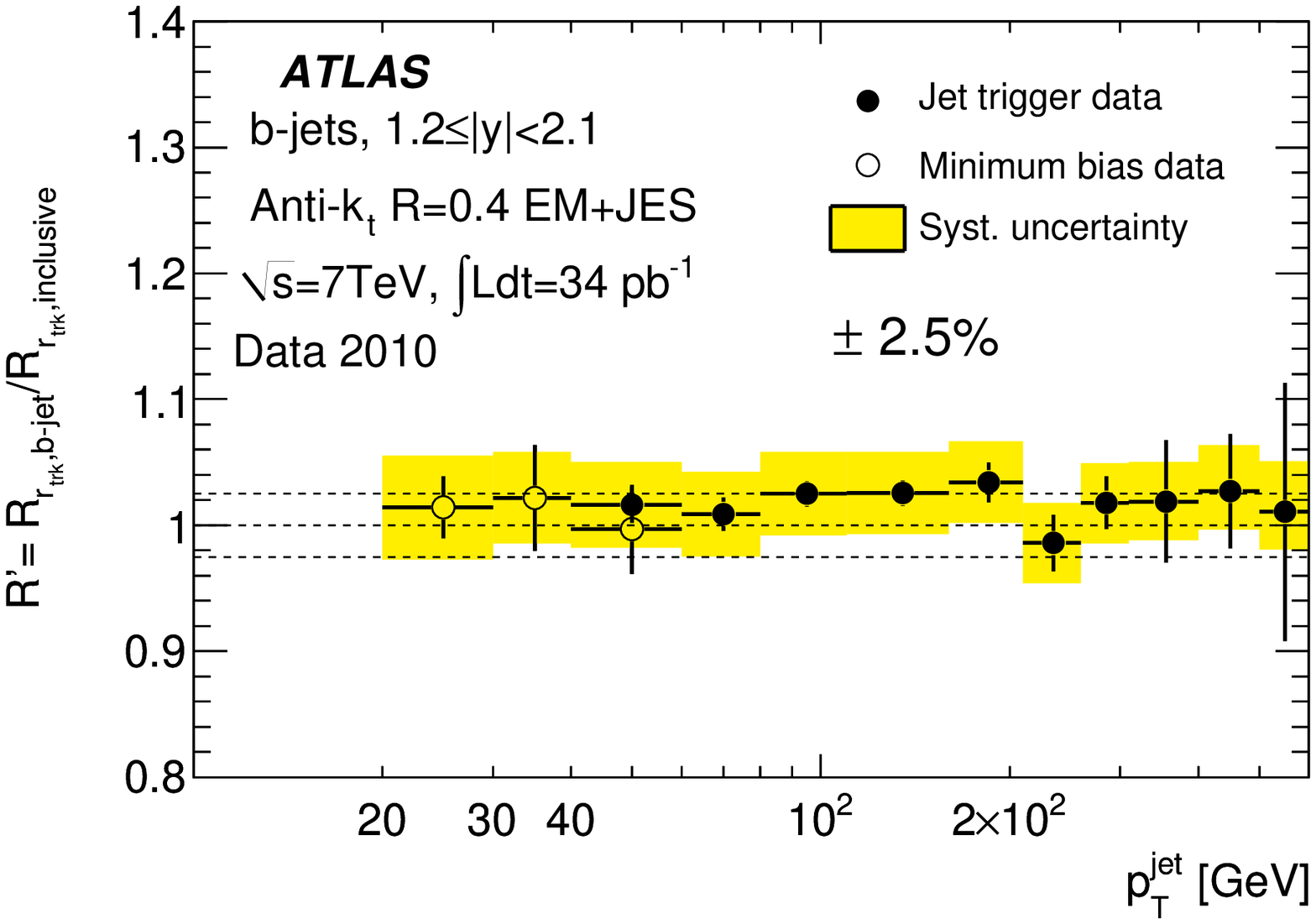}
       \includegraphics[width=0.5\textwidth]{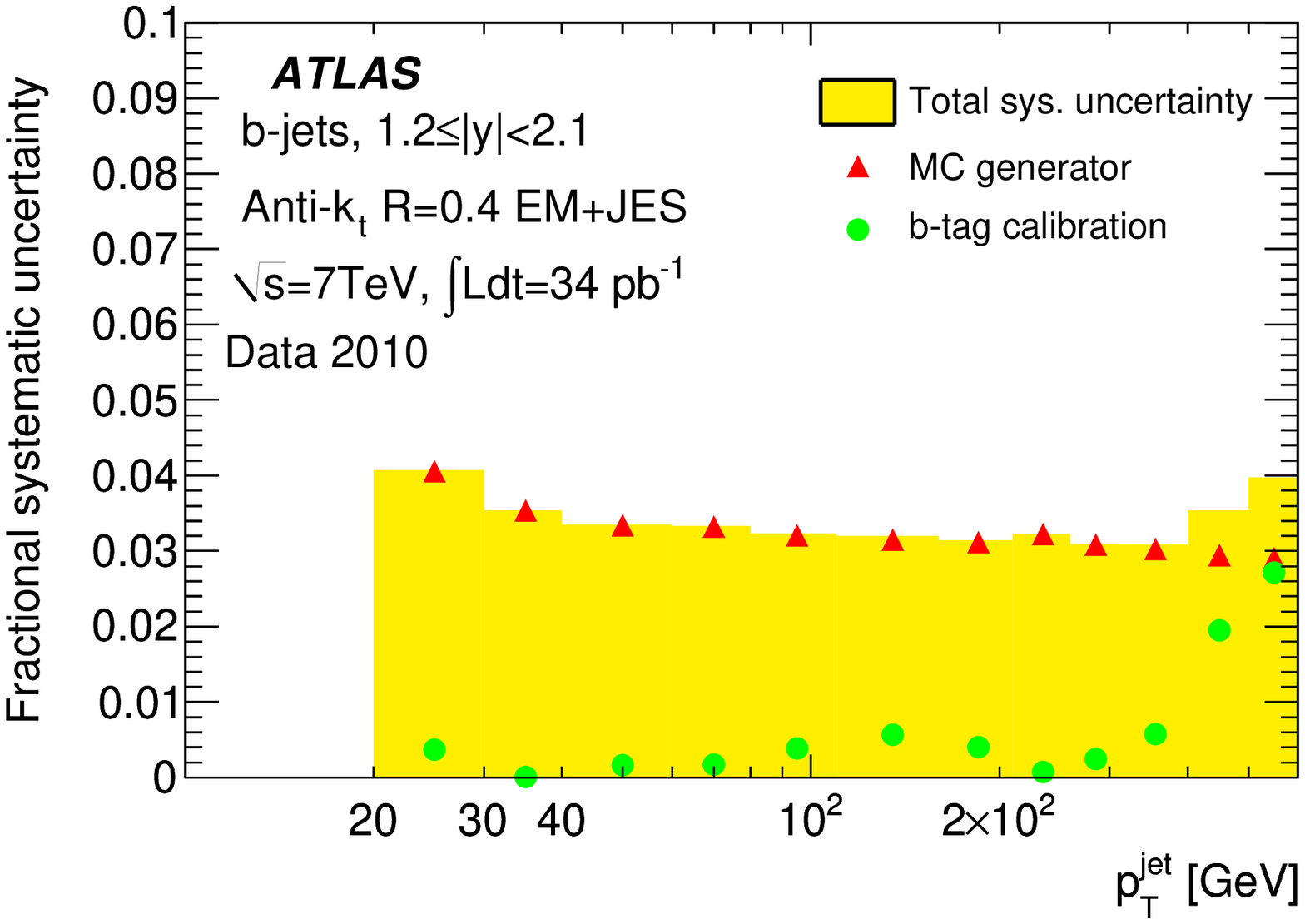}}\\
     \subfloat[$2.1 \leq |\rapjet\,| < 2.5$]{
       \includegraphics[width=0.5\textwidth]{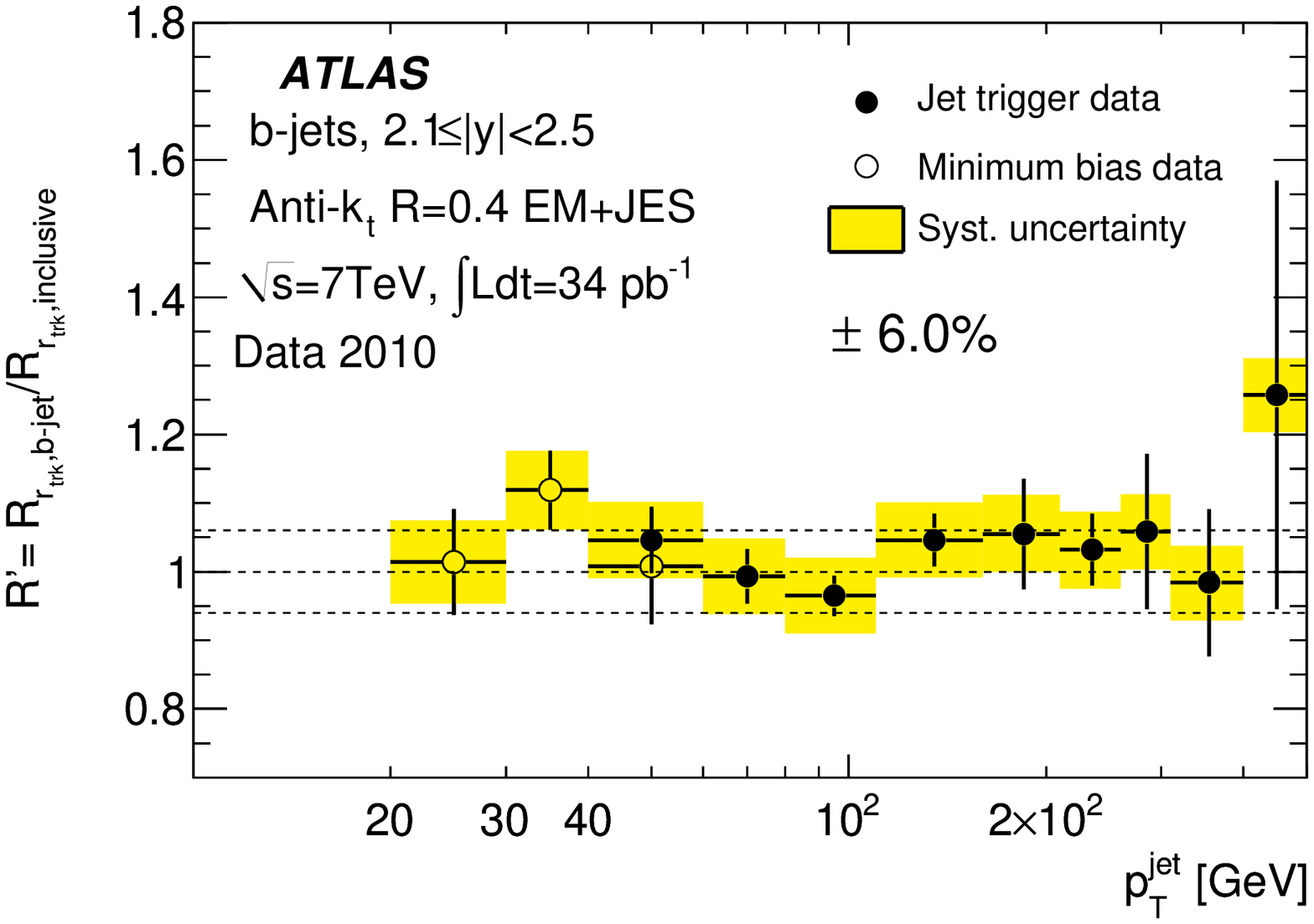}
       \includegraphics[width=0.5\textwidth]{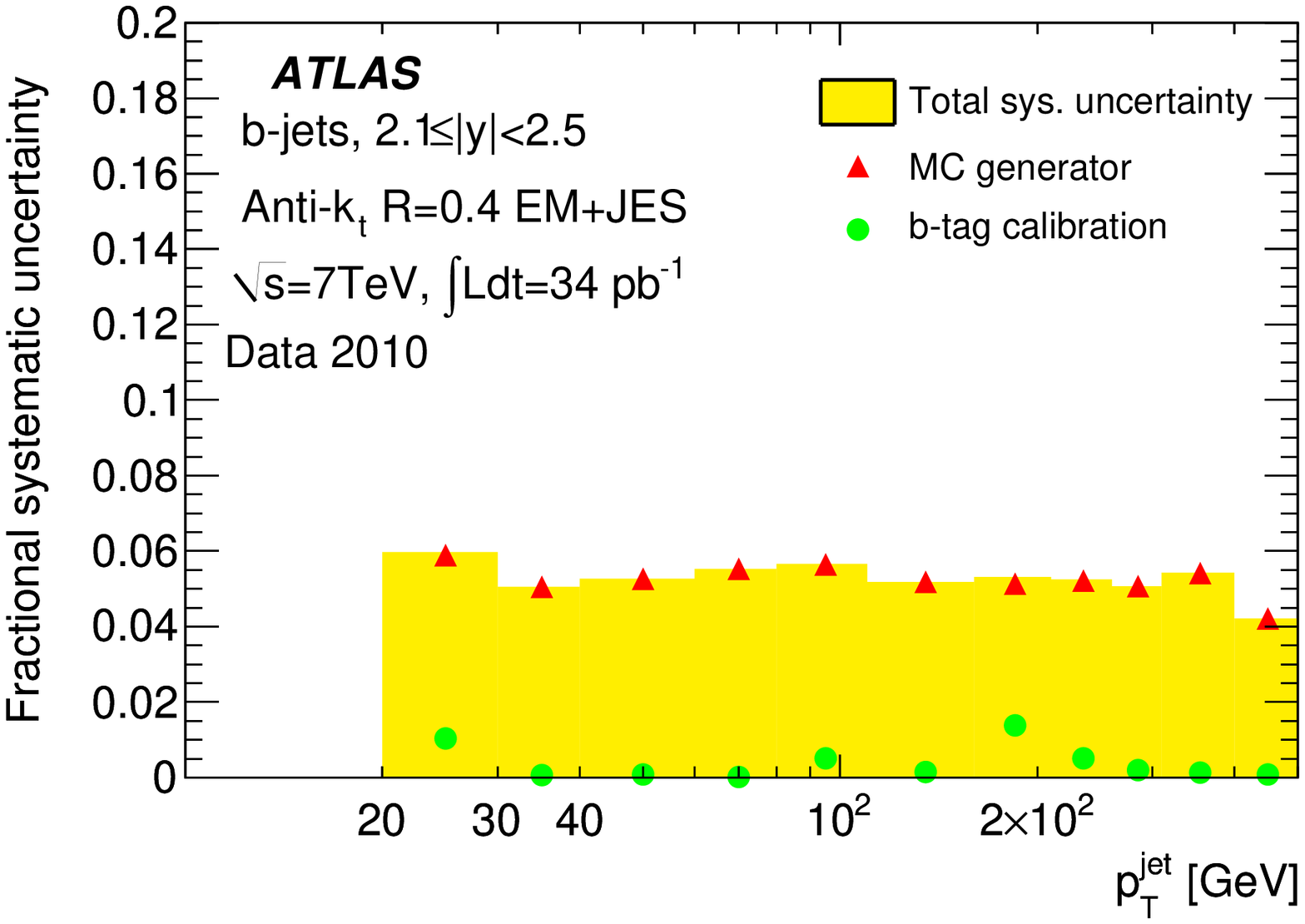}}
     \caption{The ratio $R'$ (see Equation \ref{eq:bjet-rprime}) of $R_{{\rm \rtrk,\bjet}}$ for identifed \bjets{} and $R_{{\rm \rtrk,inclusive}}$ for inclusive jets (left) and the fractional systematic uncertainty (right) as a function of \ptjet{} for 
$|\rapjet\,| < 1.2$ (a), $1.2 \leq |\rapjet\,| < 2.1$ (b) and $2.1 \leq |\rapjet\,| < 2.5$ (c).
\Antikt{} jets with $R=0.4$ calibrated with the \EMJES{} scheme are used. 
The dashed lines
indicate the estimated uncertainty from the data and Monte Carlo simulation agreement. 
Only statistical uncertainties are shown on the data points.
Note the changed axis ranges in (c). \label{fig:bjetrprime}}
   \end{center}
 \end{figure*}

\subsection{Validation of the heavy quark energy scale using tracks}
\label{ssec:datab}
The validation of the identified \bjet{} \JES{} uncertainty uses the tracks associated to the \bjet{} as reference
object and closely follows the method described in Section~\ref{sec:trackjet}.
The transverse momentum of a jet is compared to the total transverse momentum measured 
in tracks associated to the jet (see Equation \ref{eq:def-rtrk}). 

\subsubsection{Method}
The double ratio of charged-to-total momentum observed in data to that obtained in Monte Carlo simulation 
defined in Equation~\ref{eq:def-doublerat} will be referred to as $R_{\rm \rtrk, inclusive}$. 
In analogy this ratio is studied for $b$-tagged jets: \\ 
\begin{equation}
R_{{\rm \rtrk,\bjet}} \equiv \frac{{\left [ \langle {\rtrk \; \bjet}  \rangle \right ]}_{\rm Data}\; }
                             {{\left [ \langle {\rtrk \; \bjet}  \rangle \right ]}_{\rm MC} }.
\label{eq:bjet-doublerat}
\end{equation}

The \rtrk{} distributions for all \pt{} bins are calculated and the mean values of  \rtrk{} 
for data and Monte Carlo simulation are derived. 
The relative response to $b$-jets relative to inclusive jets, $R'$, is defined as
\begin{equation}
  R' \equiv \frac{R_{\rtrk,\bjet}}{R_{\rm \rtrk, inclusive}}.
  \label{eq:bjet-rprime}
\end{equation}

\subsubsection{Systematic uncertainties}
The systematic uncertainties arise from the modelling of the $b$-fragmentation, $b$-tagging calibration, 
jet resolution and tracking efficiency. They are assumed to be uncorrelated. 
The resulting fractional systematic uncertainties are shown on 
the right part of Figure~\ref{fig:bjetrtrkratio} and are determined as follows:
\begin{enumerate}
\item {\bf MC generator:} The \rtrk{} distribution is also calculated
  from \herwigpp\ samples. The shift in the distribution is fitted by
  a constant function. The variations in the data to Monte Carlo simulation ratio are
  taken as a systematic uncertainty. 
\item {\bf  $b$-tagging calibration:} The scale factors are varied correlated within their systematic uncertainty 
in the Monte Carlo simulation and the ratio is re-evaluated. 
The resulting shifts are added in quadrature to the systematic uncertainty.
\item {\bf Material description:} The knowledge of the tracking efficiency modelling in 
Monte Carlo simulation was evaluated in detail in Ref. \cite{MinBias2}.
The systematic uncertainty on the tracking efficiency for isolated tracks increases from $2 \%$ 
($|\etatrk| < 1.3$) to $7 \%$ ($2.3 \le |\etatrk| <2.5$) for tracks with $\pt > 500$ \MeV. 
The resulting effect on \rtrk{} is $2 \%$ for $|\rapjet\,| < 1.2$, $3.1 \%$ for $1.2 \leq |\rapjet\,| < 2.1$ 
and $5.5 \%$ for $2.1 \leq |\rapjet\,| < 2.5$.
\item{\bf Tracking in jet core:} High track densities in the jet core influence the tracking efficiency 
due to shared hits between tracks, fake tracks and lost tracks. 
The number of shared hits is well-described in Monte Carlo simulation. The \pt{} carried by fake tracks is negligible.
 
A relative systematic uncertainty of $50 \%$ on the loss of efficiency is assigned. 
The shift of \rtrk{} due to this uncertainty on the loss of efficiency is evaluated in Monte Carlo simulation on 
generated charged particles. 
Monte Carlo pseudo-experiments are generated according to the varied inefficiency. 
For each jet the ratio of the  \pt{} sum 
of the associated generated particles (truth tracks) with $\pttrk > 1$~\GeV{} to the  \pt{} sum of those associated truth tracks 
with $\pt > 1$~\GeV{} which also have a matched reconstructed track with $\pttrk > 1$~\GeV, is calculated. 
In this latter sample a truth track without or with a reconstructed track with $\pttrk > 1$~\GeV{} is added 
or respectively discarded according to the inefficiency uncertainty. 
The relative shift in the ratio \rtrk{} is added in quadrature to the systematic uncertainty.
\item {\bf Jet resolution:} The jet energy resolution in Monte Carlo simulation is degraded. 
A random energy that corresponds to a resolution smearing of $10 \%$ is added to each jet. 
The resulting shift of the ratio \rtrk{} is evaluated and added in quadrature to the overall systematic uncertainty.
\end{enumerate}

The two biggest contributions to the systematic uncertainty are
due to the material description and the difference between the
\rtrk{} distribution for \herwigpp{} and \pythia.

\subsubsection{Results}
Figure~\ref{fig:bjetrtrkratio} (left) shows the ratio of data to Monte Carlo simulation. 
An agreement of the calorimeter to track jet \pt{} measurements  is found
within $2 \%$ in the bin $|\rapjet\,| < 1.2$, within $4 \%$ 
for $1.2 \leq |\rapjet\,| < 2.1$ and within $6 \%$ for $2.1 \leq |\rapjet\,| < 2.5$.

The relative response $R'$ between identified \bjets{} and inclusive jets is shown in Figure~\ref{fig:bjetrprime} 
for all $\rapjet\,$-bins indicating the resulting relative \bjet{} energy scale uncertainty 
with respect to the inclusive jets sample. The uncertainty for \bjets{} is estimated 
to be $2 \%$, $2.5 \%$ and $6 \%$ 
in the range $|\rapjet\,| < 1.2$, $1.2 \leq |\rapjet\,| < 2.1$ and 
$2.1 \leq |\rapjet\,| < 2.5$, respectively. 
For the calculation of the systematic uncertainty in $R'$ it is assumed that at first order 
the uncertainty in the denominator and numerator of $R'$ from the tracking, namely tracking efficiency, 
material description, are fully correlated and cancel. 
The \ptjet{} resolution for inclusive and identified \bjets{} is considered to be similar.
Both assumptions are exactly valid for high \pt\ jets; 
for low \pt\ jets the second order deviations are estimated to be about $0.2 \%$.

The most significant systematic uncertainties on $R'$ are due to the choice of the Monte Carlo generator and the \btag\ calibration. 
Those independent uncertainties are added in quadrature. The Monte Carlo generator uncertainties from the inclusive sample 
and from the \btagged\ sample are also added in quadrature.

\subsubsection{Summary}
The jet energy scale for identified \bjets{} relative to that of inclusive jets
is evaluated for \antikt{} jets with $R = 0.4$ for the \EMJES\ calibration scheme. 
The resulting relative \bjet{} energy scale with respect to the inclusive jets sample is derived
within $2 \%$, $2.5 \%$ and $6\%$ 
in the range $|\rapjet\,| < 1.2$, $1.2 \leq |\rapjet\,| < 2.1$ and $2.1 \leq |\rapjet\,| < 2.5$, respectively. 

%
\begin{figure}[bh!]
        \centering
\includegraphics[width=0.45\textwidth]{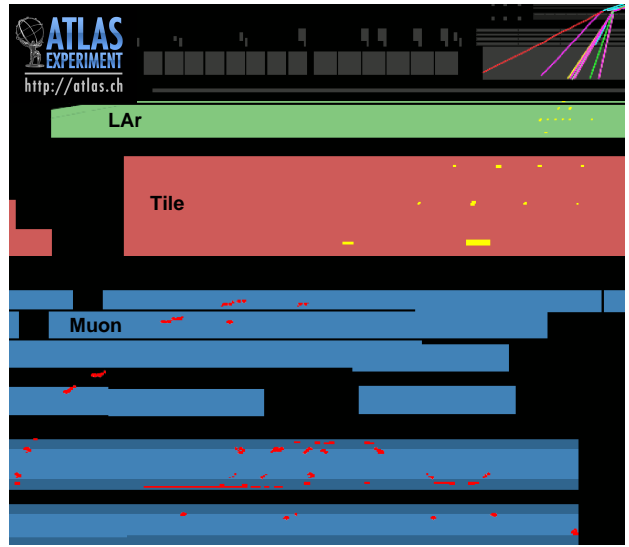}
        \caption{Graphical representation in a zoomed $x$-$y$ view
        of an event candidate with one large transverse momentum jet
        ($\ptjet=176$~\GeV) having a large activity in the last \Tile{} calorimeter layer ($82$~\GeV{} at the \EM{} scale)
        and in the muon detectors. The tracks in the inner detector are shown as lines in the top right,
        the energy deposits in the \LAr{}  and \Tile{} calorimeters are shown as
        light boxes. The hits in the muon system are shown as points.
        There are $128$ hits measured in the muon system.
        \label{fig:display}}
\end{figure}
\begin{figure*}
  \centering
    \subfloat[Leading jet]{\includegraphics[width=0.42\textwidth]{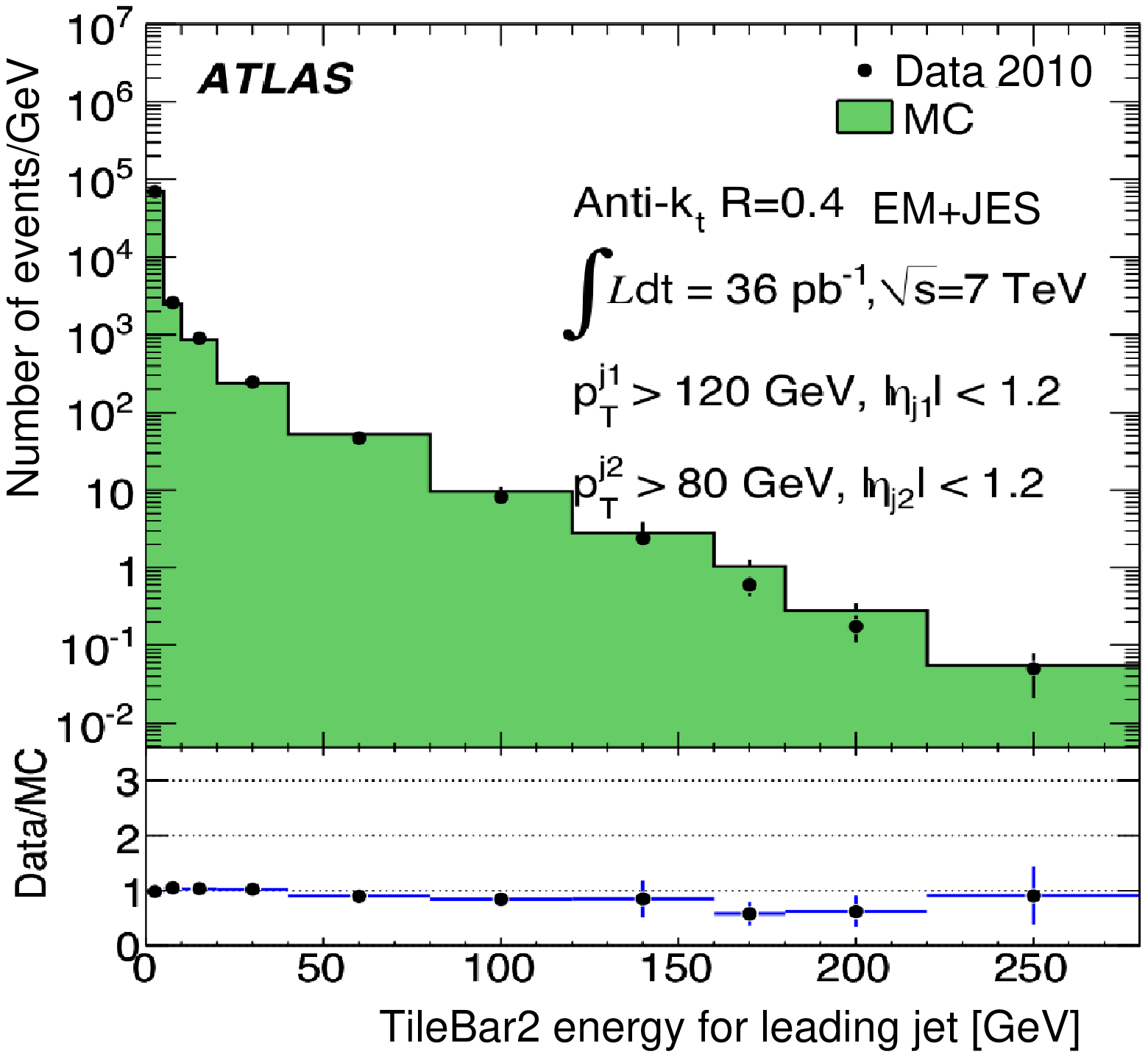}}
\hspace{1.cm}
    \subfloat[Subleading jet]{\includegraphics[width=0.445\textwidth]{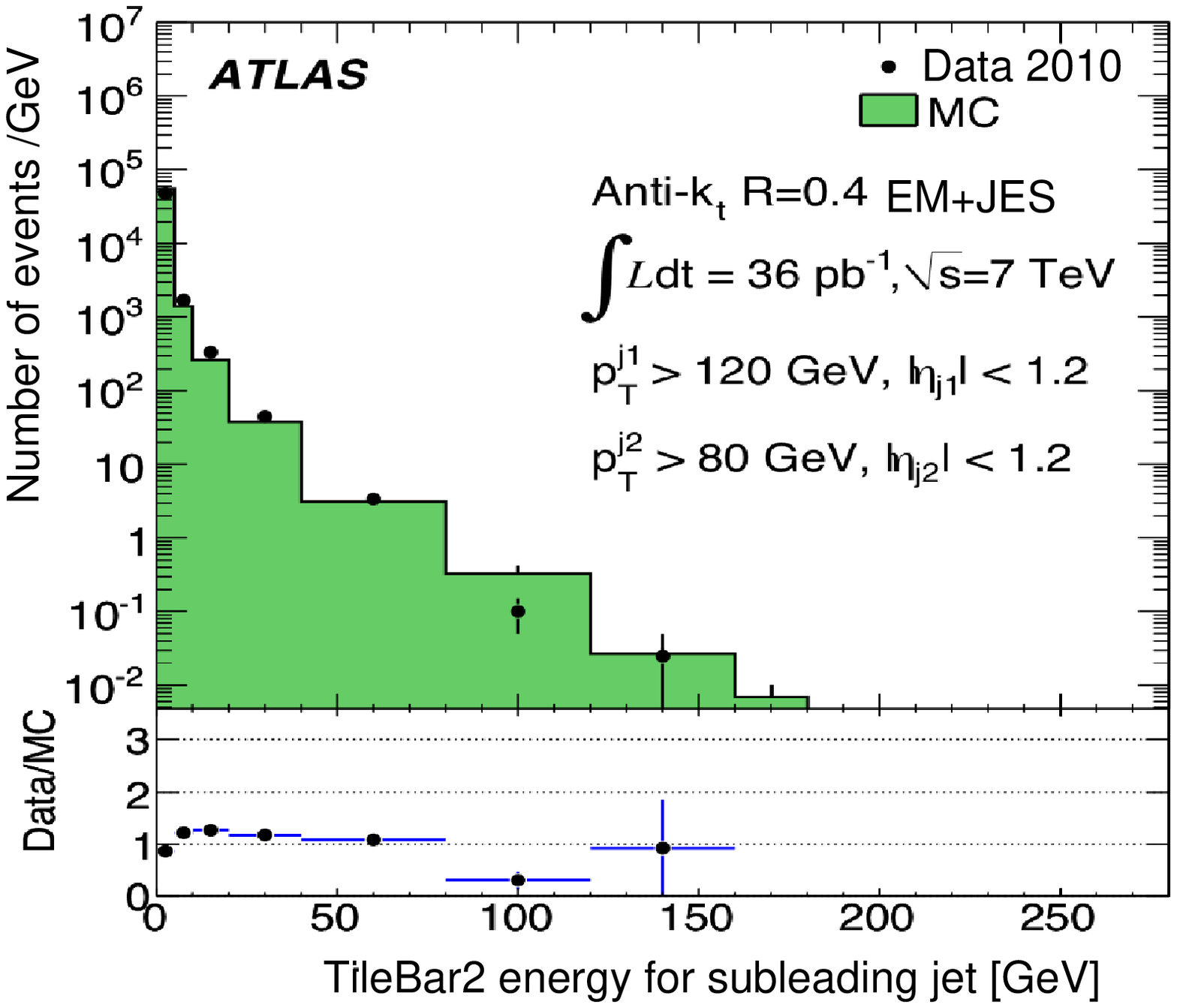}}\\
    \caption{Distribution of the deposited energy in the outermost layer 
             of the \Tile{} barrel calorimeter measured at the \EM-scale for the leading (a) and the subleading (b) jet.
    \Antikt{} jets with $R = 0.4$ within $|\etajet|<1.2$ and calibrated with the \EMJES{} scheme are used. 
    The leading jet is required to be above $\ptjet > 120$~\GeV{} 
    the subleading jet is required to be above $\ptjet > 80$~\GeV.
             Only statistical uncertainties are shown.
\label{fig:Tile_plots}}
\end{figure*}

\begin{figure*}[th!]
  \centering
    \subfloat[Leading jet]{\includegraphics[width=0.46\textwidth]{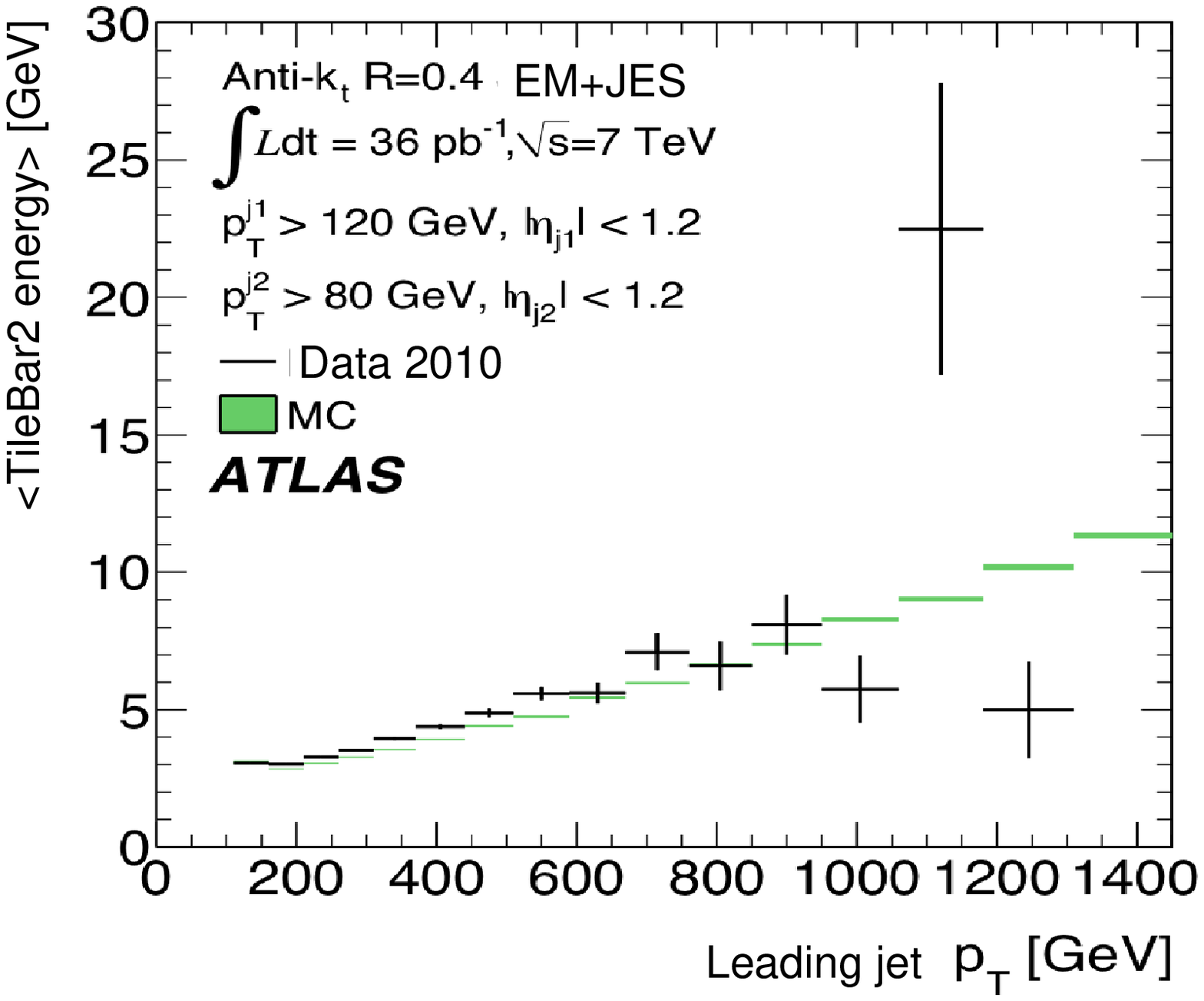}}
    \subfloat[Subleading jet]{\includegraphics[width=0.45\textwidth]{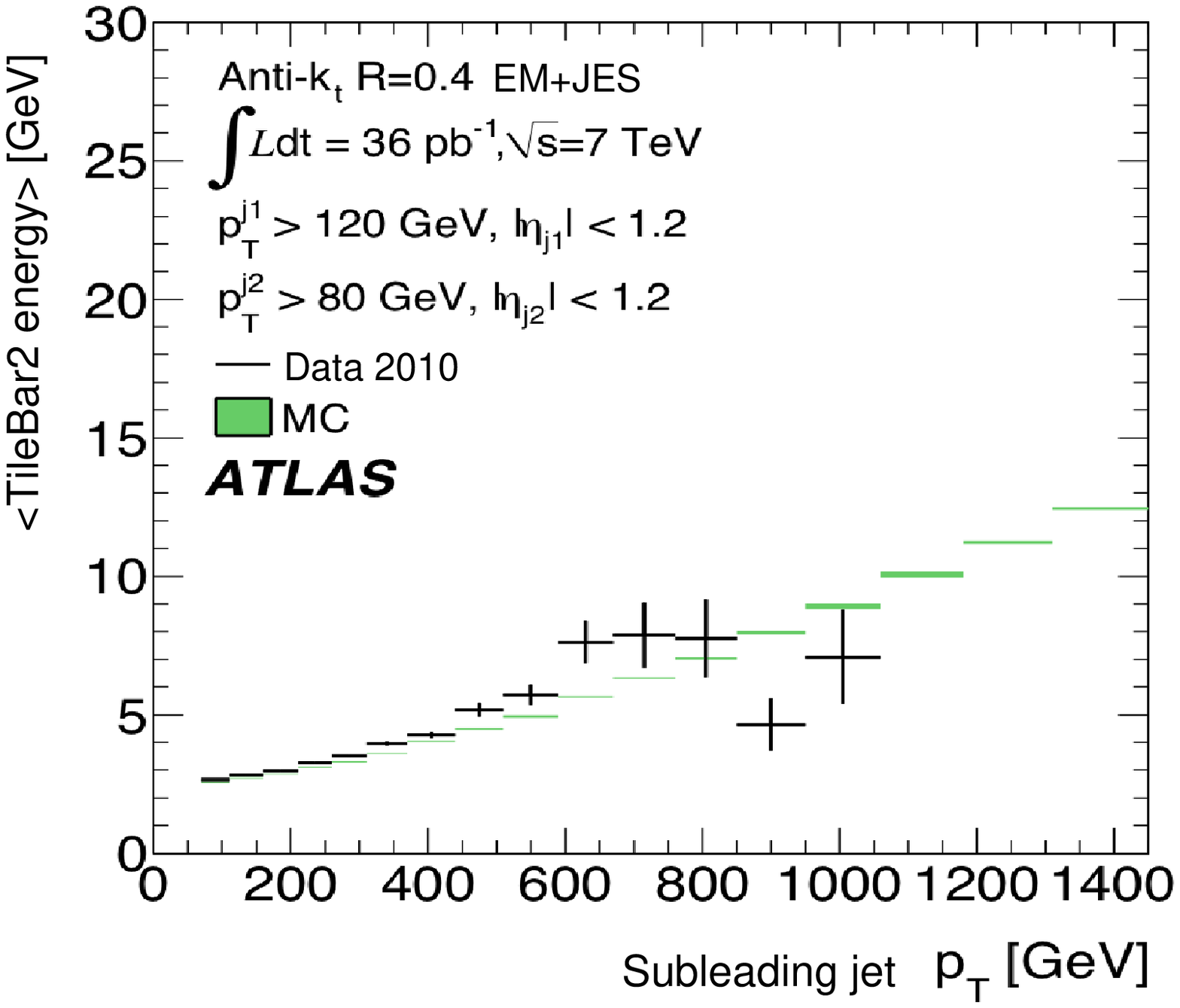}}\\
    \caption{Average energy deposited in the outermost layer of the  
    \Tile{} barrel calorimeter at the \EM-scale for the leading (a) and the subleading jet (b) as a function
    of the jet transverse momentum. 
    \Antikt{} jets with $R = 0.4$ within $|\etajet|<1.2$ and calibrated with the \EMJES{} scheme are used. 
    The leading jet is required to be above $\ptjet > 120$~\GeV{} 
    the subleading jet is required to be above $\ptjet > 80$~\GeV.
    Only statistical uncertainties are shown.
\label{fig:Tile_vs_pT}}
\end{figure*}
%
\begin{figure*}[th!]
  \centering
    \subfloat[Inclusive events]{\includegraphics[width=0.45\textwidth]{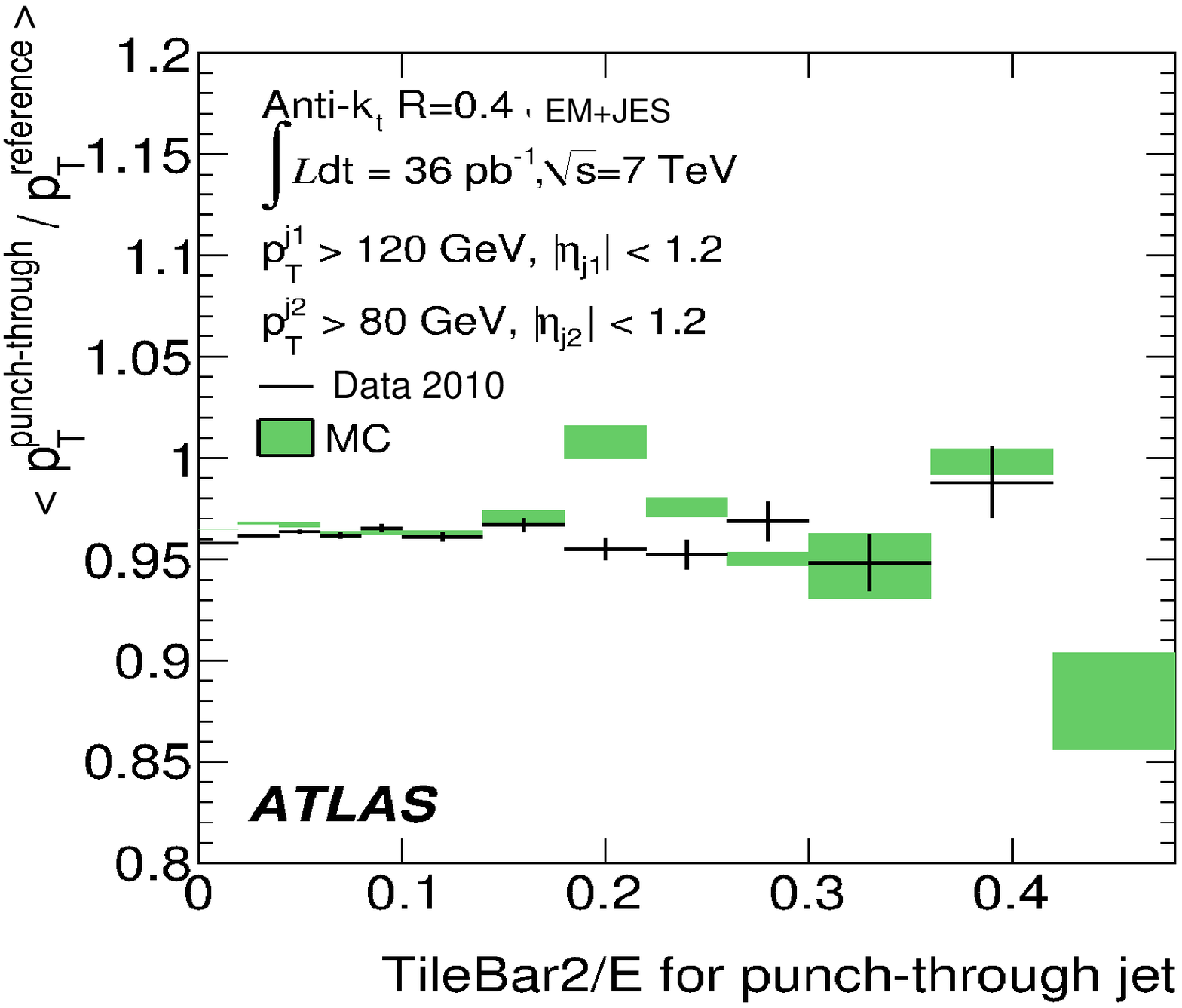}}
\hspace{1.cm}
    \subfloat[Events with large \Etmiss{}]{\includegraphics[width=0.45\textwidth]{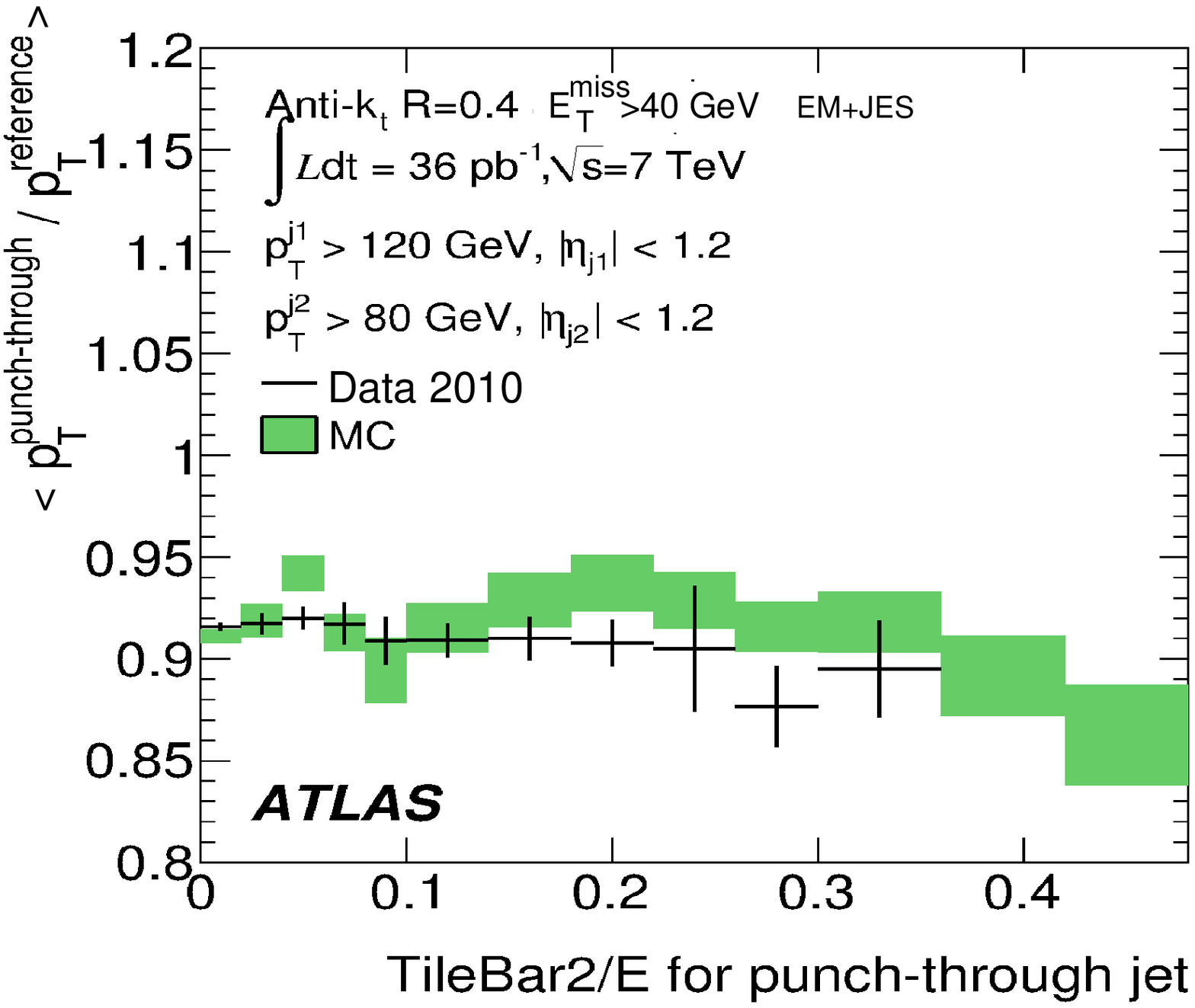}}
    \caption{Average relative jet response as a function of the energy deposited in the outermost layer of the \Tile{} 
    barrel calorimeter at the \EM-scale divided by the total jet energy.
    In (a) the standard event selection is employed, whilst in (b) an extra requirement in placed upon 
    the missing transverse momentum in the event.
    \Antikt{} jets with $R = 0.4$ within $|\etajet|<1.2$ and calibrated with the \EMJES{} scheme are used. 
    The leading jet is required to be above $\ptjet > 120$~\GeV{} 
    the subleading jet is required to be above $\ptjet > 80$~\GeV.
    Only statistical uncertainties are shown (errors bars for data, shaded band for Monte Carlo simulation).
\label{fig:tagprobe_met_vs_pT}
}
\end{figure*}

\section{Study of jet punch-through}
\label{sec:punchthrough}
For jets at very high transverse momentum it is possible that part of the energy is not deposited
in the calorimeter, but leaks out to the detector components beyond the calorimeter.
This leads to a systematic reduction in the measured jet energy.

Jets that  deposit energy beyond the hadronic \Tile{} calorimeter and in the muon system are 
called {\it punch-through} jets.
A graphical representation of a candidate for a punch-through jet in data is shown in Figure~\ref{fig:display}.

In this section the Monte Carlo simulation of energy deposits in the outermost calorimeter layer is tested.
Quantitative estimates of the energy lost beyond the calorimeter are obtained
using a tag-and-probe technique.

\subsection{Event selection for punch-through analysis}
\Antikt{} jets with $R = 0.4$ calibrated with the \EMJES{} scheme are used in this study.
Jets in the barrel of the \Tile{} calorimeter with $|\etajet| < 1.2$ are used.
Events with at least two jets are retained, if the highest \pt{} jet
satisfies $\pt^{j1} > 120$~\GeV{} and the second highest \pt{} jet satisfies $\pt^{j2} > 80$~\GeV.
The two leading jets are required to be back-to-back requiring $\Delta \phi > 170^{\circ}$.

\subsection{Energy depositions in the hadronic calorimeter}
The energy deposits in the outermost layer of the barrel 
of the \Tile{} calorimeter are a good indicator 
of the jet energy depositions beyond the calorimeter.
These are shown in Figure~\ref{fig:Tile_plots} for the leading and the sub-leading jet.
Most jets deposit only about $3$ to $7$~\GeV{} energy in the outermost calorimeter layer. 
The Monte Carlo simulation gives a good description of the data for $\ptjet < 80$~\GeV.
For higher \ptjet{} the data distribution is below the Monte Carlo simulation, but
the statistical uncertainties are large.

Figure~\ref{fig:Tile_vs_pT} shows
the dependence of the energy deposition in the outermost layer of the \Tile{} calorimeter 
measured at the \EM{} scale for the leading and sub-leading jet \pt.
The energy in the third layer of the \Tile{} calorimeter increases with rising jet \pt. 
The data are well described by the Monte Carlo simulation
in the low \ptjet{} region. Starting from about $400$~\GeV{} the data tend to
be $5 - 10 \%$ above the Monte Carlo simulation. 
For high \ptjet{} the statistical uncertainties are large.

\subsection{Dijet balance as an indication of punch-through}
The relative calorimeter response between the two jets in a dijet event can be measured 
using the dijet \pt{} balance method. 
In Section~\ref{sec:etaintercalibration} the reference jet is chosen as a well-measured object in the
central detector region that is used to assess the \JES{} uncertainty of the probe jet in the forward region.
However, in the context of punch-through such a distinction cannot be made.
Jet punch-through can occur in any detector pseudo\-rapidity region.
Fluctuations in the particle composition or in the hadronic shower occur
with equal probability for both jets and it is not possible to know {\it a priori} which of the
jets will be affected. 

A different approach is therefore employed.
The energy lost beyond the calorimeter will create a component of the missing transverse energy \Etmiss{} 
in the direction of the punch-through jet.
The punch-through jet can therefore be defined as the one that is closest to the \Etmiss{} 
$\phi$-direction.
The asymmetry between the transverse momentum of the reference jet ($p_{\rm T}^{\rm reference}$) 
away from the \Etmiss{} direction and the punch-through jet, acting as 
probe jet ($p_{\rm T}^{\rm punch-through}$), can then be measured as a function of 
the energy deposition of the jet that is the candidate for punch-through.

Figure~\ref{fig:tagprobe_met_vs_pT}a 
show the average ratio of the transverse momentum of the punch-through jet to the one of the
reference jet
as a function of the energy depositions in the last \Tile{} calorimeter layer measured at the \EM-scale
with respect to the total jet energy.
Figure~\ref{fig:tagprobe_met_vs_pT}b show the average jet response where $\Etmiss > 40$~\GeV{} is required.
This selection enhances possible punch-through effects.
The transverse momentum of the punch-through jet is lower than that of the reference jet.
This is a bias due to the selection of the  punch-through jet as the one pointing to the
direction of the missing transverse momentum. There is no dependence on the energy fraction in the
outermost layer, indicating that energy losses due to punch-through are small for the jet \pt{}
range considered in this study. The Monte Carlo simulation describes the data within the
statistical uncertainties.

\subsection{Summary of the jet punch-through study}
The energy deposition in the outer layer of the \Tile{} calorimeter and its dependence on the
transverse jet energy is described by the Monte Carlo simulation.
This indicates that the Monte Carlo simulation 
is able to  describe energy deposited beyond the calorimeter.
For the few jets that are potentially affected by punch-through
no additional uncertainty due to punch-through effects is assigned.

\section{Summary}
\label{sec:conclusions}
The jet energy scale (\JES) and its systematic uncertainty for various jet calibration schemes
are determined for jets measured  with the \ATLAS{} detector in the $2010$ data set 
corresponding to an integrated luminosity of \mylumi.
Jets produced in proton-proton collisions at the \LHC{} 
with a centre-of-mass energy of $\sqrt{s}=7$~\TeV{}
are reconstructed with the \antikt{} algorithm with distance parameters $R=0.4$ or $R=0.6$.
The energy and the direction of the jets
are calibrated with simple factors derived from Monte Carlo simulations
for transverse jet momenta $\ptjet \geq 20$~\GeV{} and pseudorapidities \AetaRange{4.5}
using various jet calibration schemes.

In the simplest calibration scheme (\EMJES), where the \JES{} correction factor is directly applied
to the calorimeter measurement at the electromagnetic scale,
the \JES{} systematic uncertainty is estimated 
using the single hadron response measured \insitu{} and in test-beams and by studying
systematic variations in Monte Carlo simulations.
The transverse momentum balance between central and forward jets in dijet events is used to
derive the \JES{} uncertainty for forward jets.

In the central region, \AetaRange{0.8}, the \EMJES{} uncertainty is lower than $4.6 \%$ for all jets 
with $\ptjet>20$~\GeV{} and less than $2.5 \%$
for jets with \ptRange{60}{800}.
Jets with transverse momenta in the \TeV-regime have a \JES{} uncertainty of $3$ to $4 \%$.
Towards the forward region the \EMJES{}  uncertainty increases, mainly because of differences
between the Monte Carlo event generators \pythia{} and \herwig{}
when deriving the relative $\eta$-intercalibration uncertainty.
The largest \JES{} uncertainty of about $14\%$ is found for low \pt{} jets with \ptRange{20}{30}
at \etaRange{3.6}{4.5}.
The jet energy scale uncertainty is found to be similar for jets
reconstructed with both jet distance parameters studied: $R=0.4$ and $R=0.6$.

The additional energy induced by superimposed events from multiple
proton-proton collisions in the same bunch crossing (pile-up) 
is determined to be $0.5$~\GeV{} per additional reconstructed vertex.
The \JES{} uncertainty after applying the pile-up correction 
is estimated as a function of the number of primary vertices.
In the case of two primary vertices per event, the uncertainty due to pile-up for jets with
$\ptjet \approx 20$~\GeV{} and \etaRange{0.3}{0.8} is about $1 \%$, while
it amounts to about $2 \%$ for jets with \etaRange{2.1}{2.8}. 
For jets with transverse momentum above $200$~\GeV, the uncertainty due to pile-up
is negligible for all jets. 

The \JES{} and its uncertainty are validated up to $\ptjet \lesssim 1$~\TeV{} to the level of a few per cent 
using several \insitu{} techniques by comparing the high \pt{} jet to a well known reference
recoiling against it. These reference objects include
the sum of the transverse track momenta associated to the jet, a system of low \pt{} jets 
or the photon \pt.
The track-based method covers the full \ptjet{} range and has the highest statistical
precision. However, the systematic uncertainty of the method is as large as $6 \%$ for very
low \pt{} jets and about $3$-$4\%$ for \ptRange{40}{800} and somewhat higher for jets 
outside this \ptjet\ range.
The $\gamma$-jet method has a systematic uncertainty of about $1 \%$, but is 
still statistically limited and reaches only $\ptjet \lesssim 300$~\GeV.
Balancing very high \pt{} jets against a recoil
system of lower \pt{} jets allows the validation of the high \pt{} jets within $5 \%$ up to $1$~\TeV.
In this range the statistical uncertainty is roughly equivalent to, or smaller than, the systematic uncertainty.

The \JES{} uncertainty derived from a combination of  \insitu{} techniques
is compatible to the one derived from the single hadron response measurements
over a wide kinematic range, but it is larger for very low and very large \ptjet,
where the uncertainties of the \insitu{} methods are large.

More sophisticated jet energy calibration schemes based on cell energy density weighting
or jet properties are studied. These provide a better jet energy resolution 
and a reduced flavour dependence of the jet response.

The global sequential jet calibration (\GS) based on global properties of the internal jet structure 
improves the energy resolution and reduces flavour dependence of the \EMJES{} calibration scheme. 
The \JES{} uncertainty for the \GS{} jet calibration scheme is given by the sum in 
quadrature of the \EMJES{} uncertainty and the uncertainty associated to the 
\GS{} corrections. The latter is conservatively taken to be 
$0.5 \%$ for $30 \le \pt < 800$~\GeV{} and $|\eta|<2.1$ and $1 \%$ for $\ptjet < 30$~\GeV{} and
$2.1 < |\eta| < 2.8$. These uncertainties are also supported by \insitu{} measurements
with the  \gammajet{} and the dijet technique.

The global cell weighting scheme (\GCW) derives cell wei\-ghts by optimising the
resolution of reconstructed jets relative to their respective truth jets. The local
cluster calibration (\LCW) derives energy corrections for calorimeter clusters
using single hadron Monte Carlo simulations.
The \JES{} uncertainty is obtained from \insitu{} techniques. 
Over a wide kinematic range the \JES{} uncertainties for the various schemes are similar,
except at very low and very high \ptjet{} where the uncertainty from the \insitu{} techniques are larger.
The \LCWJES{} and the \GCWJES{} scheme show similar performance.

For all jet calibration methods additional uncertainties are derived for close-by jet topologies
and for response differences for jets induced by quarks, gluons or heavy flavour quarks. 
A method to reduce the uncertainty due to uncertainties on the
quark and gluon composition in a given event sample is shown.
The uncertainty due to close-by jets is largest for low \pt{} jets,
but is at most below $2$ to $3 \%$. The \JES{} uncertainty of jets containing $B$-hadrons
is about $2 \%$ for low-\pt{} jets and smaller than $1 \%$ for jets with $\ptjet > 200$~\GeV.

The jet reconstruction efficiency is derived using the Monte Carlo simulation
and the systematic uncertainty  evaluated with a tag-and-probe technique using track jets.
The jet reconstruction efficiency is well-described by the Monte Carlo simulation. 
The associated systematic uncertainty is below $2 \%$ for jets
with $\ptjet < 30$~\GeV{} and negligible for higher \ptjet.

The Monte Carlo simulation gives a good description of the
main aspects of the data. 
Detailed studies show that the calorimeter
cell energy densities in jets, the calorimeter energy topology 
induced by jets and track related properties are well-described.
This includes the amount of energy deposited in the outermost
calorimeter layers from which it is inferred that the \JES{} uncertainty due
to energy leaking beyond the calorimeter is small and well-described by
the Monte Carlo simulation.
No additional uncertainty for punch-through effects is assigned for high-\pt{} jets.

In summary, the precision of the jet energy measurement with the \ATLAS{} detector
has been established using various techniques
in the first year of proton-proton collisions at the LHC. 
In the central detector the jet energy can be measured
with a precision of about $2$ to $3 \%$ over a wide transverse momentum range.

This excellent performance would not have been possible without 
a very detailed understanding of the detector and sophisticated 
calorimeter calibration procedures as well as 
the good description of the \ATLAS{} detector in the Monte Carlo simulation.

\section*{Acknowledgement}
We thank CERN for the very successful operation of the LHC, as well as the
support staff from our institutions without whom \ATLAS{} could not be
operated efficiently.

We acknowledge the support of ANPCyT, Argentina; YerPhI, Armenia; ARC,
Australia; BMWF, Austria; ANAS, Azerbaijan; SSTC, Belarus; CNPq and FAPESP,
Brazil; NSERC, NRC and CFI, Canada; CERN; CONICYT, Chile; CAS, MOST and
NSFC, China; COLCIENCIAS, Colombia; MSMT CR, MPO CR and VSC CR, Czech
Republic; DNRF, DNSRC and Lundbeck Foundation, Denmark; ARTEMIS, European
Union; IN2P3-CNRS, CEA-DSM/IRFU, France; GNAS, Georgia; BMBF, DFG, HGF, MPG
and AvH Foundation, Germany; GSRT, Greece; ISF, MINERVA, GIF, DIP and
Benoziyo Center, Israel; INFN, Italy; MEXT and JSPS, Japan; CNRST, Morocco;
FOM and NWO, Netherlands; RCN, Norway; MNiSW, Poland; GRICES and FCT,
Portugal; MERYS (MECTS), Romania; MES of Russia and ROSATOM, Russian
Federation; JINR; MSTD, Serbia; MSSR, Slovakia; ARRS and MVZT, Slovenia;
DST/NRF, South Africa; MICINN, Spain; SRC and Wallenberg Foundation,
Sweden; SER, SNSF and Cantons of Bern and Geneva, Switzerland; NSC, Taiwan;
TAEK, Turkey; STFC, the Royal Society and Leverhulme Trust, United Kingdom;
DOE and NSF, United States of America.

The crucial computing support from all WLCG partners is acknowledged
gratefully, in particular from CERN and the ATLAS Tier-1 facilities at
TRIUMF (Canada), NDGF (Denmark, Norway, Sweden), CC-IN2P3 (France),
KIT/GridKA (Germany), INFN-CNAF (Italy), NL-T1 (Netherlands), PIC (Spain),
ASGC (Taiwan), RAL (UK) and BNL (USA) and in the Tier-2 facilities
worldwide.


\bibliographystyle{atlasnote}
\bibliography{jespaper}
%

\clearpage
\onecolumn
\begin{flushleft}
{\Large The ATLAS Collaboration}

\bigskip

G.~Aad$^{\rm 48}$,
B.~Abbott$^{\rm 111}$,
J.~Abdallah$^{\rm 11}$,
A.A.~Abdelalim$^{\rm 49}$,
A.~Abdesselam$^{\rm 118}$,
O.~Abdinov$^{\rm 10}$,
B.~Abi$^{\rm 112}$,
M.~Abolins$^{\rm 88}$,
H.~Abramowicz$^{\rm 153}$,
H.~Abreu$^{\rm 115}$
% ...
\bigskip

% ...
$^{10}$ Institute of Physics, Azerbaijan Academy of Sciences, Baku, Azerbaijan\\
$^{11}$ Institut de F\'isica d'Altes Energies and Departament de F\'isica de la Universitat Aut\`onoma  de Barcelona and ICREA, Barcelona, Spain\\
% ...
$^{48}$ Fakult\"{a}t f\"{u}r Mathematik und Physik, Albert-Ludwigs-Universit\"{a}t, Freiburg i.Br., Germany\\
$^{49}$ Section de Physique, Universit\'e de Gen\`eve, Geneva, Switzerland\\
% ...
$^{88}$ Department of Physics and Astronomy, Michigan State University, East Lansing MI, United States of America\\
% ...
$^{111}$ Homer L. Dodge Department of Physics and Astronomy, University of Oklahoma, Norman OK, United States of America\\
$^{112}$ Department of Physics, Oklahoma State University, Stillwater OK, United States of America\\
% ...
$^{115}$ LAL, Univ. Paris-Sud and CNRS/IN2P3, Orsay, France\\
% ...
$^{118}$ Department of Physics, Oxford University, Oxford, United Kingdom\\
% ...
$^{153}$ Raymond and Beverly Sackler School of Physics and Astronomy, Tel Aviv University, Tel Aviv, Israel\\
% ...
\end{flushleft}
 
\end{document}